\documentclass[acmtog, authorversion]{acmart}

\newcommand{\final}{1}
\newcommand{\diff}{0} %

\usepackage[utf8]{inputenc}
\usepackage{color}
\usepackage{ifthen}
\usepackage{float}
\usepackage{alltt}
\usepackage{amsmath}
\usepackage{amsthm}
\usepackage{newlfont} %
\usepackage{floatflt}
\usepackage{wrapfig}
\usepackage{fixltx2e}
\usepackage{subfig} %
\usepackage{multirow}
\usepackage{booktabs}
\usepackage{cleveref}
\usepackage{algorithmic}
\usepackage[export]{adjustbox}
\usepackage{mathtools, cuted}
\usepackage{CJKutf8} %

\usepackage[ruled]{algorithm2e} %

\usepackage{bbm} %

\setcitestyle{square}
\def\textrightarrow{$\rightarrow$}

\SetAlFnt{\small}
\SetAlCapFnt{\small}
\SetAlCapNameFnt{\small}
\SetAlCapHSkip{0pt}
\IncMargin{-\parindent}
\newcommand{\Caption}[2]{\caption[#1]{{\em #1} #2}}

\definecolor{SithColor}{rgb}{0.7,0,0} %
\newcommand{\liyi}[1]{{\color{SithColor} Li-Yi: #1 $\qed$}}
\definecolor{ConsularColor}{rgb}{0,0.4,0} %
\definecolor{GuardianColor}{rgb}{0,0,0.8} %
\definecolor{WinduColor}{rgb}{0.56,0.34,0.62} %

\newcommand{\peihan}[1]{{\color{GuardianColor} Peihan: #1 $\qed$}}
\newcommand{\takeo}[1]{{\color{ConsularColor} Takeo: #1 $\qed$}} %
\newcommand{\koji}[1]{{\color{WinduColor} Koji: #1 $\qed$}} %
\newcommand{\matthias}[1]{{\color{WinduColor} Matthias: #1 $\qed$}} %
\newcommand{\yoda}[1]{{\color{ConsularColor} Yoda: #1 $\qed$}}
\newcommand{\warning}[1]{{\it\color{red} #1}}
\newcommand{\note}[1]{{\it\color{blue} #1}}
\newcommand{\nothing}[1]{}

\definecolor{AudioColor}{rgb}{0.56,0.34,0.62}
\newcommand{\audio}[2][]{{\color{AudioColor} Audio: #2 $\qed$}}
\definecolor{VideoColor}{rgb}{0.44,0.66,0.38}
\newcommand{\video}[1]{{\color{VideoColor} Video: #1 $\qed$}}

\definecolor{TodoColor}{rgb}{0.9,0.4,0} %
\newcommand{\todo}[1]{{\bf\color{TodoColor} ETA: #1}}
\definecolor{DoneColor}{rgb}{0.1,0.6,1.0} %

\definecolor{NewColor}{rgb}{0.9,0.4,0}
\newcommand{\new}[1]{{\color{NewColor} #1}}
\definecolor{DeleteColor}{rgb}{0.1,0.6,1.0}
\newcommand{\delete}[1]{{\color{DeleteColor} #1}}
\definecolor{MoveColor}{rgb}{0.5,0.1,0.5}
\newcommand{\update}[1]{{\color{MoveColor} #1}}
\newcommand{\replace}[2]{\delete{#1}\new{#2}}
\newcommand{\add}[1]{\new{#1}}

\definecolor{figred}{rgb}{1,0,0}
\definecolor{figgreen}{rgb}{0,0.6,0}
\definecolor{figblue}{rgb}{0,0,1}
\definecolor{figpink}{rgb}{1,0.63,0.63}

\ifthenelse{\equal{\final}{1}}
{
\renewcommand{\liyi}[1]{}
\renewcommand{\peihan}[1]{}
\renewcommand{\takeo}[1]{}
\renewcommand{\koji}[1]{}
\renewcommand{\matthias}[1]{}
\renewcommand{\yoda}[1]{}
\renewcommand{\warning}[1]{}
\renewcommand{\note}[1]{}
}
{}

\newcommand{\pseudocode}{Algorithm}
\floatstyle{plain}
\newfloat{algorithm}{tbhp}{lop}
\floatname{algorithm}{\pseudocode}

\newcommand{\funct}[1]{\texttt{#1}}

\ifthenelse{\equal{\diff}{1}}
{
}
{
\renewcommand{\new}[1]{{#1}}
\renewcommand{\delete}[1]{}
\renewcommand{\update}[1]{{#1}}
}

\newcommand{\filename}[1]{\url{#1}}
\newcommand{\foldername}[1]{\url{#1}}

\let\oldparagraph\paragraph

\renewcommand{\paragraph}[1]{\oldparagraph{{#1}.}} %

\newcommand{\umd}{University of Maryland, College Park}
\newcommand{\adobe}{Adobe Research}
\newcommand{\utokyo}{University of Tokyo}

\newcommand{\norm}[1]{\left\lVert#1\right\rVert}
\newcommand{\pluseq}{\mathrel{{+}{=}}}
\DeclareMathOperator*{\argmax}{arg\,max}
\DeclareMathOperator*{\argmin}{arg\,min}
\DeclarePairedDelimiter\abs{\lvert}{\rvert}%

\newcommand{\samplesym}{s}
\newcommand{\samplesymprime}{\samplesym^\prime}
\newcommand{\samplesymhat}{\hat{\samplesym}}
\newcommand{\samplesymhatprime}{\hat{\samplesym}^\prime}

\newcommand{\samplespace}{\mathbf{p}}
\newcommand{\sampleposition}{\samplespace} %
\newcommand{\sampleid}{q} \nothing{}
\newcommand{\sampleattributes}{\mathbf{a}}

\newcommand{\sampleconnections}{N_{\mathbf{o}}}
\newcommand{\sampleorientations}{\mathbf{o}}
\newcommand{\sampleorientationentry}{o}

\newcommand{\sampleappr}{\mathbf{a}}
\newcommand{\sampletemp}{\mathbf{t}}
\newcommand{\sampleedge}{e}
\newcommand{\sampleedgesym}[2]{\sampleedge_{#1#2}}

\newcommand{\sampleedgeset}{\boldsymbol{\mathcal{E}}}

\newcommand{\sampleedgeoutput}{\sampleedgesym{\sampleoutput}{\sampleoutputprime}}

\newcommand{\sampleedgeinput}{\sampleedgesym{\sampleinput}{\sampleinputprime}}
\newcommand{\sampleedgeoutputprime}{\sampleedge_o^{\prime}}

\newcommand{\sampleedgeinputprime}{\sampleedge_i^{\prime}}
\newcommand{\sampleedgesetinput}{\sampleedgeset_i}

\newcommand{\outputsampleedgeset}{\mathcal{E}_o}
\newcommand{\edgelength}{\ell}

\newcommand{\samplevec}{\mathbf{u}}
\newcommand{\sampleneigh}{\mathbf{n}}
\newcommand{\sampleoutput}{\samplesym_o}
\newcommand{\sampleinput}{\samplesym_i}

\newcommand{\sampleoutputprime}{\samplesym_o^\prime}
\newcommand{\sampleinputprime}{\samplesym_i^\prime}

\newcommand{\sampleoutputstar}{\samplesym_o^{*}}
\newcommand{\sampleoutputstarprime}{\samplesym_o^{*\prime}}

\newcommand{\neighinput}{\sampleneigh(\sampleinput)}
\newcommand{\neighoutput}{\sampleneigh(\sampleoutput)}

\newcommand{\samplecandidate}{\samplesym_o^c}

\newcommand{\graph}{g}

\newcommand{\reconpathinput}{p_i}
\newcommand{\reconpathoutput}{p_o}

\newcommand{\differencesym}[1]{\hat{#1}}
\newcommand{\weightappr}{\alpha}
\newcommand{\weightattributes}{\gamma}

\newcommand{\weighttemp}{\beta}
\newcommand{\weightconnections}{\beta}
	
\newcommand{\distance}[1]{\left\| #1 \right\|} 
\newcommand{\match}{m}
\newcommand{\matchneigh}{\match_n}
\newcommand{\matchorient}{\match_o}
\newcommand{\matchedge}{\match_e}
\newcommand{\matchsample}{\match_s}

\newcommand{\neighborhoodradius}{r}

\newcommand{\editopset}{O}
\newcommand{\editop}{o}
\newcommand{\editcost}{c}

\newcommand{\graphsamplesym}{\samplesym_g}

\newcommand{\hierlevel}{l}
\newcommand{\outputsym}{\mathcal{O}}

\newcommand{\energy}{E}

\newcommand{\preferencescore}{p}
\newcommand{\preferencescoremax}{\preferencescore_{max}}
\newcommand{\preferencescorethreshold}{\preferencescore_{th}}

\newcommand{\preferencefunction}{\mathcal{P}}
\newcommand{\neighvecsize}{\mathcal{N}}
\newcommand{\neighsize}{N}
\newcommand{\canvassize}{L}

\newcommand{\predelementset}{\mathcal{E}_p}
\newcommand{\userelementset}{\mathcal{E}_u}
\newcommand{\predelement}{e_p}
\newcommand{\userelement}{e_u}
\newcommand{\predelementsetsize}{n_p}
\newcommand{\userelementsetsize}{n_u}

\newcommand{\shapesimilarity}{\text{sim}}
\newcommand{\matchedsetsize}{n_m}
\newcommand{\predsample}{s_p}
\newcommand{\usersample}{s_u}

\newcommand{\maxhierlevel}{l_{max}} %
\newcommand{\boundconstraint}{\mathcal{C}}

\newcommand{\assign}{\leftarrow}
\newcommand{\lowerbound}[1]{\textrm{LB}(#1)}
\newcommand{\upperbound}[1]{\textrm{UB}(#1)}

\newcommand{\neighsizeoptim}{\neighsize_{opt}}

\newcommand{\adjacencymatrix}{A}
\newcommand{\weightedadjacencymatrix}{A_w}
\newcommand{\outputadjacencymatrix}{\adjacencymatrix_o}
\newcommand{\inputadjacencymatrix}{\adjacencymatrix_i}
\newcommand{\inputpath}{P_i}
\newcommand{\outputpath}{P_o}

\newcommand{\goldenratio}{\textrm{GR}}
\newcommand{\goldensearchneightermin}{\lambda}

\newcommand{\nummatchedneighborhoods}{N_n}
\newcommand{\numoutputsamples}{N_o}
\newcommand{\numinputsamples}{N_i}
\newcommand{\indexend}{k_e}
\newcommand{\indexstart}{k_s}

\newcommand{\constraintterm}{\Theta}
\newcommand{\domain}{\mathcal{D}}
\newcommand{\distancesym}{dist}

\newcommand{\numiteration}{T}

\newcommand{\probability}{P}
\newcommand{\probabilitybi}{\overline{\probability}} %
\newcommand{\normalizationfactor}{\mathcal{Z}}
\newcommand{\softmatchingsigma}{\sigma}

\newcommand{\existenceconfidence}{P_e}
\newcommand{\energyothers}{E_e^{\prime\prime}}
\newcommand{\energyedge}{E_e^{\prime}}

\newcommand{\threshold}{\epsilon}
\newcommand{\binaryarray}{\mathcal{B}}

\newcommand{\binaryenergyweight}{\lambda}
\newcommand{\edgeenergy}{E_e}
\newcommand{\unaryenergy}{E_u}
\newcommand{\unarycost}{D_u}
\newcommand{\binaryenergy}{E_p}
\newcommand{\binarycost}{D_p}

\newcommand{\neighboringedgeset}{\mathcal{N}}
\newcommand{\edgeconfidence}{\theta}
\newcommand{\jointedgeconfidence}{\mu}

\newcommand{\unmatched}{\texttt{NUL}}
\newcommand{\samplediff}{\differencesym{\samplevec}_{\sampleoutput\sampleinput}(\sampleoutputprime,\sampleinputprime)}
\newcommand{\sampleexistence}{i}
\newcommand{\costmatrix}{\mathcal{C}}
\newcommand{\enmatchedcost}[2]{\differencesym{\samplevec}_{\sampleoutput\sampleinput}(\sampleoutput^{\prime#1},\sampleinput^{\prime#2}) }
\newcommand{\unmatchedcost}{c}
\newcommand{\costmatrixenmatched}{\costmatrix_m}
\newcommand{\costmatrixunmatched}{\costmatrix_u}

\newcommand{\numneighsample}{N_{\sampleneigh}}
\newcommand{\numOutput}{n_\outputsym}
\newcommand{\maxiter}{I_{max}}

\newcommand{\numoutput}{N_{\sampleneigh o}}
\newcommand{\numinput}{N_{\sampleneigh i}}

\newcommand{\samplingdistance}{\delta} %
\newcommand{\samplecluster}{\mathcal{S}}

\newcommand{\setsub}{\ominus}

\newcommand{\logicor}{\vee}

\newcommand{\auto}[1]{ \textcolor{blue}{#1}}
\newcommand{\reject}[1]{ \textcolor{red}{#1}}
\newcommand{\manual}[1]{ \textcolor{blue}{#1}}

\newcommand{\bigcdot}{\boldsymbol{\cdot}}

\graphicspath{
{figs/handdrawn/}
{figs/raster/}
{../../Figures/}
}%

\acmJournal{TOG}
\acmYear{2020}
\acmVolume{39}
\acmNumber{6}
\acmArticle{168}
\acmMonth{12}
\acmDOI{10.1145/3414685.3417780}
\acmSubmissionID{papers\_168} %

\begin{document}

\title{Autocomplete Element Pattern Design}

\title{Autocomplete Zentangle Design} %

\title{Autocomplete Pattern Design}

\title{Autocomplete Pattern Design with Interactive Learning} %

\title{Autocomplete Hierarchical Pattern Design with Interactive Learning} %

\title{Autocomplete Complex Hierarchical Pattern Design with Online Parameter Learning} 

\title{Autocomplete Hierarchical Pattern Design via Online Parameter Learning} %

\title{Brushing Continuous Vector Patterns} %

\title{Synthesizing Continuous \nothing{Vector }Patterns from User Exemplars} %

\title{Continuous Curve Textures} %

\author{Peihan Tu}
\affiliation{\institution{\umd}}

\author{Li-Yi Wei}
\affiliation{\institution{\adobe}}

\author{Koji Yatani}
\author{Takeo Igarashi}
\affiliation{\institution{\utokyo}}

\author{Matthias Zwicker}
\affiliation{\institution{\umd}}

\setcopyright{acmlicensed}

\begin{abstract}

Repetitive patterns are ubiquitous in natural and \replace{man}{human}-made objects, \update{and can be created with a variety of tools and methods}.
Manual authoring provides \replace{unprecedented}{unmatched} degree of freedom and control, but can require significant artistic expertise and manual labor.
Computational methods can automate parts of the manual creation process, but are mainly tailored for discrete pixels or elements instead of more general continuous structures.
We propose an \nothing{automatic, }example-based method to synthesize continuous curve patterns from exemplars.
Our main idea is to extend prior \nothing{example-based }sample-based discrete element synthesis methods\nothing{~\cite{Ma:2011:DET}} to consider not only sample positions (geometry) but also their connections (topology).
Since continuous structures can exhibit higher complexity than discrete elements, we also propose robust, hierarchical synthesis to enhance output quality. \nothing{ and computation speed %
}%
Our algorithm can generate a variety of \nothing{high-quality }continuous curve patterns fully automatically. 
For further quality improvement and customization, we also present an autocomplete user interface to facilitate interactive creation and iterative editing.
\nothing{
Similar to existing pattern authoring systems, users can freely draw their designs.
They can specify an output domain and our system can automatically generate predicted patterns that resemble what have already been drawn.
Users can accept or modify the predictions to maintain full control, as well as manually clone from specific source regions to target regions.
The combination of our synthesis methods and authoring interface allows our system to create high-quality continuous curve patterns.
}%
We evaluate our methods and interface via different patterns, ablation studies, and comparisons with alternative methods.

\nothing{
Element pattern is composed of many small geometric elements, which is a fundamental aspect in illustrations. 
It is used to enhance visual complexity and support many artistic effects.
However, element pattern design is a tedious task usually involving manual repetitions. 

In this paper, we present an autocomplete, sketch-based system for element pattern design to alleviate user workload by predicting what the user might want to draw based on analyzing past workflows.
Users can accept, partially accept or ignore the provided suggestions and thus maintain full control and individual style.
Differently from previous autocomplete systems that are limited to simple patterns, by combing pattern grammars and example-based synthesis, our method can analyze complex (stochastic or regular, stationary or non-stationary) patterns and generate reasonable predictions. We demonstrate the quality and accessibility of our system via pattern design results and user feedbacks.
}%

\nothing{
}%

\nothing{
}%

\end{abstract}

\nothing{

}%




\begin{teaserfigure}
  \centering
\captionsetup[subfloat]{captionskip=1pt}
  \begin{tabular}{c}
	\subfloat[Output of \protect\subref{fig:teaser:string:input}]{%
          \label{fig:teaser:string:output}%
		\includegraphics[width=0.19\linewidth]{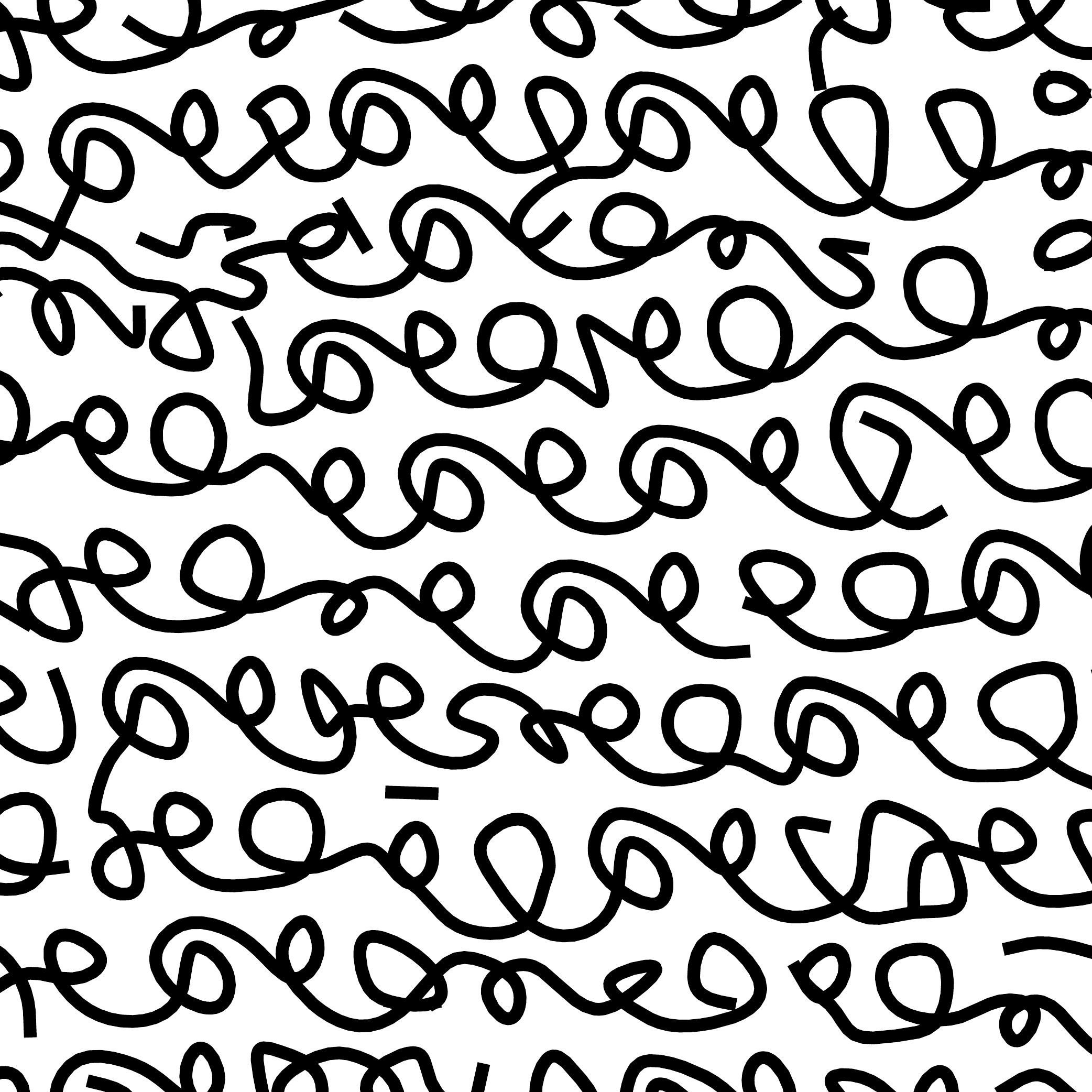}
	}%
\end{tabular}
\hspace{-1.8em}
  \begin{tabular}{cc}
  	\subfloat[String]{%
          \label{fig:teaser:string:input}%
  	\includegraphics[width=0.084\linewidth]{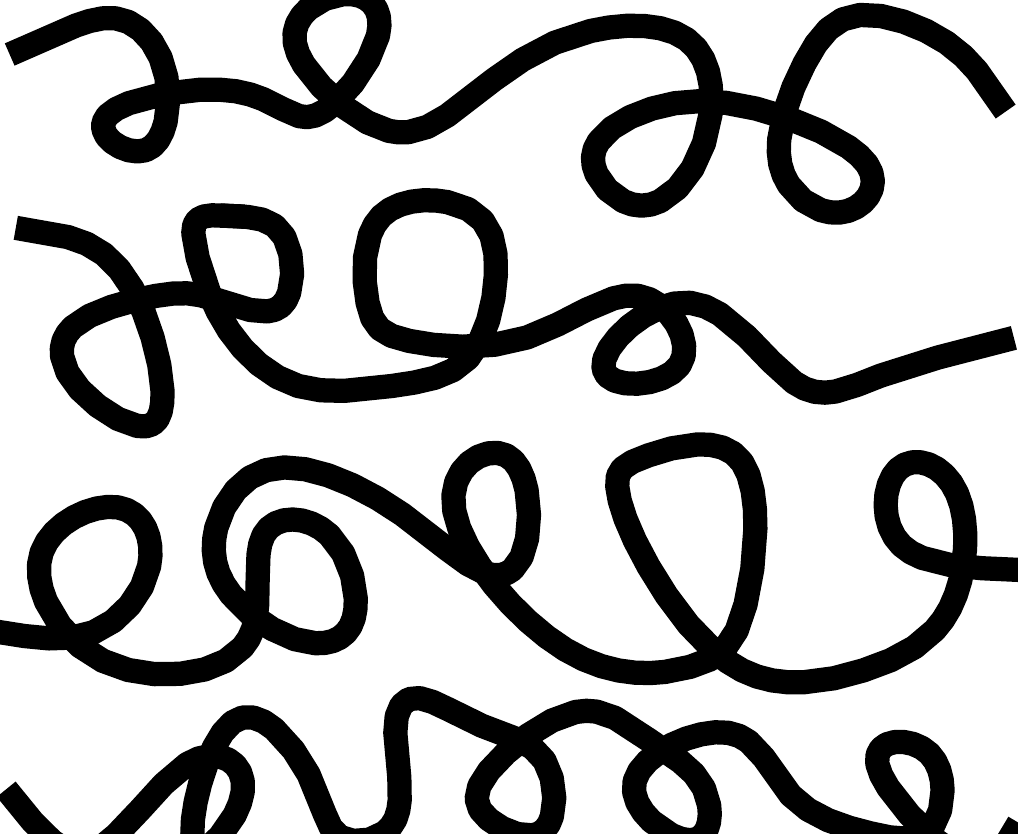}
  	}%
  \vspace{-0.8em}
  \\
	\subfloat[Strip]{%
        \label{fig:teaser:strip:input}%
	\includegraphics[width=0.105\linewidth]{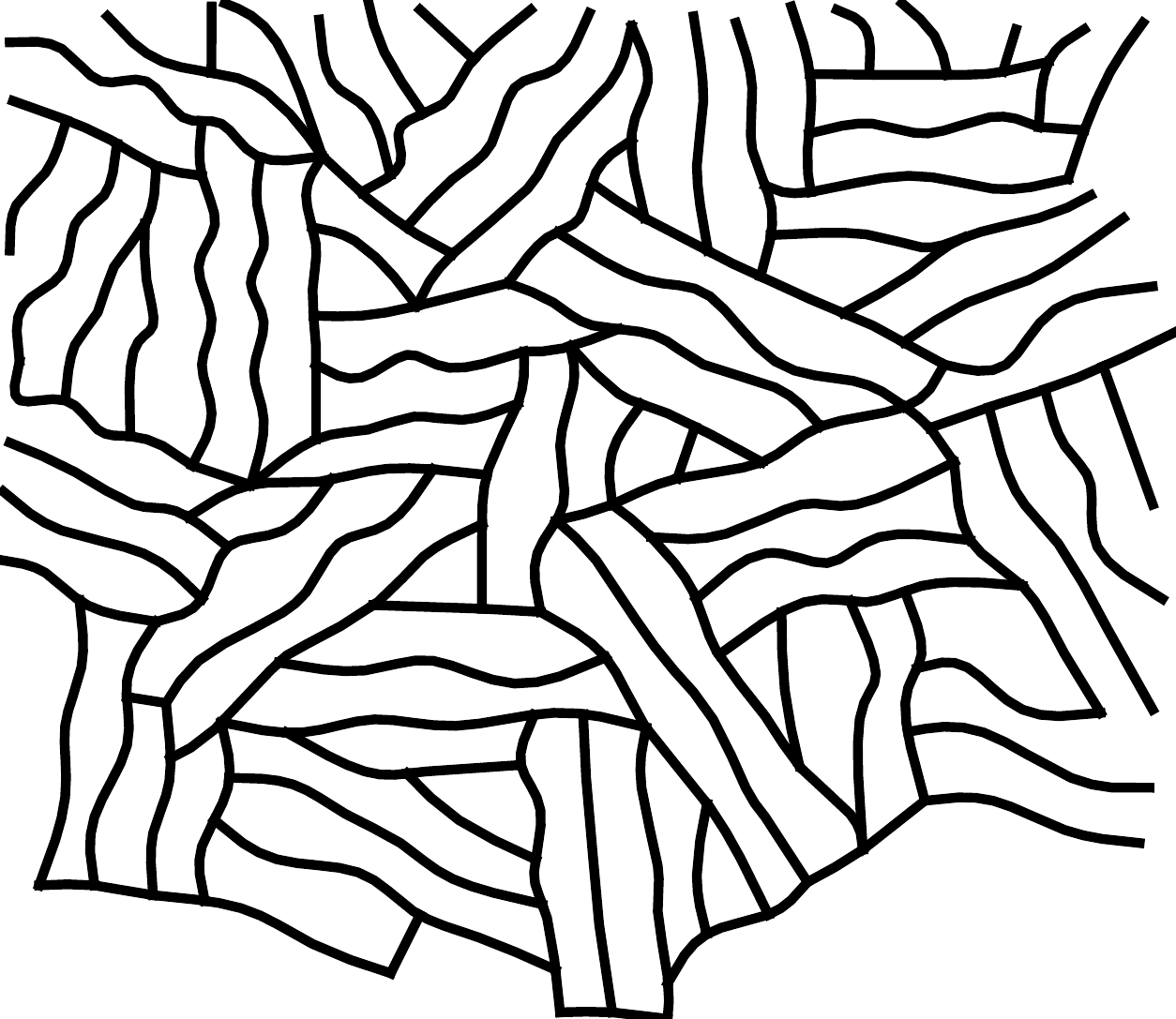}
	}
  \end{tabular}
\hspace{-1.8em}
  \begin{tabular}{c}
	\subfloat[Output of \protect\subref{fig:teaser:strip:input}]{%
        \label{fig:teaser:strip:output}%
		\includegraphics[width=0.19\linewidth]{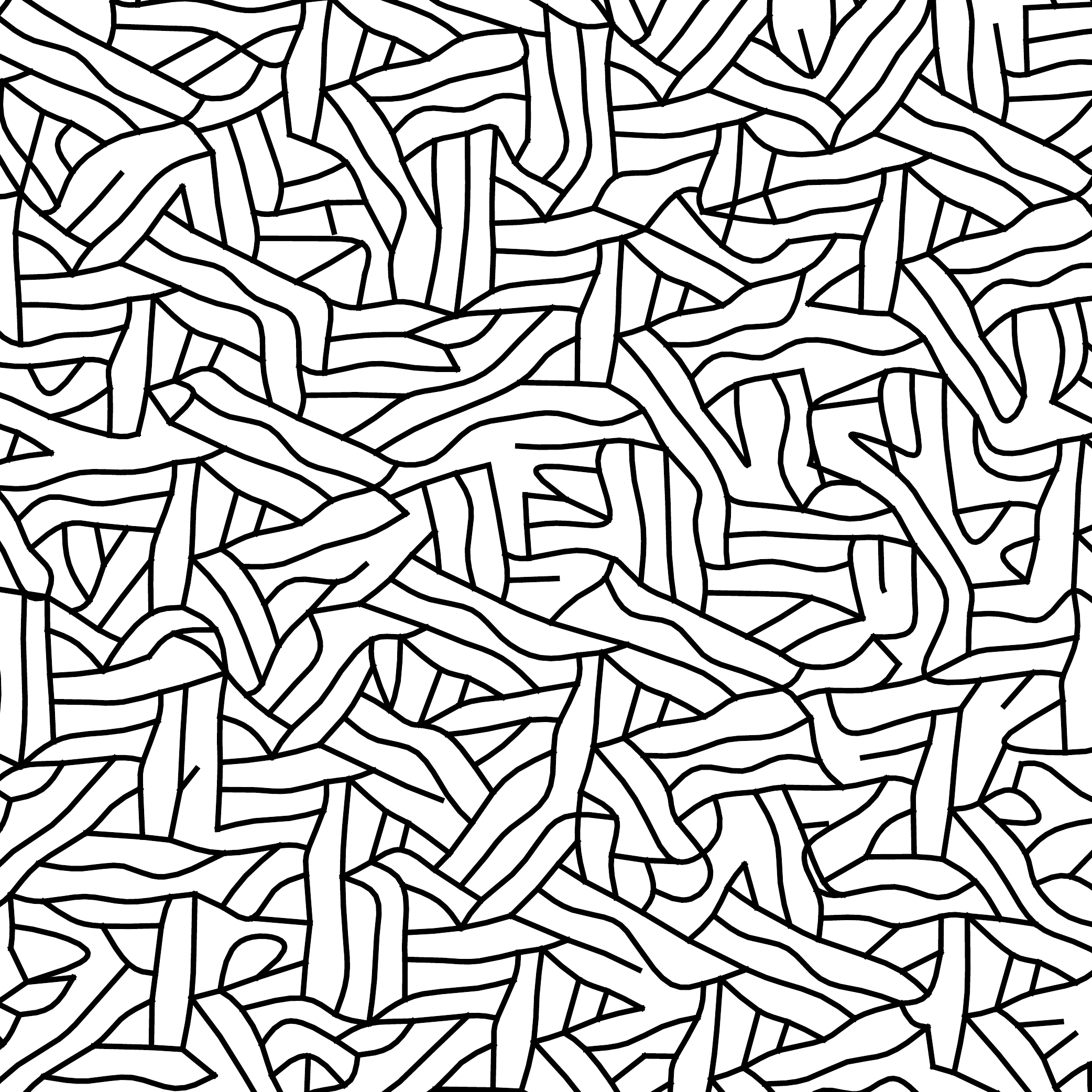}
	}%
\end{tabular}
\hspace{-1.77em}
  \begin{tabular}{c}
	\subfloat[Output of \protect\subref{fig:teaser:maze:input}]{%
        \label{fig:teaser:maze:output}%
		\includegraphics[width=0.19\linewidth]{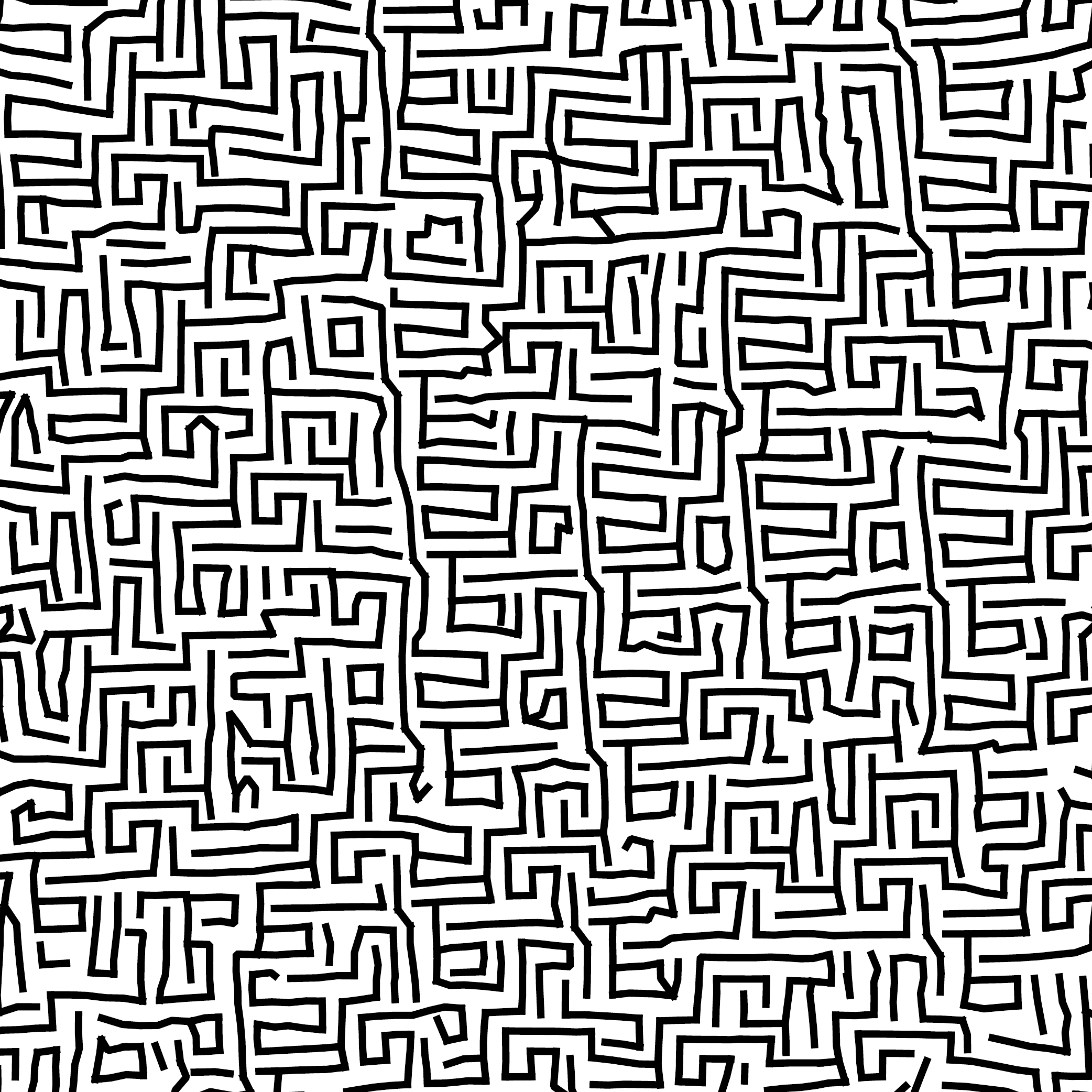}
	}%
\end{tabular}
\hspace{-1.8em}
  \begin{tabular}{cc}
\subfloat[Maze]{%
        \label{fig:teaser:maze:input}%
	\includegraphics[width=0.076\linewidth]{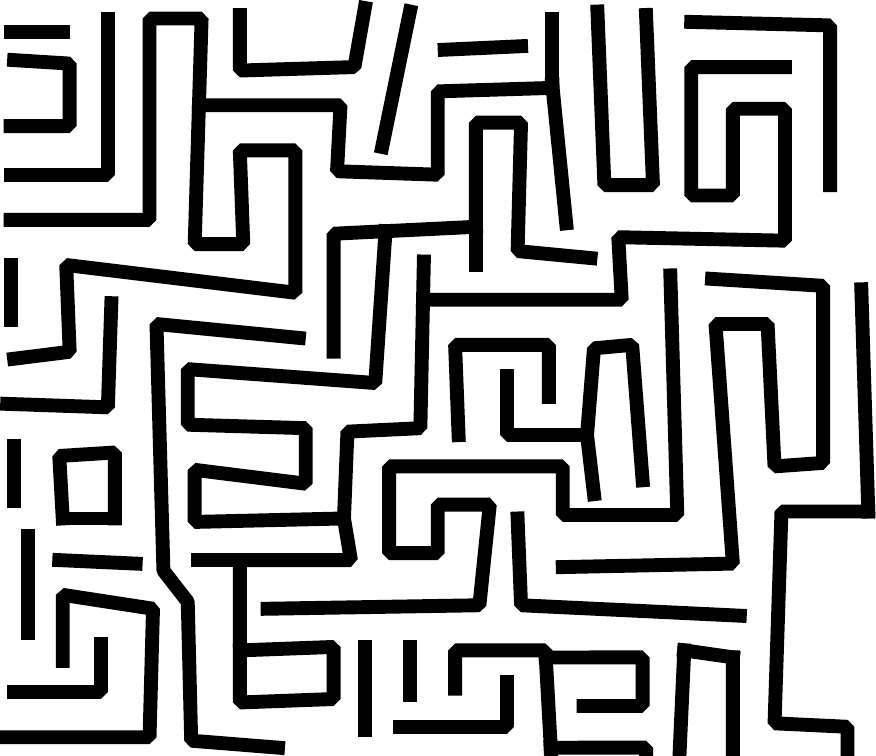}
	}
\vspace{-1.1em}
	\\
	\subfloat[Branches]{%
        \label{fig:teaser:branch:input}%
		\includegraphics[width=0.11\linewidth]{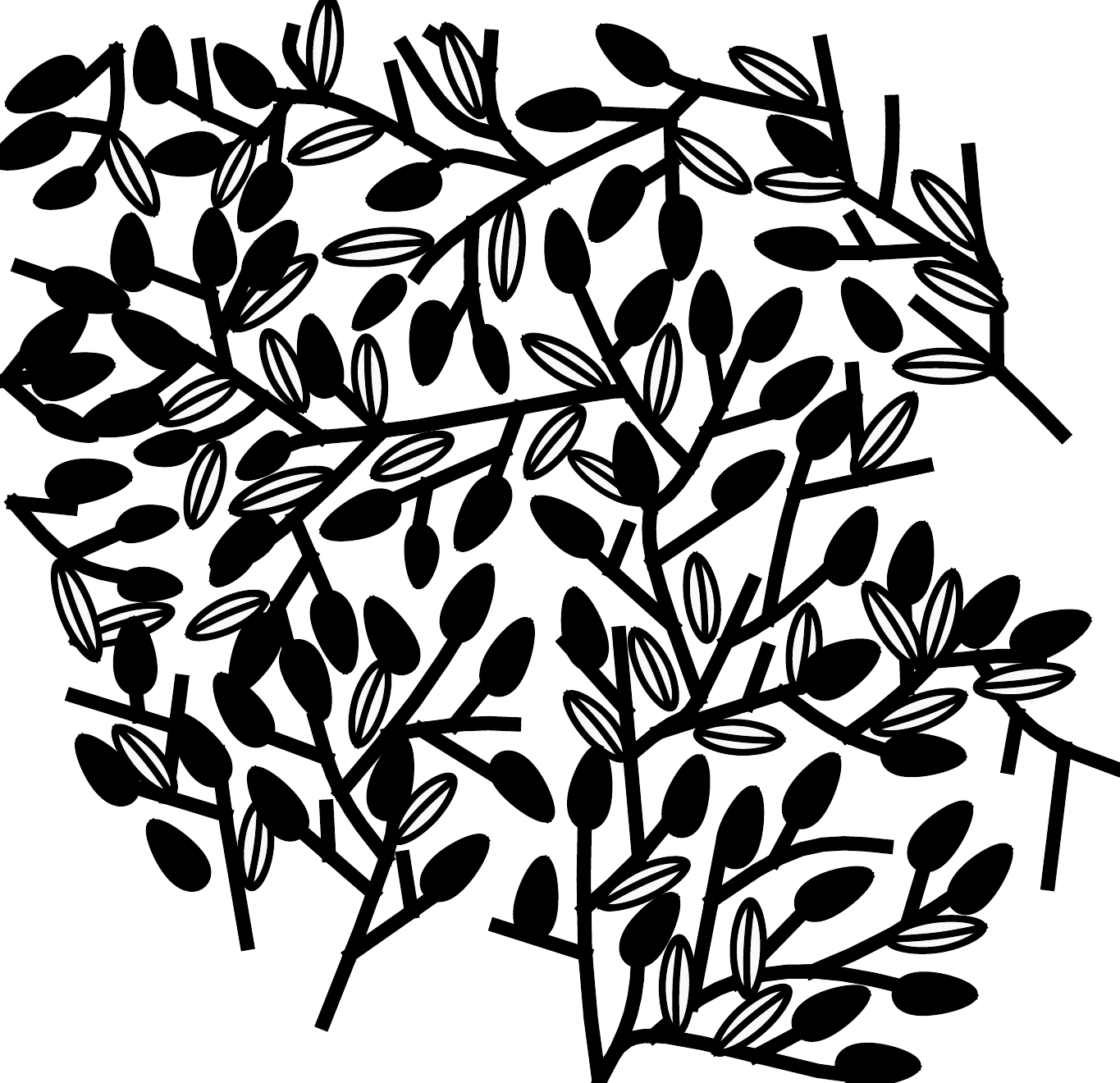}
	}%
\end{tabular}
\hspace{-1.8em}
  \begin{tabular}{c}
	\subfloat[Output of \protect\subref{fig:teaser:branch:input}]{%
        \label{fig:teaser:branch:output}%
		\includegraphics[width=0.19\linewidth]{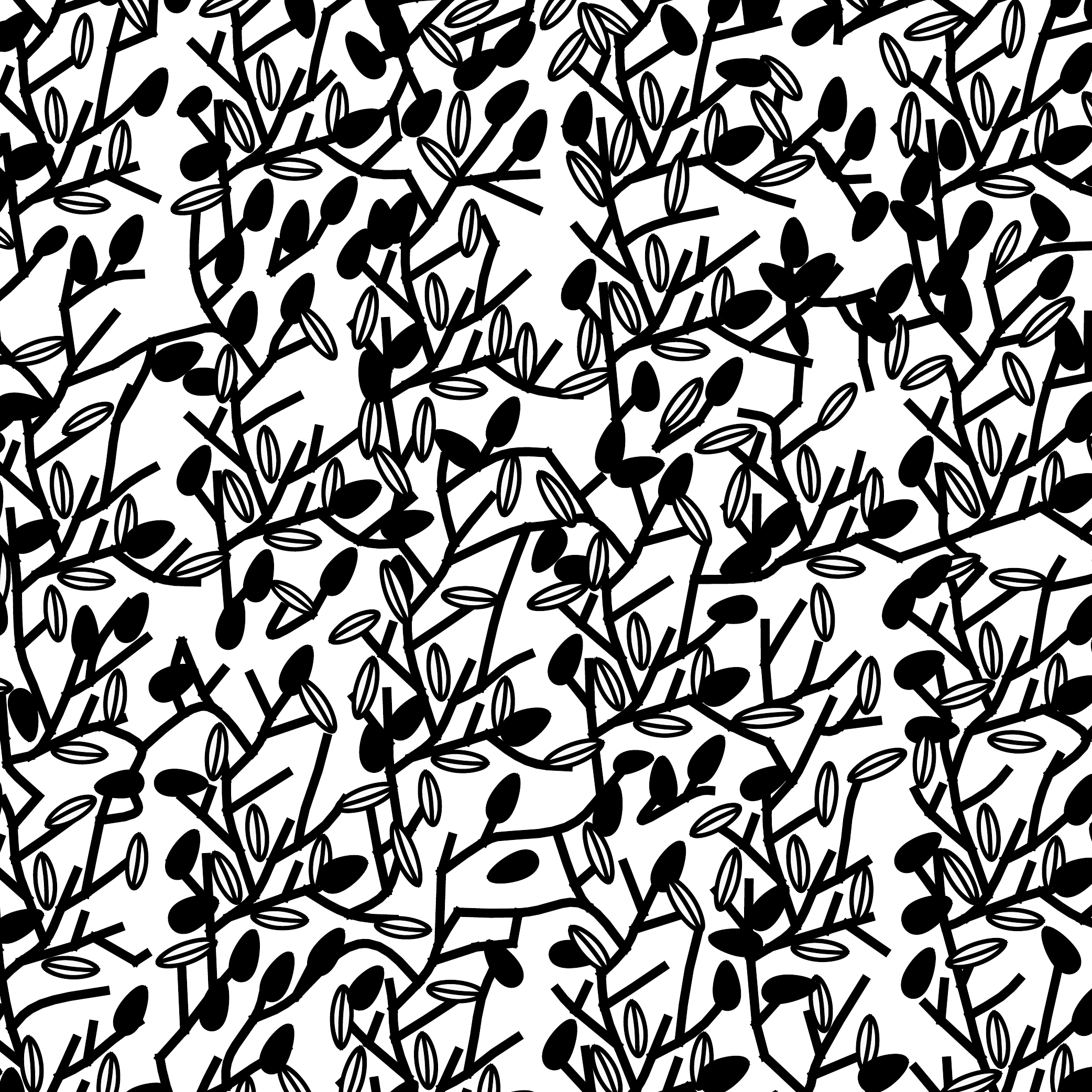}
	}%
\end{tabular}
  \nothing{
  \subfloat[elements: \auto{231}/359; path lengths: \auto{11489}/16209]{
  	\includegraphics[width=0.32\linewidth]{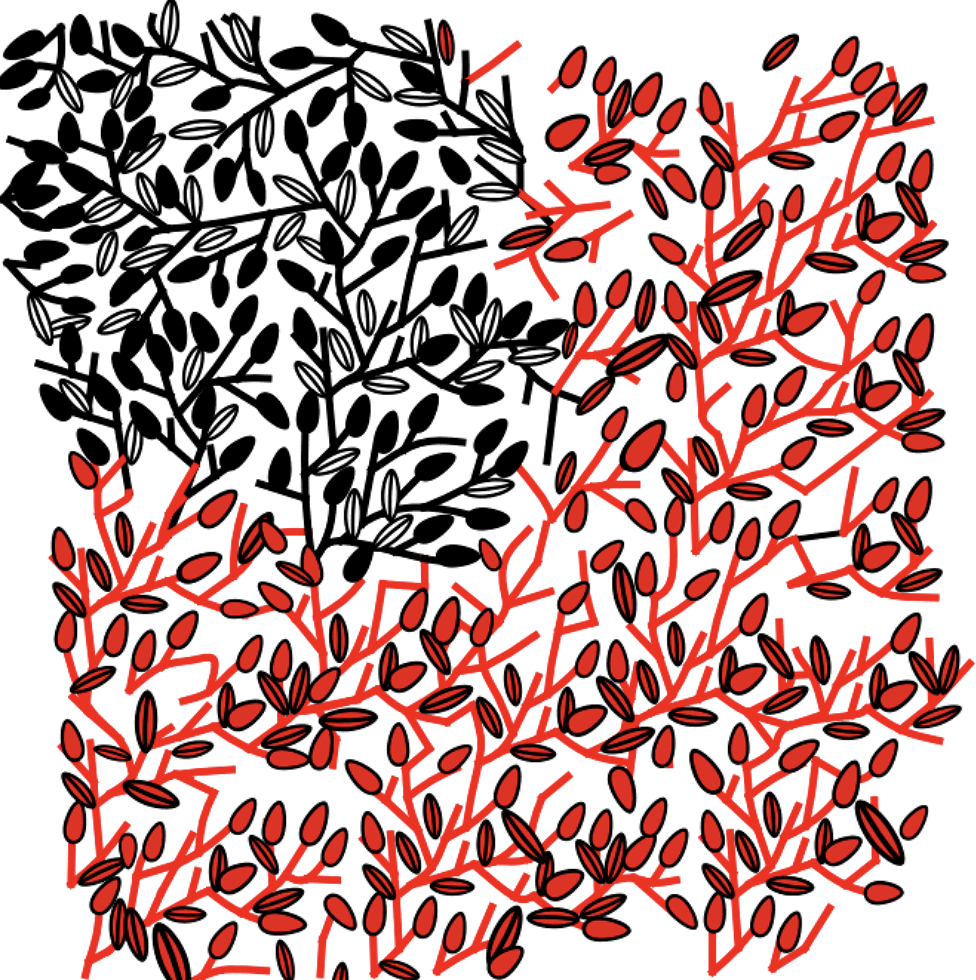}
  }%
\subfloat[elements: \auto{141}/171; path lengths: \auto{17175}/22172]{
	\includegraphics[width=0.32\linewidth]{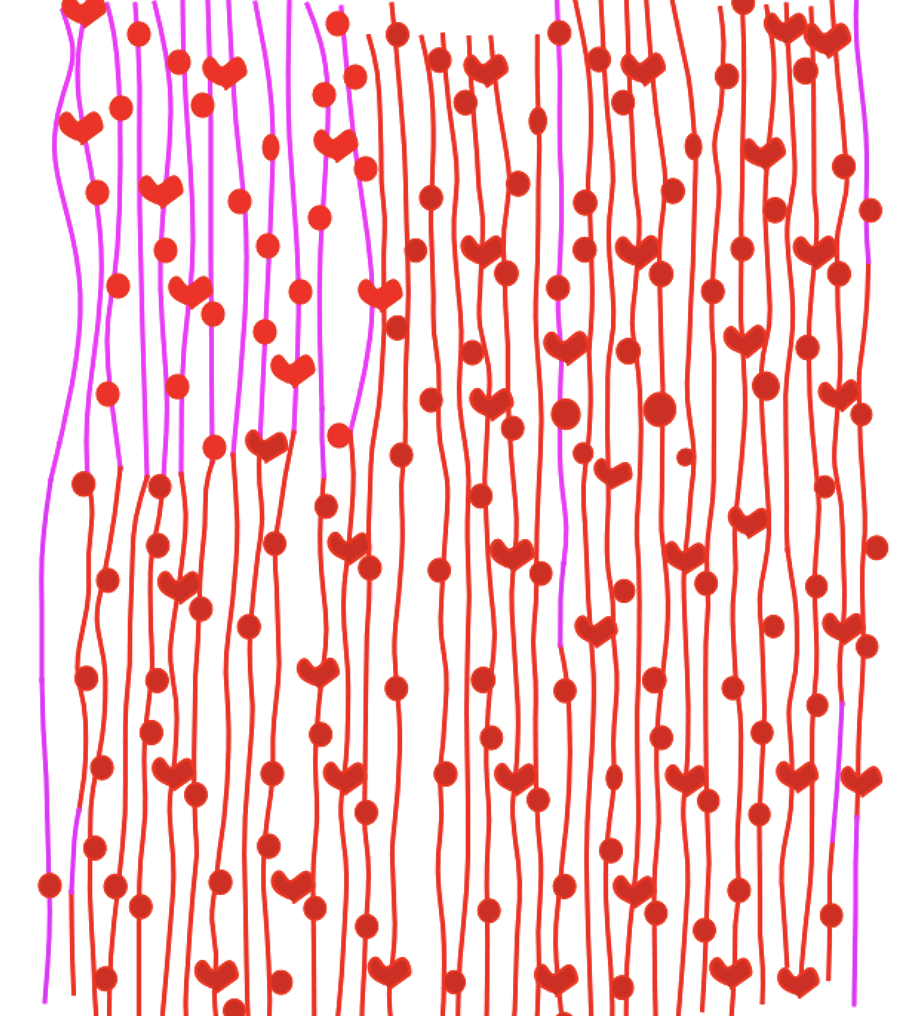}
}%
  \subfloat[elements: \auto{411}/564; path lengths: \textcolor{blue}{11420}/16450]{
	\includegraphics[width=0.32\linewidth]{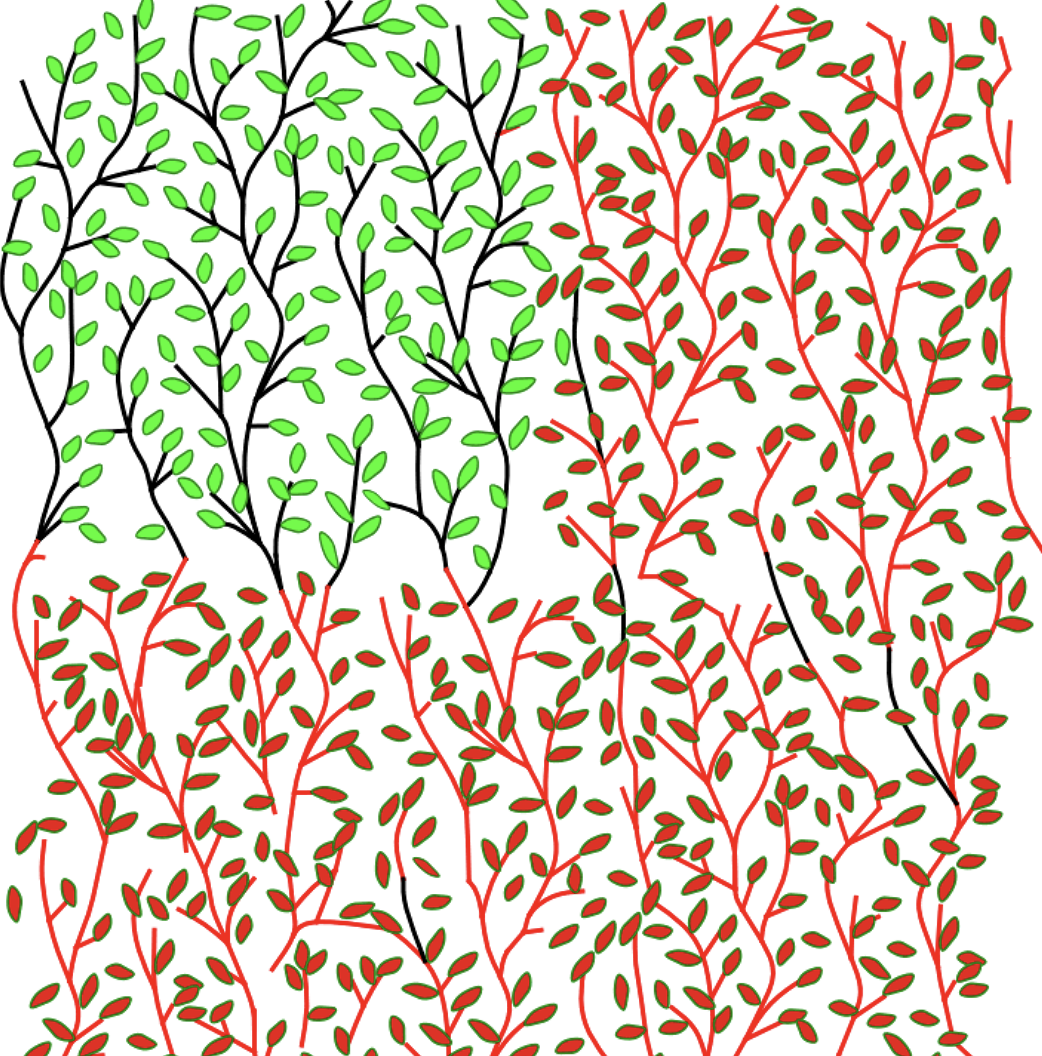}
}%
}%
  \Caption{\update{Example inputs and outputs of our method.}}
  {%
  	Given a small input exemplar, our algorithm can synthesize various types of continuous curve textures. 
  	Our results are shown in vector format. Please zoom in to see the details.
\nothing{
}%
\nothing{
}%
\nothing{
  Our system can be used to author various patterns. The user draw a smaller exemplars on the top-left. 
  Our system can generate patterns (in red) that resemble the initial exemplars.
  The users can do further edits on the predicted patterns (e.g. black paths within red paths, which are user edits).
  The numbers below are the numbers of \auto{autocomplete} and total discrete elements (e.g. leaves), and \auto{autocomplete} and total lengths of continuous paths (e.g. branches).
}%
  \nothing{
}%
 }
  \label{fig:teaser}
\end{teaserfigure}

\maketitle

\section{Introduction}
\label{sec:introduction}

\begin{figure*}[tbh]
	\centering
	\setlength{\tabcolsep}{0pt}
\begin{tabular}{ccccccc}
	\includegraphics[width=0.064\linewidth]{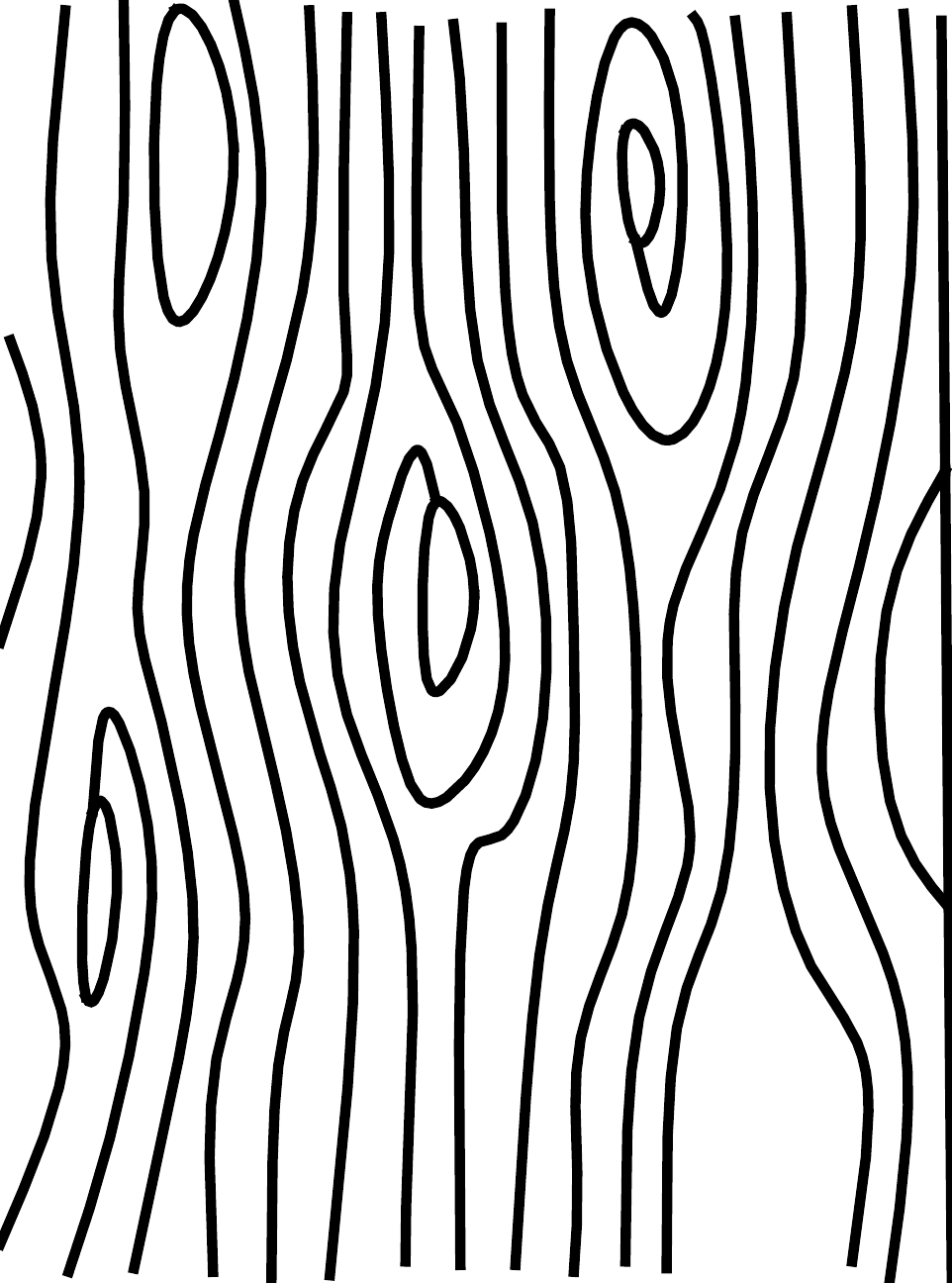}
	&\includegraphics[width=0.064\linewidth]{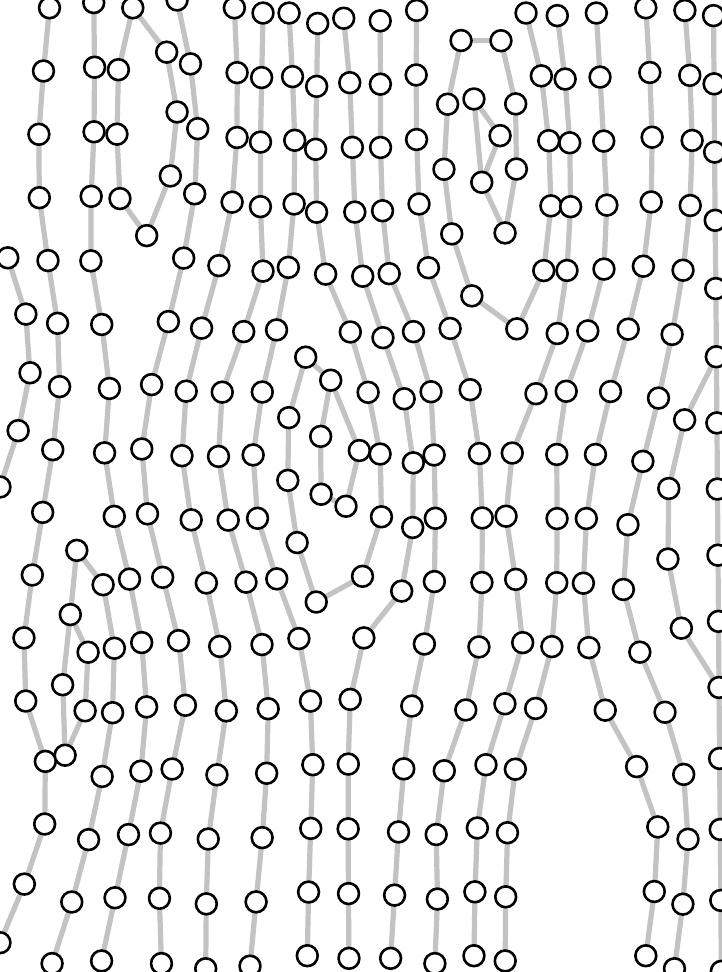}
	&\includegraphics[width=0.14\linewidth]{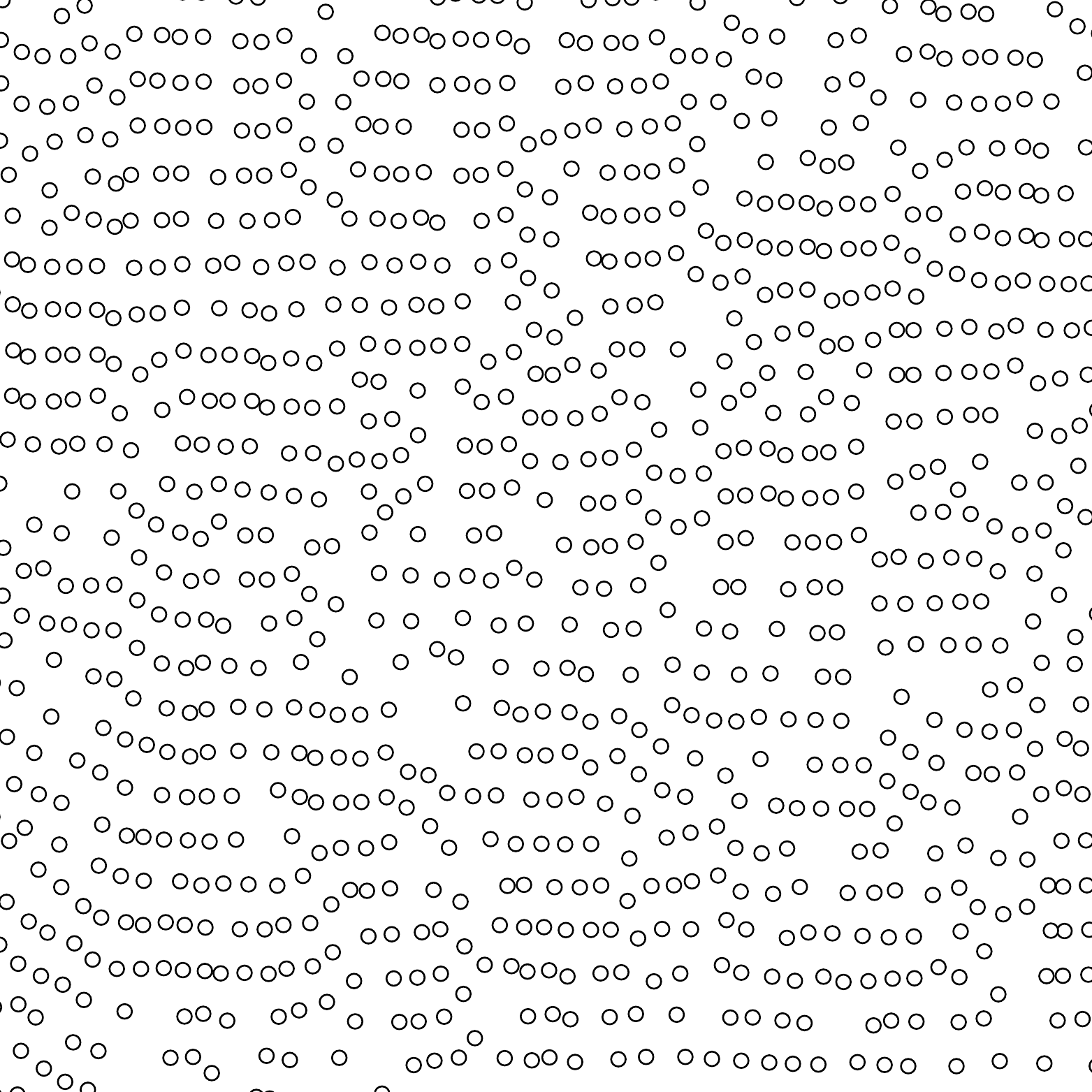}
	&\includegraphics[width=0.14\linewidth]{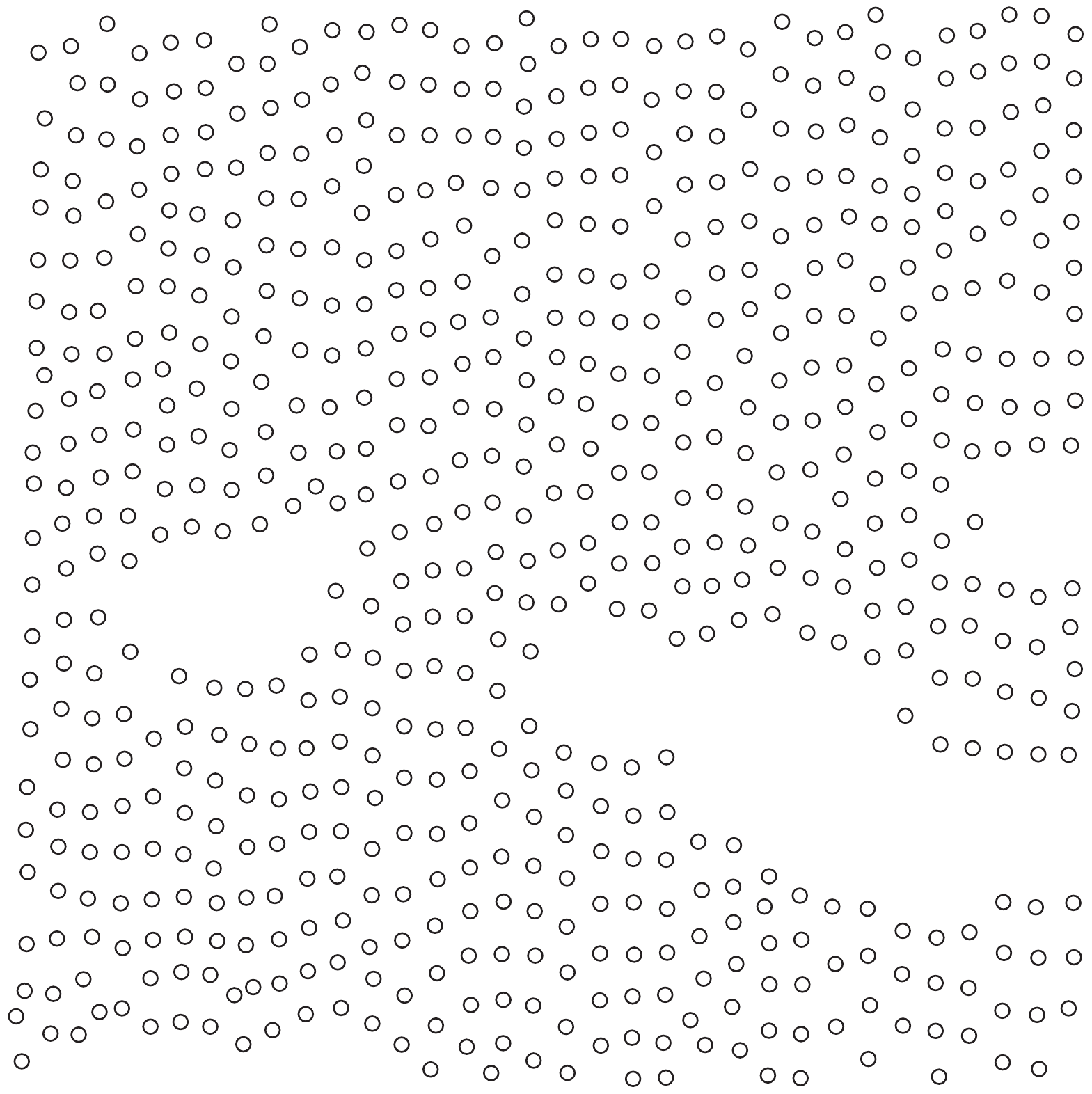}
	&\includegraphics[width=0.14\linewidth]{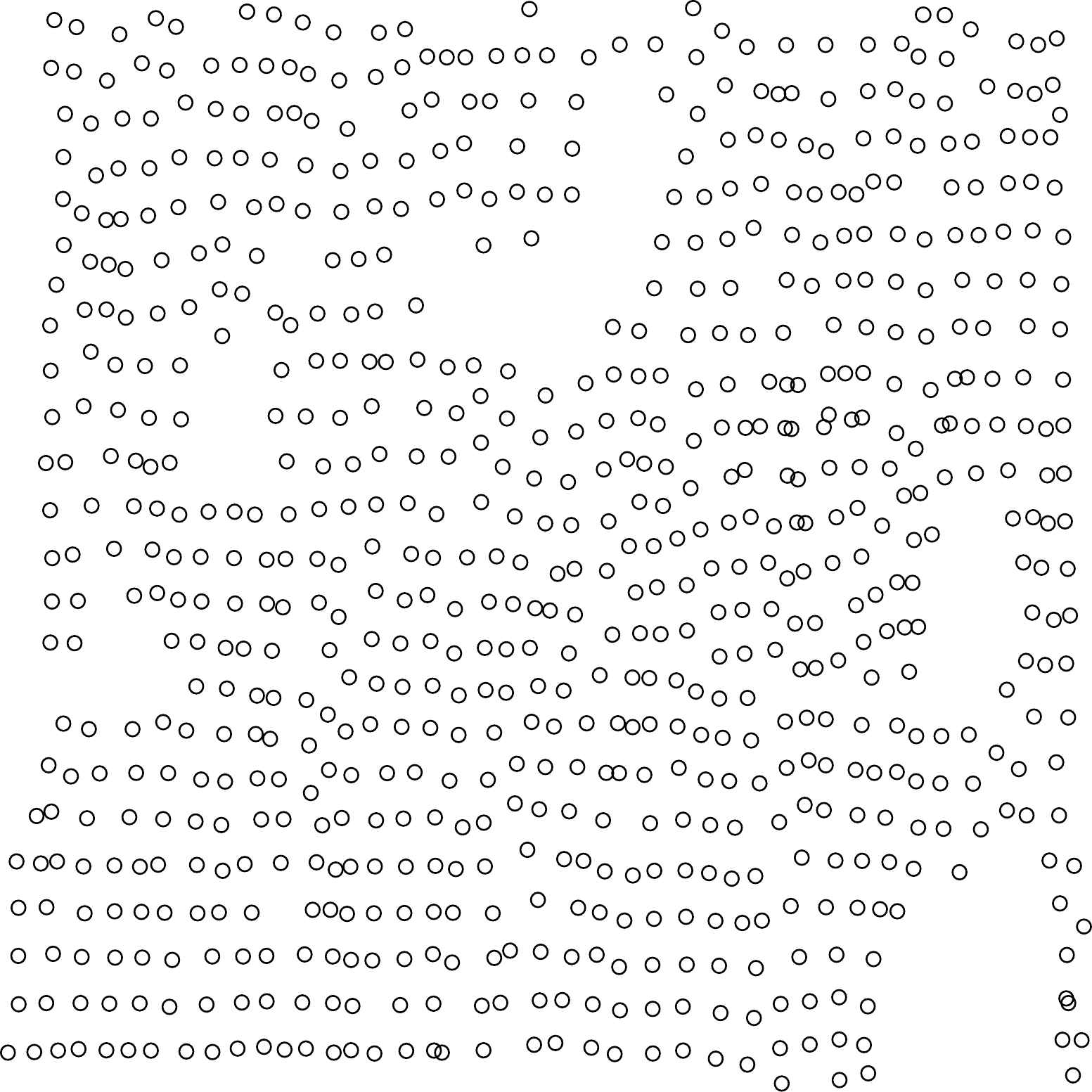}
	&\includegraphics[width=0.14\linewidth]{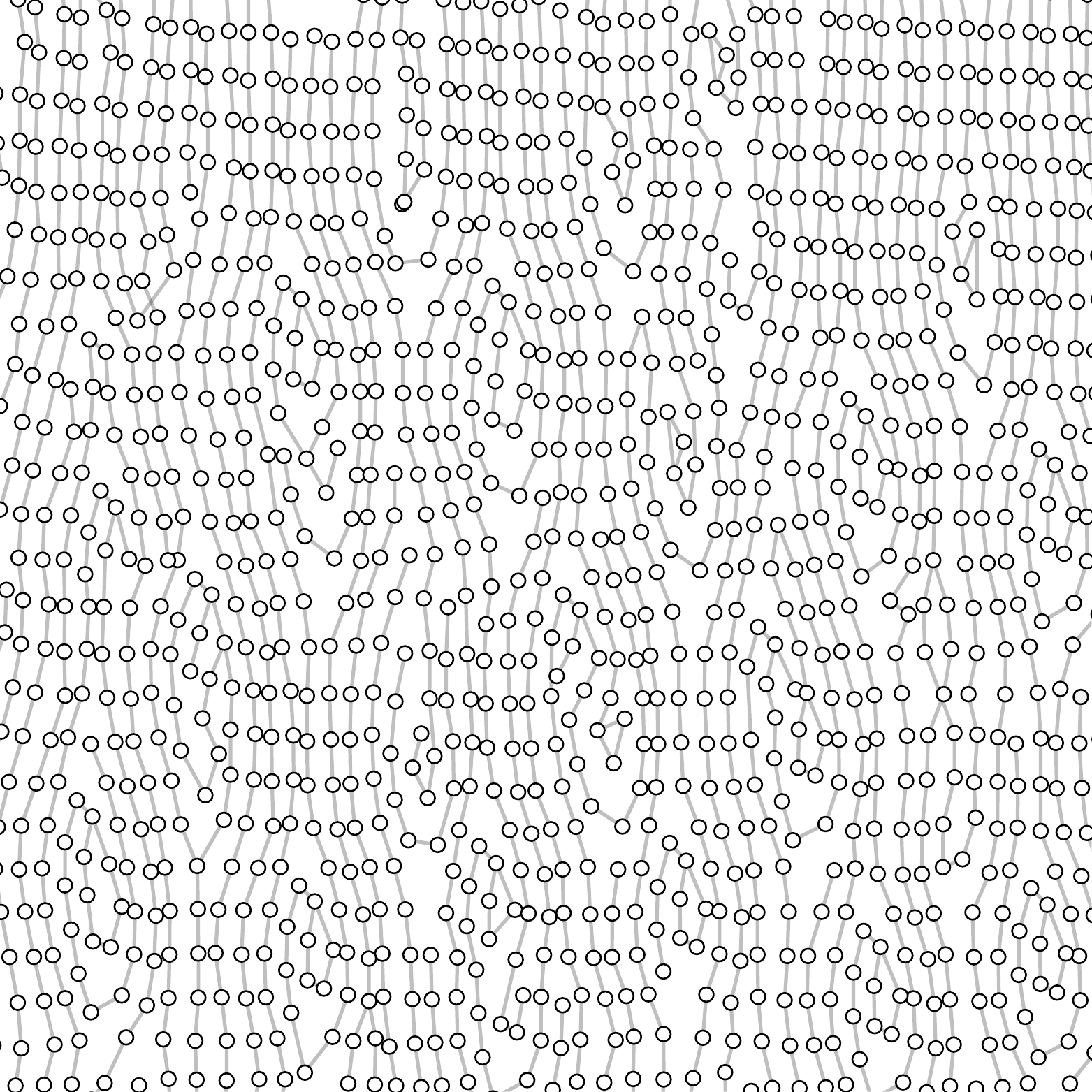}
	&\includegraphics[width=0.14\linewidth]{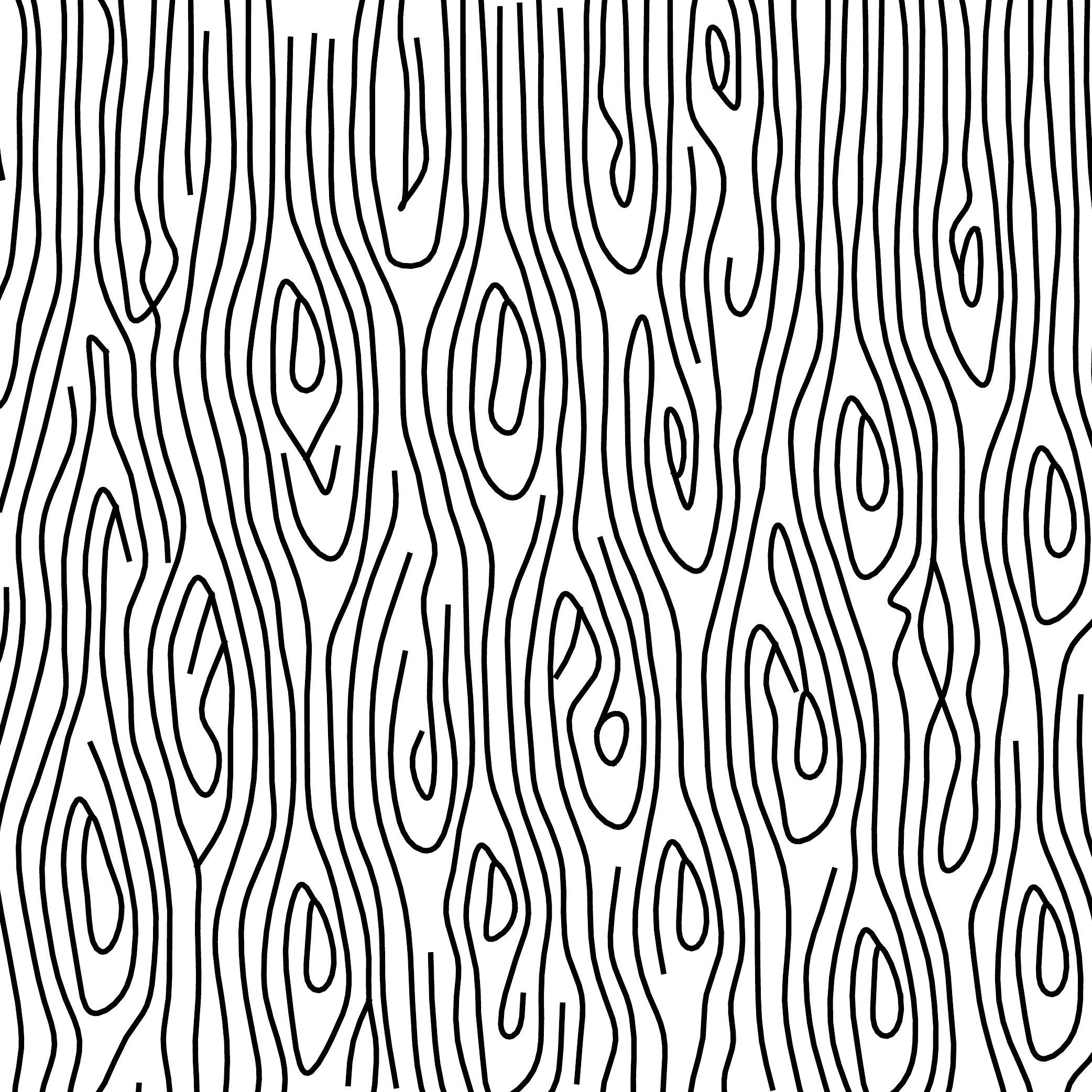}
\vspace{-1em}
   \\
   \subfloat[Exemplar]{
	\label{fig:comparison_sample_synthesis:exemplar}
	\includegraphics[width=0.096\linewidth]{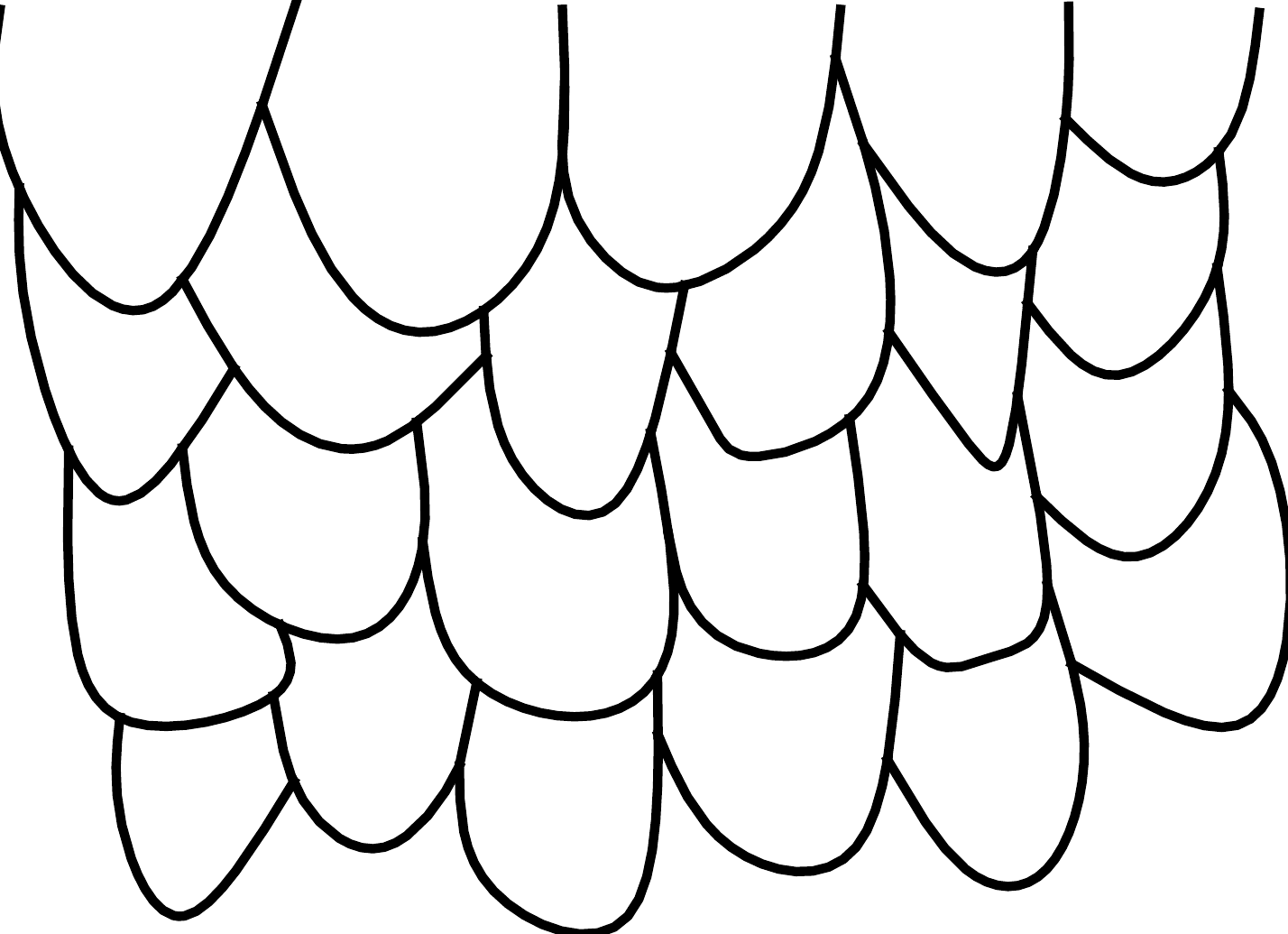}
}%
   &\subfloat[Samples of \protect\subref{fig:comparison_sample_synthesis:exemplar}]{
	\label{fig:comparison_sample_synthesis:exemplar_samples}
	\includegraphics[width=0.096\linewidth]{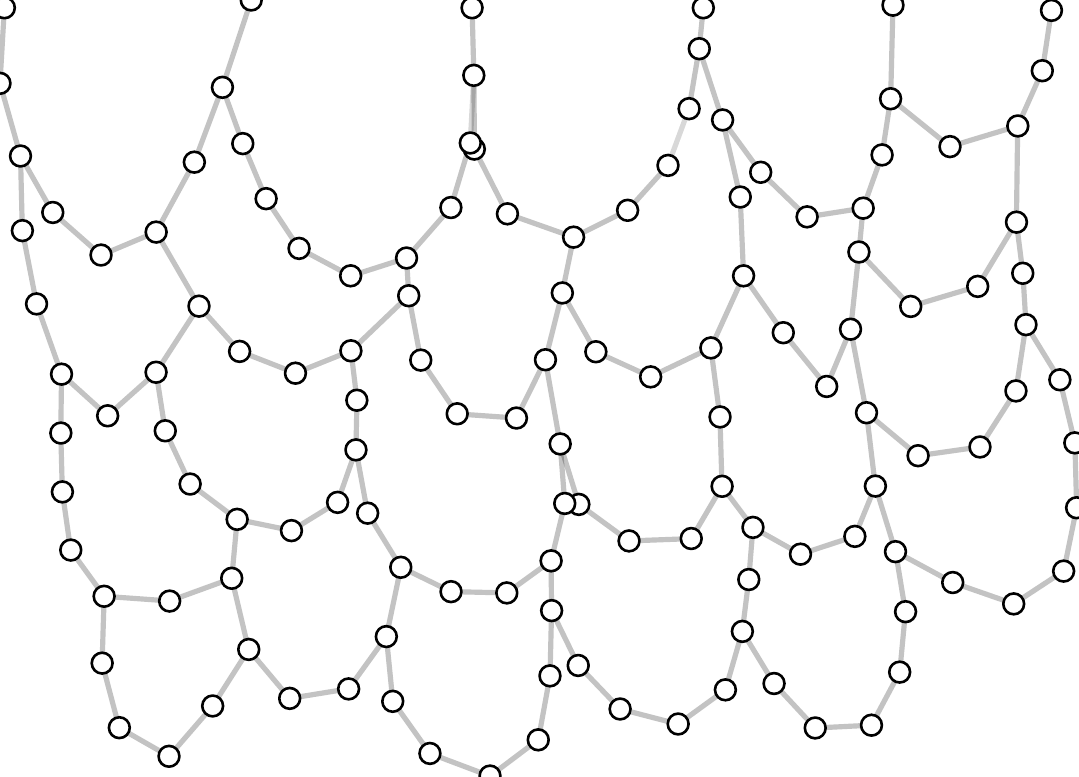}
}%
&\subfloat[\cite{Ma:2011:DET}]{
	\label{fig:comparison_sample_synthesis:ma}
	\includegraphics[width=0.14\linewidth]{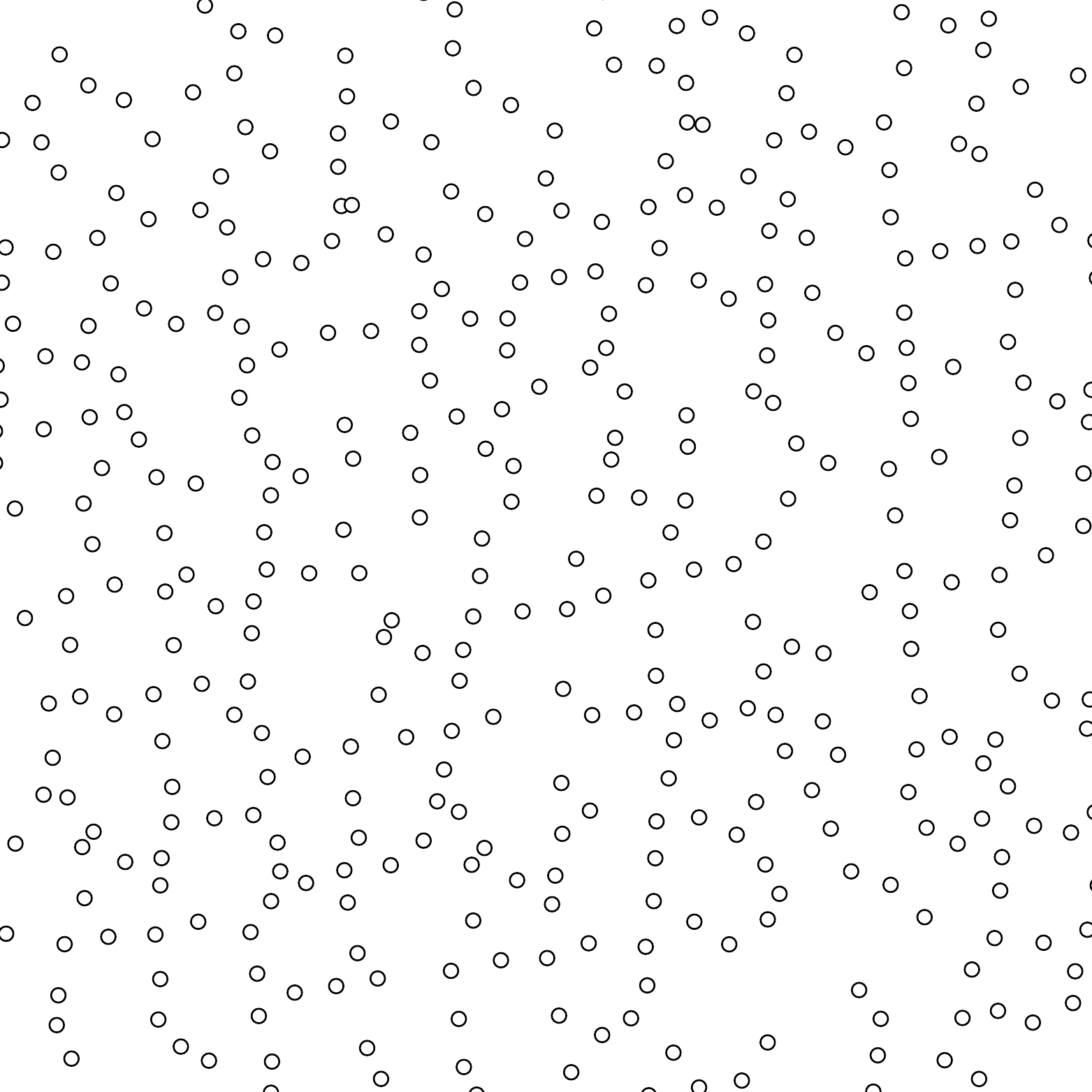}
}%
&\subfloat[\cite{Roveri:2015:EBR}]{
	\label{fig:comparison_sample_synthesis:roveri}
	\includegraphics[width=0.14\linewidth]{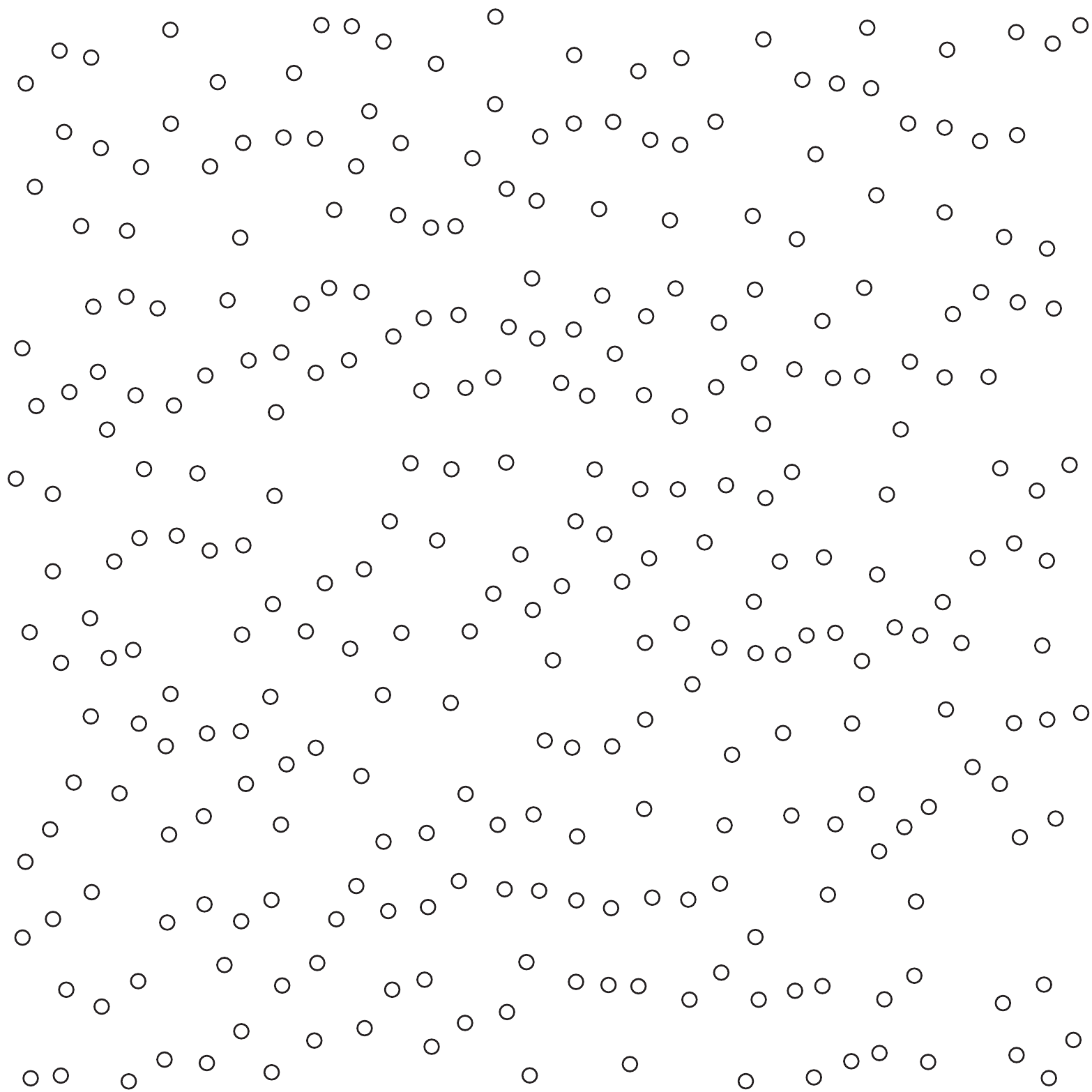}
}%
&\subfloat[\cite{Tu:2019:PPS}]{
	\label{fig:comparison_sample_synthesis:tu}
	\includegraphics[width=0.14\linewidth]{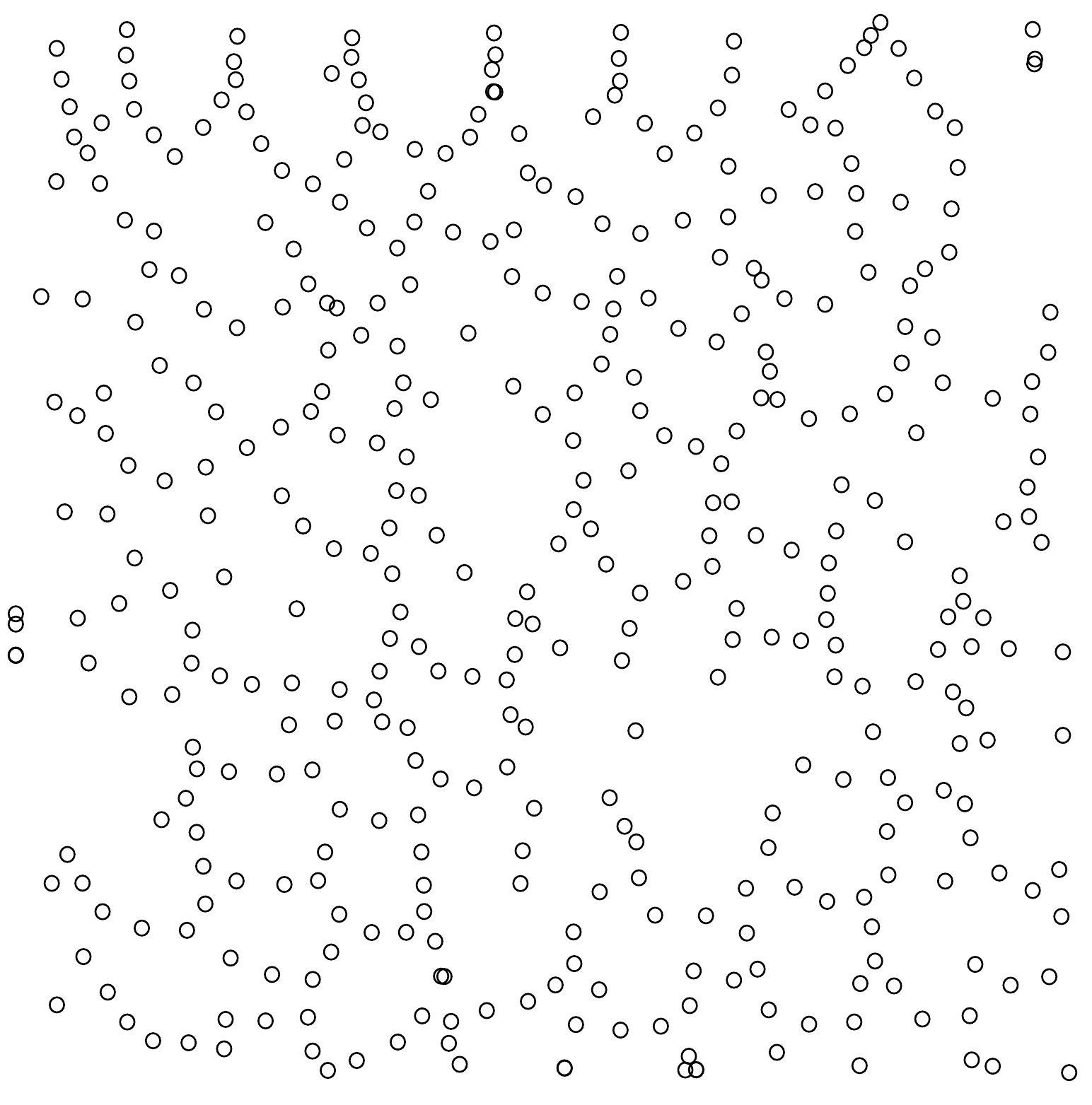}
}%
&\subfloat[Our output samples\nothing{ synthesis}]{
	\label{fig:comparison_sample_synthesis:ours:sample_synthesis}
	\includegraphics[width=0.14\linewidth]{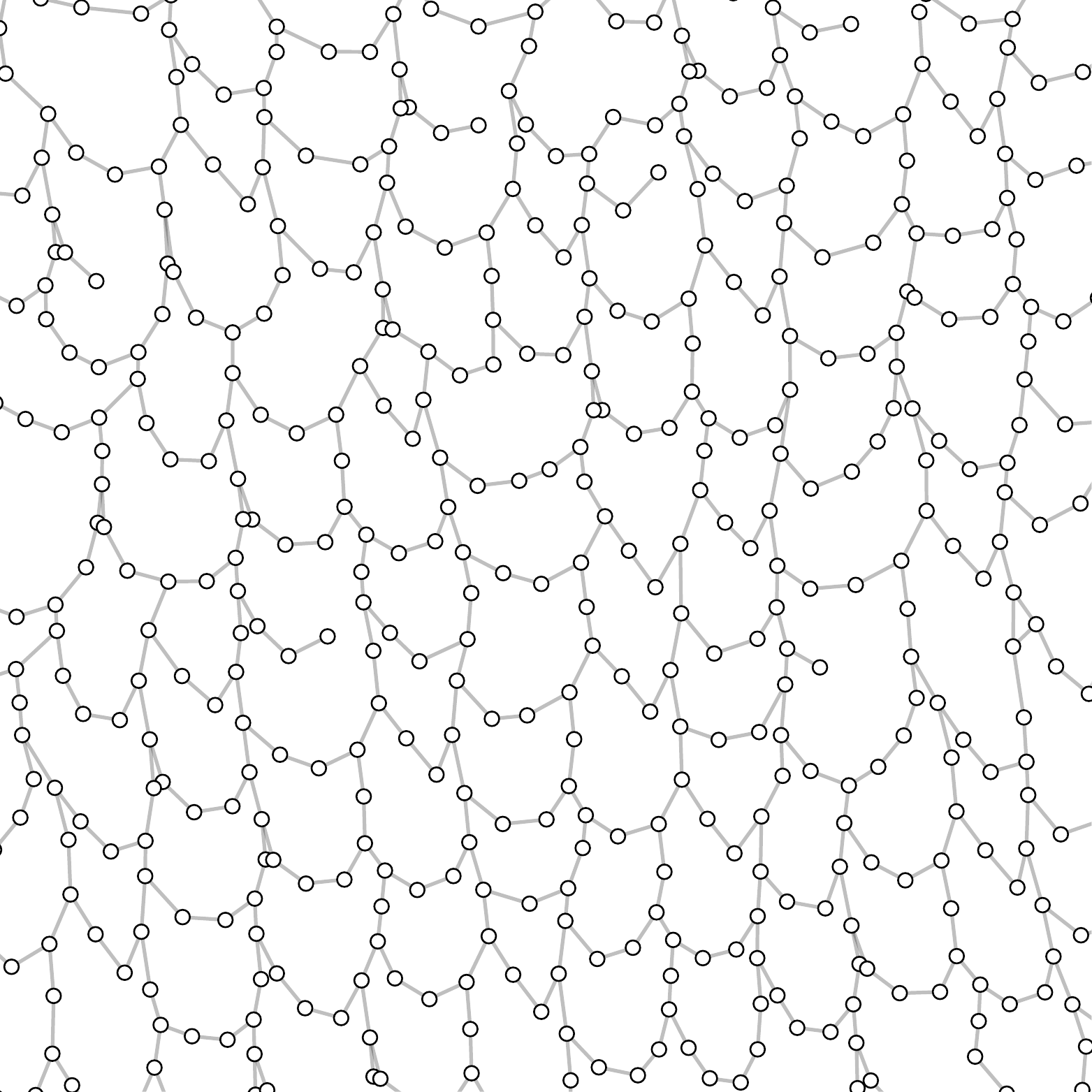}
}%
&\subfloat[Reconstruction of \protect\subref{fig:comparison_sample_synthesis:ours:sample_synthesis}]{
	\label{fig:comparison_sample_synthesis:ours:curve_recon}
	\includegraphics[width=0.14\linewidth]{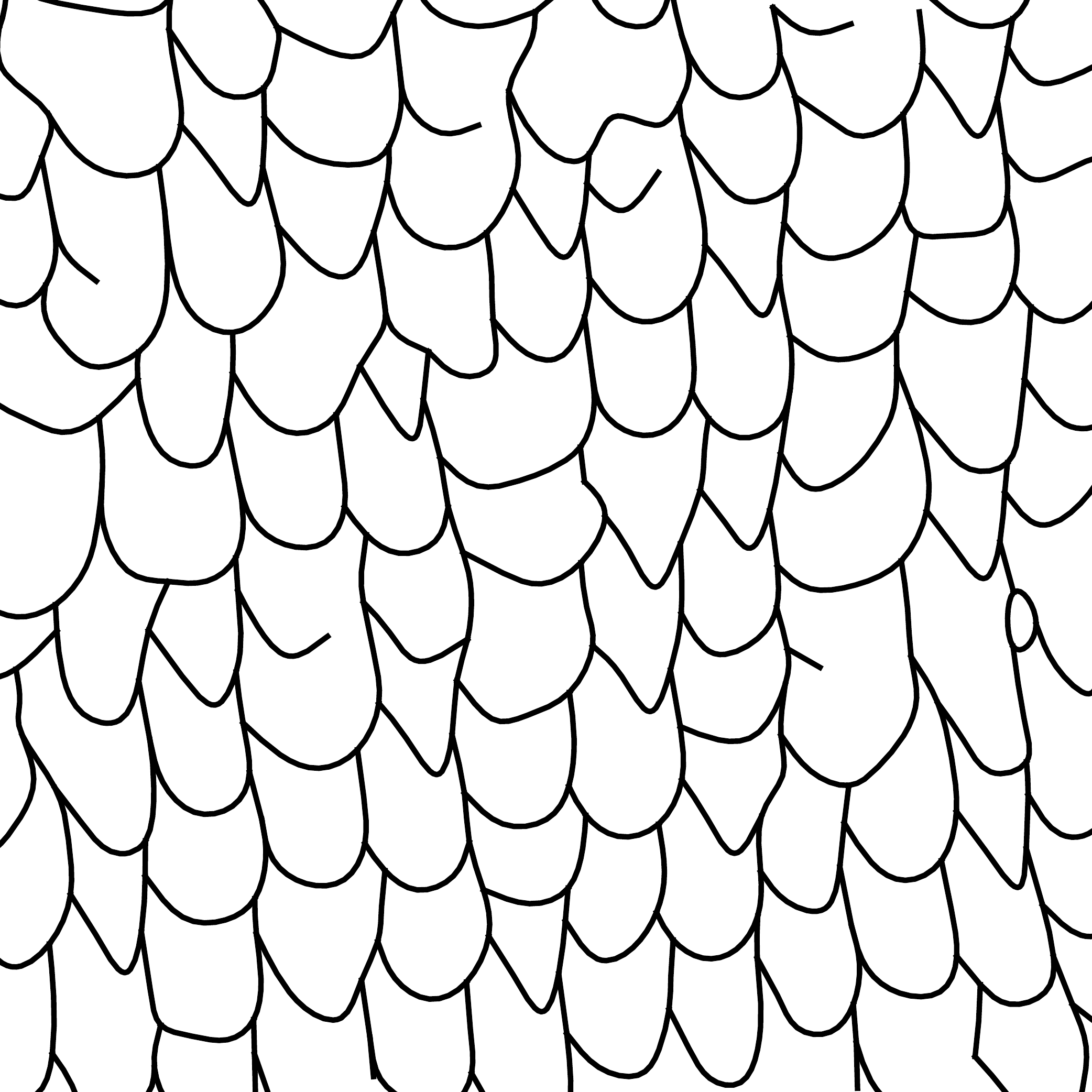}
}%
\end{tabular}
	\Caption{Comparison with prior point/sample synthesis algorithms.}
	{%
		The methods in \cite{Ma:2011:DET,Roveri:2015:EBR,Tu:2019:PPS} generate samples without considering their connections in the exemplars \subref{fig:comparison_sample_synthesis:exemplar_samples}.
                Thus, their results, as shown in \subref{fig:comparison_sample_synthesis:ma}, \add{\subref{fig:comparison_sample_synthesis:roveri}}, and \subref{fig:comparison_sample_synthesis:tu}, preserve the sample distributions less well than ours in \subref{fig:comparison_sample_synthesis:ours:sample_synthesis}.
                 It is also unclear how to reconstruct continuous curve patterns from \subref{fig:comparison_sample_synthesis:ma}, \add{\subref{fig:comparison_sample_synthesis:roveri}}, and \subref{fig:comparison_sample_synthesis:tu}. 
                \subref{fig:comparison_sample_synthesis:ours:curve_recon} shows the curve reconstruction from \subref{fig:comparison_sample_synthesis:ours:sample_synthesis}.
\nothing{
		They could not be applied to generate patterns we target \subref{fig:comparison_sample_synthesis:ours:curve_recon}.
		Post-processing is required to reconstruct curve patterns from synthesized sample distributions, which is notoriously difficult and unrobust especially when the samples are sparse.
		Instead, our method generates samples with connectivities \subref{fig:comparison_sample_synthesis:ours:sample_synthesis}, which are used to reconstruct continuous curves patterns \subref{fig:comparison_sample_synthesis:ours:curve_recon}.
}%
\nothing{
}%
}%
	\label{fig:comparison_sample_synthesis}
\end{figure*}

 %

\nothing{
}%

Repetitive patterns are fundamental for a variety of tasks in design \cite{Kazi:2012:VIT,Lu:2014:DDS} and engineering \cite{Zhou:2014:TSV,Martinez:2015:SAO,Zehnder:2016:DSO,Chen:2016:SFD,Schumacher:2016:SDS}.
Manually creating these patterns provides high degrees of individual freedom, but can also require significant technical/artistic expertise and manual labor.
\nothing{Element pattern is composed of many identifiable geometric elements with specific local element interactions and global arrangements \cite{Loi:2017:PAE}.
Element pattern is used to enhance visual complexity and support many artistic effects \cite{Chen:2016:SFD}.
However, designing element pattern is a rather tedious task usually involving intensive manual repetitions.
}%
\nothing{
.
}%
These usability barriers can be reduced by automatic methods that can synthesize patterns similar to user-supplied exemplars \cite{Ma:2011:DET,Ma:2013:DET,Ijiri:2008:EBP,Hurtut:2009:ASE,Barla:2006:SPA,Kazi:2012:VIT,Lu:2014:DDS,\nothing{Xing:2014:APR,Xing:2015:AHA,Peng:2018:A3S,}Suzuki:2017:ATP,Hsu:2018:BEF,Hsu:2020:AEF}.
However, existing techniques mainly focus on discrete patterns consisting of image pixels or shape elements, and might not apply to general patterns consisting of continuous \replace{curved patterns}{curves}\add{, which can be connected or intersected with one another}.

\nothing{

}%

\nothing{
Another important direction is to generate patterns \cite{Loi:2017:PAE,Reas:2019:P} using procedural modeling that has been successfully applied to create high-quality content for modeling virtual environment \cite{Smelik:2014:SPM}. 
However, procedural models require users to manipulate complex rules and parameters to generate desired content. This process requires high-level user expertise.
To address this issue, inverse procedural modeling attempts to seek (semi-)automatic solutions to deduce procedural representations of provided content or its abstract specifications \cite{Aliaga:2016:IPM}. 
However, there is little attention to inverse procedural modeling of element patterns.

}%

We propose an example-based method that can automatically synthesize continuous curve patterns from user-supplied exemplars\nothing{ in the vector format}.
Similar to prior pixel/sample-based methods \cite{Wei:2009:SAE,Ma:2011:DET,Lu:2012:HES,Landes:2013:SAM,Ma:2013:DET,Lu:2014:DDS,Roveri:2015:EBR}, users can provide exemplars and have the algorithm automatically produce results in desired sizes and shapes.
However, different from previous methods and systems that are restricted to discrete pixels/elements or limited continuous structures, our method can handle both discrete elements and continuous curves in a variety of patterns\nothing{ in the vector graphics format} (\Cref{fig:teaser}).

\nothing{We apply an online learning method that automatically tunes the desired algorithm parameters based on user feedbacks so that our system is able to handle diverse hierarchical patterns without manual parameter tuning that needs expertise.
}%

Our main idea is to extend prior \nothing{example-based }sample-based element synthesis methods \cite{Ma:2011:DET,Kazi:2012:VIT,Ma:2013:DET,Hsu:2018:BEF} to consider not only sample positions (geometry) but also their connections (topology) in all major algorithm components, including pattern representation, neighborhood similarity, and synthesis optimization consisting of search and assignment steps.
\nothing{
Our main idea is to combine procedural pattern modeling \cite{Loi:2017:PAE} and example-based pattern synthesis \cite{Ma:2011:DET} to analyze prior workflows and synthesize predictions that autocomplete user operations. 
While the procedural modeling cannot well control local element interactions but may generate interesting global arrangement of elements in a pattern, the example-based synthesis can faithfully reproduce local element interactions but may not preserve global arrangement. Such a combination allows the system to analyze and synthesize complex patterns online during user interactions.
}%
Our algorithm uses a graph representation for both topology synthesis and geometric path reconstruction for general continuous patterns, in contrast to the graph representations in \cite{Hsu:2018:BEF,Hsu:2020:AEF} that only apply to discrete elements.
Since continuous patterns can exhibit higher complexity than discrete elements, we propose robust, hierarchical synthesis\nothing{ and online learning (of synthesis parameters)} \cite{Wei:2000:FTS,Wei:2001:TSO} to enhance output quality. \nothing{ and computation speed.}%
\nothing{%
Analogous to multi-resolution image texture synthesis \cite{Wei:2000:FTS,Wei:2001:TSO}, our hierarchical synthesis proceeds from sparser samples and larger neighborhoods, which serve as initialization for the next steps, towards finer samples and smaller neighborhoods.
}%
\nothing{
Since hierarchical patterns might be more sensitive to synthesis parameters, such as number of hierarchies, neighborhood sizes, and neighborhood samples, which can be challenging to pick and tune manually for different patterns, we incorporate an online parameter learning method that automatically tunes the desired synthesis parameters based on interactive user feedbacks (e.g., accept or modification).
The combination of new synthesis algorithms and online learning allows us to create more complex patterns that better match user intentions.
}%

\nothing{

Autocomplete systems have been presented \cite{Xing:2014:APR,Xing:2015:AHA,Peng:2018:A3S,Suzuki:2017:ATP} which allow the user to accept, modify, or ignore predictions provided by the systems. These systems are more natural to use as it does not require the users to provide an exemplar beforehand.
But they are limited to simple patterns that are either highly stochastic or regular. It's desired to extend the scope of existing pattern design systems.

Generating more complex patterns requires more complex algorithms with more parameters. However, different patterns can require very different parameters to produce desired results. 
These parameters can be difficult to predict a priori automatically. Existing systems \cite{Xing:2014:APR,Peng:2018:A3S} targets simple patterns in which case a single set of parameters might suffice. These systems do not learn from interactive user responses, which reflects user preferences towards predictions.

We formulate the online parameter learning problem as a {\em derivative-free optimization problem} with respect an underlying (unknown) user preference function. 
Based on past experience about texture synthesis \cite{Wei:2009:SAE}, we heuristically decompose the optimization problem involving multiple parameters into several one-dimensional problem where the objective is unimodal. 
By analyzing user feedbacks on predictions, we can compare and evaluate the user preference towards different parameter settings, and find the optimal parameters using simple Golden-section search \cite{Kiefer:1953:SMS}.
}%

\nothing{
On the synthesis side, the element pattern grammar introduced in \cite{Loi:2017:PAE} is programming-based and complex.
It may not be directly adopted into a sketch-based pattern design system. We propose a pattern grammar that is suitable to our system.
Our grammar first procedurally generates structured or unstructured auxiliary guides that specify the global arrangement of elements \cite{Loi:2017:PAE}.
The example-based synthesis algorithm is employed to generate larger patterns with these guides \cite{Ma:2011:DET} and thus maintain global element arrangements.

On the analysis side, our method automatically predicts the pattern grammar from user drawings via inverse procedural modeling. 

}%

Automatically generated outputs, although convenient, might not \add{have sufficient quality or} fit what users have in mind for their particular applications.
To facilitate \replace{individual}{further} \add{editing and} customization, we also propose an interactive autocomplete authoring interface \cite{Xing:2014:APR,Hsu:2020:AEF} built upon our synthesis algorithm components.
Similar to existing \nothing{pattern }design tools, users can create various free-style patterns. 
\nothing{
Meanwhile, our system analyzes what they have done and predict what they should do next and thus reduce input workload.
}%
When \delete{they feel that }they have \delete{created }sufficient exemplar\delete{ pattern}s and would like to reduce further manual repetitions, they can specify an output domain to be automatically filled\nothing{ by our system} \cite{Kazi:2012:VIT,Xing:2014:APR}.
The synthesized patterns resemble and seamlessly connect with what has already been drawn.
If not satisfied, users can accept or modify the predictions, or ask for re-synthesis to maintain full control.
They can further designate specific source regions for cloning to target regions.

We analyze our algorithm via ablation studies, compare it with alternative methods, and demonstrate the quality and accessibility of our system via pattern design results\nothing{ and user feedback}.
We plan to share our code along with the publication of this paper to facilitate reproduction.

In sum, the contributions of this work are:
\begin{itemize}

\item A hierarchical \nothing{graph }representation and a synthesis method for both geometry and topology of continuous and discrete patterns.

\nothing{
\item An optimization process that can produce continuous curve patterns with intricate structures\nothing{ in high quality and fast speed}.
}%

\item An interactive authoring system \nothing{for continuous and discrete patterns }with autocomplete functions to reduce manual workloads and facilitate user control.

\nothing{
\item An online parameter learning method for autocomplete functions such as hint and clone.
}%
\end{itemize}

\section{Related Work}
\label{sec:prior}

Our work is inspired by prior art in vector patterns, image textures, and interactive workflows. 
Procedural methods \cite{Pedersen:2006:OLM,Loi:2017:PAE,Santoni:2016:GGP} can produce intricate structures, but are limited in scope and difficult to generalize for different types of patterns.
Example-based methods are general, but existing work predominantly focuses on image textures \cite{Wei:2009:SAE,Lu:2014:DDS,Gatys:2016:IST} rather than vector patterns.
Below, we survey methods most related to our work.

\vspace{-0.8em}

\subsection{Example-based Pattern Generation}
\label{subsec:examplepattern}

Example-based methods \replace{have been very successful}{are designed} to generate large patterns from small exemplars with an optional control provided by the users \cite{Ijiri:2008:EBP,Ma:2011:DET,Ma:2013:DET,Hurtut:2009:ASE,Barla:2006:SPA,\nothing{Kazi:2012:VIT,}Landes:2013:SAM,Roveri:2015:EBR,Hsu:2018:BEF,\nothing{Xing:2014:APR,Xing:2015:AHA,Peng:2018:A3S,Hsu:2020:AEF,}Tu:2019:PPS,Bhat:2004:GTS,Zhou:2006:MQG,Zhou:2007:TSD}.
However, these methods target discrete elements or samples \cite{Ijiri:2008:EBP,Ma:2011:DET,Hurtut:2009:ASE,\nothing{Kazi:2012:VIT,}Landes:2013:SAM,Hsu:2018:BEF,Hsu:2020:AEF,Tu:2019:PPS} and treat continuous structures as special cases via curve/surface reconstruction from point samples \cite{Ma:2011:DET,Roveri:2015:EBR}.
\delete{Thus, they might not preserve the pattern structures.
For example, broken elements can be seen in Figure 5b of \cite{Zhou:2006:MQG}.
}%
\nothing{
}%
\add{
Roveri et al. \shortcite{Roveri:2015:EBR} reconstructs the output surface from synthesized point samples via their associated surface normals without considering sample connections\nothing{ (graphs)}, and thus can only be applied to surfaces relatively smooth to the underlying sampling density\delete{ without complex topology}.
Tu et al.~\shortcite{Tu:2019:PPS} extends neural point synthesis by treating a graph edge as a line of points, which is essentially point synthesis.
The neural optimization method does not provide the \add{same} flexibility and efficiency as in our method and it is unstable when the points contain attributes beyond positions.
\nothing{
Our method generates higher-quality graphs than \cite{Tu:2019:PPS} does.
}%
}%
\nothing{
}%
Relatively few works focus on curves, such as enriching details of given coarse curves \cite{Hertzmann:2002:CA} or growing L-system-like curves \cite{Merrell:2010:ECS}.
Our system is inspired by these prior sample-based algorithms \cite{Ma:2011:DET,Roveri:2015:EBR,Hsu:2020:AEF}\nothing{ and autocomplete interactions \cite{Xing:2014:APR}}, but we explicitly incorporate both samples and their connectivity into our representation and optimization to synthesize more general continuous structures, as shown in \Cref{fig:comparison_sample_synthesis}.

\nothing{
}%
\nothing{These methods require the users to prepare the exemplars beforehand and thus lack natural user interactions. 
To address this issue,
some systems predict what users might want to draw based on analyzing prior workflows using example-based approaches \cite{Xing:2014:APR,Xing:2015:AHA,Peng:2018:A3S}. The users can either accept or ignore the suggestions and thus keep their natural interactions.
}%

\nothing{On the other hand, however, since the example-based  approaches either assume commonly used Markov-Random-Field \cite{Ijiri:2008:EBP,Ma:2011:DET,Roveri:2015:EBR,Barla:2006:SPA} or heuristically defined statistical pattern model \cite{Landes:2013:SAM,Hurtut:2009:ASE}, the synthesis quality of these approaches might be significantly compromised when the exemplar is highly structured or non-stationary.

However, previous methods synthesize patterns on a single resolution, which limits the scope of these methods. 
We propose hierarchical element representation and synthesis method. 
Our method can extend the synthesis scope and generate diverse, complex patterns.
}%

\nothing{
Ma et al \shortcite{Ma:2011:DET} and Roveri et al \shortcite{Roveri:2015:EBR} are two main previous works that target optimization-based pattern synthesis which are derived from \cite{Kwatra:2005:TOE}. These two algorithms maximize the texture/pattern similarity defined on neighborhoods by iterating two steps: search for the most similar input neighorhood for each output neighborhood and assign the information from the former to the latter. 
Ma et al \shortcite{Ma:2011:DET} adopt sample-based representation which is limited to synthesize discrete patterns, as discussed in \cite{Roveri:2015:EBR},
while Roveri et al \shortcite{Roveri:2015:EBR} seperate structure representation and sample matching to enforce all neighborhoods are matched equally well. To synthesize discrete patterns where elements are anisotropic and have to be encoded as multiple samples, however, it is necessary to consider the structure representation during the matching \cite{Ma:2011:DET}. Thus, Roveri et al 's method is only able to synthesize continuous and simple, single-sample isotropic element but most interesting discrete patterns.

Our system can generate both discrete and continuous patterns and their mixtures by applying graph-based representation to continuous structures. Note Hsu et al \cite{Hsu:2018:BEF} has applied graph to represent an element to better encode its topology, but not extended to continuous structures as we do. To achieve our goal, we propose a greedy graph matching algorithm which is effective and efficient for interactive pattern authoring. 
}

\nothing{
Our system can generate both highly structured and non-stationary patterns by combing the benefits of controlling global element arrangements via grammars and local element interactions by examples.
}%

\delete{
\subsection{Continuous Structure Synthesis}

\nothing{
}%

In addition to curves \cite{Hertzmann:2002:CA,Merrell:2010:ECS}, exemplar-based synthesis has also been applied to geometry such as surface patterns and terrains \cite{Bhat:2004:GTS,Zhou:2006:MQG,Zhou:2007:TSD}.
However, these methods might not be able to handle discrete elements, e.g., broken elements in Figure 5b of \cite{Zhou:2006:MQG}.
The method in \cite{Roveri:2015:EBR} handles continuous structures via surface/curve reconstruction of optimized point samples without explicit considering connections among sample.
Thus, the method can only deal with \nothing{isolated, isotropic }discrete elements\nothing{ that can be represented as a single sample,} and simple \nothing{3D }watertight continuous surfaces without complex topologies\nothing{, or their mixtures}.
Our method can preserve both complex discrete elements and continuous structures, and can be deployed for interactive authoring of diverse 2D patterns. 
\nothing{
}%
}%

\nothing{

\subsection{Stylized Fabrication}

Generating patterns has also been studied in stylized fabrication but with focuses on pattern structures and fabricability \cite{Chen:2017:FTD,Chen:2016:SFD,Dumas:2015:BSS,Martinez:2015:SAO,Zhou:2014:TSV,Li:2019:AQP}.
These methods strive to provide fully automatic solutions and consume significant amount of time to generate large patterns and mainly allow indirect control over the pattern appearance by proving examples. In particular, Zehnder et al \shortcite{Zehnder:2016:DSO} presented an interface to provide interactive user control but mainly for structural reasons.
The interface in \cite{Bian:2018:TPD} allows the users to author topology-constrained patterns by drawing a source tile set, and the output pattern is constructed by tiling \cite{ Cohen:2003:WTI}. The tiling process prevents more direct control from the users.
On the contrary, our interface provides direct and full control over the pattern generation process during authoring. 
Our synthesized continuous patterns can be naturally connected and thus suitable to fabricate.
\nothing{
}%
}%

\nothing{
	However, these methods are restrictive by the quality of the synthesis algorithm. For instance, \cite{Ijiri:2008:EBP} cannot be adapted to strongly anisotropic elements owing to that it is based on a centroidal element model. 
	Vignette might be one of the most related works to ours \cite{Kazi:2012:VIT}. It is an interactive texture design system for pen-and-ink illustration which adopts texture synthesis method in \cite{Ma:2011:DET}. Vignette lets users define an exemplar and then use gesture to control texture growth and thus its spatial arrangement. The system would \textit{auto-complete} the textures. However, it does not consider complex element interaction such as overlap, contact and stroke sharing which are very common in element textures. Also, Vignette does not explicitly consider the variations of elements within textures. The users have to define a large enough exemplar to encode element variations. Our approach address both defects.
	Auto-completion technique has been adapted to paintings \cite{Kazi:2012:VIT,Xing:2014:APR}, animations \cite{Xing:2015:AHA}, and 3D texture design \cite{Suzuki:2017:ATP}. However, their scope of resulting textures is rather limited, as they do not combine the benefits of procedural and example-based method, while our system relies on procedural method to synthesize the global arrangement.
	Another related work is WorldBrush \cite{Emilien:2015:WIE} which is an interactive system for synthesizing virtual worlds. The synthesized world is represented by element textures. For example, tree and river object is represented by point and graph respectively. This system learns point distributions or graph interaction parameters from examples and synthesize virtual worlds using learned distributions \cite{Oztireli:2012:ASP}. However, as mentioned in \cite{Ma:2011:DET}, single-sample element representation is not sufficient to capture element shape and other complicated element interactions such as overlap and contact. 
}%

\nothing{
\subsection{Inverse Procedural Modeling}
Procedural methods, which are not limited to pattern generation, still present challenges for providing users good control. They require users to manipulate complex and abstract procedural rules and numerous parameters.
Inverse procedural modeling can address some of the challenges by discovering procedural representation from a given input model or its abstract specifications (semi-)automatically. 
Talton et al. \shortcite{Talton:2011:MPM,Talton:2012:LDP} proposed general but slow statistical solutions for learning the parameters of existing procedural models by Metropolis procedural modeling and Bayesian networks.
More specific inverse methods have been proposed for proceduralizing trees \cite{Stava:2014:IPM}, vector graphics \cite{Stava:2010:IPM}, facade layouts \cite{Wu:2014:IPM}, cities \cite{Vanegas:2012:IDU} and textures \cite{Liu:2004:NRT,Hays:2006:DTR}.

However, inverse procedural modeling of repetitive element patterns with complex arrangement has received little attention. Our method is designed to learn element arrangement and deduce pattern grammar by analyzing workflow during user interactions.
}%

\nothing{
\subsection{Preference Learning}
\label{subsec:preferencelearning}
Many design tasks require artists to tweak numerous parameters or rules. The purpose of such parameter tweaking is to produce a design that maximizes the artists' preference.
Learning a preference function thus becomes very important for assisting the artists in a tedious design process. Prior methods have been proposed for learning user preference functions in various tasks, including material modeling \cite{Zsolnai:2018:GMS}, shape design \cite{Dang:2015:IDP,Talton:2009:EMC}, image enhancement \cite{Shapira:2009:IAE,Koyama:2017:SLS} and animation design \cite{Brochu:2010:BIO}. These preference functions are either learned via crowd sourcing \cite{Koyama:2014:CPA,Koyama:2017:SLS,Talton:2009:EMC} or interactive user scoring \cite{Zsolnai:2018:GMS,Dang:2015:IDP,Shapira:2009:IAE,Koyama:2016:SPL,Brochu:2010:BIO}.

In pattern synthesis, we can readily assume the preference function is unimodal. Our method optimizes the unimodal function via a derivative-free optimization algorithm with only function value comparisons \cite{Kiefer:1953:SMS}.
We integrate machine learning into our system with guidelines presented in \cite{Amershi:2019:GHI}.

}

\subsection{Interactive Authoring}

Workflow analysis has been investigated to assist various content creation task \cite{Nancel:2014:CCM}. 
Examples includes static and animated sketches \cite{Xing:2014:APR,Xing:2015:AHA}, 3D sculpting \cite{Peng:2018:A3S,Peng:2020:AAS}, texture design \cite{Suzuki:2017:ATP}, hand-writing beautification \cite{Zitnick:2013:HBU}, and image editing \cite{Chen:2011:NRC,Koyama:2016:SPL}.
Our work is inspired by prior autocomplete \cite{Xing:2014:APR,Xing:2015:AHA,Peng:2018:A3S,Suzuki:2017:ATP} and interactive systems \cite{Kazi:2012:VIT,Bian:2018:TPD,Hsu:2020:AEF}.
However, these systems can automate only relatively simple patterns (e.g., repetitive hatches \add{or strokes})\nothing{ and require significant user manual inputs for more complex patterns}.
Our method can automatically generate diverse and complex continuous curve patterns to facilitate iterative design with reduced input workload. \nothing{and enhanced output quality. 
}%

\nothing{
\subsection{Learning Materials}

Another totally different direction is to estimate Spatially Varying Bidirectional Reflectance Distribution Function (SVBRDF) from the 2D texture samples \cite{Blender:2018:BFO,PixPlant:2018:PPC,Allegorithmic:2018:ABM}. By manipulating estimated SVBRDF , 3D rendering system can generate new 2D textures with different appearance, but the flexibility and editing operations are very limited. This type of manipulation involving 3D rendering system is more relevant to image-based material modeling \cite{Dong:2011:AIM} instead of texture design.

Recently, a learning-based system is proposed for improving the efficiency for material design \cite{Zsolnai:2018:GMS}. This system is composed by three components: user preferences modeling, parameters dimensionality reduction and real-time material visualization. However, this system can only deal with design process with limited hyper-parameters and where the user preference function is relatively smooth with respect to these parameters. For procedural texture design, the texture appearance is decided by hundreds or even up to thousands of parameters and might be altered drastically with a little change on those parameters.

\subsection{Constrained procedural modeling and interactive design exploration}
\cite{Talton:2011:MPM}
\cite{Huang:2017:SSS}
\cite{Yeh:2012:SOW}
\cite{Stava:2014:IPM}
\cite{Ritchie:2016:NPM}

\cite{Yumer:2015:PMU}
\cite{Dang:2015:IDP}
\cite{Zsolnai:2018:GMS}
\cite{Bao:2013:GEG}
\cite{Guerrero:2016:PEP}
}%

\section{User Interface}
\label{sec:interface}

Our system can be used for both automatic synthesis and interactive editing.
Similar to prior work like traditional texture synthesis \cite{Wei:2009:SAE}, the user provides an exemplar pattern and lets our system automatically produce the output with desired size and shape.
The exemplar is represented via B\'{e}zier curves. Inputs in other formats can be converted to B\'{e}zier curves (for example, by vectorizing a raster image).
\nothing{
}%

\begin{figure}[t]
	\centering
	\includegraphics[width=0.98\linewidth]{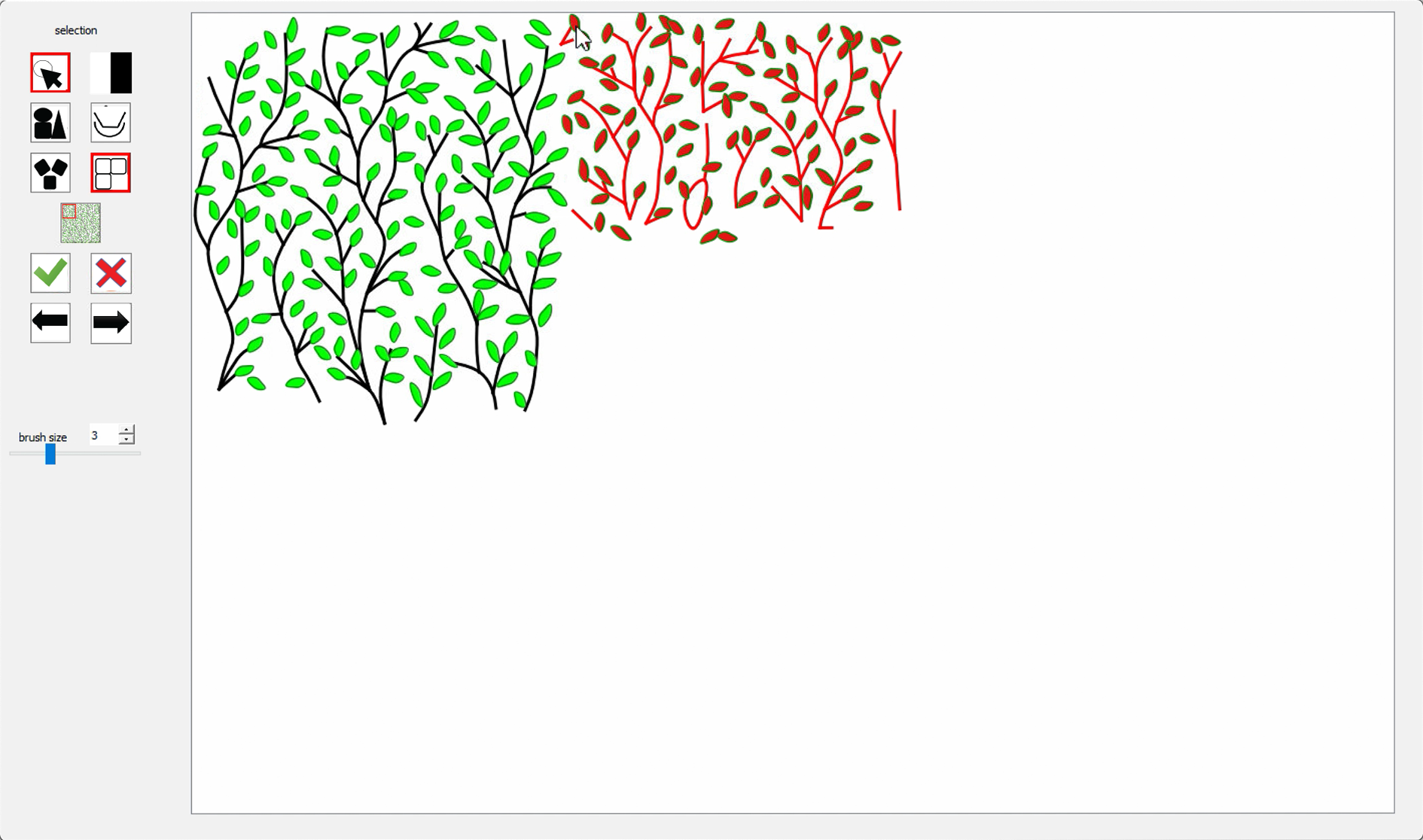}

	\Caption{User Interface.}
	{%
		Our interface has a widget panel and a canvas.
		The widget panel provides basic tools, such as selection in a vector editor, as well as controls to the graphics parameters, including color and pen width, and modes unique to our autocomplete system.
		
		\nothing{
}%
}%
	\label{fig:ui}
\end{figure}

\begin{figure}[t]
	\centering
	\subfloat[Autocomplete: before]{
	\label{subfig:autocomplete:before}
	\includegraphics[width=0.46\linewidth]{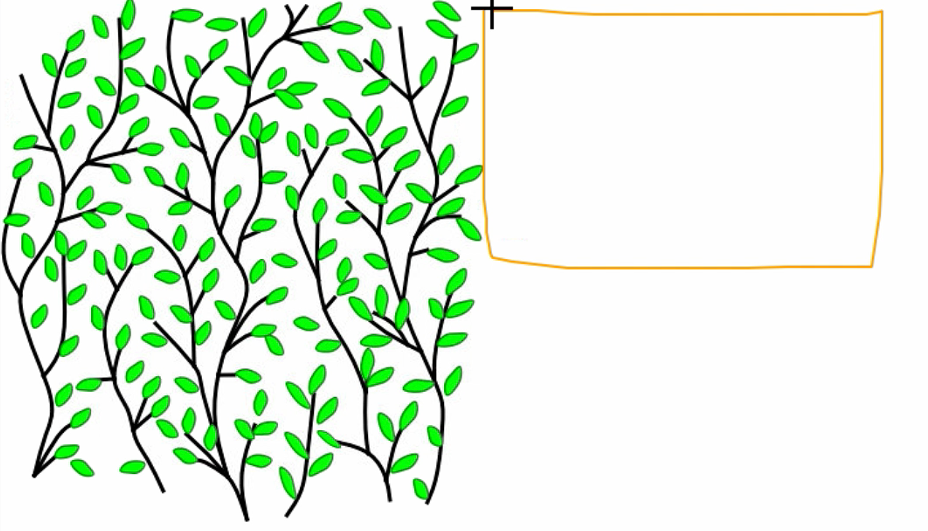}
	}
	\subfloat[Autocomplete: after]{
	\label{subfig:autocomplete:after}
	\includegraphics[width=0.46\linewidth]{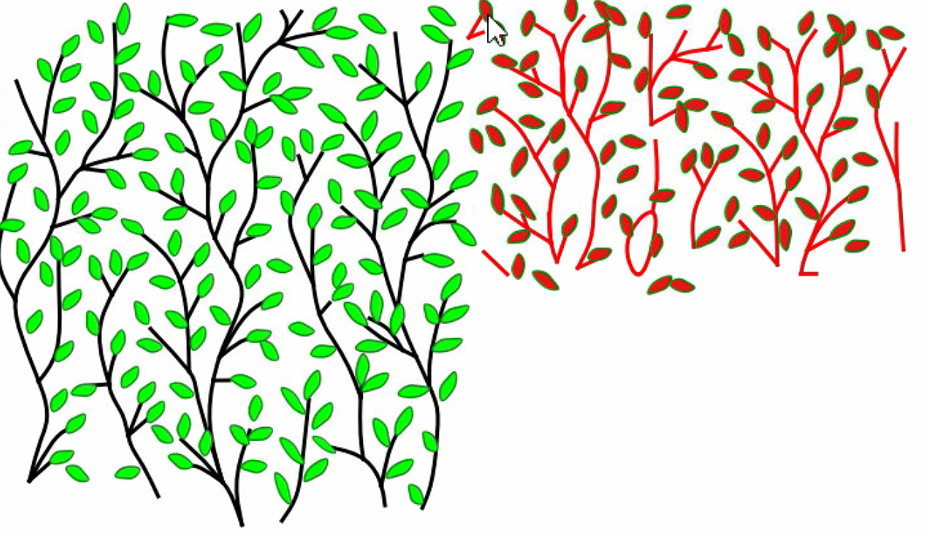}
	}

	\subfloat[Clone: before]{
	\label{subfig:workflow_clone:before}
	\includegraphics[width=0.48\linewidth]{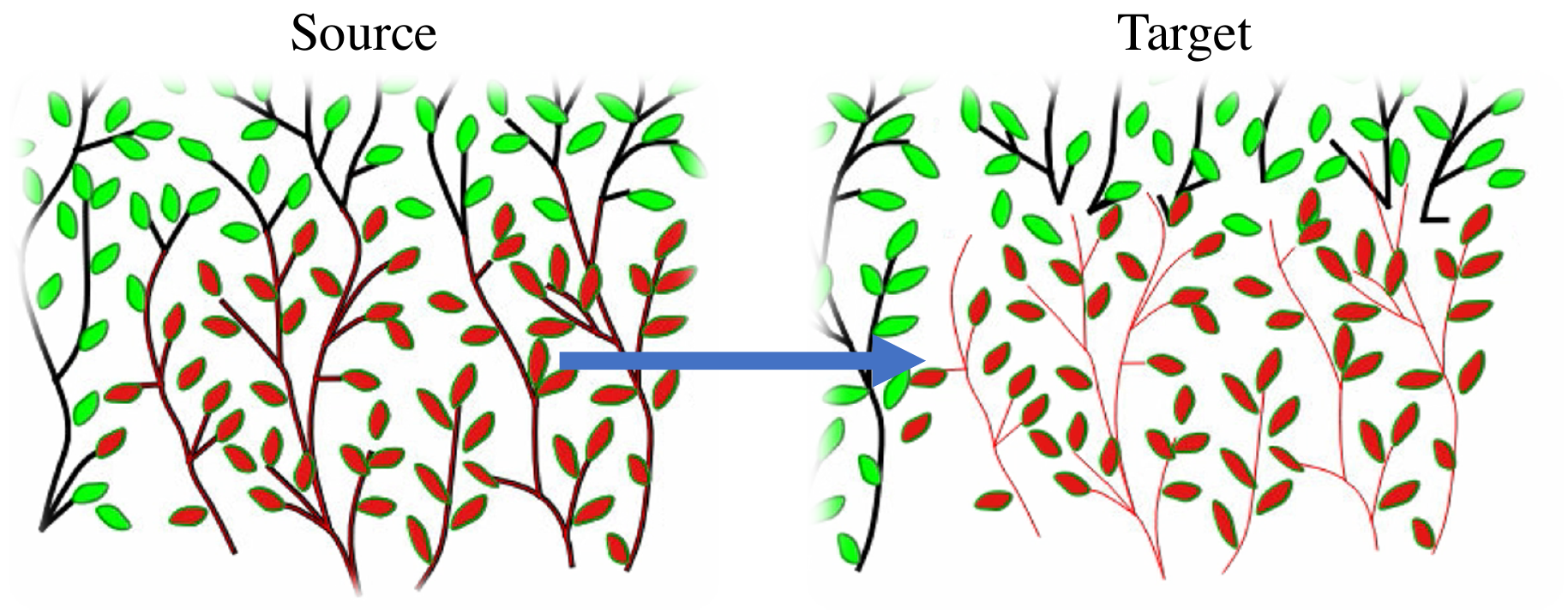}
	}
	\subfloat[Clone: after]{
	\label{subfig:workflow_clone:synthesized}
	\includegraphics[width=0.22\linewidth]{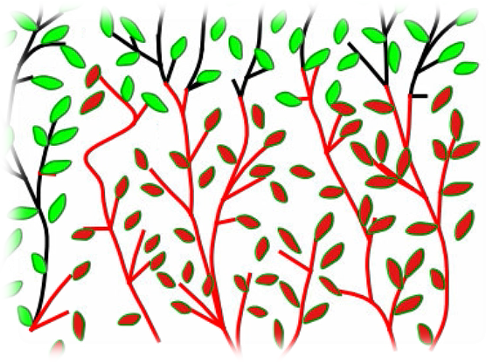}
	}
	\subfloat[User edit]{
	\label{subfig:user_editing}
	\includegraphics[width=0.22\linewidth]{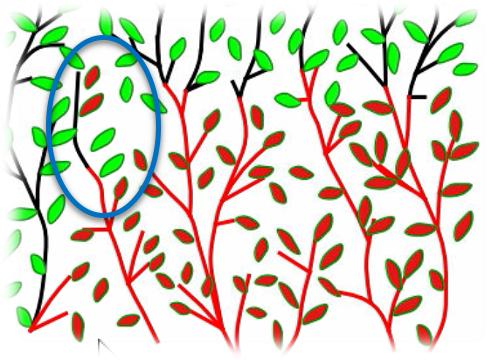}
}
	
	\Caption{Autocomplete and clone.}
	{%
	  In the autocomplete mode, the user can specify an output region (shown in yellow) \subref{subfig:autocomplete:before} and let our system generate predicted patterns \subref{subfig:autocomplete:after}.
	  In the \nothing{workflow }clone mode, the user can specify a source region (in red) and clone it to a target region \subref{subfig:workflow_clone:before}.
	  Our system can generate predictions adaptive to the existing patterns \subref{subfig:workflow_clone:synthesized}, upon which users can perform further refinements \subref{subfig:user_editing}. 
	  \subref{subfig:user_editing} is generated by editing the top left corner (in blue) of the predictions, including 1) partially rejecting several paths, 2) copy-pasting two elements, and 3) adding a path.
}
	\label{fig:ui_autocomplete}
\end{figure}

Since the automatic synthesis results might not be\delete{~exactly} what the users want and they might need to create new patterns manually, we also provide an interface built upon our automatic synthesis algorithms for users to author patterns interactively\nothing{ for further quality improvement and customization}.
Through the interface, users can specify an output domain in desired size and shape (\Cref{subfig:autocomplete:before}) and let our system predict patterns that resemble what the users have already drawn ({\em autocomplete} mode, \Cref{subfig:autocomplete:before,subfig:autocomplete:after}).
The users can also explicitly control the prediction by copying-pasting from an input region to an output region ({\em clone} mode, \Cref{subfig:workflow_clone:before,subfig:workflow_clone:synthesized}).
\nothing{%
The users can perform further edits on the predictions (\Cref{subfig:user_editing}).
}%
They can accept, partially accept, or reject the predictions via keyboard shortcuts and mouse selections.
\nothing{
We also provide the users shortcuts to accept or reject all predictions at the same time.
}%
They can also perform further edits\nothing{as visualized in the black path in \Cref{subfig:user_editing}}, such as selecting regions for re-synthesis or adding paths in the predictions.
Please refer to the supplementary video for live actions.
Since continuous patterns often contain complex structures beyond fine-grained autocomplete of individual strokes \cite{Xing:2014:APR}, we design our current interface to focus on autocomplete pattern regions instead of strokes.

\nothing{
}%

\nothing{

If there are multiple predictions, the users can choose which one is better and perform further edits on it, and our system refine the synthesis parameters in the background.
}%

\nothing{

}%

 %
\section{Method}
\label{sec:method}

\newcommand{\unifydiscretecontinuous}{1} %

Our method extends the sample-based element texture synthesis method in \cite{Ma:2011:DET} to consider not only individual point samples but also their curved connections via graphs \cite{Hsu:2018:BEF,Hsu:2020:AEF}.
\delete{
We also generalize our representation and synthesis for continuous patterns beyond discrete elements in these prior methods (\Cref{subsec:representation}\nothing{, we will introduce sample-based \cite{Ma:2011:DET} and our novel graph-based pattern representations for continuous curve pattern synthesis.}).
}%
We describe our pattern \new{representation in \Cref{subsec:representation}}, similarity measures in \Cref{subsec:similarity_measure}, and the corresponding synthesis and reconstruction algorithms in \Cref{subsec:sample_synthesis,subsec:pattern_reconstruction}.
\ifdefined\unifydiscretecontinuous
Our method can handle discrete elements, continuous structures, and their combinations.
We will describe when and how our algorithms treat them similarly or differently.
\else
\fi

\nothing{
In \Cref{subsec:sample_synthesis,subsec:pattern_reconstruction}, we detail how to synthesize samples and finally reconstruct patterns from samples.
}%

\subsection{Representation}
\label{subsec:representation}

\begin{figure}[t]
	\centering
	\subfloat[Single-res\nothing{olution element representation}]{
		\label{fig:single_resolution_repres}
		\includegraphics[width=0.22\linewidth]{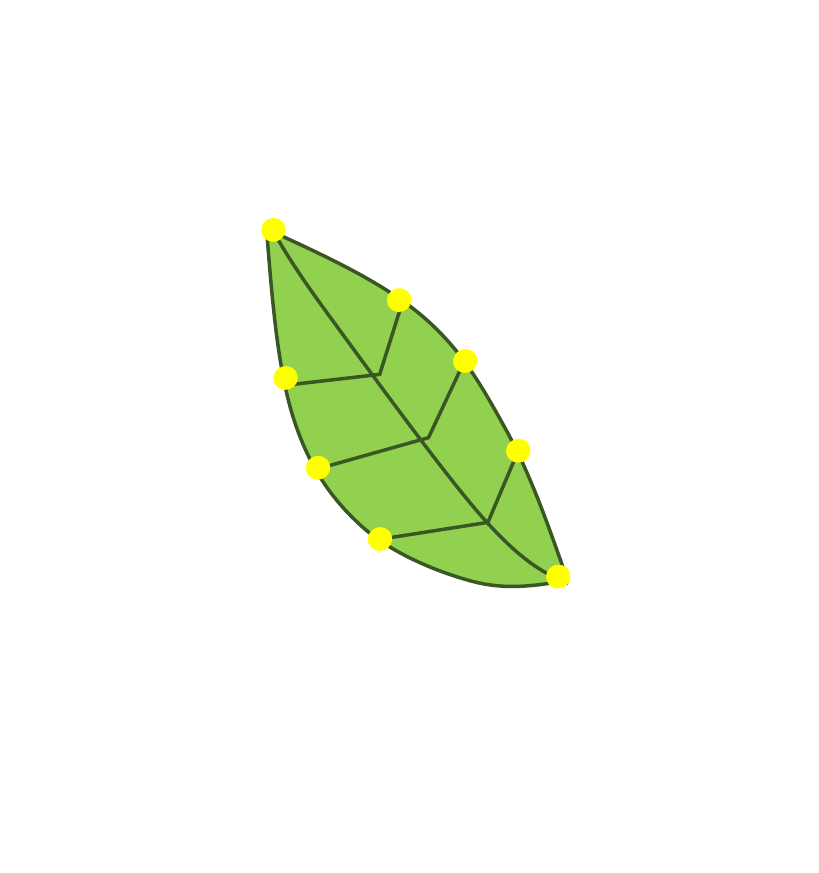}
	}%
	\subfloat[Multi-res\nothing{olutioinhierarchical element representation}]{
	\label{fig:hierarchical_repres}
		\includegraphics[width=0.22\linewidth]{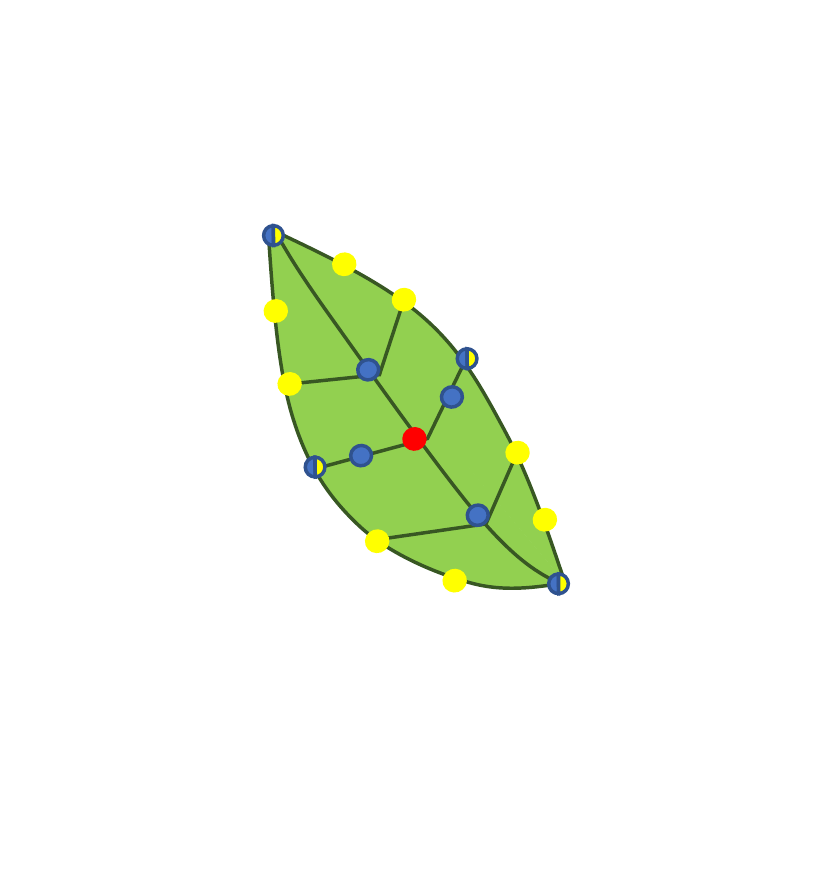}
	}%
	\subfloat[\nothing{continuous }Pattern]{
	\label{fig:continuous_pattern}
	\includegraphics[width=0.22\linewidth]{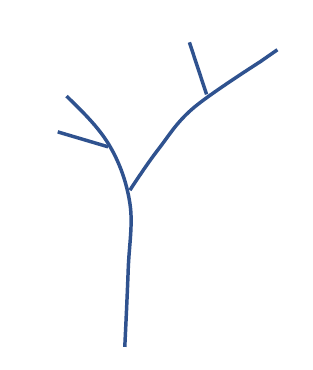}
}%
\subfloat[Graph\nothing{ representation}]{
	\label{fig:graph_repres}
	\includegraphics[width=0.22\linewidth]{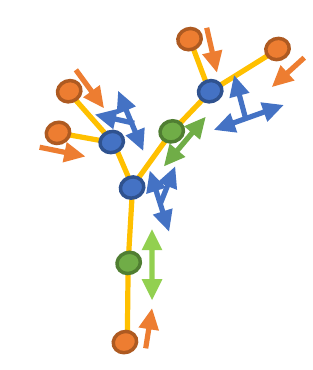}
}%

	\Caption{Pattern representation.}
	{%
		Existing algorithms \cite{Ma:2011:DET} only use a single-resolution element representation \subref{fig:single_resolution_repres}. 
		We propose a hierarchical element representation to improve the synthesis quality \subref{fig:hierarchical_repres}.
		Our synthesis proceeds from red, blue, to yellow samples in coarse to fine levels.
                \nothing{
                }%
	        A pattern \subref{fig:continuous_pattern} is represented by a graph \subref{fig:graph_repres}, where we record connections (yellow edges) and local path orientations (arrows) in addition to point samples. 
		The color of a sample indicates the number of connections $|\sampleedgeset(\samplesym)|$ associated with it; orange, green, and blue indicate 1, 2, and 3.
                \nothing{

                }%
		\nothing{
		We apply  \subref{fig:sample_representation} sample and \subref{fig:graph_representation} graph-based representations for discrete and continuous patterns, respectively. Samples are not sufficient to encode necessary information for synthesizing continuous patterns, as shown in \subref{fig:sample_matching} and \subref{fig:graph_matching}. In \subref{fig:graph_representation}, red, green and blue nodes indicate joint, end and path nodes, respectively.
	}%
\nothing{
}%
	}
		
	\label{fig:pattern_representation}
	\label{fig:method_continuous_matching}
\end{figure}

We represent patterns and elements via point samples and graphs (\Cref{fig:pattern_representation}) \cite{Ma:2011:DET,Roveri:2015:EBR,Hsu:2018:BEF}.
\nothing{
The synthesis requires our pattern representation to be simple and thus synthesizable but compact without losing necessary information for reconstruction.
}%
Each sample $\samplesym$ records its \nothing{global }position $\samplespace(\samplesym)$\replace{~and}{,} attributes $\sampleattributes(\samplesym)$\nothing{ that may vary in terms of types of patterns} (\Cref{tab:possible_sample_attr})\replace{. 
To optimize the number of samples, we also include}{, and} $\sampleexistence(\samplesym) \in [0, 1]$ to indicate the confidence of \replace{sample}{its} existence\new{~to optimize the number of samples}\replace{.
Formally, each sample is represented as
}{:}
\begin{align}
\samplevec(\samplesym)=\left(\samplespace(\samplesym),\sampleattributes(\samplesym), \sampleexistence(\samplesym)  \right).
\label{eqn:sample_representation}
\end{align}

\begin{table}[tbh]
	\begin{center}
	\Caption{Sample attributes.}
        {Each sample $\samplesym$ records its attributes $\sampleattributes(\samplesym)$ that may vary in terms of types of patterns (discrete element or continuous structure). 
		$\sampleid(\samplesym) \geq 0$ indicates uniqueness of a sample \new{relative} to other samples within a discrete element%
\ifdefined\unifydiscretecontinuous
, and $\sampleid(\samplesym) = -1$ for continuous structures%
\else
\fi
.
\ifdefined\unifydiscretecontinuous
\else
		In continuous structures,
\fi
$\sampleedgeset(\samplesym)= \{\sampleedgesym{\samplesym}{\samplesymprime}\}$ records all edges associated with $\samplesym$, 
		where $\sampleedgesym{\samplesym}{\samplesymprime}$ is an edge between $\samplesym$ and $\samplesymprime$.
		$\sampleorientations(\samplesym)$ records the local orientations of the paths intersecting at sample $\samplesym$.
        }%
		\label{tab:possible_sample_attr}
		\begin{tabular}{|l | c|c c|} %
\ifdefined\unifydiscretecontinuous
\else
			\hline 
			 & Discrete & \multicolumn{2}{|c|}{Continuous} \\ 
\fi
			\hline 
			\textbf{Attributes} & Sample id & Connectivity & Orientation  \\
			\hline
			$\sampleattributes(\samplesym)$ & $\sampleid(\samplesym)$ & $\sampleedgeset(\samplesym) = \{\sampleedgesym{\samplesym}{\samplesymprime}\} $ & $\sampleorientations(\samplesym)$  \\
                        \hline
		\end{tabular}
	\end{center}
\end{table}

\nothing{
}%

\nothing{
}%

\paragraph{Discrete elements}
\label{subsubsec:sample}

Following \cite{Ma:2011:DET,Ma:2013:DET}, for discrete elements, sample attributes include a sample id $\sampleid(\samplesym)$ that indicates the uniqueness of $\samplesym$ to other samples within its containing element.

\nothing{

}%

\nothing{
They are spatial parameters $\samplespace(\samplesym)$ which include position, direction at $\samplesym$ as well as sample id that records its relative position within a discrete element, appearance parameters $\sampleappr(\samplesym)$ that include pressure and size and temporal parameters $\sampletemp(\samplesym)$ that records the global time stamp.

}%

\nothing{
In some cases, we can identify each individual element by examining whether it has closed shape.  Optionally, users can trivially specify only one element and the remaining elements can be automatically and robustly extracted from the workflow, by analyzing the shape similarity defined via calculating the differences between sample positions, id and normals.
}%

\paragraph{Continuous structures}
\label{subsubsec:continuous_structures}
In discrete element synthesis, it is sufficient to use only samples with id $\sampleid(\samplesym)$ (\Cref{fig:single_resolution_repres,fig:hierarchical_repres}) to encode element shape because every element has the same topology \cite{Ma:2011:DET}.
On the other hand, continuous structures are composed of paths and have more flexibilities.
In our paper, we represent paths by linear and quadratic B\'{e}zier curves, even though other vector curve formats can be easily added.
The paths may be connected to each other with complex topologies (\Cref{fig:teaser}).
Thus, samples alone are not sufficient to disambiguate matching and reconstruction of continuous structures.
Therefore, we also consider connectivity among samples, leading to a graph-based representation (\Cref{fig:graph_repres}). 
Note the method in \cite{Hsu:2018:BEF,Hsu:2020:AEF} also adopts a graph-based representation, but it is for discrete elements only \nothing{but it is used for shape deformation }without explicitly modeling paths in a continuous pattern.

Specifically, we record the connectivity $\sampleedgeset(\samplesym)$ for $\samplesym$ within $\sampleattributes(\samplesym)$.
$\sampleedgeset(\samplesym)=\{\sampleedgesym{\samplesym}{\samplesymprime}\}$ is the set of edges associated with $\samplesym$,
where $\sampleedgesym{\samplesym} {\samplesymprime}$ represents the edge between the two samples $\samplesym$ and $\samplesymprime$.
\nothing{
}%
We use $\sampleexistence(\sampleedge) \in [0\ 1]$ to indicate the confidence of an edge existence;
$\sampleexistence(\sampleedgesym{\samplesym}{\samplesymprime})=1/0$ indicates the presence/absence of an edge between $\samplesym$ and $\samplesymprime$.
We will relax the binary $\sampleexistence(\sampleedgesym{\samplesym}{\samplesymprime})$ to be within the range $[0\ 1]$ during optimization-based synthesis.
\nothing{
}%
While $\sampleedgeset(\samplesym)$ records pattern topology,
we also record the tangent angles\nothing{ in opposite directions of local tangent lines} at $\samplesym$ on a path via an orientation attribute $\sampleorientations \in \mathcal{R}^{\sampleconnections}$ as part of $\sampleattributes(\samplesym)$\nothing{~the other sample attribute $\sampleattributes(\samplesym)=\left(\sampleedgeset(\samplesym), \sampleorientations (\samplesym)\right)$}, where $\sampleconnections$ is the number of entries in $\sampleorientations (\samplesym)$.
\nothing{
Intuitively, $\sampleedgeset(\samplesym)$ and $\sampleorientations (\samplesym)$ record the topology and geometry information of edges.
}%
\nothing{
and $\{\samplesymprime\}$ is the set of all other samples except $\samplesym$.
}%
\nothing{
}%
Each entry $\sampleorientationentry$ of $\sampleorientations$ is within $[0, 2\pi)$.
\nothing{%
Note the angle of tangent line might not equal to that of graph edge orientation. There is small numerical difference between them.
}%
We record \replace{tangent angles}{$\sampleorientations$} to facilitate pattern reconstruction from graphs (\Cref{subsubsec:recon_continuous_structures,fig:ablation:orientation}). 
Note that, for any input samples $\sampleinput$, we always have $| \sampleedgeset(\sampleinput) | = \sampleconnections(\sampleinput)$ (\Cref{fig:graph_repres}), where $|\sampleedgeset|$ is the size of the edge set $\sampleedgeset$.
However, this strict constraint is relaxed for the output during the synthesis to facilitate faster convergence.

\subsubsection{Hierarchical Pattern Sampling}
\label{sec:representation:hierarchical}

We adopt a multi-resolution representation of sample graphs to handle patterns with \delete{potential~}complex structures, analogous to prior multi-resolution algorithms for color texture synthesis \cite{Wei:2000:FTS,Wei:2001:TSO}.
The representation is sparser with less samples at coarser resolutions and becomes denser with more samples at finer resolutions.
By default, we use three level of hierarchies, which suffice in our experiments.
Users can decide to use fewer levels if needed.

\paragraph{Discrete elements}
We generate element samples using a simple approach (\Cref{fig:single_resolution_repres,fig:hierarchical_repres}). 
\nothing{

}%
The finest\nothing{ (third)} level of samples are generated by sampling the element polygon. 
The coarest level contains only one sample centered at each element.
The middle level of samples are located at the midpoints of each downsampled finest-level samples and the coarest-level element centers.
\nothing{
When necessary, element can also be sampled manually \cite{Hsu:2020:AEF}  or automatically with a more advanced approach \cite{Wei:2020:SMR}.
}%

\nothing{
}

\paragraph{Continuous structures}
For continuous patterns (\Cref{fig:continuous_pattern}), 
we sample the intersections (blue samples in \Cref{fig:graph_repres}) and ends of paths (orange samples) and uniformly place samples (green samples) along paths with spacing $\samplingdistance$.   
\new{We discuss the parameter $\samplingdistance$ values in \Cref{sec:result}.}

\nothing{
After synthesis, the output patterns can be reconstructed from the synthesized samples by utilizing the recorded sample attributes and connections (graphs).
}%

\nothing{
}%

\nothing{
}%

\nothing{
\subsubsection{Connectivity}
\label{subsubsec:connectivity}
To better handle both discrete elements and continuous patterns, 
}%

\nothing{ (\Cref{subsubsec:recon_continuous_structures}) as well as local matchings between input and output.}

\nothing{
\nothing{%
To address this issue, we instead apply an approximated graph representation without recording the exact sample connectivity.

\begin{equation}
\samplevec(\graphsamplesym)=\big(\samplespace(\samplesym),\sampleappr(\samplesym), \sampletemp(\samplesym),\sampleedge(\samplesym)\big)
\end{equation}

}%

}%

\nothing{
}%

\nothing{
For continuous structures, there exists long and short strokes. We segment or combine strokes into unit strokes, and synthesize the continuous structures as discrete element patterns in which each unit stroke is an element.
}%

\subsection{Similarity Measure}
\label{subsec:similarity}
\label{subsec:similarity_measure}

\nothing{
We follow the approach presented in \cite{Ma:2011:DET} to synthesize patterns by maximizing similarity between patterns.
}%

A core part of pattern synthesis is a measure of similarity between local regions \cite{Wei:2009:SAE,Ma:2011:DET,Roveri:2015:EBR}.
Here, we describe our similarity measure for continuous patterns via their sample-graph representation (\Cref{subsec:representation}), which, in turn, will form the basis for our synthesis optimization (\Cref{sec:synthesis}).
\nothing{
We formulate the pattern synthesis as an optimization problem of maximizing similarity between patterns.
}%

\subsubsection{Sample Similarity}
\label{subsubsec:sample_similarity}

\nothing{
}%

The difference between two samples  $\samplesym$ and $\samplesymprime$, 
which includes the differences in the global position $\samplespace$ and attributes $\sampleattributes$, is defined as follows:
\ifdefined\unifydiscretecontinuous
\begin{align}
&\differencesym{\samplespace}(\samplesym,\samplesymprime)=\samplespace(\samplesym)-\samplespace(\samplesymprime),
\\
&\differencesym{\sampleattributes}(\samplesym,\samplesymprime) =
\left(\differencesym{\sampleid}(\samplesym,\samplesymprime), \differencesym{\sampleedgeset}(\samplesym,\samplesymprime)\right)
\label{eq:sample_similiarity}
\end{align}
\else
\begin{align}
&\differencesym{\samplespace}(\samplesym,\samplesymprime)=\samplespace(\samplesym)-\samplespace(\samplesymprime),
\\
&\differencesym{\sampleattributes}(\samplesym,\samplesymprime) =
\begin{cases}
\differencesym{\sampleid}(\samplesym,\samplesymprime) & \text{if discrete element}
\\
\differencesym{\sampleedgeset }(\samplesym,\samplesymprime) & \text{if continuous pattern}
\end{cases}
.
\label{eq:sample_similiarity}
\end{align}
\fi
\nothing{
}%
The differences in sample id attribute $\sampleid$ is
\begin{align}
\differencesym{\sampleid}(\samplesym,\samplesymprime)=\mathbbm{1}\left\{ \sampleid(\samplesym)  \not= \sampleid(\samplesymprime)\right\},
\end{align}
where $\mathbbm{1}(\bigcdot)$ is an indicator function that equals to one if its condition $\bigcdot$ holds, and zero otherwise.
\nothing{
}%
The edge set difference is 
\begin{align}
\differencesym{\sampleedgeset}(\samplesym,\samplesymprime) =   \left(\sum_{
\substack{
\sampleedgesym{\samplesym}{\samplesymhat}\in \sampleedgeset(\samplesym)
\nothing{\sampleedgesym{\samplesymprime}{\samplesymhatprime} \in \sampleedgeset(\samplesymprime)}
\nothing{
\\ \sampleedgesym{\samplesymprime}{\samplesymhatprime} = \matchedge(\sampleedgesym{\samplesym}{\samplesymhat})  \nothing{\in \sampleedgeset(\samplesymprime)}
}%
}
}^{\nothing{b}} \distancesym\left(\sampleedgesym{\samplesym}{\samplesymhat}, \sampleedgesym{ \samplesymprime}{\samplesymhatprime}\right) \right)  + \weightconnections \left| |\sampleedgeset(\samplesym)| - |\sampleedgeset(\samplesymprime)|\right|,
\label{eq:sample_edge_similarity}
\end{align}
where $\distancesym\left(\sampleedgesym{\samplesym}{\samplesymhat}, \sampleedgesym{ \samplesymprime}{\samplesymhatprime}\right)= \|\differencesym{\samplespace}(\samplesym,\samplesymhat) - \differencesym{\samplespace}(\samplesymprime,\samplesymhatprime) \|$ is the difference between $\sampleedgesym{\samplesym}{\samplesymhat}$ and $\sampleedgesym{\samplesymprime}{\samplesymhatprime}$,
and $\sampleedgesym{\samplesymprime}{\samplesymhatprime} = \matchedge(\sampleedgesym{\samplesym}{\samplesymhat}) \in \sampleedgeset(\samplesymprime)$ is the matching edge for $\sampleedgesym{\samplesym}{\samplesymhat}$ via the Hungarian algorithm \add{(which solves the one-to-one matching relationship between edges)} \cite{Kuhn:1955:HMA} to minimize the first term in \Cref{eq:sample_edge_similarity} ($\match$ indicates matching relationship).
\nothing{
}%
$\weightconnections$ is a weighting parameter set to sampling distance $\samplingdistance$ (\Cref{sec:representation:hierarchical}) in our experiments.

For newly added output samples that do not have any edges, either by initialization or existence assignment (\Cref{sec:synthesis}), we want them to be useful and connected to existing output samples.
To this end, they should be encouraged (\replace{have}{via} low\add{er} cost) to match with input samples during the search step (\Cref{subsubsec:search_step}).
We set $\weightconnections = 0$ for these samples and thus \Cref{eq:sample_edge_similarity} becomes 0, as the first term is also 0 since newly created samples do not have edges. 
\nothing{

}%

\nothing{
}%
We do not include $\sampleorientations$ within the attribute similarity term (\Cref{eq:sample_similiarity}) since \replace{$\sampleedge$}{$\sampleedgeset$} already contain similar information in $\sampleorientations$.
However, we still need to update $\sampleorientations$ during synthesis (assignment step, \Cref{subsubsec:assignment_step}) and reconstruction (\Cref{subsec:pattern_reconstruction}).
This requires us to match orientation entries $\sampleorientationentry$ within $\sampleorientations$ from $\samplesym$ and $\samplesymprime$ respectively. 
The matching $\sampleorientationentry(\samplesymprime) = \matchorient\left(\sampleorientationentry(\samplesym)\right) \in \sampleorientations(\samplesymprime)$ is computed via the Hungarian algorithm \cite{Kuhn:1955:HMA} by minimizing the sum of smallest absolute differences between matched $\sampleorientationentry(\samplesym)$ and $\sampleorientationentry(\samplesymprime)$
\nothing{
}%
\begin{align}
 \differencesym{\sampleorientations}(\samplesym,\samplesymprime) =
\sum_{
 	\substack{
\sampleorientationentry(\samplesym)\in \sampleorientations(\samplesym)\nothing{\\
\matchorient\left(\sampleorientationentry(\samplesym)\right)\in \sampleorientations(\samplesymprime)}
}
  }^{} \differencesym{\sampleorientationentry}(\samplesym,\samplesymprime),
 \label{eq:sample_orien_similarity}
\end{align}
where $\differencesym{\sampleorientationentry}(\samplesym,\samplesymprime)=\min\left( |\sampleorientationentry(\samplesym)- \matchorient\left( \sampleorientationentry(\samplesym)\right)|,2\pi-|\sampleorientationentry(\samplesym)-\match_o\left(\sampleorientationentry(\samplesym)\right)|\right)$.
\nothing{

}%
\nothing{
}%

\nothing{
The second term in \Cref{eq:sample_orien_similarity} computes the differences between number of local sample connections $\sampleconnections$, weighted by $\weightconnections$.
(Note $\differencesym{\sampleorientations}(\samplesym,\samplesymprime)$ is a number not a vector.)
}%

\nothing{
Note we do not directly take connectivity $\sampleedge(\samplesym,\samplesymprime)$ into account in the sample similarity for three reasons. 
First, $\sampleedge(\samplesym,\samplesymprime)$ involves a pair of samples, which requires to solve a NP-hard quadratic assignment problem to match samples from two neighborhoods.
Second, pattern paths are uniformly sampled so that the length of $\sampleedge(\samplesym,\samplesymprime)$ are approximately equal.
The variation among $\sampleedge(\samplesym,\samplesymprime)$ is their orientations that can be approximated by $\sampleorientations$.
Third, we find our approximation produce synthesis results of good quality.
}%

We also have not found it necessary to include $\sampleexistence$ in the sample similarity measure (\Cref{eq:sample_similiarity}). 
Instead, $\sampleexistence$ will be used to optimize the number of samples.
\nothing{and
$\sampleedge$ will be used during the reconstruction step.
}%
\nothing{
}%

\nothing{
For $\sampleorientations \in \mathcal{R}^{\sampleconnections}$, we use the minimum assignment cost computed by Hungarian algorithm~\cite{Kuhn:1955:HMA} (thus $\differencesym{\sampleorientations}(\samplesym,\samplesymprime)$ is a number not an vector). To compute the assignment, the cost between entries $\sampleorientationentry(\samplesym)$ and $\sampleorientationentry(\samplesymprime)$  is the small angle between two angles, namely .
}%

\nothing{
\subsubsection{Connectivity similarity}

In addition to per sample positions $\sampleposition$ and attributes $\sampleattributes$, we also explicitly compute similarity in sample connectivity.
Specifically, from the recorded $\sampleedge(\samplesym, \samplesymprime)$ between two samples $\samplesym$ and $\samplesymprime$, we can compute their edge length:
\begin{align}
\edgelength(\samplesym, \samplesymprime) =
\sampleedge(\samplesym, \samplesymprime)
|\differencesym{\samplespace}(\samplesym,\samplesymprime)|
\end{align}
Note that $\edgelength$ might not be the actual length of the curve connecting the two samples, but suffice to serve as a proxy of their connectivity difference.
}%

\nothing{
}%

\subsubsection{Neighborhood Similarity}
We define $\sampleneigh(\samplesym)$, the neighborhood of $\samplesym$, as a set of samples around $\samplesym$'s spatial vicinity within a certain radius $\neighborhoodradius$.
The neighborhood similarity is defined as
\nothing{
\begin{align}
\distance{\neighoutput-\neighinput}= 
\sum_{\substack{\sampleoutputprime \in \matchneigh \left(\neighinput \right)}}
\distance{\differencesym{\samplevec}(\sampleoutput,\sampleoutputprime)-\differencesym{\samplevec}(\sampleinput,\sampleinputprime)}^2 
+
\sum_{\substack{\sampleoutputprime \in \neighoutput \setsub  \matchneigh}}^{} \unmatchedcost(\sampleoutputprime)   
\label{eq:robust_neighborhood_similarity}
\end{align}
}%
\begin{align}
\distance{\neighoutput-\neighinput}= 
\sum_{\substack{\sampleoutputprime \in \matchneigh \left(\neighinput \right)}}
\samplediff  +
\sum_{\substack{\sampleoutputprime \in \neighoutput \setsub  \matchneigh}}^{} \unmatchedcost(\sampleoutputprime),   
\label{eq:robust_neighborhood_similarity}
\end{align}
where
\begin{equation}
\samplediff = \|\differencesym{\sampleposition}(\sampleoutput,\sampleoutputprime)-\differencesym{\sampleposition}(\sampleinput,\sampleinputprime)\|
+ \weightattributes
\|\differencesym{\sampleattributes}(\sampleoutputprime, \sampleinputprime) \|
\end{equation}
is the sample similarity between $\sampleoutputprime$ and $\sampleinputprime$ within neighborhoods centered at $\sampleoutput$ and $\sampleinput$ respectively.
\nothing{
}%
$\sampleinputprime = \matchsample(\sampleoutputprime)$ is the matching input sample for $\sampleoutputprime$.
We discuss how to match samples within $\neighoutput$ and $\neighinput$ in \replace{the next section}{\Cref{sec:neighborhood_matching:robust}}.
The positional differences $\|\differencesym{\sampleposition}(\sampleoutput,\sampleoutputprime)-\differencesym{\sampleposition}(\sampleinput,\sampleinputprime)\|$ are computed in local neighborhood coordinate systems centered at $\sampleoutput$ or $\sampleinput$.
\new{The two terms in \Cref{eq:robust_neighborhood_similarity} partition $\neighoutput$ into two sets.}
\add{In the first term of \Cref{eq:robust_neighborhood_similarity},}
$\matchneigh \left(\neighinput\right) $ is the subset of $\neighoutput$ matched with samples \nothing{$\sampleinputprime$ }within $\neighinput$.
\add{In the second term of \Cref{eq:robust_neighborhood_similarity}}, $\unmatchedcost(\sampleoutputprime)$ is the cost resulting from unmatched output samples $\sampleoutputprime$.
In our implementation, $\weightattributes=0.5$.
\Cref{eq:robust_neighborhood_similarity} is designed for our robust neighborhood matching, described next.

\nothing{
}%

\subsubsection{Robust Neighborhood Matching}
\label{sec:neighborhood_matching:robust}

\begin{figure*}[tbh!p]
	\centering
	\subfloat[Hard neighborhood matching]{
		\label{fig:hard_neighborhood_matching}
		\includegraphics[width=0.32\linewidth]{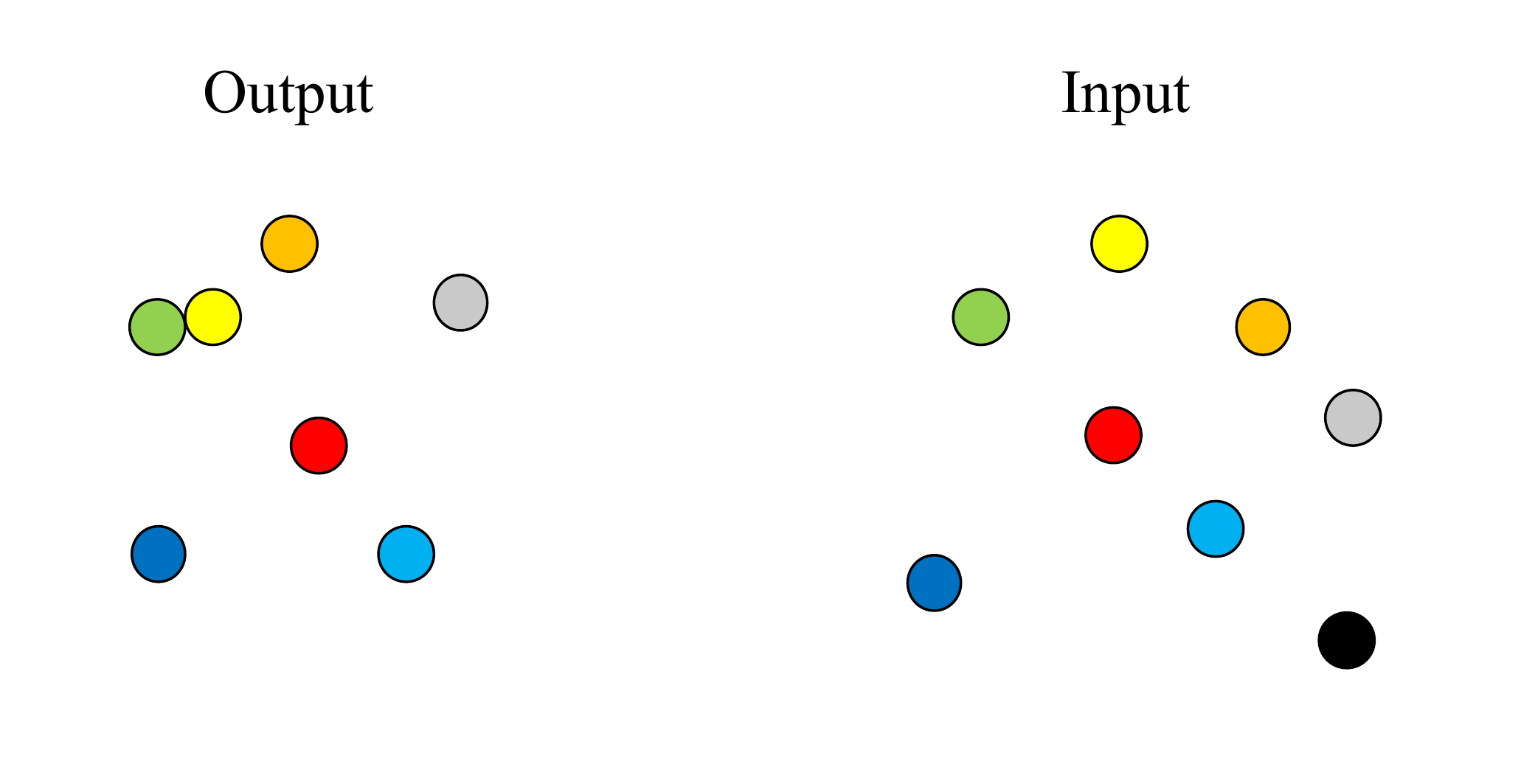}
	}%
	\subfloat[Soft neighborhood matching]{
		\label{fig:soft_neighborhood_matching}
		\includegraphics[width=0.32\linewidth]{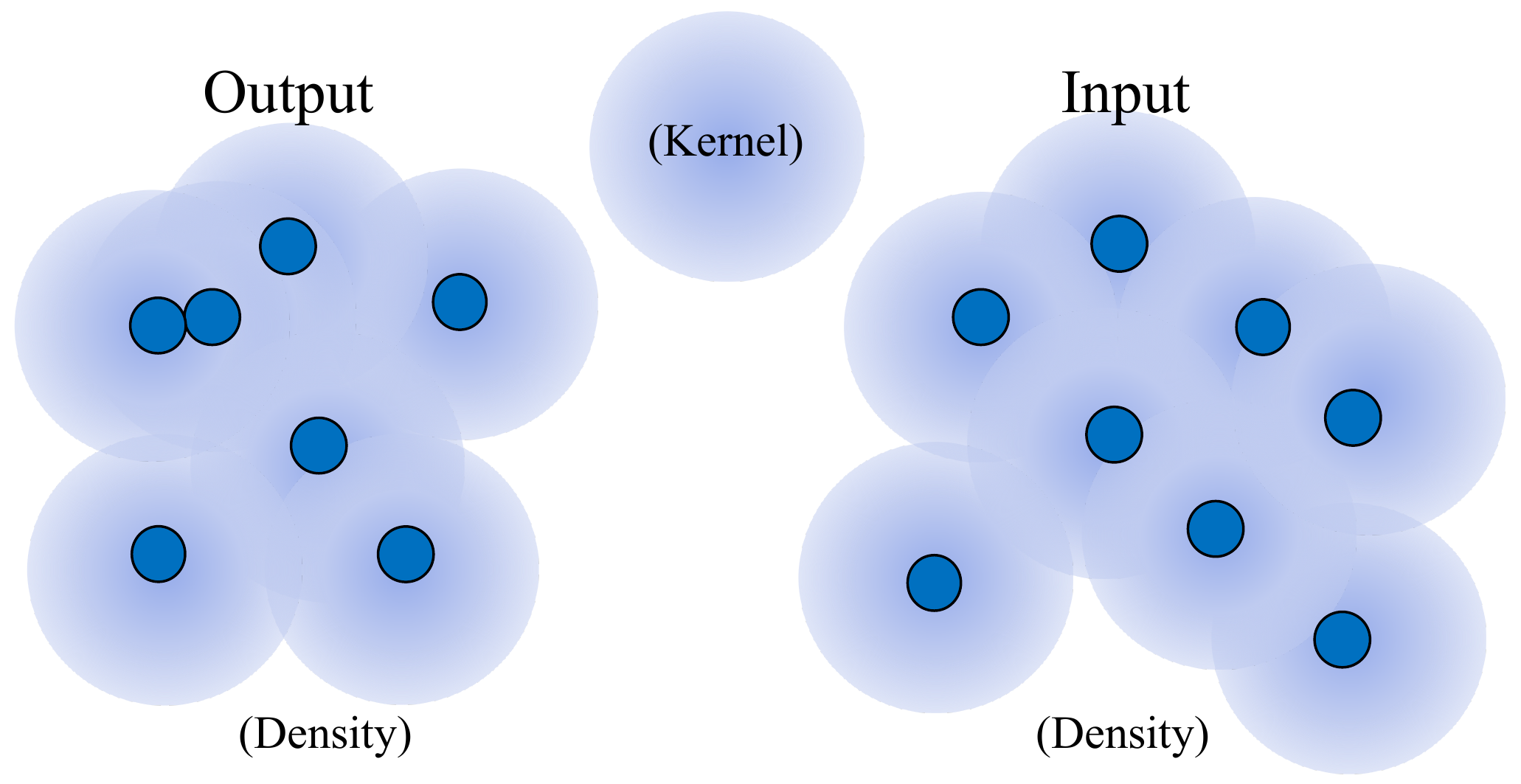}
	}%
	\subfloat[Robust neighborhood matching]{
	\label{fig:robust_neighborhood_matching}
	\includegraphics[width=0.32\linewidth]{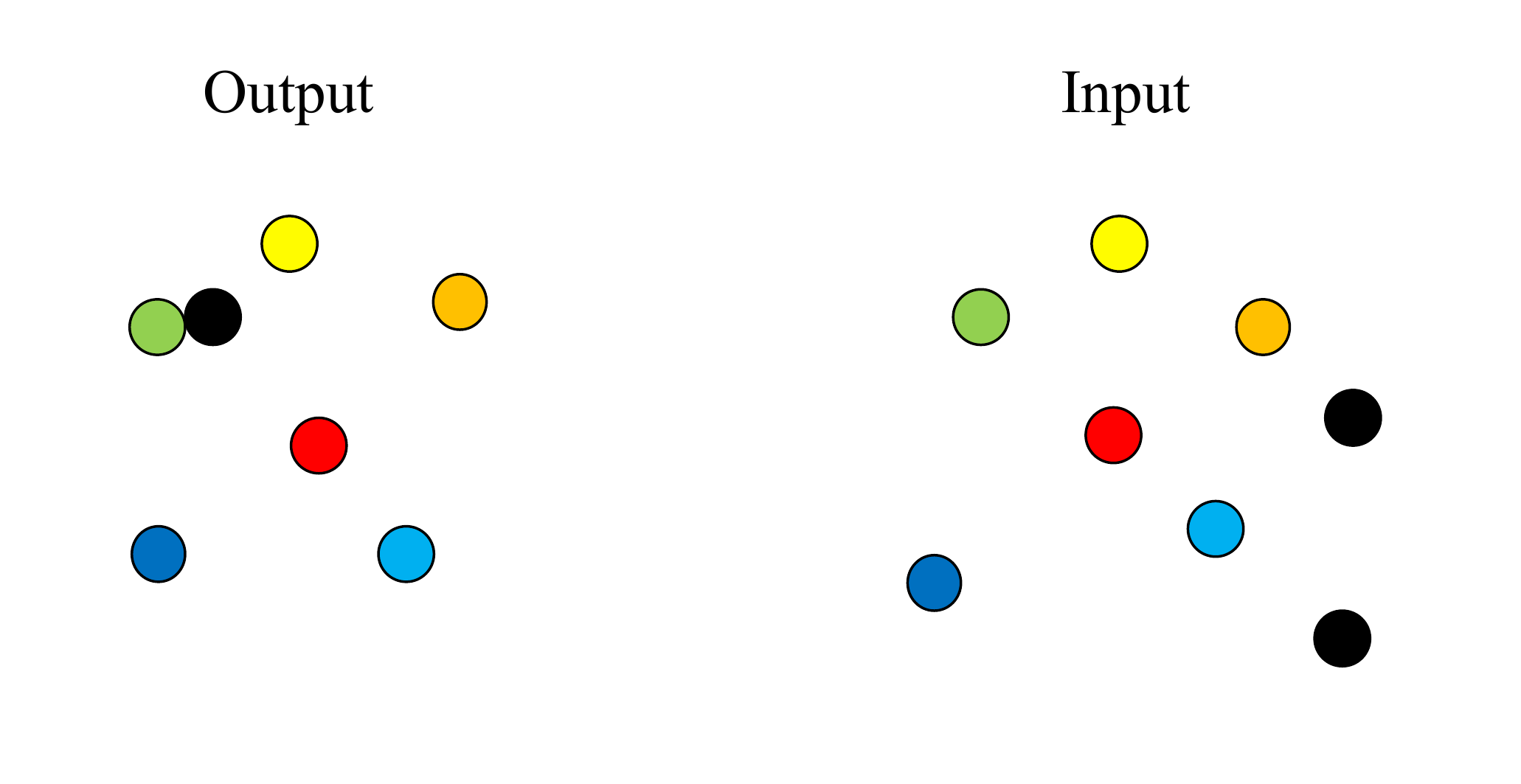}
	}%

	\Caption{Different neighborhood matching methods.}
	{%
		\nothing{

}%
		In \subref{fig:hard_neighborhood_matching} each sample in the output (left) is forced to match with another one in the input (right), which could be problematic since some output samples are outliers and should not be matched to any input samples, and some output samples might be missing in the current iteration of optimization.
		Matched samples are with the same color. Black indicates unmatched.
		In \subref{fig:soft_neighborhood_matching} samples are not explicitly matched but the neighborhood is matched as a whole, by transforming the samples with density kernels \cite{Roveri:2015:EBR}. 
		There is no one-to-one correspondence between samples, and thus it is not easy to integrate with various sample attributes. 
		Instead, we propose robust neighborhood matching \subref{fig:robust_neighborhood_matching}  to generate high-quality sample distributions to accommodate various types of patterns. We allow output samples to be unmatched if it can result in high matching cost increase.
		In the above example, the yellow output sample in \subref{fig:hard_neighborhood_matching} destroys the hard neighborhood matching and forces other (orange, and gray) samples to be less matched, while our robust matching leaves the bad sample unmatched.
		\nothing{
		Each output sample is probabilistically matched with input samples within a neighborhood and vice versa. Here, the yellow output sample is probabilistically matched with input samples colored with yellows of different lightness. A light color indicates high probability.}%
		\nothing{
	}%
	}
	\label{fig:neighborhood_matching}
\end{figure*}

In \cite{Ma:2011:DET}, each output sample is forced to match with another sample in the input, which could be problematic since some output samples are outliers and should not be matched to any input samples.
Some output samples might be missing in the current iteration of optimization.
But this forced matching allows to easily define sample similarity for various sample attributes, as shown in \Cref{subsubsec:sample_similarity}, since we have one-to-one sample correspondence.
We call this {\em hard neighborhood matching}
(\Cref{fig:hard_neighborhood_matching}).
In \cite{Roveri:2015:EBR}, the neighborhoods are matched via comparing their density fields estimated with Gaussian kernels. 
This similarity criterion is computed with the neighborhoods as a whole. 
{\em There is no one-to-one correspondence between samples}. 
We call it {\em soft neighborhood matching} (\Cref{fig:soft_neighborhood_matching}).
The method \cite{Roveri:2015:EBR} "smears the sample attributes into their neighborhood" by encoding them as the height of the density kernel, which could unnecessarily couple the position\delete{al} and attribute information.
\nothing{

}%
It is not easy to integrate \add{soft matching} with various sample attributes, which can include edges. 
\nothing{
This assumes the attributes must be continuous, summable and spatially coherent, as kernels perform density estimation weighted by attributes which unnecessarily couples the positional and attribute information.
This method could not achieve high accuracy and robustness in terms of attributed sample synthesis.
}%
\nothing{
}%

Instead, we propose to use a robust neighborhood matching that explicitly considers outliers in the output to address these issues (\Cref{fig:robust_neighborhood_matching}).
An output sample is either matched with an input sample or unmatched as an outlier with additional cost $\unmatchedcost$. 
\replace{Still, w}{W}e apply the Hungarian algorithm to compute the matchings between input $\neighinput$ and output neighborhoods $\neighoutput$. 
The input of the Hungarian algorithm is a cost matrix where each entry indicates the matching cost between an output and an input sample. 
\nothing{
In our case, we allow output samples to be unmatched.
}%
Inspired by \cite{Riesen:2009:AGE}, we define our cost matrix $\costmatrix \in \mathbb{R}^{\numoutput \times (\numinput + \numoutput)}$ as:
\begin{flalign}
\costmatrix & =
\left[
\begin{array}{c|c}
\costmatrixenmatched
 & 
\costmatrixunmatched
 \nothing{
  \\
\hline
 \begin{matrix}
\unmatchedcost & \infty & \cdots & \infty \\
\infty & \unmatchedcost  & \cdots & \infty \\
\vdots & \infty & \ddots & \vdots \\
\infty & \infty & \cdots &  \unmatchedcost
\end{matrix}
& 
\begin{matrix}
0 & 0 & \cdots & 0 \\
0 & 0 & \cdots & \vdots \\
0 & 0 & \ddots & 0 \\
0 & 0 & \cdots & 0 
\end{matrix}
}
\end{array}
\right]
\\
\costmatrixenmatched & =
\left[
\arraycolsep=1.4pt\def\arraystretch{1}
\begin{matrix}
\enmatchedcost{1}{1} & 
\enmatchedcost{1}{2} &
\cdots &
\enmatchedcost{1}{\numinput}  \\
\enmatchedcost{2}{1}  & 
\enmatchedcost{2}{2} & 
\cdots &  
\enmatchedcost{2}{\numinput}\\
\vdots & \vdots& \ddots & \vdots\\
\enmatchedcost{\numoutput}{1} & \enmatchedcost{\numoutput}{2} & \cdots & \enmatchedcost{\numoutput}{\numinput}\\
\end{matrix}
\right]
\\
\costmatrixunmatched & =
\left[
 \begin{matrix}
 \unmatchedcost_1 & \unmatchedcost_1 & \cdots & \unmatchedcost_1 \\
\unmatchedcost_2 & \unmatchedcost_2  & \cdots &  \unmatchedcost_2 \\
\vdots & \vdots & \vdots &  \vdots \\
\unmatchedcost_{\numoutput} & \unmatchedcost_{\numoutput} &\cdots  &  \unmatchedcost_{\numoutput}
 \end{matrix}
\right],
\end{flalign}
where the superscripts of $\sampleoutputprime$ or  $\sampleinputprime$ represent the index of a sample within $\neighoutput$ or $\neighinput$.
There are $\numinput$ and $\numoutput$ samples within $\neighinput$ and $\neighoutput$, respectively. 
In \cite{Ma:2013:DET}, the sample matching is computed via only the $\costmatrixenmatched$ part of $\costmatrix$, in which case every output sample should be matched. 
Our cost matrix is augmented with the $\costmatrixunmatched$ side, where each entry represents the cost of unmatched outliers in $\neighoutput$. 
We make sure there are enough samples in the input neighborhood so that an output sample would not be matched only when it is an outlier that would result in a high cost increase in matching.
For the same reason, we do not take missing output samples into account in the cost matrix formulation.
\nothing{
In \Cref{subsubsec:adaptive_sampling}, we will talk about how to adaptively add missing samples back and remove outliers, based on robustly matched neighborhood pairs.
}%
\nothing{
}%

In our implementation, $\unmatchedcost_k$ is set as $\min(2, 1.2 + 0.4|\sampleedgeset(\sampleoutput^k)| )\samplingdistance$ if the output sample $\sampleoutput^k$ is from a continuous pattern,
and $1.5 \times$ average nearest neighbor distance if $\sampleoutput^k$ is from a discrete element.
In an interactive system, we may synthesize predictions near the provided exemplars.
If $\sampleoutput^k$ is from the exemplars, $\unmatchedcost_k = \infty$ because none of the  samples from the exemplars are outliers and all of them should be matched.

In a neighborhood, there might be samples from both discrete elements and continuous structures. 
We only match samples from the same type of patterns and with the same id, i.e. continuous structures only match with continuous structures (negative id), and discrete elements only match the same discrete elements and their samples with the same (non-negative) ids.

\nothing{

\nothing{
Matching samples from continuous structures are constrained by graph edges which is a classic, NP-hard (for global optimal solution) graph matching problem \cite{West:1996:IGT}. We propose a greedy algorithm for deciding the matchings. 
For purposes of algorithm clarification, graph nodes with one, two and more than two neighbors are called \textit{end node}, \textit{path node}, \textit{joint node}, as illustrated in \Cref{fig:graph_representation}.
The basic idea is as follows, as illustrated in 1) We match each pair of nodes in a greedy fashion, the next one or several matchings are constrained by existing matchings and graph edge.  2)  We first match two central nodes $\sampleinput$ and $\sampleoutput$, and then proceed to match joint and end nodes in an order of traversing the graph (which excludes path nodes) by breadth-first search, and finally match path nodes.
For matching neighbors of a pair of matched node, we apply Hungarian algorithm \cite{Kuhn:1955:HMA}. This process is repeated until there is no joint or end node that can be matched in $\neighoutput$. 3) Finally, we only match path nodes from the same graph edge (a graph excluding path nodes).
}

}%

\nothing{
}%

\subsection{Pattern Synthesis}
\label{sec:synthesis}
\label{subsec:sample_synthesis}

Based on our pattern representation (\Cref{subsec:representation}) and similarity measures (\Cref{subsec:similarity}), we now describe how to synthesize an output similar to a given input.
\nothing{
In this step, the algorithm takes an input pattern representation (samples and their edges) and generates a similar output.
}%

\subsubsection{Optimization Objective}
We synthesize output predictions $\outputsym$ via optimizing the following objective:
\begin{align}
\energy(\outputsym)= \sum_{
\substack{
\sampleinput = \match(\sampleoutput),
\sampleoutput \in \outputsym}
}^{}\distance{\neighoutput-\neighinput} + \constraintterm(\outputsym,\domain).
\label{eq:pattern_energy}
\end{align}
where $\sampleoutput$ is matched with $\sampleinput$\delete{ ($\sampleinput = \matchsample(\sampleoutput)$)}.
This energy sums up the similarity between every $\neighoutput$ in $\outputsym$ and its most similar $\neighinput$ via \Cref{eq:robust_neighborhood_similarity}.
$\constraintterm(\outputsym,\domain)$ is the domain constraint term \cite{Dumas:2018:PAE} to encourage the synthesized samples to stay within the user-specified domain $\domain$.

\nothing{
}%

The pattern optimization framework adopts an EM-like strategy to minimize \Cref{eq:pattern_energy}, by iterating the search and assignment steps as detailed below.

\subsubsection{Initialization}
\label{sec:initialization}

Similar to prior patch-based texture synthesis methods \cite{Efros:2001:IQT,Liang:2001:RTS}, we copy new patches one-by-one with similar boundary \add{patterns} to existing patches for initialization.
\replace{
The nearby patches are overlapped and pairs of patches have high similarity as \replace{evaluated}{measured} by \Cref{eq:robust_neighborhood_similarity} in the overlapping regions.
}
{
Each next patch is selected to ensure high similarity (as evaluated by \Cref{eq:robust_neighborhood_similarity}) in the overlapped boundary regions with existing patches.
}%
In the overlapping regions, we only copy samples unmatched with any sample in the existing patches.
We make sure that the initialized samples are within the output domain $\domain$ by removing samples outside it.
We copy discrete elements in wholes like \cite{Ma:2011:DET}.
In \Cref{sec:study}, we will show the robustness of our method to random sample initializations (\Cref{fig:initialization_robustness}). 
But patch-based initialization makes the algorithm converge faster, contributing to the responsiveness of the interface.
For simplicity, we do not copy edges in the initialization step.

\nothing{
}

\begin{figure*}[tbh!p]
	\centering
	\subfloat[Matched neighborhood pair 1]{
		\label{fig:existence_assign:n1}
		\frame{
	\includegraphics[width=0.27\linewidth]{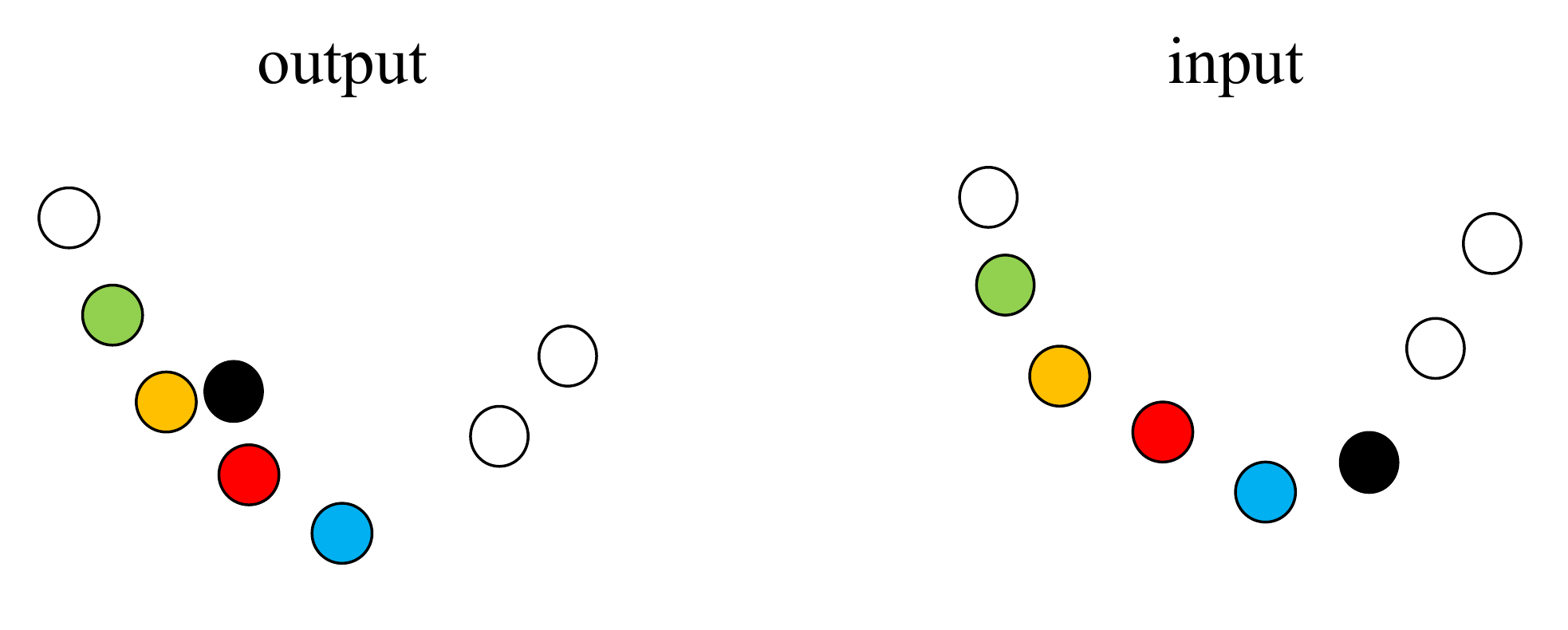}}
}%
	\subfloat[Matched neighborhood pair 2]{
	\label{fig:existence_assign:n2}
			\frame{
		\includegraphics[width=0.27\linewidth]{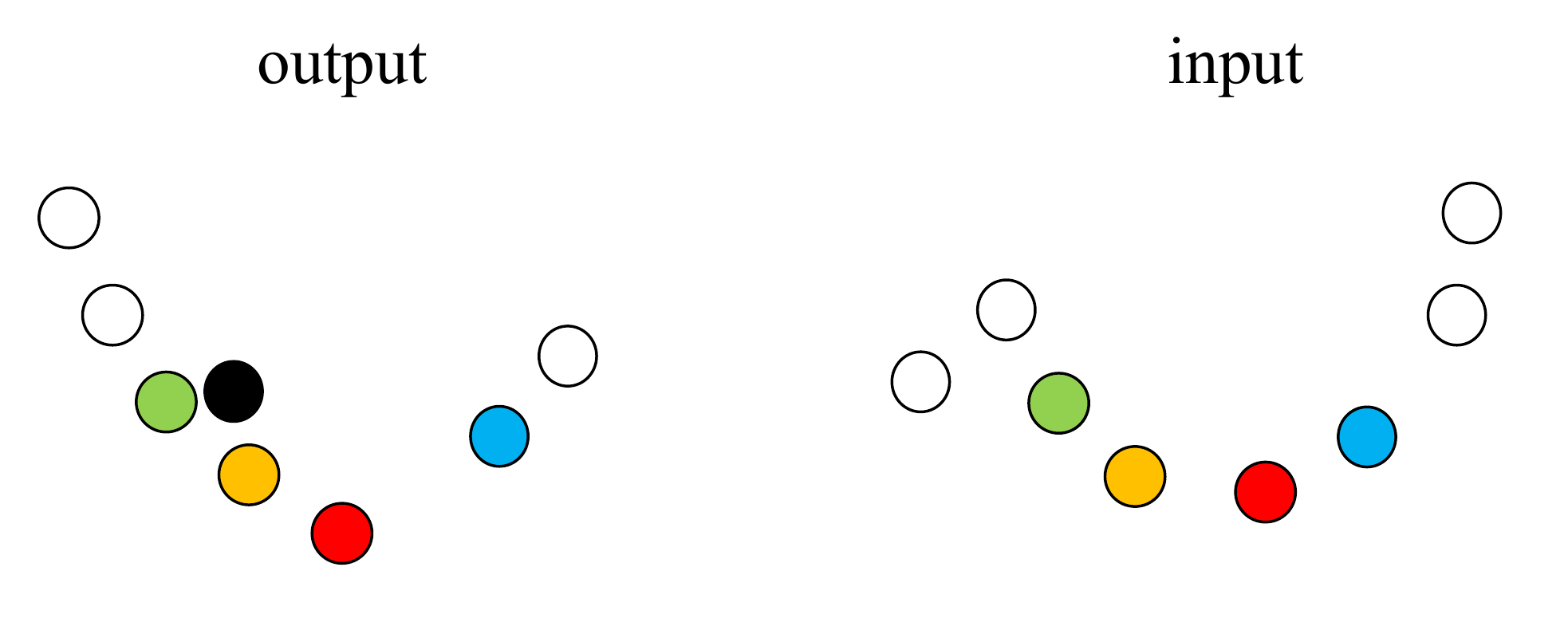}
	}
	}%
	\subfloat[Matched neighborhood pair 3]{
	\label{fig:existence_assign:n3}
			\frame{
		\includegraphics[width=0.27\linewidth]{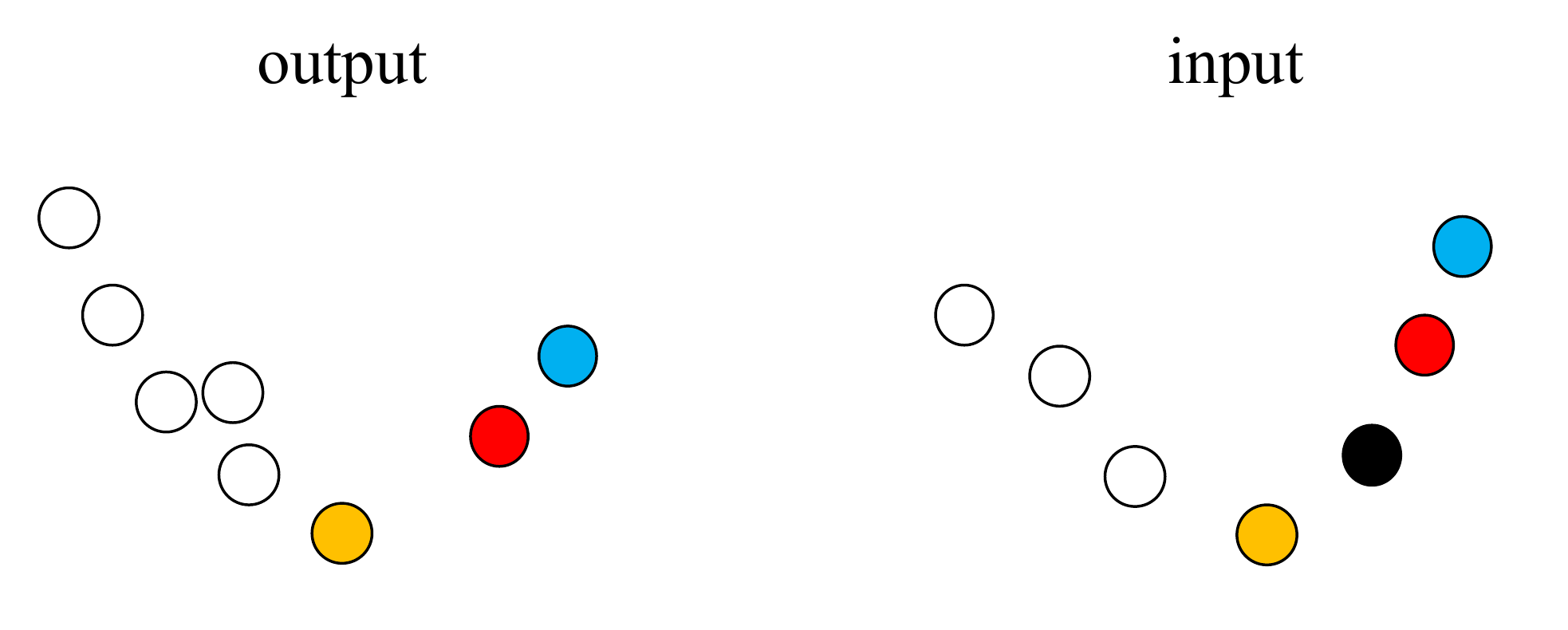}
	}
	}%
	\subfloat[Assign existence]{
	\label{fig:existence_assign:merged}
		\includegraphics[width=0.12\linewidth]{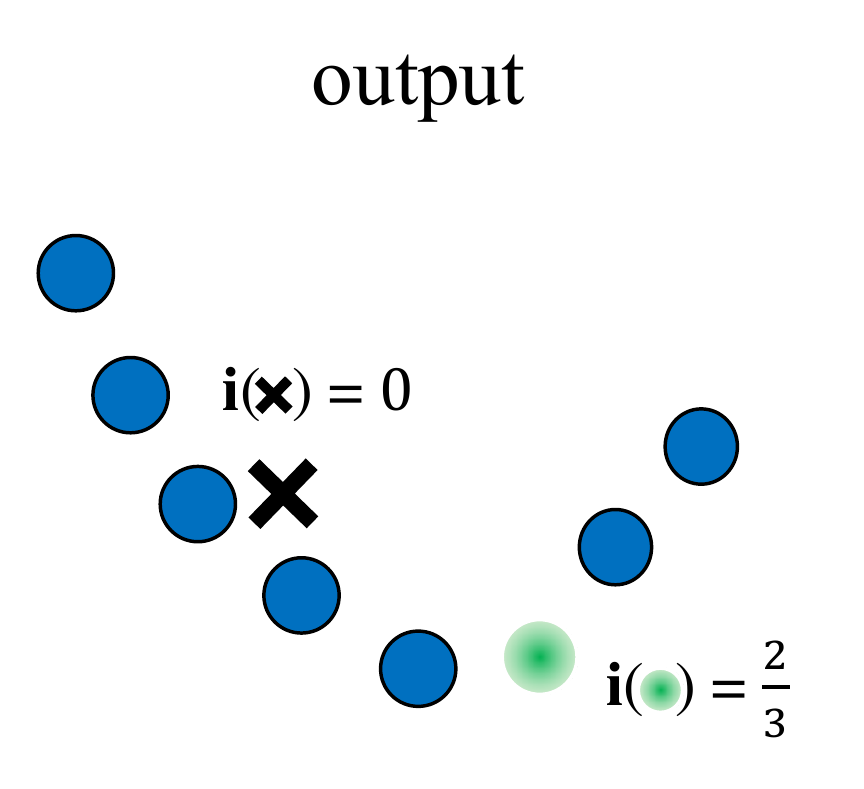}
}%

	\Caption{Existence assignment example.} 
	{%
\nothing{

}%
	We visualize how to compute confidences of existence of output samples $\sampleexistence(\sampleoutput)$ (\Cref{eqn:assignment:existence,alg:generate_new_output_samples}) from a set of matched input and output neighborhoods.
		\subref{fig:existence_assign:n1} \subref{fig:existence_assign:n2} \subref{fig:existence_assign:n3} show three pairs of matched input and output neighborhoods centered at different samples (shown in red) over the same set of samples.
		Matched samples are in the same color.
                Empty black circles indicate samples outside a\nothing{n output} neighborhood.
		Solid black circles indicate samples within a\nothing{n output} neighborhood but unmatched.
                \nothing{
		In this example, there are eight output samples, which can produce eight pairs of input and output neighborhoods.
		For the sake of explanation, we only use the above three pairs \subref{fig:existence_assign:n1} \subref{fig:existence_assign:n2} \subref{fig:existence_assign:n3} to perform the existence assignment step to compute $\sampleexistence(\sampleoutput)$.
                }%
                \nothing{
            }%
The black cross sample in \subref{fig:existence_assign:merged} has $\sampleexistence(\sampleoutput)=0$, since it is unmatched with any $\sampleinput$ ($\sampleexistence(\sampleinput)=0$) in \subref{fig:existence_assign:n1} and \subref{fig:existence_assign:n2}.
		The green sample in \subref{fig:existence_assign:merged} has confidence $\sampleexistence(\sampleoutput)=\frac{2}{3}$: in the three pairs of neighborhoods, there are two pairs \subref{fig:existence_assign:n1} and \subref{fig:existence_assign:n3} where each has an unmatched input sample, which indicates there could be a missing sample in the output located at approximately the same location relative to its neighborhood center;
\nothing{
}%
		the two unmatched input samples are merged to generate the green output sample in \subref{fig:existence_assign:merged}.
                \nothing{
                }%
	}
	\label{fig:existence_assign}
\end{figure*}

\begin{figure*}[tbh]
	\centering
	\subfloat[Initialization]{
	\label{fig:existence_assign:init}
	\includegraphics[width=0.18\linewidth]{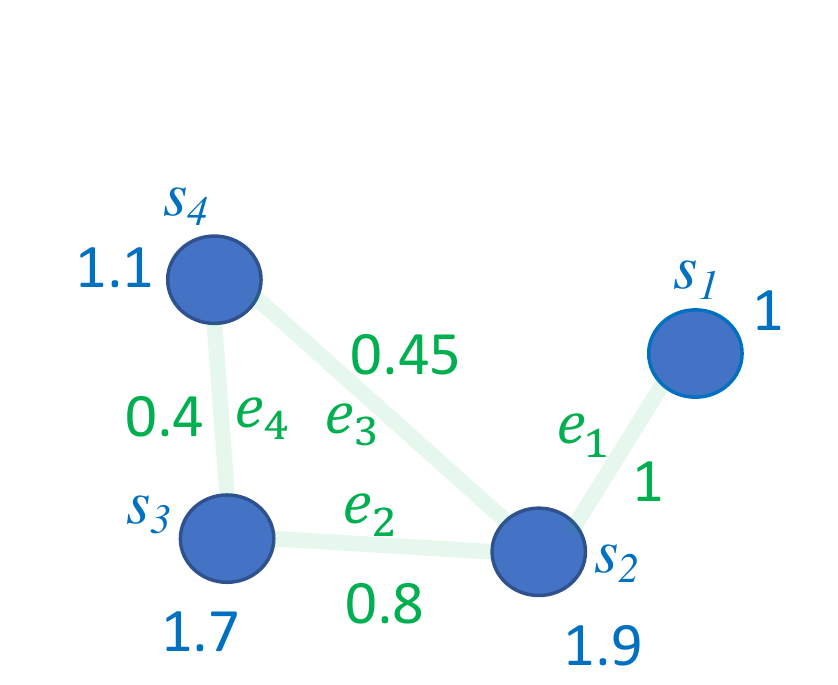}
	}%
	\subfloat[Step 1]{
	\label{fig:existence_assign:1}
	\includegraphics[width=0.18\linewidth]{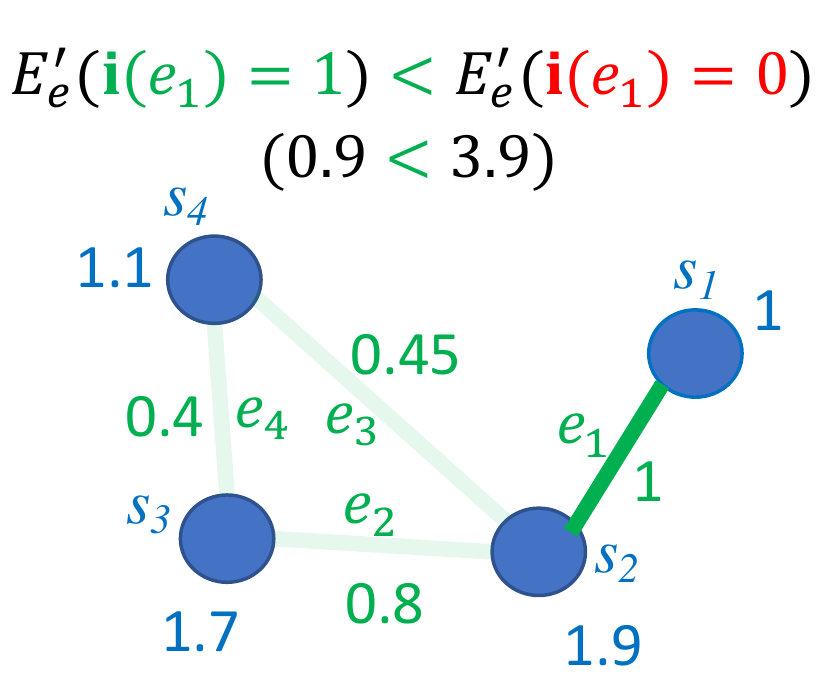}
}
	\subfloat[Step 2]{
	\label{fig:existence_assign:2}
		\includegraphics[width=0.18\linewidth]{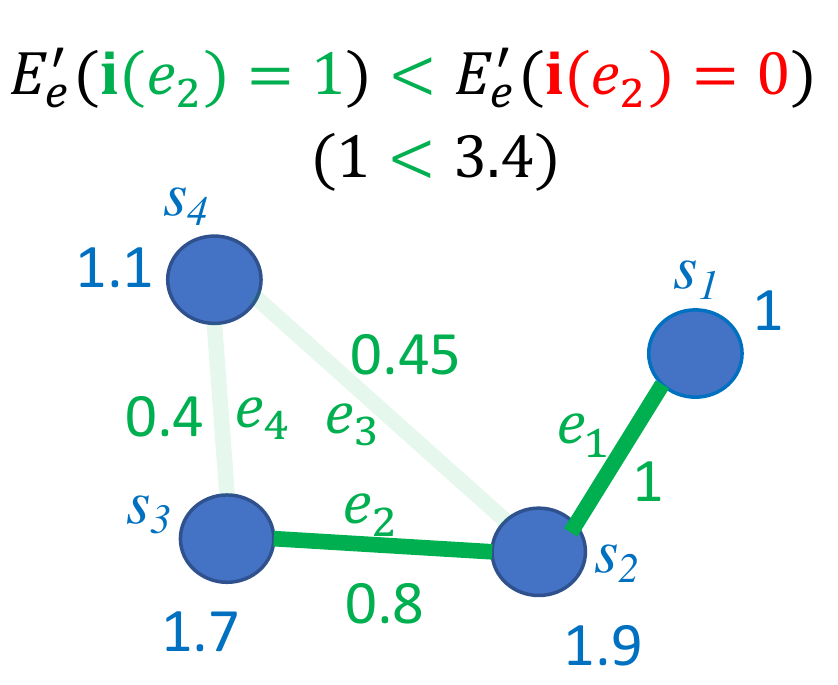}
	}%
	\subfloat[Step 3]{
	\label{fig:existence_assign:3}
		\includegraphics[width=0.18\linewidth]{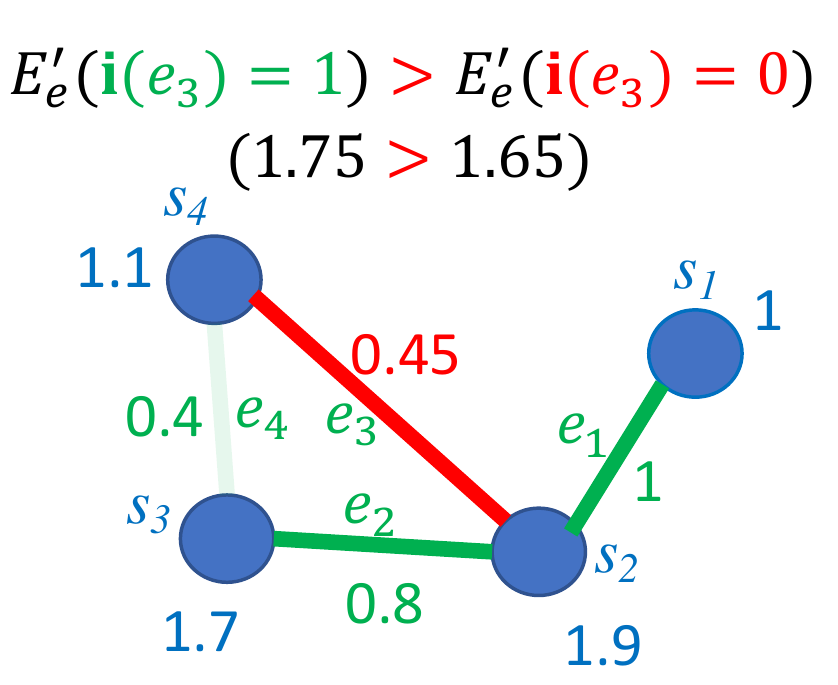}
	}%
	\subfloat[Step 4]{
	\label{fig:existence_assign:4}
		\includegraphics[width=0.18\linewidth]{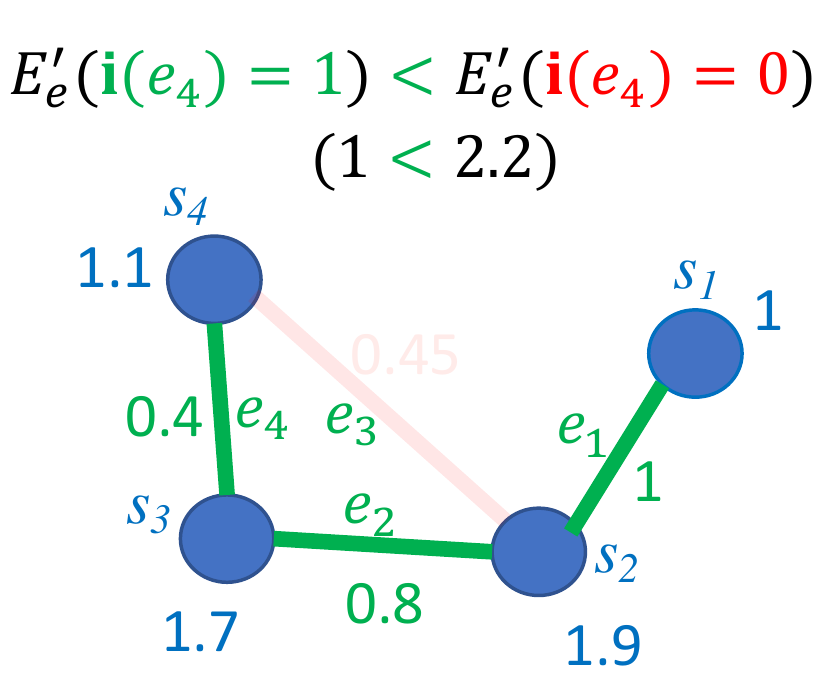}
}%

	\Caption{Edge assignment example.}
	{%
		This example illustrates how to solve \Cref{eqn:assignment:edge} for edge assignment by solving \Cref{eqn:assignment:edge:univar:final} in a loop. 
		Each step \subref{fig:existence_assign:1}-\subref{fig:existence_assign:4} solves \Cref{eqn:assignment:edge:univar:final} once.
		There are four samples and four potential edges (with $\overline{\sampleexistence}(\sampleedge)> 0 $).
		 The numbers near edges or samples indicate
		 the expected confidence of existence $\overline{\sampleexistence}(\sampleedge)$ of the edges or the expected number of edges $|\overline{\sampleedgeset}(\samplesym)|$ associated with samples  (e.g. $\overline{\sampleexistence}(\sampleedge_2)=0.8$,  $|\overline{\sampleedgeset}(\samplesym_2)|=1.9$).
		 In \subref{fig:existence_assign:init}, we initialize all $\sampleexistence(\sampleedge)=0$ and sort all potential edges by its $\overline{\sampleexistence}$ from the largest to the smallest: $\sampleedge_1$ is the one with the largest $\overline{\sampleexistence}$ and $\sampleedge_4$ is the one with the smallest.
		 The optimization loops from $\sampleedge_1$ \subref{fig:existence_assign:1} to $\sampleedge_4$ \subref{fig:existence_assign:4} in decreasing $\overline{\sampleexistence}(\sampleedge)$ values.
                 The number below each $\energyedge\left(\sampleexistence(\sampleedge)\right)$ is its computed value (e.g. $\energyedge\left(\sampleexistence(\sampleedge_1)=1\right)= |\sampleexistence(\sampleedge_1)-\overline{\sampleexistence}(\sampleedge_1)| + |\sampleedgeset(\samplesym_1)-\overline{\sampleedgeset}(\samplesym_1)| + |\sampleedgeset(\samplesym_2)-\overline{\sampleedgeset}(\samplesym_2)| =  
		 |1-1|+|1-1|+|1-1.9|=0.9$).
		 The light green edges are not optimized with initial values $\sampleexistence(\sampleedge)=0$.
		 The dark green edges are optimized with $\sampleexistence(\sampleedge)=1$.
		 The red edge is optimized with  $\sampleexistence(\sampleedge)=0$.
		 The light red indicates there is no edge after optimization. 
		 \nothing{
	}
}%
	\label{fig:edge_assign}
\end{figure*}

\subsubsection{Search Step}
\label{subsubsec:search_step}

We adopt PatchMatch~\cite{Barnes:2009:PRC,Chen:2012:NPT} to compute approximate nearest neighbors (ANN) for each output sample. 
The standard PatchMatch algorithm 1) randomly generates the initial nearest neighbor field, and 2) alternates between propagation and search steps by traversing the regular image grid in a scanline order.
Initially, we generate the ANN by randomly assigning an output sample to an input sample (with identical sample id for discrete elements).
One issue is how to choose a sample traversal order.
We follow the steps from \cite{Chen:2012:NPT} which works on meshes.
We build a simple graph by connecting each sample with its $k$-nearest neighbors ($k=8$ in our implemenetation), and perform breadth-first search. In the next iteration, the traversal starts from the last sample in the most recent sequence.

In our implementation, for the random search step, the maximum window size is 150, and the minimum size is 25, and the search window is exponentially decreased with factor 2.
In each pattern optimization step, we need to compute an ANN. In two consecutive steps, the output sample distributions are similar. So the previous ANN is used to initialize the subsequent patch match algorithm.
Since the initialization is close to the converged ANN, a small number (2) of Patch Match iterations is used, except for the initial step at each level of hierarchical synthesis (\Cref{sec:method:sample_synthesis:hierarchical_synthesis}), which uses 5 iterations.
Our patch match implementation is parallelized by equally dividing the output domain into regions, the number of which equals to that of threads.
The search step consumes most of the computation time needed by the synthesis. 
The computational complexity of the search step in an optimization step is  $O(\numOutput \numneighsample^3)$, where $\numOutput$ is the total number of output samples and $\numneighsample$ is the average number of samples within neighborhoods.
Please refer to \Cref{sec:performance} for more details.
 
\nothing{
In the search step, for each output sample,  we find the input sample with the most similar neighborhood with K-coherence search for acceleration~\cite{Tong:2002:SBT}.
In the pre-processing, K-coherence search builds a similarity set for each input sample by brute force search, during which the input exemplar is fixed. 
Due to the interactive and dynamic nature of our system, the input exemplar is changing over time with added new samples, when the user copy and paste a new element, draw a continuous curve or simply accept predictions provided by the system.

We dynamically construct the similarity set for each new input sample. There are two types of new input samples: created sample and synthesized sample. The created sample is completely new and associated with newly created element or continuous curve. 
The synthesized sample is associated with newly synthesized predictions. For the former, we build the similarity set via brute-force search. For the latter, we simply collect similar neighborhoods of that sample during synthesis.

}

\subsubsection{Assignment Step}
\label{subsubsec:assignment_step}

Here, we describe how to determine the values of sample positions $\samplespace$, attributes including edge $\sampleedgeset$ and orientation $\sampleorientations$, as well as sample existence $\sampleexistence$.
\nothing{
}%
The assignments of these different quantities are extended from the assignment step of pixel colors \cite{Kwatra:2005:TOE} and sample positions \cite{Ma:2011:DET} by taking votes from overlapping output neighborhoods at the same entity (such as sample or edge).
In particular, discrete samples only have sample id attributes $\sampleid$, which is used in the search step to make sure only samples with the same $\sampleid$ are matched. 
Thus, only position assignment is deployed for discrete samples.

\paragraph{Position assignment}

For each output sample $\sampleoutput$ and its neighbor\delete{s} $\sampleoutputprime$, there is a set of \nothing{a few }matched input sample pairs $\left(\sampleinput, \sampleinputprime \right)$ provided by the previous search step.
The estimated distance $\differencesym{\samplespace}(\sampleoutput,\sampleoutputprime)$ between $\sampleoutput$ and $\sampleoutputprime$ is
\begin{align}
\differencesym{\samplespace}(\sampleoutput,\sampleoutputprime) \approx\samplespace(\sampleinput)-\samplespace(\sampleinputprime).
\label{eq:estimated_distance}
\end{align}
We use least squares \cite{Ma:2011:DET} to estimate $\samplespace(\sampleoutput)$ \nothing{ and $\samplespace(\sampleoutputprime)$,} by
\begin{align}
\argmin_{\{\samplespace(\sampleoutput)\}}
\sum_{\sampleoutput\in\outputsym} \sum_{\sampleoutputprime\in\neighoutput}
^{} \left\|\differencesym{\samplespace}(\sampleoutput,\sampleoutputprime)-\left(\samplespace(\sampleinput)-\samplespace(\sampleinputprime)\right)\right\|^2 + \sum_{\sampleoutput \notin \domain}^{} \|\differencesym{\samplespace}(\sampleoutput,\domain)\|^2.
\label{eqn:assignment:position}
\end{align}
The second term in \Cref{eqn:assignment:position} is the domain constraint to encourage output samples to stay within $\domain$,
where $\differencesym{\samplespace}(\sampleoutput,\domain)$ is the shortest vector from $\sampleoutput$ to the boundary of $\domain$,
and $\sampleoutput \notin \domain$ indicates $\sampleoutput$ is outside $\domain$. 

\nothing{
\Cref{eqn:assignment:position} is minimized with iterative reweighed least squares (IRLS). In an assignment step, the number of IRLS iterations is . \Cref{fig:lpnorm_comparison} shows comparison of using $l_1$ and $l_2$ norms.
}%

\nothing{
\begin{figure}[htb]
	\centering
	\subfloat[$l_1$-norm regression]{
		\label{fig:l_1norm_regression}
		\includegraphics[width=0.45\linewidth]{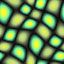}
	}%
	\subfloat[$l_2$-norm regression]{
		\label{fig:l_2norm_regression}
		\includegraphics[width=0.45\linewidth]{figs/raster/161.jpg}
	}%

	\Caption{Comparison of using $l_1$ and $l_2$ norms in sample position assignment.}
	{%
		$l_1$ norm is more robust to outliers and encourage sparsity in residuals. 
	}
	\label{fig:lpnorm_comparison}
\end{figure}
}%

\nothing{
}%

\nothing{
\begin{align}
\constraintterm(\outputsym,\domain)=\sum_{\sampleoutput \in \outputsym}^{}\max(0,\distancesym(\sampleoutput,\domain)) + 
\end{align}
}%

\nothing{
}%

\paragraph{Existence assignment}

\nothing{
}%

\nothing{
}%
Our method adjusts the number of samples within local regions during the synthesis process for better quality.
\nothing{
If there are more and less samples than required, the pattern structure will be compromised.
}%
The number of samples is optimized via existence $\sampleexistence$ assignment, again via a \nothing{simple }voting scheme\nothing{.
We collect not only a set of input samples that are matched with $\sampleoutput$ but also record the input sample as $\unmatched$  when $\sampleoutput$ is unmatched with any samples in a input neighborhood, in a set $\{\sampleinput\}$}:
\begin{align}
\argmin_{\sampleexistence(\sampleoutput) \in [0\ 1] } \sum_{\sampleinput\in\{\sampleinput\} }^{} \left|\sampleexistence(\sampleoutput)-\sampleexistence(\sampleinput)\right|^2,
\label{eqn:assignment:existence}
\end{align}
where $\sampleinput$ runs through the set $\{\sampleinput\}$ we collect during neighborhood matching, i.e. the corresponding input samples of an output sample from overlapping input neighborhoods.
$\sampleexistence(\sampleinput)=0$ if $\sampleoutput$ is not matched with any $\sampleinput$ in a pair of matched input and output neighborhoods, and $\sampleexistence(\sampleinput)=1$ otherwise.
\Cref{eqn:assignment:existence} computes the confidence of existence $\sampleexistence(\sampleoutput) \in [0\ 1]$ of an output sample $\sampleoutput$.
\nothing{
}%
\nothing{
If $\sampleoutput$ is unmatched with any $\sampleinput$ $\sampleinput=\unmatched$),
}%
Every iteration , we remove output samples $\sampleoutput$ whose $\sampleexistence(\sampleoutput)<0.5$.
The above assignment step is applied to samples that are already in the output sample distribution.
To add back missing output samples, we first generate candidate samples\nothing{ $\samplecandidate$}, merge them as output samples\nothing{ $\sampleoutput$},
and pick those with $\sampleexistence(\sampleoutput)>0.5$ as added output samples.
\add{
\nothing{We find the existence assignment heuristics works well.}
The energy in \Cref{eqn:assignment:existence} is not guaranteed to decrease immediately after sample addition or removal, but it will generally decrease through iterations.
}%
Please refer to \Cref{fig:existence_assign} for an example and \Cref{sec:appendix:existence_assignment} for more algorithm details.

\nothing{
}%

\nothing{
When a new $\samplecandidate$ is added, we match it with the nearest, existing candidate sample if their distance is within $0.5\samplingdistance$.
This indicates the two samples are probably associated to an identical missing output sample.
Otherwise, it means the old candidate is matched to a special sample \texttt{None}.
Finally, the algorithm produce many clusters of candidate samples. Each cluster contains samples that may be associated with one missing output sample. 
We merge all samples within a cluster by averaging their position, attributes and existence in the same way as the assignment step. 
Every 10 iterations, we upgrade the merged candidate samples with $\sampleexistence(\sampleoutput)>0.5$ as output samples.
}%

\nothing{
What's a little different here is the way to compute  $\sampleexistence$ for $\samplecandidate$. For a $\samplecandidate$, we collect a set of matched input and output neighborhoods pairs where the output is overlapped over $\samplecandidate$. If there 
}%

\paragraph{Edge assignment}

\nothing{
Edge assignment is done by, again, a voting scheme from overlapping matched neighborhoods.
Specifically, to determine whether pairs of output samples $\sampleoutput, \sampleoutputprime$ should be connected, we can optimize the binary connectivity $\sampleedgeoutput(\sampleoutput, \sampleoutputprime)$ via:
\nothing{%
\begin{align}
\argmin_{\sampleedgeoutput(\sampleoutput, \sampleoutputprime)} 
&
\sum_{\sampleinput\in\{\sampleinput(\sampleoutput)\} }^{} \left(\sampleedgeoutput(\sampleoutput, \sampleoutputprime)-\sampleedgeinput(\sampleinput, \sampleinputprime)\right)^2
\\
+
&
\sum_{\sampleinputprime\in\{\sampleinputprime(\sampleoutputprime)\} }^{} \left(\sampleedgeoutput(\sampleoutput, \sampleoutputprime)-\sampleedgeinput(\sampleinput, \sampleinputprime)\right)^2
\label{eqn:assignment:edge:neighborhood}
\end{align}
}%
\begin{align}
\argmin_{\sampleedgeoutput(\sampleoutput, \sampleoutputprime)} 
\sum_{\sampleedgeinput \in \sampleedgesetinput(\sampleedgeoutput) }^{} \left(\sampleedgeoutput(\sampleoutput, \sampleoutputprime)-\sampleedgeinput(\sampleinput, \sampleinputprime)\right)^2
\label{eqn:assignment:edge:neighborhood}
\end{align}
,
where $\sampleedgesetinput(\sampleedgeoutput)$ is the set of input edges that overlap $\sampleedgeoutput$:
\begin{align}
\sampleedgesetinput\left(\sampleedgeoutput(\sampleoutput, \sampleoutputprime)\right)
=
\{\sampleedgeinput(\sampleinput, \sampleinputprime) \; | \; \sampleinput\in\{\sampleinput(\sampleoutput)\} \logicor \sampleinputprime\in\{\sampleinputprime(\sampleoutputprime)\} \}
\label{eqn:assignment:edge:setinput}
\end{align}
,
where $\sampleinput(\sampleoutput)$/$\sampleinputprime(\sampleoutputprime)$ is a matching input sample for output sample $\sampleoutput$/$\sampleoutputprime$ and $\sampleinputprime$/$\sampleinput$ is the neighbor sample of $\sampleinput$/$\sampleinputprime$ matched to $\sampleoutputprime$/$\sampleoutputprime$\nothing{ for the first/second term}.
}%
\nothing{
In our experiments, we found out that we can simplify \Cref{eqn:assignment:edge:neighborhood} as follows:
}%

We assign edges by optimizing the following objective:
\begin{align}
\argmin_{\{\sampleexistence(\sampleedgesym{\sampleoutput}{ \sampleoutputprime}) \in \{0,1\} \}}
\sum_{\{\sampleedgesym{\sampleoutput}{\sampleoutputprime} \} }^{} \left|\sampleexistence(\sampleedgesym{\sampleoutput}{ \sampleoutputprime})-\overline{\sampleexistence}( \sampleedgesym{\sampleoutput}{\sampleoutputprime})\right|
+
\sum^{}_{\nothing{\sampleoutput\in}\{\sampleoutput\}} \left| |\sampleedgeset(\sampleoutput)|- |\overline{\sampleedgeset}(\sampleoutput) |\right|,
\label{eqn:assignment:edge}
\end{align}
\nothing{
}%
\nothing{
,
in the first term of which $(\sampleinput,\sampleinputprime)$ run through all input sample pairs matched to $(\sampleoutput,\sampleoutputprime)$, and $\sampleoutput$ runs though all output samples $\{\sampleoutput\}$ in the second term.
}%
\nothing{, $\binaryarray$ is the set of binary sets}
where the first term computes the difference between the actual and expected edge confidences $\overline{\sampleexistence}( \sampleedgesym{\sampleoutput}{\sampleoutputprime}) \in [0\ 1]$.
\delete{Intuitively, }$\overline{\sampleexistence}$ is the vote by overlapping input neighborhoods on the same edge, computed using least squares by replacing samples in \Cref{eqn:assignment:existence} with edges.
$\{\sampleedgesym{\sampleoutput}{ \sampleoutputprime} \} $ is the set of edges that have $\overline{\sampleexistence}(\sampleedgesym{\sampleoutput}{ \sampleoutputprime}) > 0$, and there is no edge between $\sampleoutput$ and $\sampleoutputprime$ if
$\overline{\sampleexistence}(\sampleedgesym{\sampleoutput}{ \sampleoutputprime}) = 0$.
The second term computes the differences between the optimized number of edges $|\sampleedgeset(\sampleoutput)|$ and the expected number of edges $|\overline{\sampleedgeset}(\sampleoutput)|$  connected to $\sampleoutput$.
$|\overline{\sampleedgeset}(\sampleoutput)|$ is similarly computed by voting from overlapping output neighborhoods on the same sample $\sampleoutput$:
\begin{align}
\argmin_{|\overline{\sampleedgeset}(\sampleoutput)|} \sum_{\sampleinput\in\{\sampleinput\} }^{} \left||\overline{\sampleedgeset}(\sampleoutput)|-|\sampleedgeset(\sampleinput)|\right|^2.
\label{eqn:assignment:numedges}
\end{align}
Basically, we compute \nothing{the nearest integer} the average of $\{|\sampleedgeset(\sampleinput)|\}$.
In sum, the first term is edge-centric while the second is sample-centric.

It is non-trivial to optimize \Cref{eqn:assignment:edge}, where the optimization variables $\{\sampleexistence(\sampleedgesym{\sampleoutput}{ \sampleoutputprime})\}$ are binary.
Thus, we solve it in a greedy fashion.
\delete{First, }We initialize all $\sampleexistence(\sampleedgesym{\sampleoutput}{\sampleoutputprime})=0$.
\delete{Second, we only consider the influence of the edge-centric term in \Cref{eqn:assignment:edge}.}
\nothing{
We compute the confidence of existence of output edges $\sampleexistence(\sampleedgesym{\sampleoutput}{\sampleoutputprime})$ by
\begin{align}
\argmin_{\sampleexistence(\sampleedgeoutput) \in [0,1] } \sum_{(\sampleinput,\sampleinputprime)\in\{(\sampleinput,\sampleinputprime)\} }^{} \left|\sampleexistence(\sampleedgeoutput)-\sampleexistence( \sampleedgeinput)\right|^2
\end{align}
, where we relax the constraint of $\sampleedgeoutput$  from $\sampleedgeoutput\in \{0,1\}$ to $\sampleedgeoutput\in [0,1]$. 
}%
We sort output edges $\{\sampleedgeoutput\}$ by its expected confidence of existence $\overline{\sampleexistence}(\sampleedgeoutput)$,
and optimize $\sampleexistence(\sampleedgeoutput)$ greedily by looping over the sorted $\{\sampleedgeoutput\}$ in decreasing confidence.
For each $\sampleedgeoutput$, we decide whether $\sampleexistence(\sampleedgeoutput) =$ $0$ or $1$ by choosing the one that minimizes \Cref{eqn:assignment:edge}.
In other words, the multivariate optimization problem (\Cref{eqn:assignment:edge}) is optimized by solving univariate optimization problems in a loop. 
By decomposing the optimization variables in \Cref{eqn:assignment:edge} from a set of edges $\{\sampleexistence(\sampleedgesym{\sampleoutput}{ \sampleoutputprime}) \} $ to a single edge $\sampleexistence( \sampleedgesym{\sampleoutputstar}{\sampleoutputstarprime})$ to be optimized and the rest,
the univariate version of \Cref{eqn:assignment:edge} can be written as:
\begin{align}
\argmin_{\sampleexistence(\sampleedgesym{\sampleoutputstar}{\sampleoutputstarprime}) \in \{0,1\} }
\energyedge + \energyothers,
\label{eqn:assignment:edge:univar}
\end{align}
where
\begin{align}
\begin{split}
\energyedge\left(\sampleexistence(\sampleedgesym{\sampleoutputstar}{\sampleoutputstarprime})\right) & = \left|\sampleexistence(\sampleedgesym{\sampleoutputstar}{ \sampleoutputstarprime})-\overline{\sampleexistence}( \sampleedgesym{\sampleoutputstar}{\sampleoutputstarprime})\right|
+
\\
&\left| |\sampleedgeset(\sampleoutputstar)|- |\overline{\sampleedgeset}(\sampleoutputstar) |\right|+
\left| |\sampleedgeset(\sampleoutputstarprime)|- |\overline{\sampleedgeset}(\sampleoutputstarprime) |\right|,
\end{split}
\\
\begin{split}
\energyothers\left(\sampleexistence(\sampleedgesym{\sampleoutputstar}{\sampleoutputstarprime})\right) &=
\sum_{\{\sampleedgesym{\sampleoutput}{\sampleoutputprime} \} \setsub \sampleedgesym{\sampleoutputstar}{\sampleoutputstarprime} }^{} \left|\sampleexistence(\sampleedgesym{\sampleoutput}{ \sampleoutputprime})-\overline{\sampleexistence}( \sampleedgesym{\sampleoutput}{\sampleoutputprime})\right|
+
\\
&\sum^{}_{\{\sampleoutput\} \setsub \{\sampleoutputstar,\sampleoutputstarprime \}  } \left| |\sampleedgeset(\sampleoutput)|- |\overline{\sampleedgeset}(\sampleoutput) |\right|.
\end{split}
\end{align}
\nothing{
The summations $\sum$ in \Cref{eqn:assignment:edge} are decomposed into $\energyedge$ and $\energyothers$.
}%
Since $\energyothers$ is a constant in \Cref{eqn:assignment:edge:univar}, it is equivalent to:
\begin{align}
\argmin_{\sampleexistence(\sampleedgesym{\sampleoutputstar}{\sampleoutputstarprime}) \in \{0,1\} }
\energyedge.
\label{eqn:assignment:edge:univar:final}
\end{align}
\Cref{eqn:assignment:edge:univar:final} can be solved with brute-force search. The search space is 2 ($\{0, 1\}$). 
\Cref{fig:edge_assign} illustrates how to solve \Cref{eqn:assignment:edge} by solving \Cref{eqn:assignment:edge:univar:final} in a loop.

\nothing{
The loop is ended once the univariate optimization minimizer is 0.
}%

\nothing{
}

\nothing{
\Cref{eqn:assignment:edge} is an approximation of \Cref{eqn:assignment:edge:neighborhood} because it does not consider the possibility that even though $\sampleinput$ and $\sampleinputprime$ are not connected in the input, there might be another input sample $\hat{\sampleinput}$ that is similar to $\sampleinput$ (measured in terms of neighborhood and such) and is connected to $\sampleinputprime$.
However, our experiments indicate such situations happen only rarely, as exemplified in \Cref{fig:explain_edge_assignment}.
The sample matching in the two neighborhoods is recomputed with a tighter unmatched cost $\unmatchedcost=1$ to discourage misconnections in the edge assignment. 
We decide if there is an edge between two samples $\sampleoutput$ and $\sampleoutputprime$ by analyzing all the  pairs of matched input samples $\sampleinput$ and $\sampleinputprime$.
Essentially, what \Cref{eqn:assignment:edge} does is the following:
if there are 50\% percent of input sample pairs that have edges,  there will be an edge between $\sampleoutput$  and $\sampleoutputprime$.  
}%

\nothing{
}%

\nothing{

\nothing{
We optimize the \Cref{eq:edge_energy} using graph cuts \cite{Kolmogorov:2004:WEF,Boykov:2001:FAE}.
}%

\nothing{
There are a number of matched input neighborhoods over an output edge.  We can compute the probability of the existence of the edges. The algorithm is described as follows.
We choose the edge with the highest probability of existence. 
For every edge connected to it, we compute the conditional probability of edge existence given it. 
The edge with the maximum conditional probability is generated. This process is repeated until the conditional probability is lower than a threshold $\threshold$.
}%
}%

\nothing{
Regarding the assignment of $\sampleconnections$, we adopts a simple voting scheme. For an output sample $\sampleoutput$, we have a set of input samples $\{\sampleinput\}$ that are matched to it. 
Each input sample has an associated $\sampleconnections$ and we gather a vote set $\{\sampleconnections(\sampleinput)\}$.
We choose the one that has the closest to the weighted arithmetic mean of the votes by $\probabilitybi({\sampleoutputprime,\sampleinputprime})$.
\nothing{
\begin{align}
\argmin_{\sampleconnections(\sampleoutput)} \sum_{}^{}
\end{align}
}%

}%

\nothing{
For each output sample $\sampleoutputprime$, we have a number of matched overlapping input neighborhoods $\{\neighinput\}$.
In each input neighborhood, there are input samples and we have the a set of probabilities of matching each input sample to the output sample $\probability(\sampleinputprime, \sampleoutputprime)$.
We can define the quantity below.
\begin{align}
\existenceconfidence(\sampleoutputprime) = \frac{1}{\nummatchedneighborhoods} \sum_{\neighinput\in\{\neighinput\}}\sum_{\sampleinputprime\in\neighinput}\probability(\sampleinputprime, \sampleoutputprime)
\label{eq:existence_confidence}
\end{align}
,
where $\nummatchedneighborhoods $ is number of overlapping input neighborhoods $\{\neighinput\}$ over $\sampleoutput$.
\Cref{eq:existence_confidence} suggests the existence confidence of $\sampleoutputprime$. 
If $\existenceconfidence(\sampleoutputprime)$ is low, it suggests that $\sampleoutputprime$ is less likely to be matched with input samples, and thus $\sampleoutputprime$ is likely to be removed.
If $\existenceconfidence(\sampleoutputprime)$ is larger than one, it will suggest $\sampleoutputprime$ is likely to be matched with more than one input sample, and thus there should be another output sample near $\sampleoutputprime$.
We adaptively remove and add samples into the output during optimization.
}%

\nothing{

}%

\nothing{
Edge assignment adopts simple voting scheme. For each pair of output samples $\sampleoutput$ and $\sampleoutputprime$ , if there are more than half number of matched input pairs that have edges connecting them, then we assign an edge connecting $\sampleoutput$ and $\sampleoutputprime$
}%

\nothing{
\begin{align}
\differencesym{\samplespace}(\sampleoutput,\sampleoutputprime)=\samplespace(\sampleinput)-\samplespace(\sampleinputprime)
\end{align}
}%

\paragraph{Orientation assignment}

\begin{figure}[tbh]
	\centering
	\captionsetup[subfigure]{justification=centering}
		
	\subfloat[\nothing{Input }$\{\sampleorientations(\sampleinput)\}$]{
		\label{fig:orientation_assign:input}
		\includegraphics[width=0.42\linewidth]{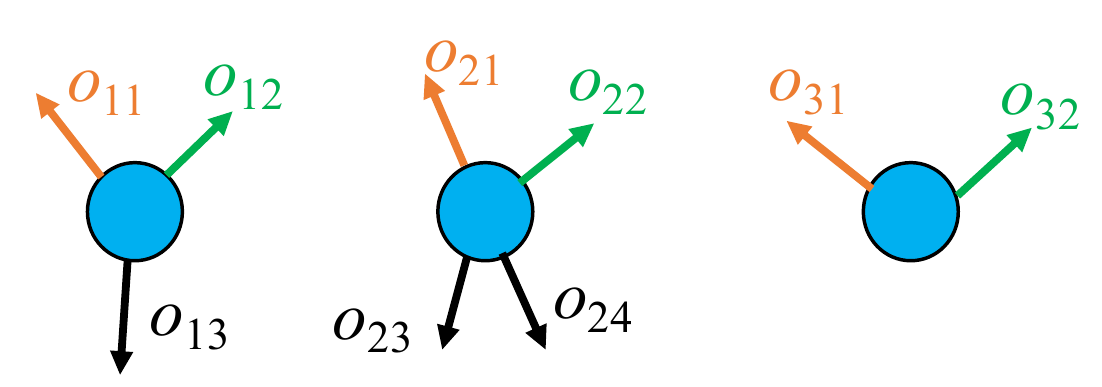}
	}%
	\subfloat[\nothing{Output }$\sampleorientations(\sampleoutput)$]{
		\label{fig:orientation_assign:output}
		\includegraphics[width=0.12\linewidth]{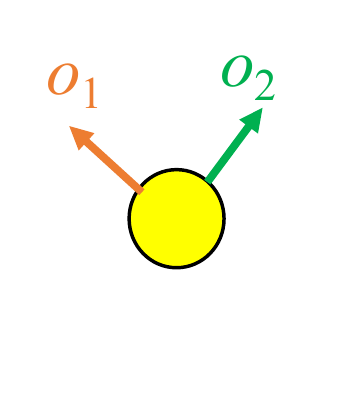}
	}%
	\subfloat[Updated \protect\subref{fig:orientation_assign:output}]{
		\label{fig:orientation_assign:assign}
		\includegraphics[width=0.37\linewidth]{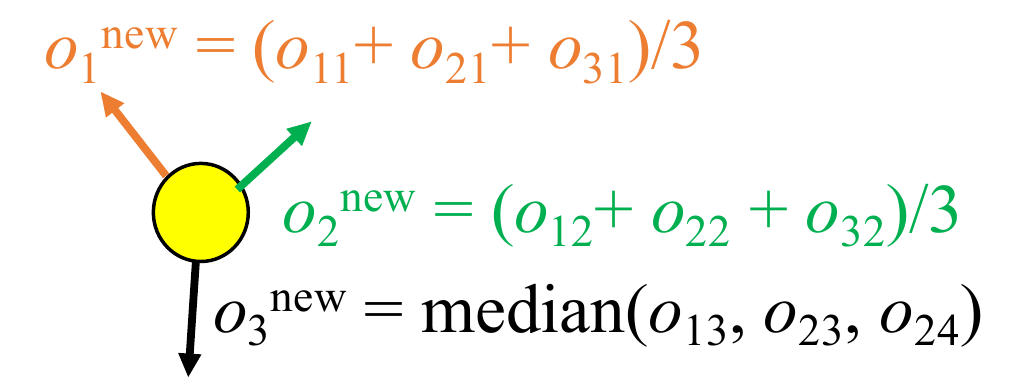}
	}%
	\Caption{Orientation assignment example.}
	{%
	        This example illustrates the orientation assignment step when $\sampleconnections(\sampleoutput)$ is increased from $2$ to $3$.
                The output sample \subref{fig:orientation_assign:output} is matched with the three input samples \subref{fig:orientation_assign:input}. 
		Matched orientations $\sampleorientationentry$ (arrows) are visualized in the same colors.
		Unmatched inputs $\sampleorientationentry$ are in black.
		\subref{fig:orientation_assign:assign} shows the updated orientations of the output sample. 
		The orange $\sampleorientationentry_1^{\text{new}}$ and green orientations $\sampleorientationentry_2^{\text{new}}$ are updated by averaging matched input orientations.
		The black output orientation $\sampleorientationentry_3^{\text{new}}$ is newly added by choosing the median from three unmatched orientations ($\sampleorientationentry_{13}$,$\sampleorientationentry_{23}$,$\sampleorientationentry_{24}$) in \subref{fig:orientation_assign:input}.
	}
	\label{fig:orientation_assign}
\end{figure}

In the search step (\Cref{subsubsec:search_step}), each $\sampleorientations(\sampleoutput)$ is matched with a set of $\{\sampleorientations(\sampleinput)\}$ associated with input samples coming from different input neighborhoods,
and each entry $\sampleorientationentry(\sampleoutput) \in \sampleorientations(\sampleoutput)$ has been matched with a $\sampleorientationentry(\sampleinput) \in \sampleorientations(\sampleinput)$.
The local orientation attribute $\sampleorientations(\sampleoutput)$ is updated by a voting scheme among $\{\sampleorientations(\sampleinput)\}$, where $\{\sampleorientations(\sampleinput)\}$ could have different lengths across different $\sampleinput$.

We optimize both dimension $\sampleconnections$ and value of entries $\sampleorientationentry$ in order.
\nothing{
First, we apply simple voting scheme to compute $\sampleconnections$.
\begin{align}
\argmin_{\sampleconnections(\sampleoutput)} \sum_{\sampleinput\in\{\sampleinput\} }^{} \left|\sampleconnections(\sampleoutput)-\sampleconnections(\sampleinput)\right|^2
\label{eqn:assignment:numconnections}
\end{align}
where $\sampleinput$ runs through all matched input samples of $\sampleoutput$. 
}%
In the input exemplar, the number of orientation entries $\sampleconnections(\sampleinput)$ equals $|\sampleedgeset(\sampleinput)|$.
Thus $\sampleconnections$ can be computed like in \Cref{eqn:assignment:numedges} and rounding the result as integers.
Essentially, we are trying to find an integer $\sampleconnections(\sampleoutput)$ that is the closest to the arithmetic average of $\{ \sampleconnections(\sampleinput)\}$\nothing{ \cite{Ma:2011:DET}}.

Similarly, we can update the values $\sampleorientationentry(\sampleoutput)$ in $\sampleorientations(\sampleoutput)$ using the same voting scheme to \Cref{eqn:assignment:existence,eqn:assignment:numedges}\nothing{ because each entry $\sampleorientationentry(\sampleoutput)$ is also matched with a set of votes  $\{\sampleorientationentry(\sampleinput)\}$ from the input samples}.
A special case is when $\sampleconnections(\sampleoutput)$ is updated to a new value (changing $\sampleorientations(\sampleoutput)$ vector length). 
In this case, we will need to add or remove one or several entries\nothing{ $\sampleorientationentry(\sampleoutput)$ } to or from the original $\sampleorientations(\sampleoutput)$.
To remove an entry\nothing{ $\sampleorientationentry(\sampleoutput)$} from $\sampleorientations(\sampleoutput)$, we pick the one\nothing{$\sampleorientationentry(\sampleoutput)$} whose matched set of input votes $\{\sampleorientationentry(\sampleinput)\}$ has the largest variance.
(We have experimented with another strategy that removes $\sampleorientationentry(\sampleoutput)$ whose matched set of input votes $\{\sampleorientationentry(\sampleinput)\}$ has the least number of entries, but have not found visible differences to the maximum variance strategy above.)
\nothing{
}%
\iftrue
To add an entry to $\sampleorientations(\sampleoutput)$, we collect orientation entries $\{\sampleorientationentry^{\prime}(\sampleinput)\}$ from the input samples that remain unmatched to any orientation entries $\sampleorientations(\sampleoutput)$ of the matched output sample, and add a new entry $\sampleorientationentry(\sampleoutput)$ into $\sampleorientations(\sampleoutput)$ as the median from the unmatched set $\{\sampleorientationentry^{\prime}(\sampleinput)\}$.
An example is illustrated in \Cref{fig:orientation_assign}.
\else
We may also need to add a new entry to $\sampleorientations(\sampleoutput)$, when $\sampleconnections$ is updated to a bigger value compared to the original one.
To this end, 
for each output sample $\sampleoutput$, we generate a set of extra orientation entries $\{\sampleorientationentry^{\prime}(\sampleinput)\}$ from the input samples.
}%
In \Cref{subsec:similarity}, Hungarian algorithm is used to compute one-to-one correspondences between two set of orientation entries from input $\sampleinput$ and output $\sampleoutput$ samples respectively, where
extra orientation entries on one of the two samples are left unmatched.
$\{\sampleorientationentry^{\prime}(\sampleinput)\}$ are collected when an output sample is matched with an input sample and the input sample $\sampleinput$ has more orientation entries $\sampleorientationentry(\sampleinput)$ within $\sampleorientations(\sampleinput)$ than $\sampleoutput$ does. 
The extra input orientation entry $\sampleorientationentry(\sampleinput)$ is added \nothing{ }%
into the set $\{\sampleorientationentry^{\prime}(\sampleinput)\}$. 
$\{\sampleorientationentry^{\prime}(\sampleinput)\}$ allows repeated entries.
When adding $\sampleorientationentry(\sampleoutput)$ into $\sampleorientations(\sampleoutput)$, we add the median from the unmatched set $\{\sampleorientationentry^{\prime}(\sampleinput)\}$.
\fi
In the rare case where we need to add more than one entry\nothing{$\sampleorientationentry(\sampleoutput)$} to $\sampleorientations(\sampleoutput)$, we randomly choose\nothing{ one} from $\{\sampleorientationentry^{\prime}(\sampleinput)\}$ after the median is used for the first add-on. 
\nothing{

}%

\subsubsection{Hierarchical Synthesis}
\label{sec:method:sample_synthesis:hierarchical_synthesis}

\begin{figure*}[htb]
	
	\centering
	\captionsetup[subfigure]{labelformat=empty}
	\setlength{\tabcolsep}{1pt}
\begin{tabular}{cccccc}
	\subfloat[Leaves]{
	\includegraphics[width=0.106\linewidth]{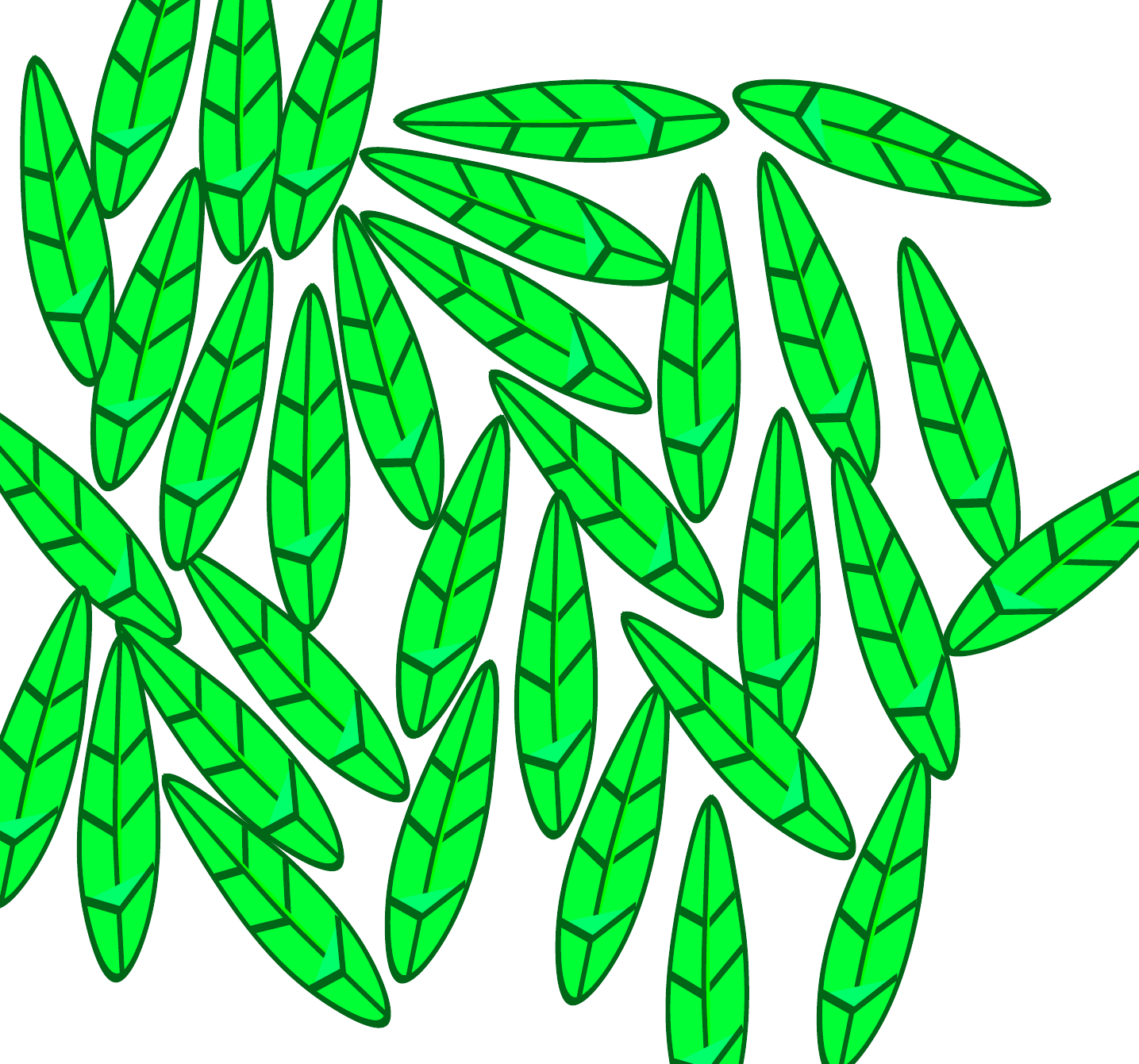}
}
&
\subfloat[]{
	\includegraphics[width=0.155\linewidth]{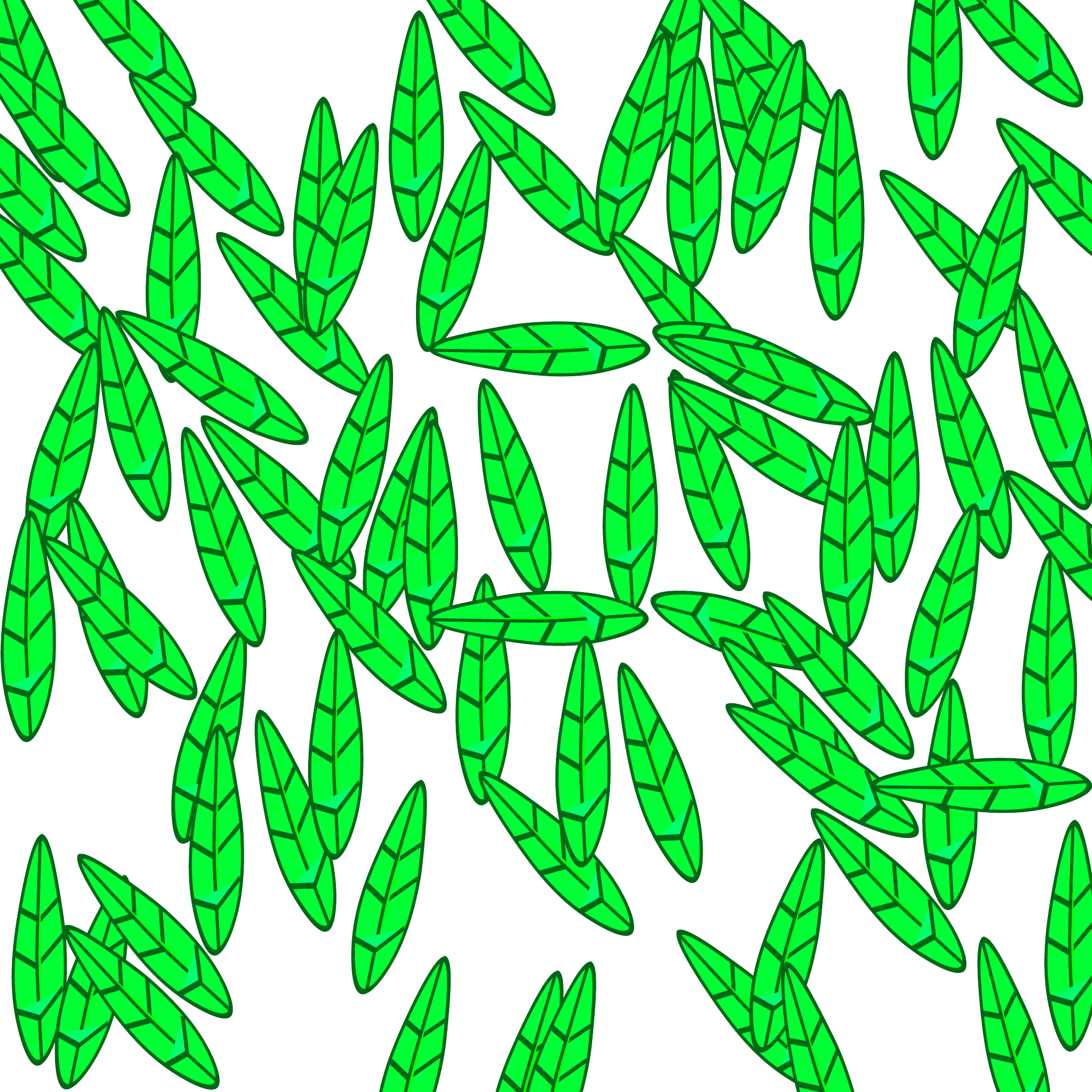}
}
&
\subfloat[]{
	\includegraphics[width=0.155\linewidth]{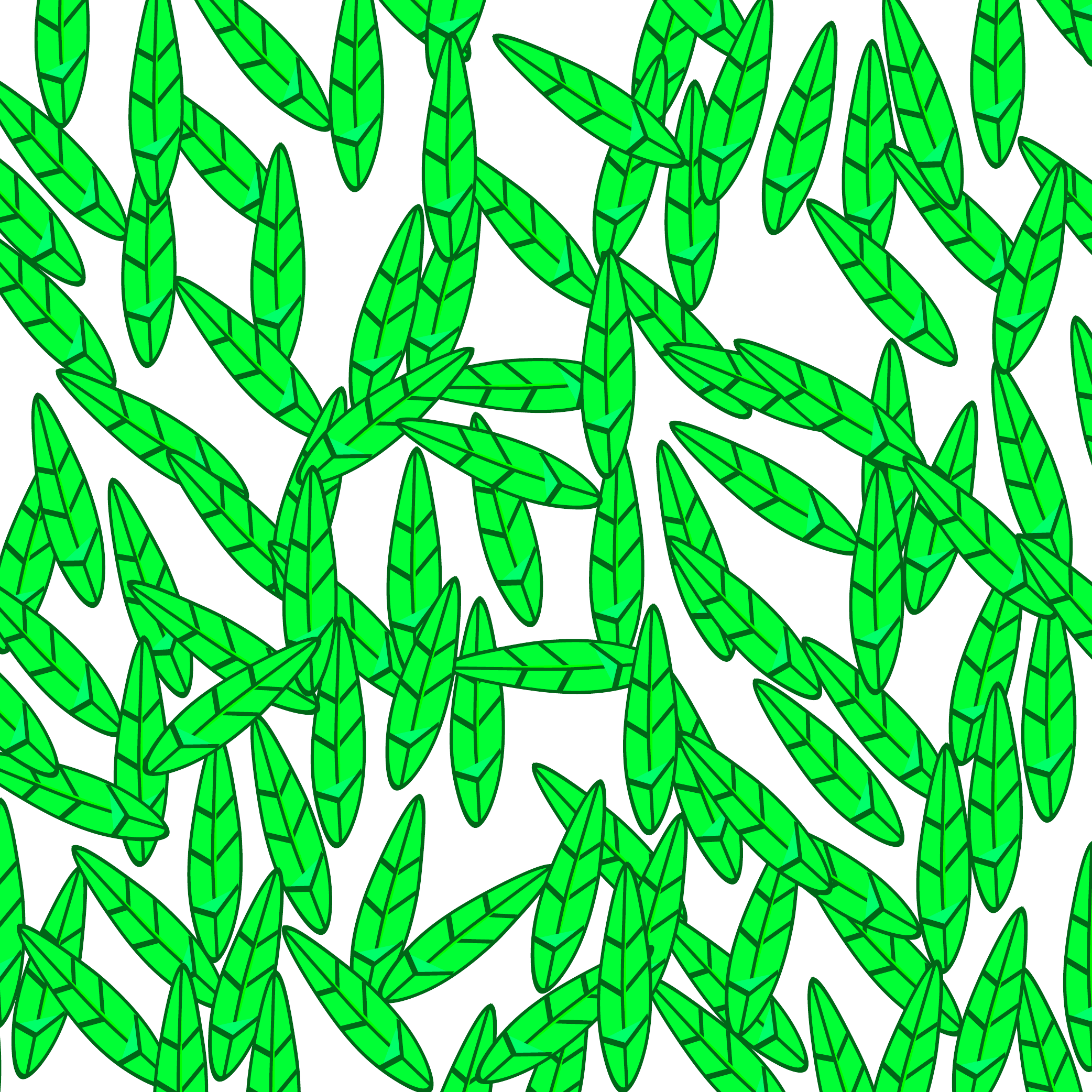}
}
&
\subfloat[]{
	\includegraphics[width=0.155\linewidth]{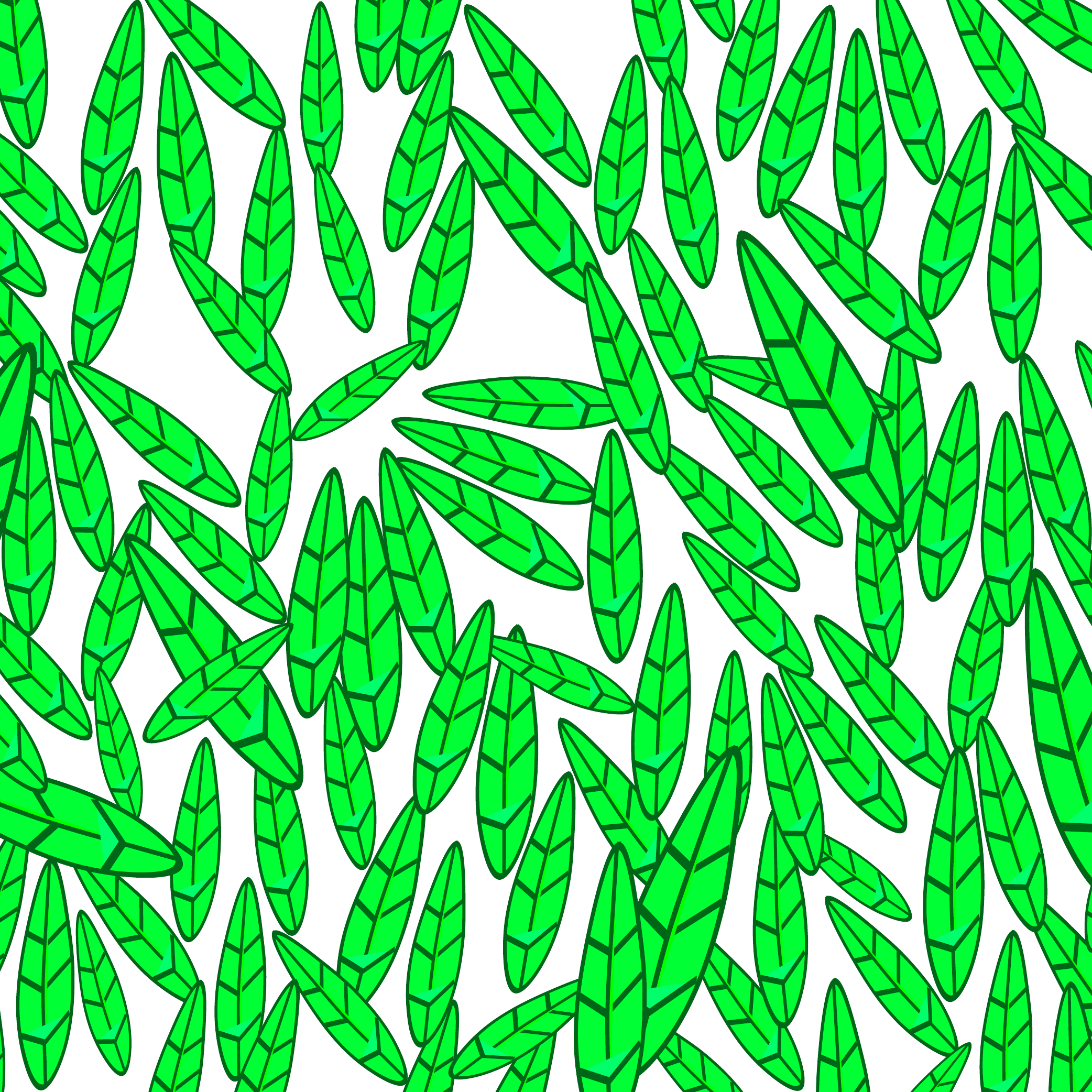}
}
&
\subfloat[]{
	\includegraphics[width=0.155\linewidth]{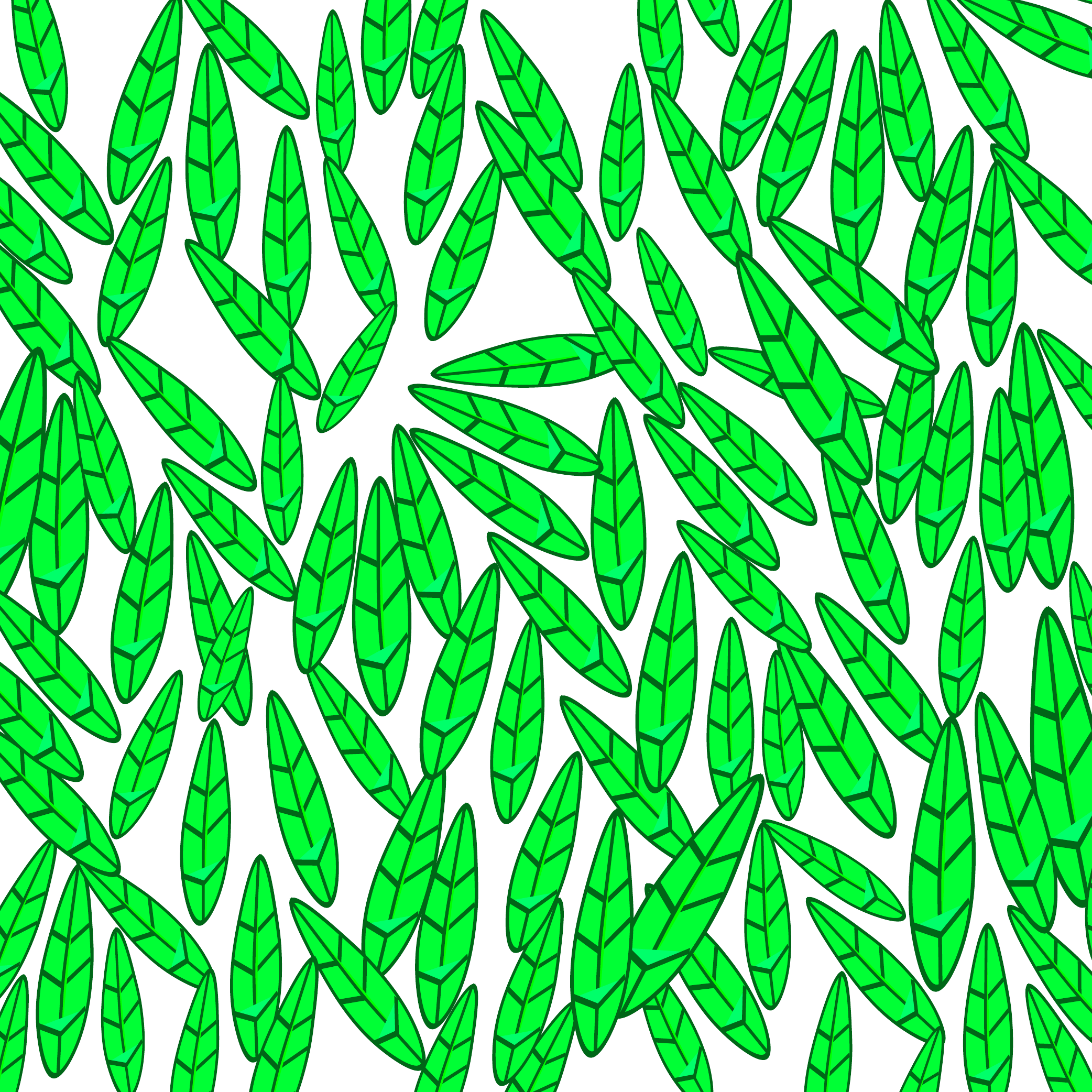}
}
&
\subfloat[]{
	\includegraphics[width=0.155\linewidth]{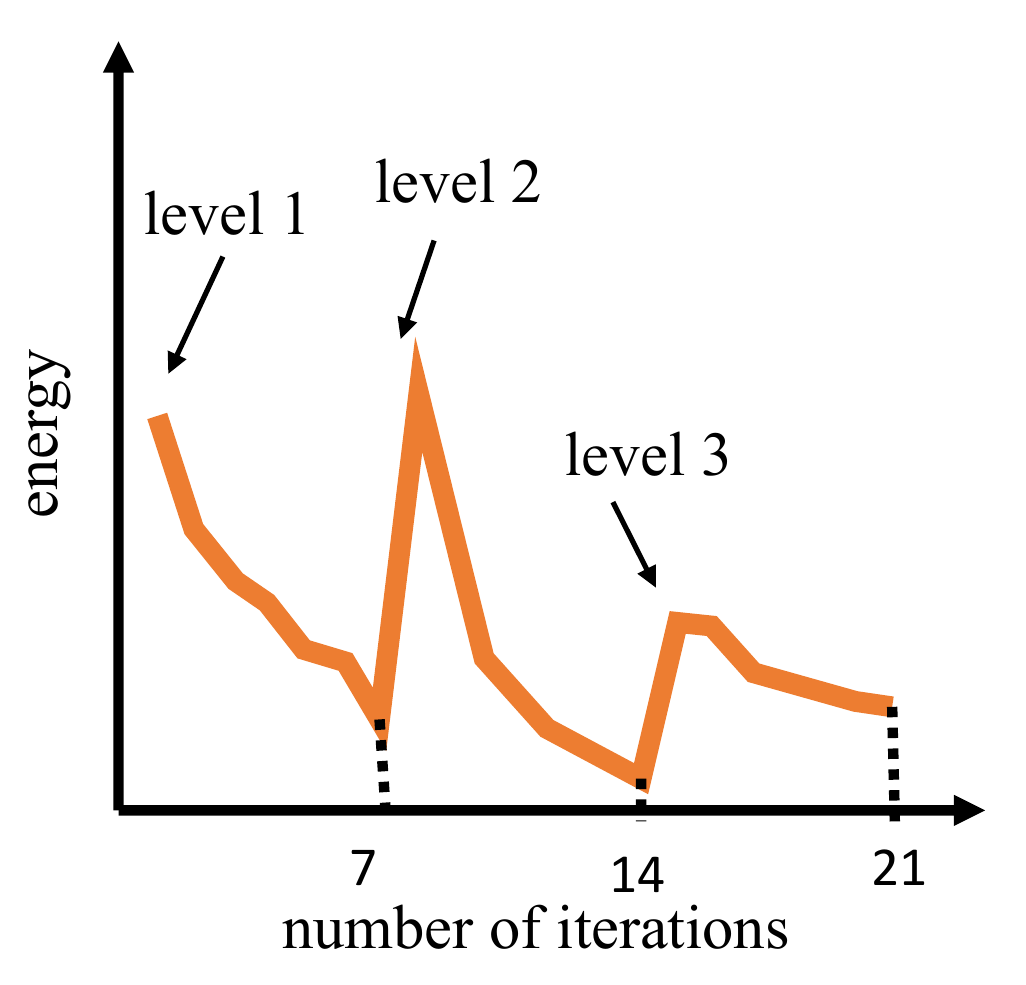}
}
\vspace{-2em}
\\
\subfloat[Exemplar]{
	\label{fig:hier_exemplar}
	\includegraphics[width=0.128\linewidth]{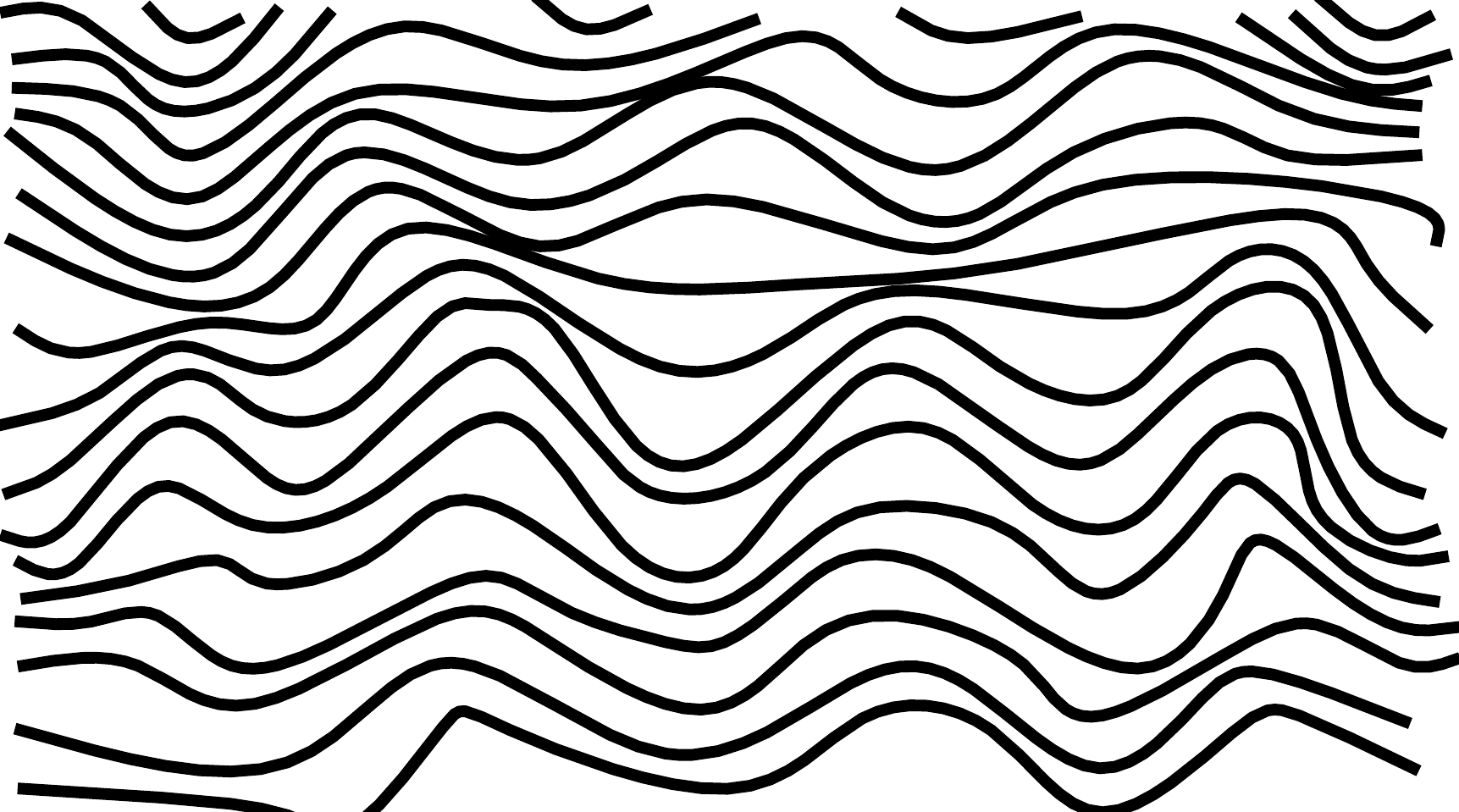}
}%
&
\subfloat[Initial step]{
	\label{fig:hier_init}
	\includegraphics[width=0.155\linewidth]{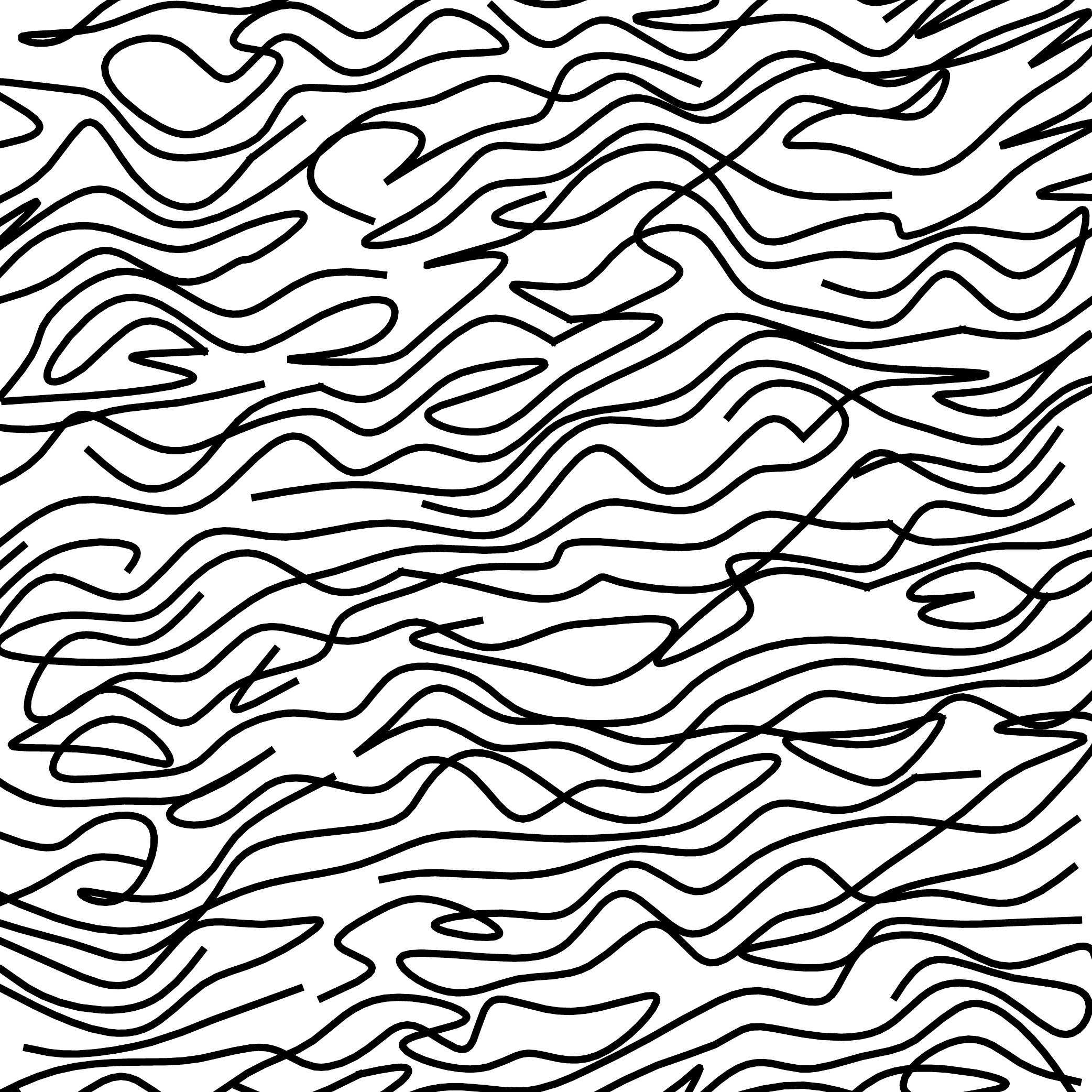}
}%
&
\subfloat[First level]{
	\label{fig:hier_1}
	\includegraphics[width=0.155\linewidth]{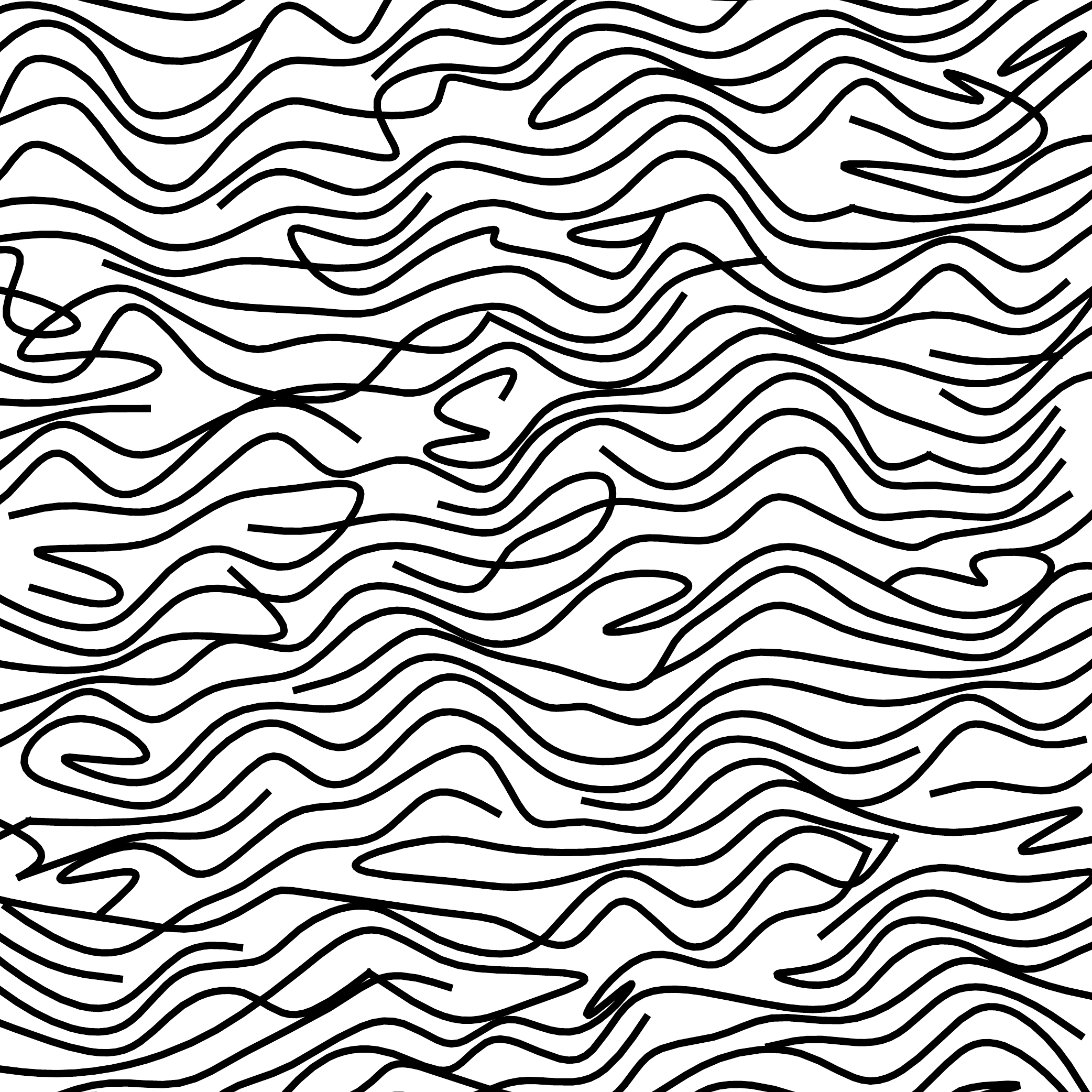}
}%
&
\subfloat[Second level]{
	\label{fig:hier_2}
	\includegraphics[width=0.155\linewidth]{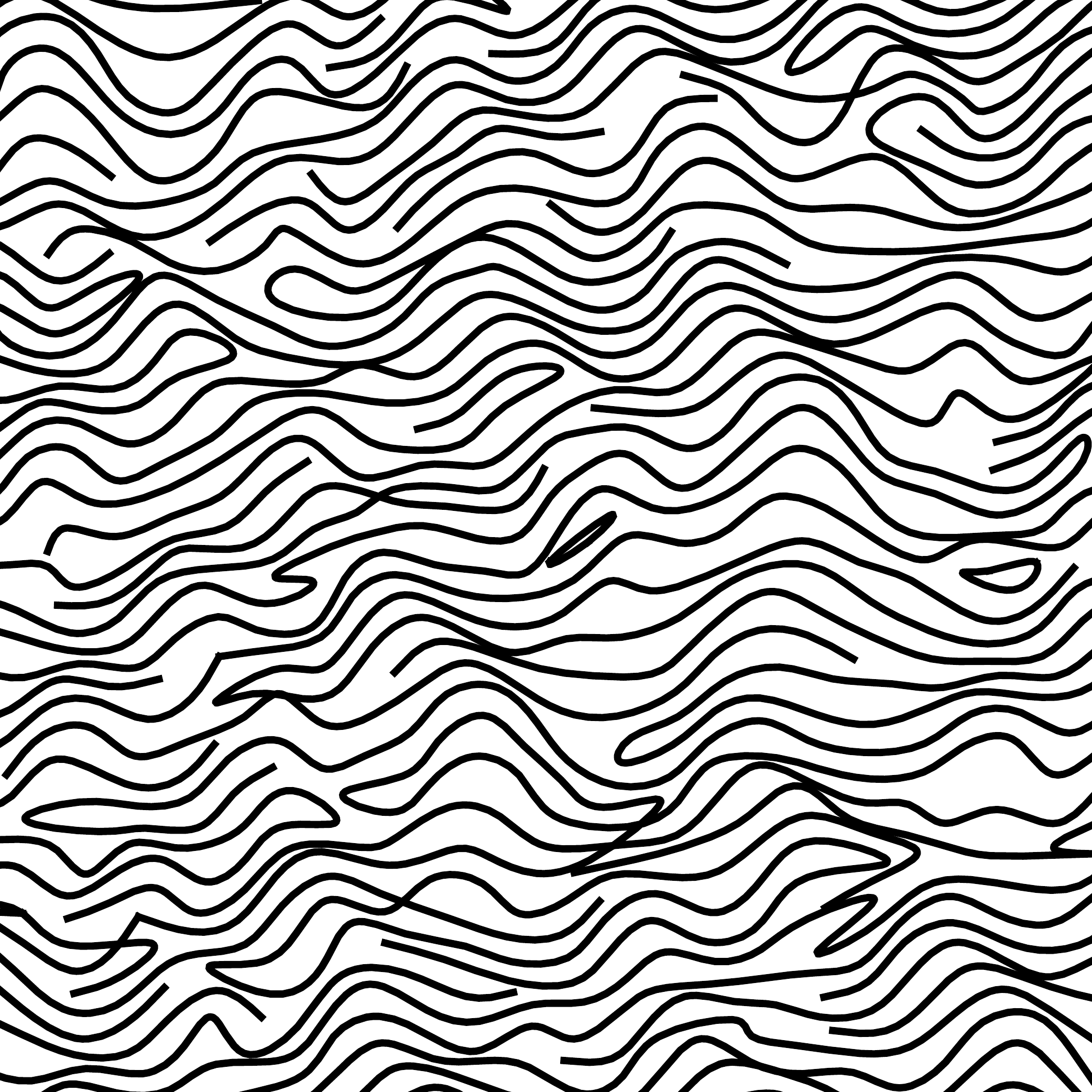}
}%
&
\subfloat[Third level]{
	\label{fig:hier_3}
	\includegraphics[width=0.155\linewidth]{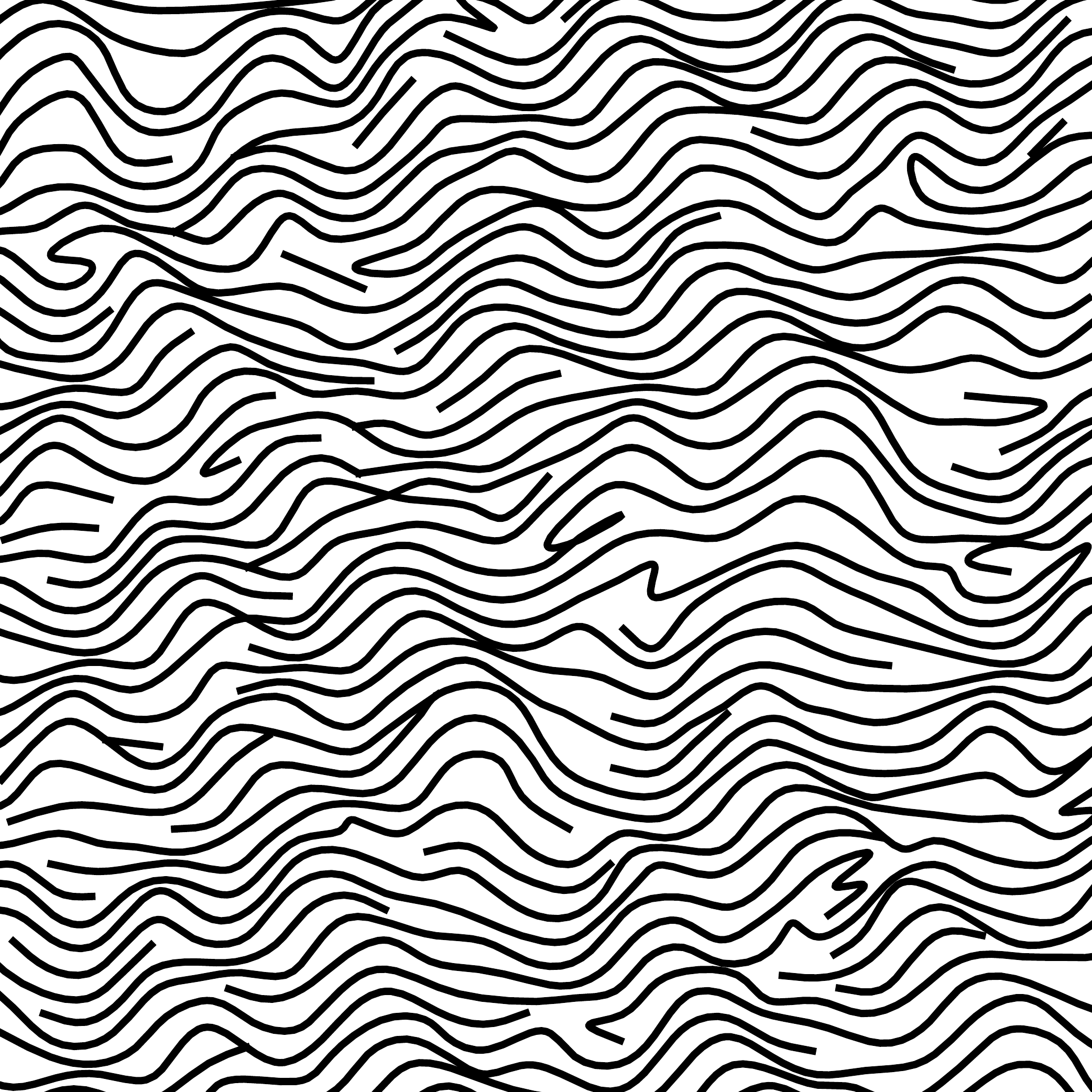}
}%
&
\subfloat[Energy plot]{
	\includegraphics[width=0.155\linewidth]{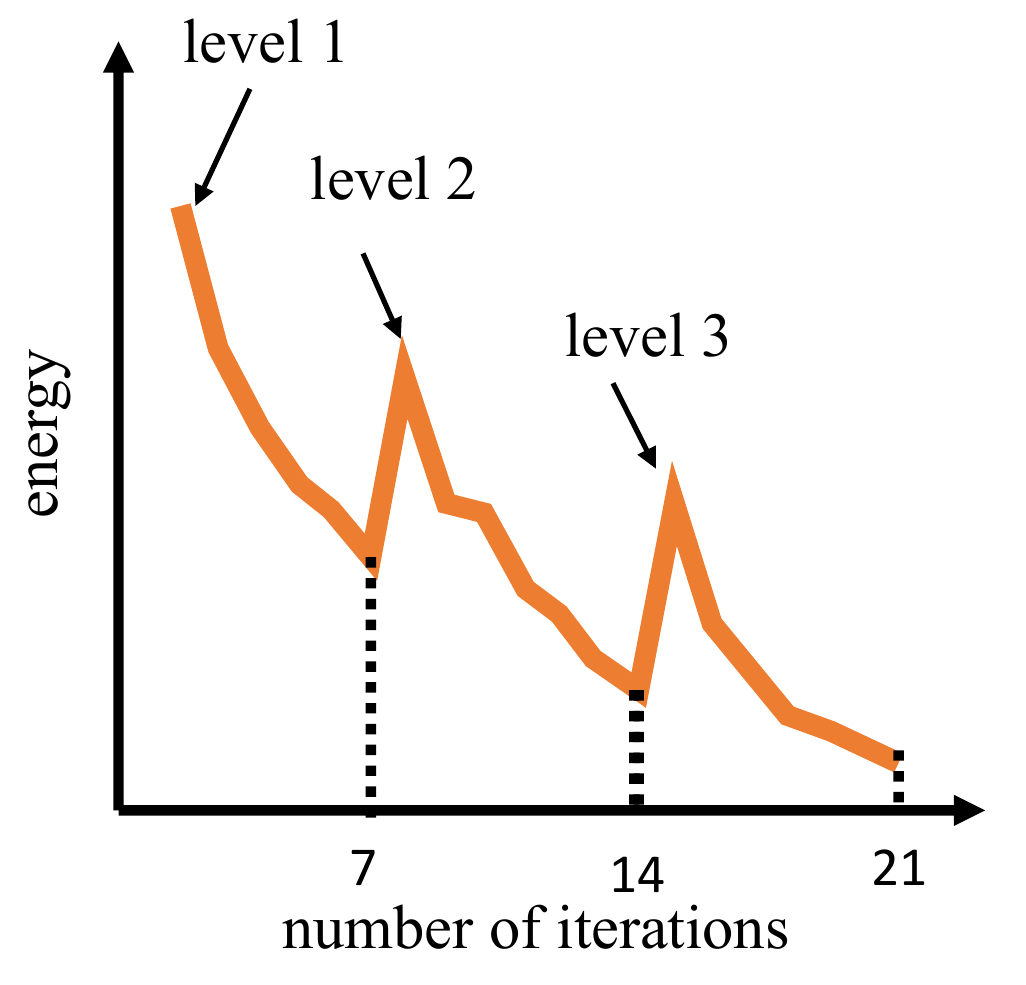}
}%
\end{tabular}

	\Caption{Hierarchical synthesis visualization.}
	{%
		The hierarchical synthesis proceeds from the coarsest (first level) to the finest levels (third level), with upsampling sample representations (\Cref{fig:hierarchical_repres}). 
		The hierarchical synthesis gradually refines the pattern from large to small scale structures.
		Note that the initialization has fewer elements than required (top row), hence our existence optimization adaptively controls the total number of elements. 
\nothing{
}%
\nothing{

}%
	}
	\label{fig:hierarchical_synthesis}
        \label{fig:synthesis}
\end{figure*}

Instead of using a single-resolution representation \cite{Ma:2011:DET}, we apply a hierarchical representation (\Cref{sec:representation:hierarchical}) for multi-resolution synthesis. 
We first synthesize the predictions at a coarse level using sparse representation, and then reconstruct the patterns based on sparse samples. 
We continue this process with a denser and denser pattern representation. \Cref{fig:hierarchical_repres} shows an example of multi-resolution element representation. 
For continuous structures, the sampling distance $\samplingdistance$ of continuous pattern is gradually increasing with respect to the level of hierarchy.
During synthesis, we use multi-scale neighborhood sizes to keep both large and local structures. 
The neighborhood size is gradually reduced at different hierarchies.
In our implementation, at each hierarchy,  there are 7 search-assignment iterations.
See \Cref{fig:hierarchical_synthesis} for an example.

\nothing{
Hierarchical synthesis is also more efficient and suitable to interactive systems.

}%

\nothing{

}%

\subsection{Pattern Reconstruction}
\label{subsec:pattern_reconstruction}

\nothing{

}%

The reconstruction step takes a synthesized pattern representation as input to generate output patterns that may consist of discrete elements and continuous structures.

\subsubsection{Discrete Elements}
For discrete elements, each sample is uniquely associated with an element. The reconstruction is to transform the element shapes by treating samples as control points. 
Specifically, we assume similarity transform to reconstruct the elements.

\subsubsection{Continuous Structures}
\label{subsubsec:recon_continuous_structures}
\nothing{
}%

\begin{figure*}[tbh]
	\centering

	\subfloat[2-neighbor $\samplesym$]{
	\label{fig:curve_recon:graph:1}
	\includegraphics[width=0.12\linewidth]{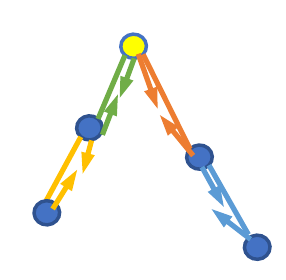}
       }%
        \subfloat[2 paths from \protect\subref{fig:curve_recon:graph:1}]{
	\label{fig:curve_recon:good_recon:1}
	\includegraphics[width=0.12\linewidth]{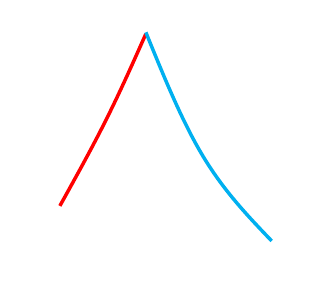}
        }%
        \subfloat[1 path from \protect\subref{fig:curve_recon:graph:1}]{
	\label{fig:curve_recon:bad_recon:1}
	\includegraphics[width=0.12\linewidth]{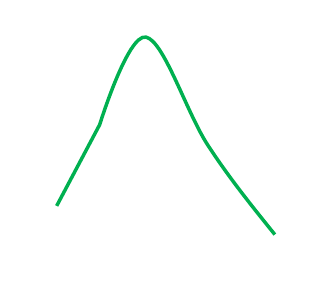}
        }%
	\subfloat[$>$2 neighbor $\samplesym$]{
	\label{fig:curve_recon:graph}
	\includegraphics[width=0.12\linewidth]{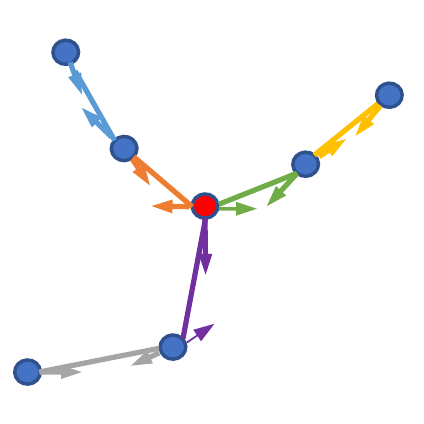}
	}%
	\subfloat[Recon from \protect\subref{fig:curve_recon:graph}]{
	\label{fig:curve_recon:good_recon}
	\includegraphics[width=0.12\linewidth]{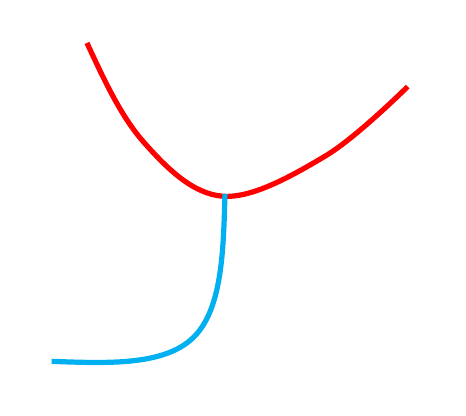}
	}%
	\subfloat[Other possible reconstructions from \protect\subref{fig:curve_recon:graph}]{
	\label{fig:curve_recon:bad_recon}
	\includegraphics[width=0.34\linewidth]{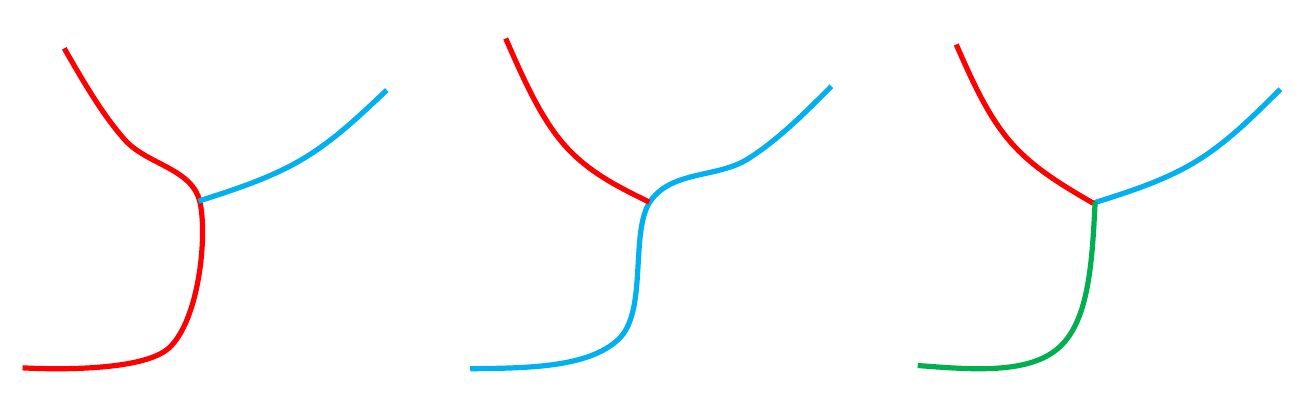}
}%

	\Caption{Curve reconstruction from a graph using orientation attributes.}
	{%
 If we only consider the different statuses of the yellow sample in \subref{fig:curve_recon:graph:1},
 there are two possible reconstructions \subref{fig:curve_recon:good_recon:1}\subref{fig:curve_recon:bad_recon:1}, depending on whether it is a junction \subref{fig:curve_recon:good_recon:1} or path \subref{fig:curve_recon:bad_recon:1} sample.
\nothing{
		Nearby blue samples are connected via edges without branches \subref{fig:curve_recon:graph}. So a pair of blue samples are identified within a path.
	}%
If we only consider the different statuses of the red sample in \subref{fig:curve_recon:graph}, 
there are four possible reconstructions \subref{fig:curve_recon:good_recon}\subref{fig:curve_recon:bad_recon},
depending on the red sample is included within which two or three paths.
\nothing{
Due to the ambiguity in the red junction sample, there are four possible path reconstructions. 
The red sample may be included in two or three paths.
}%
For the yellow (in \subref{fig:curve_recon:graph:1})  and red samples (in \subref{fig:curve_recon:graph}), we decide the reconstruction by examining its associated local orientation attribute and the fact a pair of local orientations of a sample should be opposite if the sample is included within the path. Our algorithm will reconstruct \subref{fig:curve_recon:good_recon:1} from \subref{fig:curve_recon:graph:1} and \subref{fig:curve_recon:good_recon} from \subref{fig:curve_recon:graph}.
\nothing{

}%
\nothing{
		First, we match local direction vectors with edges using Hungarian algorithm by minimizing the difference between local direction and graph edge angles. Matched direction vectors and edges are in the same colors.
		Second, we identify pairs of local path directions (if any) that are opposite, for example, the green arrows in \subref{fig:curve_recon:graph}. Then we include the edges and connected samples associated with the pair of local direction within a single path  \subref{fig:curve_recon:good_recon}. Other possible reconstructions from  \subref{fig:curve_recon:graph} are shown in \subref{fig:curve_recon:bad_recon}.
}%
	}
	\label{fig:curve_recon}
\end{figure*}

We have synthesized a graph whose sample positions and edge connections represent the topology of the output\nothing{ continuous patterns} (\Cref{sec:synthesis}).
However, these edges are piecewise linear, and they thus capture only connectivity/topology, but not shape/geometry information.
The original continuous patterns can be composed of smooth paths (e.g. quadratic B\'{e}zier curves).
Therefore, we need to reconstruct paths from the graph samples and edges. The samples are used as control points of B\'{e}zier curves.

\nothing{
}%

\nothing{
We will need to reconstruct continuous paths from a synthesized output graph, where each graph sample record several local path direction vectors.
}%
\nothing{

}%

Next, we talk about how to identify which sample and which edge are included within which path. 
This process relies on the synthesized sample orientation attributes $\sampleorientations$.

\nothing{
It is straightforward to include successively edge-connected samples (e.g. the blue samples in \Cref{fig:curve_recon:graph}) with no more than two graph neighbors within a path.
}%

\nothing{

}%

Samples with only one neighbor are unambiguous and thus only included within one path.
Samples with only two neighbors could be included in one path (all blue samples with two neighbors in \Cref{fig:curve_recon}) as path samples, or two paths as junction samples (e.g. the yellow sample in \Cref{fig:curve_recon:graph:1}).
Samples with more than two neighbors are junction samples (e.g. the red sample in \Cref{fig:curve_recon:graph}) that are included in multiple potential paths.

To disambiguate these cases, we examine a sample's local orientations $\sampleorientations(\samplesym)$.
Since $\sampleorientationentry \in \sampleorientations(\samplesym)$ should be tangent to the sample's local path, if a sample $\samplesym$ has a pair of orientations $\sampleorientationentry(\samplesym)$ that are almost opposite ($ 8\pi/9<$ absolute orientation difference $< 10\pi/9$), it will suggest that the sample is included inside a path as opposed to at the ends of a path. 
Therefore, there are three steps to reconstruct a pattern without ambiguity.
First, we identify pairs of local path orientations $\sampleorientationentry(\samplesym)$ (if any) that are opposite (e.g., a pair of arrows associated with 2-neighbor blue samples in \Cref{fig:curve_recon:graph:1,fig:curve_recon:graph}, or the orange and green ones associated with the red junction sample in \Cref{fig:curve_recon:graph}).
Second, we match local orientations $\sampleorientationentry(\samplesym)$ with edges $\sampleedge \in \sampleedgeset(\samplesym)$ connected to the sample using the Hungarian algorithm by minimizing the sum of absolute difference between local orientation and edge angles.
The arrows and graph edges in  \Cref{fig:curve_recon:graph:1,fig:curve_recon:graph} with the same colors are matched.
Third, we generate a path by including edges that are connected together and matched with opposite orientations. 
A B\'{e}zier curve is generated by interpolating samples along a path.
This reconstruction strategy using $\sampleorientations$ can help
preserving the original curve shapes, as demonstrated in \Cref{fig:ablation:orientation}.

\nothing{
we find samples that are connected to the junction sample that are also in that path.
Then we include the edges and connected samples matched with the pair of local directions within a single path (\Cref{fig:curve_recon:good_recon}).
}%

\nothing{
}%

\nothing{%

}%

\nothing{
}%

\nothing{

On the boundary between output and input patterns, we also need to compute connections between output and input samples, which we do not include in the algorithm description for notation simplification.
}%

\nothing{
\paragraph{Initial graph reconstruction}
\label{subsubsec:recon_continuous_structures:init}

In sample synthesis, we have the correspondence between a few pair of output and input neighborhoods $\{\neighoutput,\neighinput,\match\}$.
For input samples, we already record the edge connections among them in adjacency matrix $\inputadjacencymatrix$.
The initial graph reconstruction directly transfers the input edges $\inputadjacencymatrix$ to the output $\outputadjacencymatrix$ based on $\{\neighoutput,\neighinput,\match\}$.

For a pair of output samples $\sampleoutput$ and $\sampleoutputprime$, we have a pair of matched input samples $\sampleinput$ and $\sampleinputprime$. 
If there is a direct connection between $\sampleinput$ and $\sampleinputprime$, in other words, $\inputadjacencymatrix^{\sampleinput,\sampleinputprime}$=1, we add a vote between $\sampleoutput$ and $\sampleoutputprime$: $\outputadjacencymatrix^{\sampleoutput,\sampleoutputprime} \pluseq 1$.
Finally, $\outputadjacencymatrix^{\sampleoutput,\sampleoutputprime}$ is normalized with the number of overlapping neighborhoods on the edge. The entries in $\outputadjacencymatrix$ represent the existence confidence of the corresponding edges (\Cref{fig:initial_graph_reconstruction}).

\nothing{
Using breadth-first search, the shortest path between $\sampleinput$ and $\sampleinputprime$ on input graph, is computed.
For every pair of $\sampleinput$, $\sampleinputprime$, we find the shortest paths $\reconpathinput$ (on input example graph) between them.  For every $\reconpathinput$, we can find the matched path among output samples $\reconpathoutput$.
Thus, there are many overlapped paths the may compose of the output continuous graph structure.
We then {\em merge} these overlapped paths into a continuous graph and greedily remove unreliable noisy edges based on each sample's $\sampleconnections$.
The pseudocode of our algorithm is shown in \Cref{alg:continuous_recon}.
}%

\paragraph{Pattern graph denoising with minimum spanning tree (MST)}
\label{subsubsec:recon_continuous_structures:denoise}

$\outputadjacencymatrix$ is noisy. We observe the noisy pattern graph has many {\em cycle noise}. 
That is, a good pattern graph  (\Cref{fig:example_graph_init}) usually does not contain unnecessary small graph cycles and thus is smoother. %
To get rid of such noise, we first remove outlier edges via thresholding $\outputadjacencymatrix^{\sampleoutput,\sampleoutputprime} < 0.2$.
Then we remove cycles by detecting them with breadth-first search and computing minimum spanning trees on nodes that are included in cycles, as trees are naturally free of cycles and smoother.
The edges that are not in the MST is removed. In the MST computation, the edge weight is  $1/\outputadjacencymatrix^{\sampleoutput,\sampleoutputprime}$, which says, the edges with low-confidence have more cost to stay in the final reconstruction.
Graph denoising (\Cref{fig:denoising}) resembles image denoising with image filters.

\nothing{
We can remove edges that have less votes. We normalize $\outputadjacencymatrix^{\sampleoutput,\sampleoutputprime}$ by diving each edge with the number of overlapped neighborhoods of it as $\outputadjacencymatrix^{\prime\sampleoutput,\sampleoutputprime}$.
We filter out edges with $\outputadjacencymatrix^{\prime\sampleoutput,\sampleoutputprime} < 0.1$. 
Then we greedily remove edges that have smallest $\outputadjacencymatrix^{\prime\sampleoutput,\sampleoutputprime}$, if all of $\sampleconnections$ of its connected nodes is smaller than the number of edges associated with it, which is recorded in $\outputadjacencymatrix^{\prime\sampleoutput,\sampleoutputprime}$.

We observe a good pattern graph does not contain small cycles. To further refine the graph reconstruction results, the small cycles are detected with breadth-first search. 
A MST is computed with nodes on the cycles. The edges that are not in the MST is removed. In the MST computation, the edge weight is  $1/\outputadjacencymatrix^{\prime\sampleoutput,\sampleoutputprime}$.
}

}%

\nothing{
In the synthesized samples, each sample records an orientation vector.
Using only this information, which is noisy, it's not easy to reconstruct the continuous patterns. 
Thus, We also incorporate information from the exemplar. For a pair of output samples $\sampleoutput$ and $\sampleoutputprime$, we have a pair of matched input samples $\sampleinput$ and $\sampleinputprime$. Ideally, the geodesic distance between $\sampleinput$ and $\sampleinputprime$ should be equal to that between $\sampleoutput$ and $\sampleoutputprime$.

The problem is abstracted as follows. $\{\sampleoutput\}$ is a set of graph nodes. We want to construct a graph from these nodes with the following conditions; For each node, we know the number of connections $\{\sampleconnections(\sampleoutput)\}$ it has, and their orientations $\{\sampleorientations(\sampleoutput)\}$. The expected geodesic distances between pair of nodes in the constructed graph are also known. Note these information might be noisy.
}

\nothing{
}

\nothing{
\subsection{Element Extraction}
In previous sections, we assume 1) discrete elements and continuous structures are classified from the workflow; 2) Continuous structures are also segmented as 
Here, we present the method of detecting individual elements. The undetected are treated as continuous structures.
}

\nothing{
Ahuja et al \shortcite{Ahuja:2007:ETN} extracts texels from raster textures by performing repetition detection on tree structures constructed by multi-scale image segmentation, instead of on raster images themselves, which reduces the algorithm search space. 
Inspired by their work, we segment recent user workflow hierarchically within a temporal window to obtain a segmentation tree. 

After extracting individual elements, we compute a mean shape for element of the same categories.

To determine the sampling strategy, we compute circular variance.

We apply curvature-based sampling strategy to sample the elements. 

}

\nothing{
\subsubsection{Continuous graph construction}
Remaining workflows are identified as belonging to continuous structures. Since strokes are, in nature, discrete, we merge  stroke samples that are sufficiently close as joint node.
}

\section{Evaluation}
\label{sec:evaluation}
\label{sec:study}

We evaluate our method with sample results, ablation studies, and comparisons with existing art.
We will make our code repository \cite{Tu:2020:CCTC} public to facilitate future research.

\subsection{Results}
\label{sec:result}

\nothing{

}%
\nothing{
We share the executable of the system in the supplementary material for reviewers to evaluate it.
}%

\nothing{
}%
\nothing{

}%

\begin{figure*}[tb!p]
	\centering
	\setlength{\tabcolsep}{-2.5pt}
	\captionsetup[subfigure]{justification=centering}
\newcommand{\figuresize}{0.155}

\begin{tabular}{cccccccccc}
	\centering
\subfloat[Zentangle]{
	\label{fig:zentangle:exemplar}
	\includegraphics[width=0.0723\linewidth]{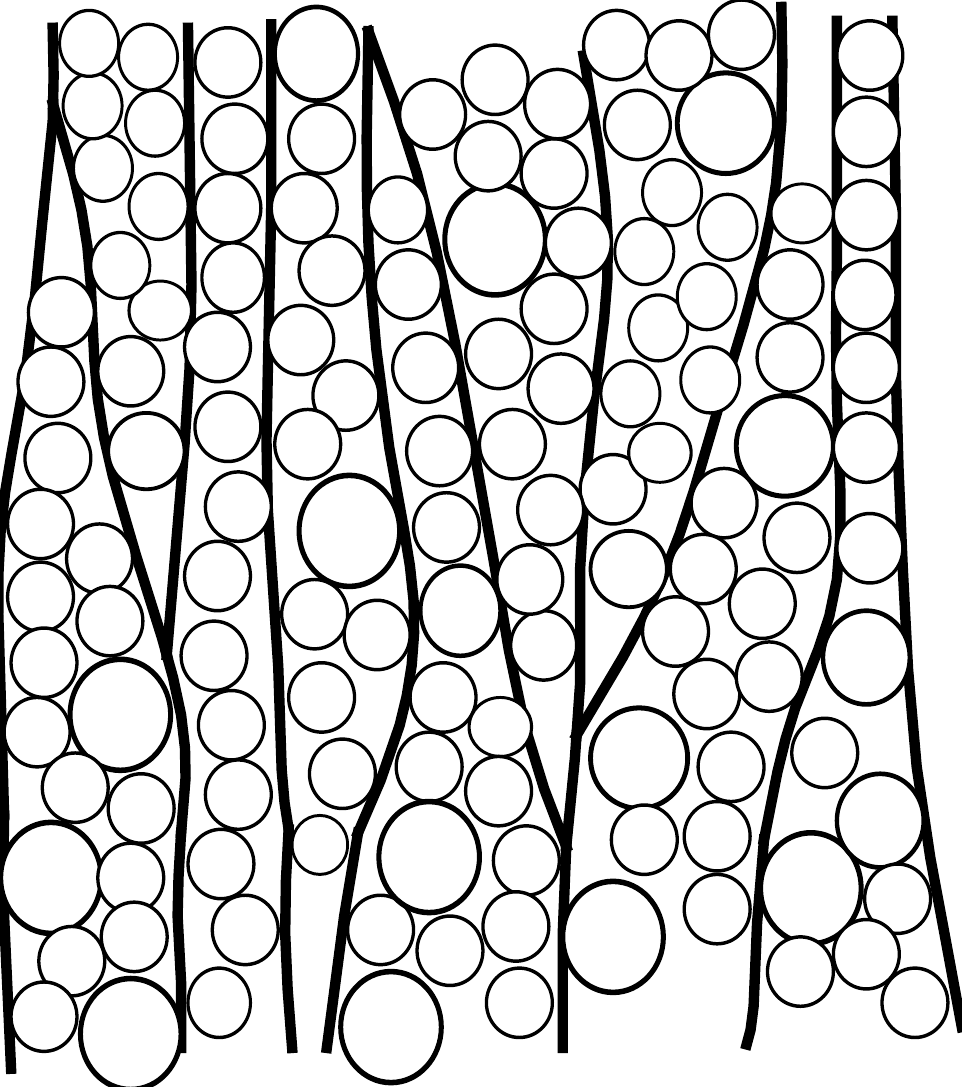}
}%
&\subfloat[]{
	\label{fig:zentangle:automatic_synthesis}
	\includegraphics[width=\figuresize\linewidth]{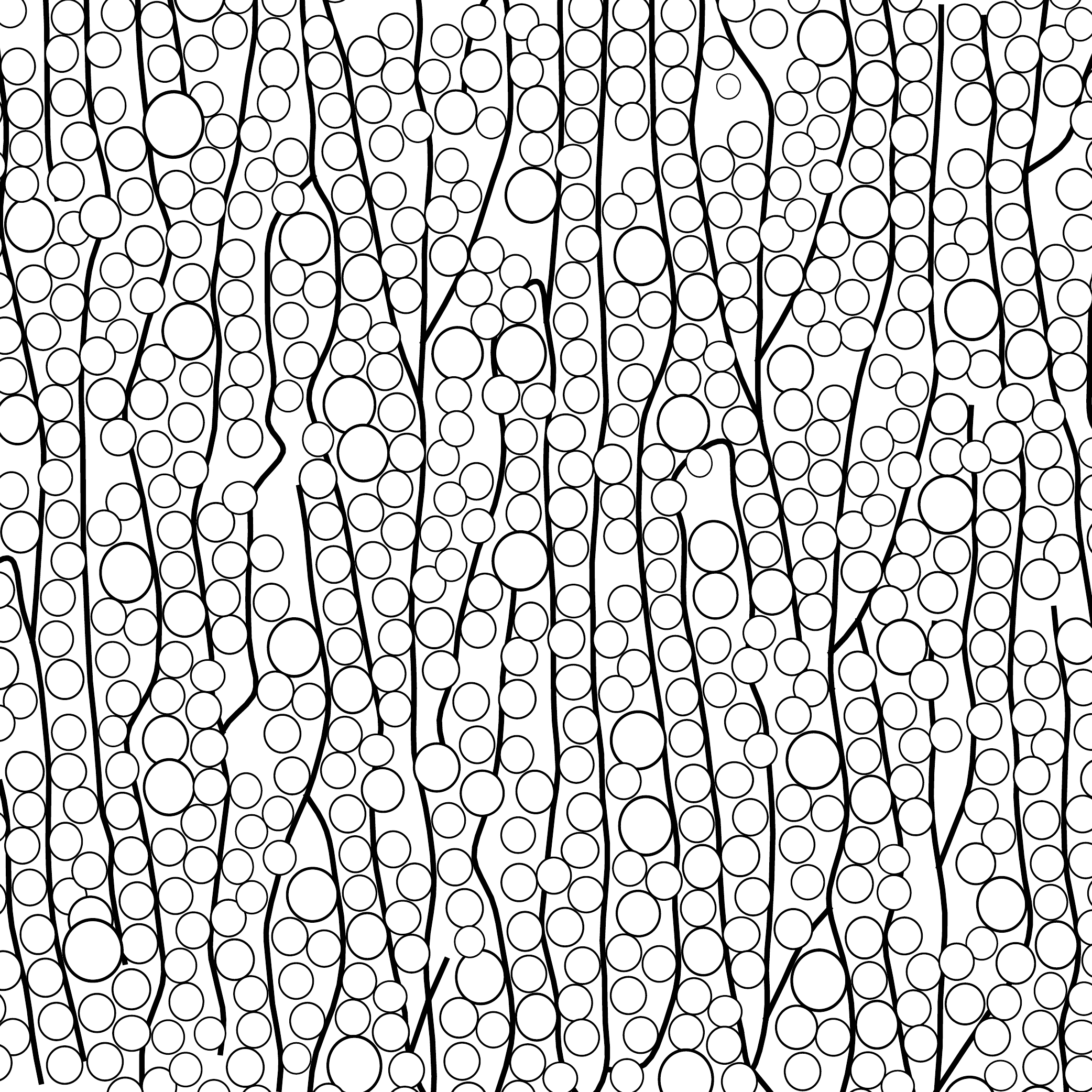}
}%
&	\subfloat[Diffusion]{
	\label{fig:diffusion:exemplar}
	\includegraphics[width=0.097\linewidth]{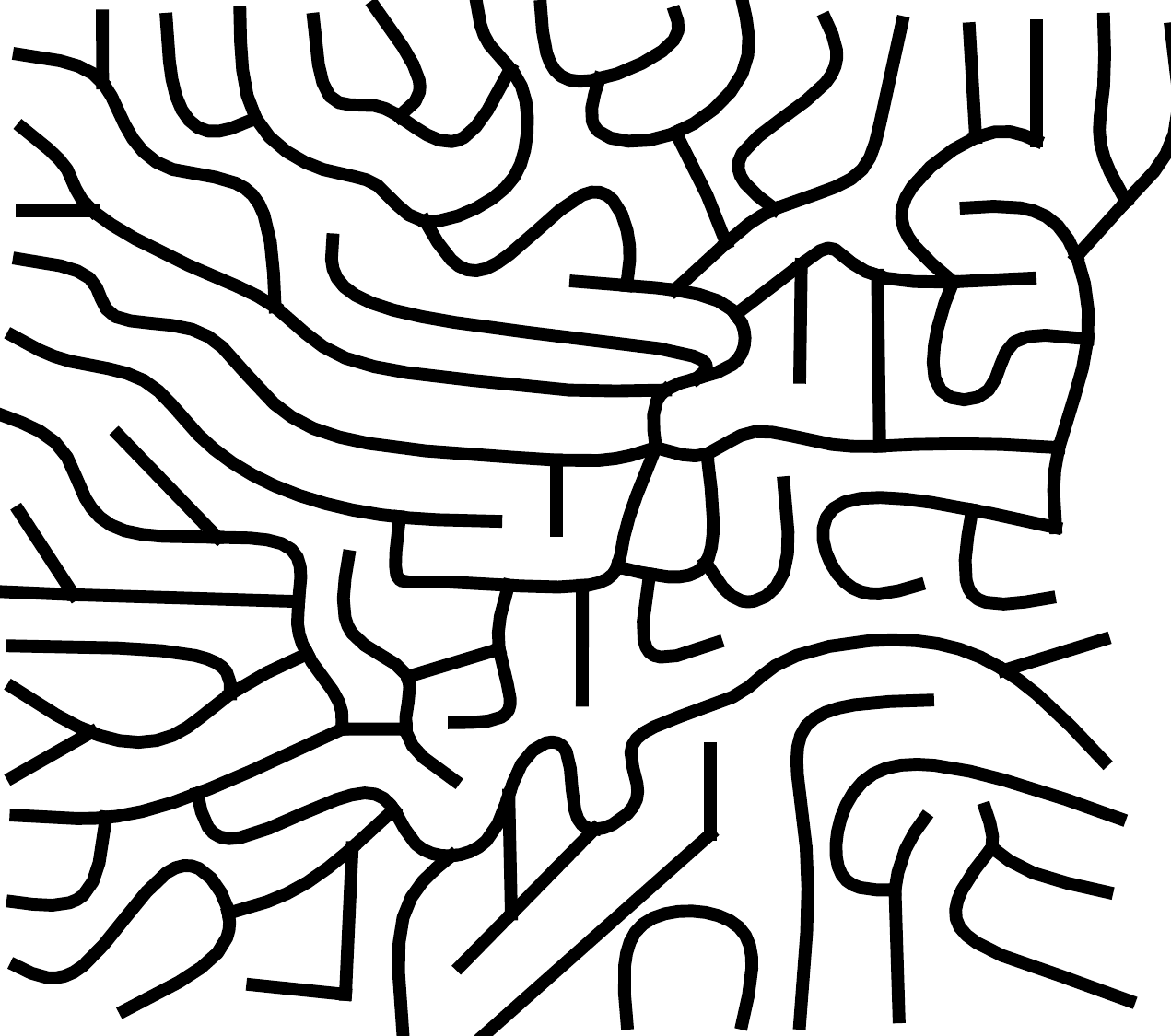}
}%
&\subfloat[]{
	\label{fig:diffusion:automatic_synthesis}
	\includegraphics[width=\figuresize\linewidth]{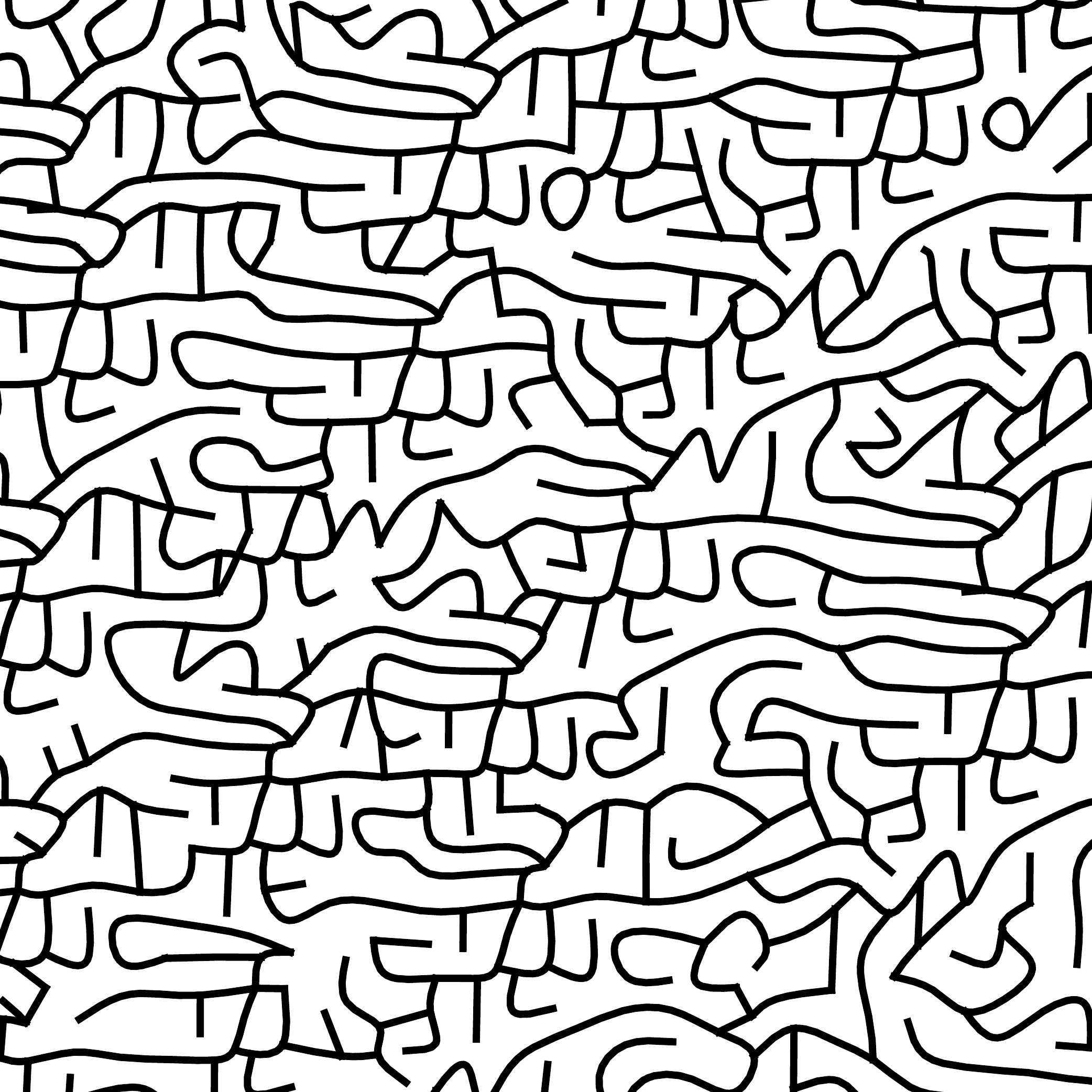}
}%
&\subfloat[Wet flow]{
	\label{fig:distorted_blocks:exemplar}
	\includegraphics[width=0.0831\linewidth]{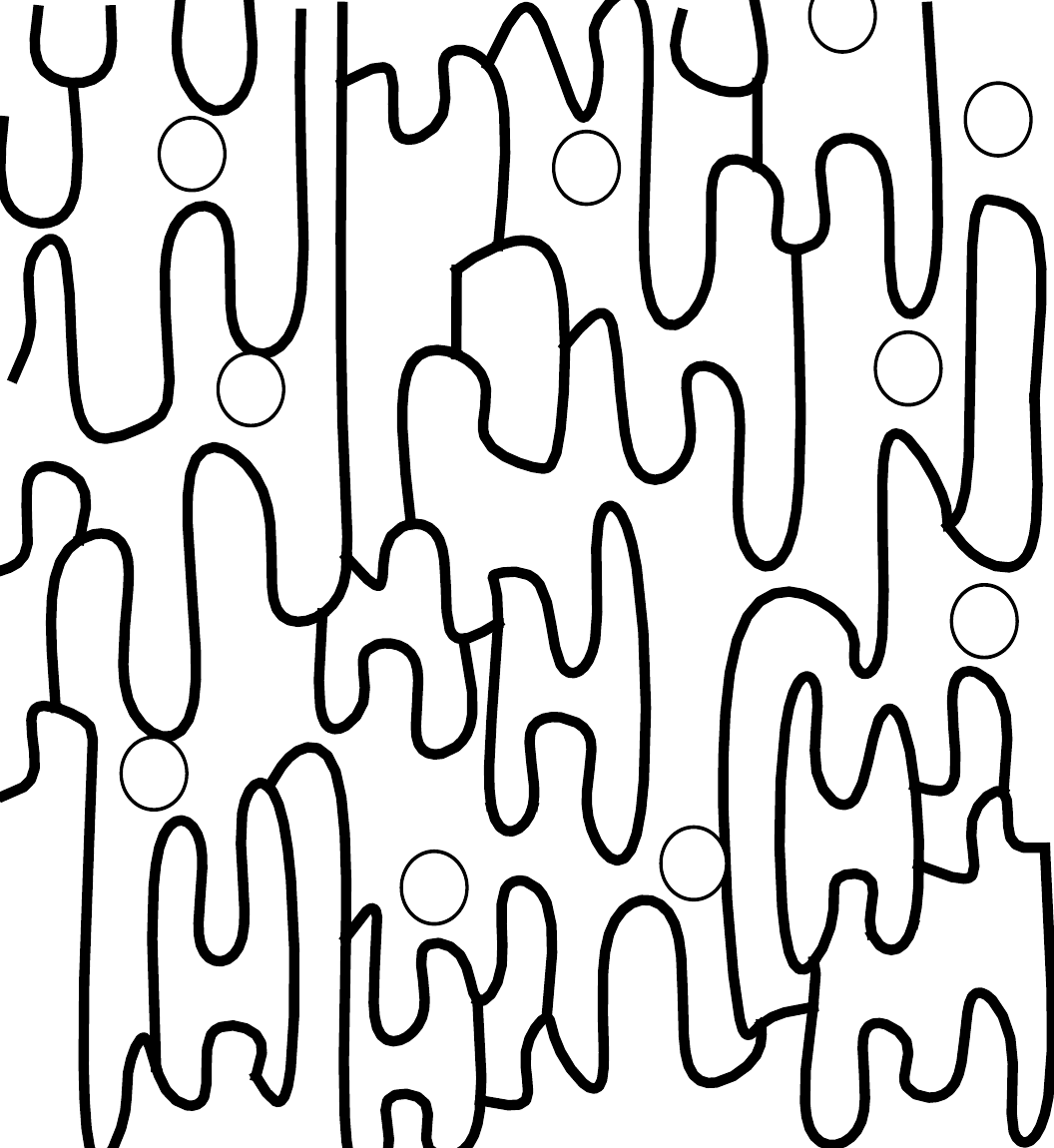}
}%
&\subfloat[]{
	\label{fig:distorted_blocks:automatic_synthesis}
	\includegraphics[width=\figuresize\linewidth]{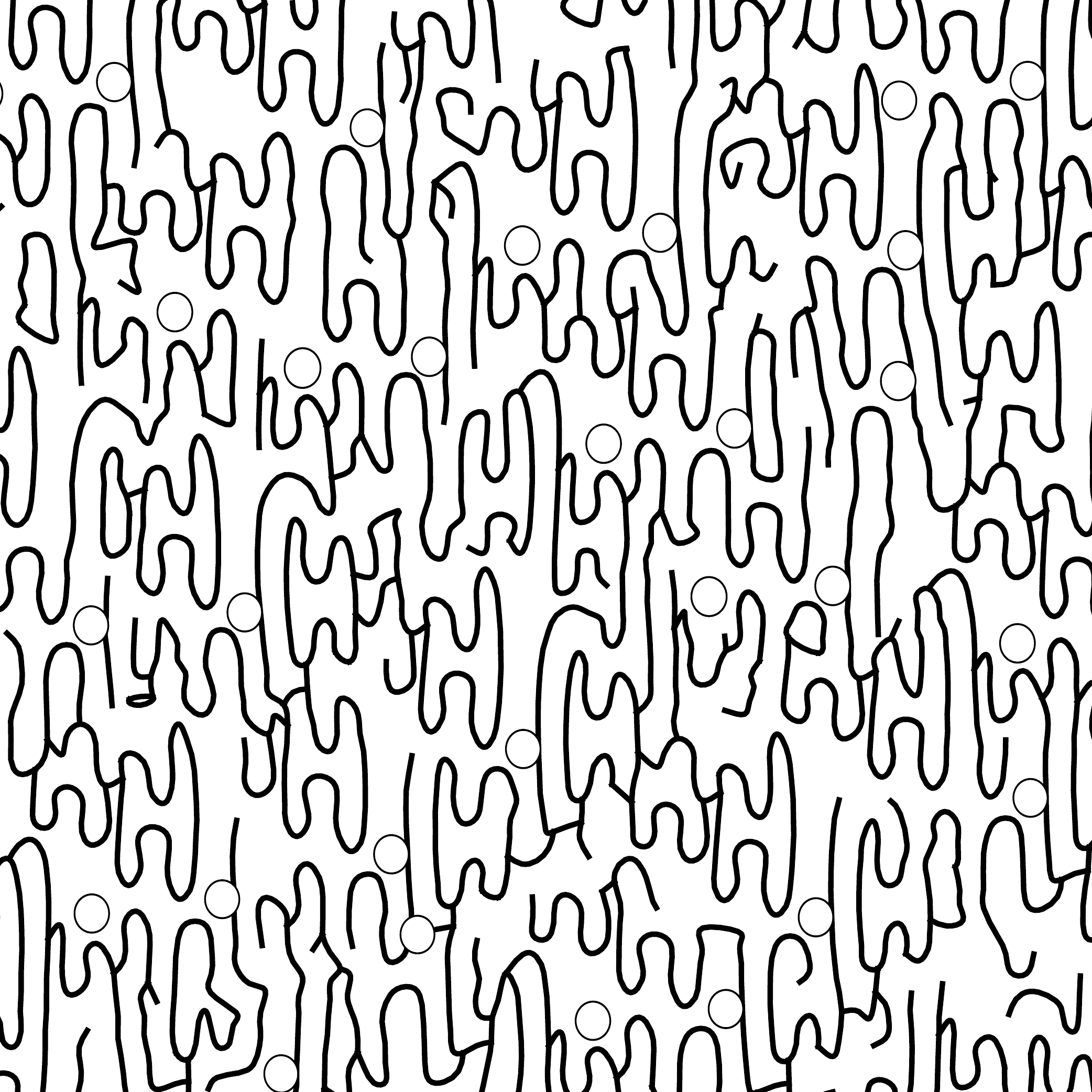}
}%
&
	\subfloat[Brick wall]{
	\label{fig:regular_brick_wall:exemplar}
	\includegraphics[width=0.0898\linewidth]{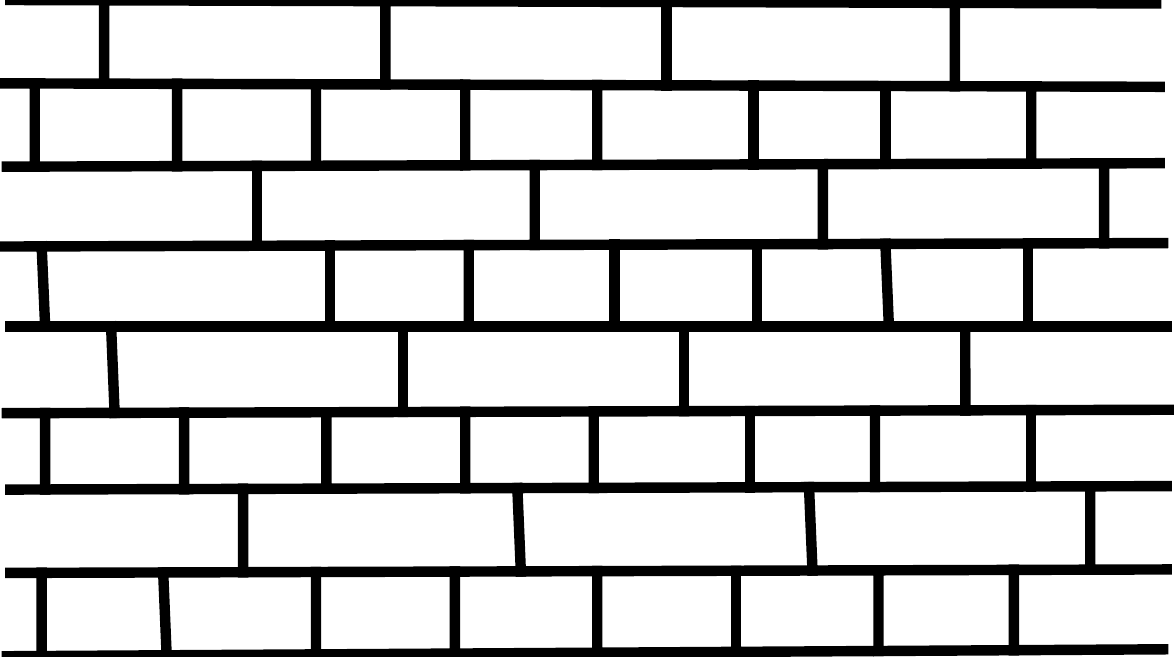}
}%
&\subfloat[]{
	\label{fig:regular_brick_wall:automatic_synthesis}
	\includegraphics[width=\figuresize\linewidth]{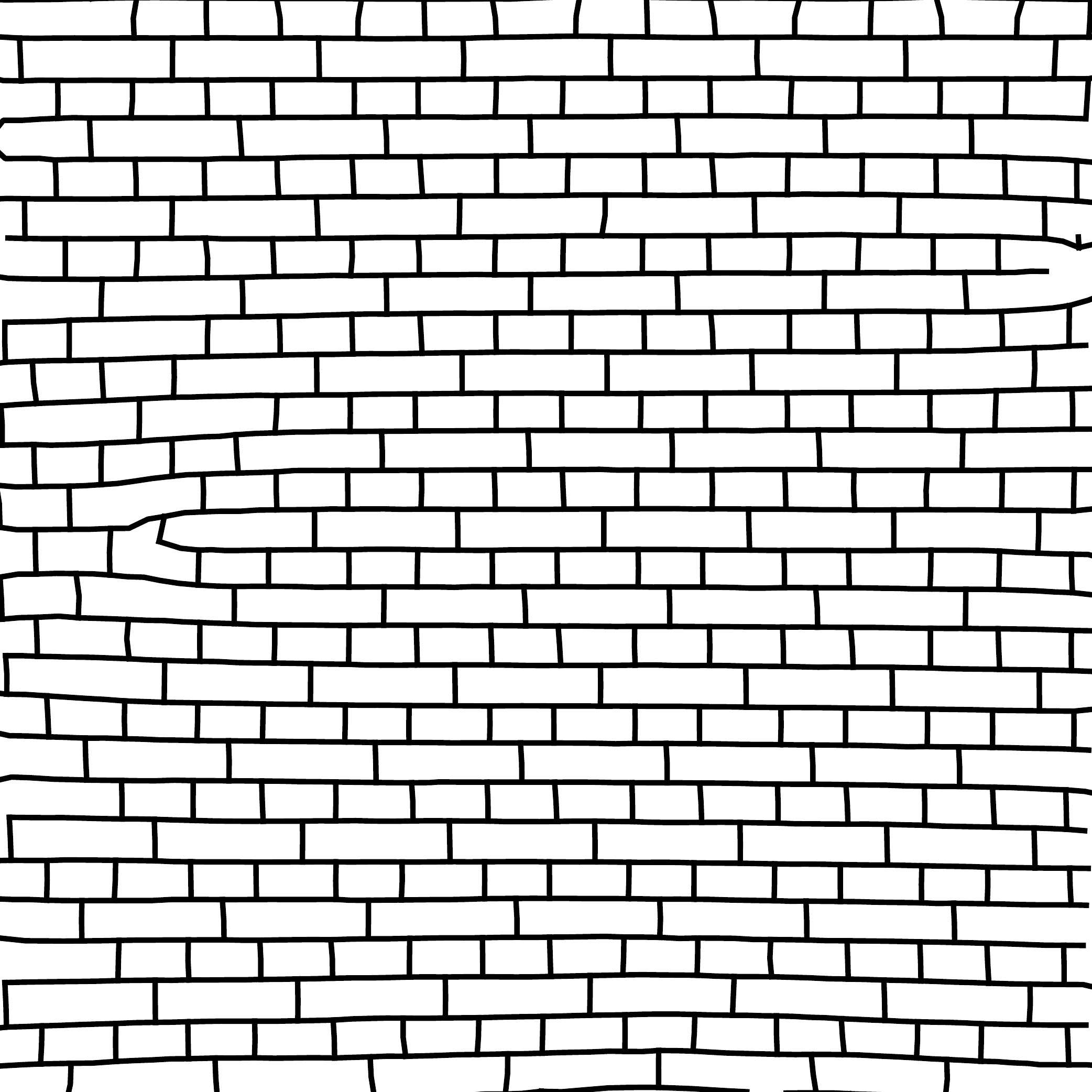}
}%
\\
	\subfloat[Fence]{
	\label{fig:fence:exemplar}
	\includegraphics[width=0.0505\linewidth]{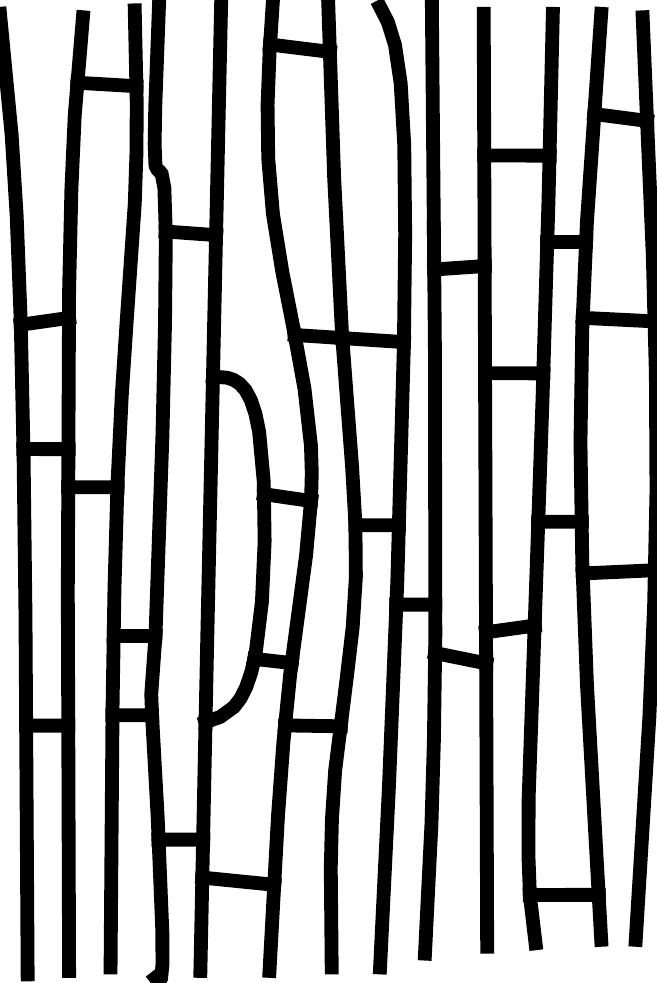}
}%
&\subfloat[]{
	\label{fig:fence:automatic_synthesis}
	\includegraphics[width=\figuresize\linewidth]{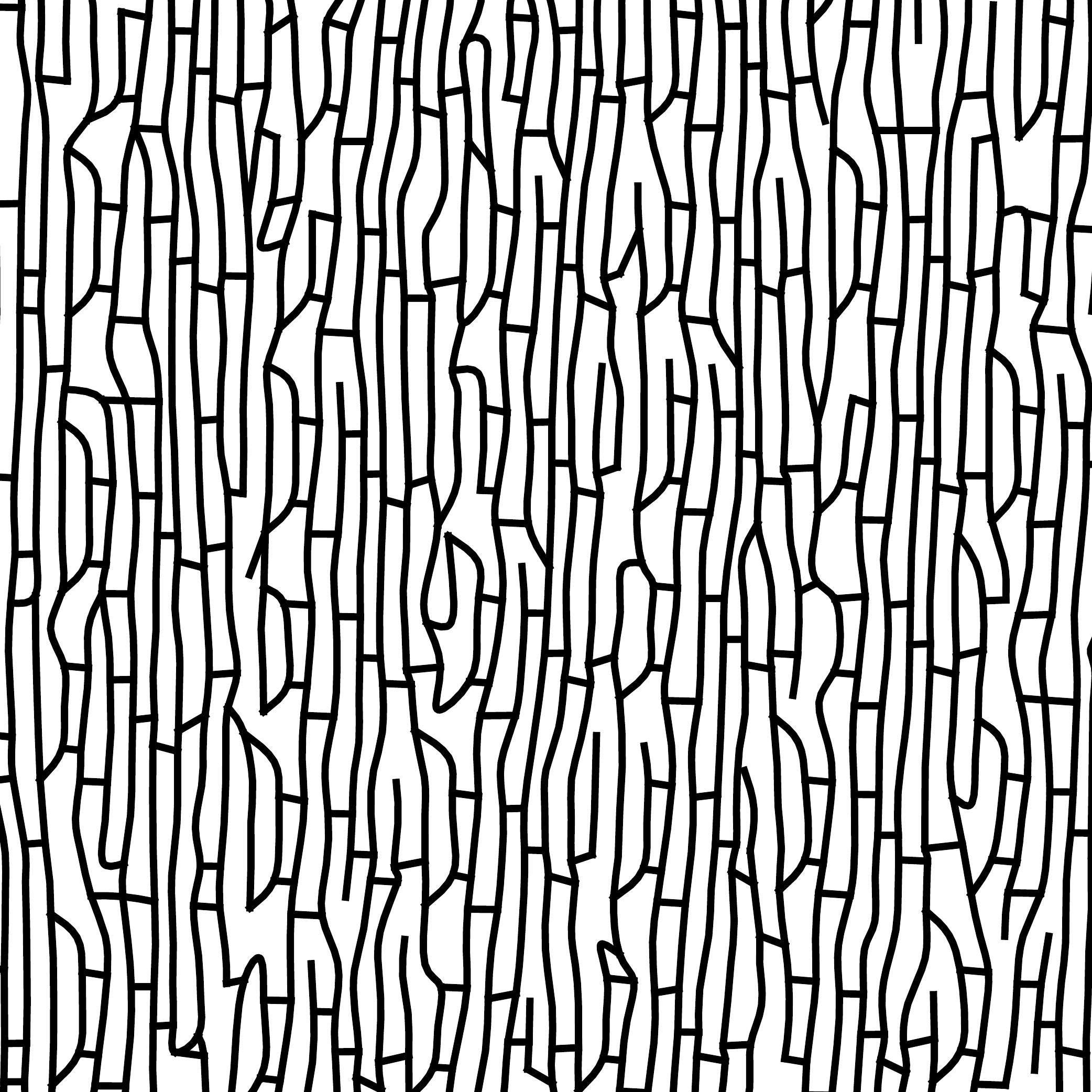}
}%
&
\subfloat[Blocks]{
	\label{fig:blocks:exemplar}
	\includegraphics[width=0.0911\linewidth]{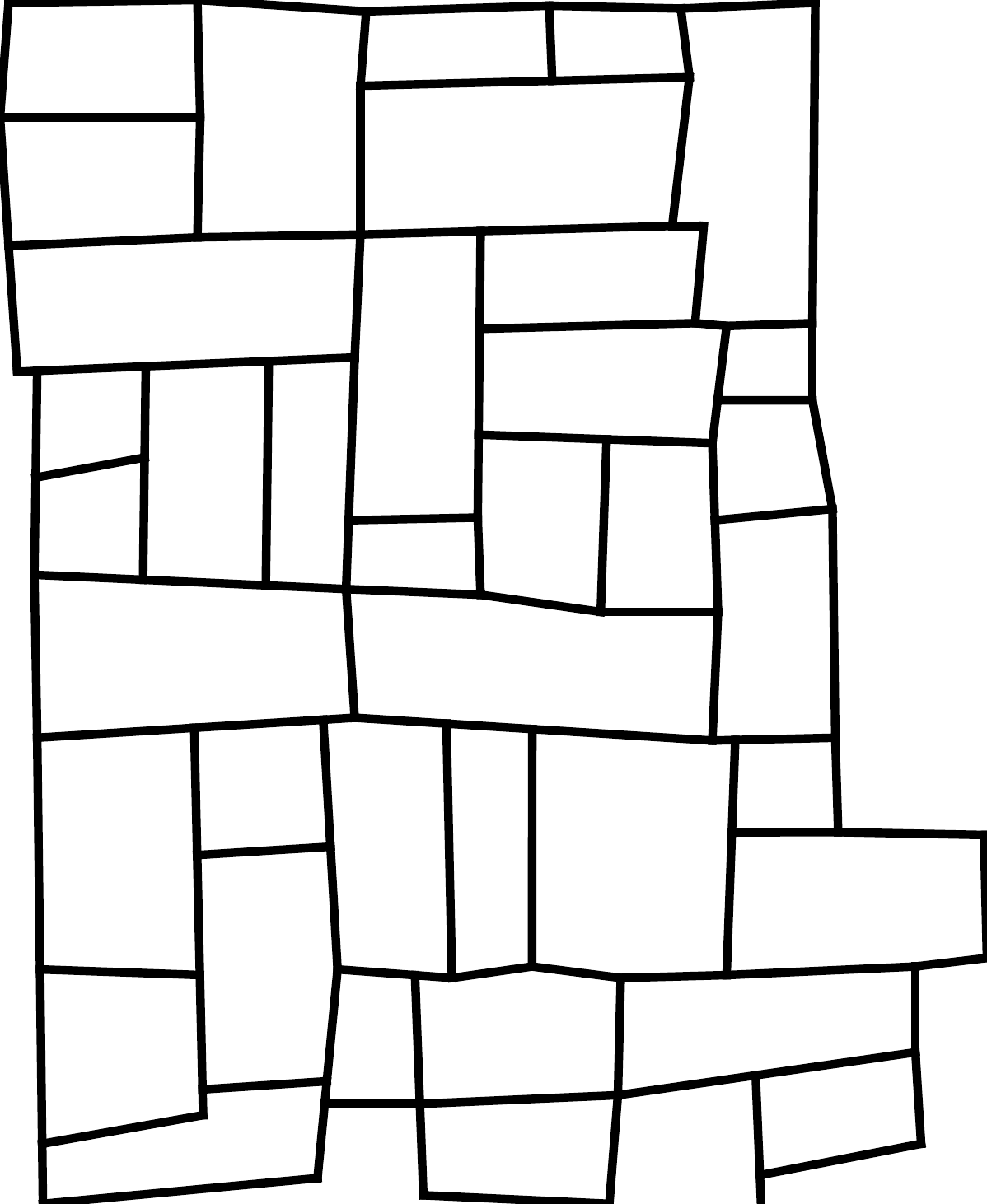}
}%
&\subfloat[]{
	\label{fig:blocks:automatic_synthesis}
	\includegraphics[width=\figuresize\linewidth]{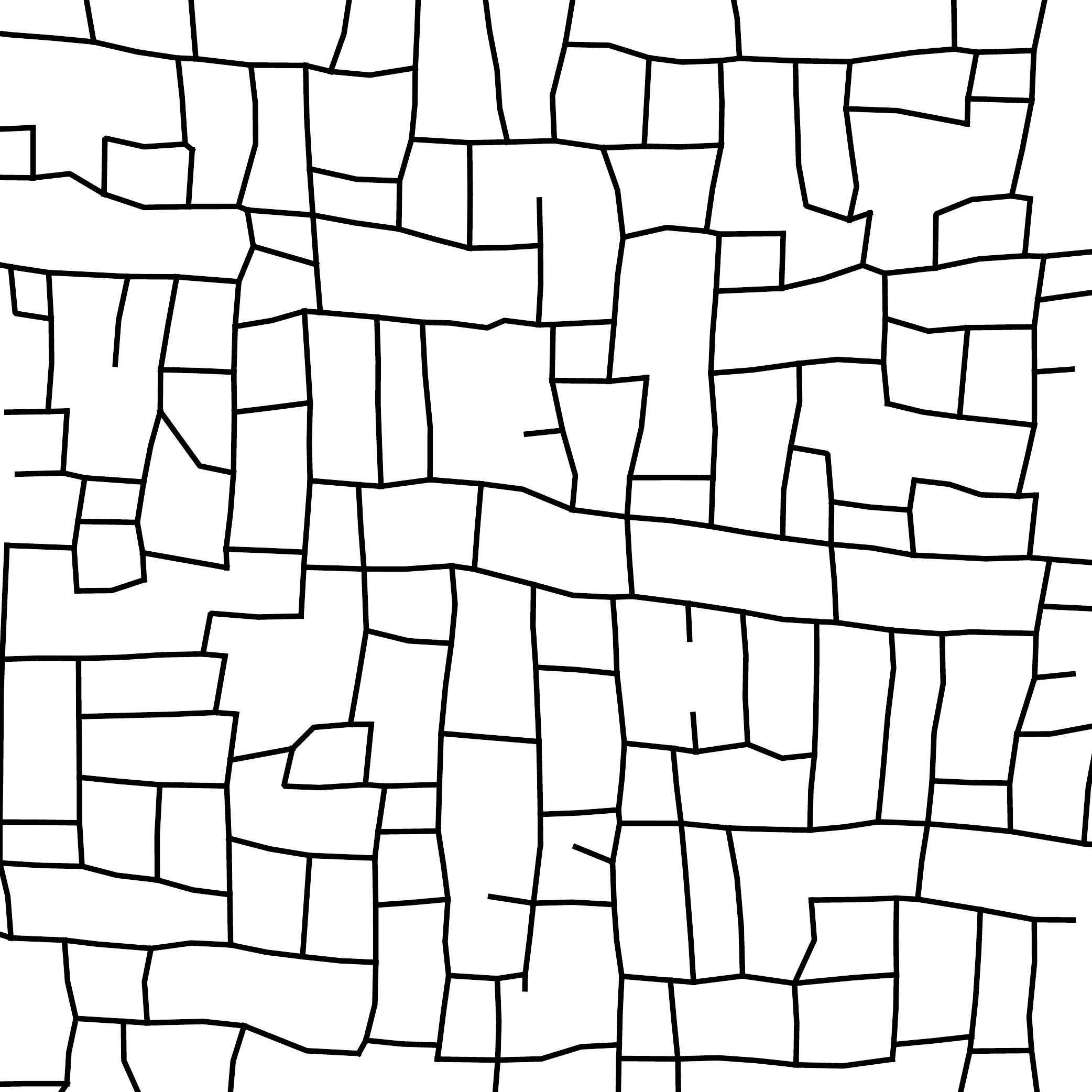}
}%
&
	\subfloat[Heart]{
	\label{fig:heart:exemplar}
	\includegraphics[width=0.0587\linewidth]{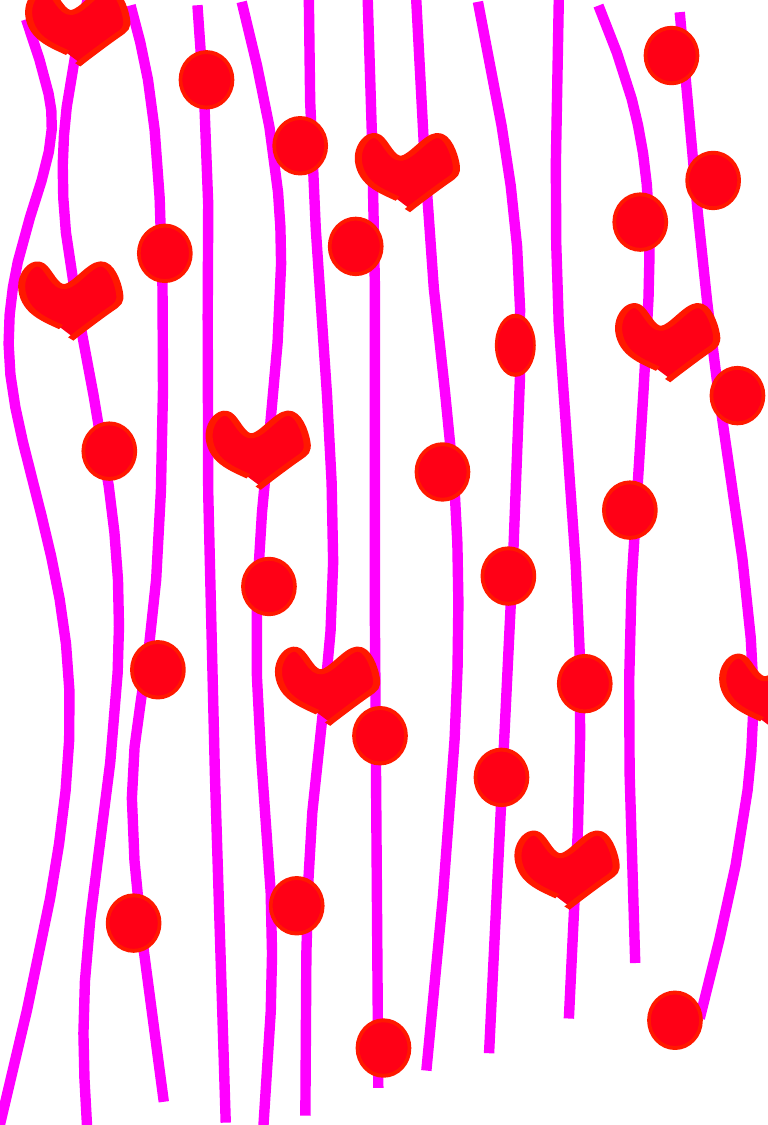}
}%
&\subfloat[]{
	\label{fig:heart:automatic_synthesis}
	\includegraphics[width=\figuresize\linewidth]{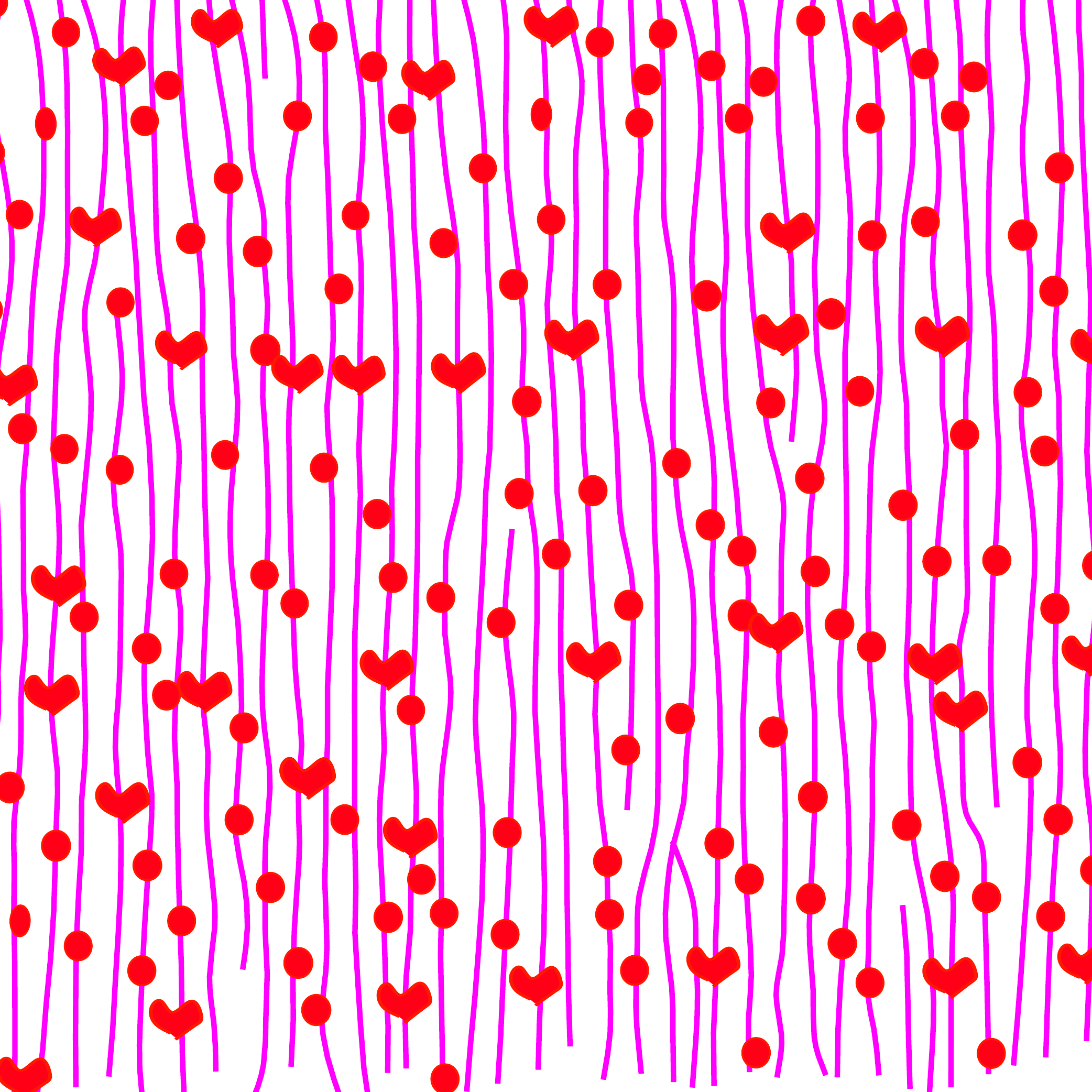}
}%
&	\subfloat[Curve grid]{
	\label{fig:distorted_grid:exemplar}
	\includegraphics[width=0.088\linewidth]{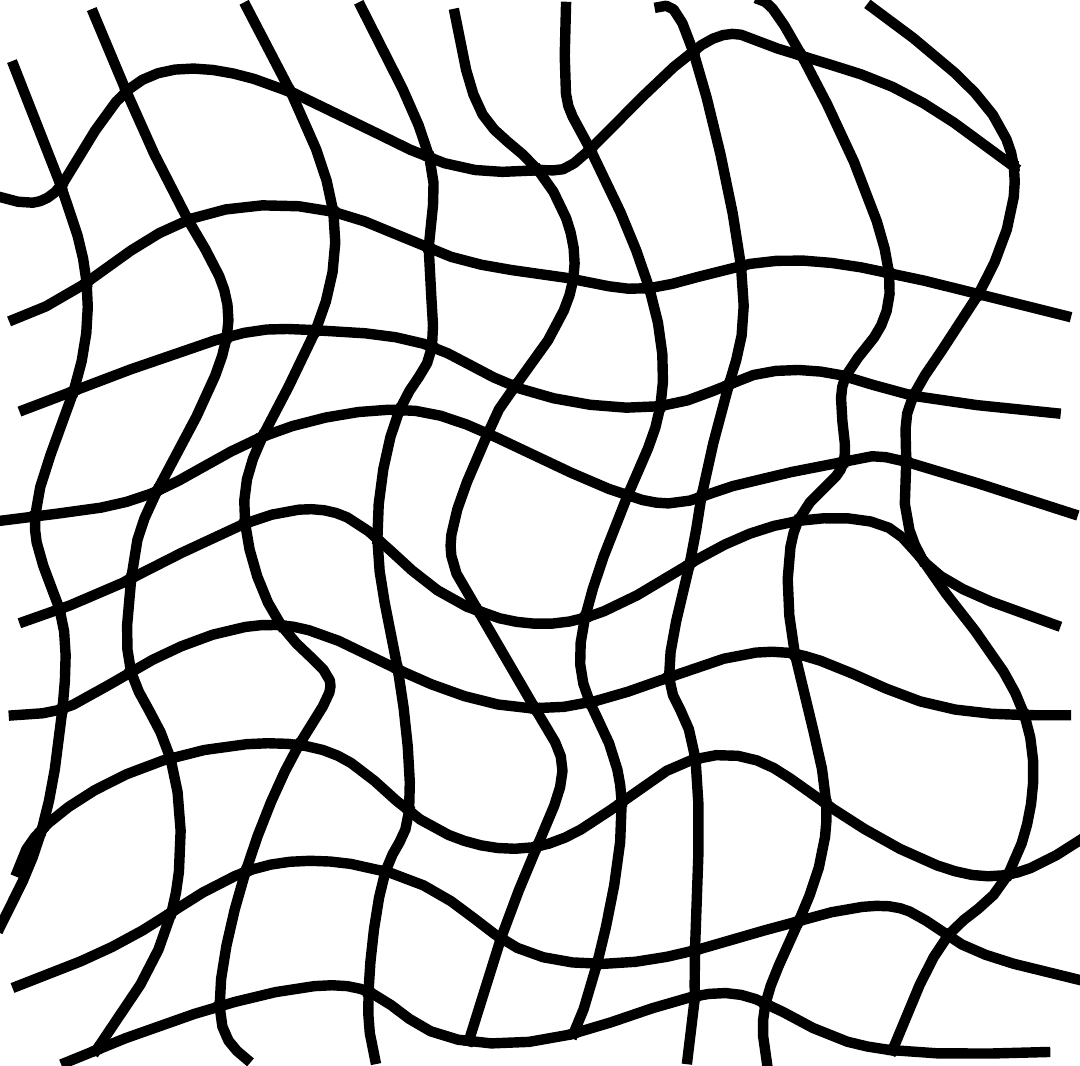}
}%
&\subfloat[]{
	\label{fig:distorted_grid:automatic_synthesis}
	\includegraphics[width=\figuresize\linewidth]{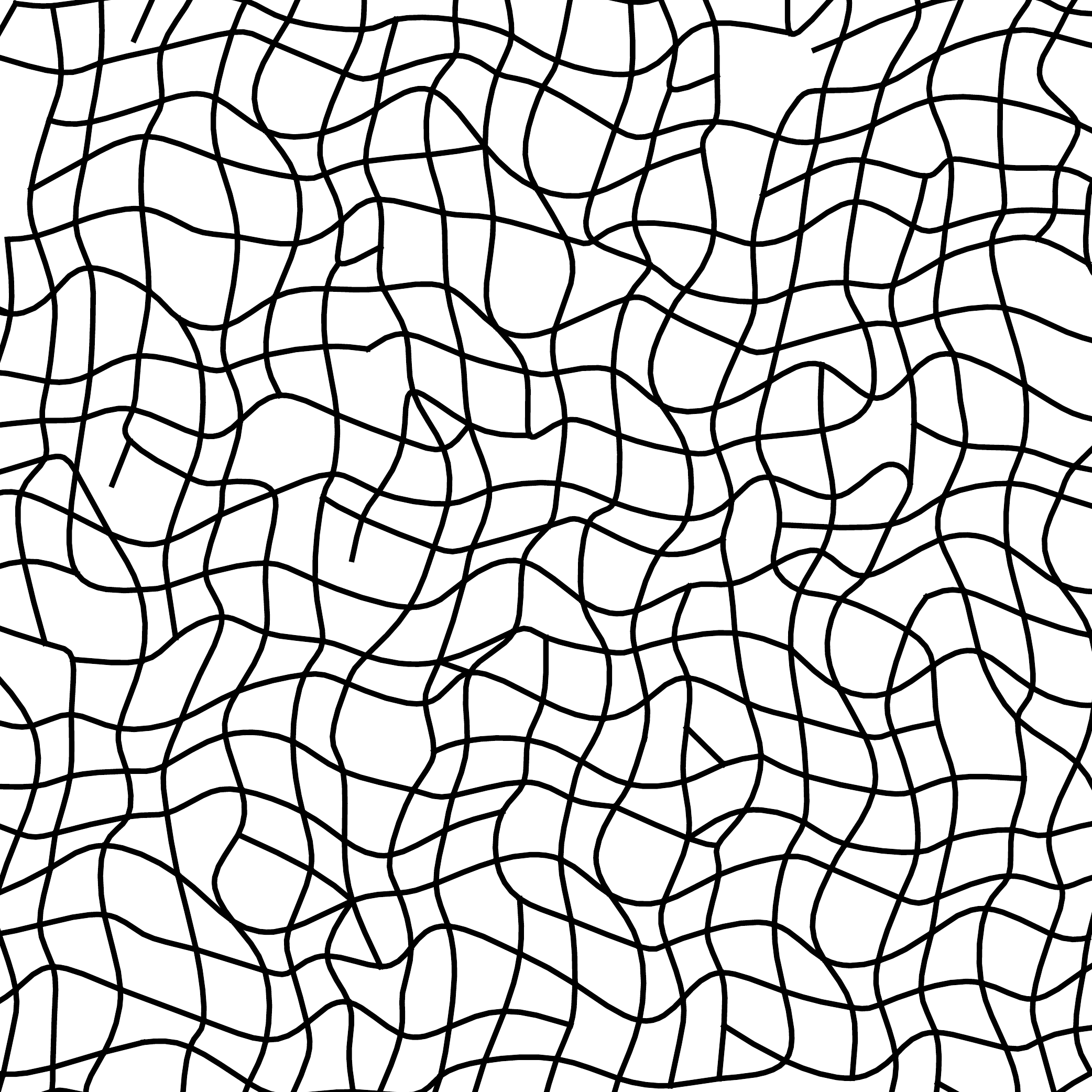}
}%
\\
\subfloat[\nothing{topographic map}Isocontour]{
	\label{fig:topographic_map:exemplar}
	\includegraphics[width=0.0815\linewidth]{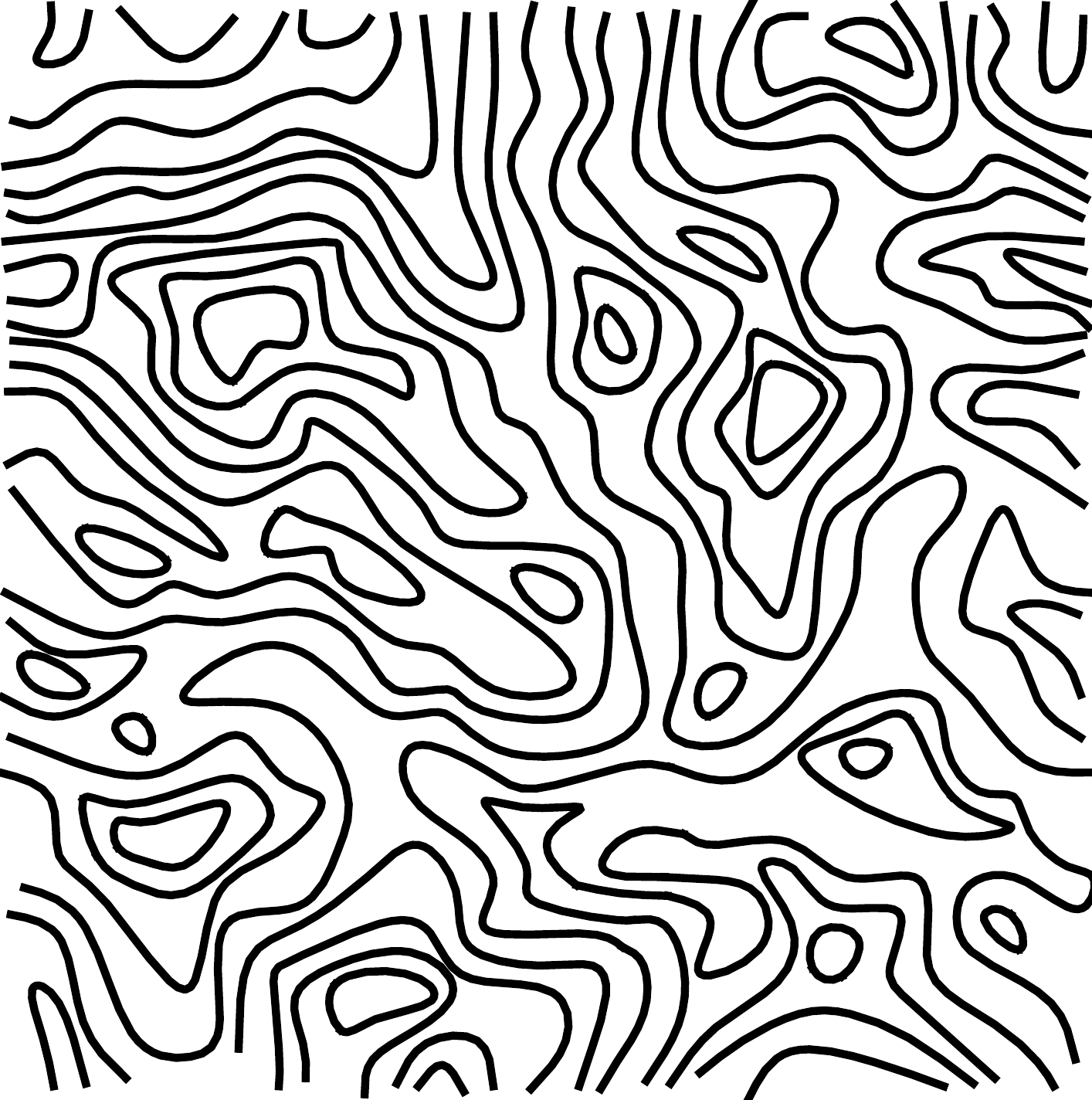}
}%
&\subfloat[]{
	\label{fig:topographic_map:automatic_synthesis}
	\includegraphics[width=\figuresize\linewidth]{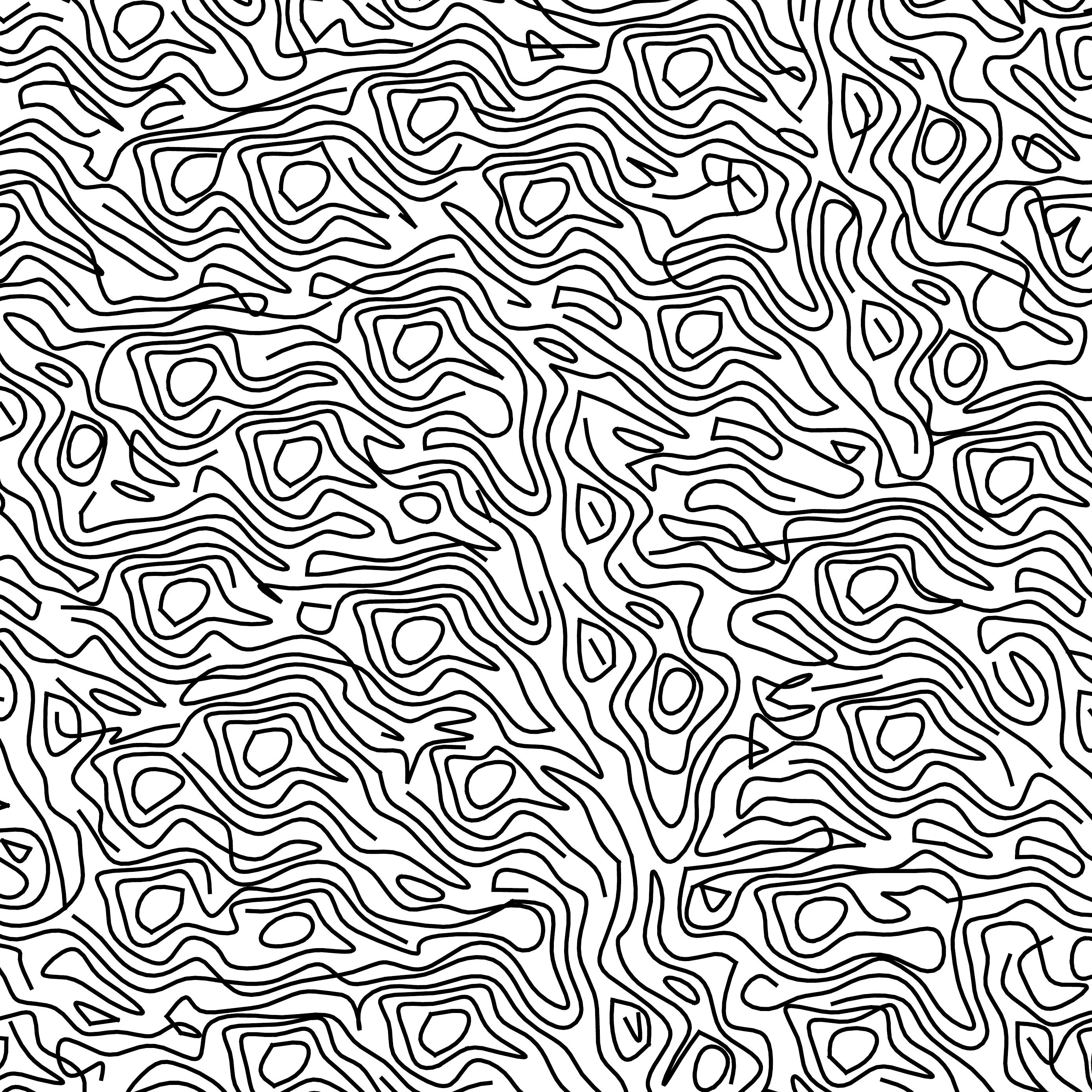}
}%
&
\subfloat[Circuit\nothing{ board}]{
	\label{fig:circuit_board:exemplar}
	\includegraphics[width=0.08\linewidth]{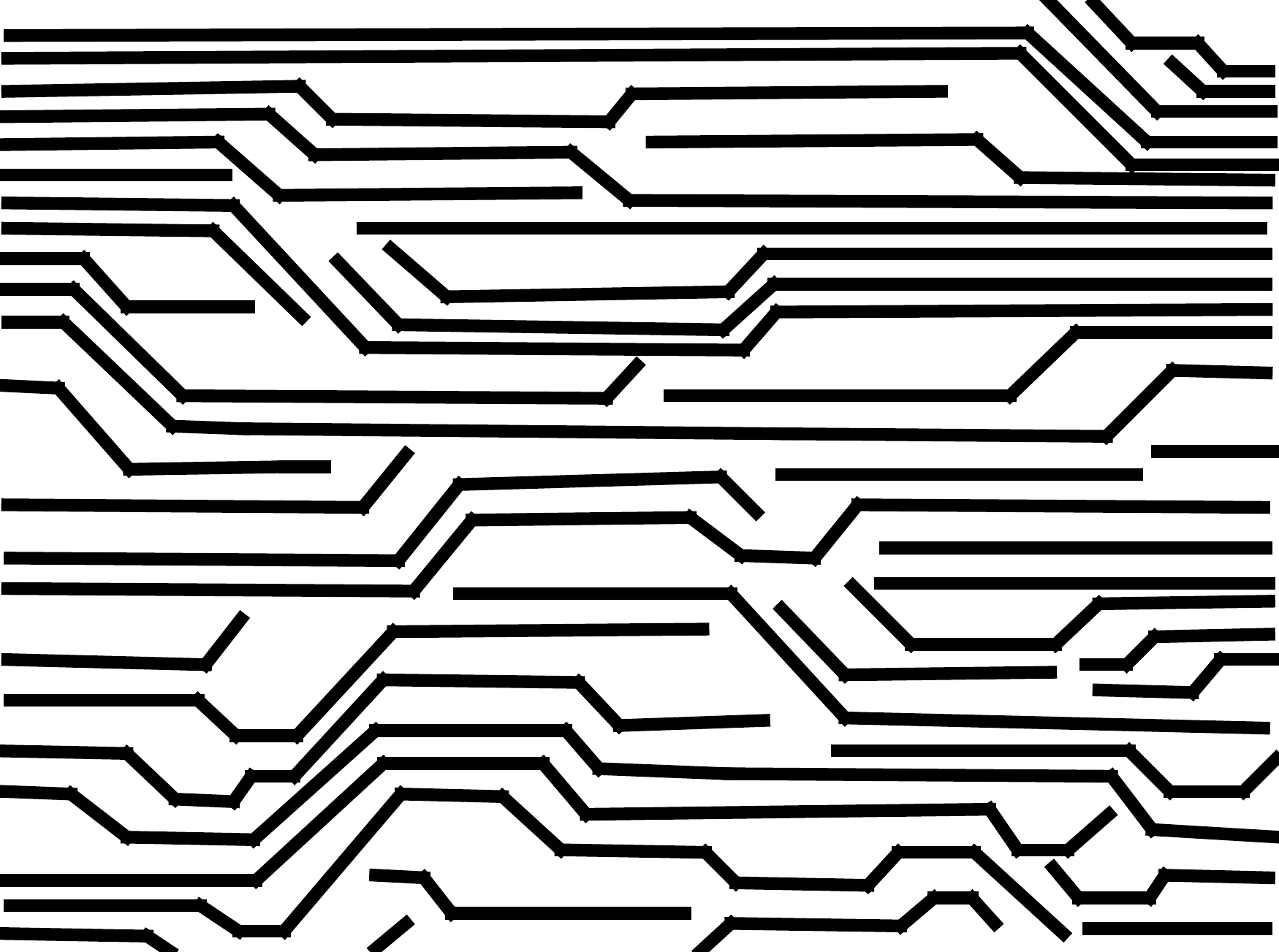}
}%
&\subfloat[]{
	\label{fig:circuit_board:automatic_synthesis}
	\includegraphics[width=\figuresize\linewidth]{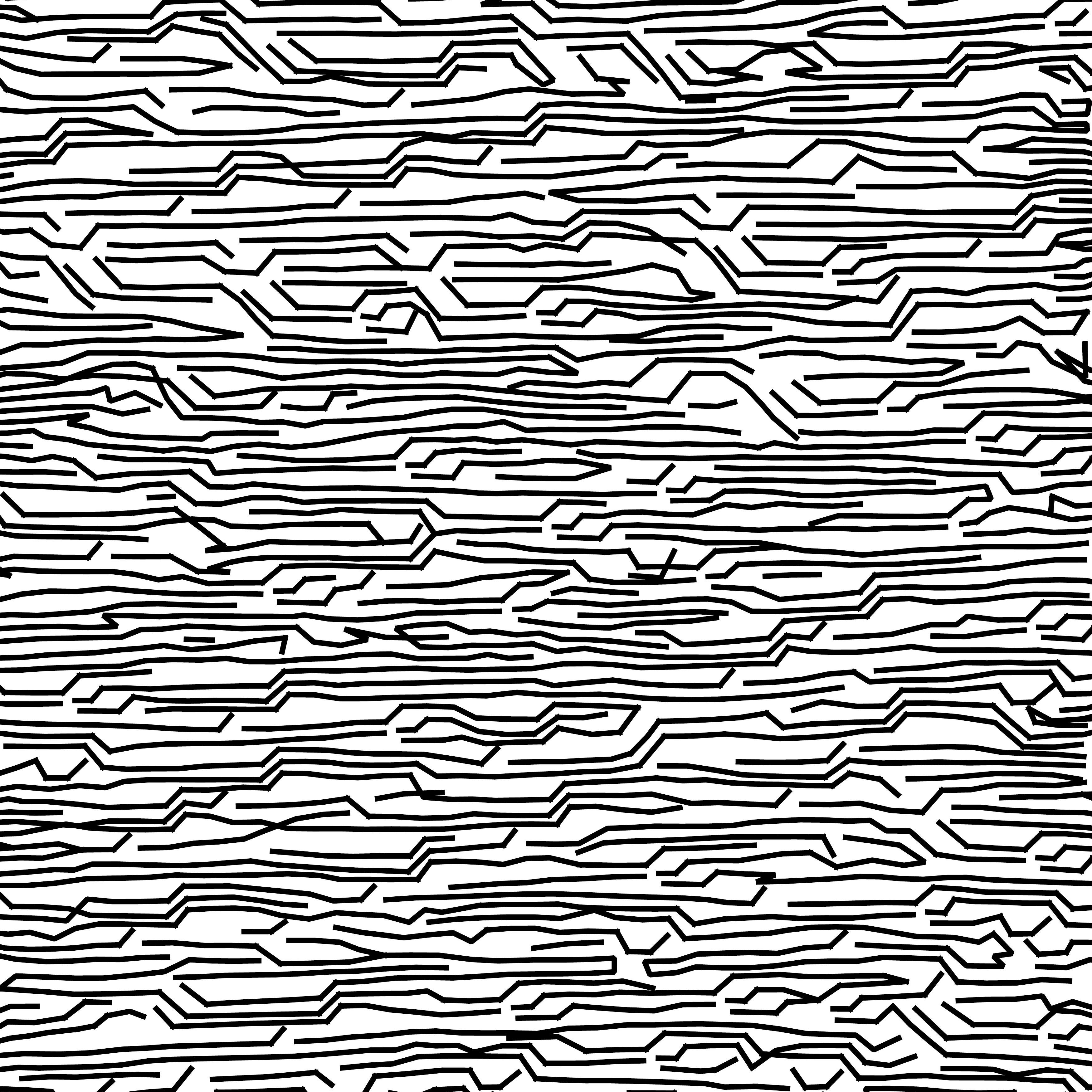}
}%
&
\subfloat[Chain]{
	\label{fig:chain:exemplar}
	\includegraphics[width=0.063\linewidth]{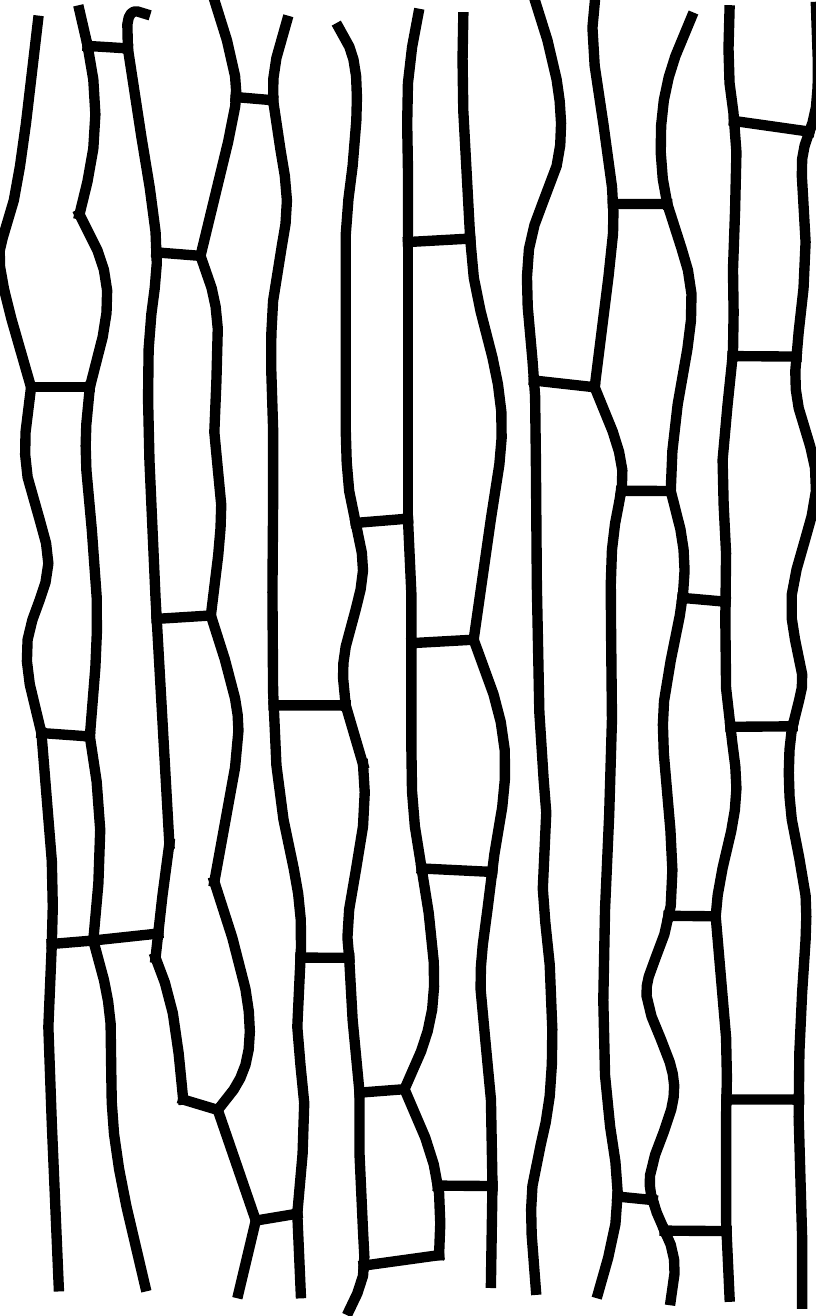}
}%
&\subfloat[]{
	\label{fig:chain:automatic_synthesis}
	\includegraphics[width=\figuresize\linewidth]{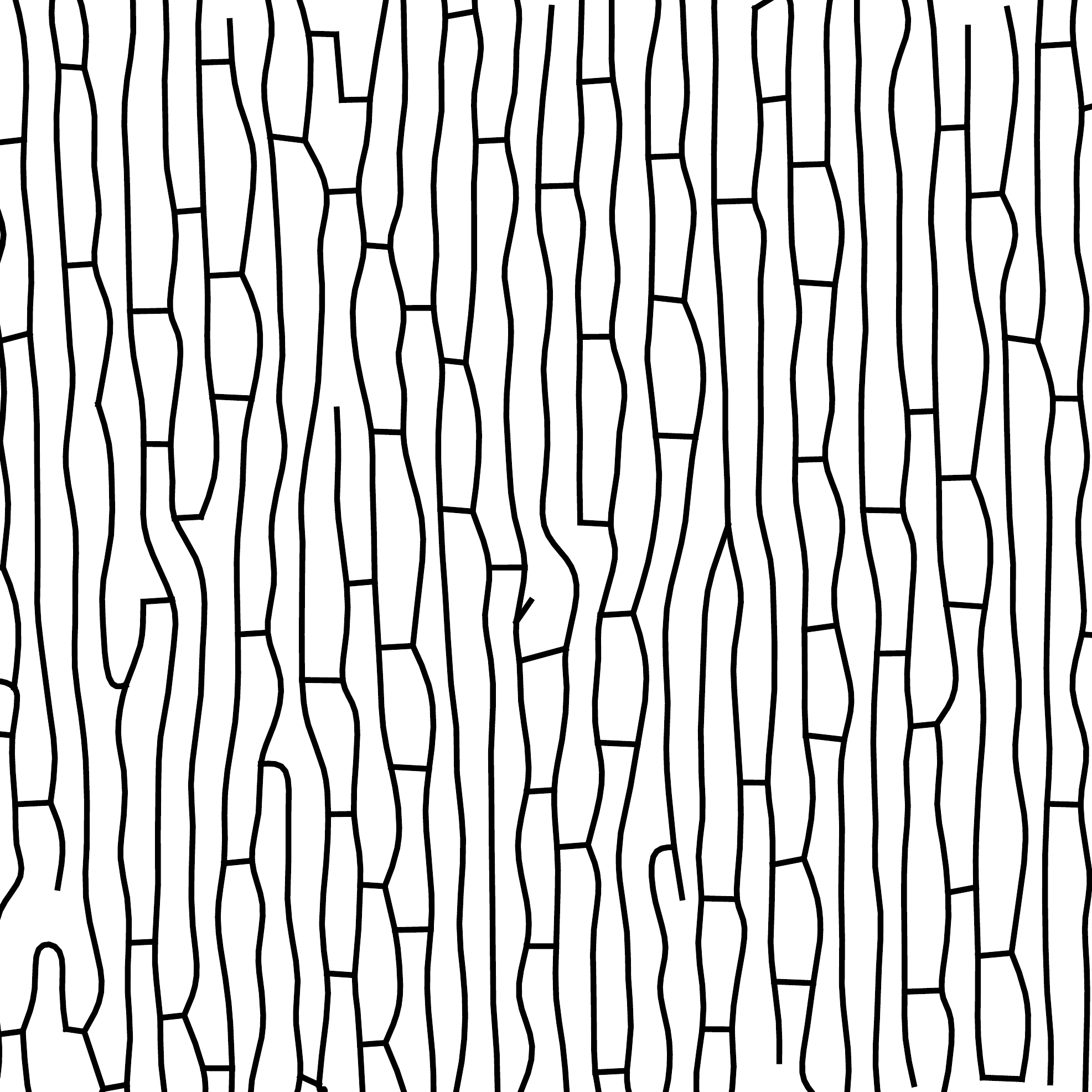}
}%
&	\subfloat[Fabric]{
		\label{fig:fabric:exemplar}
		\includegraphics[width=0.0788\linewidth]{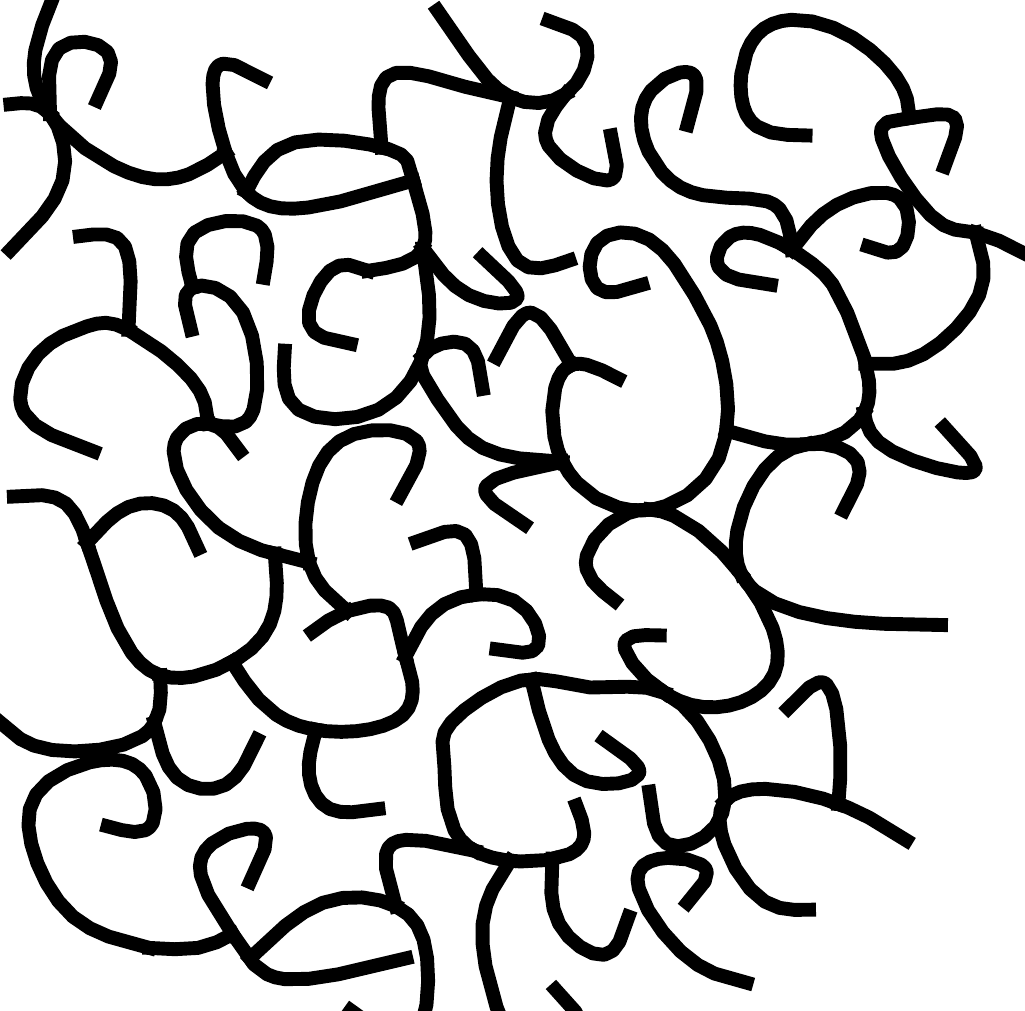}
	}%
&\subfloat[]{
		\label{fig:fabric:automatic_synthesis}
		\includegraphics[width=\figuresize\linewidth]{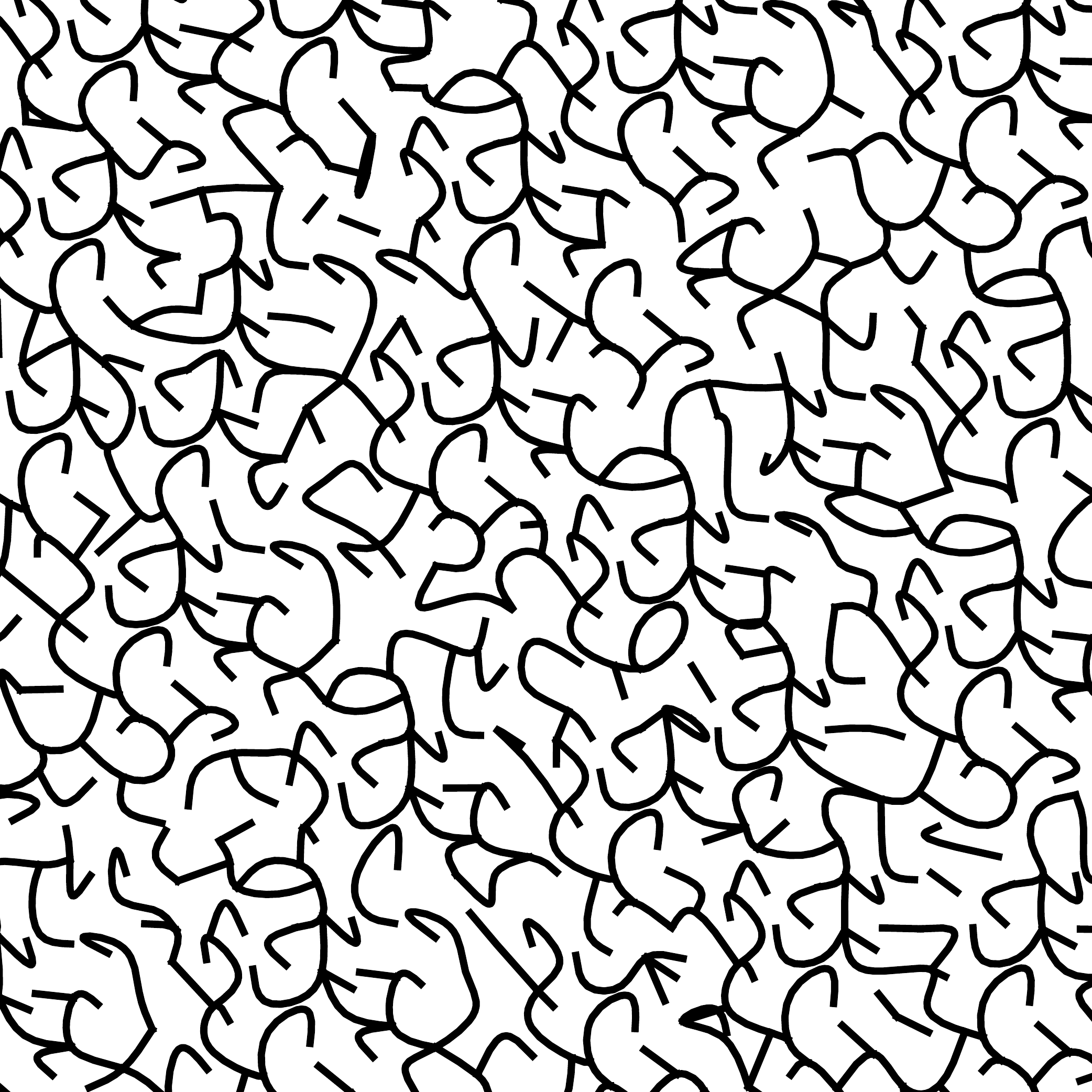}
	}%
\end{tabular}

\nothing{
	\subfloat[branches]{
	\label{fig:branches:exemplar}
	\includegraphics[width=0.105\linewidth]{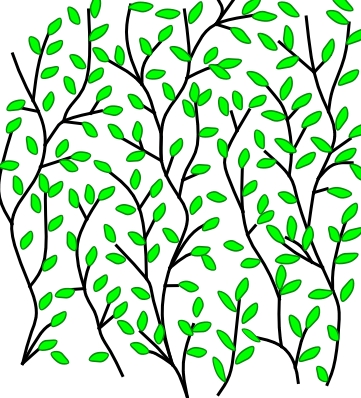}
}%
\subfloat[]{
	\label{fig:branches:automatic_synthesis}
	\includegraphics[width=\figuresize\linewidth]{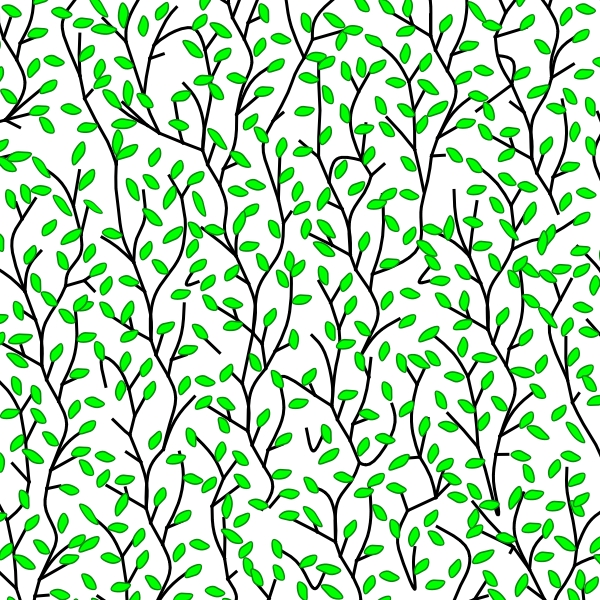}
}%
}%

	\Caption{Automatic synthesis results by our method.}
	{%
	  Within each pair of images, the input exemplar is smaller and shown on the left, the automatic synthesis result is bigger and shown on the right.
          \nothing{
}%
                \nothing{
		The number of autocomplete and total graphical elements. 
		A graphical element is either a discrete element or a continuous path. 
		Autocompleted are visualized in red. 
		The patterns on the top-left corners are initial exemplars.
                }%
                \nothing{
                }%
		\nothing{
		$\neighsize_1$, $\neighsize_2$ and $\neighsize_3$ indicates the synthesis parameters learned by our learning algorithms.  $\numiteration$ is the number of iterations to get these parameters in our experiments.
		}%
	}
	\label{fig:full_auto_outputs}
\end{figure*}

\begin{figure*}[tb!p]
	\centering
	\captionsetup[subfigure]{labelformat=empty}
	\captionsetup[subfigure]{justification=centering}
		\setlength{\tabcolsep}{-2.5pt}

\newcommand{\semiautofiguresize}{0.135}

\begin{tabular}{cccccccc}
\centering
\subfloat[Uneven bricks]{
	\label{fig:irregularbrickwall:exemplar}
	\includegraphics[width=0.06\linewidth]{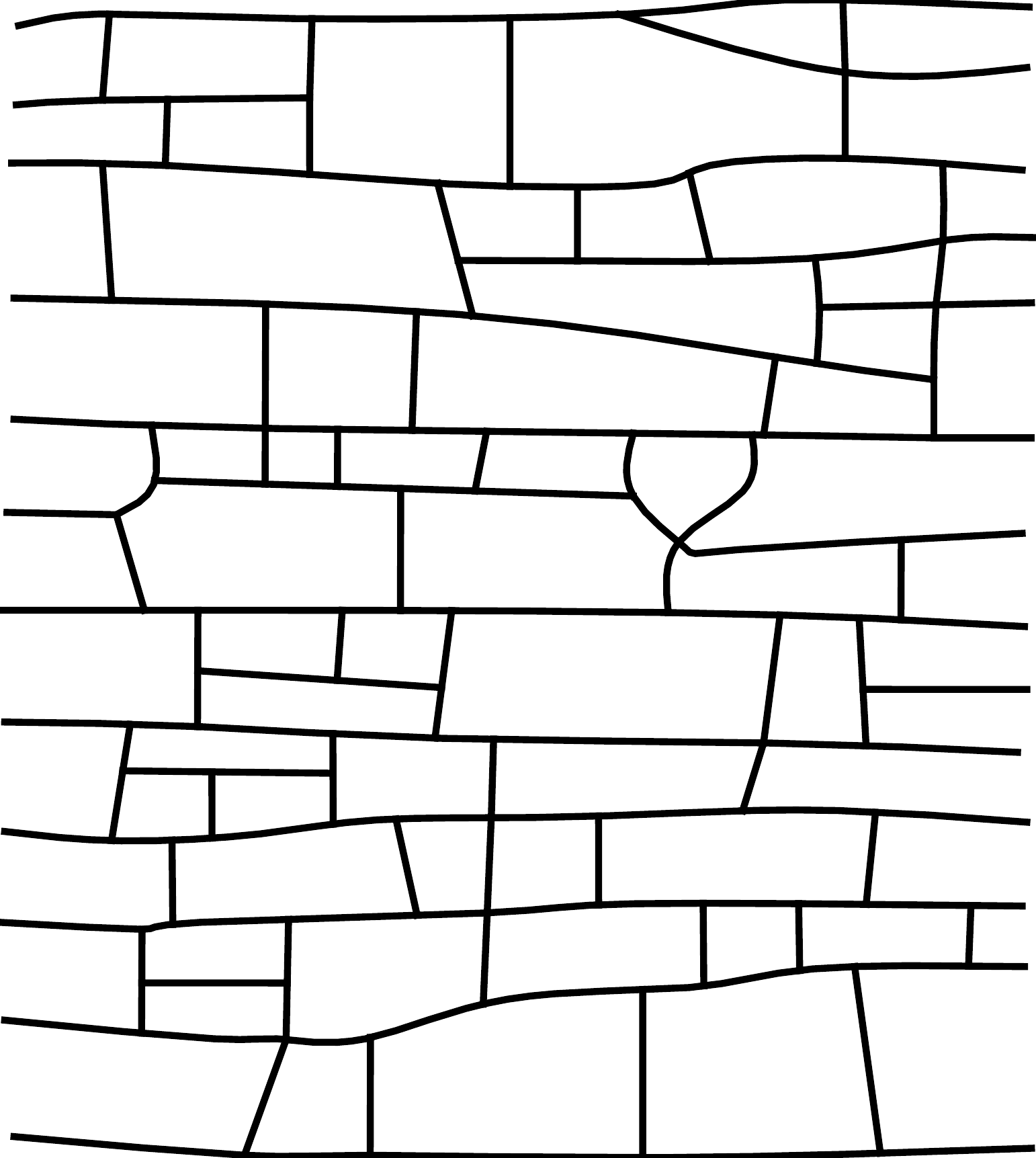}
}%
&\subfloat[Auto]{
	\label{fig:irregularbrickwall:auto}
	\includegraphics[width=\semiautofiguresize\linewidth]{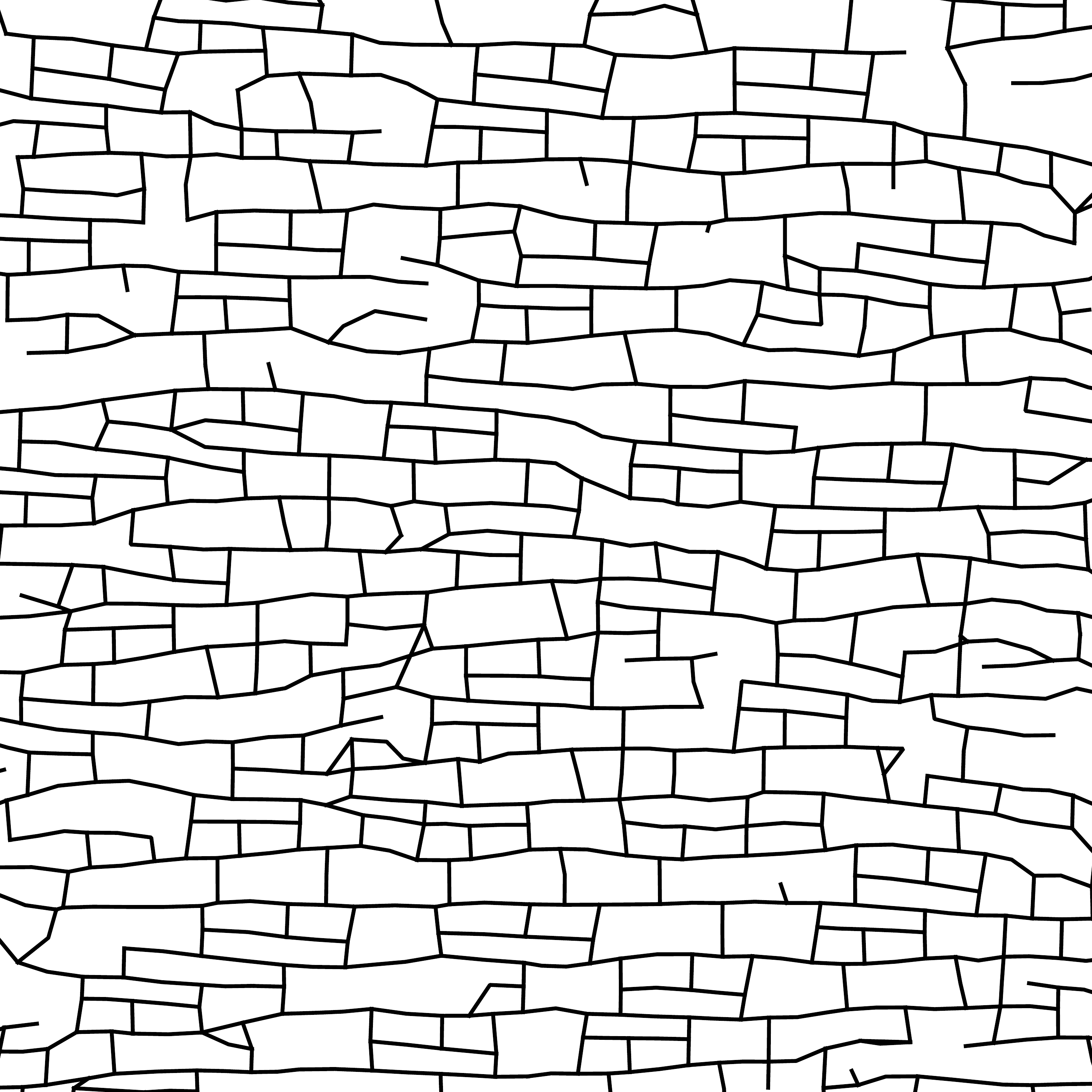}
}%
&\subfloat[User \reject{16} + \manual{11}]{
	\label{fig:irregularbrickwall:user}
	\includegraphics[width=\semiautofiguresize\linewidth]{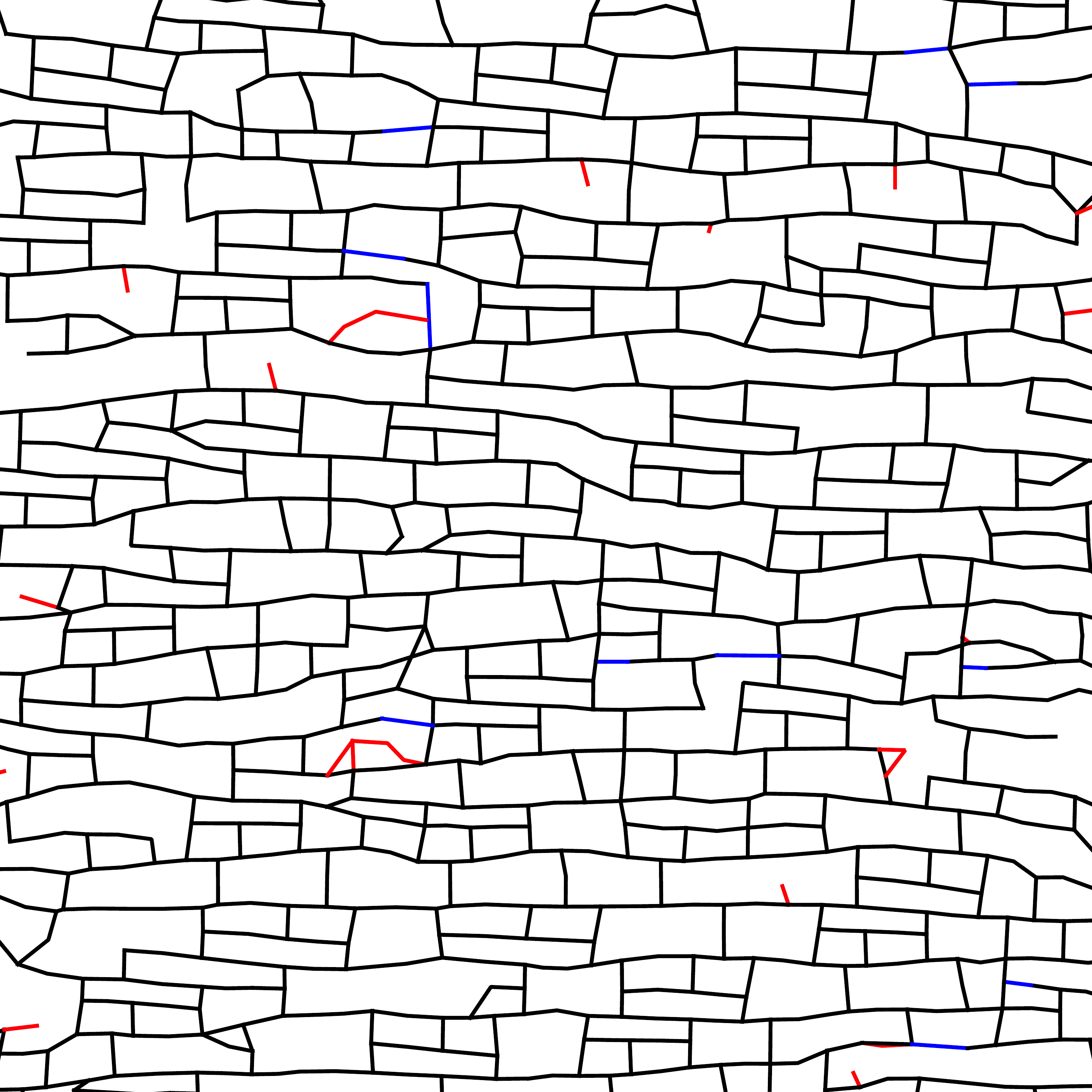}
}%
&\subfloat[Edited]{
	\label{fig:irregularbrickwall:edited}
	\includegraphics[width=\semiautofiguresize\linewidth]{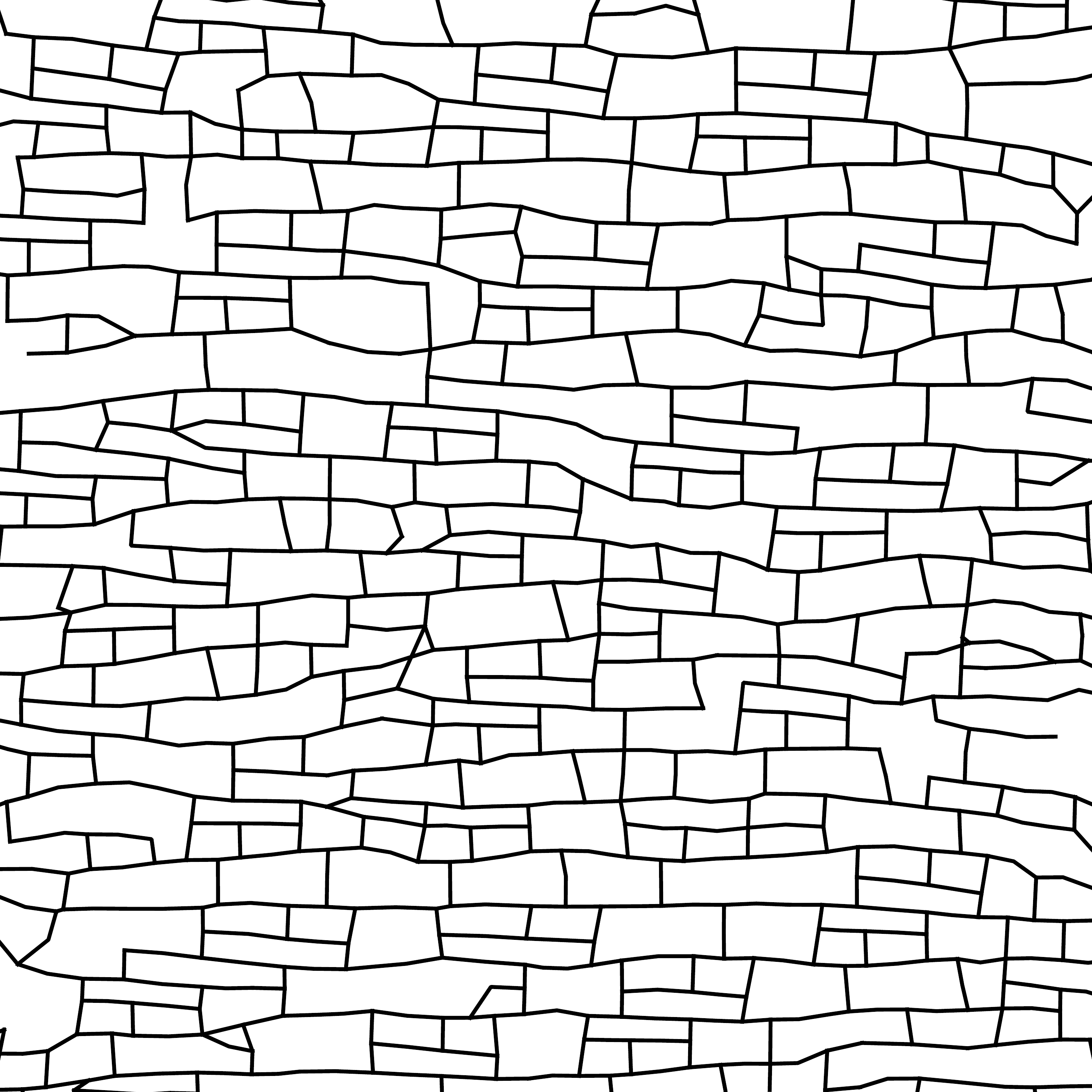}
}%
&
	\subfloat[Zigzag]{
	\label{fig:zigzag:exemplar}
	\includegraphics[width=0.0723\linewidth]{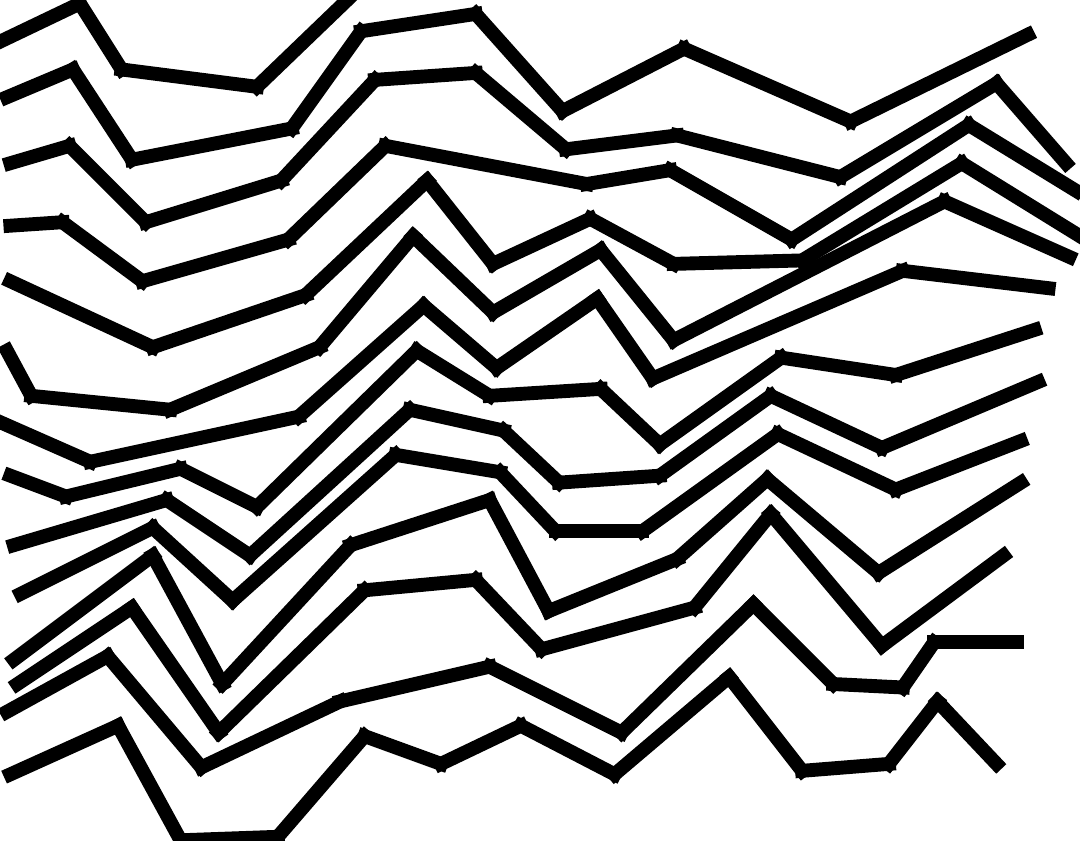}
}%
&\subfloat[Auto]{
	\label{fig:zigzag:auto}
	\includegraphics[width=\semiautofiguresize\linewidth]{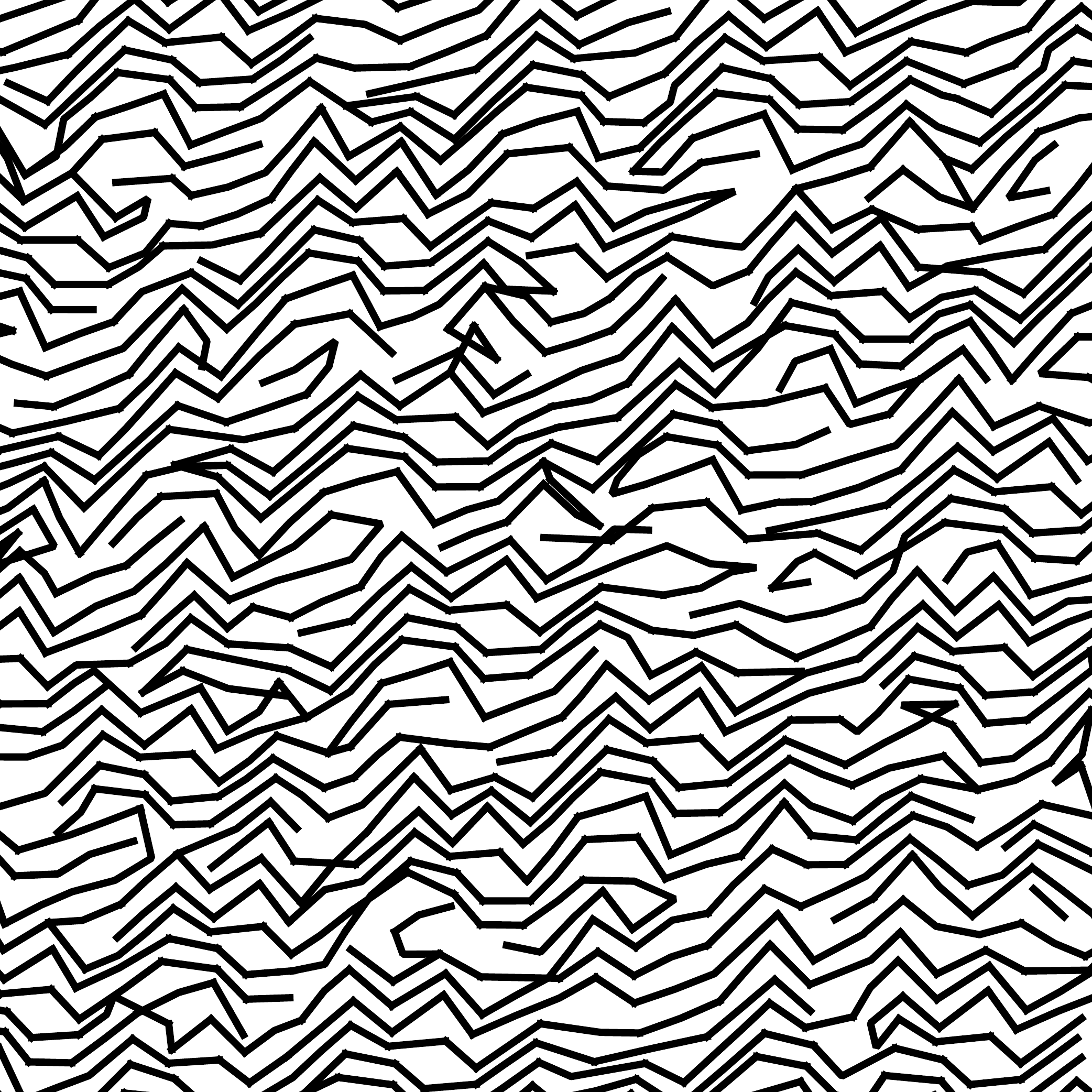}
}%
&\subfloat[User  \reject{11} + \manual{2}]{
	\label{fig:zigzag:user}
	\includegraphics[width=\semiautofiguresize\linewidth]{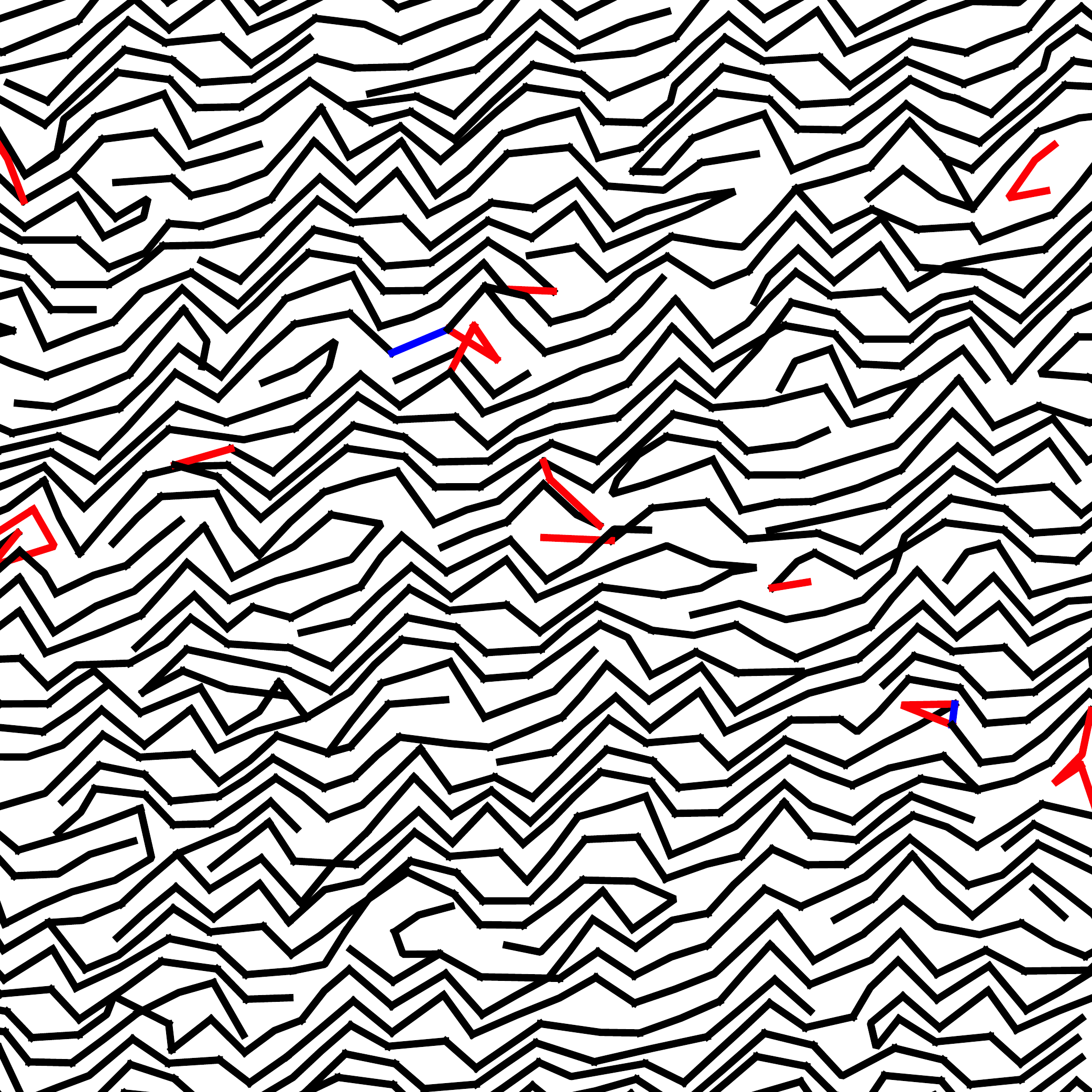}
}%
&\subfloat[Edited]{
	\label{fig:zigzag:edited}
	\includegraphics[width=\semiautofiguresize\linewidth]{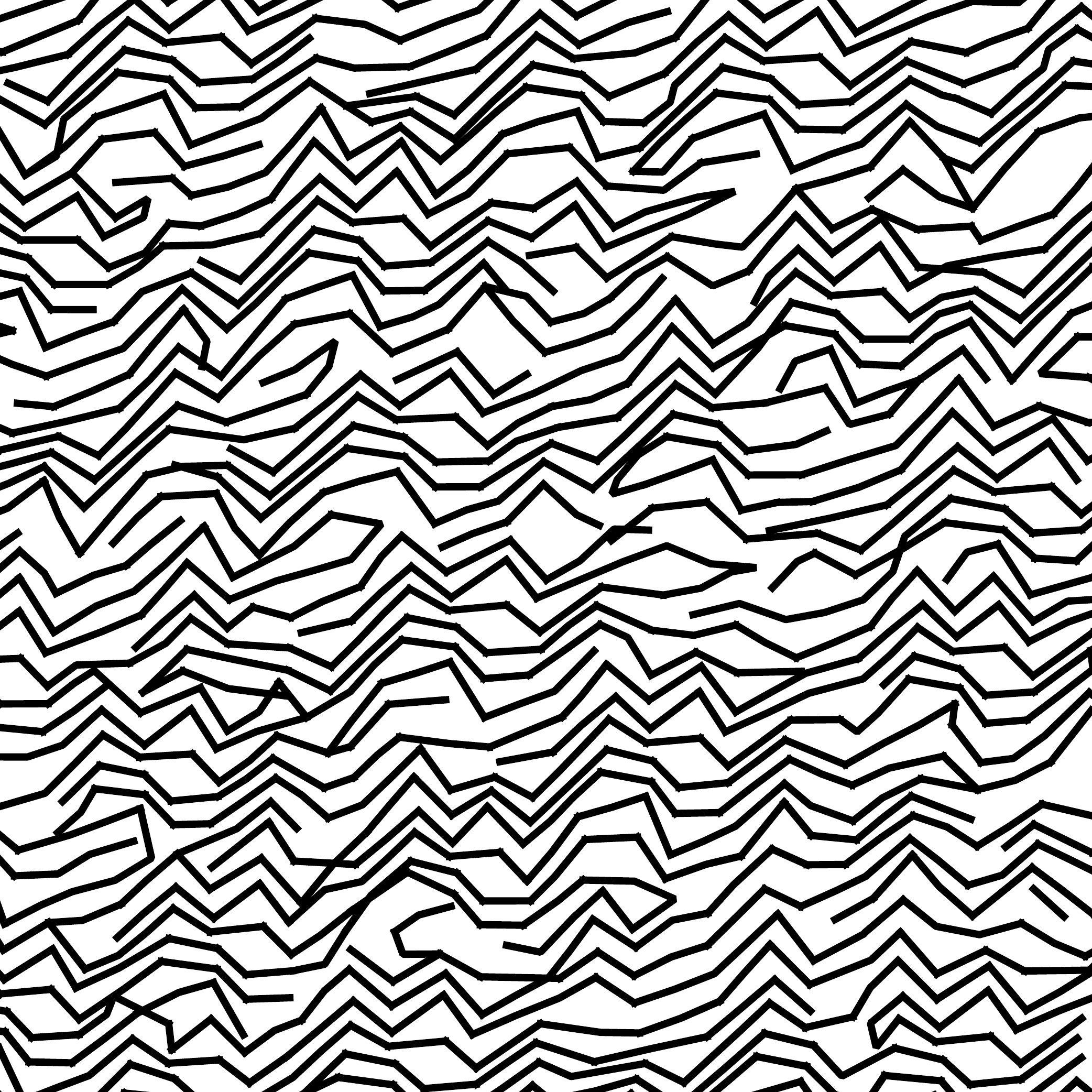}
}%
\\
	\subfloat[Prisma]{
	\label{fig:prisma:exemplar}
	\includegraphics[width=0.0595\linewidth]{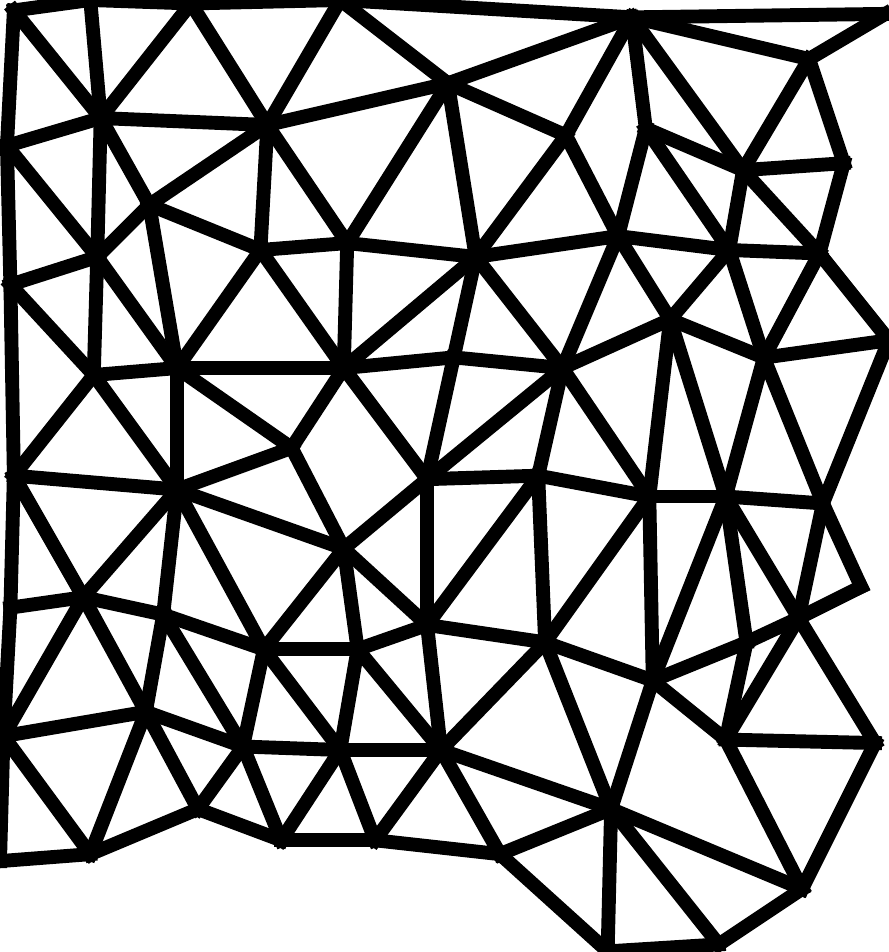}
}%
&\subfloat[Auto]{
	\label{fig:prisma:auto}
	\includegraphics[width=\semiautofiguresize\linewidth]{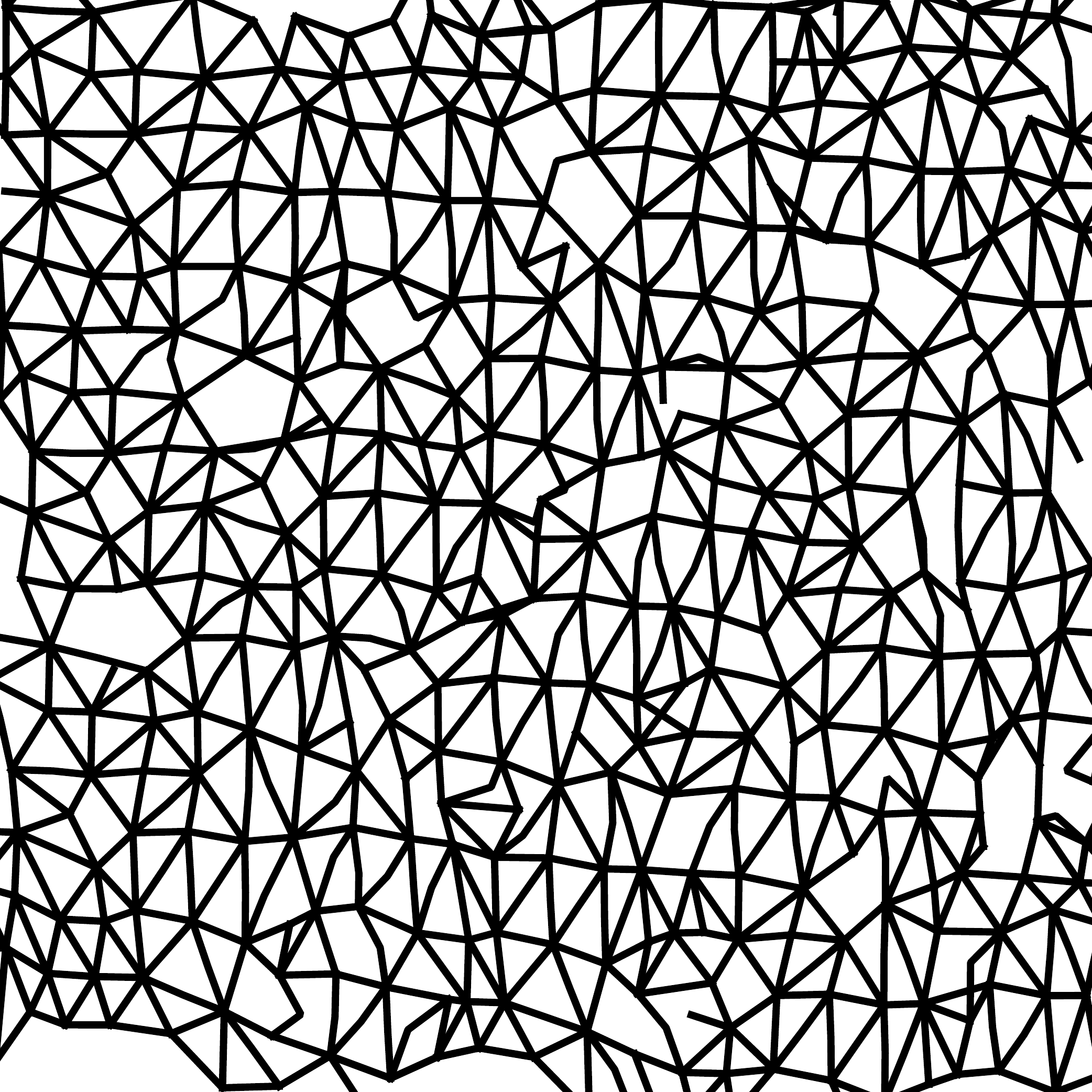}
}%
&\subfloat[User \reject{5} + \manual{28}]{
	\label{fig:prisma:user}
	\includegraphics[width=\semiautofiguresize\linewidth]{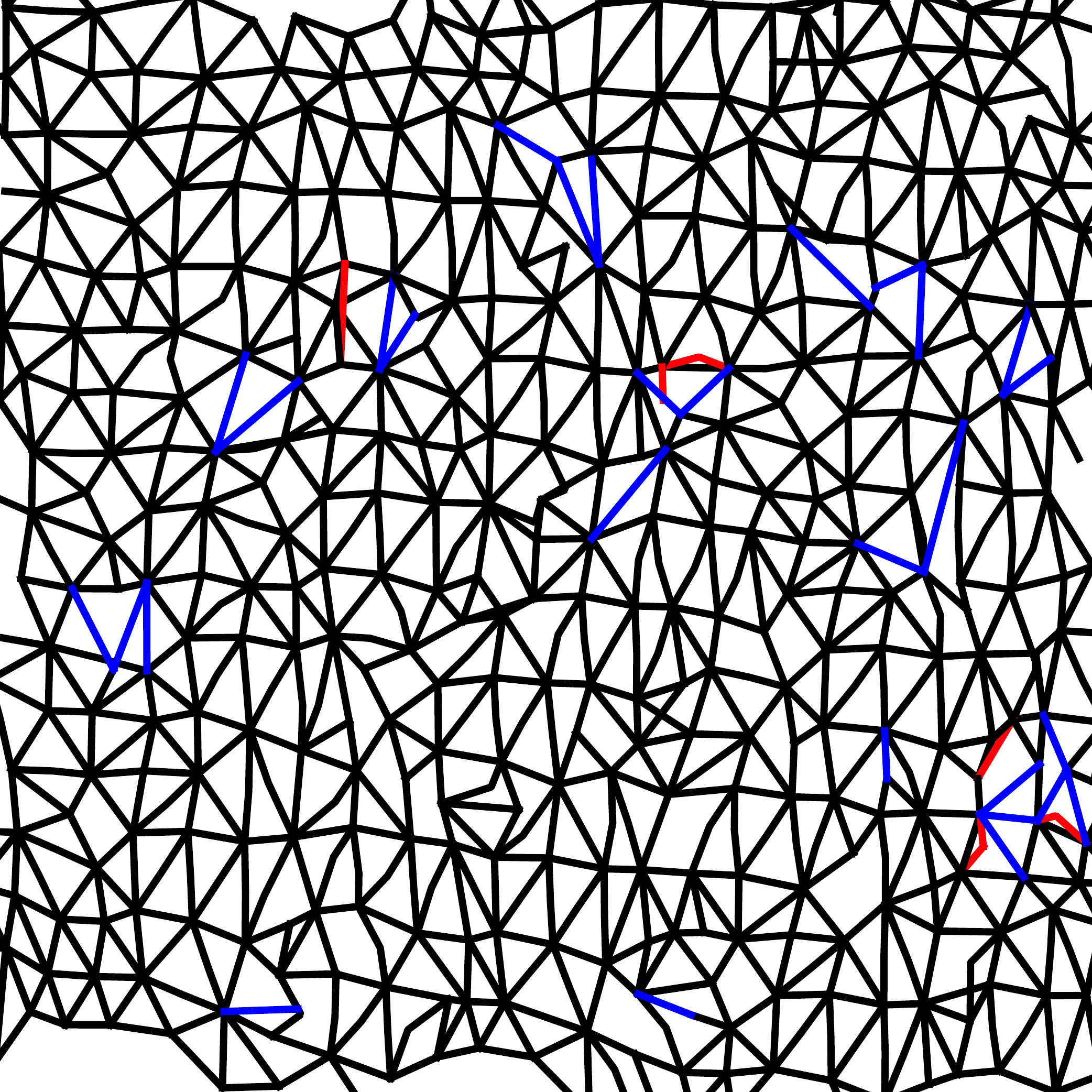}
}%
&\subfloat[Edited]{
	\label{fig:prisma:edited}
	\includegraphics[width=\semiautofiguresize\linewidth]{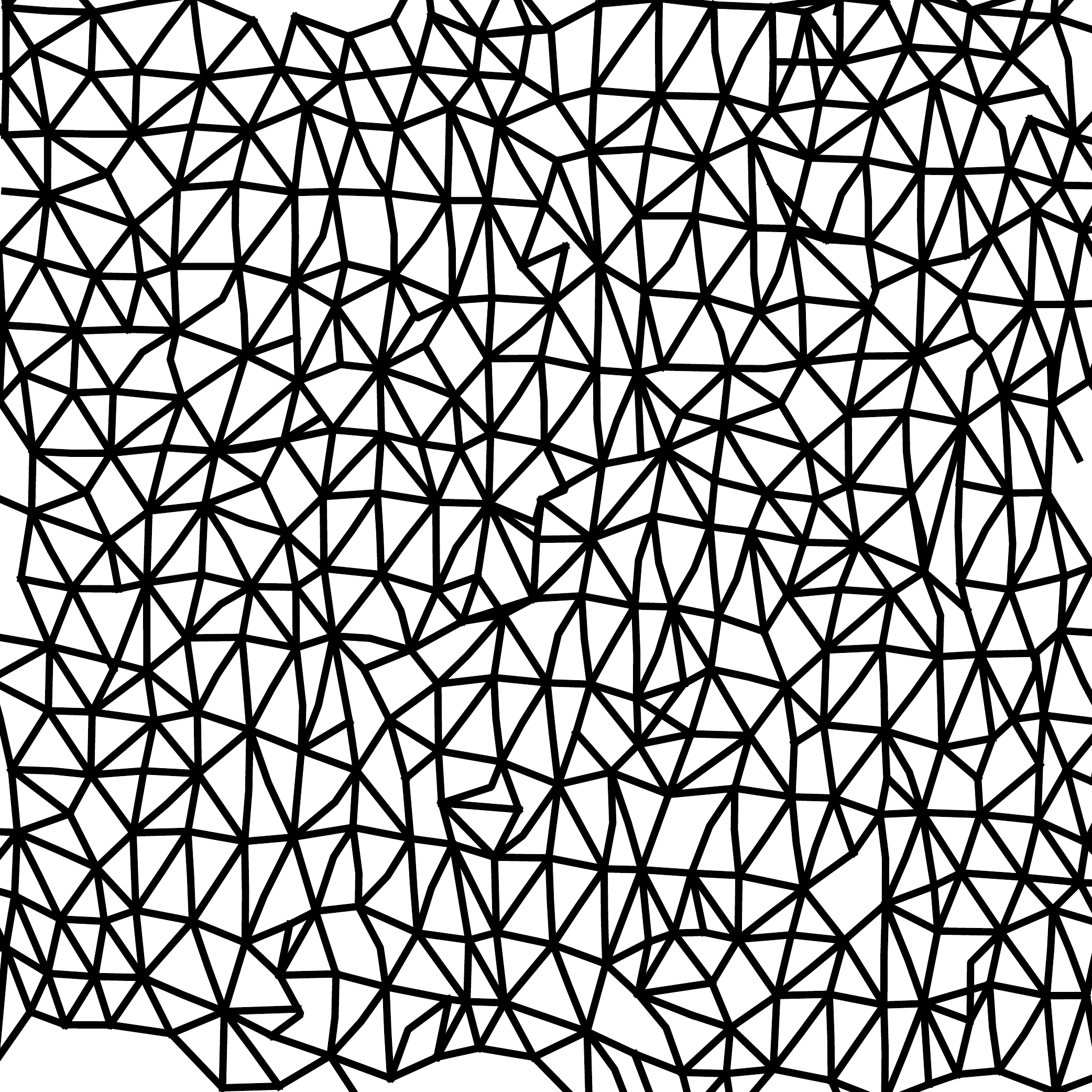}
}%
&
	\subfloat[Waves (small)]{
	\label{fig:waves:exemplar}
	\includegraphics[width=0.0875\linewidth]{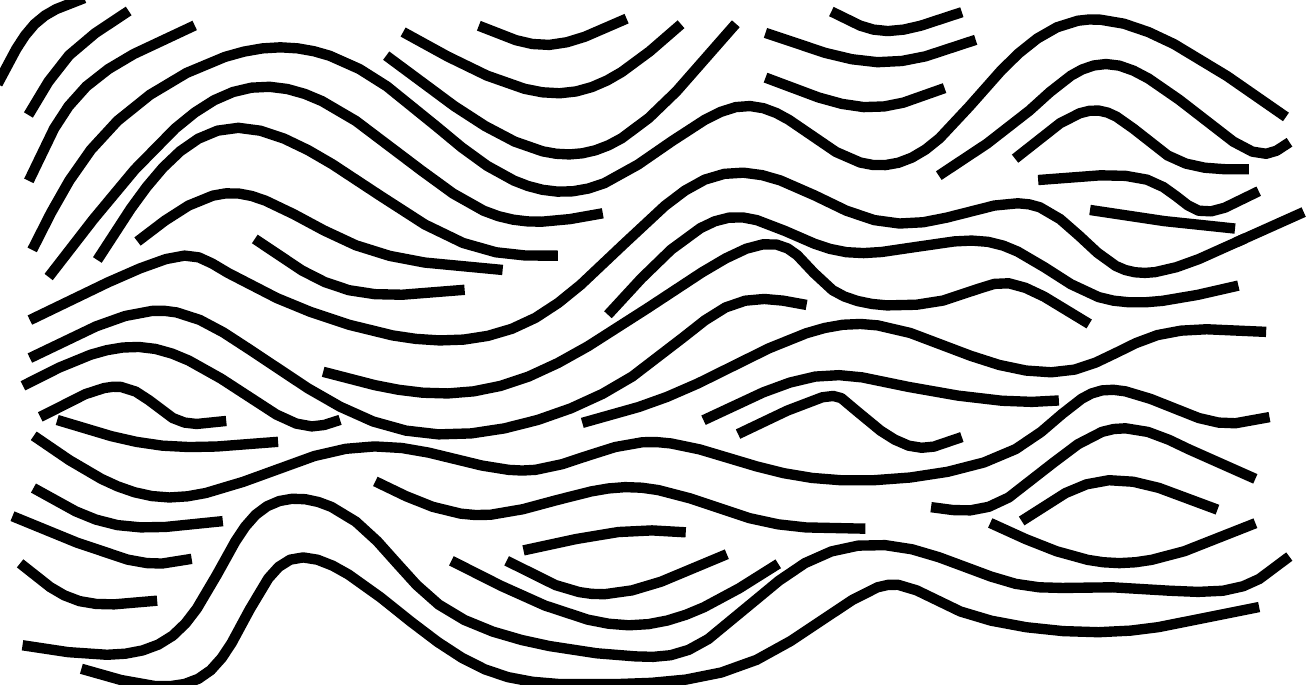}
}%
&\subfloat[Auto]{
	\label{fig:waves:automatic_synthesis}
	\includegraphics[width=\semiautofiguresize\linewidth]{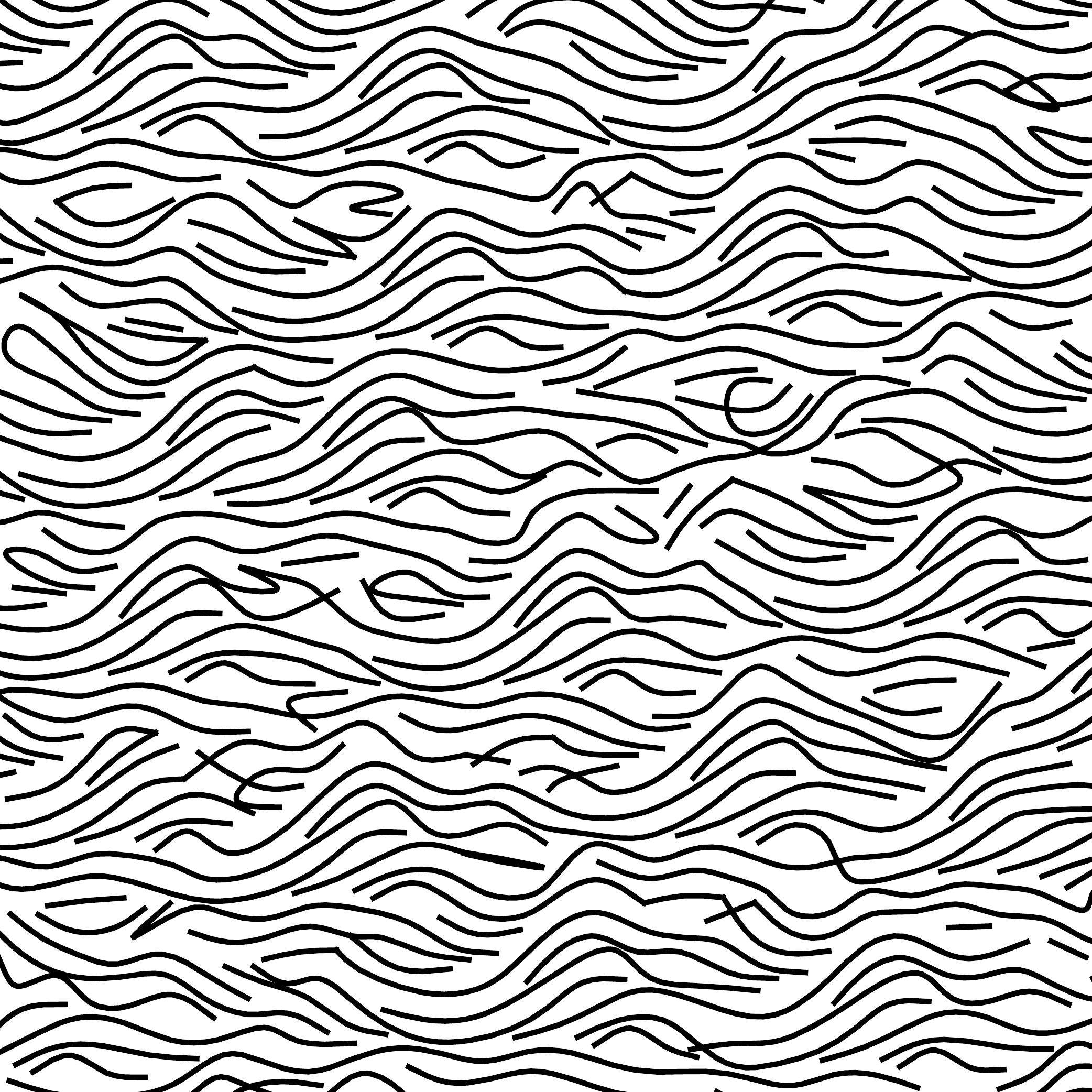}
}%
&\subfloat[User \reject{21} + \manual{0}]{
	\label{fig:waves:user}
	\includegraphics[width=\semiautofiguresize\linewidth]{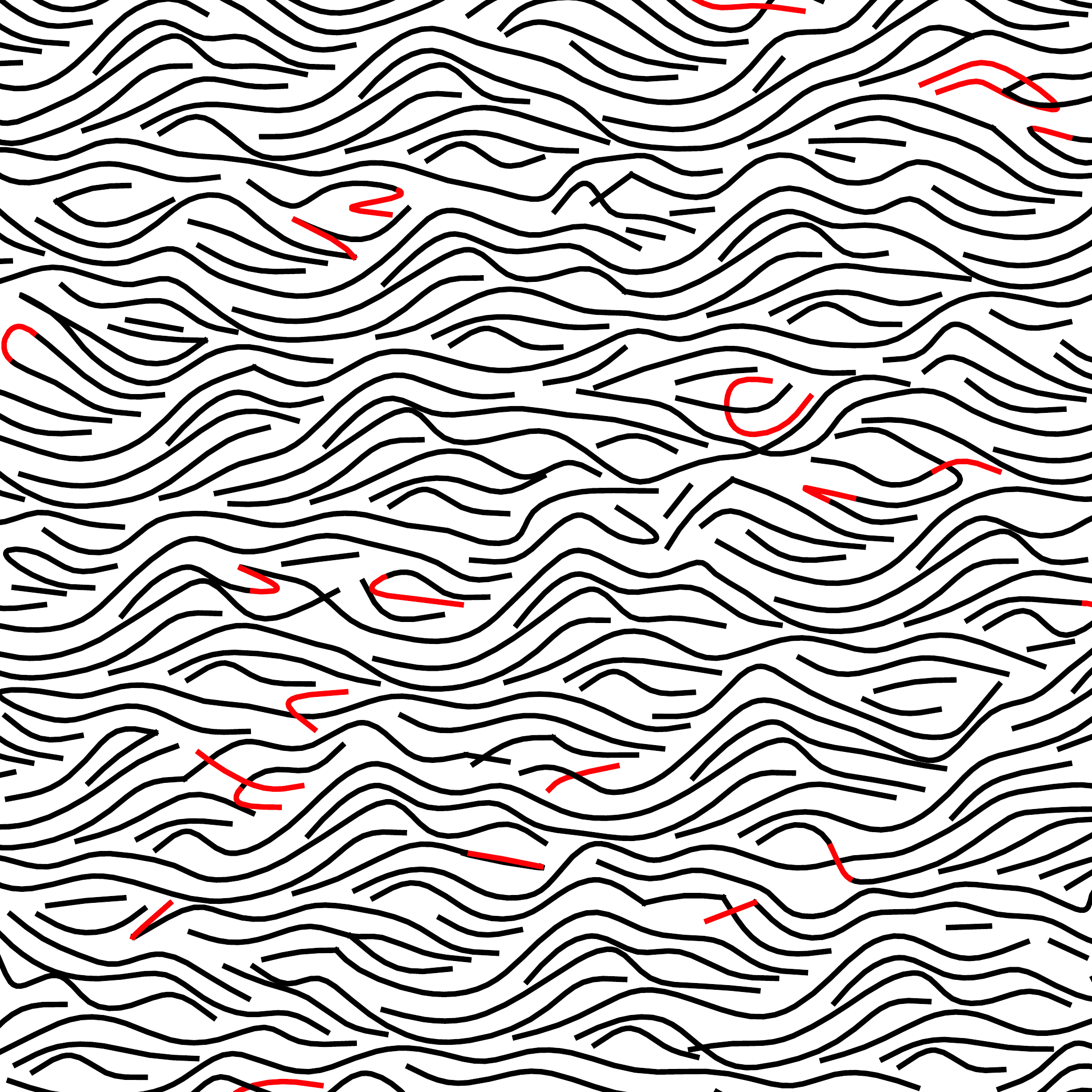}
}%
&\subfloat[Edited]{
	\label{fig:waves:edited}
	\includegraphics[width=\semiautofiguresize\linewidth]{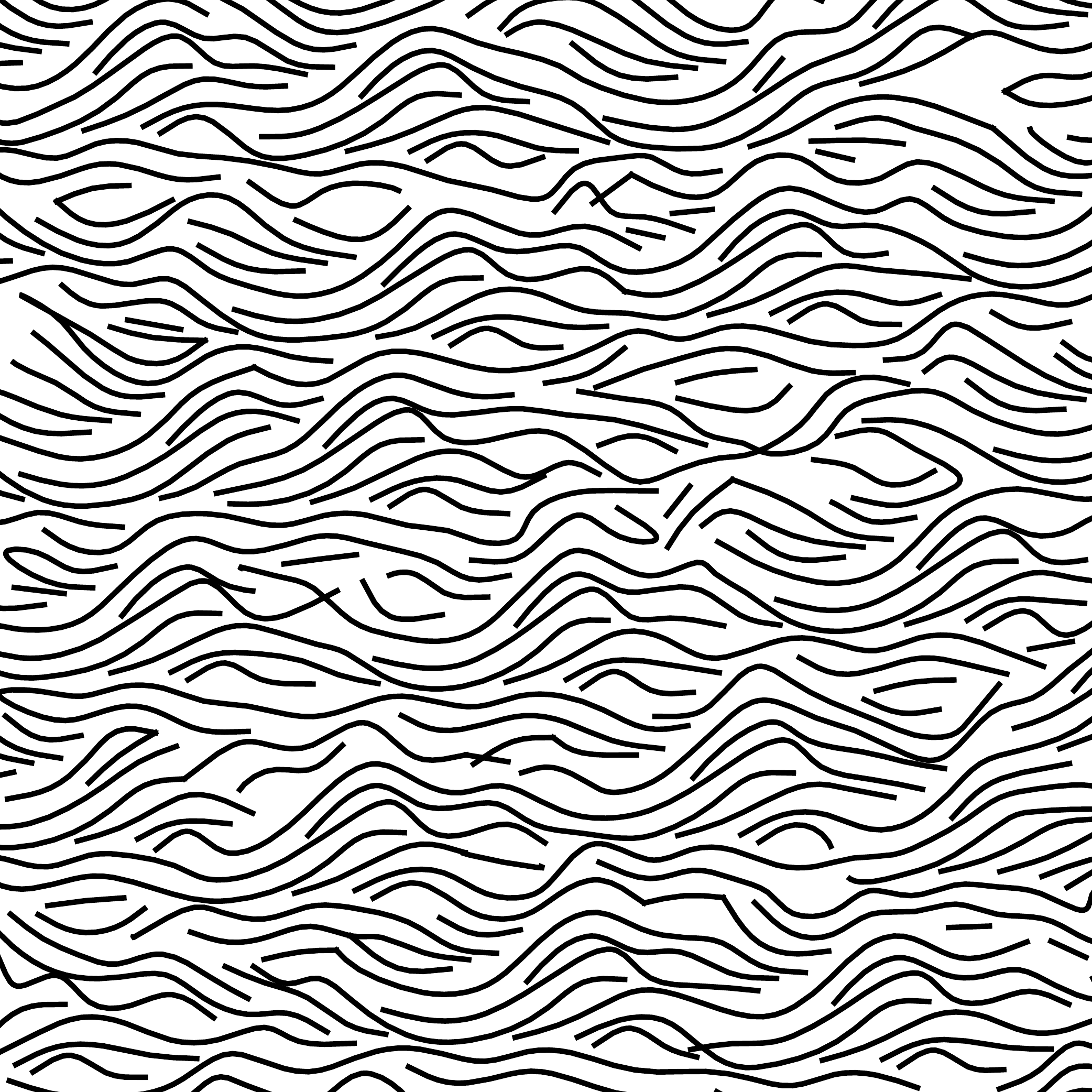}
}%

\end{tabular}

	\Caption{Automatic synthesis and user-assisted results.}
	{%
\nothing{%
}%
		We count the number of user operations needed for correcting artifacts. \reject{Red} indicates the number of rejection and \manual{blue} the number of manual path drawing. \nothing{or element copy\&paste}%
		\nothing{
}%
\nothing{
}%
	}
	\label{fig:semi_auto_outputs}
\end{figure*}

\nothing{
\paragraph{Automatic synthesis}
}%

\nothing{
Here we show fully automatic and interactive editing results.
}%
Our method can automatically synthesize satisfactory results for a variety of patterns without user intervention, as exemplified in \Cref{fig:teaser,fig:full_auto_outputs,fig:semi_auto_outputs} and our (full) results in \Cref{fig:synthesis,fig:initialization_robustness,fig:ablation,fig:ablation:orientation,fig:comparison_sample_synthesis}.
\new{
However, like existing techniques, our method might not always produce what users would like to have, and some artifacts can be visible in local regions (such as unfinished or dangling components in \replace{\Cref{fig:teaser:string:output,fig:comparison_sample_synthesis:ours:curve_recon}}{\Cref{fig:teaser:string:output} and \Cref{fig:comparison_sample_synthesis:ours:curve_recon}} or inconsistent curvatures in \Cref{fig:hier_3} bottom) and global structures (such as the regular and warped grids in \Cref{fig:regular_brick_wall:automatic_synthesis,fig:distorted_grid:automatic_synthesis}, the rectangular blocks in \Cref{fig:blocks:automatic_synthesis}, and the straight lines in \Cref{fig:teaser:maze:output,fig:circuit_board:automatic_synthesis}).
}%
\nothing{
\paragraph{Interactive editing}
}%
For further quality improvement and customization, users can also interactively edit the system suggestions via our system interface, as demonstrated in \Cref{fig:semi_auto_outputs}.
\nothing{
\paragraph{Parameters}
}%
Unless otherwise noted, all our results are produced with three hierarchies using neigborhood radii $\neighborhoodradius \in  \{60, 50, 40\}$ with sampling distance $\samplingdistance \in \{40, 30, 25\}$, while the longer side of bounding box of exemplars are varying between 250 and 500. 
Our method is robust to variations of neighborhood radii.
See \Cref{sec:appendix:parameters} for more details about our parameter settings.

\nothing{
}%

\nothing{
\subsubsection{Direct discrete choice versus analyzed discrete choice}

}

\nothing{
}%

\subsection{Ablation Study}

\begin{figure*}[htb]
	\centering
	\captionsetup[subfigure]{labelformat=empty}
	\captionsetup[subfigure]{justification=centering}
   \setlength{\tabcolsep}{0pt}
\begin{tabular}{ccccc}
    \subfloat[Voronoi]{
   \label{fig:robustness_exemplar:1}
	\includegraphics[width=0.123\linewidth]{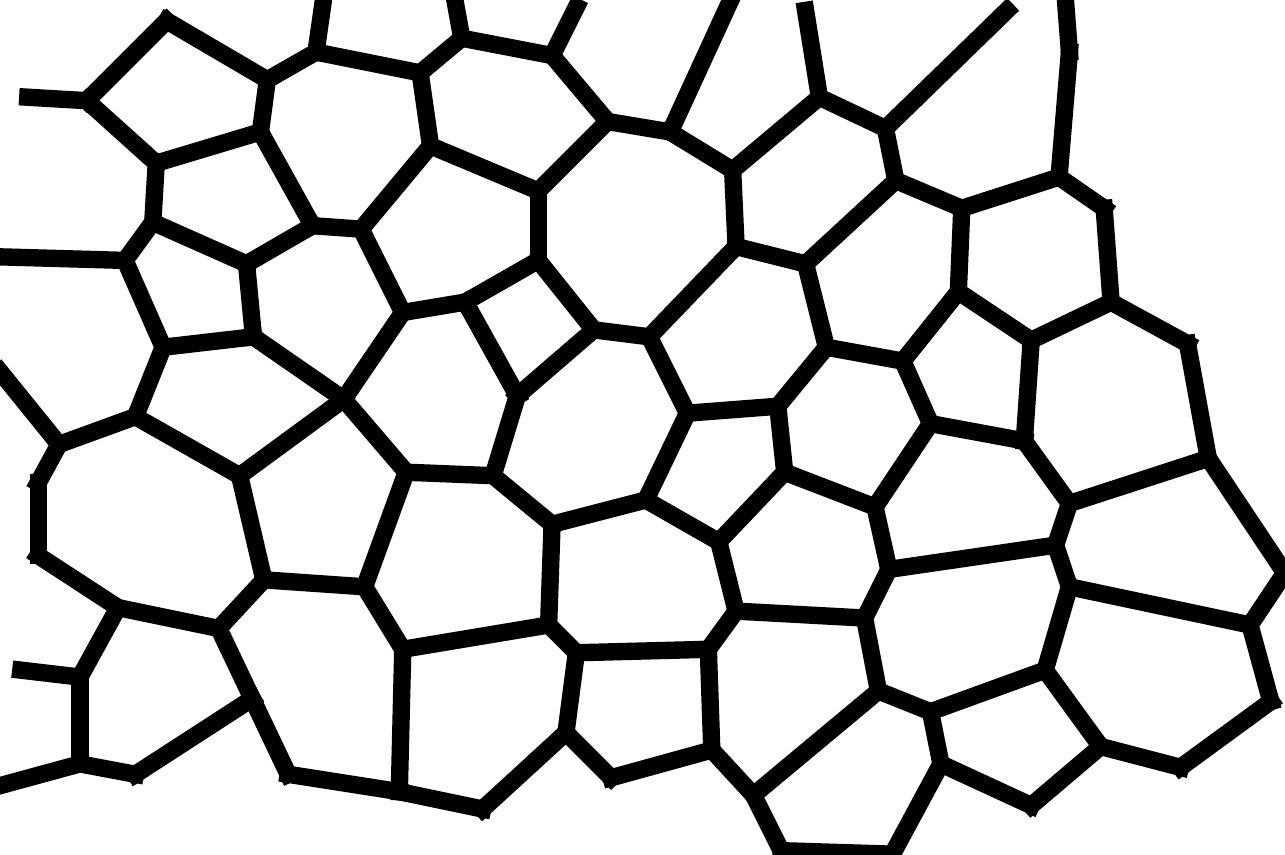}
	}%
&\subfloat[ ]{
	\label{fig:random_patch_copy_init:1}
	\includegraphics[width=0.20\linewidth]{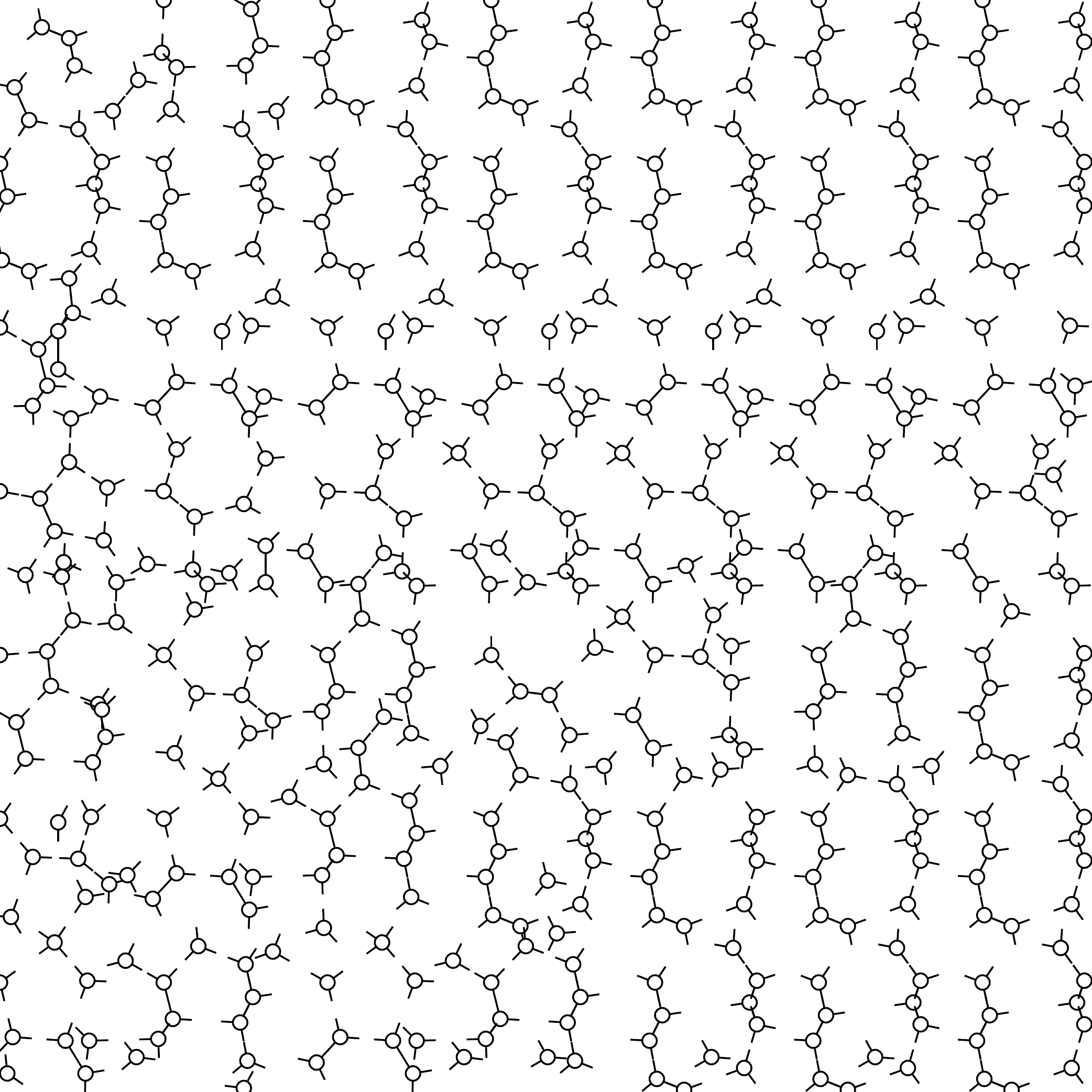}
	}%
&\subfloat[]{
	\label{fig:random_patch_copy_converged:1}
	\includegraphics[width=0.20\linewidth]{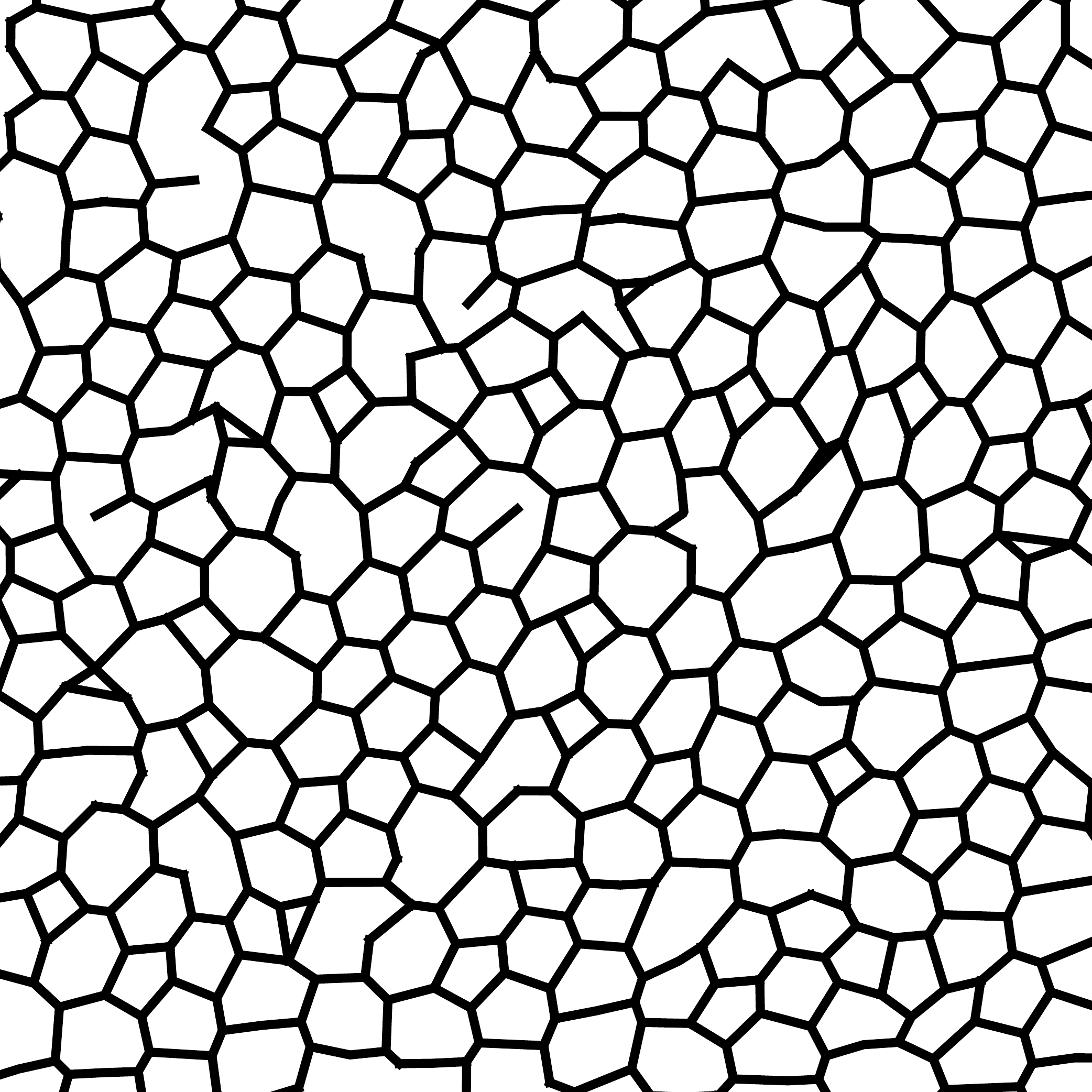}
	}%
&\subfloat[]{
	\label{fig:random_sample_copy_init:1}
	\includegraphics[width=0.20\linewidth]{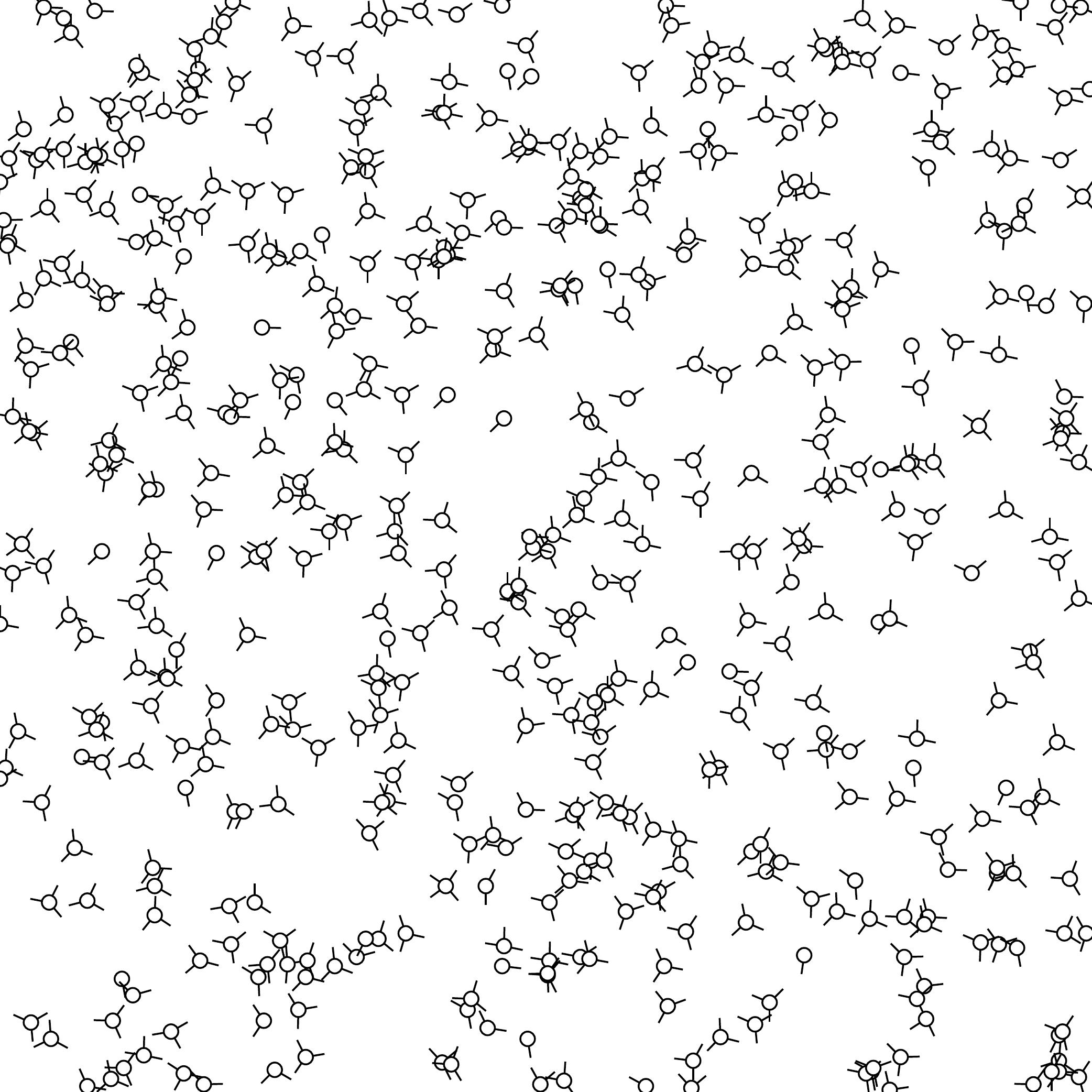}
	}%
&\subfloat[]{
	\label{fig:random_sample_copy_converged:1}
	\includegraphics[width=0.20\linewidth]{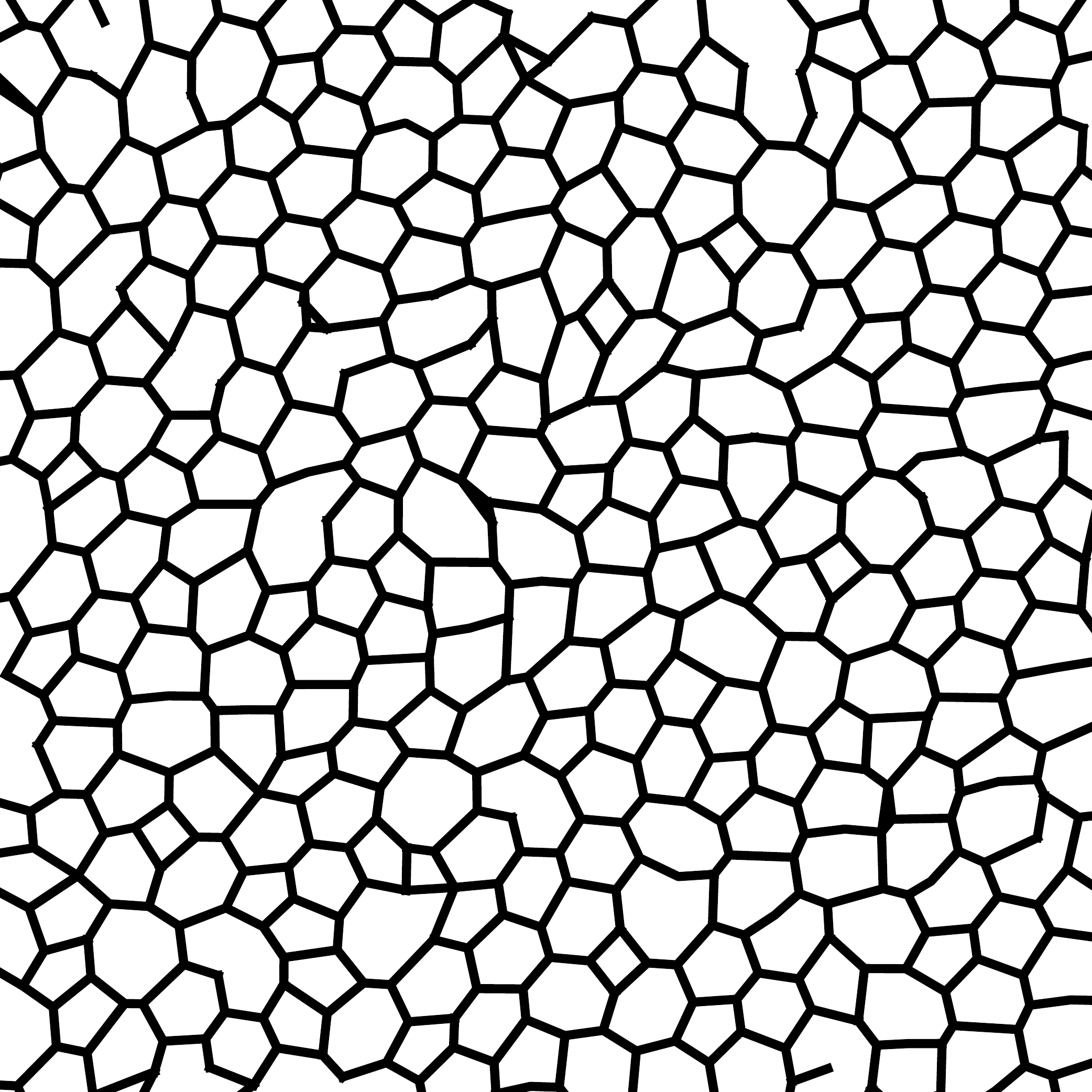}
	}%
\vspace{-2em}
\\
\subfloat[Tree]{
	\label{fig:robustness_exemplar}
	\includegraphics[width=0.120\linewidth]{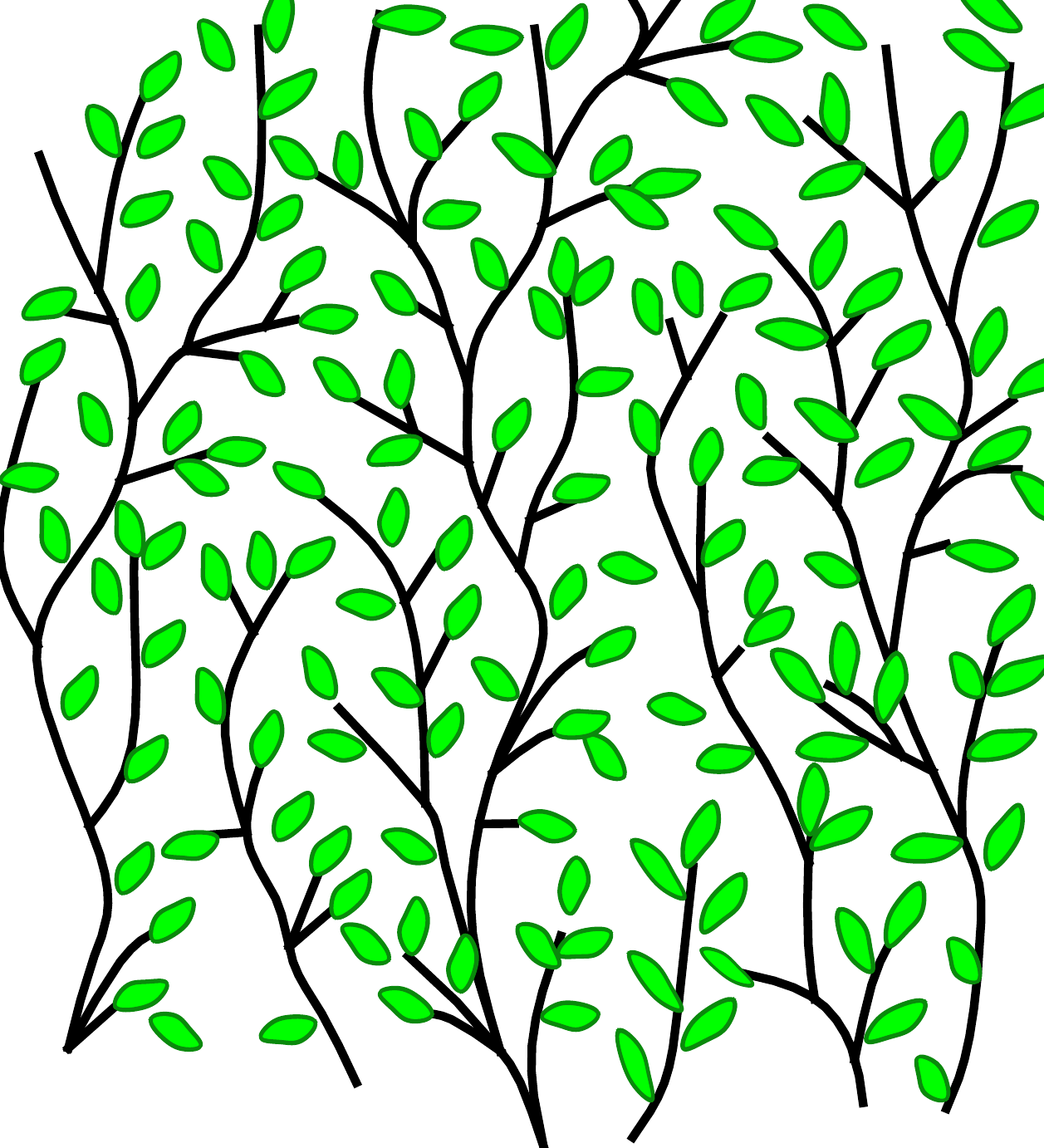}
	}%
&\subfloat[Patch-based  initialization]{
		\label{fig:random_patch_copy_init}
	\includegraphics[width=0.20\linewidth]{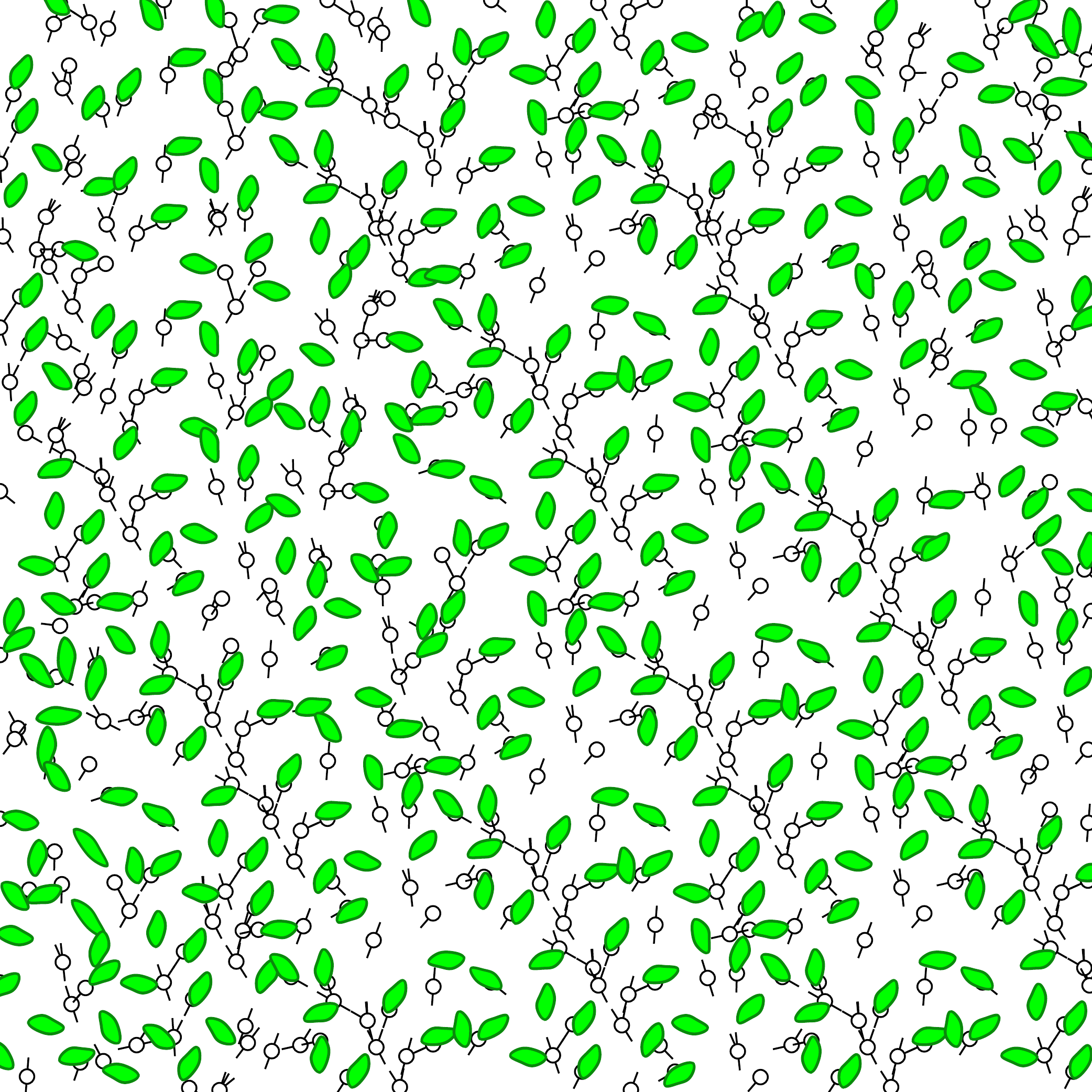}
	}%
&\subfloat[Patch-based \nothing{converged}final output]{
	\label{fig:random_patch_copy_converged:tree}
	\includegraphics[width=0.20\linewidth]{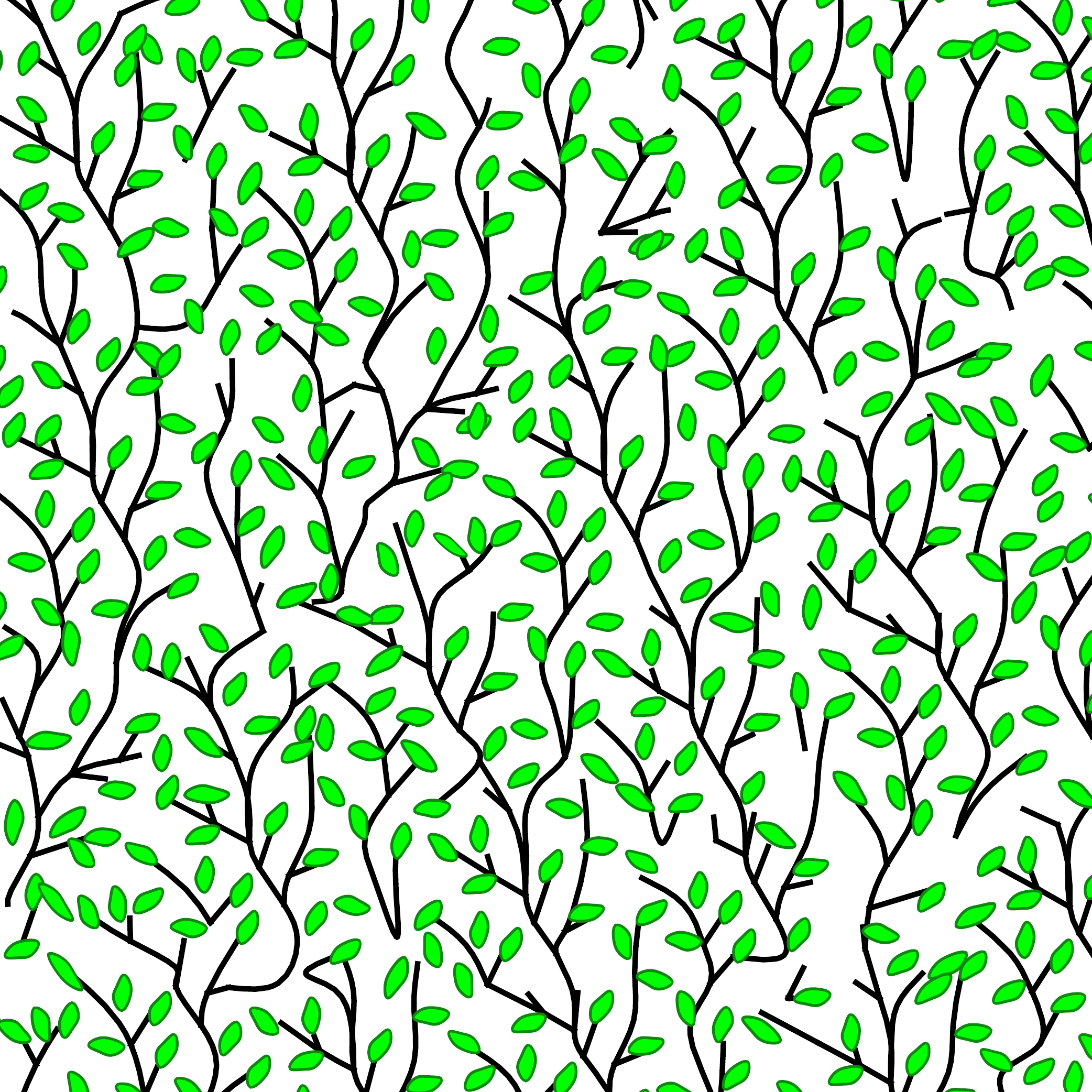}
}%
&\subfloat[Random sample initialization]{
		\label{fig:random_sample_copy_init}
	\includegraphics[width=0.20\linewidth]{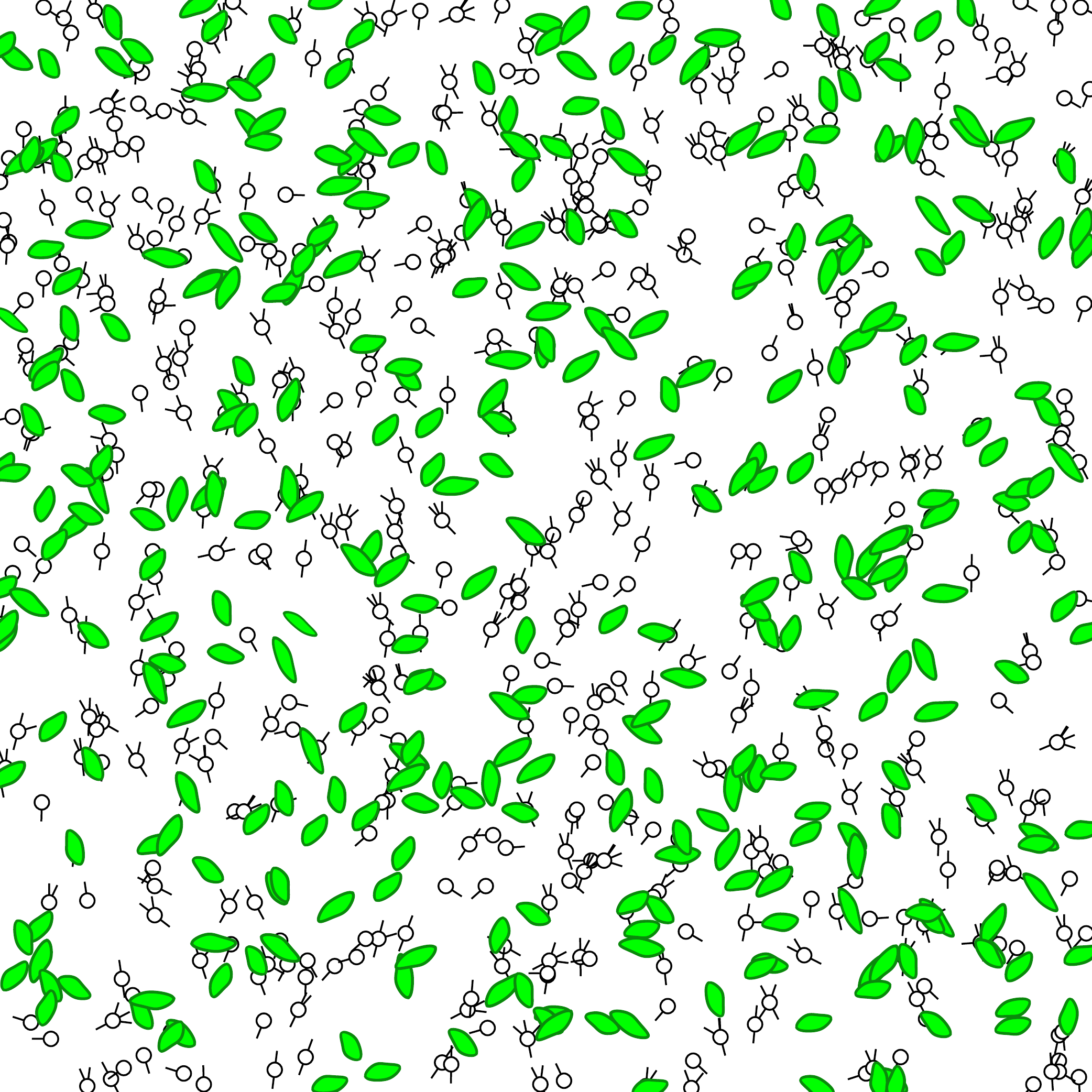}
}
&\subfloat[Random sample final output]{
		\label{fig:random_sample_copy_converged}
	\includegraphics[width=0.20\linewidth]{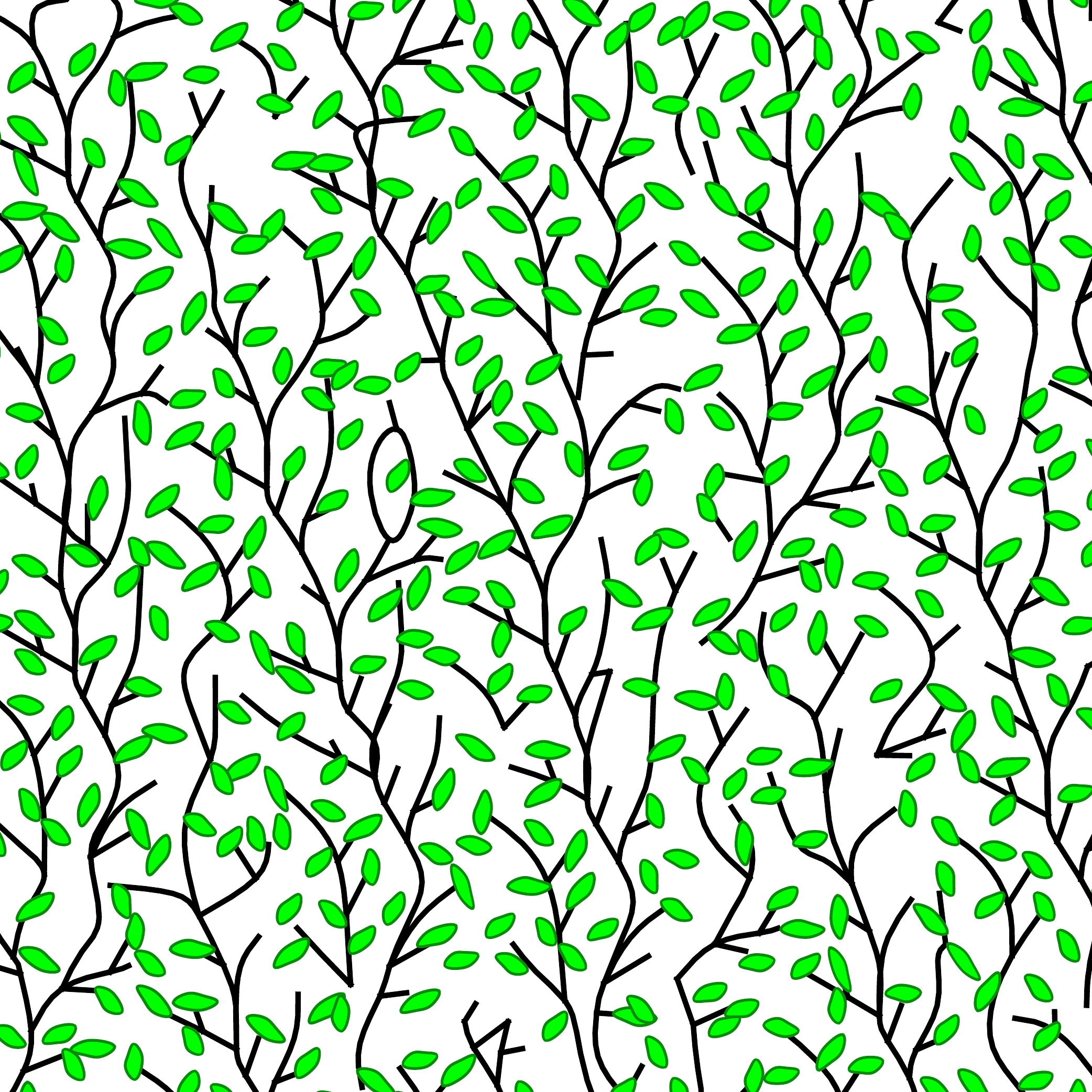}
}
\end{tabular}
	\Caption{Algorithm robustness to different initialization.}
	{%
\nothing{
		\subref{fig:robustness_exemplar} shows the input exemplars; \subref{fig:random_patch_copy_init} shows the initializations by our patch-based initialization method;
		\subref{fig:random_sample_copy_converged} is the converged results of patch-based initialization.
		\subref{fig:random_sample_copy_init}
		is the initialization by randomly copying samples (white noise). \subref{fig:random_sample_copy_converged} shows the converged results of random sample initialization.
}%
		Our algorithm can generate similar results with both patch-based or random initialization. 
		\nothing{
	}%
	}
	\label{fig:initialization_robustness}
\end{figure*}

\nothing{

}%

\begin{figure*}[tbh]
  \centering

\begin{tabular}{cc}
	\subfloat[Crocodile skin]{
	\label{fig:curve_recon:exemplar:1}
	\includegraphics[width=0.114\linewidth]{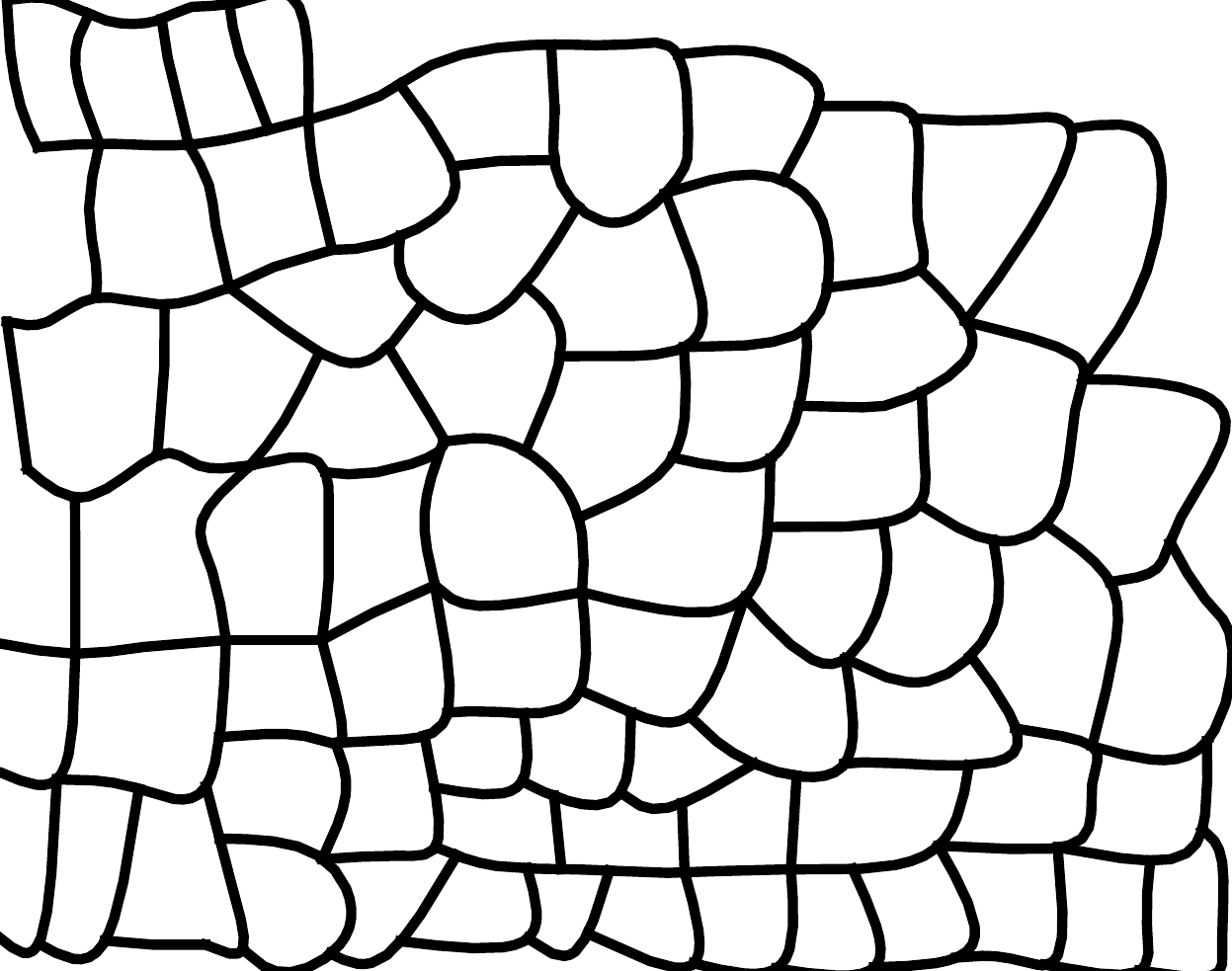}
}%
&
\subfloat[With $\sampleorientations$]{
	\label{fig:curve_recon:ours:1}
	\includegraphics[width=0.19\linewidth]{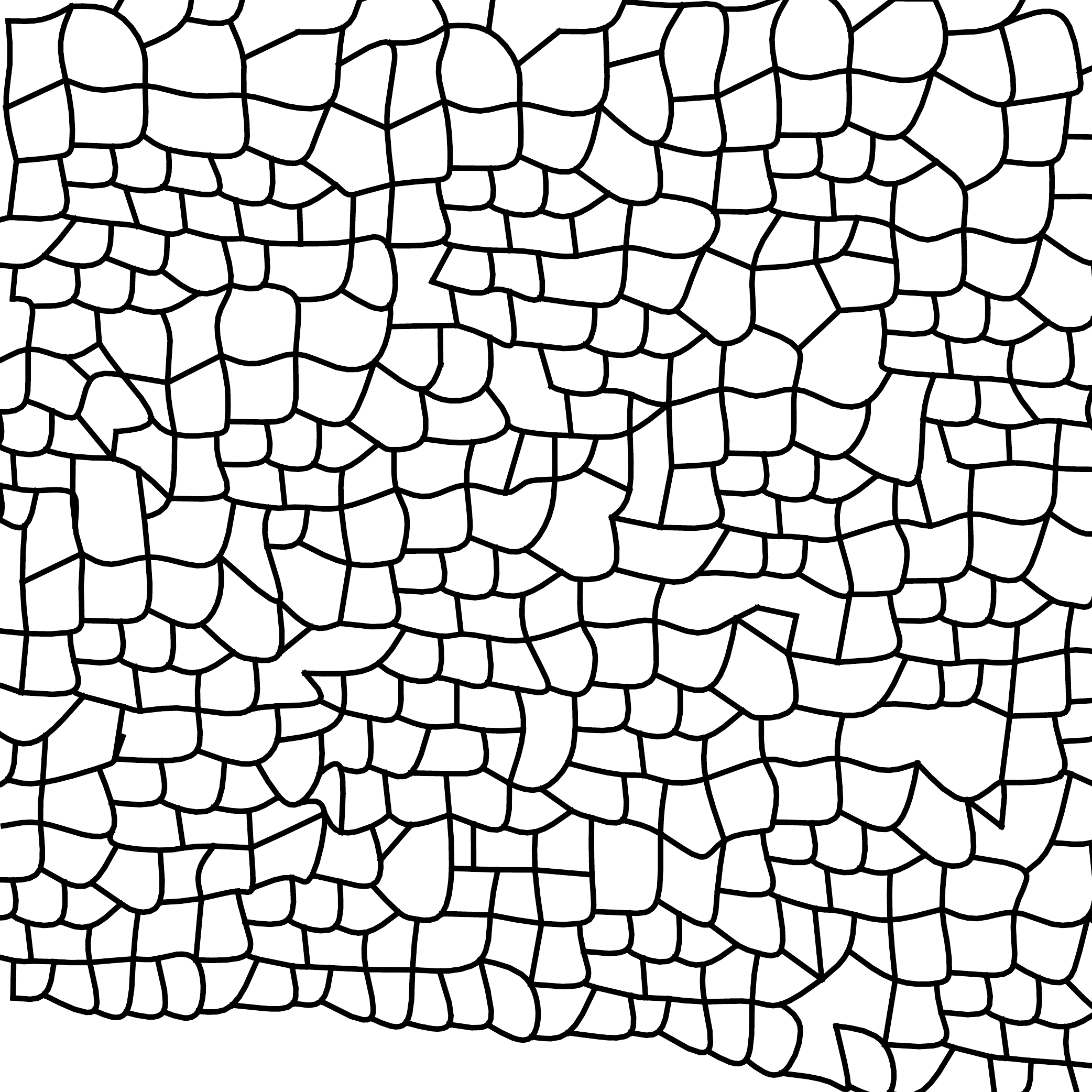}
}%
\end{tabular}
\begin{tabular}{c}
\subfloat[Zoom-in of \protect\subref{fig:curve_recon:ours:1}]{
	\label{fig:curve_recon:with:1}
	\includegraphics[width=0.10\linewidth]{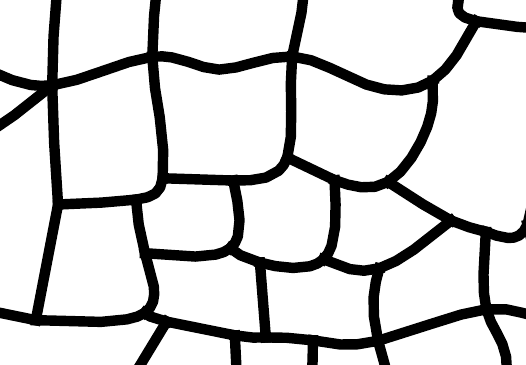}
}%
\vspace{0.2em}
\\
\subfloat[Without $\sampleorientations$\nothing{ (close-up)}]{
	\label{fig:curve_recon:without:1}
	\includegraphics[width=0.10\linewidth]{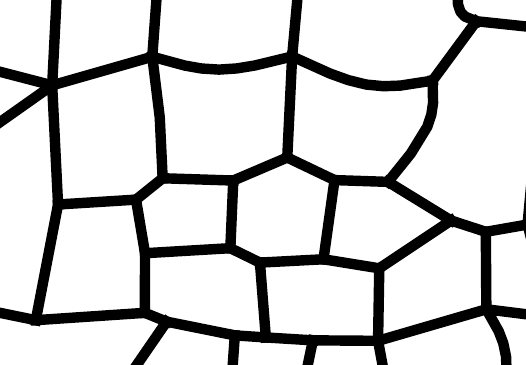}
}%
\end{tabular}
\begin{tabular}{cc}
\subfloat[Flame]{
	\label{fig:curve_recon:exemplar:2}
	\includegraphics[width=0.094\linewidth]{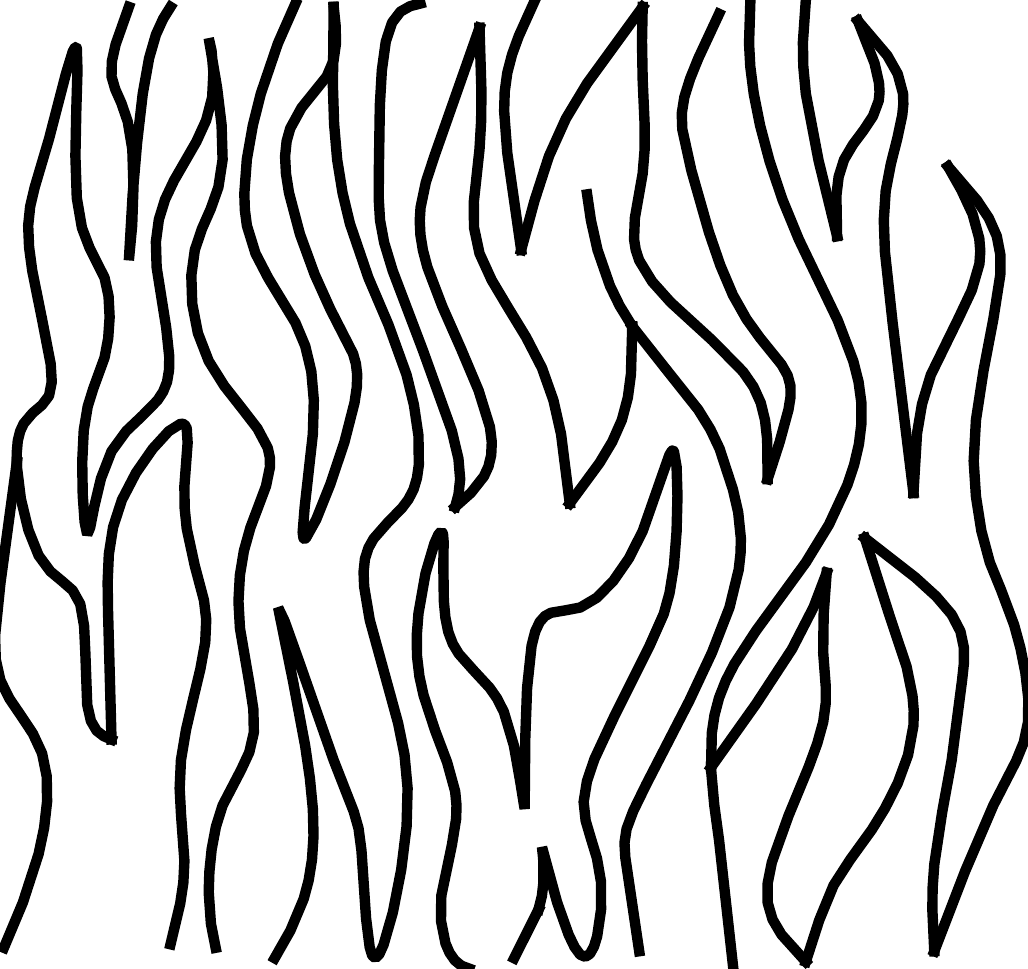}
}%
&
\subfloat[With $\sampleorientations$]{
	\label{fig:curve_recon:ours:2}
	\includegraphics[width=0.19\linewidth]{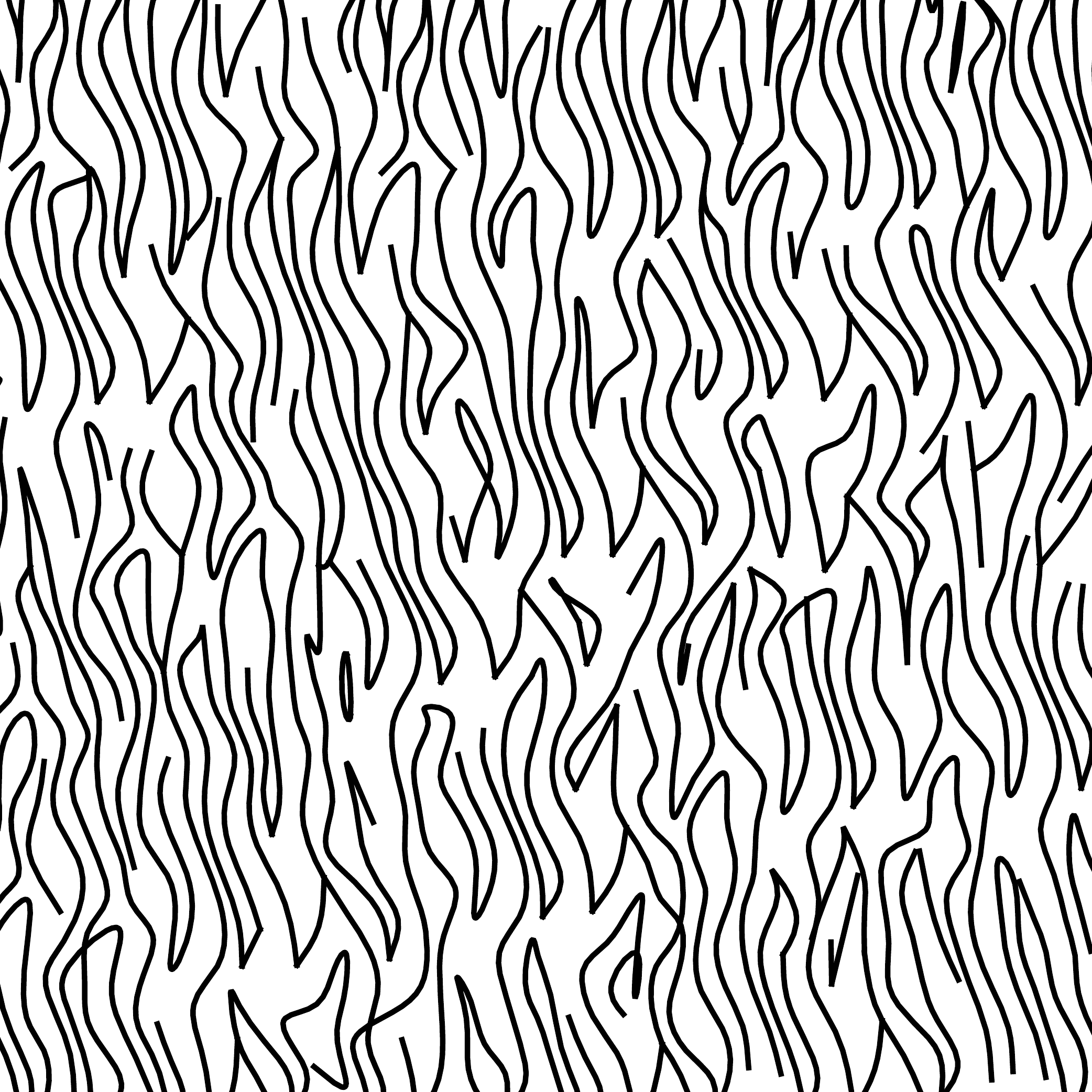}
}%
\end{tabular}
\begin{tabular}{c}
\addlinespace[2ex]
\subfloat[Zoom-in of \protect\subref{fig:curve_recon:ours:2}]{
	\label{fig:curve_recon:with}
	\includegraphics[width=0.10\linewidth]{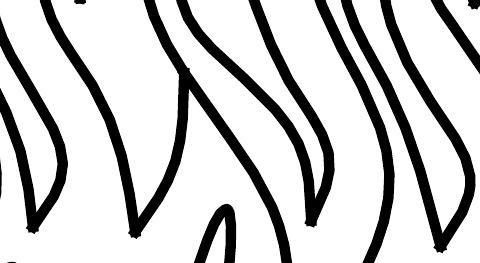}
}%
\vspace{1em}
\\
\subfloat[Without $\sampleorientations$\nothing{ (close-up)}]{
	\label{fig:curve_recon:without:2}
\includegraphics[width=0.10\linewidth]{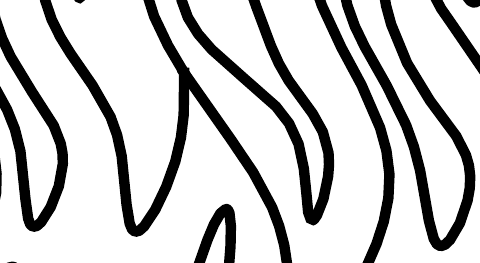}
}
\end{tabular}
\Caption{Ablation study for $\sampleorientations$.}
{%
		The orientation attribute $\sampleorientations$ is useful to faithfully recover curve appearance in the exemplars.
		In the "crocodile skin" example, the curves should be smooth at junction; in the "flame" example, the curves should be sharp at the flame tip.
\nothing{
		Please zoom in the see the differences of results between with or without $\sampleorientations$.
}%
}
\label{fig:ablation:orientation}
\end{figure*}

\begin{figure*}[tbh]
	\centering
  \captionsetup[subfigure]{labelformat=empty}
   	\captionsetup[subfigure]{justification=centering}
   \setlength{\tabcolsep}{0pt}
\begin{tabular}{ccccc}
   \subfloat[Fish scale]{
	\label{fig:ablation:exemplar:1}
	\includegraphics[width=0.137\linewidth]{figs/results/vector_results/exemplars/g_exemplar_circles_pattern_corrected.pdf}
	}%
	&\subfloat[]{
	\label{fig:ablation:search_without_edge:1}
	\includegraphics[width=0.20\linewidth]{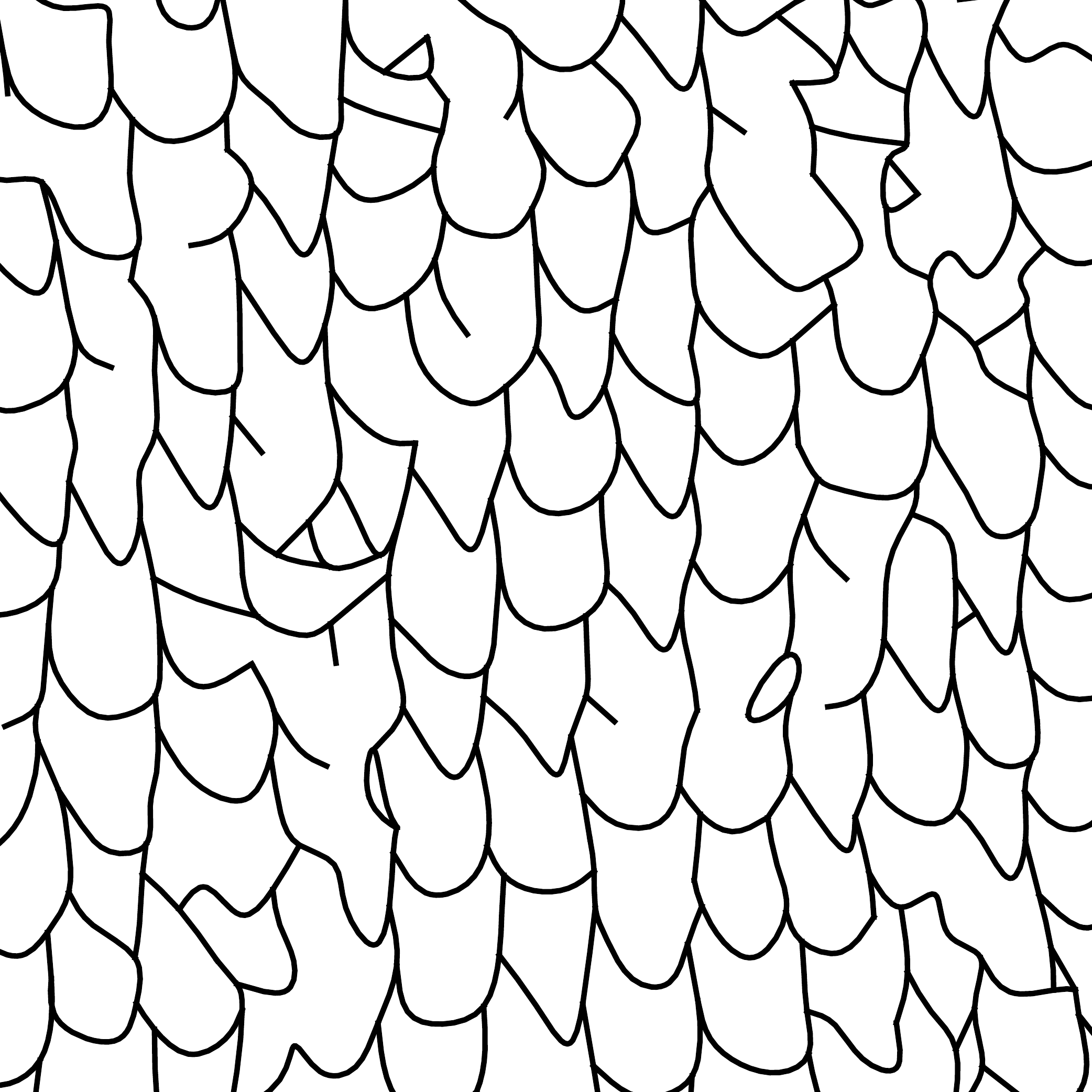}
	}%
	&\subfloat[]{
	\label{fig:ablation:robust_neigh:1}
	\includegraphics[width=0.20\linewidth]{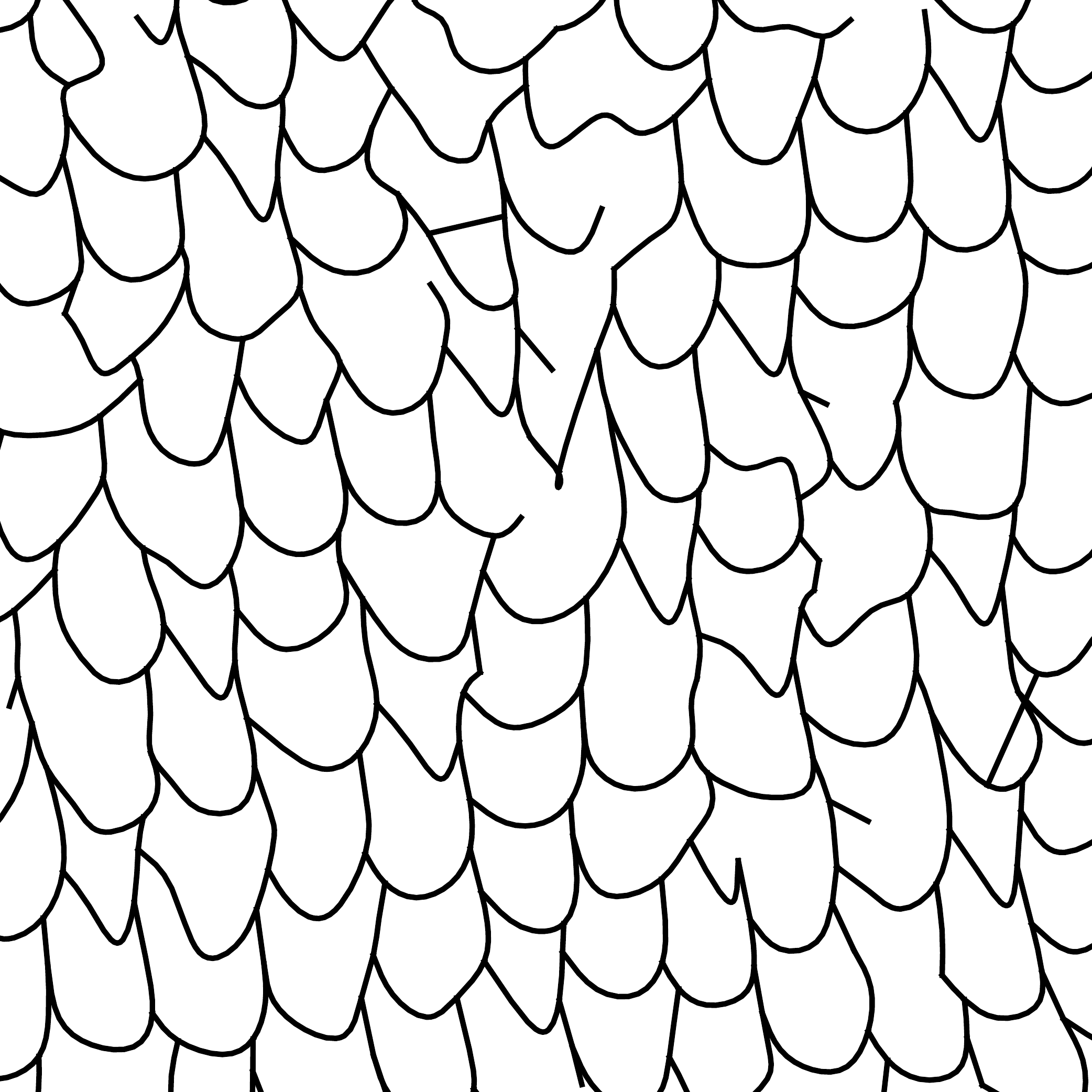}
}%
	&\subfloat[]{
	\label{fig:ablation:without_existence_assignment:1}
	\includegraphics[width=0.20\linewidth]{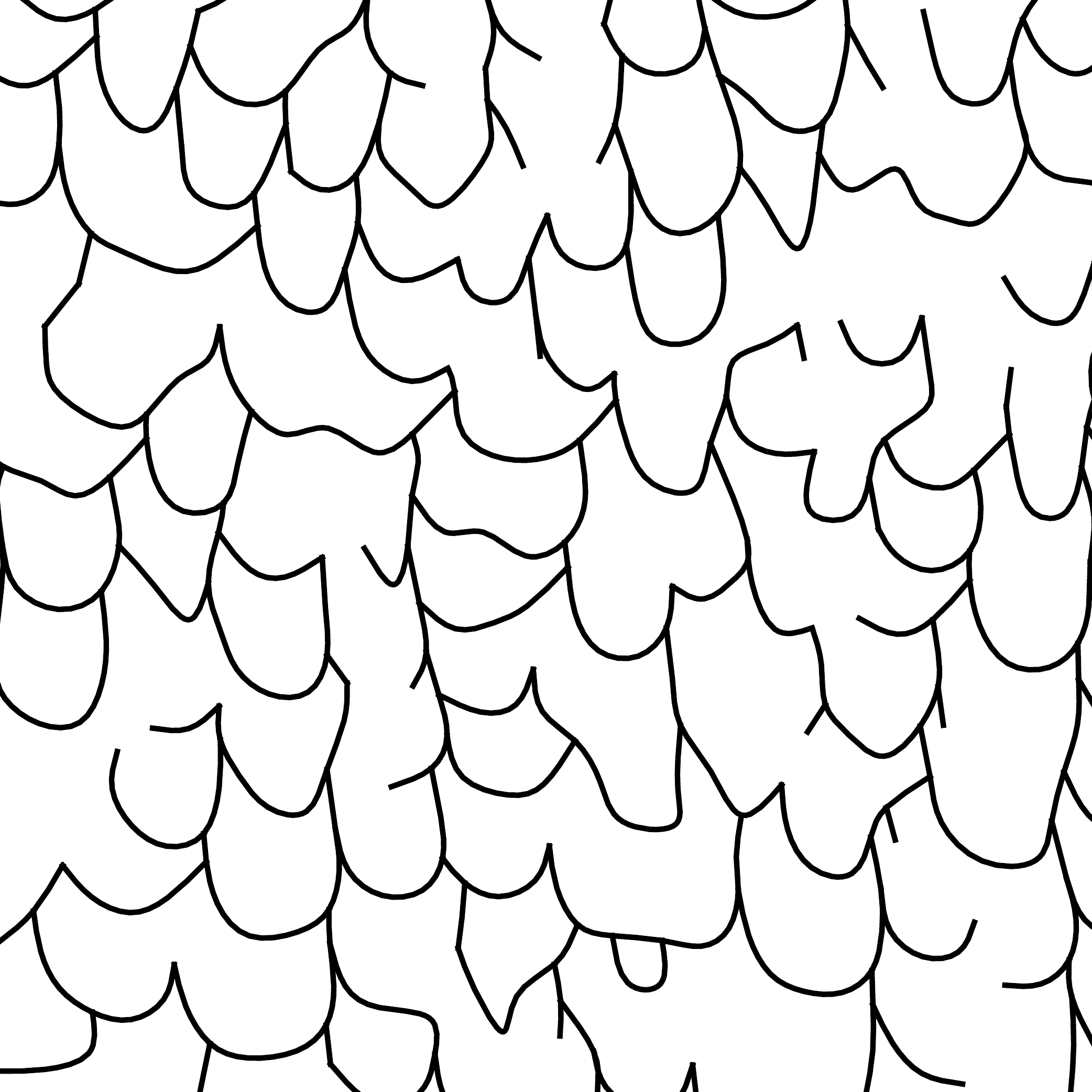}
}%
	&\subfloat[]{
	\label{fig:ablation:full:1}
	\includegraphics[width=0.20\linewidth]{figs/results/vector_results/ours/g_curve_recon_circles_pattern_corrected_n0_60_n1_45_n2_40_level_2_iter_10.pdf}
}%
\vspace{-2em}
\\
   \subfloat[Roof tiles]{
	\label{fig:ablation:exemplar:2}
	\includegraphics[width=0.127\linewidth]{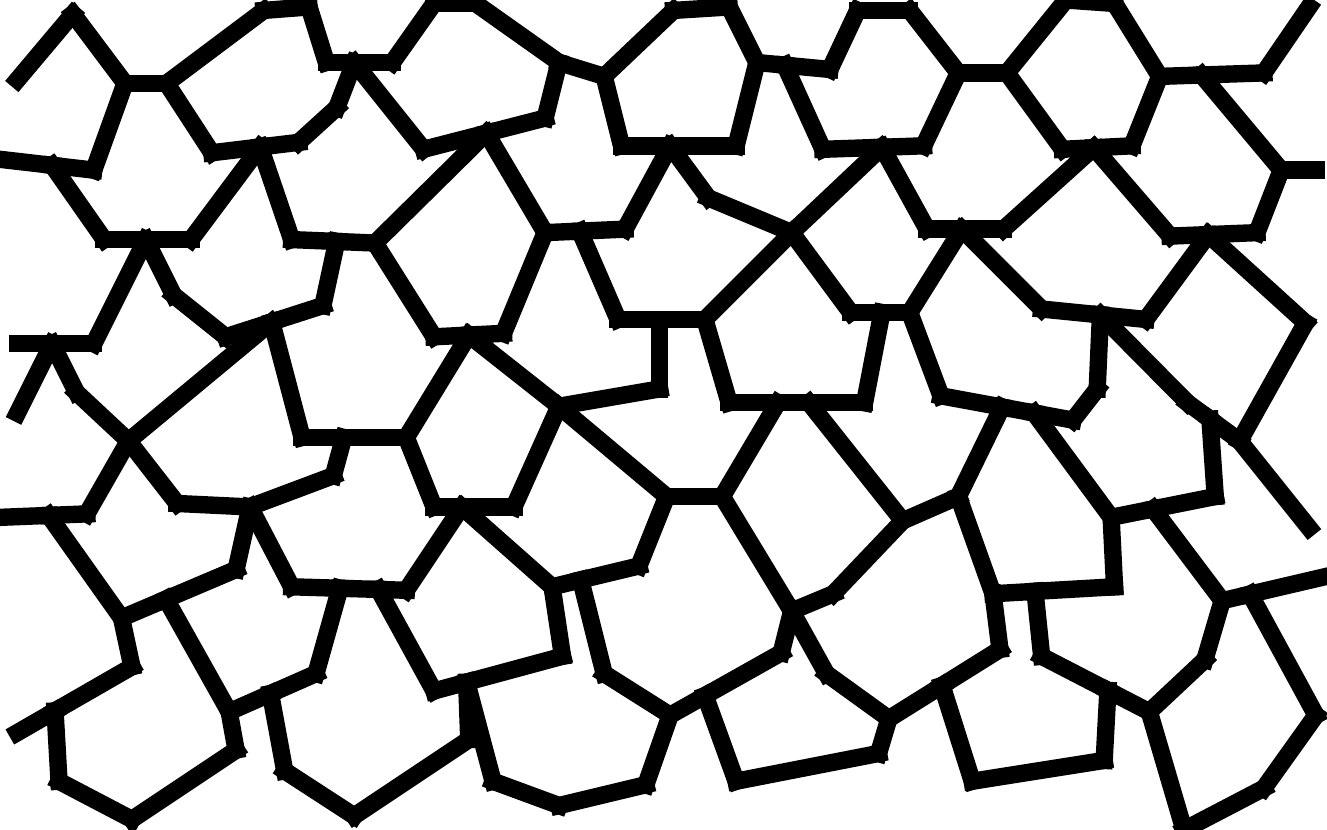}
}%
&\subfloat[Without edge search]{
	\label{fig:ablation:search_without_edge:2}
	\includegraphics[width=0.20\linewidth]{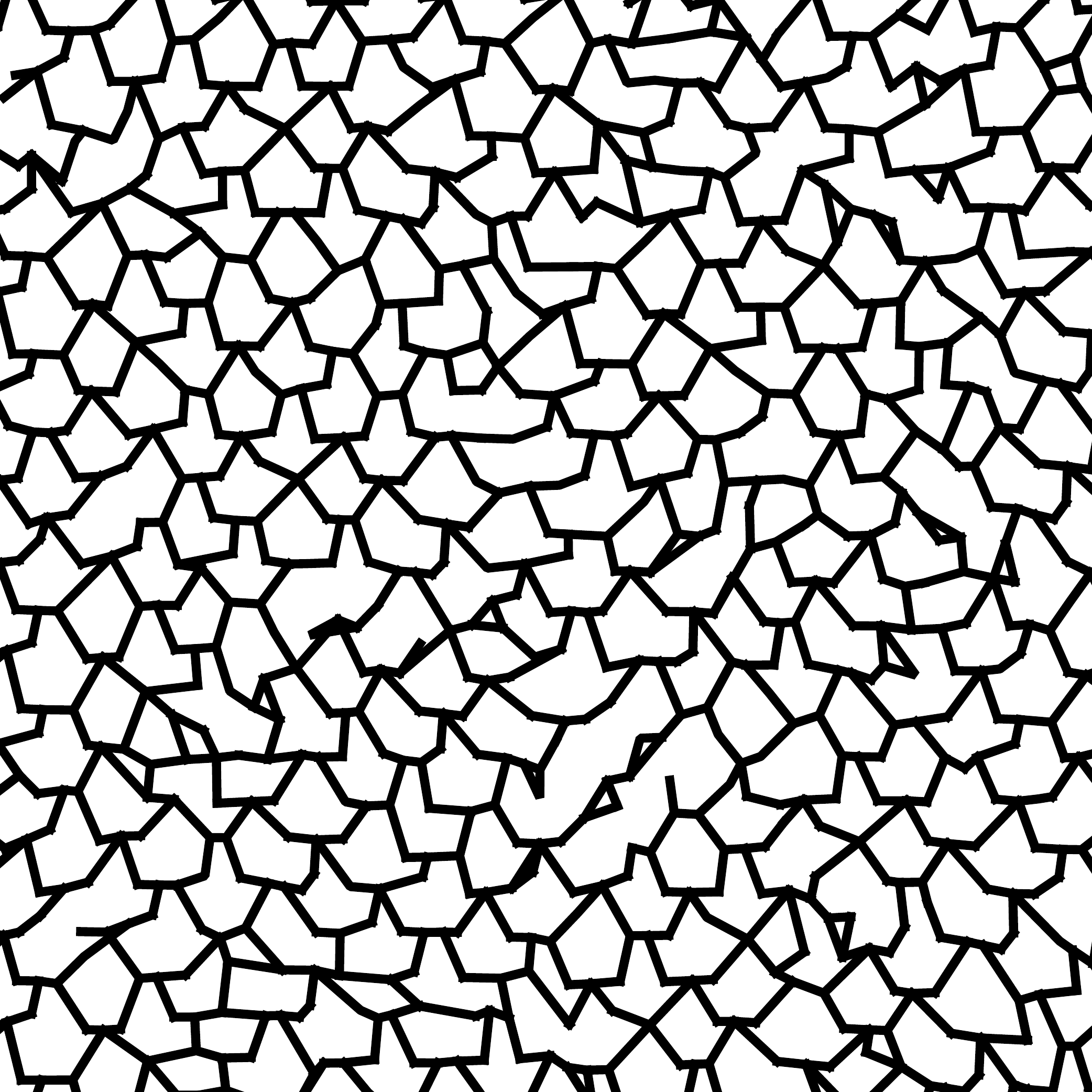}
}%
&\subfloat[Without robust matching]{
	\label{fig:ablation:robust_neigh:2}
	\includegraphics[width=0.20\linewidth]{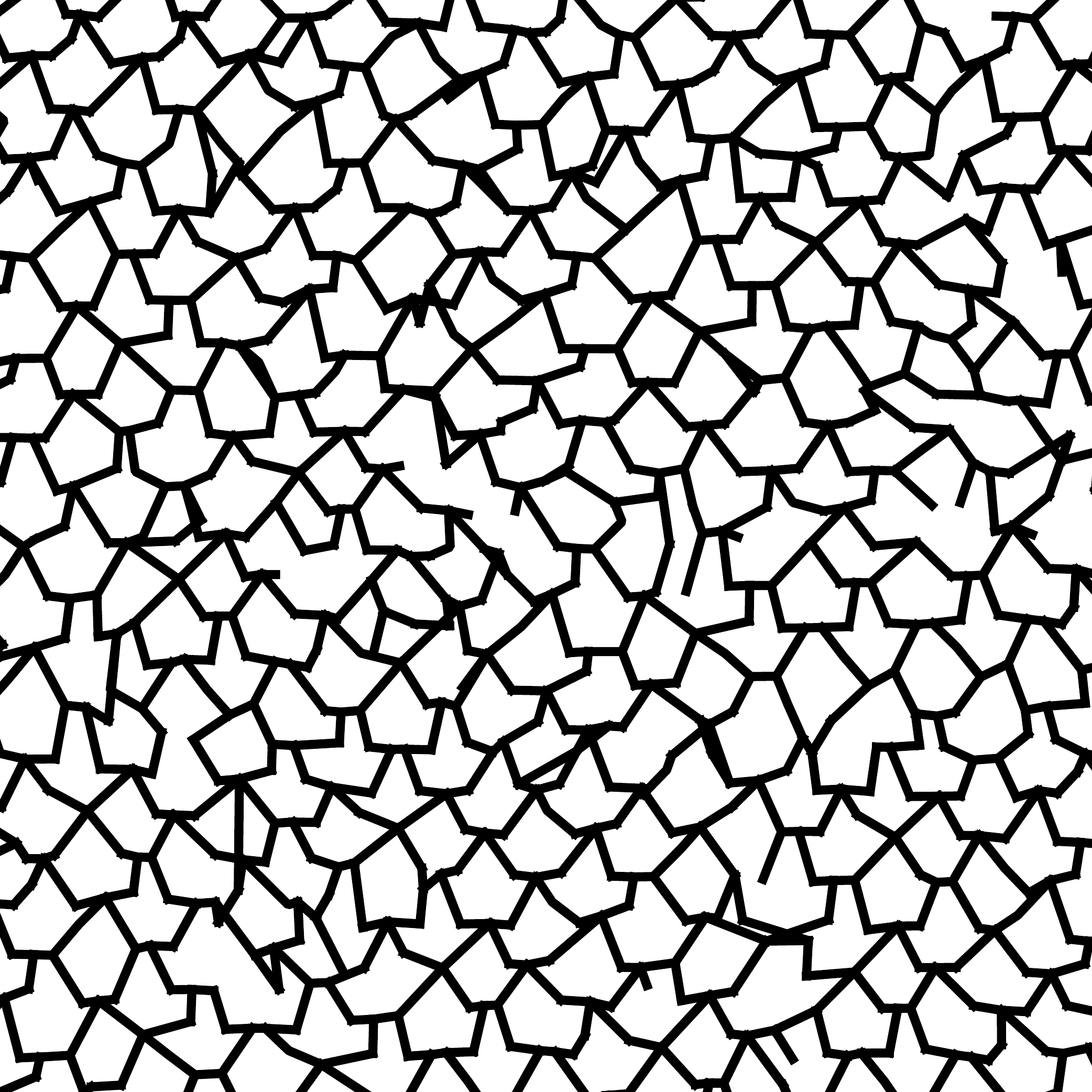}
}%
&\subfloat[Without existence assignment]{
	\label{fig:ablation:without_existence_assignment:2}
	\includegraphics[width=0.20\linewidth]{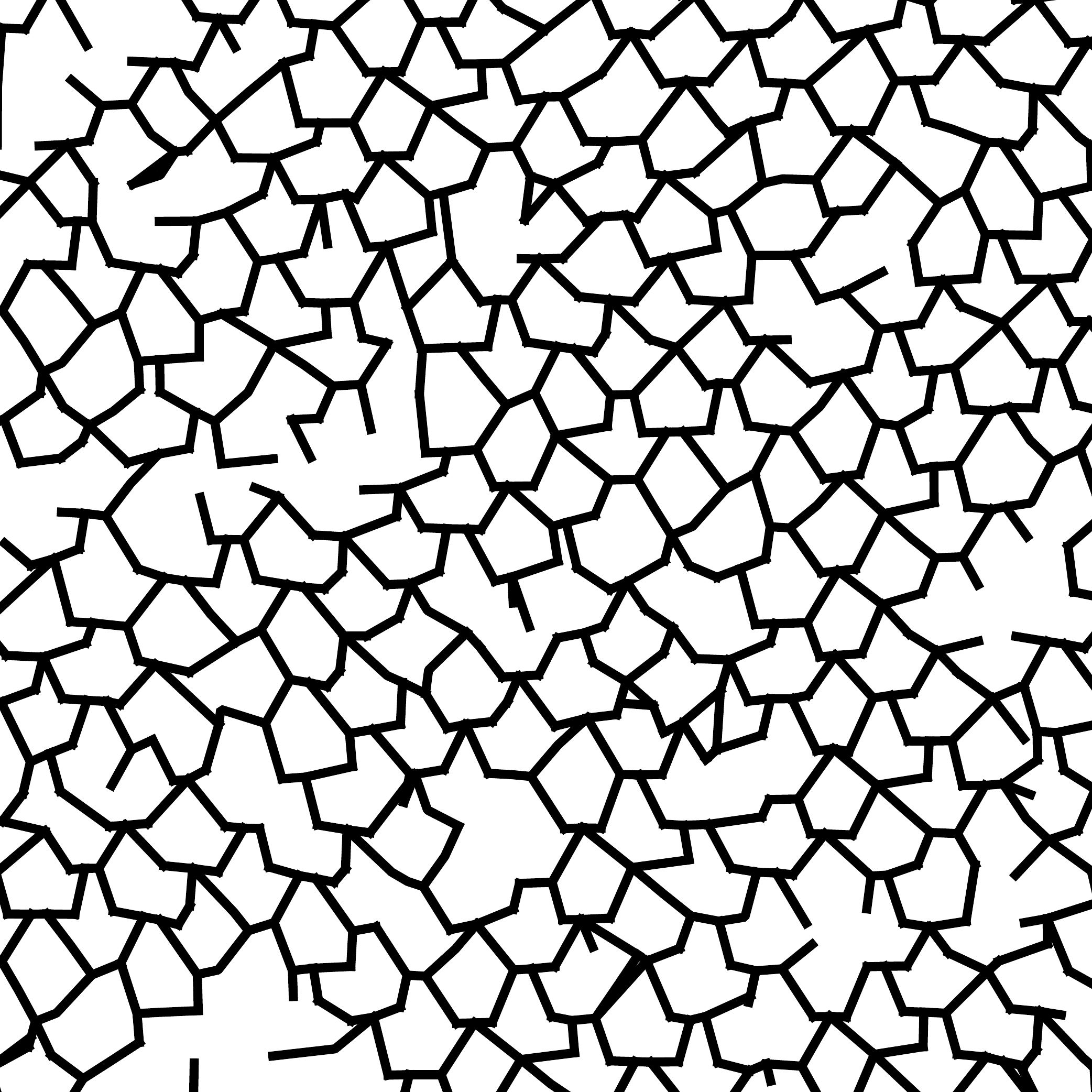}
}%
&\subfloat[Full result]{
	\label{fig:ablation:full:2}
	\includegraphics[width=0.20\linewidth]{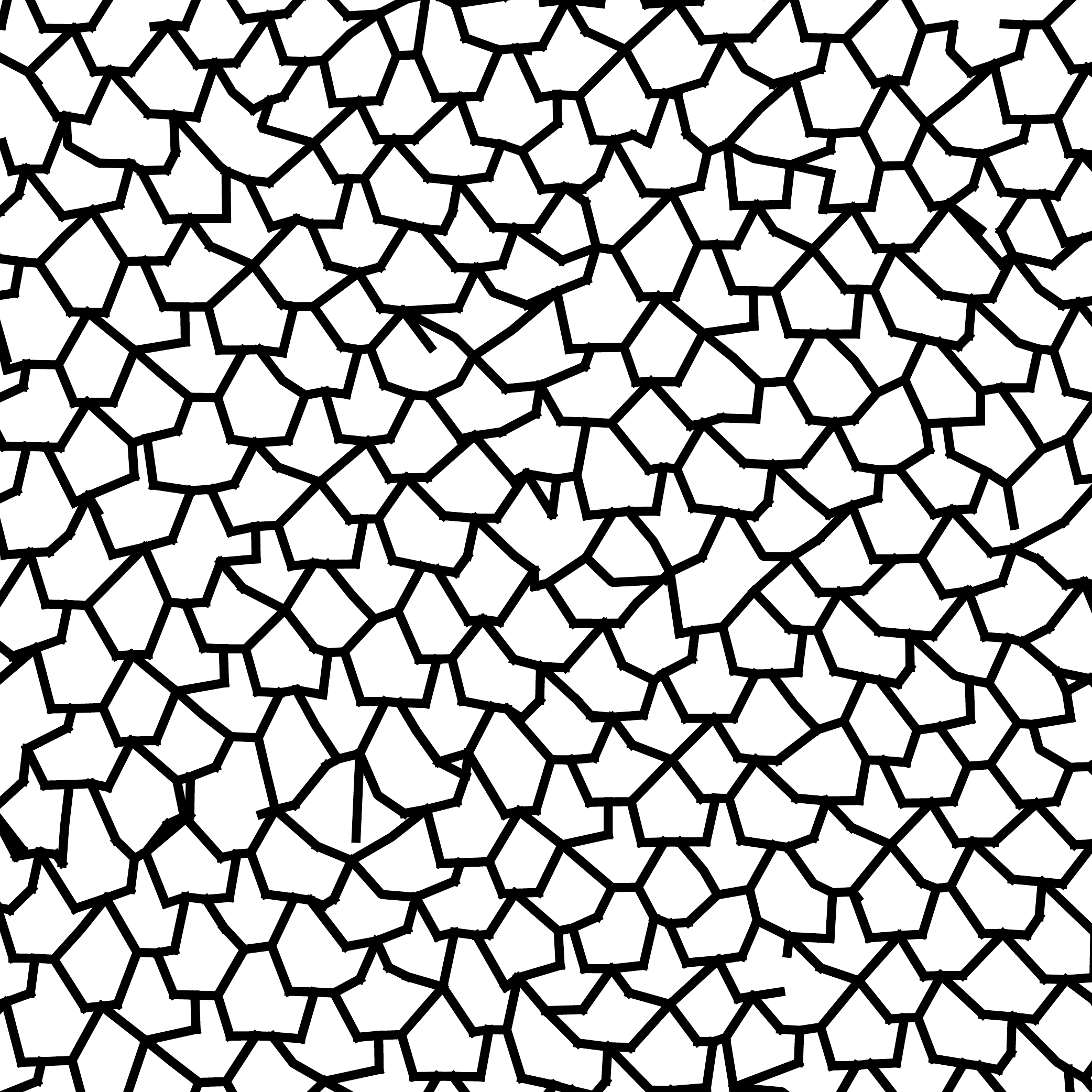}
}%
\end{tabular}

	\Caption{Ablation study.}
	{%
		Without using edges (\Cref{eq:sample_edge_similarity}) in the search step (the second column) or robust matching (\Cref{sec:neighborhood_matching:robust}) that considers outliers (the third column),  the algorithm produces lower quality results.
		Without existence assignment (the third paragraph in \Cref{subsubsec:assignment_step}), the algorithm produces broken curves and empty space due to outliers and missing samples (the forth column). Our results are shown in the last column.
		\nothing{
	        }%
	}
	\label{fig:ablation}
\end{figure*}

Although we use patch-based methods for initialization in our implementation, 
our algorithm is robust to different initial conditions (\Cref{fig:initialization_robustness}), even if the initial sample distribution is randomly distributed (white noise).
\nothing{
The white noise initialization can produce similar results to patch-based one's.
}%
\Cref{fig:ablation:orientation} is the ablation study for the orientation attribute $\sampleorientations$.
\Cref{fig:ablation} shows other components of our algorithm. 
Without the edge term (\Cref{eq:sample_edge_similarity}) or robust matching (\Cref{sec:neighborhood_matching:robust}) in the search step, our algorithm produces lower quality results with obvious artifacts. 
Without existence assignment (the  third paragraph in \Cref{subsubsec:assignment_step}), the algorithm cannot automatically adjust the number of samples within local regions and can produce empty space or extra broken curves.

\nothing{
}%

\nothing{
\subsection{Sampling rate}
Results produced with various sampling rates

\begin{figure*}[htb]
	\centering
	\captionsetup[subfigure]{labelformat=empty}
	
   \subfloat[]{
	\label{fig:sampling_rate:exemplar:1}
	\includegraphics[width=0.13\linewidth]{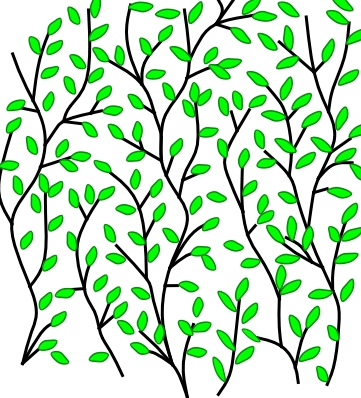}
	}%
	\subfloat[]{
		\label{fig:sampling_rate:40:1}
	\includegraphics[width=0.20\linewidth]{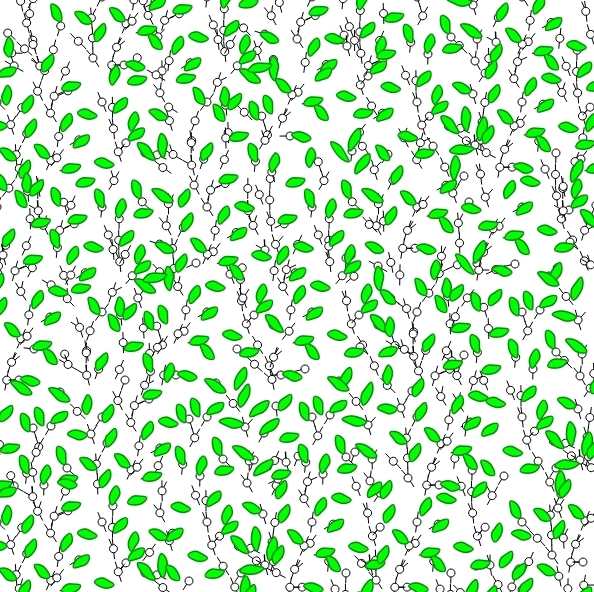}
	}%
	\subfloat[]{
		\label{fig:sampling_rate:30:1}
	\includegraphics[width=0.20\linewidth]{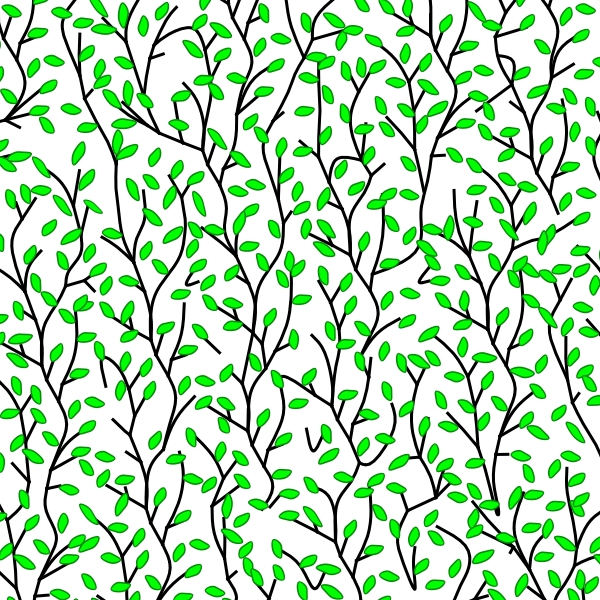}
}%
	\subfloat[]{
		\label{fig:sampling_rate:20:1}
	\includegraphics[width=0.20\linewidth]{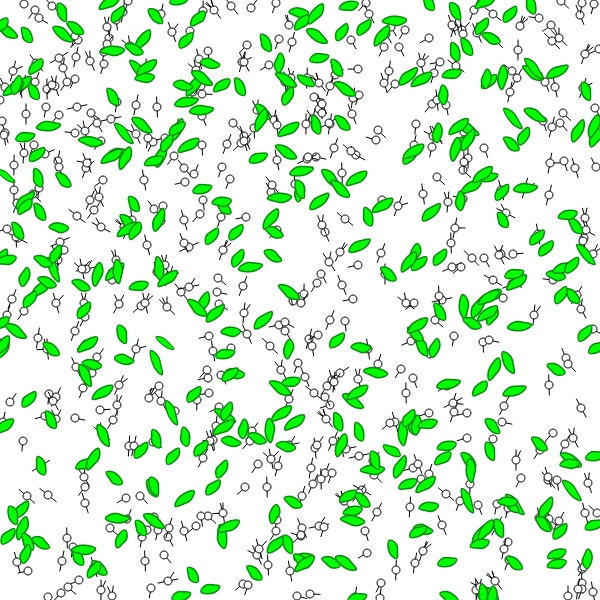}
}
	\subfloat[]{
	\label{fig:sampling_rate:10:1}
	\includegraphics[width=0.20\linewidth]{figs/method/robustness/sample_init_mixed_tree_random_sample.jpg}
}

   \subfloat[examplar]{
	\label{fig:sampling_rate:exemplar:2}
	\includegraphics[width=0.13\linewidth]{figs/method/robustness/exemplar_mixed_tree.jpg}
}%
\subfloat[40]{
	\label{fig:sampling_rate:40:2}
	\includegraphics[width=0.20\linewidth]{figs/method/robustness/sample_init_mixed_tree_patch.jpg}
}%
\subfloat[30]{
	\label{fig:sampling_rate:30:2}
	\includegraphics[width=0.20\linewidth]{figs/method/robustness/curve_recon_mixed_tree_patch.jpg}
}%
\subfloat[20]{
	\label{fig:sampling_rate:20:2}
	\includegraphics[width=0.20\linewidth]{figs/method/robustness/sample_init_mixed_tree_random_sample.jpg}
}
\subfloat[10]{
	\label{fig:sampling_rate:10:2}
	\includegraphics[width=0.20\linewidth]{figs/method/robustness/sample_init_mixed_tree_random_sample.jpg}
}

	\Caption{Results generated with various sampling rate.}
	{%

	}
	\label{fig:sampling_rate}
\end{figure*}

}%

\nothing{
}%

\subsection{Comparison to Previous Methods}

To our knowledge, there is no previous example-based method\nothing{ except \cite{Bian:2018:TPD}} that can generate the types of patterns we target.
\nothing{
}%
\nothing{
However, \cite{Bian:2018:TPD} requires manual authoring as it lacks an automatic synthesis method for 2D patterns.
}%
The sample-based methods in \cite{Ma:2011:DET,Roveri:2015:EBR,Tu:2019:PPS} are the most related.
\nothing{
The key to synthesize continuous patterns is our novel graph-based representation and other new techniques in the sample synthesis.
}%
We compare against \cite{Ma:2011:DET,Roveri:2015:EBR} and a state-of-art point distribution synthesis method in \cite{Tu:2019:PPS} which applies convolutional neural networks to preserve both local and global structures\nothing{~for synthesized samples}.
As shown in \Cref{fig:comparison_sample_synthesis}, our method can produce better spatial sample distributions\nothing{ by considering their connections} than \cite{Ma:2011:DET,Roveri:2015:EBR,Tu:2019:PPS}. 
Note that we compare only sample distributions in \Cref{fig:comparison_sample_synthesis} since it is unclear how to reconstruct continuous curve patterns from synthesized samples without connectivity \cite{Ma:2011:DET,Roveri:2015:EBR,Tu:2019:PPS}.

\nothing{
Without recording the connectivity information, recover it by post-processing the synthesized sample distributions using curve reconstruction algorithms.
}
\nothing{
Thus, we compare only sample positions but without curve connections.
}%
\nothing{
}%

\begin{figure*}[thb]
	\centering
	\captionsetup[subfigure]{labelformat=empty}
	\captionsetup[subfigure]{justification=centering}
		\setlength{\tabcolsep}{-1.25pt}
\begin{tabular}{cccccc}
   \subfloat[Waves (large)]{
	\includegraphics[width=0.145\linewidth]{figs/results/vector_results/exemplars/g_exemplar_global_waves.pdf}
}%
 &\subfloat[]{
	\includegraphics[width=0.18\linewidth]{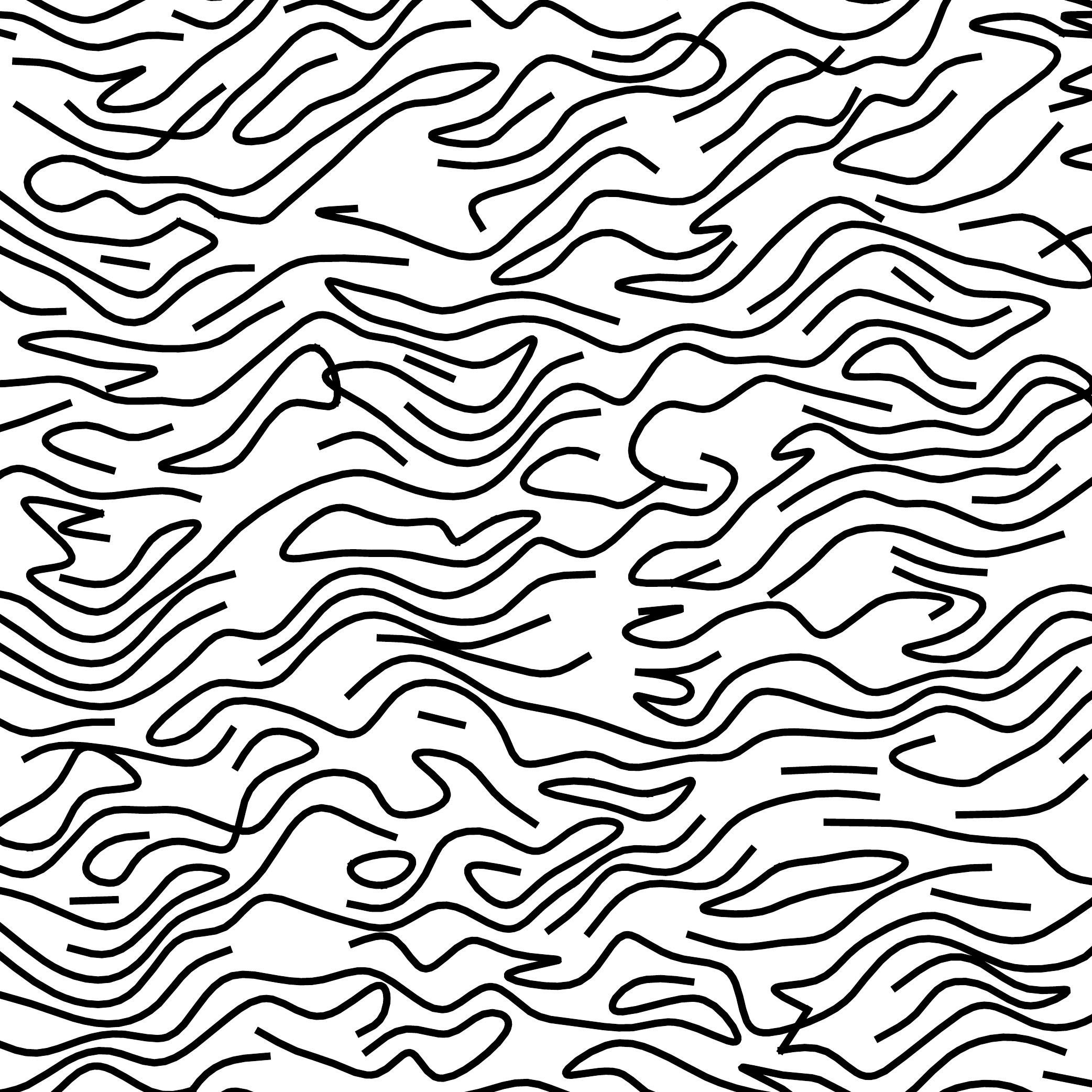}
}%
 &\subfloat[]{
     \label{fig:comparison_pattern_synthesis:wave:our}
	\includegraphics[width=0.18\linewidth]{figs/results/vector_results/hier/g_curve_recon_global_waves_n0_60_n1_50_n2_40_level_2_iter_6.pdf}
}%
& \subfloat[Roof tiles]{
	\includegraphics[width=0.114\linewidth]{figs/results/vector_results/exemplars/g_exemplar_roof_tiles.pdf}
}%
&\subfloat[]{
	\includegraphics[width=0.18\linewidth]{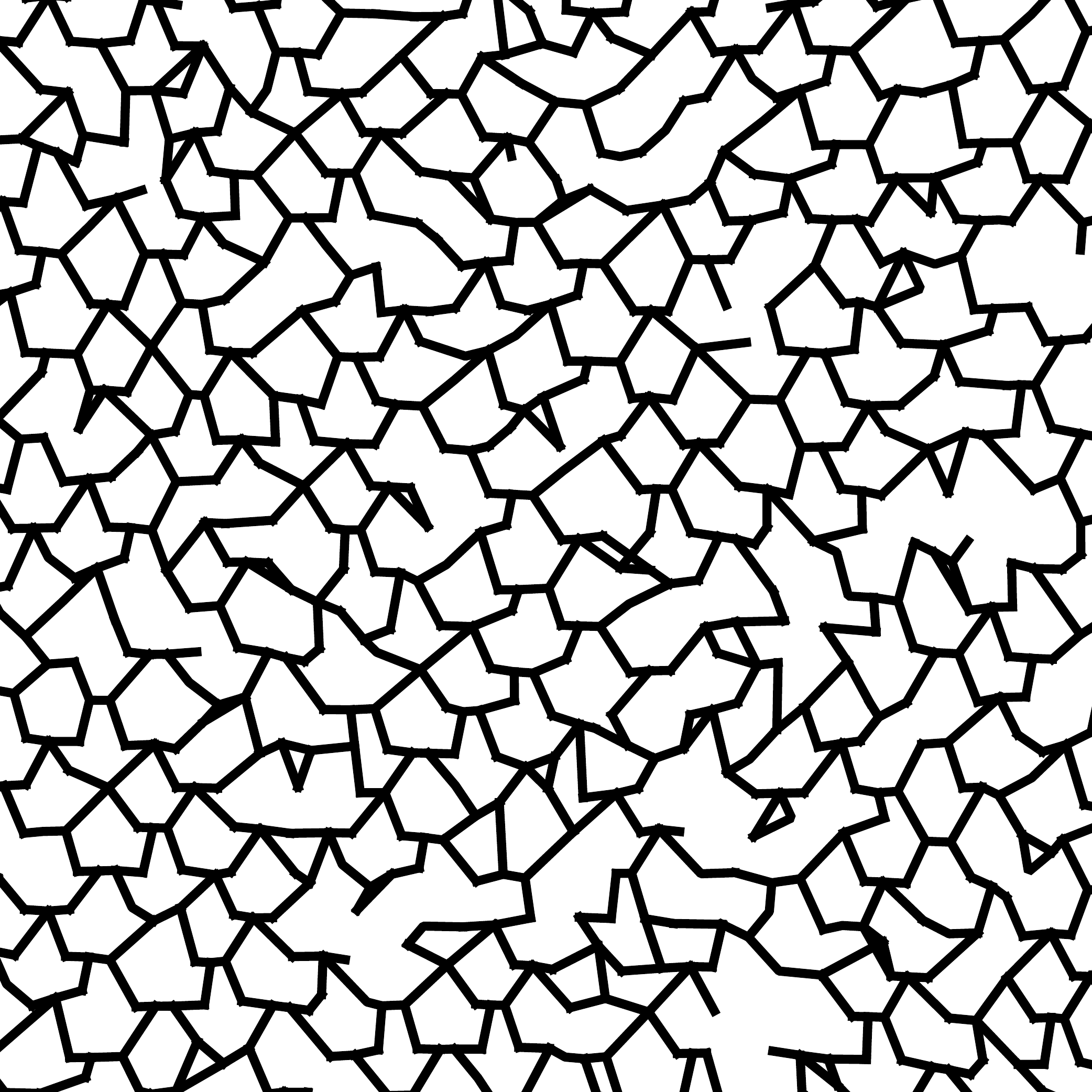}
}%
&\subfloat[]{
	\includegraphics[width=0.18\linewidth]{figs/results/vector_results/ours/g_curve_recon_roof_tiles_n0_60_n1_50_n2_40_level_2_iter_6.pdf}
}%
	\vspace{-2em}
	\\
   \subfloat[Fish scale]{
	\includegraphics[width=0.1235\linewidth]{figs/results/vector_results/exemplars/g_exemplar_circles_pattern_corrected.pdf}
}%
	&\subfloat[Enhanced \protect\cite{Ma:2011:DET}]{
	\includegraphics[width=0.18\linewidth]{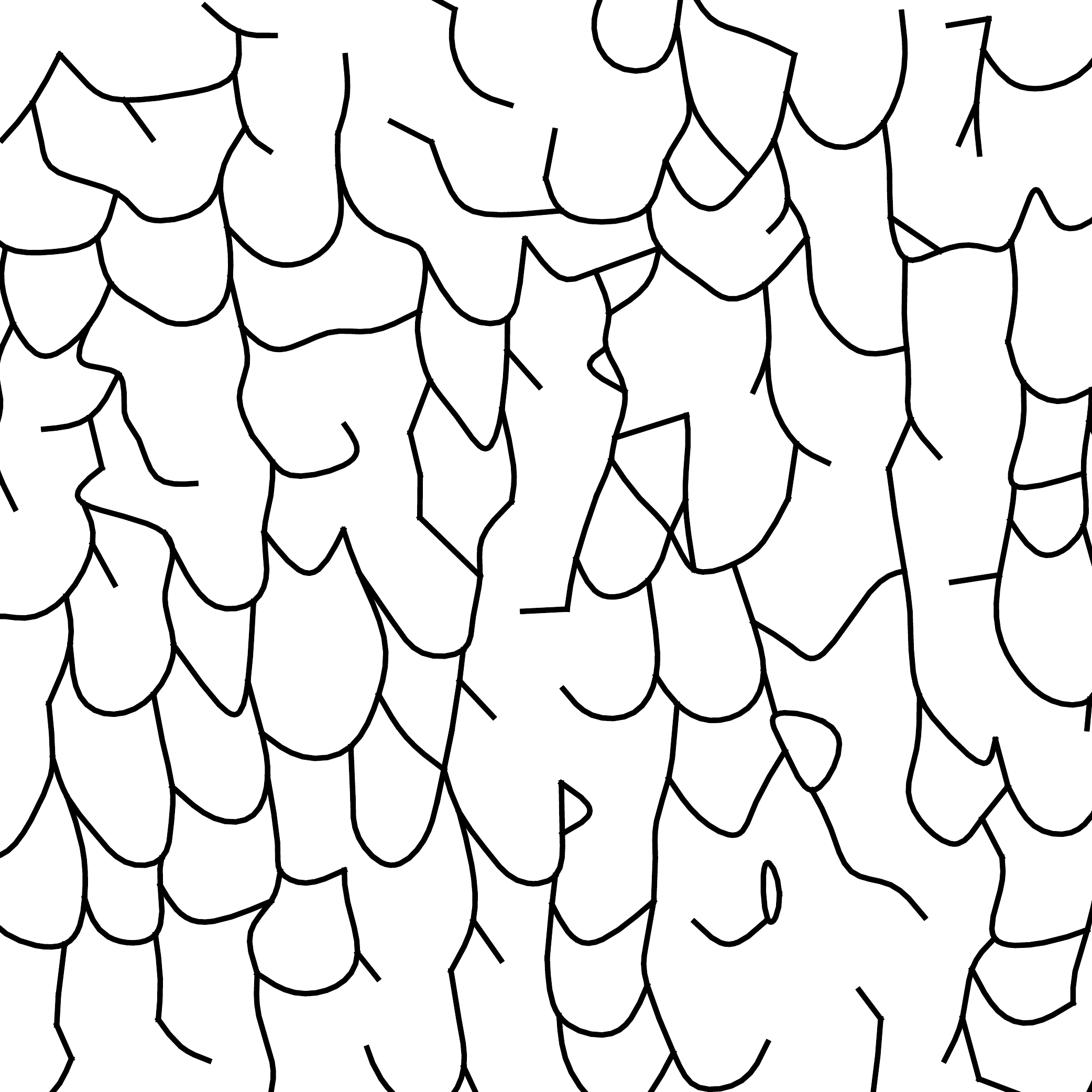}
	}%
	&\subfloat[Ours]{
	\includegraphics[width=0.18\linewidth]{figs/results/vector_results/ours/g_curve_recon_circles_pattern_corrected_n0_60_n1_45_n2_40_level_2_iter_10.pdf}
	}%
&
   \subfloat[Wood ring]{
	\includegraphics[width=0.083\linewidth]{figs/results/vector_results/exemplars/g_exemplar_wood_ring.pdf}
}%
	&\subfloat[\nothing{connectivity + edge assignment+}Enhanced \protect\cite{Ma:2011:DET}]{
	\includegraphics[width=0.18\linewidth]{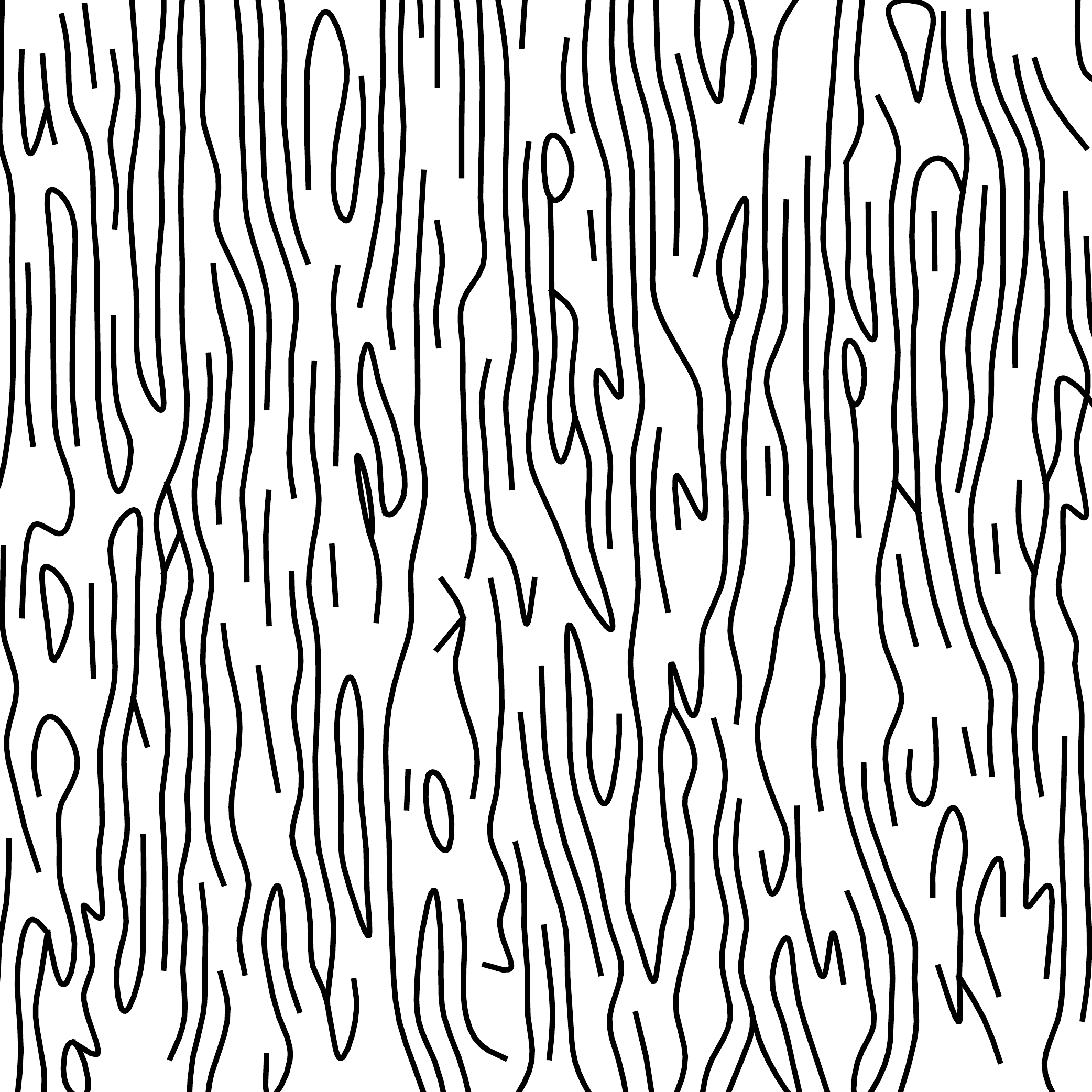}
	}%
	&\subfloat[Ours]{
	\label{fig:wood_ring:automatic_synthesis}
	\includegraphics[width=0.18\linewidth]{figs/results/vector_results/ours/g_curve_recon_wood_ring_n0_60_n1_50_n2_40_level_2_iter_6.pdf}
	}%
\end{tabular}
	\Caption{Comparison of our algorithm with enhanced \protect\cite{Ma:2011:DET}.}
	{%
		\nothing{
		Previous methods \cite{Ma:2011:DET,Roveri:2015:EBR} could not be applied to generate patterns we target.
	}%
		We compare our methods to an enhanced version of \cite{Ma:2011:DET} that incorporates our ideas, including the sample connectivity and edge assignment.		
		\nothing{
		Our method produces higher-quality sample distribution than \cite{Ma:2011:DET}. 
		This is because our method adopts a more robust neighborhood similarity criterion as well as novel existence assignment. 
	}%
\nothing{
}%
}
	\label{fig:comparison_pattern_synthesis}
\end{figure*}

We also {\em enhance} \cite{Ma:2011:DET} for comparisons, by incorporating it with the sample connectivity (\Cref{fig:graph_repres}) \nothing{ (edge)} and edge assignment step (\Cref{eqn:assignment:edge}), but without the edge set difference (\Cref{eq:sample_edge_similarity}) and robust matching (\Cref{sec:neighborhood_matching:robust}) in the search step, as well as without the existence assignment step (the third paragraph in \Cref{subsubsec:assignment_step}).
As shown in \Cref{fig:comparison_pattern_synthesis}, our method can generate better results than the enhanced version of \cite{Ma:2011:DET}.
\iftrue
Unlike for \cite{Ma:2011:DET}, we are unable to enhance \cite{Tu:2019:PPS,Roveri:2015:EBR} due to the lack of one-to-one sample correspondences which are needed for the edge assignment step.
\nothing{
}%
\nothing{
}%
\else
For \cite{Roveri:2015:EBR}, it focuses on spatial distribution synthesis and could be problematic when dealing with attributed samples due to its mixing of spatial and attribute dimensions.
The use of soft neighborhood matching (\Cref{fig:soft_neighborhood_matching}) also prohibits it to incorporate edge assignment.
\fi

\nothing{
}%

\nothing{
In terms of sample synthesis, the similarity formulation in \cite{Roveri:2015:EBR} only allows it handle attributes that are summable and spatially coherent. 
This limits its ability of handle any sample distributions in our paper, as the samples may include discrete attribute such as sample id, or non-summable attributes, such as orientations and connectivity.
Here,  we compare our methods to \cite{Ma:2011:DET}.

Our method produces higher-quality sample distribution than \cite{Ma:2011:DET}. 
This is because our method adopts a more robust neighborhood similarity criterion as well as novel existence assignment.
}%

\nothing{

}%

\section{Conclusions, Limitations, and Future Work}
\label{sec:conclusion}

\nothing{
Our system development focuses on discrete pattern generation but continuous patterns \cite{Roveri:2015:EBR}. In the future, we are happy to extend the system for continuous patterns.
}%

\nothing{
It's the first time online learning is integrated into the example-pattern design system.
The online learning method is based on Golden-section search algorithm.
The search algorithm operates by successively narrowing the search range, which makes it slower but more robust. 
In the future, we would like to apply different kinds of search algorithm and study how to balance the speed and robustness in the context of user interactions. 
}%

Repetitive patterns have many applications, whose creation has been a main focus of research in computer graphics and interactive techniques.
This work focuses on methods and interfaces to help users author continuous curve patterns.
Analysis and results of diverse patterns have demonstrated the promise of our approach.

\begin{figure}[ht]
	\centering
		\captionsetup[subfigure]{justification=centering}
		\setlength{\tabcolsep}{-2pt}
		
\begin{tabular}{cccc}
	\subfloat[Input]{
	\label{fig:failure:dna:exemplar}
	\includegraphics[width=0.234\linewidth]{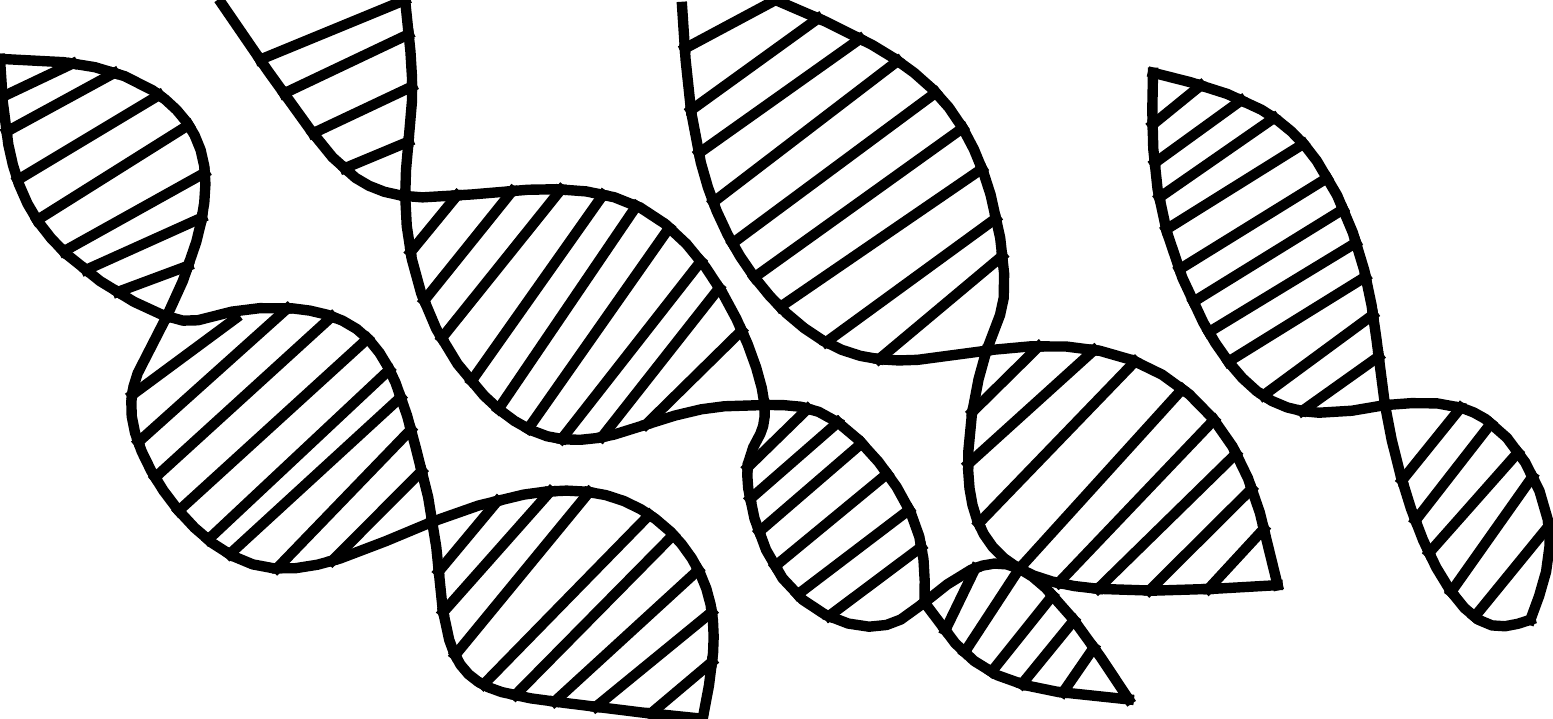}
}
	&
	\subfloat[Output]{
	\label{fig:failure:dna:output}
	\includegraphics[width=0.3\linewidth]{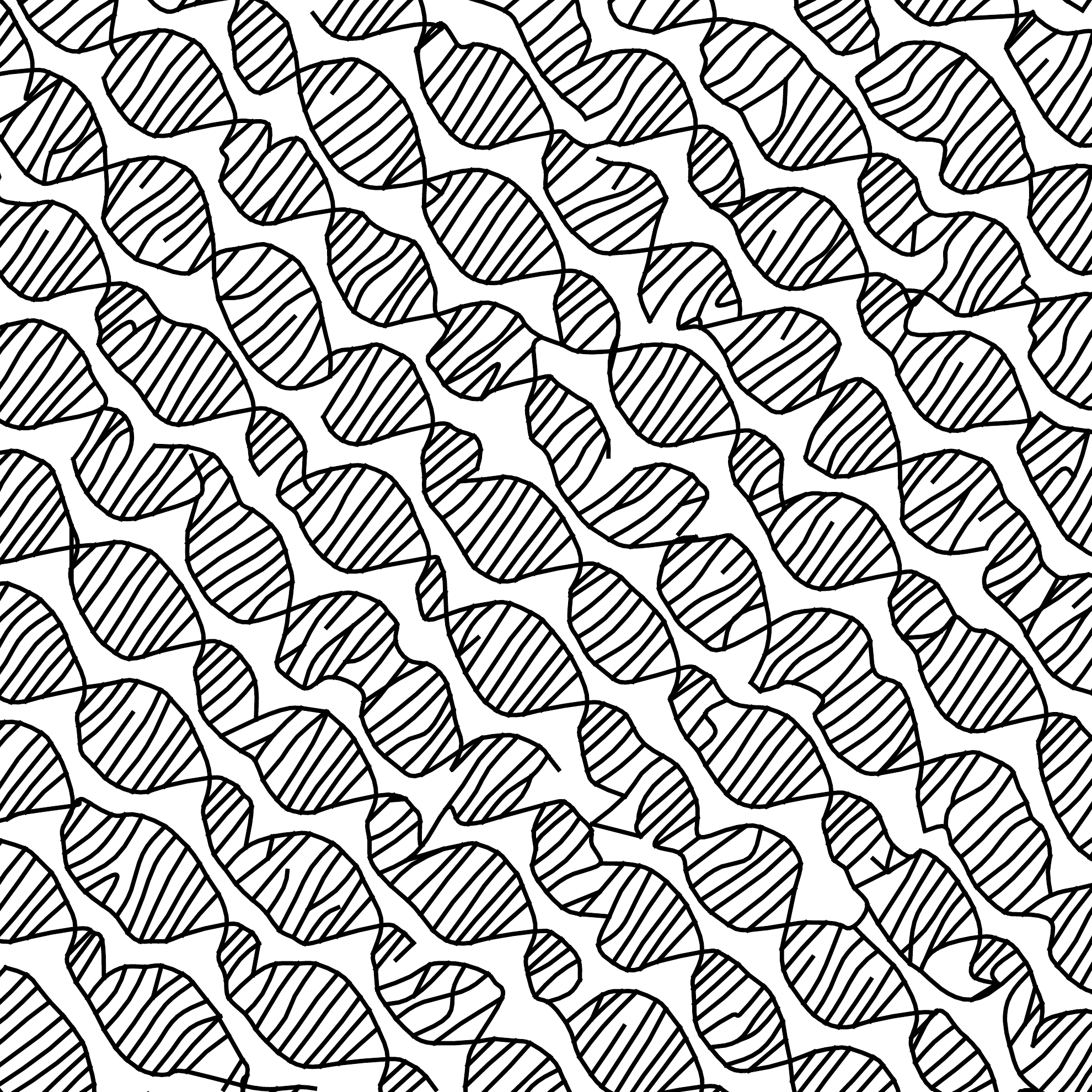}
}
	&
	\subfloat[Input]{
	\label{fig:failure:leaves:exemplar}
	\includegraphics[width=0.132\linewidth]{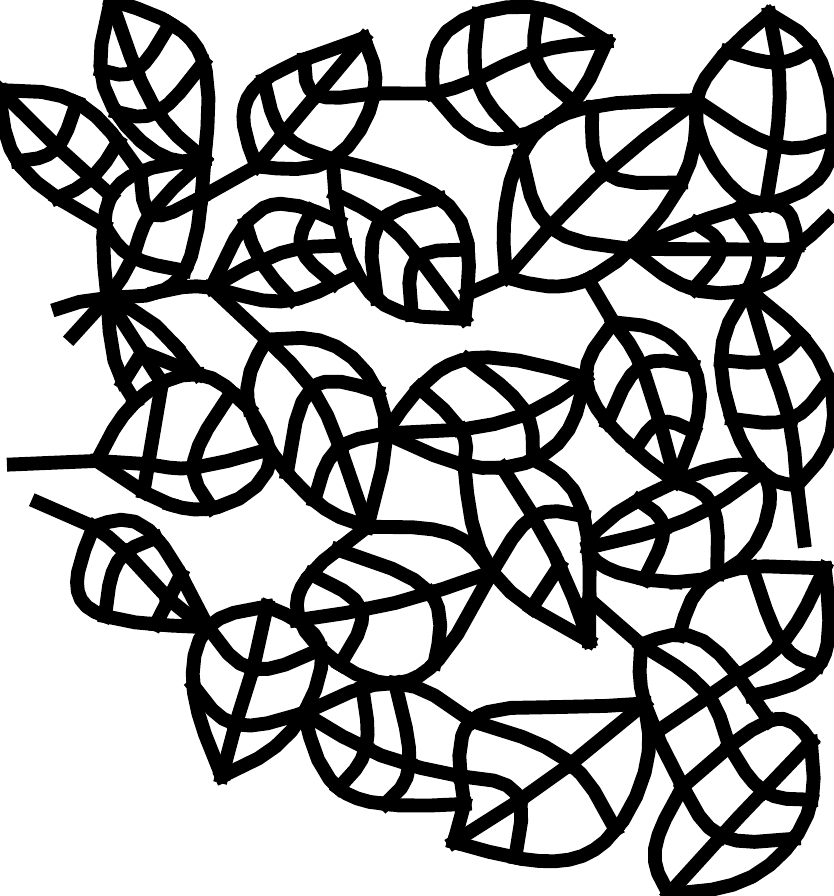}
	}%
	&\subfloat[Ouptut]{
	\label{fig:failure:leaves:output}
	\includegraphics[width=0.3\linewidth]{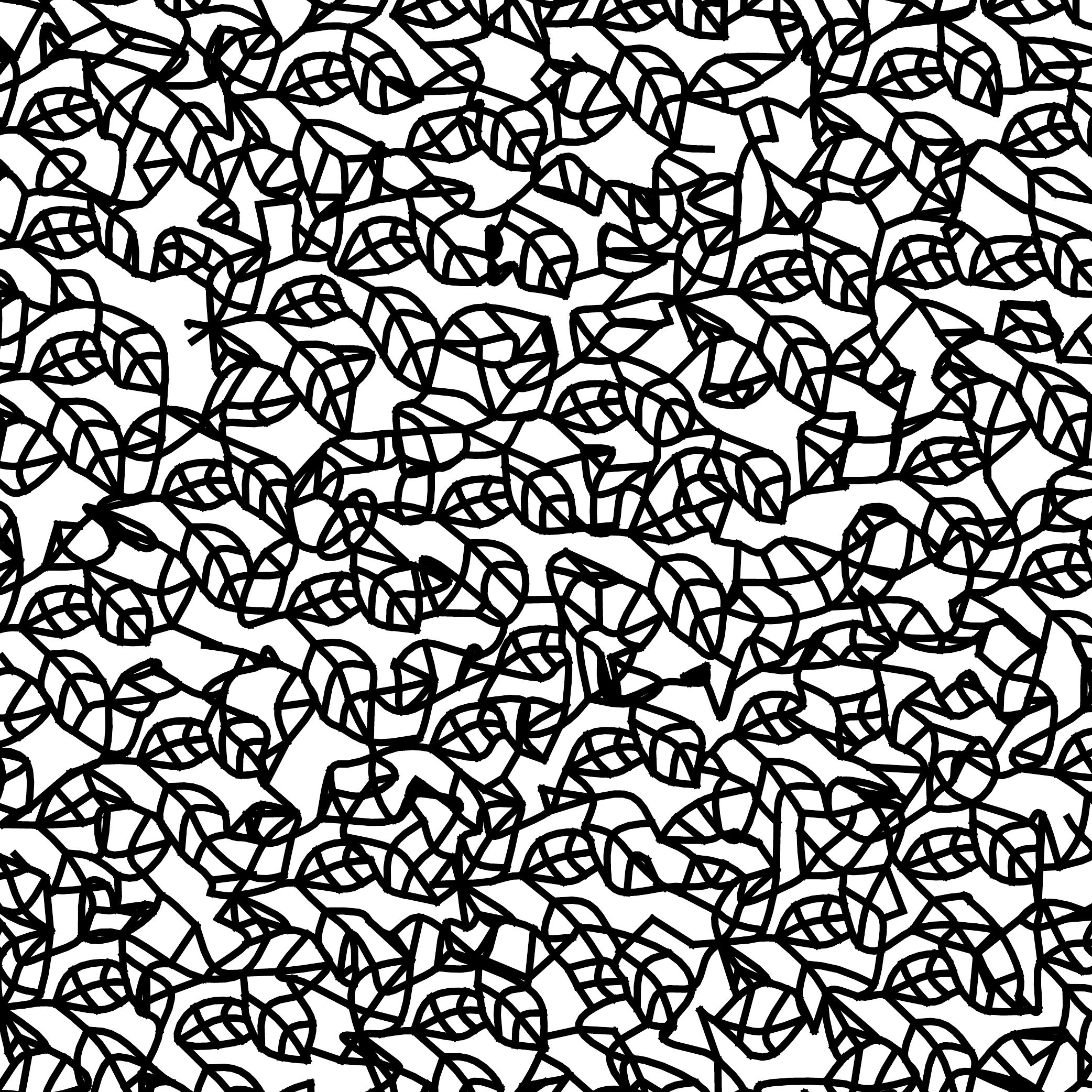}
	}%
\end{tabular}
	\Caption{Failure case.}
	{%
	  Our algorithm fails to preserve the identifiable DNA-segment and tree-leaf elements within the continuous structures.
	  \nothing{
  }%
	}
	\label{fig:failure}
\end{figure}

\add{
Like other neighborhood-based texture/pattern synthesis methods, our algorithm also assumes local \nothing{Markov random field (MRF) }properties and thus cannot capture global structures and may introduce stochastic variations, such as broken and distorted curves, as shown in \Cref{sec:result}.
These artifacts can be reduced by other improvements, such as bidirectional similarity, additional feature masks, and smart initialization \cite{Kaspar:2015:STT}.
}%

\add{
Our current reconstruction algorithm is based on B\'{e}zier curve interpolation, which might not preserve the exemplar curves.
One possibility is to treat each curve segment like a discrete element and reconstruct via sample-based warping \cite{Ma:2011:DET,Hsu:2020:AEF}, while ensuring that curve segments sharing common samples are well connected. 
}%

Our current algorithm treats discrete elements and continuous structures separately and thus might not preserve identifiable elements within continuous structures, as exemplified in \Cref{fig:failure}.
\delete{
These continuous patterns, unlike discrete elements \cite{Ma:2011:DET}, can share common vertices, boundaries, or even structures.
}%
A potential future work is to find a unified representation and approach for both discrete and continuous patterns.

The pattern synthesis requires nearest neighborhood searching for output samples, which can become computationally expensive for large outputs.
This neighborhood searching process can be readily parallelized \cite{Huang:2007:APT}. 

We focus on curves as the first step to handle continuous vector patterns.
A next step is to incorporate more vector graphics features as parts of the sample/edge attributes, such as color and thickness, as well as higher dimensional primitives including 2D regions and 3D volumes \cite{Takayama:2010:VMD,Wang:2010:VST,Wang:2011:MVV}.
\nothing{
A future direction is to extend the current algorithm, which is about synthesis of pattern consisting of curves, to generate such different vector textures.
}%
More controls can also be added to facilitate more diverse authoring effects such as local variations in scales and orientations \cite{Hsu:2020:AEF}.
In addition to optimizing pattern appearance as in this work, adding mechanical structures constraints can facilitate the application of curve structures for rapid manufacturing \cite{Zehnder:2016:DSO,Chen:2017:FTD,Chen:2016:SFD,Zhou:2014:TSV,Bian:2018:TPD,Li:2019:AQP}.

\nothing{
}%

\nothing{
Disclose all limitations, and describe how potential future works can address these and lead to more interesting and ground breaking stuff.

I am continuously updating this document based on your questions and feedbacks.

I believe in the (near?) future all research papers, at least in computer science, will be published in a form that contains everything in one place, including math, code, and data, so that readers can parse the descriptions and repeat the experiments, like what we can do with Jupyter/Ipython notebooks today.
}%

\begin{acks}

We would like to thank the anonymous reviewers for their valuable feedback.
Peihan Tu conducted parts of this research as a visiting student with the \utokyo{} and an intern with \adobe{}.
This work has been partially supported by an Adobe gift funding and JSPS KAKENHI Grant Number 17H00752.

\end{acks}

\bibliographystyle{acmart}
{
\bibliography{paper,misc}

%%% -*-BibTeX-*-
%%% Do NOT edit. File created by BibTeX with style
%%% ACM-Reference-Format-Journals [18-Jan-2012].

\begin{thebibliography}{191}

%%% ====================================================================
%%% NOTE TO THE USER: you can override these defaults by providing
%%% customized versions of any of these macros before the \bibliography
%%% command.  Each of them MUST provide its own final punctuation,
%%% except for \shownote{}, \showDOI{}, and \showURL{}.  The latter two
%%% do not use final punctuation, in order to avoid confusing it with
%%% the Web address.
%%%
%%% To suppress output of a particular field, define its macro to expand
%%% to an empty string, or better, \unskip, like this:
%%%
%%% \newcommand{\showDOI}[1]{\unskip}   % LaTeX syntax
%%%
%%% \def \showDOI #1{\unskip}           % plain TeX syntax
%%%
%%% ====================================================================

\ifx \showCODEN    \undefined \def \showCODEN     #1{\unskip}     \fi
\ifx \showDOI      \undefined \def \showDOI       #1{#1}\fi
\ifx \showISBNx    \undefined \def \showISBNx     #1{\unskip}     \fi
\ifx \showISBNxiii \undefined \def \showISBNxiii  #1{\unskip}     \fi
\ifx \showISSN     \undefined \def \showISSN      #1{\unskip}     \fi
\ifx \showLCCN     \undefined \def \showLCCN      #1{\unskip}     \fi
\ifx \shownote     \undefined \def \shownote      #1{#1}          \fi
\ifx \showarticletitle \undefined \def \showarticletitle #1{#1}   \fi
\ifx \showURL      \undefined \def \showURL       {\relax}        \fi
% The following commands are used for tagged output and should be
% invisible to TeX
\providecommand\bibfield[2]{#2}
\providecommand\bibinfo[2]{#2}
\providecommand\natexlab[1]{#1}
\providecommand\showeprint[2][]{arXiv:#2}

\bibitem[\protect\citeauthoryear{Adobe}{Adobe}{2020}]%
        {Adobe:2020:CCT:PP}
\bibfield{author}{\bibinfo{person}{Adobe}.} \bibinfo{year}{2020}\natexlab{}.
\newblock \bibinfo{title}{Continuous Curve Textures Project Page}.
\newblock
\newblock
\urldef\tempurl%
\url{https://research.adobe.com/publication/continuous-curve-textures/}
\showURL{%
\tempurl}


\bibitem[\protect\citeauthoryear{Agarwal, Foster, Hsu, Kakade, and
  Rakhlin}{Agarwal et~al\mbox{.}}{2011}]%
        {Agarwal:2011:SCO}
\bibfield{author}{\bibinfo{person}{Alekh Agarwal}, \bibinfo{person}{Dean~P
  Foster}, \bibinfo{person}{Daniel~J Hsu}, \bibinfo{person}{Sham~M Kakade},
  {and} \bibinfo{person}{Alexander Rakhlin}.} \bibinfo{year}{2011}\natexlab{}.
\newblock \showarticletitle{Stochastic convex optimization with bandit
  feedback}. In \bibinfo{booktitle}{\emph{Advances in Neural Information
  Processing Systems}}. \bibinfo{pages}{1035--1043}.
\newblock


\bibitem[\protect\citeauthoryear{Ahuja and Todorovic}{Ahuja and
  Todorovic}{2007}]%
        {Ahuja:2007:ETN}
\bibfield{author}{\bibinfo{person}{Narendra Ahuja} {and}
  \bibinfo{person}{Sinisa Todorovic}.} \bibinfo{year}{2007}\natexlab{}.
\newblock \showarticletitle{Extracting texels in 2.1 D natural textures}. In
  \bibinfo{booktitle}{\emph{Computer Vision, 2007. ICCV 2007. IEEE 11th
  International Conference on}}. IEEE, \bibinfo{pages}{1--8}.
\newblock


\bibitem[\protect\citeauthoryear{Allegorithmic}{Allegorithmic}{2018a}]%
        {Allegorithmic:2018:ASD}
\bibfield{author}{\bibinfo{person}{Allegorithmic}.}
  \bibinfo{year}{2018}\natexlab{a}.
\newblock \bibinfo{title}{Allegorithmic Substance Designer}.
\newblock
\newblock
\newblock
\shownote{\url{https://www.allegorithmic.com/products/substance-designer}.}


\bibitem[\protect\citeauthoryear{Allegorithmic}{Allegorithmic}{2018b}]%
        {Allegorithmic:2016:SDG}
\bibfield{author}{\bibinfo{person}{Allegorithmic}.}
  \bibinfo{year}{2018}\natexlab{b}.
\newblock \bibinfo{title}{Substance Designer Getting Started: 01 - Creating a
  base material part One}.
\newblock
\newblock
\newblock
\shownote{\url{https://youtu.be/NzKtubDsC9o?list=PLB0wXHrWAmCwLRTzdb-RxadGk_xBBQKar&t=1149
  }.}


\bibitem[\protect\citeauthoryear{Allegorithmic:Youtube}{Allegorithmic:Youtube}{2018}]%
        {Allegorithmic:2018:SDO}
\bibfield{author}{\bibinfo{person}{Allegorithmic:Youtube}.}
  \bibinfo{year}{2018}\natexlab{}.
\newblock \bibinfo{title}{Substance Designer: Overview}.
\newblock
\newblock
\newblock
\shownote{\url{https://www.youtube.com/watch?v=ScttSShgXlw}.}


\bibitem[\protect\citeauthoryear{AlMeraj, Kaplan, and Asente}{AlMeraj
  et~al\mbox{.}}{2013a}]%
        {AlMeraj:2013:PGT}
\bibfield{author}{\bibinfo{person}{Zainab AlMeraj}, \bibinfo{person}{Craig~S.
  Kaplan}, {and} \bibinfo{person}{Paul Asente}.}
  \bibinfo{year}{2013}\natexlab{a}.
\newblock \showarticletitle{Patch-Based Geometric Texture Synthesis}. In
  \bibinfo{booktitle}{\emph{CAE ’13}}. \bibinfo{pages}{15--19}.
\newblock
\showISBNx{9781450322034}
\urldef\tempurl%
\url{https://doi.org/10.1145/2487276.2487278}
\showDOI{\tempurl}


\bibitem[\protect\citeauthoryear{AlMeraj, Kaplan, and Asente}{AlMeraj
  et~al\mbox{.}}{2013b}]%
        {AlMeraj:2013:TEE}
\bibfield{author}{\bibinfo{person}{Zainab AlMeraj}, \bibinfo{person}{Craig~S.
  Kaplan}, {and} \bibinfo{person}{Paul Asente}.}
  \bibinfo{year}{2013}\natexlab{b}.
\newblock \showarticletitle{Towards Effective Evaluation of Geometric Texture
  Synthesis Algorithms}. In \bibinfo{booktitle}{\emph{NPAR ’13}}.
  \bibinfo{pages}{5--14}.
\newblock
\urldef\tempurl%
\url{https://doi.org/10.1145/2486042.2486043}
\showDOI{\tempurl}


\bibitem[\protect\citeauthoryear{Barla, Breslav, Thollot, Sillion, and
  Markosian}{Barla et~al\mbox{.}}{2006}]%
        {Barla:2006:SPA}
\bibfield{author}{\bibinfo{person}{Pascal Barla}, \bibinfo{person}{Simon
  Breslav}, \bibinfo{person}{Jo{\"e}lle Thollot},
  \bibinfo{person}{Fran{\c{c}}ois Sillion}, {and} \bibinfo{person}{Lee
  Markosian}.} \bibinfo{year}{2006}\natexlab{}.
\newblock \showarticletitle{Stroke pattern analysis and synthesis}. In
  \bibinfo{booktitle}{\emph{Computer Graphics Forum}},
  Vol.~\bibinfo{volume}{25}. Wiley Online Library, \bibinfo{pages}{663--671}.
\newblock


\bibitem[\protect\citeauthoryear{Barnes, Shechtman, Finkelstein, and
  Goldman}{Barnes et~al\mbox{.}}{2009}]%
        {Barnes:2009:PRC}
\bibfield{author}{\bibinfo{person}{Connelly Barnes}, \bibinfo{person}{Eli
  Shechtman}, \bibinfo{person}{Adam Finkelstein}, {and} \bibinfo{person}{Dan~B
  Goldman}.} \bibinfo{year}{2009}\natexlab{}.
\newblock \showarticletitle{PatchMatch: A Randomized Correspondence Algorithm
  for Structural Image Editing}.
\newblock \bibinfo{journal}{\emph{ACM Trans. Graph.}} \bibinfo{volume}{28},
  \bibinfo{number}{3}, Article \bibinfo{articleno}{24} (\bibinfo{date}{July}
  \bibinfo{year}{2009}), \bibinfo{numpages}{11}~pages.
\newblock
\showISSN{0730-0301}
\urldef\tempurl%
\url{https://doi.org/10.1145/1531326.1531330}
\showDOI{\tempurl}


\bibitem[\protect\citeauthoryear{Bergmann, Jetchev, and Vollgraf}{Bergmann
  et~al\mbox{.}}{2017}]%
        {Bergmann:2017:LTM}
\bibfield{author}{\bibinfo{person}{Urs Bergmann}, \bibinfo{person}{Nikolay
  Jetchev}, {and} \bibinfo{person}{Roland Vollgraf}.}
  \bibinfo{year}{2017}\natexlab{}.
\newblock \showarticletitle{Learning texture manifolds with the periodic
  spatial GAN}.
\newblock \bibinfo{journal}{\emph{arXiv preprint arXiv:1705.06566}}
  (\bibinfo{year}{2017}).
\newblock


\bibitem[\protect\citeauthoryear{Bhat, Ingram, and Turk}{Bhat
  et~al\mbox{.}}{2004}]%
        {Bhat:2004:GTS}
\bibfield{author}{\bibinfo{person}{Pravin Bhat}, \bibinfo{person}{Stephen
  Ingram}, {and} \bibinfo{person}{Greg Turk}.} \bibinfo{year}{2004}\natexlab{}.
\newblock \showarticletitle{Geometric texture synthesis by example}. In
  \bibinfo{booktitle}{\emph{SGP '04}}. \bibinfo{pages}{41--44}.
\newblock


\bibitem[\protect\citeauthoryear{Bian, Wei, and Lefebvre}{Bian
  et~al\mbox{.}}{2018}]%
        {Bian:2018:TPD}
\bibfield{author}{\bibinfo{person}{Xiaojun Bian}, \bibinfo{person}{Li-Yi Wei},
  {and} \bibinfo{person}{Sylvain Lefebvre}.} \bibinfo{year}{2018}\natexlab{}.
\newblock \showarticletitle{Tile-based Pattern Design with Topology Control}.
\newblock \bibinfo{journal}{\emph{Proc. ACM Comput. Graph. Interact. Tech.}}
  \bibinfo{volume}{1}, \bibinfo{number}{1}, Article \bibinfo{articleno}{23}
  (\bibinfo{date}{July} \bibinfo{year}{2018}), \bibinfo{numpages}{15}~pages.
\newblock
\showISSN{2577-6193}
\urldef\tempurl%
\url{https://doi.org/10.1145/3203204}
\showDOI{\tempurl}


\bibitem[\protect\citeauthoryear{Brochu, Cora, and De~Freitas}{Brochu
  et~al\mbox{.}}{2010}]%
        {Brochu:2010:TBO}
\bibfield{author}{\bibinfo{person}{Eric Brochu}, \bibinfo{person}{Vlad~M Cora},
  {and} \bibinfo{person}{Nando De~Freitas}.} \bibinfo{year}{2010}\natexlab{}.
\newblock \showarticletitle{A tutorial on Bayesian optimization of expensive
  cost functions, with application to active user modeling and hierarchical
  reinforcement learning}.
\newblock \bibinfo{journal}{\emph{arXiv preprint arXiv:1012.2599}}
  (\bibinfo{year}{2010}).
\newblock


\bibitem[\protect\citeauthoryear{Brooks and Dodgson}{Brooks and
  Dodgson}{2002}]%
        {Brooks:2002:SBT}
\bibfield{author}{\bibinfo{person}{Stephen Brooks} {and} \bibinfo{person}{Neil
  Dodgson}.} \bibinfo{year}{2002}\natexlab{}.
\newblock \showarticletitle{Self-similarity Based Texture Editing}.
\newblock \bibinfo{journal}{\emph{ACM Trans. Graph.}} \bibinfo{volume}{21},
  \bibinfo{number}{3} (\bibinfo{date}{July} \bibinfo{year}{2002}),
  \bibinfo{pages}{653--656}.
\newblock
\showISSN{0730-0301}
\urldef\tempurl%
\url{https://doi.org/10.1145/566654.566632}
\showDOI{\tempurl}


\bibitem[\protect\citeauthoryear{Chan and Zhu}{Chan and Zhu}{2005}]%
        {Chan:2005:LSB}
\bibfield{author}{\bibinfo{person}{Tony Chan} {and} \bibinfo{person}{Wei Zhu}.}
  \bibinfo{year}{2005}\natexlab{}.
\newblock \showarticletitle{Level set based shape prior segmentation}. In
  \bibinfo{booktitle}{\emph{2005 IEEE Computer Society Conference on Computer
  Vision and Pattern Recognition (CVPR'05)}}, Vol.~\bibinfo{volume}{2}. IEEE,
  \bibinfo{pages}{1164--1170}.
\newblock


\bibitem[\protect\citeauthoryear{Chen, Wei, and Chang}{Chen
  et~al\mbox{.}}{2011}]%
        {Chen:2011:NRC}
\bibfield{author}{\bibinfo{person}{Hsiang-Ting Chen}, \bibinfo{person}{Li-Yi
  Wei}, {and} \bibinfo{person}{Chun-Fa Chang}.}
  \bibinfo{year}{2011}\natexlab{}.
\newblock \showarticletitle{Nonlinear Revision Control for Images}.
\newblock \bibinfo{journal}{\emph{ACM Trans. Graph.}} \bibinfo{volume}{30},
  \bibinfo{number}{4}, Article \bibinfo{articleno}{105} (\bibinfo{date}{July}
  \bibinfo{year}{2011}), \bibinfo{numpages}{10}~pages.
\newblock
\showISSN{0730-0301}
\urldef\tempurl%
\url{https://doi.org/10.1145/2010324.1965000}
\showDOI{\tempurl}


\bibitem[\protect\citeauthoryear{Chen and Wild}{Chen and Wild}{2015}]%
        {Chen:2015:RDF}
\bibfield{author}{\bibinfo{person}{Ruobing Chen} {and} \bibinfo{person}{Stefan
  Wild}.} \bibinfo{year}{2015}\natexlab{}.
\newblock \showarticletitle{Randomized derivative-free optimization of noisy
  convex functions}.
\newblock \bibinfo{journal}{\emph{arXiv preprint arXiv:1507.03332}}
  (\bibinfo{year}{2015}).
\newblock


\bibitem[\protect\citeauthoryear{Chen, Ma, Lefebvre, Xin, Mart\'{\i}nez, and
  wang}{Chen et~al\mbox{.}}{2017}]%
        {Chen:2017:FTD}
\bibfield{author}{\bibinfo{person}{Weikai Chen}, \bibinfo{person}{Yuexin Ma},
  \bibinfo{person}{Sylvain Lefebvre}, \bibinfo{person}{Shiqing Xin},
  \bibinfo{person}{Jon\`{a}s Mart\'{\i}nez}, {and} \bibinfo{person}{wenping
  wang}.} \bibinfo{year}{2017}\natexlab{}.
\newblock \showarticletitle{Fabricable Tile Decors}.
\newblock \bibinfo{journal}{\emph{ACM Trans. Graph.}} \bibinfo{volume}{36},
  \bibinfo{number}{6}, Article \bibinfo{articleno}{175} (\bibinfo{date}{Nov.}
  \bibinfo{year}{2017}), \bibinfo{numpages}{15}~pages.
\newblock
\showISSN{0730-0301}
\urldef\tempurl%
\url{https://doi.org/10.1145/3130800.3130817}
\showDOI{\tempurl}


\bibitem[\protect\citeauthoryear{Chen, Zhang, Xin, Xia, Lefebvre, and
  Wang}{Chen et~al\mbox{.}}{2016b}]%
        {Chen:2016:SFD}
\bibfield{author}{\bibinfo{person}{Weikai Chen}, \bibinfo{person}{Xiaolong
  Zhang}, \bibinfo{person}{Shiqing Xin}, \bibinfo{person}{Yang Xia},
  \bibinfo{person}{Sylvain Lefebvre}, {and} \bibinfo{person}{Wenping Wang}.}
  \bibinfo{year}{2016}\natexlab{b}.
\newblock \showarticletitle{Synthesis of Filigrees for Digital Fabrication}.
\newblock \bibinfo{journal}{\emph{ACM Trans. Graph.}} \bibinfo{volume}{35},
  \bibinfo{number}{4}, Article \bibinfo{articleno}{98} (\bibinfo{date}{July}
  \bibinfo{year}{2016}), \bibinfo{numpages}{13}~pages.
\newblock
\showISSN{0730-0301}
\urldef\tempurl%
\url{https://doi.org/10.1145/2897824.2925911}
\showDOI{\tempurl}


\bibitem[\protect\citeauthoryear{Chen, Funkhouser, Goldman, and Shechtman}{Chen
  et~al\mbox{.}}{2012}]%
        {Chen:2012:NPT}
\bibfield{author}{\bibinfo{person}{Xiaobai Chen}, \bibinfo{person}{Tom
  Funkhouser}, \bibinfo{person}{Dan~B Goldman}, {and} \bibinfo{person}{Eli
  Shechtman}.} \bibinfo{year}{2012}\natexlab{}.
\newblock \showarticletitle{Non-parametric texture transfer using meshmatch}.
\newblock \bibinfo{journal}{\emph{Adobe Technical Report}} \bibinfo{number}{2}
  (\bibinfo{year}{2012}).
\newblock


\bibitem[\protect\citeauthoryear{Chen, Fu, and Au}{Chen et~al\mbox{.}}{2016a}]%
        {Chen:2016:MSI}
\bibfield{author}{\bibinfo{person}{Yilan Chen}, \bibinfo{person}{Hongbo Fu},
  {and} \bibinfo{person}{Kin~Chung Au}.} \bibinfo{year}{2016}\natexlab{a}.
\newblock \showarticletitle{A Multi-level Sketch-based Interface for Decorative
  Pattern Exploration}. In \bibinfo{booktitle}{\emph{SIGGRAPH ASIA 2016
  Technical Briefs}} \emph{(\bibinfo{series}{SA '16})}. Article
  \bibinfo{articleno}{26}, \bibinfo{numpages}{4}~pages.
\newblock
\showISBNx{978-1-4503-4541-5}
\urldef\tempurl%
\url{https://doi.org/10.1145/3005358.3005362}
\showDOI{\tempurl}


\bibitem[\protect\citeauthoryear{Cheng, Zhang, Mitra, Huang, and Hu}{Cheng
  et~al\mbox{.}}{2010}]%
        {Cheng:2010:RFA}
\bibfield{author}{\bibinfo{person}{Ming-Ming Cheng}, \bibinfo{person}{Fang-Lue
  Zhang}, \bibinfo{person}{Niloy~J. Mitra}, \bibinfo{person}{Xiaolei Huang},
  {and} \bibinfo{person}{Shi-Min Hu}.} \bibinfo{year}{2010}\natexlab{}.
\newblock \showarticletitle{RepFinder: Finding Approximately Repeated Scene
  Elements for Image Editing}. In \bibinfo{booktitle}{\emph{ACM SIGGRAPH 2010
  Papers}} \emph{(\bibinfo{series}{SIGGRAPH '10})}. \bibinfo{publisher}{ACM},
  \bibinfo{address}{New York, NY, USA}, Article \bibinfo{articleno}{83},
  \bibinfo{numpages}{8}~pages.
\newblock
\showISBNx{978-1-4503-0210-4}
\urldef\tempurl%
\url{https://doi.org/10.1145/1833349.1778820}
\showDOI{\tempurl}


\bibitem[\protect\citeauthoryear{Cohen, Shade, Hiller, and Deussen}{Cohen
  et~al\mbox{.}}{2003}]%
        {Cohen:2003:WTI}
\bibfield{author}{\bibinfo{person}{Michael~F. Cohen}, \bibinfo{person}{Jonathan
  Shade}, \bibinfo{person}{Stefan Hiller}, {and} \bibinfo{person}{Oliver
  Deussen}.} \bibinfo{year}{2003}\natexlab{}.
\newblock \showarticletitle{Wang Tiles for Image and Texture Generation}.
\newblock \bibinfo{journal}{\emph{ACM Trans. Graph.}} \bibinfo{volume}{22},
  \bibinfo{number}{3} (\bibinfo{date}{July} \bibinfo{year}{2003}),
  \bibinfo{pages}{287--294}.
\newblock
\showISSN{0730-0301}
\urldef\tempurl%
\url{https://doi.org/10.1145/882262.882265}
\showDOI{\tempurl}


\bibitem[\protect\citeauthoryear{committee}{committee}{2020}]%
        {Tu:2020:GRS}
\bibfield{author}{\bibinfo{person}{Replicability committee}.}
  \bibinfo{year}{2020}\natexlab{}.
\newblock \bibinfo{title}{Graphics Replicability Stamp Initiative}.
\newblock
\newblock
\newblock
\shownote{\url{http://www.replicabilitystamp.org/}.}


\bibitem[\protect\citeauthoryear{Conn, Scheinberg, and Vicente}{Conn
  et~al\mbox{.}}{2009}]%
        {Conn:2009:IDF}
\bibfield{author}{\bibinfo{person}{Andrew~R Conn}, \bibinfo{person}{Katya
  Scheinberg}, {and} \bibinfo{person}{Luis~N Vicente}.}
  \bibinfo{year}{2009}\natexlab{}.
\newblock \bibinfo{booktitle}{\emph{Introduction to derivative-free
  optimization}}. Vol.~\bibinfo{volume}{8}.
\newblock \bibinfo{publisher}{Siam}.
\newblock


\bibitem[\protect\citeauthoryear{Conte, Foggia, Sansone, and Vento}{Conte
  et~al\mbox{.}}{2004}]%
        {Conte:2004:TYG}
\bibfield{author}{\bibinfo{person}{Donatello Conte}, \bibinfo{person}{Pasquale
  Foggia}, \bibinfo{person}{Carlo Sansone}, {and} \bibinfo{person}{Mario
  Vento}.} \bibinfo{year}{2004}\natexlab{}.
\newblock \showarticletitle{Thirty years of graph matching in pattern
  recognition}.
\newblock \bibinfo{journal}{\emph{International journal of pattern recognition
  and artificial intelligence}} \bibinfo{volume}{18}, \bibinfo{number}{03}
  (\bibinfo{year}{2004}), \bibinfo{pages}{265--298}.
\newblock


\bibitem[\protect\citeauthoryear{Cornet and Rouquier}{Cornet and
  Rouquier}{2004}]%
        {Cornet:2004:GTP}
\bibfield{author}{\bibinfo{person}{Emmanuel Cornet} {and}
  \bibinfo{person}{Jean-Baptiste Rouquier}.} \bibinfo{year}{2004}\natexlab{}.
\newblock \bibinfo{title}{GIMP Texturize plugin}.
\newblock
\newblock
\newblock
\shownote{\url{https://lmanul.github.io/gimp-texturize/}.}


\bibitem[\protect\citeauthoryear{Dalstein, Ronfard, and van~de Panne}{Dalstein
  et~al\mbox{.}}{2015}]%
        {Dalstein:2015:VGA}
\bibfield{author}{\bibinfo{person}{Boris Dalstein}, \bibinfo{person}{R\'{e}mi
  Ronfard}, {and} \bibinfo{person}{Michiel van~de Panne}.}
  \bibinfo{year}{2015}\natexlab{}.
\newblock \showarticletitle{Vector Graphics Animation with Time-Varying
  Topology}.
\newblock \bibinfo{journal}{\emph{ACM Trans. Graph.}} \bibinfo{volume}{34},
  \bibinfo{number}{4}, Article \bibinfo{articleno}{145} (\bibinfo{date}{July}
  \bibinfo{year}{2015}), \bibinfo{numpages}{12}~pages.
\newblock
\showISSN{0730-0301}
\urldef\tempurl%
\url{https://doi.org/10.1145/2766913}
\showDOI{\tempurl}


\bibitem[\protect\citeauthoryear{De~Goes, Cohen-Steiner, Alliez, and
  Desbrun}{De~Goes et~al\mbox{.}}{2011}]%
        {DeGoes:2011:OTA}
\bibfield{author}{\bibinfo{person}{Fernando De~Goes}, \bibinfo{person}{David
  Cohen-Steiner}, \bibinfo{person}{Pierre Alliez}, {and}
  \bibinfo{person}{Mathieu Desbrun}.} \bibinfo{year}{2011}\natexlab{}.
\newblock \showarticletitle{An optimal transport approach to robust
  reconstruction and simplification of 2d shapes}. In
  \bibinfo{booktitle}{\emph{Computer Graphics Forum}},
  Vol.~\bibinfo{volume}{30}. \bibinfo{pages}{1593--1602}.
\newblock


\bibitem[\protect\citeauthoryear{Dekel, Gan, Krishnan, Liu, and Freeman}{Dekel
  et~al\mbox{.}}{2018}]%
        {Dekel:2018:SSC}
\bibfield{author}{\bibinfo{person}{Tali Dekel}, \bibinfo{person}{Chunag Gan},
  \bibinfo{person}{Dilip Krishnan}, \bibinfo{person}{Ce Liu}, {and}
  \bibinfo{person}{William Freeman}.} \bibinfo{year}{2018}\natexlab{}.
\newblock \showarticletitle{Sparse, Smart Contours to Represent and Edit
  Images}. In \bibinfo{booktitle}{\emph{CVPR '18}}.
\newblock


\bibitem[\protect\citeauthoryear{Dekel, Michaeli, Irani, and Freeman}{Dekel
  et~al\mbox{.}}{2015}]%
        {Dekel:2015:RMN}
\bibfield{author}{\bibinfo{person}{Tali Dekel}, \bibinfo{person}{Tomer
  Michaeli}, \bibinfo{person}{Michal Irani}, {and} \bibinfo{person}{William~T.
  Freeman}.} \bibinfo{year}{2015}\natexlab{}.
\newblock \showarticletitle{Revealing and Modifying Non-local Variations in a
  Single Image}.
\newblock \bibinfo{journal}{\emph{ACM Trans. Graph.}} \bibinfo{volume}{34},
  \bibinfo{number}{6}, Article \bibinfo{articleno}{227} (\bibinfo{date}{Oct.}
  \bibinfo{year}{2015}), \bibinfo{numpages}{11}~pages.
\newblock
\showISSN{0730-0301}
\urldef\tempurl%
\url{https://doi.org/10.1145/2816795.2818113}
\showDOI{\tempurl}


\bibitem[\protect\citeauthoryear{Dey, Dey, and Kumar}{Dey
  et~al\mbox{.}}{1999}]%
        {Dey:1999:SPA}
\bibfield{author}{\bibinfo{person}{Tamal~K. Dey}, \bibinfo{person}{Tamal~K.
  Dey}, {and} \bibinfo{person}{Piyush Kumar}.} \bibinfo{year}{1999}\natexlab{}.
\newblock \showarticletitle{A Simple Provable Algorithm for Curve
  Reconstruction}. In \bibinfo{booktitle}{\emph{Proceedings of the Tenth Annual
  ACM-SIAM Symposium on Discrete Algorithms}} \emph{(\bibinfo{series}{SODA
  '99})}. \bibinfo{publisher}{Society for Industrial and Applied Mathematics},
  \bibinfo{address}{Philadelphia, PA, USA}, \bibinfo{pages}{893--894}.
\newblock
\showISBNx{0-89871-434-6}
\urldef\tempurl%
\url{http://dl.acm.org/citation.cfm?id=314500.315073}
\showURL{%
\tempurl}


\bibitem[\protect\citeauthoryear{Drive}{Drive}{2019a}]%
        {Tu:2019:IL}
\bibfield{author}{\bibinfo{person}{Google Drive}.}
  \bibinfo{year}{2019}\natexlab{a}.
\newblock \bibinfo{title}{Interactive Learning}.
\newblock
\newblock
\newblock
\shownote{\url{https://docs.google.com/presentation/d/1Us3pZ7Iz7wb3-7sXLrgWLUUD6xI3a8MGIttjfjh6WJU/}.}


\bibitem[\protect\citeauthoryear{Drive}{Drive}{2019b}]%
        {Tu:2019:HSR}
\bibfield{author}{\bibinfo{person}{Google Drive}.}
  \bibinfo{year}{2019}\natexlab{b}.
\newblock \bibinfo{title}{Results: Hierarchical Synthesis}.
\newblock
\newblock
\newblock
\shownote{\url{https://docs.google.com/presentation/d/1-e6kGPpDPMrEVHmZLYToFMvIK3WC8ul0GogP4PUXdO8/}.}


\bibitem[\protect\citeauthoryear{Dumas, Mart{\'\i}nez, Lefebvre, and Wei}{Dumas
  et~al\mbox{.}}{2018}]%
        {Dumas:2018:PAE}
\bibfield{author}{\bibinfo{person}{J{\'e}r{\'e}mie Dumas},
  \bibinfo{person}{Jon{\`a}s Mart{\'\i}nez}, \bibinfo{person}{Sylvain
  Lefebvre}, {and} \bibinfo{person}{Li-Yi Wei}.}
  \bibinfo{year}{2018}\natexlab{}.
\newblock \showarticletitle{Printable Aggregate Elements}.
\newblock \bibinfo{journal}{\emph{arXiv preprint arXiv:1811.02626}}
  (\bibinfo{year}{2018}).
\newblock


\bibitem[\protect\citeauthoryear{Efros and Freeman}{Efros and Freeman}{2001}]%
        {Efros:2001:IQT}
\bibfield{author}{\bibinfo{person}{Alexei~A. Efros} {and}
  \bibinfo{person}{William~T. Freeman}.} \bibinfo{year}{2001}\natexlab{}.
\newblock \showarticletitle{Image Quilting for Texture Synthesis and Transfer}.
  In \bibinfo{booktitle}{\emph{SIGGRAPH '01}}. \bibinfo{pages}{341--346}.
\newblock
\showISBNx{1-58113-374-X}
\urldef\tempurl%
\url{https://doi.org/10.1145/383259.383296}
\showDOI{\tempurl}


\bibitem[\protect\citeauthoryear{Ellis, Ritchie, Solar-Lezama, and
  Tenenbaum}{Ellis et~al\mbox{.}}{2018}]%
        {Ellis:2018:LIG}
\bibfield{author}{\bibinfo{person}{Kevin Ellis}, \bibinfo{person}{Daniel
  Ritchie}, \bibinfo{person}{Armando Solar-Lezama}, {and} \bibinfo{person}{Josh
  Tenenbaum}.} \bibinfo{year}{2018}\natexlab{}.
\newblock \showarticletitle{Learning to infer graphics programs from hand-drawn
  images}. In \bibinfo{booktitle}{\emph{NIPS}}. \bibinfo{pages}{6062--6071}.
\newblock


\bibitem[\protect\citeauthoryear{Emilien, Vimont, Cani, Poulin, and
  Benes}{Emilien et~al\mbox{.}}{2015}]%
        {Emilien:2015:WIE}
\bibfield{author}{\bibinfo{person}{Arnaud Emilien}, \bibinfo{person}{Ulysse
  Vimont}, \bibinfo{person}{Marie-Paule Cani}, \bibinfo{person}{Pierre Poulin},
  {and} \bibinfo{person}{Bedrich Benes}.} \bibinfo{year}{2015}\natexlab{}.
\newblock \showarticletitle{WorldBrush: Interactive Example-based Synthesis of
  Procedural Virtual Worlds}.
\newblock \bibinfo{journal}{\emph{ACM Trans. Graph.}} \bibinfo{volume}{34},
  \bibinfo{number}{4}, Article \bibinfo{articleno}{106} (\bibinfo{date}{July}
  \bibinfo{year}{2015}), \bibinfo{numpages}{11}~pages.
\newblock
\showISSN{0730-0301}
\urldef\tempurl%
\url{https://doi.org/10.1145/2766975}
\showDOI{\tempurl}


\bibitem[\protect\citeauthoryear{Eric, Freitas, and Ghosh}{Eric
  et~al\mbox{.}}{2008}]%
        {Brochu:2008:APL}
\bibfield{author}{\bibinfo{person}{Brochu Eric}, \bibinfo{person}{Nando~D
  Freitas}, {and} \bibinfo{person}{Abhijeet Ghosh}.}
  \bibinfo{year}{2008}\natexlab{}.
\newblock \showarticletitle{Active preference learning with discrete choice
  data}. In \bibinfo{booktitle}{\emph{Advances in neural information processing
  systems}}. \bibinfo{pages}{409--416}.
\newblock


\bibitem[\protect\citeauthoryear{Fails and Olsen}{Fails and Olsen}{2003}]%
        {Fails:2003:IML}
\bibfield{author}{\bibinfo{person}{Jerry~Alan Fails} {and}
  \bibinfo{person}{Dan~R. Olsen, Jr.}} \bibinfo{year}{2003}\natexlab{}.
\newblock \showarticletitle{Interactive Machine Learning}. In
  \bibinfo{booktitle}{\emph{IUI '03}}. \bibinfo{pages}{39--45}.
\newblock
\showISBNx{1-58113-586-6}
\urldef\tempurl%
\url{https://doi.org/10.1145/604045.604056}
\showDOI{\tempurl}


\bibitem[\protect\citeauthoryear{Galerne, Lagae, Lefebvre, and
  Drettakis}{Galerne et~al\mbox{.}}{2012}]%
        {Galerne:2012:GNE}
\bibfield{author}{\bibinfo{person}{Bruno Galerne}, \bibinfo{person}{Ares
  Lagae}, \bibinfo{person}{Sylvain Lefebvre}, {and} \bibinfo{person}{George
  Drettakis}.} \bibinfo{year}{2012}\natexlab{}.
\newblock \showarticletitle{Gabor Noise by Example}.
\newblock \bibinfo{journal}{\emph{ACM Trans. Graph.}} \bibinfo{volume}{31},
  \bibinfo{number}{4}, Article \bibinfo{articleno}{73} (\bibinfo{date}{July}
  \bibinfo{year}{2012}), \bibinfo{numpages}{9}~pages.
\newblock
\showISSN{0730-0301}
\urldef\tempurl%
\url{https://doi.org/10.1145/2185520.2185569}
\showDOI{\tempurl}


\bibitem[\protect\citeauthoryear{Gatys, Ecker, and Bethge}{Gatys
  et~al\mbox{.}}{2016}]%
        {Gatys:2016:IST}
\bibfield{author}{\bibinfo{person}{Leon~A. Gatys},
  \bibinfo{person}{Alexander~S. Ecker}, {and} \bibinfo{person}{Matthias
  Bethge}.} \bibinfo{year}{2016}\natexlab{}.
\newblock \showarticletitle{Image Style Transfer Using Convolutional Neural
  Networks}. In \bibinfo{booktitle}{\emph{CVPR '16}}.
  \bibinfo{pages}{2414--2423}.
\newblock


\bibitem[\protect\citeauthoryear{Ghazanfarpour and Dischler}{Ghazanfarpour and
  Dischler}{1995}]%
        {Ghazanfarpour:1995:SAA}
\bibfield{author}{\bibinfo{person}{Djamchid Ghazanfarpour} {and}
  \bibinfo{person}{Jean-Michel Dischler}.} \bibinfo{year}{1995}\natexlab{}.
\newblock \showarticletitle{Spectral analysis for automatic 3-d texture
  generation}.
\newblock \bibinfo{journal}{\emph{Computers \& Graphics}} \bibinfo{volume}{19},
  \bibinfo{number}{3} (\bibinfo{year}{1995}), \bibinfo{pages}{413--422}.
\newblock


\bibitem[\protect\citeauthoryear{Gilet, Sauvage, Vanhoey, Dischler, and
  Ghazanfarpour}{Gilet et~al\mbox{.}}{2014}]%
        {Gilet:2014:LRN}
\bibfield{author}{\bibinfo{person}{Guillaume Gilet}, \bibinfo{person}{Basile
  Sauvage}, \bibinfo{person}{Kenneth Vanhoey}, \bibinfo{person}{Jean-Michel
  Dischler}, {and} \bibinfo{person}{Djamchid Ghazanfarpour}.}
  \bibinfo{year}{2014}\natexlab{}.
\newblock \showarticletitle{Local Random-phase Noise for Procedural Texturing}.
\newblock \bibinfo{journal}{\emph{ACM Trans. Graph.}} \bibinfo{volume}{33},
  \bibinfo{number}{6}, Article \bibinfo{articleno}{195} (\bibinfo{date}{Nov.}
  \bibinfo{year}{2014}), \bibinfo{numpages}{11}~pages.
\newblock
\showISSN{0730-0301}
\urldef\tempurl%
\url{https://doi.org/10.1145/2661229.2661249}
\showDOI{\tempurl}


\bibitem[\protect\citeauthoryear{Guerin, Digne, Galin, Peytavie, Wolf, Benes,
  and Martinez}{Guerin et~al\mbox{.}}{2017}]%
        {Guerin:2017:IET}
\bibfield{author}{\bibinfo{person}{Eric Guerin}, \bibinfo{person}{Julie Digne},
  \bibinfo{person}{Eric Galin}, \bibinfo{person}{Adrien Peytavie},
  \bibinfo{person}{Christian Wolf}, \bibinfo{person}{Bedrich Benes}, {and}
  \bibinfo{person}{Benoit Martinez}.} \bibinfo{year}{2017}\natexlab{}.
\newblock \showarticletitle{Interactive Example-Based Terrain Authoring with
  Conditional Generative Adversarial Networks}.
\newblock \bibinfo{journal}{\emph{ACM Transactions on Graphics (proceedings of
  Siggraph Asia 2017)}} \bibinfo{volume}{36}, \bibinfo{number}{6}
  (\bibinfo{year}{2017}).
\newblock


\bibitem[\protect\citeauthoryear{Guerrero, Bernstein, Li, and Mitra}{Guerrero
  et~al\mbox{.}}{2016a}]%
        {Guerrero:2016:PEP}
\bibfield{author}{\bibinfo{person}{Paul Guerrero}, \bibinfo{person}{Gilbert
  Bernstein}, \bibinfo{person}{Wilmot Li}, {and} \bibinfo{person}{Niloy~J.
  Mitra}.} \bibinfo{year}{2016}\natexlab{a}.
\newblock \showarticletitle{PATEX: Exploring Pattern Variations}.
\newblock \bibinfo{journal}{\emph{ACM Trans. Graph.}} \bibinfo{volume}{35},
  \bibinfo{number}{4}, Article \bibinfo{articleno}{48} (\bibinfo{date}{July}
  \bibinfo{year}{2016}), \bibinfo{numpages}{13}~pages.
\newblock
\showISSN{0730-0301}
\urldef\tempurl%
\url{https://doi.org/10.1145/2897824.2925950}
\showDOI{\tempurl}


\bibitem[\protect\citeauthoryear{Guerrero, Mitra, and Wonka}{Guerrero
  et~al\mbox{.}}{2016b}]%
        {Guerrero:2016:RRI}
\bibfield{author}{\bibinfo{person}{Paul Guerrero}, \bibinfo{person}{Niloy~J.
  Mitra}, {and} \bibinfo{person}{Peter Wonka}.}
  \bibinfo{year}{2016}\natexlab{b}.
\newblock \showarticletitle{RAID: A Relation-augmented Image Descriptor}.
\newblock \bibinfo{journal}{\emph{ACM Trans. Graph.}} \bibinfo{volume}{35},
  \bibinfo{number}{4}, Article \bibinfo{articleno}{46} (\bibinfo{date}{July}
  \bibinfo{year}{2016}), \bibinfo{numpages}{12}~pages.
\newblock
\showISSN{0730-0301}
\urldef\tempurl%
\url{https://doi.org/10.1145/2897824.2925939}
\showDOI{\tempurl}


\bibitem[\protect\citeauthoryear{Ha and Eck}{Ha and Eck}{2017}]%
        {Ha:2017:NRS}
\bibfield{author}{\bibinfo{person}{David Ha} {and} \bibinfo{person}{Douglas
  Eck}.} \bibinfo{year}{2017}\natexlab{}.
\newblock \showarticletitle{A Neural Representation of Sketch Drawings}.
\newblock \bibinfo{journal}{\emph{ArXiv e-prints}} (\bibinfo{date}{April}
  \bibinfo{year}{2017}).
\newblock
\showeprint[arxiv]{1704.03477}


\bibitem[\protect\citeauthoryear{Habib}{Habib}{2020}]%
        {Habib:2020:SSP}
\bibfield{author}{\bibinfo{person}{Rubaiat Habib}.}
  \bibinfo{year}{2020}\natexlab{}.
\newblock \bibinfo{title}{Seminar slides: pen-and-ink illustration}.
\newblock
\newblock
\newblock
\shownote{\url{https://www.dropbox.com/s/xng5ltm28oqz2nq/seminar\%20slides.pptx?dl=0}.}


\bibitem[\protect\citeauthoryear{Han, Zhou, Wei, Gong, Bao, Zhang, and Guo}{Han
  et~al\mbox{.}}{2006}]%
        {Han:2006:FES}
\bibfield{author}{\bibinfo{person}{Jianwei Han}, \bibinfo{person}{Kun Zhou},
  \bibinfo{person}{Li-Yi Wei}, \bibinfo{person}{Minmin Gong},
  \bibinfo{person}{Hujun Bao}, \bibinfo{person}{Xinming Zhang}, {and}
  \bibinfo{person}{Baining Guo}.} \bibinfo{year}{2006}\natexlab{}.
\newblock \showarticletitle{Fast example-based surface texture synthesis via
  discrete optimization}.
\newblock \bibinfo{journal}{\emph{The Visual Computer}} \bibinfo{volume}{22},
  \bibinfo{number}{9-11} (\bibinfo{year}{2006}), \bibinfo{pages}{918--925}.
\newblock


\bibitem[\protect\citeauthoryear{Hays, Leordeanu, Efros, and Liu}{Hays
  et~al\mbox{.}}{2006}]%
        {Hays:2006:DTR}
\bibfield{author}{\bibinfo{person}{James Hays}, \bibinfo{person}{Marius
  Leordeanu}, \bibinfo{person}{Alexei~A. Efros}, {and} \bibinfo{person}{Yanxi
  Liu}.} \bibinfo{year}{2006}\natexlab{}.
\newblock \showarticletitle{Discovering Texture Regularity As a Higher-order
  Correspondence Problem}. In \bibinfo{booktitle}{\emph{ECCV'06}}.
  \bibinfo{pages}{522--535}.
\newblock
\showISBNx{3-540-33834-9, 978-3-540-33834-5}
\urldef\tempurl%
\url{https://doi.org/10.1007/11744047_40}
\showDOI{\tempurl}


\bibitem[\protect\citeauthoryear{Hertzmann, Oliver, Curless, and
  Seitz}{Hertzmann et~al\mbox{.}}{2002}]%
        {Hertzmann:2002:CA}
\bibfield{author}{\bibinfo{person}{Aaron Hertzmann}, \bibinfo{person}{Nuria
  Oliver}, \bibinfo{person}{Brian Curless}, {and} \bibinfo{person}{Steven~M.
  Seitz}.} \bibinfo{year}{2002}\natexlab{}.
\newblock \showarticletitle{Curve Analogies}. In \bibinfo{booktitle}{\emph{EGRW
  '02}}. \bibinfo{pages}{233--246}.
\newblock
\showISBNx{1581135343}


\bibitem[\protect\citeauthoryear{hikeart}{hikeart}{2018}]%
        {hikeart:2018:HRS}
\bibfield{author}{\bibinfo{person}{hikeart}.} \bibinfo{year}{2018}\natexlab{}.
\newblock \bibinfo{title}{How to Repeat a Shape Along Any Path in Adobe
  Illustrator}.
\newblock
\newblock
\newblock
\shownote{\url{https://youtu.be/d-0aSsuFGCY}.}


\bibitem[\protect\citeauthoryear{Hopcroft and Wong}{Hopcroft and Wong}{1974}]%
        {Hopcroft:1974:LTA}
\bibfield{author}{\bibinfo{person}{J.~E. Hopcroft} {and} \bibinfo{person}{J.~K.
  Wong}.} \bibinfo{year}{1974}\natexlab{}.
\newblock \showarticletitle{Linear Time Algorithm for Isomorphism of Planar
  Graphs (Preliminary Report)}. In \bibinfo{booktitle}{\emph{Proceedings of the
  Sixth Annual ACM Symposium on Theory of Computing}}
  \emph{(\bibinfo{series}{STOC '74})}. \bibinfo{publisher}{ACM},
  \bibinfo{address}{New York, NY, USA}, \bibinfo{pages}{172--184}.
\newblock
\urldef\tempurl%
\url{https://doi.org/10.1145/800119.803896}
\showDOI{\tempurl}


\bibitem[\protect\citeauthoryear{Hsu, Wei, You, and Zhang}{Hsu
  et~al\mbox{.}}{2018}]%
        {Hsu:2018:BEF}
\bibfield{author}{\bibinfo{person}{Chen-Yuan Hsu}, \bibinfo{person}{Li-Yi Wei},
  \bibinfo{person}{Lihua You}, {and} \bibinfo{person}{Jian~Jun Zhang}.}
  \bibinfo{year}{2018}\natexlab{}.
\newblock \showarticletitle{Brushing Element Fields}. In
  \bibinfo{booktitle}{\emph{SIGGRAPH Asia 2018 Technical Briefs}}
  \emph{(\bibinfo{series}{SA '18})}. Article \bibinfo{articleno}{6},
  \bibinfo{numpages}{4}~pages.
\newblock
\showISBNx{978-1-4503-6062-3}
\urldef\tempurl%
\url{https://doi.org/10.1145/3283254.3283274}
\showDOI{\tempurl}


\bibitem[\protect\citeauthoryear{Hsu, Wei, You, and Zhang}{Hsu
  et~al\mbox{.}}{2020}]%
        {Hsu:2020:AEF}
\bibfield{author}{\bibinfo{person}{Chen-Yuan Hsu}, \bibinfo{person}{Li-Yi Wei},
  \bibinfo{person}{Lihua You}, {and} \bibinfo{person}{Jian~Jun Zhang}.}
  \bibinfo{year}{2020}\natexlab{}.
\newblock \showarticletitle{Autocomplete Element Fields}. In
  \bibinfo{booktitle}{\emph{CHI '20}}. \bibinfo{pages}{1--13}.
\newblock
\showISBNx{9781450367080}
\urldef\tempurl%
\url{https://doi.org/10.1145/3313831.3376248}
\showDOI{\tempurl}


\bibitem[\protect\citeauthoryear{Huang, Tong, and Wang}{Huang
  et~al\mbox{.}}{2007}]%
        {Huang:2007:APT}
\bibfield{author}{\bibinfo{person}{Hao-Da Huang}, \bibinfo{person}{Xin Tong},
  {and} \bibinfo{person}{Wen-Cheng Wang}.} \bibinfo{year}{2007}\natexlab{}.
\newblock \showarticletitle{Accelerated parallel texture optimization}.
\newblock \bibinfo{journal}{\emph{Journal of Computer Science and Technology}}
  \bibinfo{volume}{22}, \bibinfo{number}{5} (\bibinfo{year}{2007}),
  \bibinfo{pages}{761--769}.
\newblock


\bibitem[\protect\citeauthoryear{Hurtut, Landes, Thollot, Gousseau, Drouillhet,
  and Coeurjolly}{Hurtut et~al\mbox{.}}{2009}]%
        {Hurtut:2009:ASE}
\bibfield{author}{\bibinfo{person}{T. Hurtut}, \bibinfo{person}{P.-E. Landes},
  \bibinfo{person}{J. Thollot}, \bibinfo{person}{Y. Gousseau},
  \bibinfo{person}{R. Drouillhet}, {and} \bibinfo{person}{J.-F. Coeurjolly}.}
  \bibinfo{year}{2009}\natexlab{}.
\newblock \showarticletitle{Appearance-guided Synthesis of Element Arrangements
  by Example}. In \bibinfo{booktitle}{\emph{NPAR '09}}.
  \bibinfo{pages}{51--60}.
\newblock
\showISBNx{978-1-60558-604-5}
\urldef\tempurl%
\url{https://doi.org/10.1145/1572614.1572623}
\showDOI{\tempurl}


\bibitem[\protect\citeauthoryear{Ijiri, Mech, Igarashi, and Miller}{Ijiri
  et~al\mbox{.}}{2008}]%
        {Ijiri:2008:EBP}
\bibfield{author}{\bibinfo{person}{Takashi Ijiri},
  \bibinfo{person}{Radom{\'\i}r Mech}, \bibinfo{person}{Takeo Igarashi}, {and}
  \bibinfo{person}{Gavin Miller}.} \bibinfo{year}{2008}\natexlab{}.
\newblock \showarticletitle{An Example-based Procedural System for Element
  Arrangement}. In \bibinfo{booktitle}{\emph{Computer Graphics Forum}},
  Vol.~\bibinfo{volume}{27}. Wiley Online Library, \bibinfo{pages}{429--436}.
\newblock


\bibitem[\protect\citeauthoryear{Jacobs, Brandt, Mech, and Resnick}{Jacobs
  et~al\mbox{.}}{2018}]%
        {Jacobs:2018:EMD}
\bibfield{author}{\bibinfo{person}{Jennifer Jacobs}, \bibinfo{person}{Joel
  Brandt}, \bibinfo{person}{Radom\'{\i}r Mech}, {and} \bibinfo{person}{Mitchel
  Resnick}.} \bibinfo{year}{2018}\natexlab{}.
\newblock \showarticletitle{Extending Manual Drawing Practices with
  Artist-Centric Programming Tools}. In \bibinfo{booktitle}{\emph{Proceedings
  of the 2018 CHI Conference on Human Factors in Computing Systems}}
  \emph{(\bibinfo{series}{CHI '18})}. \bibinfo{publisher}{ACM},
  \bibinfo{address}{New York, NY, USA}, Article \bibinfo{articleno}{590},
  \bibinfo{numpages}{13}~pages.
\newblock
\showISBNx{978-1-4503-5620-6}
\urldef\tempurl%
\url{https://doi.org/10.1145/3173574.3174164}
\showDOI{\tempurl}


\bibitem[\protect\citeauthoryear{Jacobs, Gogia, M\u{e}ch, and Brandt}{Jacobs
  et~al\mbox{.}}{2017}]%
        {Jacobs:2017:SEP}
\bibfield{author}{\bibinfo{person}{Jennifer Jacobs}, \bibinfo{person}{Sumit
  Gogia}, \bibinfo{person}{Radom\'{\i}r M\u{e}ch}, {and}
  \bibinfo{person}{Joel~R. Brandt}.} \bibinfo{year}{2017}\natexlab{}.
\newblock \showarticletitle{Supporting Expressive Procedural Art Creation
  Through Direct Manipulation}. In \bibinfo{booktitle}{\emph{Proceedings of the
  2017 CHI Conference on Human Factors in Computing Systems}}
  \emph{(\bibinfo{series}{CHI '17})}. \bibinfo{publisher}{ACM},
  \bibinfo{address}{New York, NY, USA}, \bibinfo{pages}{6330--6341}.
\newblock
\showISBNx{978-1-4503-4655-9}
\urldef\tempurl%
\url{https://doi.org/10.1145/3025453.3025927}
\showDOI{\tempurl}


\bibitem[\protect\citeauthoryear{Jamieson, Nowak, and Recht}{Jamieson
  et~al\mbox{.}}{2012}]%
        {Jamieson:2012:QCD}
\bibfield{author}{\bibinfo{person}{Kevin~G Jamieson}, \bibinfo{person}{Robert
  Nowak}, {and} \bibinfo{person}{Ben Recht}.} \bibinfo{year}{2012}\natexlab{}.
\newblock \showarticletitle{Query complexity of derivative-free optimization}.
  In \bibinfo{booktitle}{\emph{Advances in Neural Information Processing
  Systems}}. \bibinfo{pages}{2672--2680}.
\newblock


\bibitem[\protect\citeauthoryear{Jian and Vemuri}{Jian and Vemuri}{2010}]%
        {Jian:2010:RPS}
\bibfield{author}{\bibinfo{person}{Bing Jian} {and} \bibinfo{person}{Baba~C
  Vemuri}.} \bibinfo{year}{2010}\natexlab{}.
\newblock \showarticletitle{Robust point set registration using gaussian
  mixture models}.
\newblock \bibinfo{journal}{\emph{IEEE transactions on pattern analysis and
  machine intelligence}} \bibinfo{volume}{33}, \bibinfo{number}{8}
  (\bibinfo{year}{2010}), \bibinfo{pages}{1633--1645}.
\newblock


\bibitem[\protect\citeauthoryear{Kaspar, Neubert, Lischinski, Pauly, and
  Kopf}{Kaspar et~al\mbox{.}}{2015}]%
        {Kaspar:2015:STT}
\bibfield{author}{\bibinfo{person}{Alexandre Kaspar}, \bibinfo{person}{Boris
  Neubert}, \bibinfo{person}{Dani Lischinski}, \bibinfo{person}{Mark Pauly},
  {and} \bibinfo{person}{Johannes Kopf}.} \bibinfo{year}{2015}\natexlab{}.
\newblock \showarticletitle{Self Tuning Texture Optimization}.
\newblock \bibinfo{journal}{\emph{Comput. Graph. Forum}} \bibinfo{volume}{34},
  \bibinfo{number}{2} (\bibinfo{date}{May} \bibinfo{year}{2015}),
  \bibinfo{pages}{349--359}.
\newblock
\showISSN{0167-7055}
\urldef\tempurl%
\url{https://doi.org/10.1111/cgf.12565}
\showDOI{\tempurl}


\bibitem[\protect\citeauthoryear{Kazi, Chevalier, Grossman, Zhao, and
  Fitzmaurice}{Kazi et~al\mbox{.}}{2014}]%
        {Kazi:2014:DBL}
\bibfield{author}{\bibinfo{person}{Rubaiat~Habib Kazi}, \bibinfo{person}{Fanny
  Chevalier}, \bibinfo{person}{Tovi Grossman}, \bibinfo{person}{Shengdong
  Zhao}, {and} \bibinfo{person}{George Fitzmaurice}.}
  \bibinfo{year}{2014}\natexlab{}.
\newblock \showarticletitle{Draco: bringing life to illustrations with kinetic
  textures}. In \bibinfo{booktitle}{\emph{Proceedings of the SIGCHI Conference
  on Human Factors in Computing Systems}}. \bibinfo{pages}{351--360}.
\newblock


\bibitem[\protect\citeauthoryear{Kazi, Igarashi, Zhao, and Davis}{Kazi
  et~al\mbox{.}}{2012}]%
        {Kazi:2012:VIT}
\bibfield{author}{\bibinfo{person}{Rubaiat~Habib Kazi}, \bibinfo{person}{Takeo
  Igarashi}, \bibinfo{person}{Shengdong Zhao}, {and} \bibinfo{person}{Richard
  Davis}.} \bibinfo{year}{2012}\natexlab{}.
\newblock \showarticletitle{Vignette: Interactive Texture Design and
  Manipulation with Freeform Gestures for Pen-and-ink Illustration}. In
  \bibinfo{booktitle}{\emph{CHI '12}}. \bibinfo{pages}{1727--1736}.
\newblock
\showISBNx{978-1-4503-1015-4}
\urldef\tempurl%
\url{https://doi.org/10.1145/2207676.2208302}
\showDOI{\tempurl}


\bibitem[\protect\citeauthoryear{Kiefer}{Kiefer}{1953}]%
        {Kiefer:1953:SMS}
\bibfield{author}{\bibinfo{person}{Jack Kiefer}.}
  \bibinfo{year}{1953}\natexlab{}.
\newblock \showarticletitle{Sequential minimax search for a maximum}.
\newblock \bibinfo{journal}{\emph{Proceedings of the American mathematical
  society}} \bibinfo{volume}{4}, \bibinfo{number}{3} (\bibinfo{year}{1953}),
  \bibinfo{pages}{502--506}.
\newblock


\bibitem[\protect\citeauthoryear{Kira}{Kira}{2008}]%
        {Kira:2008:EBP}
\bibfield{author}{\bibinfo{person}{Yoshikage Kira}.}
  \bibinfo{year}{2008}\natexlab{}.
\newblock \bibinfo{title}{An example-based procedural system for element
  arrangement}.
\newblock
\newblock
\newblock
\shownote{\url{https://www.youtube.com/watch?v=GfOxdlZV6_k}.}


\bibitem[\protect\citeauthoryear{Kopf, Cohen, and Szeliski}{Kopf
  et~al\mbox{.}}{2014}]%
        {Kopf:2014:FPH}
\bibfield{author}{\bibinfo{person}{Johannes Kopf}, \bibinfo{person}{Michael~F.
  Cohen}, {and} \bibinfo{person}{Richard Szeliski}.}
  \bibinfo{year}{2014}\natexlab{}.
\newblock \showarticletitle{First-person Hyper-lapse videos}.
\newblock \bibinfo{journal}{\emph{Transactions on Graphics (Proceedings of
  SIGGRAPH)}} \bibinfo{volume}{33}, \bibinfo{number}{4} (\bibinfo{year}{2014}),
  \bibinfo{pages}{article no. 78}.
\newblock


\bibitem[\protect\citeauthoryear{Kopf, Cohen-Or, Deussen, and Lischinski}{Kopf
  et~al\mbox{.}}{2006}]%
        {Kopf:2006:RWT}
\bibfield{author}{\bibinfo{person}{Johannes Kopf}, \bibinfo{person}{Daniel
  Cohen-Or}, \bibinfo{person}{Oliver Deussen}, {and} \bibinfo{person}{Dani
  Lischinski}.} \bibinfo{year}{2006}\natexlab{}.
\newblock \showarticletitle{Recursive Wang Tiles for Real-time Blue Noise}. In
  \bibinfo{booktitle}{\emph{ACM SIGGRAPH 2006 Papers}}
  \emph{(\bibinfo{series}{SIGGRAPH '06})}. \bibinfo{publisher}{ACM},
  \bibinfo{address}{New York, NY, USA}, \bibinfo{pages}{509--518}.
\newblock
\showISBNx{1-59593-364-6}
\urldef\tempurl%
\url{https://doi.org/10.1145/1179352.1141916}
\showDOI{\tempurl}


\bibitem[\protect\citeauthoryear{Kopf, Fu, Cohen-Or, Deussen, Lischinski, and
  Wong}{Kopf et~al\mbox{.}}{2007}]%
        {Kopf:2007:STS}
\bibfield{author}{\bibinfo{person}{Johannes Kopf}, \bibinfo{person}{Chi-Wing
  Fu}, \bibinfo{person}{Daniel Cohen-Or}, \bibinfo{person}{Oliver Deussen},
  \bibinfo{person}{Dani Lischinski}, {and} \bibinfo{person}{Tien-Tsin Wong}.}
  \bibinfo{year}{2007}\natexlab{}.
\newblock \showarticletitle{Solid Texture Synthesis from 2D Exemplars}.
\newblock \bibinfo{journal}{\emph{ACM Transactions on Graphics (Proceedings of
  SIGGRAPH 2007)}} \bibinfo{volume}{26}, \bibinfo{number}{3}
  (\bibinfo{year}{2007}), \bibinfo{pages}{2:1--2:9}.
\newblock


\bibitem[\protect\citeauthoryear{Koyama, Sakamoto, and Igarashi}{Koyama
  et~al\mbox{.}}{2016}]%
        {Koyama:2016:SPL}
\bibfield{author}{\bibinfo{person}{Yuki Koyama}, \bibinfo{person}{Daisuke
  Sakamoto}, {and} \bibinfo{person}{Takeo Igarashi}.}
  \bibinfo{year}{2016}\natexlab{}.
\newblock \showarticletitle{SelPh: Progressive Learning and Support of Manual
  Photo Color Enhancement}. In \bibinfo{booktitle}{\emph{CHI '16}}.
  \bibinfo{pages}{2520--2532}.
\newblock
\showISBNx{978-1-4503-3362-7}
\urldef\tempurl%
\url{https://doi.org/10.1145/2858036.2858111}
\showDOI{\tempurl}


\bibitem[\protect\citeauthoryear{Koyama, Sato, Sakamoto, and Igarashi}{Koyama
  et~al\mbox{.}}{2017}]%
        {Koyama:2017:SLS}
\bibfield{author}{\bibinfo{person}{Yuki Koyama}, \bibinfo{person}{Issei Sato},
  \bibinfo{person}{Daisuke Sakamoto}, {and} \bibinfo{person}{Takeo Igarashi}.}
  \bibinfo{year}{2017}\natexlab{}.
\newblock \showarticletitle{Sequential Line Search for Efficient Visual Design
  Optimization by Crowds}.
\newblock \bibinfo{journal}{\emph{ACM Trans. Graph.}} \bibinfo{volume}{36},
  \bibinfo{number}{4}, Article \bibinfo{articleno}{48} (\bibinfo{date}{July}
  \bibinfo{year}{2017}), \bibinfo{numpages}{11}~pages.
\newblock
\showISSN{0730-0301}
\urldef\tempurl%
\url{https://doi.org/10.1145/3072959.3073598}
\showDOI{\tempurl}


\bibitem[\protect\citeauthoryear{Kuhn}{Kuhn}{1955}]%
        {Kuhn:1955:HMA}
\bibfield{author}{\bibinfo{person}{Harold~W Kuhn}.}
  \bibinfo{year}{1955}\natexlab{}.
\newblock \showarticletitle{The Hungarian method for the assignment problem}.
\newblock \bibinfo{journal}{\emph{Naval research logistics quarterly}}
  \bibinfo{volume}{2}, \bibinfo{number}{1-2} (\bibinfo{year}{1955}),
  \bibinfo{pages}{83--97}.
\newblock


\bibitem[\protect\citeauthoryear{Kwatra, Essa, Bobick, and Kwatra}{Kwatra
  et~al\mbox{.}}{2005}]%
        {Kwatra:2005:TOE}
\bibfield{author}{\bibinfo{person}{Vivek Kwatra}, \bibinfo{person}{Irfan Essa},
  \bibinfo{person}{Aaron Bobick}, {and} \bibinfo{person}{Nipun Kwatra}.}
  \bibinfo{year}{2005}\natexlab{}.
\newblock \showarticletitle{Texture Optimization for Example-based Synthesis}.
\newblock \bibinfo{journal}{\emph{ACM Trans. Graph.}} \bibinfo{volume}{24},
  \bibinfo{number}{3} (\bibinfo{date}{July} \bibinfo{year}{2005}),
  \bibinfo{pages}{795--802}.
\newblock
\showISSN{0730-0301}
\urldef\tempurl%
\url{https://doi.org/10.1145/1073204.1073263}
\showDOI{\tempurl}


\bibitem[\protect\citeauthoryear{Kwatra, Sch\"{o}dl, Essa, Turk, and
  Bobick}{Kwatra et~al\mbox{.}}{2003}]%
        {Kwatra:2003:GTI}
\bibfield{author}{\bibinfo{person}{Vivek Kwatra}, \bibinfo{person}{Arno
  Sch\"{o}dl}, \bibinfo{person}{Irfan Essa}, \bibinfo{person}{Greg Turk}, {and}
  \bibinfo{person}{Aaron Bobick}.} \bibinfo{year}{2003}\natexlab{}.
\newblock \showarticletitle{Graphcut Textures: Image and Video Synthesis Using
  Graph Cuts}. In \bibinfo{booktitle}{\emph{SIGGRAPH '03}}.
  \bibinfo{pages}{277--286}.
\newblock
\showISBNx{1-58113-709-5}
\urldef\tempurl%
\url{https://doi.org/10.1145/1201775.882264}
\showDOI{\tempurl}


\bibitem[\protect\citeauthoryear{Laboratory}{Laboratory}{2015}]%
        {Youtube:2015:EBR}
\bibfield{author}{\bibinfo{person}{Computer~Graphics Laboratory}.}
  \bibinfo{year}{2015}\natexlab{}.
\newblock \bibinfo{title}{Example Based Repetitive Structure Synthesis}.
\newblock
\newblock
\newblock
\shownote{\url{https://youtu.be/wVWSGUS5D5A}.}


\bibitem[\protect\citeauthoryear{Lagae, Lefebvre, Cook, DeRose, Drettakis,
  Ebert, Lewis, Perlin, and Zwicker}{Lagae et~al\mbox{.}}{2010a}]%
        {Lagae:2010:SPN}
\bibfield{author}{\bibinfo{person}{Ares Lagae}, \bibinfo{person}{Sylvain
  Lefebvre}, \bibinfo{person}{Rob Cook}, \bibinfo{person}{Tony DeRose},
  \bibinfo{person}{George Drettakis}, \bibinfo{person}{David~S Ebert},
  \bibinfo{person}{John~P Lewis}, \bibinfo{person}{Ken Perlin}, {and}
  \bibinfo{person}{Matthias Zwicker}.} \bibinfo{year}{2010}\natexlab{a}.
\newblock \showarticletitle{A survey of procedural noise functions}. In
  \bibinfo{booktitle}{\emph{Computer Graphics Forum}},
  Vol.~\bibinfo{volume}{29}. Wiley Online Library, \bibinfo{pages}{2579--2600}.
\newblock


\bibitem[\protect\citeauthoryear{Lagae, Vangorp, Lenaerts, and Dutr{\'e}}{Lagae
  et~al\mbox{.}}{2010b}]%
        {Lagae:2010:PIS}
\bibfield{author}{\bibinfo{person}{Ares Lagae}, \bibinfo{person}{Peter
  Vangorp}, \bibinfo{person}{Toon Lenaerts}, {and} \bibinfo{person}{Philip
  Dutr{\'e}}.} \bibinfo{year}{2010}\natexlab{b}.
\newblock \showarticletitle{Procedural Isotropic Stochastic Textures by
  Example}.
\newblock \bibinfo{journal}{\emph{Comput. Graph.}} \bibinfo{volume}{34},
  \bibinfo{number}{4} (\bibinfo{date}{Aug.} \bibinfo{year}{2010}),
  \bibinfo{pages}{312--321}.
\newblock
\showISSN{0097-8493}
\urldef\tempurl%
\url{https://doi.org/10.1016/j.cag.2010.05.004}
\showDOI{\tempurl}


\bibitem[\protect\citeauthoryear{Landes, Galerne, and Hurtut}{Landes
  et~al\mbox{.}}{2013}]%
        {Landes:2013:SAM}
\bibfield{author}{\bibinfo{person}{Pierre-Edouard Landes},
  \bibinfo{person}{Bruno Galerne}, {and} \bibinfo{person}{Thomas Hurtut}.}
  \bibinfo{year}{2013}\natexlab{}.
\newblock \showarticletitle{A Shape-Aware Model for Discrete Texture
  Synthesis}.
\newblock \bibinfo{journal}{\emph{Computer Graphics Forum}}
  \bibinfo{volume}{32}, \bibinfo{number}{4} (\bibinfo{year}{2013}),
  \bibinfo{pages}{67--76}.
\newblock


\bibitem[\protect\citeauthoryear{Lee, Zitnick, and Cohen}{Lee
  et~al\mbox{.}}{2011}]%
        {Lee:2011:SRU}
\bibfield{author}{\bibinfo{person}{Yong~Jae Lee}, \bibinfo{person}{C.~Lawrence
  Zitnick}, {and} \bibinfo{person}{Michael~F. Cohen}.}
  \bibinfo{year}{2011}\natexlab{}.
\newblock \showarticletitle{ShadowDraw: Real-time User Guidance for Freehand
  Drawing}.
\newblock \bibinfo{journal}{\emph{ACM Trans. Graph.}} \bibinfo{volume}{30},
  \bibinfo{number}{4}, Article \bibinfo{articleno}{27} (\bibinfo{date}{July}
  \bibinfo{year}{2011}), \bibinfo{numpages}{10}~pages.
\newblock
\showISSN{0730-0301}
\urldef\tempurl%
\url{https://doi.org/10.1145/2010324.1964922}
\showDOI{\tempurl}


\bibitem[\protect\citeauthoryear{Lefebvre and Hoppe}{Lefebvre and
  Hoppe}{2006}]%
        {Lefebvre:2006:ATS}
\bibfield{author}{\bibinfo{person}{Sylvain Lefebvre} {and}
  \bibinfo{person}{Hugues Hoppe}.} \bibinfo{year}{2006}\natexlab{}.
\newblock \showarticletitle{Appearance-space Texture Synthesis}. In
  \bibinfo{booktitle}{\emph{ACM SIGGRAPH 2006 Papers}}
  \emph{(\bibinfo{series}{SIGGRAPH '06})}. \bibinfo{publisher}{ACM},
  \bibinfo{address}{New York, NY, USA}, \bibinfo{pages}{541--548}.
\newblock
\showISBNx{1-59593-364-6}
\urldef\tempurl%
\url{https://doi.org/10.1145/1179352.1141921}
\showDOI{\tempurl}


\bibitem[\protect\citeauthoryear{Li, Lafarge, and Marlet}{Li
  et~al\mbox{.}}{2020}]%
        {Li:2020:ASI}
\bibfield{author}{\bibinfo{person}{Muxingzi Li}, \bibinfo{person}{Florent
  Lafarge}, {and} \bibinfo{person}{Renaud Marlet}.}
  \bibinfo{year}{2020}\natexlab{}.
\newblock \showarticletitle{Approximating shapes in images with low-complexity
  polygons}. In \bibinfo{booktitle}{\emph{The IEEE/CVF Conference on Computer
  Vision and Pattern Recognition (CVPR)}}.
\newblock


\bibitem[\protect\citeauthoryear{Li, Breen, McCann, and Hodgins}{Li
  et~al\mbox{.}}{2019}]%
        {Li:2019:AQP}
\bibfield{author}{\bibinfo{person}{Yifei Li}, \bibinfo{person}{David~E. Breen},
  \bibinfo{person}{James McCann}, {and} \bibinfo{person}{Jessica Hodgins}.}
  \bibinfo{year}{2019}\natexlab{}.
\newblock \showarticletitle{Algorithmic Quilting Pattern Generation for Pieced
  Quilts}. In \bibinfo{booktitle}{\emph{Proceedings of the 45th Graphics
  Interface Conference on Proceedings of Graphics Interface 2019}}
  \emph{(\bibinfo{series}{GI\&\#39;19})}. Article \bibinfo{articleno}{13},
  \bibinfo{numpages}{9}~pages.
\newblock
\showISBNx{978-0-9947868-4-5}
\urldef\tempurl%
\url{https://doi.org/10.20380/GI2019.13}
\showDOI{\tempurl}


\bibitem[\protect\citeauthoryear{Liang, Liu, Xu, Guo, and Shum}{Liang
  et~al\mbox{.}}{2001}]%
        {Liang:2001:RTS}
\bibfield{author}{\bibinfo{person}{Lin Liang}, \bibinfo{person}{Ce Liu},
  \bibinfo{person}{Ying-Qing Xu}, \bibinfo{person}{Baining Guo}, {and}
  \bibinfo{person}{Heung-Yeung Shum}.} \bibinfo{year}{2001}\natexlab{}.
\newblock \showarticletitle{Real-time Texture Synthesis by Patch-based
  Sampling}.
\newblock \bibinfo{journal}{\emph{ACM Trans. Graph.}} \bibinfo{volume}{20},
  \bibinfo{number}{3} (\bibinfo{date}{July} \bibinfo{year}{2001}),
  \bibinfo{pages}{127--150}.
\newblock
\showISSN{0730-0301}
\urldef\tempurl%
\url{https://doi.org/10.1145/501786.501787}
\showDOI{\tempurl}


\bibitem[\protect\citeauthoryear{Loi, Hurtut, Vergne, and Thollot}{Loi
  et~al\mbox{.}}{2017}]%
        {Loi:2017:PAE}
\bibfield{author}{\bibinfo{person}{Hugo Loi}, \bibinfo{person}{Thomas Hurtut},
  \bibinfo{person}{Romain Vergne}, {and} \bibinfo{person}{Joelle Thollot}.}
  \bibinfo{year}{2017}\natexlab{}.
\newblock \showarticletitle{Programmable 2D Arrangements for Element Texture
  Design}.
\newblock \bibinfo{journal}{\emph{ACM Trans. Graph.}} \bibinfo{volume}{36},
  \bibinfo{number}{4}, Article \bibinfo{articleno}{105a} (\bibinfo{date}{May}
  \bibinfo{year}{2017}).
\newblock
\showISSN{0730-0301}
\urldef\tempurl%
\url{https://doi.org/10.1145/3072959.2983617}
\showDOI{\tempurl}


\bibitem[\protect\citeauthoryear{Loviscach}{Loviscach}{2009}]%
        {Loviscach:2009:CSR}
\bibfield{author}{\bibinfo{person}{J{\"o}rn Loviscach}.}
  \bibinfo{year}{2009}\natexlab{}.
\newblock \showarticletitle{Continuous Search and Replace in Vector Graphics.}.
  In \bibinfo{booktitle}{\emph{Eurographics (Short Papers)}}.
  \bibinfo{pages}{17--20}.
\newblock


\bibitem[\protect\citeauthoryear{Lu, Barnes, Wan, Asente, Mech, and
  Finkelstein}{Lu et~al\mbox{.}}{2014}]%
        {Lu:2014:DDS}
\bibfield{author}{\bibinfo{person}{Jingwan Lu}, \bibinfo{person}{Connelly
  Barnes}, \bibinfo{person}{Connie Wan}, \bibinfo{person}{Paul Asente},
  \bibinfo{person}{Radomir Mech}, {and} \bibinfo{person}{Adam Finkelstein}.}
  \bibinfo{year}{2014}\natexlab{}.
\newblock \showarticletitle{DecoBrush: Drawing Structured Decorative Patterns
  by Example}.
\newblock \bibinfo{journal}{\emph{ACM Trans. Graph.}} \bibinfo{volume}{33},
  \bibinfo{number}{4}, Article \bibinfo{articleno}{90} (\bibinfo{date}{July}
  \bibinfo{year}{2014}), \bibinfo{numpages}{9}~pages.
\newblock
\showISSN{0730-0301}
\urldef\tempurl%
\url{https://doi.org/10.1145/2601097.2601190}
\showDOI{\tempurl}


\bibitem[\protect\citeauthoryear{Lu, Yu, Finkelstein, and DiVerdi}{Lu
  et~al\mbox{.}}{2012}]%
        {Lu:2012:HES}
\bibfield{author}{\bibinfo{person}{Jingwan Lu}, \bibinfo{person}{Fisher Yu},
  \bibinfo{person}{Adam Finkelstein}, {and} \bibinfo{person}{Stephen DiVerdi}.}
  \bibinfo{year}{2012}\natexlab{}.
\newblock \showarticletitle{HelpingHand: Example-based Stroke Stylization}.
\newblock \bibinfo{journal}{\emph{ACM Trans. Graph.}} \bibinfo{volume}{31},
  \bibinfo{number}{4}, Article \bibinfo{articleno}{46} (\bibinfo{date}{July}
  \bibinfo{year}{2012}), \bibinfo{numpages}{10}~pages.
\newblock
\showISSN{0730-0301}
\urldef\tempurl%
\url{https://doi.org/10.1145/2185520.2185542}
\showDOI{\tempurl}


\bibitem[\protect\citeauthoryear{Luk\'{a}\v{c}, S\'{y}kora, Sunkavalli,
  Shechtman, Jamri\v{s}ka, Carr, and Pajdla}{Luk\'{a}\v{c}
  et~al\mbox{.}}{2017}]%
        {Lukac:2017:NRR}
\bibfield{author}{\bibinfo{person}{Michal Luk\'{a}\v{c}},
  \bibinfo{person}{Daniel S\'{y}kora}, \bibinfo{person}{Kalyan Sunkavalli},
  \bibinfo{person}{Eli Shechtman}, \bibinfo{person}{Ond\v{r}ej Jamri\v{s}ka},
  \bibinfo{person}{Nathan Carr}, {and} \bibinfo{person}{Tom\'{a}\v{s} Pajdla}.}
  \bibinfo{year}{2017}\natexlab{}.
\newblock \showarticletitle{Nautilus: Recovering Regional Symmetry
  Transformations for Image Editing}.
\newblock \bibinfo{journal}{\emph{ACM Trans. Graph.}} \bibinfo{volume}{36},
  \bibinfo{number}{4}, Article \bibinfo{articleno}{108} (\bibinfo{date}{July}
  \bibinfo{year}{2017}), \bibinfo{numpages}{11}~pages.
\newblock
\showISSN{0730-0301}
\urldef\tempurl%
\url{https://doi.org/10.1145/3072959.3073661}
\showDOI{\tempurl}


\bibitem[\protect\citeauthoryear{Lun, Zou, Huang, Kalogerakis, Tan, Cani, and
  Zhang}{Lun et~al\mbox{.}}{2017}]%
        {Lun:2017:LGD}
\bibfield{author}{\bibinfo{person}{Zhaoliang Lun}, \bibinfo{person}{Changqing
  Zou}, \bibinfo{person}{Haibin Huang}, \bibinfo{person}{Evangelos
  Kalogerakis}, \bibinfo{person}{Ping Tan}, \bibinfo{person}{Marie-Paule Cani},
  {and} \bibinfo{person}{Hao Zhang}.} \bibinfo{year}{2017}\natexlab{}.
\newblock \showarticletitle{Learning to Group Discrete Graphical Patterns}.
\newblock \bibinfo{journal}{\emph{ACM Transactions on Graphics}}
  \bibinfo{volume}{36}, \bibinfo{number}{6} (\bibinfo{year}{2017}),
  \bibinfo{pages}{Article 225}.
\newblock


\bibitem[\protect\citeauthoryear{Ma, Wei, Lefebvre, and Tong}{Ma
  et~al\mbox{.}}{2013}]%
        {Ma:2013:DET}
\bibfield{author}{\bibinfo{person}{Chongyang Ma}, \bibinfo{person}{Li-Yi Wei},
  \bibinfo{person}{Sylvain Lefebvre}, {and} \bibinfo{person}{Xin Tong}.}
  \bibinfo{year}{2013}\natexlab{}.
\newblock \showarticletitle{Dynamic Element Textures}.
\newblock \bibinfo{journal}{\emph{ACM Trans. Graph.}} \bibinfo{volume}{32},
  \bibinfo{number}{4}, Article \bibinfo{articleno}{90} (\bibinfo{date}{July}
  \bibinfo{year}{2013}), \bibinfo{numpages}{10}~pages.
\newblock
\showISSN{0730-0301}
\urldef\tempurl%
\url{https://doi.org/10.1145/2461912.2461921}
\showDOI{\tempurl}


\bibitem[\protect\citeauthoryear{Ma, Wei, and Tong}{Ma et~al\mbox{.}}{2011}]%
        {Ma:2011:DET}
\bibfield{author}{\bibinfo{person}{Chongyang Ma}, \bibinfo{person}{Li-Yi Wei},
  {and} \bibinfo{person}{Xin Tong}.} \bibinfo{year}{2011}\natexlab{}.
\newblock \showarticletitle{Discrete Element Textures}.
\newblock \bibinfo{journal}{\emph{ACM Trans. Graph.}} \bibinfo{volume}{30},
  \bibinfo{number}{4}, Article \bibinfo{articleno}{62} (\bibinfo{date}{July}
  \bibinfo{year}{2011}), \bibinfo{numpages}{10}~pages.
\newblock
\showISSN{0730-0301}
\urldef\tempurl%
\url{https://doi.org/10.1145/2010324.1964957}
\showDOI{\tempurl}


\bibitem[\protect\citeauthoryear{Mart\'{\i}nez, Dumas, Lefebvre, and
  Wei}{Mart\'{\i}nez et~al\mbox{.}}{2015}]%
        {Martinez:2015:SAO}
\bibfield{author}{\bibinfo{person}{Jon\`{a}s Mart\'{\i}nez},
  \bibinfo{person}{J{\'e}r{\'e}mie Dumas}, \bibinfo{person}{Sylvain Lefebvre},
  {and} \bibinfo{person}{Li-Yi Wei}.} \bibinfo{year}{2015}\natexlab{}.
\newblock \showarticletitle{Structure and Appearance Optimization for
  Controllable Shape Design}.
\newblock \bibinfo{journal}{\emph{ACM Trans. Graph.}} \bibinfo{volume}{34},
  \bibinfo{number}{6}, Article \bibinfo{articleno}{229} (\bibinfo{date}{Oct.}
  \bibinfo{year}{2015}), \bibinfo{numpages}{11}~pages.
\newblock
\showISSN{0730-0301}
\urldef\tempurl%
\url{https://doi.org/10.1145/2816795.2818101}
\showDOI{\tempurl}


\bibitem[\protect\citeauthoryear{Matusik, Zwicker, and Durand}{Matusik
  et~al\mbox{.}}{2005}]%
        {Matusik:2005:TDU}
\bibfield{author}{\bibinfo{person}{Wojciech Matusik}, \bibinfo{person}{Matthias
  Zwicker}, {and} \bibinfo{person}{Fr{\'e}do Durand}.}
  \bibinfo{year}{2005}\natexlab{}.
\newblock \showarticletitle{Texture Design Using a Simplicial Complex of
  Morphable Textures}.
\newblock \bibinfo{journal}{\emph{ACM Trans. Graph.}} \bibinfo{volume}{24},
  \bibinfo{number}{3} (\bibinfo{date}{July} \bibinfo{year}{2005}),
  \bibinfo{pages}{787--794}.
\newblock
\showISSN{0730-0301}
\urldef\tempurl%
\url{https://doi.org/10.1145/1073204.1073262}
\showDOI{\tempurl}


\bibitem[\protect\citeauthoryear{Merrell and Manocha}{Merrell and
  Manocha}{2010}]%
        {Merrell:2010:ECS}
\bibfield{author}{\bibinfo{person}{Paul Merrell} {and} \bibinfo{person}{Dinesh
  Manocha}.} \bibinfo{year}{2010}\natexlab{}.
\newblock \showarticletitle{Example-based curve synthesis}.
\newblock \bibinfo{journal}{\emph{Computers \& Graphics}} \bibinfo{volume}{34},
  \bibinfo{number}{4} (\bibinfo{year}{2010}), \bibinfo{pages}{304--311}.
\newblock


\bibitem[\protect\citeauthoryear{Myronenko and Song}{Myronenko and
  Song}{2010}]%
        {Myronenko:2010:PSR}
\bibfield{author}{\bibinfo{person}{Andriy Myronenko} {and}
  \bibinfo{person}{Xubo Song}.} \bibinfo{year}{2010}\natexlab{}.
\newblock \showarticletitle{Point set registration: Coherent point drift}.
\newblock \bibinfo{journal}{\emph{IEEE transactions on pattern analysis and
  machine intelligence}} \bibinfo{volume}{32}, \bibinfo{number}{12}
  (\bibinfo{year}{2010}), \bibinfo{pages}{2262--2275}.
\newblock


\bibitem[\protect\citeauthoryear{Nancel and Cockburn}{Nancel and
  Cockburn}{2014}]%
        {Nancel:2014:CCM}
\bibfield{author}{\bibinfo{person}{Mathieu Nancel} {and} \bibinfo{person}{Andy
  Cockburn}.} \bibinfo{year}{2014}\natexlab{}.
\newblock \showarticletitle{Causality: A Conceptual Model of Interaction
  History}. In \bibinfo{booktitle}{\emph{CHI '14}}.
  \bibinfo{pages}{1777--1786}.
\newblock
\showISBNx{978-1-4503-2473-1}
\urldef\tempurl%
\url{https://doi.org/10.1145/2556288.2556990}
\showDOI{\tempurl}


\bibitem[\protect\citeauthoryear{Nesterov and Spokoiny}{Nesterov and
  Spokoiny}{2017}]%
        {Nesterov:2017:RGF}
\bibfield{author}{\bibinfo{person}{Yurii Nesterov} {and}
  \bibinfo{person}{Vladimir Spokoiny}.} \bibinfo{year}{2017}\natexlab{}.
\newblock \showarticletitle{Random gradient-free minimization of convex
  functions}.
\newblock \bibinfo{journal}{\emph{Foundations of Computational Mathematics}}
  \bibinfo{volume}{17}, \bibinfo{number}{2} (\bibinfo{year}{2017}),
  \bibinfo{pages}{527--566}.
\newblock


\bibitem[\protect\citeauthoryear{Nishida, Garcia-Dorado, Aliaga, Benes, and
  Bousseau}{Nishida et~al\mbox{.}}{2016}]%
        {Nishida:2016:ISU}
\bibfield{author}{\bibinfo{person}{Gen Nishida}, \bibinfo{person}{Ignacio
  Garcia-Dorado}, \bibinfo{person}{Daniel~G. Aliaga}, \bibinfo{person}{Bedrich
  Benes}, {and} \bibinfo{person}{Adrien Bousseau}.}
  \bibinfo{year}{2016}\natexlab{}.
\newblock \showarticletitle{Interactive Sketching of Urban Procedural Models}.
\newblock \bibinfo{journal}{\emph{ACM Trans. Graph.}} \bibinfo{volume}{35},
  \bibinfo{number}{4}, Article \bibinfo{articleno}{130} (\bibinfo{date}{July}
  \bibinfo{year}{2016}), \bibinfo{numpages}{11}~pages.
\newblock
\showISSN{0730-0301}
\urldef\tempurl%
\url{https://doi.org/10.1145/2897824.2925951}
\showDOI{\tempurl}


\bibitem[\protect\citeauthoryear{Ohrhallinger and Wimmer}{Ohrhallinger and
  Wimmer}{2019}]%
        {Ohrhallinger:2019:FCN}
\bibfield{author}{\bibinfo{person}{Stefan Ohrhallinger} {and}
  \bibinfo{person}{Michael Wimmer}.} \bibinfo{year}{2019}\natexlab{}.
\newblock \showarticletitle{Fitconnect: Connecting noisy 2d samples by fitted
  neighbourhoods}. In \bibinfo{booktitle}{\emph{Computer Graphics Forum}},
  Vol.~\bibinfo{volume}{38}. Wiley Online Library, \bibinfo{pages}{126--137}.
\newblock


\bibitem[\protect\citeauthoryear{Palacios, Roy, Kumar, Hsu, Chen, Ma, Wei, and
  Zhang}{Palacios et~al\mbox{.}}{2017}]%
        {Palacios:2017:TFD}
\bibfield{author}{\bibinfo{person}{Jonathan Palacios},
  \bibinfo{person}{Lawrence Roy}, \bibinfo{person}{Prashant Kumar},
  \bibinfo{person}{Chen-Yuan Hsu}, \bibinfo{person}{Weikai Chen},
  \bibinfo{person}{Chongyang Ma}, \bibinfo{person}{Li-Yi Wei}, {and}
  \bibinfo{person}{Eugene Zhang}.} \bibinfo{year}{2017}\natexlab{}.
\newblock \showarticletitle{Tensor Field Design in Volumes}.
\newblock \bibinfo{journal}{\emph{ACM Trans. Graph.}} \bibinfo{volume}{36},
  \bibinfo{number}{6}, Article \bibinfo{articleno}{188} (\bibinfo{date}{Nov.}
  \bibinfo{year}{2017}), \bibinfo{numpages}{15}~pages.
\newblock
\showISSN{0730-0301}
\urldef\tempurl%
\url{https://doi.org/10.1145/3130800.3130844}
\showDOI{\tempurl}


\bibitem[\protect\citeauthoryear{Pedersen and Singh}{Pedersen and
  Singh}{2006}]%
        {Pedersen:2006:OLM}
\bibfield{author}{\bibinfo{person}{Hans Pedersen} {and} \bibinfo{person}{Karan
  Singh}.} \bibinfo{year}{2006}\natexlab{}.
\newblock \showarticletitle{Organic Labyrinths and Mazes}. In
  \bibinfo{booktitle}{\emph{NPAR '06}}. \bibinfo{pages}{79--86}.
\newblock
\showISBNx{1595933573}
\urldef\tempurl%
\url{https://doi.org/10.1145/1124728.1124742}
\showDOI{\tempurl}


\bibitem[\protect\citeauthoryear{Peng, Wei, Kazi, and Kim}{Peng
  et~al\mbox{.}}{2020}]%
        {Peng:2020:AAS}
\bibfield{author}{\bibinfo{person}{Mengqi Peng}, \bibinfo{person}{Li-Yi Wei},
  \bibinfo{person}{Rubaiat~Habib Kazi}, {and} \bibinfo{person}{Vladimir~G.
  Kim}.} \bibinfo{year}{2020}\natexlab{}.
\newblock \showarticletitle{Autocomplete Animated Sculpting}. In
  \bibinfo{booktitle}{\emph{UIST '20}}.
\newblock
\urldef\tempurl%
\url{https://doi.org/10.1145/3379337.3415884}
\showDOI{\tempurl}


\bibitem[\protect\citeauthoryear{Peng, Xing, and Wei}{Peng
  et~al\mbox{.}}{2018}]%
        {Peng:2018:A3S}
\bibfield{author}{\bibinfo{person}{Mengqi Peng}, \bibinfo{person}{Jun Xing},
  {and} \bibinfo{person}{Li-Yi Wei}.} \bibinfo{year}{2018}\natexlab{}.
\newblock \showarticletitle{Autocomplete {3D} Sculpting}.
\newblock \bibinfo{journal}{\emph{ACM Trans. Graph.}} \bibinfo{volume}{37},
  \bibinfo{number}{4}, Article \bibinfo{articleno}{132} (\bibinfo{date}{July}
  \bibinfo{year}{2018}), \bibinfo{numpages}{15}~pages.
\newblock
\showISSN{0730-0301}
\urldef\tempurl%
\url{https://doi.org/10.1145/3197517.3201297}
\showDOI{\tempurl}


\bibitem[\protect\citeauthoryear{Pinterest}{Pinterest}{2020a}]%
        {Pinterest:2020:CPC}
\bibfield{author}{\bibinfo{person}{Pinterest}.}
  \bibinfo{year}{2020}\natexlab{a}.
\newblock \bibinfo{title}{Continuous Pattern: Continuity in 2D}.
\newblock
\newblock
\newblock
\shownote{\url{https://pin.it/5iFuMTv}.}


\bibitem[\protect\citeauthoryear{Pinterest}{Pinterest}{2020b}]%
        {Pinterest:2020:CPL}
\bibfield{author}{\bibinfo{person}{Pinterest}.}
  \bibinfo{year}{2020}\natexlab{b}.
\newblock \bibinfo{title}{Continuous Pattern: leaves}.
\newblock
\newblock
\newblock
\shownote{\url{https://pin.it/5Ig5mC4}.}


\bibitem[\protect\citeauthoryear{Pinterest}{Pinterest}{2020c}]%
        {Pinterest:2020:CPS}
\bibfield{author}{\bibinfo{person}{Pinterest}.}
  \bibinfo{year}{2020}\natexlab{c}.
\newblock \bibinfo{title}{Continuous Pattern: spider web}.
\newblock
\newblock
\newblock
\shownote{\url{https://pin.it/ssJIHxC}.}


\bibitem[\protect\citeauthoryear{Pinterest}{Pinterest}{2020d}]%
        {Pinterest:2020:CPT}
\bibfield{author}{\bibinfo{person}{Pinterest}.}
  \bibinfo{year}{2020}\natexlab{d}.
\newblock \bibinfo{title}{Continuous Pattern: Topology-varying discrete
  elements}.
\newblock
\newblock
\newblock
\shownote{\url{https://www.pinterest.com/peihantu/continuous-pattern-design/topology-varying-discrete-elements/}.}


\bibitem[\protect\citeauthoryear{Pinterest}{Pinterest}{2020e}]%
        {Pinterest:2020:HCC}
\bibfield{author}{\bibinfo{person}{Pinterest}.}
  \bibinfo{year}{2020}\natexlab{e}.
\newblock \bibinfo{title}{How To Create Custom Fancy Type Letters - A - in
  Adobe Illustrator}.
\newblock
\newblock
\newblock
\shownote{\url{https://pin.it/5j1lG6b}.}


\bibitem[\protect\citeauthoryear{Pinterest}{Pinterest}{2020f}]%
        {Pinterest:2020:HCS}
\bibfield{author}{\bibinfo{person}{Pinterest}.}
  \bibinfo{year}{2020}\natexlab{f}.
\newblock \bibinfo{title}{How to Create Simple Art with Paintbrush Tool in
  Illustrator}.
\newblock
\newblock
\newblock
\shownote{\url{https://pin.it/3eCbGr1}.}


\bibitem[\protect\citeauthoryear{Pinterest}{Pinterest}{2020g}]%
        {Pinterest:2020:ITA}
\bibfield{author}{\bibinfo{person}{Pinterest}.}
  \bibinfo{year}{2020}\natexlab{g}.
\newblock \bibinfo{title}{Illustrator Trick : Art Brush Lettering}.
\newblock
\newblock
\newblock
\shownote{\url{https://pin.it/go2La09}.}


\bibitem[\protect\citeauthoryear{Pinterest}{Pinterest}{2020h}]%
        {Pinterest:2020:RSS}
\bibfield{author}{\bibinfo{person}{Pinterest}.}
  \bibinfo{year}{2020}\natexlab{h}.
\newblock \bibinfo{title}{Research: Shape Synthesis}.
\newblock
\newblock
\newblock
\shownote{\url{https://www.pinterest.com/peihantu/research/shape-synthesis/}.}


\bibitem[\protect\citeauthoryear{Pinterest}{Pinterest}{2020i}]%
        {Pinterest:2020:WDD}
\bibfield{author}{\bibinfo{person}{Pinterest}.}
  \bibinfo{year}{2020}\natexlab{i}.
\newblock \bibinfo{title}{What does the dual graph look like for this pattern?}
\newblock
\newblock
\newblock
\shownote{\url{https://pin.it/1WUQmf3}.}


\bibitem[\protect\citeauthoryear{Reas and Fry}{Reas and Fry}{2019}]%
        {Reas:2019:P}
\bibfield{author}{\bibinfo{person}{Casey Reas} {and} \bibinfo{person}{Ben
  Fry}.} \bibinfo{year}{2019}\natexlab{}.
\newblock \bibinfo{title}{Processing}.
\newblock
\newblock
\newblock
\shownote{\url{https://processing.org/}.}


\bibitem[\protect\citeauthoryear{Red-Portal}{Red-Portal}{2018}]%
        {Red-Portal:2018:DFC}
\bibfield{author}{\bibinfo{person}{Red-Portal}.}
  \bibinfo{year}{2018}\natexlab{}.
\newblock \bibinfo{title}{Derivative Free Convex Optimization Algorithms in
  C++}.
\newblock
\newblock
\newblock
\shownote{\url{https://github.com/Red-Portal/dfopt}.}


\bibitem[\protect\citeauthoryear{Riesen and Bunke}{Riesen and Bunke}{2009}]%
        {Riesen:2009:AGE}
\bibfield{author}{\bibinfo{person}{Kaspar Riesen} {and} \bibinfo{person}{Horst
  Bunke}.} \bibinfo{year}{2009}\natexlab{}.
\newblock \showarticletitle{Approximate graph edit distance computation by
  means of bipartite graph matching}.
\newblock \bibinfo{journal}{\emph{Image and Vision computing}}
  \bibinfo{volume}{27}, \bibinfo{number}{7} (\bibinfo{year}{2009}),
  \bibinfo{pages}{950--959}.
\newblock


\bibitem[\protect\citeauthoryear{Rios and Sahinidis}{Rios and
  Sahinidis}{2013}]%
        {Rios:2013:DFO}
\bibfield{author}{\bibinfo{person}{Luis~Miguel Rios} {and}
  \bibinfo{person}{Nikolaos~V Sahinidis}.} \bibinfo{year}{2013}\natexlab{}.
\newblock \showarticletitle{Derivative-free optimization: a review of
  algorithms and comparison of software implementations}.
\newblock \bibinfo{journal}{\emph{Journal of Global Optimization}}
  \bibinfo{volume}{56}, \bibinfo{number}{3} (\bibinfo{year}{2013}),
  \bibinfo{pages}{1247--1293}.
\newblock


\bibitem[\protect\citeauthoryear{Rother, Kolmogorov, Lempitsky, and
  Szummer}{Rother et~al\mbox{.}}{2007}]%
        {Rother:2007:OBM}
\bibfield{author}{\bibinfo{person}{Carsten Rother}, \bibinfo{person}{Vladimir
  Kolmogorov}, \bibinfo{person}{Victor Lempitsky}, {and}
  \bibinfo{person}{Martin Szummer}.} \bibinfo{year}{2007}\natexlab{}.
\newblock \showarticletitle{Optimizing binary MRFs via extended roof duality}.
  In \bibinfo{booktitle}{\emph{2007 IEEE Conference on Computer Vision and
  Pattern Recognition}}. IEEE, \bibinfo{pages}{1--8}.
\newblock


\bibitem[\protect\citeauthoryear{Roveri, {\"O}ztireli, Martin, Solenthaler, and
  Gross}{Roveri et~al\mbox{.}}{2015}]%
        {Roveri:2015:EBR}
\bibfield{author}{\bibinfo{person}{Riccardo Roveri}, \bibinfo{person}{A~Cengiz
  {\"O}ztireli}, \bibinfo{person}{Sebastian Martin}, \bibinfo{person}{Barbara
  Solenthaler}, {and} \bibinfo{person}{Markus Gross}.}
  \bibinfo{year}{2015}\natexlab{}.
\newblock \showarticletitle{Example based repetitive structure synthesis}.
\newblock \bibinfo{journal}{\emph{Computer Graphics Forum}}
  \bibinfo{volume}{34}, \bibinfo{number}{5} (\bibinfo{year}{2015}),
  \bibinfo{pages}{39--52}.
\newblock


\bibitem[\protect\citeauthoryear{Roy, Zhang, and Vogel}{Roy
  et~al\mbox{.}}{2019}]%
        {Roy:2019:AAG}
\bibfield{author}{\bibinfo{person}{Quentin Roy}, \bibinfo{person}{Futian
  Zhang}, {and} \bibinfo{person}{Daniel Vogel}.}
  \bibinfo{year}{2019}\natexlab{}.
\newblock \showarticletitle{Automation Accuracy Is Good, but High
  Controllability May Be Better}. In \bibinfo{booktitle}{\emph{CHI '19}}.
  Article \bibinfo{articleno}{520}, \bibinfo{numpages}{8}~pages.
\newblock
\showISBNx{978-1-4503-5970-2}
\urldef\tempurl%
\url{https://doi.org/10.1145/3290605.3300750}
\showDOI{\tempurl}


\bibitem[\protect\citeauthoryear{Santoni and Pellacini}{Santoni and
  Pellacini}{2016}]%
        {Santoni:2016:GGP}
\bibfield{author}{\bibinfo{person}{Christian Santoni} {and}
  \bibinfo{person}{Fabio Pellacini}.} \bibinfo{year}{2016}\natexlab{}.
\newblock \showarticletitle{gTangle: A Grammar for the Procedural Generation of
  Tangle Patterns}.
\newblock \bibinfo{journal}{\emph{ACM Trans. Graph.}} \bibinfo{volume}{35},
  \bibinfo{number}{6}, Article \bibinfo{articleno}{182} (\bibinfo{date}{Nov.}
  \bibinfo{year}{2016}), \bibinfo{numpages}{11}~pages.
\newblock
\showISSN{0730-0301}
\urldef\tempurl%
\url{https://doi.org/10.1145/2980179.2982417}
\showDOI{\tempurl}


\bibitem[\protect\citeauthoryear{Saputra, Kaplan, and Asente}{Saputra
  et~al\mbox{.}}{2020}]%
        {Saputra:2020:APE}
\bibfield{author}{\bibinfo{person}{Reza~Adhitya Saputra},
  \bibinfo{person}{Craig~S. Kaplan}, {and} \bibinfo{person}{Paul Asente}.}
  \bibinfo{year}{2020}\natexlab{}.
\newblock \showarticletitle{AnimationPak: Packing Elements with Scripted
  Animations}. In \bibinfo{booktitle}{\emph{Graphics Interface '20}}.
\newblock
\newblock
\shownote{\url{https://openreview.net/forum?id=sr89orrDo-o}.}


\bibitem[\protect\citeauthoryear{Schumacher, Thomaszewski, and
  Gross}{Schumacher et~al\mbox{.}}{2016}]%
        {Schumacher:2016:SDS}
\bibfield{author}{\bibinfo{person}{Christian Schumacher},
  \bibinfo{person}{Bernhard Thomaszewski}, {and} \bibinfo{person}{Markus
  Gross}.} \bibinfo{year}{2016}\natexlab{}.
\newblock \showarticletitle{Stenciling: Designing Structurally-Sound Surfaces
  with Decorative Patterns}.
\newblock \bibinfo{journal}{\emph{Computer Graphics Forum}}
  \bibinfo{volume}{35}, \bibinfo{number}{5} (\bibinfo{year}{2016}),
  \bibinfo{pages}{101--110}.
\newblock


\bibitem[\protect\citeauthoryear{Shahriari, Swersky, Wang, Adams, and
  De~Freitas}{Shahriari et~al\mbox{.}}{2015}]%
        {Shahriari:2015:THL}
\bibfield{author}{\bibinfo{person}{Bobak Shahriari}, \bibinfo{person}{Kevin
  Swersky}, \bibinfo{person}{Ziyu Wang}, \bibinfo{person}{Ryan~P Adams}, {and}
  \bibinfo{person}{Nando De~Freitas}.} \bibinfo{year}{2015}\natexlab{}.
\newblock \showarticletitle{Taking the human out of the loop: A review of
  Bayesian optimization}.
\newblock \bibinfo{journal}{\emph{Proc. IEEE}} \bibinfo{volume}{104},
  \bibinfo{number}{1} (\bibinfo{year}{2015}), \bibinfo{pages}{148--175}.
\newblock


\bibitem[\protect\citeauthoryear{SIGGRAPH}{SIGGRAPH}{2020}]%
        {SIGGRAPH:2020:SG}
\bibfield{author}{\bibinfo{person}{SIGGRAPH}.} \bibinfo{year}{2020}\natexlab{}.
\newblock \bibinfo{title}{Speaker Guidelines}.
\newblock
\newblock
\urldef\tempurl%
\url{https://sa2020.siggraph.org/images/pdfs/SA20_General_Speaker_Guidelines.pdf}
\showURL{%
\tempurl}


\bibitem[\protect\citeauthoryear{Slides}{Slides}{2019}]%
        {Tu:2019:HS}
\bibfield{author}{\bibinfo{person}{Google Slides}.}
  \bibinfo{year}{2019}\natexlab{}.
\newblock \bibinfo{title}{Hierarchical synthesis}.
\newblock
\newblock
\newblock
\shownote{\url{https://docs.google.com/presentation/d/190DLpB1W4OW58uSL3KM6WlsIP8ixG84oTv20mXyvlmA/edit?usp=sharing}.}


\bibitem[\protect\citeauthoryear{Stackexchange}{Stackexchange}{2011}]%
        {StackExchange:2011:HLI}
\bibfield{author}{\bibinfo{person}{Stackexchange}.}
  \bibinfo{year}{2011}\natexlab{}.
\newblock \bibinfo{title}{How to link images relatively in Inkscape?}
\newblock
\newblock
\newblock
\shownote{\url{http://graphicdesign.stackexchange.com/questions/4906/how-to-link-images-relatively-in-inkscape}.}


\bibitem[\protect\citeauthoryear{Suzuki, Yeh, Yatani, and Gross}{Suzuki
  et~al\mbox{.}}{2017}]%
        {Suzuki:2017:ATP}
\bibfield{author}{\bibinfo{person}{Ryo Suzuki}, \bibinfo{person}{Tom Yeh},
  \bibinfo{person}{Koji Yatani}, {and} \bibinfo{person}{Mark~D Gross}.}
  \bibinfo{year}{2017}\natexlab{}.
\newblock \showarticletitle{Autocomplete Textures for 3D Printing}.
\newblock \bibinfo{journal}{\emph{arXiv preprint arXiv:1703.05700}}
  (\bibinfo{year}{2017}).
\newblock


\bibitem[\protect\citeauthoryear{Takayama, Sorkine, Nealen, and
  Igarashi}{Takayama et~al\mbox{.}}{2010}]%
        {Takayama:2010:VMD}
\bibfield{author}{\bibinfo{person}{Kenshi Takayama}, \bibinfo{person}{Olga
  Sorkine}, \bibinfo{person}{Andrew Nealen}, {and} \bibinfo{person}{Takeo
  Igarashi}.} \bibinfo{year}{2010}\natexlab{}.
\newblock \showarticletitle{Volumetric Modeling with Diffusion Surfaces}. In
  \bibinfo{booktitle}{\emph{SIGGRAPH ASIA '10}}. Article
  \bibinfo{articleno}{Article 180}, \bibinfo{numpages}{8}~pages.
\newblock
\urldef\tempurl%
\url{https://doi.org/10.1145/1866158.1866202}
\showDOI{\tempurl}


\bibitem[\protect\citeauthoryear{Tu}{Tu}{2018a}]%
        {Tu:2018:PDPin}
\bibfield{author}{\bibinfo{person}{Peihan Tu}.}
  \bibinfo{year}{2018}\natexlab{a}.
\newblock \bibinfo{title}{Continuous Pattern Design}.
\newblock
\newblock
\urldef\tempurl%
\url{https://www.pinterest.com/peihantu/continuous-pattern-design/}
\showURL{%
\tempurl}


\bibitem[\protect\citeauthoryear{Tu}{Tu}{2018b}]%
        {Tu:2018:ETD}
\bibfield{author}{\bibinfo{person}{Peihan Tu}.}
  \bibinfo{year}{2018}\natexlab{b}.
\newblock \bibinfo{title}{Element Texture Design}.
\newblock
\newblock
\newblock
\shownote{\url{https://www.pinterest.com/peihantu/element-texture-design/}.}


\bibitem[\protect\citeauthoryear{Tu}{Tu}{2019a}]%
        {YouTube:2019:DIL}
\bibfield{author}{\bibinfo{person}{Peihan Tu}.}
  \bibinfo{year}{2019}\natexlab{a}.
\newblock \bibinfo{title}{Demo: interactive learning}.
\newblock
\newblock
\newblock
\shownote{\url{https://youtu.be/_-56PL46IlM}.}


\bibitem[\protect\citeauthoryear{Tu}{Tu}{2019b}]%
        {Tu:2019:EC:uYOVdHYjUOo}
\bibfield{author}{\bibinfo{person}{Peihan Tu}.}
  \bibinfo{year}{2019}\natexlab{b}.
\newblock \bibinfo{title}{Element Creation}.
\newblock
\newblock
\newblock
\shownote{\url{https://youtu.be/uYOVdHYjUOo}.}


\bibitem[\protect\citeauthoryear{Tu}{Tu}{2019c}]%
        {Tu:2019:WLP}
\bibfield{author}{\bibinfo{person}{Peihan Tu}.}
  \bibinfo{year}{2019}\natexlab{c}.
\newblock \bibinfo{title}{Wavy Lines Pattern}.
\newblock
\newblock
\newblock
\shownote{\url{https://www.pinterest.com/pin/640074165768185434/}.}


\bibitem[\protect\citeauthoryear{Tu}{Tu}{2020a}]%
        {Tu:2020:CCT:PP}
\bibfield{author}{\bibinfo{person}{Peihan Tu}.}
  \bibinfo{year}{2020}\natexlab{a}.
\newblock \bibinfo{title}{Continuous Curve Textures Project Page}.
\newblock
\newblock
\urldef\tempurl%
\url{https://tph9608.github.io/cct-siga20/}
\showURL{%
\tempurl}


\bibitem[\protect\citeauthoryear{Tu}{Tu}{2020b}]%
        {Tu:2020:CCTC}
\bibfield{author}{\bibinfo{person}{Peihan Tu}.}
  \bibinfo{year}{2020}\natexlab{b}.
\newblock \bibinfo{title}{Continuous Curve Textures Source Code}.
\newblock
\newblock
\newblock
\shownote{\url{https://github.com/tph9608/continuous-curve-texture/}.}


\bibitem[\protect\citeauthoryear{Tu}{Tu}{2020c}]%
        {Tu:2020:CCT:Talk}
\bibfield{author}{\bibinfo{person}{Peihan Tu}.}
  \bibinfo{year}{2020}\natexlab{c}.
\newblock \bibinfo{title}{Continuous Curve Textures Talk Slides}.
\newblock
\newblock
\urldef\tempurl%
\url{https://umd0-my.sharepoint.com/:p:/g/personal/phtu_umd_edu/EUHszU5tWK1PhSmuiQuiCsABVLhPNo1w_TrHHNkM_6942Q}
\showURL{%
\tempurl}


\bibitem[\protect\citeauthoryear{Tu}{Tu}{2020d}]%
        {Tu:2020:CCT:TalkVideo}
\bibfield{author}{\bibinfo{person}{Peihan Tu}.}
  \bibinfo{year}{2020}\natexlab{d}.
\newblock \bibinfo{title}{Continuous Curve Textures Talk Video}.
\newblock
\newblock
\urldef\tempurl%
\url{https://youtu.be/y1UvuqH9f8U}
\showURL{%
\tempurl}


\bibitem[\protect\citeauthoryear{Tu}{Tu}{2020e}]%
        {TU:2020:DPD}
\bibfield{author}{\bibinfo{person}{Peihan Tu}.}
  \bibinfo{year}{2020}\natexlab{e}.
\newblock \bibinfo{title}{Demo: Pattern Design}.
\newblock
\newblock
\newblock
\shownote{\url{https://youtu.be/meLok9-_Bt8}.}


\bibitem[\protect\citeauthoryear{Tu}{Tu}{2020f}]%
        {Tu:2020_03_17:IA}
\bibfield{author}{\bibinfo{person}{Peihan Tu}.}
  \bibinfo{year}{2020}\natexlab{f}.
\newblock \bibinfo{title}{Interactive authoring}.
\newblock
\newblock
\urldef\tempurl%
\url{https://youtu.be/0BUXjWND9iE}
\showURL{%
\tempurl}


\bibitem[\protect\citeauthoryear{Tu}{Tu}{2020g}]%
        {TU:2020:PVP}
\bibfield{author}{\bibinfo{person}{Peihan Tu}.}
  \bibinfo{year}{2020}\natexlab{g}.
\newblock \bibinfo{title}{Paper Video: Pattern Design}.
\newblock
\newblock
\newblock
\shownote{\url{https://youtu.be/w-cxRMel94c}.}


\bibitem[\protect\citeauthoryear{Tu}{Tu}{2020h}]%
        {Tu:2020_03_26:RCT}
\bibfield{author}{\bibinfo{person}{Peihan Tu}.}
  \bibinfo{year}{2020}\natexlab{h}.
\newblock \bibinfo{title}{Resynthesis, clone tool}.
\newblock
\newblock
\urldef\tempurl%
\url{https://youtu.be/7tAvYhJqFFQ}
\showURL{%
\tempurl}


\bibitem[\protect\citeauthoryear{Tu}{Tu}{2020i}]%
        {TU:2020:UMP}
\bibfield{author}{\bibinfo{person}{Peihan Tu}.}
  \bibinfo{year}{2020}\natexlab{i}.
\newblock \bibinfo{title}{User Manual: Pattern Design}.
\newblock
\newblock
\newblock
\shownote{\url{https://docs.google.com/document/d/1UKgxtHW3xXjTuJ-k0rfUpZluMB8AfFePxrh4UMQk1HE/edit?usp=sharing}.}


\bibitem[\protect\citeauthoryear{Tu, Lischinski, and Huang}{Tu
  et~al\mbox{.}}{2019}]%
        {Tu:2019:PPS}
\bibfield{author}{\bibinfo{person}{Peihan Tu}, \bibinfo{person}{Dani
  Lischinski}, {and} \bibinfo{person}{Hui Huang}.}
  \bibinfo{year}{2019}\natexlab{}.
\newblock \showarticletitle{Point Pattern Synthesis via Irregular Convolution}.
\newblock \bibinfo{journal}{\emph{Computer Graphics Forum}}
  \bibinfo{volume}{38}, \bibinfo{number}{5} (\bibinfo{year}{2019}),
  \bibinfo{pages}{109--122}.
\newblock
\urldef\tempurl%
\url{https://doi.org/10.1111/cgf.13793}
\showDOI{\tempurl}


\bibitem[\protect\citeauthoryear{Tu and Wei}{Tu and Wei}{2018}]%
        {Tu:2018:PDPinWC}
\bibfield{author}{\bibinfo{person}{Peihan Tu} {and} \bibinfo{person}{Li-Yi
  Wei}.} \bibinfo{year}{2018}\natexlab{}.
\newblock \bibinfo{title}{Continuous Pattern Design: Workflow clone}.
\newblock
\newblock
\urldef\tempurl%
\url{https://www.pinterest.com/peihantu/continuous-pattern-design/workflow-clone/}
\showURL{%
\tempurl}


\bibitem[\protect\citeauthoryear{Tu and Wei}{Tu and Wei}{2019}]%
        {Tu:2019:CPD:OS}
\bibfield{author}{\bibinfo{person}{Peihan Tu} {and} \bibinfo{person}{Li-Yi
  Wei}.} \bibinfo{year}{2019}\natexlab{}.
\newblock \bibinfo{title}{Continuous Pattern Design - our scope}.
\newblock
\newblock
\urldef\tempurl%
\url{https://www.pinterest.com/peihantu/continuous-pattern-design/our-scope/}
\showURL{%
\tempurl}


\bibitem[\protect\citeauthoryear{Tu, Wei, Yatani, Igarashi, and Zwicker}{Tu
  et~al\mbox{.}}{2020}]%
        {Tu:2020:CCT}
\bibfield{author}{\bibinfo{person}{Peihan Tu}, \bibinfo{person}{Li-Yi Wei},
  \bibinfo{person}{Koji Yatani}, \bibinfo{person}{Takeo Igarashi}, {and}
  \bibinfo{person}{Matthias Zwicker}.} \bibinfo{year}{2020}\natexlab{}.
\newblock \showarticletitle{Continuous Curve Textures}.
\newblock \bibinfo{journal}{\emph{ACM Trans. Graph.}} (\bibinfo{year}{2020}).
\newblock


\bibitem[\protect\citeauthoryear{Turk and Banks}{Turk and Banks}{1996}]%
        {Turk:1996:ISP}
\bibfield{author}{\bibinfo{person}{Greg Turk} {and} \bibinfo{person}{David
  Banks}.} \bibinfo{year}{1996}\natexlab{}.
\newblock \showarticletitle{Image-Guided Streamline Placement}. In
  \bibinfo{booktitle}{\emph{SIGGRAPH '96}}. \bibinfo{pages}{453--460}.
\newblock
\showISBNx{0897917464}
\urldef\tempurl%
\url{https://doi.org/10.1145/237170.237285}
\showDOI{\tempurl}


\bibitem[\protect\citeauthoryear{Van~Kaick, Zhang, Hamarneh, and
  Cohen-Or}{Van~Kaick et~al\mbox{.}}{2011}]%
        {Van:2011:SSC}
\bibfield{author}{\bibinfo{person}{Oliver Van~Kaick}, \bibinfo{person}{Hao
  Zhang}, \bibinfo{person}{Ghassan Hamarneh}, {and} \bibinfo{person}{Daniel
  Cohen-Or}.} \bibinfo{year}{2011}\natexlab{}.
\newblock \showarticletitle{A survey on shape correspondence}. In
  \bibinfo{booktitle}{\emph{Computer Graphics Forum}},
  Vol.~\bibinfo{volume}{30}. Wiley Online Library, \bibinfo{pages}{1681--1707}.
\newblock


\bibitem[\protect\citeauthoryear{Wang, Yu, Zhou, and Guo}{Wang
  et~al\mbox{.}}{2011}]%
        {Wang:2011:MVV}
\bibfield{author}{\bibinfo{person}{Lvdi Wang}, \bibinfo{person}{Yizhou Yu},
  \bibinfo{person}{Kun Zhou}, {and} \bibinfo{person}{Baining Guo}.}
  \bibinfo{year}{2011}\natexlab{}.
\newblock \showarticletitle{Multiscale vector volumes}.
\newblock \bibinfo{journal}{\emph{ACM Transactions on Graphics (TOG)}}
  \bibinfo{volume}{30}, \bibinfo{number}{6} (\bibinfo{year}{2011}),
  \bibinfo{pages}{1--8}.
\newblock


\bibitem[\protect\citeauthoryear{Wang, Zhou, Yu, and Guo}{Wang
  et~al\mbox{.}}{2010}]%
        {Wang:2010:VST}
\bibfield{author}{\bibinfo{person}{Lvdi Wang}, \bibinfo{person}{Kun Zhou},
  \bibinfo{person}{Yizhou Yu}, {and} \bibinfo{person}{Baining Guo}.}
  \bibinfo{year}{2010}\natexlab{}.
\newblock \showarticletitle{Vector solid textures}.
\newblock \bibinfo{journal}{\emph{ACM Transactions on Graphics (TOG)}}
  \bibinfo{volume}{29}, \bibinfo{number}{4} (\bibinfo{year}{2010}),
  \bibinfo{pages}{1--8}.
\newblock


\bibitem[\protect\citeauthoryear{Wang, Liu, Zhu, Tao, Kautz, and
  Catanzaro}{Wang et~al\mbox{.}}{2017}]%
        {Wang:2017:HIS}
\bibfield{author}{\bibinfo{person}{Ting-Chun Wang}, \bibinfo{person}{Ming-Yu
  Liu}, \bibinfo{person}{Jun-Yan Zhu}, \bibinfo{person}{Andrew Tao},
  \bibinfo{person}{Jan Kautz}, {and} \bibinfo{person}{Bryan Catanzaro}.}
  \bibinfo{year}{2017}\natexlab{}.
\newblock \showarticletitle{High-Resolution Image Synthesis and Semantic
  Manipulation with Conditional GANs}.
\newblock \bibinfo{journal}{\emph{arXiv preprint arXiv:1711.11585}}
  (\bibinfo{year}{2017}).
\newblock


\bibitem[\protect\citeauthoryear{Wang, Du, Balakrishnan, and Singh}{Wang
  et~al\mbox{.}}{2018}]%
        {Wang:2018:SZO}
\bibfield{author}{\bibinfo{person}{Yining Wang}, \bibinfo{person}{Simon Du},
  \bibinfo{person}{Sivaraman Balakrishnan}, {and} \bibinfo{person}{Aarti
  Singh}.} \bibinfo{year}{2018}\natexlab{}.
\newblock \showarticletitle{Stochastic Zeroth-order Optimization in High
  Dimensions}. In \bibinfo{booktitle}{\emph{International Conference on
  Artificial Intelligence and Statistics}}. \bibinfo{pages}{1356--1365}.
\newblock


\bibitem[\protect\citeauthoryear{Wei}{Wei}{2002}]%
        {Wei:2002:TSF}
\bibfield{author}{\bibinfo{person}{Li-Yi Wei}.}
  \bibinfo{year}{2002}\natexlab{}.
\newblock \emph{\bibinfo{title}{Texture Synthesis by Fixed Neighborhood
  Searching}}.
\newblock \bibinfo{thesistype}{Ph.D. Dissertation}. \bibinfo{address}{Stanford,
  CA, USA}.
\newblock Advisor(s) Levoy, Marc.
\newblock
\showISBNx{0-493-52003-1}
\newblock
\shownote{AAI3038169.}


\bibitem[\protect\citeauthoryear{Wei}{Wei}{2008}]%
        {Wei:2008:PPD}
\bibfield{author}{\bibinfo{person}{Li-Yi Wei}.}
  \bibinfo{year}{2008}\natexlab{}.
\newblock \showarticletitle{Parallel Poisson Disk Sampling}.
\newblock \bibinfo{journal}{\emph{ACM Trans. Graph.}} \bibinfo{volume}{27},
  \bibinfo{number}{3}, Article \bibinfo{articleno}{20} (\bibinfo{date}{Aug.}
  \bibinfo{year}{2008}), \bibinfo{numpages}{9}~pages.
\newblock
\showISSN{0730-0301}
\urldef\tempurl%
\url{https://doi.org/10.1145/1360612.1360619}
\showDOI{\tempurl}


\bibitem[\protect\citeauthoryear{Wei}{Wei}{2010}]%
        {Wei:2010:HDP}
\bibfield{author}{\bibinfo{person}{Li-Yi Wei}.}
  \bibinfo{year}{2010}\natexlab{}.
\newblock \bibinfo{title}{How to deal with paper deadlines}.
\newblock
\newblock
\newblock
\shownote{\url{https://blog.liyiwei.org/?p=264}.}


\bibitem[\protect\citeauthoryear{Wei}{Wei}{2012}]%
        {Wei:2012:RIS}
\bibfield{author}{\bibinfo{person}{Li-Yi Wei}.}
  \bibinfo{year}{2012}\natexlab{}.
\newblock \bibinfo{title}{Representative image for SIGGRAPH submission}.
\newblock
\newblock
\newblock
\shownote{\url{https://blog.liyiwei.org/?p=546}.}


\bibitem[\protect\citeauthoryear{Wei}{Wei}{2013}]%
        {Wei:2013:PFF}
\bibfield{author}{\bibinfo{person}{Li-Yi Wei}.}
  \bibinfo{year}{2013}\natexlab{}.
\newblock \bibinfo{title}{How to do a paper fast-forward}.
\newblock
\newblock
\urldef\tempurl%
\url{https://blog.liyiwei.org/?p=1242}
\showURL{%
\tempurl}


\bibitem[\protect\citeauthoryear{Wei}{Wei}{2016}]%
        {Wei:2016:TS}
\bibfield{author}{\bibinfo{person}{Li-Yi Wei}.}
  \bibinfo{year}{2016}\natexlab{}.
\newblock \bibinfo{title}{Texture Synthesis}.
\newblock
\newblock
\urldef\tempurl%
\url{https://github.com/1iyiwei/texture}
\showURL{%
\tempurl}


\bibitem[\protect\citeauthoryear{Wei}{Wei}{2017}]%
        {Wei:2017:MRM}
\bibfield{author}{\bibinfo{person}{Li-Yi Wei}.}
  \bibinfo{year}{2017}\natexlab{}.
\newblock \bibinfo{title}{Sharing and managing research materials}.
\newblock
\newblock
\newblock
\shownote{\url{https://blog.liyiwei.org/?p=2605}.}


\bibitem[\protect\citeauthoryear{Wei}{Wei}{2020a}]%
        {Wei:2020:SPS}
\bibfield{author}{\bibinfo{person}{Li-Yi Wei}.}
  \bibinfo{year}{2020}\natexlab{a}.
\newblock \bibinfo{title}{Sharing paper source with publisher}.
\newblock
\newblock
\newblock
\shownote{\url{https://blog.liyiwei.org/?p=4022}.}


\bibitem[\protect\citeauthoryear{Wei}{Wei}{2020b}]%
        {Wei:2020:SMR}
\bibfield{author}{\bibinfo{person}{Li-Yi Wei}.}
  \bibinfo{year}{2020}\natexlab{b}.
\newblock \bibinfo{title}{Simple Methods to Represent Shapes with Sample
  Spheres}.
\newblock
\newblock
\urldef\tempurl%
\url{https://adobe.my.salesforce.com/0691O00000G7Lch}
\showURL{%
\tempurl}


\bibitem[\protect\citeauthoryear{Wei, Han, Zhou, Bao, Guo, and Shum}{Wei
  et~al\mbox{.}}{2008}]%
        {Wei:2008:ITS}
\bibfield{author}{\bibinfo{person}{Li-Yi Wei}, \bibinfo{person}{Jianwei Han},
  \bibinfo{person}{Kun Zhou}, \bibinfo{person}{Hujun Bao},
  \bibinfo{person}{Baining Guo}, {and} \bibinfo{person}{Heung-Yeung Shum}.}
  \bibinfo{year}{2008}\natexlab{}.
\newblock \showarticletitle{Inverse Texture Synthesis}.
\newblock \bibinfo{journal}{\emph{ACM Trans. Graph.}} \bibinfo{volume}{27},
  \bibinfo{number}{3}, Article \bibinfo{articleno}{52} (\bibinfo{date}{Aug.}
  \bibinfo{year}{2008}), \bibinfo{numpages}{9}~pages.
\newblock
\showISSN{0730-0301}
\urldef\tempurl%
\url{https://doi.org/10.1145/1360612.1360651}
\showDOI{\tempurl}


\bibitem[\protect\citeauthoryear{Wei, Lefebvre, Kwatra, and Turk}{Wei
  et~al\mbox{.}}{2009}]%
        {Wei:2009:SAE}
\bibfield{author}{\bibinfo{person}{Li-Yi Wei}, \bibinfo{person}{Sylvain
  Lefebvre}, \bibinfo{person}{Vivek Kwatra}, {and} \bibinfo{person}{Greg
  Turk}.} \bibinfo{year}{2009}\natexlab{}.
\newblock \showarticletitle{State of the Art in Example-based Texture
  Synthesis}. In \bibinfo{booktitle}{\emph{Eurographics 2009, State of the Art
  Report, EG-STAR}}. \bibinfo{publisher}{Eurographics Association}.
\newblock
\urldef\tempurl%
\url{http://www-sop.inria.fr/reves/Basilic/2009/WLKT09}
\showURL{%
\tempurl}


\bibitem[\protect\citeauthoryear{Wei and Levoy}{Wei and Levoy}{2000}]%
        {Wei:2000:FTS}
\bibfield{author}{\bibinfo{person}{Li-Yi Wei} {and} \bibinfo{person}{Marc
  Levoy}.} \bibinfo{year}{2000}\natexlab{}.
\newblock \showarticletitle{Fast Texture Synthesis Using Tree-structured Vector
  Quantization}. In \bibinfo{booktitle}{\emph{SIGGRAPH '00}}.
  \bibinfo{pages}{479--488}.
\newblock
\showISBNx{1-58113-208-5}
\urldef\tempurl%
\url{https://doi.org/10.1145/344779.345009}
\showDOI{\tempurl}


\bibitem[\protect\citeauthoryear{Wei and Levoy}{Wei and Levoy}{2001}]%
        {Wei:2001:TSO}
\bibfield{author}{\bibinfo{person}{Li-Yi Wei} {and} \bibinfo{person}{Marc
  Levoy}.} \bibinfo{year}{2001}\natexlab{}.
\newblock \showarticletitle{Texture Synthesis over Arbitrary Manifold
  Surfaces}. In \bibinfo{booktitle}{\emph{SIGGRAPH '01}}.
  \bibinfo{pages}{355--360}.
\newblock
\showISBNx{1-58113-374-X}
\urldef\tempurl%
\url{https://doi.org/10.1145/383259.383298}
\showDOI{\tempurl}


\bibitem[\protect\citeauthoryear{Wikipedia}{Wikipedia}{2019a}]%
        {Wiki:2019:GSS}
\bibfield{author}{\bibinfo{person}{Wikipedia}.}
  \bibinfo{year}{2019}\natexlab{a}.
\newblock \bibinfo{title}{Golden section search}.
\newblock
\newblock
\newblock
\shownote{\url{https://en.wikipedia.org/wiki/Golden-section_search}.}


\bibitem[\protect\citeauthoryear{Wikipedia}{Wikipedia}{2019b}]%
        {Wikipedia:2019:GM}
\bibfield{author}{\bibinfo{person}{Wikipedia}.}
  \bibinfo{year}{2019}\natexlab{b}.
\newblock \bibinfo{title}{Graph matching}.
\newblock
\newblock
\newblock
\shownote{\url{https://en.wikipedia.org/wiki/Graph_matching}.}


\bibitem[\protect\citeauthoryear{Wikipedia}{Wikipedia}{2019c}]%
        {Wikipedia:2019:PG}
\bibfield{author}{\bibinfo{person}{Wikipedia}.}
  \bibinfo{year}{2019}\natexlab{c}.
\newblock \bibinfo{title}{Planar Graph}.
\newblock
\newblock
\newblock
\shownote{\url{https://en.wikipedia.org/wiki/Planar_graph}.}


\bibitem[\protect\citeauthoryear{Xian, Sangkloy, Lu, Fang, Yu, and Hays}{Xian
  et~al\mbox{.}}{2018}]%
        {Xian:2018:TGC}
\bibfield{author}{\bibinfo{person}{Wenqi Xian}, \bibinfo{person}{Patsorn
  Sangkloy}, \bibinfo{person}{Jingwan Lu}, \bibinfo{person}{Chen Fang},
  \bibinfo{person}{Fisher Yu}, {and} \bibinfo{person}{James Hays}.}
  \bibinfo{year}{2018}\natexlab{}.
\newblock \showarticletitle{Texturegan: Controlling deep image synthesis with
  texture patches}. In \bibinfo{booktitle}{\emph{CVPR '18}}.
\newblock


\bibitem[\protect\citeauthoryear{Xing, Chen, and Wei}{Xing
  et~al\mbox{.}}{2014}]%
        {Xing:2014:APR}
\bibfield{author}{\bibinfo{person}{Jun Xing}, \bibinfo{person}{Hsiang-Ting
  Chen}, {and} \bibinfo{person}{Li-Yi Wei}.} \bibinfo{year}{2014}\natexlab{}.
\newblock \showarticletitle{Autocomplete Painting Repetitions}.
\newblock \bibinfo{journal}{\emph{ACM Trans. Graph.}} \bibinfo{volume}{33},
  \bibinfo{number}{6}, Article \bibinfo{articleno}{172} (\bibinfo{date}{Nov.}
  \bibinfo{year}{2014}), \bibinfo{numpages}{11}~pages.
\newblock
\showISSN{0730-0301}
\urldef\tempurl%
\url{https://doi.org/10.1145/2661229.2661247}
\showDOI{\tempurl}


\bibitem[\protect\citeauthoryear{Xing, Kazi, Grossman, Wei, Stam, and
  Fitzmaurice}{Xing et~al\mbox{.}}{2016}]%
        {Xing:2016:EBI}
\bibfield{author}{\bibinfo{person}{Jun Xing}, \bibinfo{person}{Rubaiat~Habib
  Kazi}, \bibinfo{person}{Tovi Grossman}, \bibinfo{person}{Li-Yi Wei},
  \bibinfo{person}{Jos Stam}, {and} \bibinfo{person}{George Fitzmaurice}.}
  \bibinfo{year}{2016}\natexlab{}.
\newblock \showarticletitle{Energy-Brushes: Interactive Tools for Illustrating
  Stylized Elemental Dynamics}. In \bibinfo{booktitle}{\emph{Proceedings of the
  29th Annual Symposium on User Interface Software and Technology}}
  \emph{(\bibinfo{series}{UIST ’16})}. \bibinfo{publisher}{Association for
  Computing Machinery}, \bibinfo{address}{New York, NY, USA},
  \bibinfo{pages}{755–766}.
\newblock
\showISBNx{9781450341899}
\urldef\tempurl%
\url{https://doi.org/10.1145/2984511.2984585}
\showDOI{\tempurl}


\bibitem[\protect\citeauthoryear{Xing, Wei, Shiratori, and Yatani}{Xing
  et~al\mbox{.}}{2015}]%
        {Xing:2015:AHA}
\bibfield{author}{\bibinfo{person}{Jun Xing}, \bibinfo{person}{Li-Yi Wei},
  \bibinfo{person}{Takaaki Shiratori}, {and} \bibinfo{person}{Koji Yatani}.}
  \bibinfo{year}{2015}\natexlab{}.
\newblock \showarticletitle{Autocomplete Hand-drawn Animations}.
\newblock \bibinfo{journal}{\emph{ACM Trans. Graph.}} \bibinfo{volume}{34},
  \bibinfo{number}{6}, Article \bibinfo{articleno}{169} (\bibinfo{date}{Oct.}
  \bibinfo{year}{2015}), \bibinfo{numpages}{11}~pages.
\newblock
\showISSN{0730-0301}
\urldef\tempurl%
\url{https://doi.org/10.1145/2816795.2818079}
\showDOI{\tempurl}


\bibitem[\protect\citeauthoryear{Xu, Yan, Xia, and Jia}{Xu
  et~al\mbox{.}}{2012}]%
        {Xu:2012:SET}
\bibfield{author}{\bibinfo{person}{Li Xu}, \bibinfo{person}{Qiong Yan},
  \bibinfo{person}{Yang Xia}, {and} \bibinfo{person}{Jiaya Jia}.}
  \bibinfo{year}{2012}\natexlab{}.
\newblock \showarticletitle{Structure Extraction from Texture via Natural
  Variation Measure}.
\newblock \bibinfo{journal}{\emph{ACM Transactions on Graphics (SIGGRAPH
  Asia)}} (\bibinfo{year}{2012}).
\newblock


\bibitem[\protect\citeauthoryear{You, Ying, Ren, Hamilton, and Leskovec}{You
  et~al\mbox{.}}{2018}]%
        {You:2018:GRG}
\bibfield{author}{\bibinfo{person}{Jiaxuan You}, \bibinfo{person}{Rex Ying},
  \bibinfo{person}{Xiang Ren}, \bibinfo{person}{William~L Hamilton}, {and}
  \bibinfo{person}{Jure Leskovec}.} \bibinfo{year}{2018}\natexlab{}.
\newblock \showarticletitle{Graphrnn: Generating realistic graphs with deep
  auto-regressive models}.
\newblock \bibinfo{journal}{\emph{arXiv preprint arXiv:1802.08773}}
  (\bibinfo{year}{2018}).
\newblock


\bibitem[\protect\citeauthoryear{YouTube}{YouTube}{2019a}]%
        {Tu:2019:FS}
\bibfield{author}{\bibinfo{person}{YouTube}.} \bibinfo{year}{2019}\natexlab{a}.
\newblock \bibinfo{title}{Fast Synthesis}.
\newblock
\newblock
\newblock
\shownote{\url{https://youtu.be/MqyEm2-F0Go}.}


\bibitem[\protect\citeauthoryear{YouTube}{YouTube}{2019b}]%
        {YouTube:2019:PDV1}
\bibfield{author}{\bibinfo{person}{YouTube}.} \bibinfo{year}{2019}\natexlab{b}.
\newblock \bibinfo{title}{Pattern Design Video 1}.
\newblock
\newblock
\newblock
\shownote{\url{https://youtu.be/XR8Q6mE9V_o}.}


\bibitem[\protect\citeauthoryear{YouTube}{YouTube}{2019c}]%
        {YouTube:2019:PDV2}
\bibfield{author}{\bibinfo{person}{YouTube}.} \bibinfo{year}{2019}\natexlab{c}.
\newblock \bibinfo{title}{Pattern Design Video 2}.
\newblock
\newblock
\newblock
\shownote{\url{https://youtu.be/ITRZ75OKrG0}.}


\bibitem[\protect\citeauthoryear{YouTube}{YouTube}{2019d}]%
        {YouTube:2019:PDV3}
\bibfield{author}{\bibinfo{person}{YouTube}.} \bibinfo{year}{2019}\natexlab{d}.
\newblock \bibinfo{title}{Pattern Design Video 3}.
\newblock
\newblock
\newblock
\shownote{\url{https://youtu.be/KEXg-4RnGRY}.}


\bibitem[\protect\citeauthoryear{YouTube}{YouTube}{2019e}]%
        {YouTube:2019:PDV4}
\bibfield{author}{\bibinfo{person}{YouTube}.} \bibinfo{year}{2019}\natexlab{e}.
\newblock \bibinfo{title}{Pattern Design Video 4}.
\newblock
\newblock
\newblock
\shownote{\url{https://youtu.be/mPNB0y0sY3Y}.}


\bibitem[\protect\citeauthoryear{YouTube}{YouTube}{2020}]%
        {YouTube:2020:MEM}
\bibfield{author}{\bibinfo{person}{YouTube}.} \bibinfo{year}{2020}\natexlab{}.
\newblock \bibinfo{title}{VJ/DJ Music equalizer moving Motion}.
\newblock
\newblock
\newblock
\shownote{\url{https://youtu.be/c3abfWsQk-M}.}


\bibitem[\protect\citeauthoryear{Zehnder, Coros, and Thomaszewski}{Zehnder
  et~al\mbox{.}}{2016}]%
        {Zehnder:2016:DSO}
\bibfield{author}{\bibinfo{person}{Jonas Zehnder}, \bibinfo{person}{Stelian
  Coros}, {and} \bibinfo{person}{Bernhard Thomaszewski}.}
  \bibinfo{year}{2016}\natexlab{}.
\newblock \showarticletitle{Designing Structurally-sound Ornamental Curve
  Networks}.
\newblock \bibinfo{journal}{\emph{ACM Trans. Graph.}} \bibinfo{volume}{35},
  \bibinfo{number}{4}, Article \bibinfo{articleno}{99} (\bibinfo{date}{July}
  \bibinfo{year}{2016}), \bibinfo{numpages}{10}~pages.
\newblock
\showISSN{0730-0301}
\urldef\tempurl%
\url{https://doi.org/10.1145/2897824.2925888}
\showDOI{\tempurl}


\bibitem[\protect\citeauthoryear{Zhang, Zhu, Isola, Geng, Lin, Yu, and
  Efros}{Zhang et~al\mbox{.}}{2017}]%
        {Zhang:2017:RTU}
\bibfield{author}{\bibinfo{person}{Richard Zhang}, \bibinfo{person}{Jun-Yan
  Zhu}, \bibinfo{person}{Phillip Isola}, \bibinfo{person}{Xinyang Geng},
  \bibinfo{person}{Angela~S Lin}, \bibinfo{person}{Tianhe Yu}, {and}
  \bibinfo{person}{Alexei~A Efros}.} \bibinfo{year}{2017}\natexlab{}.
\newblock \showarticletitle{Real-Time User-Guided Image Colorization with
  Learned Deep Priors}.
\newblock \bibinfo{journal}{\emph{ACM Transactions on Graphics (TOG)}}
  \bibinfo{volume}{9}, \bibinfo{number}{4} (\bibinfo{year}{2017}).
\newblock


\bibitem[\protect\citeauthoryear{Zheng, Chen, Cheng, Zhou, Hu, and Mitra}{Zheng
  et~al\mbox{.}}{2012}]%
        {Zheng:2012:IIC}
\bibfield{author}{\bibinfo{person}{Youyi Zheng}, \bibinfo{person}{Xiang Chen},
  \bibinfo{person}{Ming-Ming Cheng}, \bibinfo{person}{Kun Zhou},
  \bibinfo{person}{Shi-Min Hu}, {and} \bibinfo{person}{Niloy~J. Mitra}.}
  \bibinfo{year}{2012}\natexlab{}.
\newblock \showarticletitle{Interactive Images: Cuboid Proxies for Smart Image
  Manipulation}.
\newblock \bibinfo{journal}{\emph{ACM Transactions on Graphics}}
  \bibinfo{volume}{31}, \bibinfo{number}{4}, Article \bibinfo{articleno}{99}
  (\bibinfo{year}{2012}), \bibinfo{numpages}{11}~pages.
\newblock


\bibitem[\protect\citeauthoryear{Zhou, Sun, Turk, and Rehg}{Zhou
  et~al\mbox{.}}{2007}]%
        {Zhou:2007:TSD}
\bibfield{author}{\bibinfo{person}{Howard Zhou}, \bibinfo{person}{Jie Sun},
  \bibinfo{person}{Greg Turk}, {and} \bibinfo{person}{James~M. Rehg}.}
  \bibinfo{year}{2007}\natexlab{}.
\newblock \showarticletitle{Terrain Synthesis from Digital Elevation Models}.
\newblock \bibinfo{journal}{\emph{IEEE Transactions on Visualization and
  Computer Graphics}} \bibinfo{volume}{13}, \bibinfo{number}{4}
  (\bibinfo{date}{July} \bibinfo{year}{2007}), \bibinfo{pages}{834–848}.
\newblock
\showISSN{1077-2626}
\urldef\tempurl%
\url{https://doi.org/10.1109/TVCG.2007.1027}
\showDOI{\tempurl}


\bibitem[\protect\citeauthoryear{Zhou, Huang, Wang, Tong, Desbrun, Guo, and
  Shum}{Zhou et~al\mbox{.}}{2006}]%
        {Zhou:2006:MQG}
\bibfield{author}{\bibinfo{person}{Kun Zhou}, \bibinfo{person}{Xin Huang},
  \bibinfo{person}{Xi Wang}, \bibinfo{person}{Yiying Tong},
  \bibinfo{person}{Mathieu Desbrun}, \bibinfo{person}{Baining Guo}, {and}
  \bibinfo{person}{Heung-Yeung Shum}.} \bibinfo{year}{2006}\natexlab{}.
\newblock \showarticletitle{Mesh Quilting for Geometric Texture Synthesis}.
\newblock \bibinfo{journal}{\emph{ACM Trans. Graph.}} \bibinfo{volume}{25},
  \bibinfo{number}{3} (\bibinfo{date}{July} \bibinfo{year}{2006}),
  \bibinfo{pages}{690--697}.
\newblock
\showISSN{0730-0301}
\urldef\tempurl%
\url{https://doi.org/10.1145/1141911.1141942}
\showDOI{\tempurl}


\bibitem[\protect\citeauthoryear{Zhou, Jiang, and Lefebvre}{Zhou
  et~al\mbox{.}}{2014}]%
        {Zhou:2014:TSV}
\bibfield{author}{\bibinfo{person}{Shizhe Zhou}, \bibinfo{person}{Changyun
  Jiang}, {and} \bibinfo{person}{Sylvain Lefebvre}.}
  \bibinfo{year}{2014}\natexlab{}.
\newblock \showarticletitle{Topology-constrained Synthesis of Vector Patterns}.
\newblock \bibinfo{journal}{\emph{ACM Trans. Graph.}} \bibinfo{volume}{33},
  \bibinfo{number}{6}, Article \bibinfo{articleno}{215} (\bibinfo{date}{Nov.}
  \bibinfo{year}{2014}), \bibinfo{numpages}{11}~pages.
\newblock
\showISSN{0730-0301}
\urldef\tempurl%
\url{https://doi.org/10.1145/2661229.2661238}
\showDOI{\tempurl}


\bibitem[\protect\citeauthoryear{Zitnick}{Zitnick}{2013}]%
        {Zitnick:2013:HBU}
\bibfield{author}{\bibinfo{person}{C.~Lawrence Zitnick}.}
  \bibinfo{year}{2013}\natexlab{}.
\newblock \showarticletitle{Handwriting Beautification Using Token Means}.
\newblock \bibinfo{journal}{\emph{ACM Trans. Graph.}} \bibinfo{volume}{32},
  \bibinfo{number}{4}, Article \bibinfo{articleno}{53} (\bibinfo{date}{July}
  \bibinfo{year}{2013}), \bibinfo{numpages}{8}~pages.
\newblock
\showISSN{0730-0301}
\urldef\tempurl%
\url{https://doi.org/10.1145/2461912.2461985}
\showDOI{\tempurl}


\bibitem[\protect\citeauthoryear{Zsolnai-Feher, Wonka, and
  Wimmer}{Zsolnai-Feher et~al\mbox{.}}{2018}]%
        {Zsolnai:2018:GMS}
\bibfield{author}{\bibinfo{person}{Karoly Zsolnai-Feher},
  \bibinfo{person}{Peter Wonka}, {and} \bibinfo{person}{Michael Wimmer}.}
  \bibinfo{year}{2018}\natexlab{}.
\newblock \showarticletitle{Gaussian Material Synthesis}.
\newblock \bibinfo{journal}{\emph{ACM Trans. Graph.}} \bibinfo{volume}{37},
  \bibinfo{number}{4} (\bibinfo{date}{Aug.} \bibinfo{year}{2018}).
\newblock


\end{thebibliography}


%%% -*-BibTeX-*-
%%% Do NOT edit. File created by BibTeX with style
%%% ACM-Reference-Format-Journals [18-Jan-2012].

\begin{thebibliography}{59}

%%% ====================================================================
%%% NOTE TO THE USER: you can override these defaults by providing
%%% customized versions of any of these macros before the \bibliography
%%% command.  Each of them MUST provide its own final punctuation,
%%% except for \shownote{}, \showDOI{}, and \showURL{}.  The latter two
%%% do not use final punctuation, in order to avoid confusing it with
%%% the Web address.
%%%
%%% To suppress output of a particular field, define its macro to expand
%%% to an empty string, or better, \unskip, like this:
%%%
%%% \newcommand{\showDOI}[1]{\unskip}   % LaTeX syntax
%%%
%%% \def \showDOI #1{\unskip}           % plain TeX syntax
%%%
%%% ====================================================================

\ifx \showCODEN    \undefined \def \showCODEN     #1{\unskip}     \fi
\ifx \showDOI      \undefined \def \showDOI       #1{#1}\fi
\ifx \showISBNx    \undefined \def \showISBNx     #1{\unskip}     \fi
\ifx \showISBNxiii \undefined \def \showISBNxiii  #1{\unskip}     \fi
\ifx \showISSN     \undefined \def \showISSN      #1{\unskip}     \fi
\ifx \showLCCN     \undefined \def \showLCCN      #1{\unskip}     \fi
\ifx \shownote     \undefined \def \shownote      #1{#1}          \fi
\ifx \showarticletitle \undefined \def \showarticletitle #1{#1}   \fi
\ifx \showURL      \undefined \def \showURL       {\relax}        \fi
% The following commands are used for tagged output and should be
% invisible to TeX
\providecommand\bibfield[2]{#2}
\providecommand\bibinfo[2]{#2}
\providecommand\natexlab[1]{#1}
\providecommand\showeprint[2][]{arXiv:#2}

\bibitem[\protect\citeauthoryear{Barla, Breslav, Thollot, Sillion, and
  Markosian}{Barla et~al\mbox{.}}{2006}]%
        {Barla:2006:SPA}
\bibfield{author}{\bibinfo{person}{Pascal Barla}, \bibinfo{person}{Simon
  Breslav}, \bibinfo{person}{Jo{\"e}lle Thollot},
  \bibinfo{person}{Fran{\c{c}}ois Sillion}, {and} \bibinfo{person}{Lee
  Markosian}.} \bibinfo{year}{2006}\natexlab{}.
\newblock \showarticletitle{Stroke pattern analysis and synthesis}. In
  \bibinfo{booktitle}{\emph{Computer Graphics Forum}},
  Vol.~\bibinfo{volume}{25}. Wiley Online Library, \bibinfo{pages}{663--671}.
\newblock


\bibitem[\protect\citeauthoryear{Barnes, Shechtman, Finkelstein, and
  Goldman}{Barnes et~al\mbox{.}}{2009}]%
        {Barnes:2009:PRC}
\bibfield{author}{\bibinfo{person}{Connelly Barnes}, \bibinfo{person}{Eli
  Shechtman}, \bibinfo{person}{Adam Finkelstein}, {and} \bibinfo{person}{Dan~B
  Goldman}.} \bibinfo{year}{2009}\natexlab{}.
\newblock \showarticletitle{PatchMatch: A Randomized Correspondence Algorithm
  for Structural Image Editing}.
\newblock \bibinfo{journal}{\emph{ACM Trans. Graph.}} \bibinfo{volume}{28},
  \bibinfo{number}{3}, Article \bibinfo{articleno}{24} (\bibinfo{date}{July}
  \bibinfo{year}{2009}), \bibinfo{numpages}{11}~pages.
\newblock
\showISSN{0730-0301}
\urldef\tempurl%
\url{https://doi.org/10.1145/1531326.1531330}
\showDOI{\tempurl}


\bibitem[\protect\citeauthoryear{Bhat, Ingram, and Turk}{Bhat
  et~al\mbox{.}}{2004}]%
        {Bhat:2004:GTS}
\bibfield{author}{\bibinfo{person}{Pravin Bhat}, \bibinfo{person}{Stephen
  Ingram}, {and} \bibinfo{person}{Greg Turk}.} \bibinfo{year}{2004}\natexlab{}.
\newblock \showarticletitle{Geometric texture synthesis by example}. In
  \bibinfo{booktitle}{\emph{SGP '04}}. \bibinfo{pages}{41--44}.
\newblock


\bibitem[\protect\citeauthoryear{Bian, Wei, and Lefebvre}{Bian
  et~al\mbox{.}}{2018}]%
        {Bian:2018:TPD}
\bibfield{author}{\bibinfo{person}{Xiaojun Bian}, \bibinfo{person}{Li-Yi Wei},
  {and} \bibinfo{person}{Sylvain Lefebvre}.} \bibinfo{year}{2018}\natexlab{}.
\newblock \showarticletitle{Tile-based Pattern Design with Topology Control}.
\newblock \bibinfo{journal}{\emph{Proc. ACM Comput. Graph. Interact. Tech.}}
  \bibinfo{volume}{1}, \bibinfo{number}{1}, Article \bibinfo{articleno}{23}
  (\bibinfo{date}{July} \bibinfo{year}{2018}), \bibinfo{numpages}{15}~pages.
\newblock
\showISSN{2577-6193}
\urldef\tempurl%
\url{https://doi.org/10.1145/3203204}
\showDOI{\tempurl}


\bibitem[\protect\citeauthoryear{Chen, Wei, and Chang}{Chen
  et~al\mbox{.}}{2011}]%
        {Chen:2011:NRC}
\bibfield{author}{\bibinfo{person}{Hsiang-Ting Chen}, \bibinfo{person}{Li-Yi
  Wei}, {and} \bibinfo{person}{Chun-Fa Chang}.}
  \bibinfo{year}{2011}\natexlab{}.
\newblock \showarticletitle{Nonlinear Revision Control for Images}.
\newblock \bibinfo{journal}{\emph{ACM Trans. Graph.}} \bibinfo{volume}{30},
  \bibinfo{number}{4}, Article \bibinfo{articleno}{105} (\bibinfo{date}{July}
  \bibinfo{year}{2011}), \bibinfo{numpages}{10}~pages.
\newblock
\showISSN{0730-0301}
\urldef\tempurl%
\url{https://doi.org/10.1145/2010324.1965000}
\showDOI{\tempurl}


\bibitem[\protect\citeauthoryear{Chen, Ma, Lefebvre, Xin, Mart\'{\i}nez, and
  wang}{Chen et~al\mbox{.}}{2017}]%
        {Chen:2017:FTD}
\bibfield{author}{\bibinfo{person}{Weikai Chen}, \bibinfo{person}{Yuexin Ma},
  \bibinfo{person}{Sylvain Lefebvre}, \bibinfo{person}{Shiqing Xin},
  \bibinfo{person}{Jon\`{a}s Mart\'{\i}nez}, {and} \bibinfo{person}{wenping
  wang}.} \bibinfo{year}{2017}\natexlab{}.
\newblock \showarticletitle{Fabricable Tile Decors}.
\newblock \bibinfo{journal}{\emph{ACM Trans. Graph.}} \bibinfo{volume}{36},
  \bibinfo{number}{6}, Article \bibinfo{articleno}{175} (\bibinfo{date}{Nov.}
  \bibinfo{year}{2017}), \bibinfo{numpages}{15}~pages.
\newblock
\showISSN{0730-0301}
\urldef\tempurl%
\url{https://doi.org/10.1145/3130800.3130817}
\showDOI{\tempurl}


\bibitem[\protect\citeauthoryear{Chen, Zhang, Xin, Xia, Lefebvre, and
  Wang}{Chen et~al\mbox{.}}{2016}]%
        {Chen:2016:SFD}
\bibfield{author}{\bibinfo{person}{Weikai Chen}, \bibinfo{person}{Xiaolong
  Zhang}, \bibinfo{person}{Shiqing Xin}, \bibinfo{person}{Yang Xia},
  \bibinfo{person}{Sylvain Lefebvre}, {and} \bibinfo{person}{Wenping Wang}.}
  \bibinfo{year}{2016}\natexlab{}.
\newblock \showarticletitle{Synthesis of Filigrees for Digital Fabrication}.
\newblock \bibinfo{journal}{\emph{ACM Trans. Graph.}} \bibinfo{volume}{35},
  \bibinfo{number}{4}, Article \bibinfo{articleno}{98} (\bibinfo{date}{July}
  \bibinfo{year}{2016}), \bibinfo{numpages}{13}~pages.
\newblock
\showISSN{0730-0301}
\urldef\tempurl%
\url{https://doi.org/10.1145/2897824.2925911}
\showDOI{\tempurl}


\bibitem[\protect\citeauthoryear{Chen, Funkhouser, Goldman, and Shechtman}{Chen
  et~al\mbox{.}}{2012}]%
        {Chen:2012:NPT}
\bibfield{author}{\bibinfo{person}{Xiaobai Chen}, \bibinfo{person}{Tom
  Funkhouser}, \bibinfo{person}{Dan~B Goldman}, {and} \bibinfo{person}{Eli
  Shechtman}.} \bibinfo{year}{2012}\natexlab{}.
\newblock \showarticletitle{Non-parametric texture transfer using meshmatch}.
\newblock \bibinfo{journal}{\emph{Adobe Technical Report}} \bibinfo{number}{2}
  (\bibinfo{year}{2012}).
\newblock


\bibitem[\protect\citeauthoryear{Cornet and Rouquier}{Cornet and
  Rouquier}{2004}]%
        {Cornet:2004:GTP}
\bibfield{author}{\bibinfo{person}{Emmanuel Cornet} {and}
  \bibinfo{person}{Jean-Baptiste Rouquier}.} \bibinfo{year}{2004}\natexlab{}.
\newblock \bibinfo{title}{GIMP Texturize plugin}.
\newblock
\newblock
\newblock
\shownote{\url{https://lmanul.github.io/gimp-texturize/}.}


\bibitem[\protect\citeauthoryear{Dumas, Mart{\'\i}nez, Lefebvre, and Wei}{Dumas
  et~al\mbox{.}}{2018}]%
        {Dumas:2018:PAE}
\bibfield{author}{\bibinfo{person}{J{\'e}r{\'e}mie Dumas},
  \bibinfo{person}{Jon{\`a}s Mart{\'\i}nez}, \bibinfo{person}{Sylvain
  Lefebvre}, {and} \bibinfo{person}{Li-Yi Wei}.}
  \bibinfo{year}{2018}\natexlab{}.
\newblock \showarticletitle{Printable Aggregate Elements}.
\newblock \bibinfo{journal}{\emph{arXiv preprint arXiv:1811.02626}}
  (\bibinfo{year}{2018}).
\newblock


\bibitem[\protect\citeauthoryear{Efros and Freeman}{Efros and Freeman}{2001}]%
        {Efros:2001:IQT}
\bibfield{author}{\bibinfo{person}{Alexei~A. Efros} {and}
  \bibinfo{person}{William~T. Freeman}.} \bibinfo{year}{2001}\natexlab{}.
\newblock \showarticletitle{Image Quilting for Texture Synthesis and Transfer}.
  In \bibinfo{booktitle}{\emph{SIGGRAPH '01}}. \bibinfo{pages}{341--346}.
\newblock
\showISBNx{1-58113-374-X}
\urldef\tempurl%
\url{https://doi.org/10.1145/383259.383296}
\showDOI{\tempurl}


\bibitem[\protect\citeauthoryear{Gatys, Ecker, and Bethge}{Gatys
  et~al\mbox{.}}{2016}]%
        {Gatys:2016:IST}
\bibfield{author}{\bibinfo{person}{Leon~A. Gatys},
  \bibinfo{person}{Alexander~S. Ecker}, {and} \bibinfo{person}{Matthias
  Bethge}.} \bibinfo{year}{2016}\natexlab{}.
\newblock \showarticletitle{Image Style Transfer Using Convolutional Neural
  Networks}. In \bibinfo{booktitle}{\emph{CVPR '16}}.
  \bibinfo{pages}{2414--2423}.
\newblock


\bibitem[\protect\citeauthoryear{Hertzmann, Oliver, Curless, and
  Seitz}{Hertzmann et~al\mbox{.}}{2002}]%
        {Hertzmann:2002:CA}
\bibfield{author}{\bibinfo{person}{Aaron Hertzmann}, \bibinfo{person}{Nuria
  Oliver}, \bibinfo{person}{Brian Curless}, {and} \bibinfo{person}{Steven~M.
  Seitz}.} \bibinfo{year}{2002}\natexlab{}.
\newblock \showarticletitle{Curve Analogies}. In \bibinfo{booktitle}{\emph{EGRW
  '02}}. \bibinfo{pages}{233--246}.
\newblock
\showISBNx{1581135343}


\bibitem[\protect\citeauthoryear{Hsu, Wei, You, and Zhang}{Hsu
  et~al\mbox{.}}{2018}]%
        {Hsu:2018:BEF}
\bibfield{author}{\bibinfo{person}{Chen-Yuan Hsu}, \bibinfo{person}{Li-Yi Wei},
  \bibinfo{person}{Lihua You}, {and} \bibinfo{person}{Jian~Jun Zhang}.}
  \bibinfo{year}{2018}\natexlab{}.
\newblock \showarticletitle{Brushing Element Fields}. In
  \bibinfo{booktitle}{\emph{SIGGRAPH Asia 2018 Technical Briefs}}
  \emph{(\bibinfo{series}{SA '18})}. Article \bibinfo{articleno}{6},
  \bibinfo{numpages}{4}~pages.
\newblock
\showISBNx{978-1-4503-6062-3}
\urldef\tempurl%
\url{https://doi.org/10.1145/3283254.3283274}
\showDOI{\tempurl}


\bibitem[\protect\citeauthoryear{Hsu, Wei, You, and Zhang}{Hsu
  et~al\mbox{.}}{2020}]%
        {Hsu:2020:AEF}
\bibfield{author}{\bibinfo{person}{Chen-Yuan Hsu}, \bibinfo{person}{Li-Yi Wei},
  \bibinfo{person}{Lihua You}, {and} \bibinfo{person}{Jian~Jun Zhang}.}
  \bibinfo{year}{2020}\natexlab{}.
\newblock \showarticletitle{Autocomplete Element Fields}. In
  \bibinfo{booktitle}{\emph{CHI '20}}. \bibinfo{pages}{1--13}.
\newblock
\showISBNx{9781450367080}
\urldef\tempurl%
\url{https://doi.org/10.1145/3313831.3376248}
\showDOI{\tempurl}


\bibitem[\protect\citeauthoryear{Huang, Tong, and Wang}{Huang
  et~al\mbox{.}}{2007}]%
        {Huang:2007:APT}
\bibfield{author}{\bibinfo{person}{Hao-Da Huang}, \bibinfo{person}{Xin Tong},
  {and} \bibinfo{person}{Wen-Cheng Wang}.} \bibinfo{year}{2007}\natexlab{}.
\newblock \showarticletitle{Accelerated parallel texture optimization}.
\newblock \bibinfo{journal}{\emph{Journal of Computer Science and Technology}}
  \bibinfo{volume}{22}, \bibinfo{number}{5} (\bibinfo{year}{2007}),
  \bibinfo{pages}{761--769}.
\newblock


\bibitem[\protect\citeauthoryear{Hurtut, Landes, Thollot, Gousseau, Drouillhet,
  and Coeurjolly}{Hurtut et~al\mbox{.}}{2009}]%
        {Hurtut:2009:ASE}
\bibfield{author}{\bibinfo{person}{T. Hurtut}, \bibinfo{person}{P.-E. Landes},
  \bibinfo{person}{J. Thollot}, \bibinfo{person}{Y. Gousseau},
  \bibinfo{person}{R. Drouillhet}, {and} \bibinfo{person}{J.-F. Coeurjolly}.}
  \bibinfo{year}{2009}\natexlab{}.
\newblock \showarticletitle{Appearance-guided Synthesis of Element Arrangements
  by Example}. In \bibinfo{booktitle}{\emph{NPAR '09}}.
  \bibinfo{pages}{51--60}.
\newblock
\showISBNx{978-1-60558-604-5}
\urldef\tempurl%
\url{https://doi.org/10.1145/1572614.1572623}
\showDOI{\tempurl}


\bibitem[\protect\citeauthoryear{Ijiri, Mech, Igarashi, and Miller}{Ijiri
  et~al\mbox{.}}{2008}]%
        {Ijiri:2008:EBP}
\bibfield{author}{\bibinfo{person}{Takashi Ijiri},
  \bibinfo{person}{Radom{\'\i}r Mech}, \bibinfo{person}{Takeo Igarashi}, {and}
  \bibinfo{person}{Gavin Miller}.} \bibinfo{year}{2008}\natexlab{}.
\newblock \showarticletitle{An Example-based Procedural System for Element
  Arrangement}. In \bibinfo{booktitle}{\emph{Computer Graphics Forum}},
  Vol.~\bibinfo{volume}{27}. Wiley Online Library, \bibinfo{pages}{429--436}.
\newblock


\bibitem[\protect\citeauthoryear{Kaspar, Neubert, Lischinski, Pauly, and
  Kopf}{Kaspar et~al\mbox{.}}{2015}]%
        {Kaspar:2015:STT}
\bibfield{author}{\bibinfo{person}{Alexandre Kaspar}, \bibinfo{person}{Boris
  Neubert}, \bibinfo{person}{Dani Lischinski}, \bibinfo{person}{Mark Pauly},
  {and} \bibinfo{person}{Johannes Kopf}.} \bibinfo{year}{2015}\natexlab{}.
\newblock \showarticletitle{Self Tuning Texture Optimization}.
\newblock \bibinfo{journal}{\emph{Comput. Graph. Forum}} \bibinfo{volume}{34},
  \bibinfo{number}{2} (\bibinfo{date}{May} \bibinfo{year}{2015}),
  \bibinfo{pages}{349--359}.
\newblock
\showISSN{0167-7055}
\urldef\tempurl%
\url{https://doi.org/10.1111/cgf.12565}
\showDOI{\tempurl}


\bibitem[\protect\citeauthoryear{Kazi, Igarashi, Zhao, and Davis}{Kazi
  et~al\mbox{.}}{2012}]%
        {Kazi:2012:VIT}
\bibfield{author}{\bibinfo{person}{Rubaiat~Habib Kazi}, \bibinfo{person}{Takeo
  Igarashi}, \bibinfo{person}{Shengdong Zhao}, {and} \bibinfo{person}{Richard
  Davis}.} \bibinfo{year}{2012}\natexlab{}.
\newblock \showarticletitle{Vignette: Interactive Texture Design and
  Manipulation with Freeform Gestures for Pen-and-ink Illustration}. In
  \bibinfo{booktitle}{\emph{CHI '12}}. \bibinfo{pages}{1727--1736}.
\newblock
\showISBNx{978-1-4503-1015-4}
\urldef\tempurl%
\url{https://doi.org/10.1145/2207676.2208302}
\showDOI{\tempurl}


\bibitem[\protect\citeauthoryear{Koyama, Sakamoto, and Igarashi}{Koyama
  et~al\mbox{.}}{2016}]%
        {Koyama:2016:SPL}
\bibfield{author}{\bibinfo{person}{Yuki Koyama}, \bibinfo{person}{Daisuke
  Sakamoto}, {and} \bibinfo{person}{Takeo Igarashi}.}
  \bibinfo{year}{2016}\natexlab{}.
\newblock \showarticletitle{SelPh: Progressive Learning and Support of Manual
  Photo Color Enhancement}. In \bibinfo{booktitle}{\emph{CHI '16}}.
  \bibinfo{pages}{2520--2532}.
\newblock
\showISBNx{978-1-4503-3362-7}
\urldef\tempurl%
\url{https://doi.org/10.1145/2858036.2858111}
\showDOI{\tempurl}


\bibitem[\protect\citeauthoryear{Kuhn}{Kuhn}{1955}]%
        {Kuhn:1955:HMA}
\bibfield{author}{\bibinfo{person}{Harold~W Kuhn}.}
  \bibinfo{year}{1955}\natexlab{}.
\newblock \showarticletitle{The Hungarian method for the assignment problem}.
\newblock \bibinfo{journal}{\emph{Naval research logistics quarterly}}
  \bibinfo{volume}{2}, \bibinfo{number}{1-2} (\bibinfo{year}{1955}),
  \bibinfo{pages}{83--97}.
\newblock


\bibitem[\protect\citeauthoryear{Kwatra, Essa, Bobick, and Kwatra}{Kwatra
  et~al\mbox{.}}{2005}]%
        {Kwatra:2005:TOE}
\bibfield{author}{\bibinfo{person}{Vivek Kwatra}, \bibinfo{person}{Irfan Essa},
  \bibinfo{person}{Aaron Bobick}, {and} \bibinfo{person}{Nipun Kwatra}.}
  \bibinfo{year}{2005}\natexlab{}.
\newblock \showarticletitle{Texture Optimization for Example-based Synthesis}.
\newblock \bibinfo{journal}{\emph{ACM Trans. Graph.}} \bibinfo{volume}{24},
  \bibinfo{number}{3} (\bibinfo{date}{July} \bibinfo{year}{2005}),
  \bibinfo{pages}{795--802}.
\newblock
\showISSN{0730-0301}
\urldef\tempurl%
\url{https://doi.org/10.1145/1073204.1073263}
\showDOI{\tempurl}


\bibitem[\protect\citeauthoryear{Kwatra, Sch\"{o}dl, Essa, Turk, and
  Bobick}{Kwatra et~al\mbox{.}}{2003}]%
        {Kwatra:2003:GTI}
\bibfield{author}{\bibinfo{person}{Vivek Kwatra}, \bibinfo{person}{Arno
  Sch\"{o}dl}, \bibinfo{person}{Irfan Essa}, \bibinfo{person}{Greg Turk}, {and}
  \bibinfo{person}{Aaron Bobick}.} \bibinfo{year}{2003}\natexlab{}.
\newblock \showarticletitle{Graphcut Textures: Image and Video Synthesis Using
  Graph Cuts}. In \bibinfo{booktitle}{\emph{SIGGRAPH '03}}.
  \bibinfo{pages}{277--286}.
\newblock
\showISBNx{1-58113-709-5}
\urldef\tempurl%
\url{https://doi.org/10.1145/1201775.882264}
\showDOI{\tempurl}


\bibitem[\protect\citeauthoryear{Landes, Galerne, and Hurtut}{Landes
  et~al\mbox{.}}{2013}]%
        {Landes:2013:SAM}
\bibfield{author}{\bibinfo{person}{Pierre-Edouard Landes},
  \bibinfo{person}{Bruno Galerne}, {and} \bibinfo{person}{Thomas Hurtut}.}
  \bibinfo{year}{2013}\natexlab{}.
\newblock \showarticletitle{A Shape-Aware Model for Discrete Texture
  Synthesis}.
\newblock \bibinfo{journal}{\emph{Computer Graphics Forum}}
  \bibinfo{volume}{32}, \bibinfo{number}{4} (\bibinfo{year}{2013}),
  \bibinfo{pages}{67--76}.
\newblock


\bibitem[\protect\citeauthoryear{Li, Breen, McCann, and Hodgins}{Li
  et~al\mbox{.}}{2019}]%
        {Li:2019:AQP}
\bibfield{author}{\bibinfo{person}{Yifei Li}, \bibinfo{person}{David~E. Breen},
  \bibinfo{person}{James McCann}, {and} \bibinfo{person}{Jessica Hodgins}.}
  \bibinfo{year}{2019}\natexlab{}.
\newblock \showarticletitle{Algorithmic Quilting Pattern Generation for Pieced
  Quilts}. In \bibinfo{booktitle}{\emph{Proceedings of the 45th Graphics
  Interface Conference on Proceedings of Graphics Interface 2019}}
  \emph{(\bibinfo{series}{GI\&\#39;19})}. Article \bibinfo{articleno}{13},
  \bibinfo{numpages}{9}~pages.
\newblock
\showISBNx{978-0-9947868-4-5}
\urldef\tempurl%
\url{https://doi.org/10.20380/GI2019.13}
\showDOI{\tempurl}


\bibitem[\protect\citeauthoryear{Liang, Liu, Xu, Guo, and Shum}{Liang
  et~al\mbox{.}}{2001}]%
        {Liang:2001:RTS}
\bibfield{author}{\bibinfo{person}{Lin Liang}, \bibinfo{person}{Ce Liu},
  \bibinfo{person}{Ying-Qing Xu}, \bibinfo{person}{Baining Guo}, {and}
  \bibinfo{person}{Heung-Yeung Shum}.} \bibinfo{year}{2001}\natexlab{}.
\newblock \showarticletitle{Real-time Texture Synthesis by Patch-based
  Sampling}.
\newblock \bibinfo{journal}{\emph{ACM Trans. Graph.}} \bibinfo{volume}{20},
  \bibinfo{number}{3} (\bibinfo{date}{July} \bibinfo{year}{2001}),
  \bibinfo{pages}{127--150}.
\newblock
\showISSN{0730-0301}
\urldef\tempurl%
\url{https://doi.org/10.1145/501786.501787}
\showDOI{\tempurl}


\bibitem[\protect\citeauthoryear{Loi, Hurtut, Vergne, and Thollot}{Loi
  et~al\mbox{.}}{2017}]%
        {Loi:2017:PAE}
\bibfield{author}{\bibinfo{person}{Hugo Loi}, \bibinfo{person}{Thomas Hurtut},
  \bibinfo{person}{Romain Vergne}, {and} \bibinfo{person}{Joelle Thollot}.}
  \bibinfo{year}{2017}\natexlab{}.
\newblock \showarticletitle{Programmable 2D Arrangements for Element Texture
  Design}.
\newblock \bibinfo{journal}{\emph{ACM Trans. Graph.}} \bibinfo{volume}{36},
  \bibinfo{number}{4}, Article \bibinfo{articleno}{105a} (\bibinfo{date}{May}
  \bibinfo{year}{2017}).
\newblock
\showISSN{0730-0301}
\urldef\tempurl%
\url{https://doi.org/10.1145/3072959.2983617}
\showDOI{\tempurl}


\bibitem[\protect\citeauthoryear{Lu, Barnes, Wan, Asente, Mech, and
  Finkelstein}{Lu et~al\mbox{.}}{2014}]%
        {Lu:2014:DDS}
\bibfield{author}{\bibinfo{person}{Jingwan Lu}, \bibinfo{person}{Connelly
  Barnes}, \bibinfo{person}{Connie Wan}, \bibinfo{person}{Paul Asente},
  \bibinfo{person}{Radomir Mech}, {and} \bibinfo{person}{Adam Finkelstein}.}
  \bibinfo{year}{2014}\natexlab{}.
\newblock \showarticletitle{DecoBrush: Drawing Structured Decorative Patterns
  by Example}.
\newblock \bibinfo{journal}{\emph{ACM Trans. Graph.}} \bibinfo{volume}{33},
  \bibinfo{number}{4}, Article \bibinfo{articleno}{90} (\bibinfo{date}{July}
  \bibinfo{year}{2014}), \bibinfo{numpages}{9}~pages.
\newblock
\showISSN{0730-0301}
\urldef\tempurl%
\url{https://doi.org/10.1145/2601097.2601190}
\showDOI{\tempurl}


\bibitem[\protect\citeauthoryear{Lu, Yu, Finkelstein, and DiVerdi}{Lu
  et~al\mbox{.}}{2012}]%
        {Lu:2012:HES}
\bibfield{author}{\bibinfo{person}{Jingwan Lu}, \bibinfo{person}{Fisher Yu},
  \bibinfo{person}{Adam Finkelstein}, {and} \bibinfo{person}{Stephen DiVerdi}.}
  \bibinfo{year}{2012}\natexlab{}.
\newblock \showarticletitle{HelpingHand: Example-based Stroke Stylization}.
\newblock \bibinfo{journal}{\emph{ACM Trans. Graph.}} \bibinfo{volume}{31},
  \bibinfo{number}{4}, Article \bibinfo{articleno}{46} (\bibinfo{date}{July}
  \bibinfo{year}{2012}), \bibinfo{numpages}{10}~pages.
\newblock
\showISSN{0730-0301}
\urldef\tempurl%
\url{https://doi.org/10.1145/2185520.2185542}
\showDOI{\tempurl}


\bibitem[\protect\citeauthoryear{Ma, Wei, Lefebvre, and Tong}{Ma
  et~al\mbox{.}}{2013}]%
        {Ma:2013:DET}
\bibfield{author}{\bibinfo{person}{Chongyang Ma}, \bibinfo{person}{Li-Yi Wei},
  \bibinfo{person}{Sylvain Lefebvre}, {and} \bibinfo{person}{Xin Tong}.}
  \bibinfo{year}{2013}\natexlab{}.
\newblock \showarticletitle{Dynamic Element Textures}.
\newblock \bibinfo{journal}{\emph{ACM Trans. Graph.}} \bibinfo{volume}{32},
  \bibinfo{number}{4}, Article \bibinfo{articleno}{90} (\bibinfo{date}{July}
  \bibinfo{year}{2013}), \bibinfo{numpages}{10}~pages.
\newblock
\showISSN{0730-0301}
\urldef\tempurl%
\url{https://doi.org/10.1145/2461912.2461921}
\showDOI{\tempurl}


\bibitem[\protect\citeauthoryear{Ma, Wei, and Tong}{Ma et~al\mbox{.}}{2011}]%
        {Ma:2011:DET}
\bibfield{author}{\bibinfo{person}{Chongyang Ma}, \bibinfo{person}{Li-Yi Wei},
  {and} \bibinfo{person}{Xin Tong}.} \bibinfo{year}{2011}\natexlab{}.
\newblock \showarticletitle{Discrete Element Textures}.
\newblock \bibinfo{journal}{\emph{ACM Trans. Graph.}} \bibinfo{volume}{30},
  \bibinfo{number}{4}, Article \bibinfo{articleno}{62} (\bibinfo{date}{July}
  \bibinfo{year}{2011}), \bibinfo{numpages}{10}~pages.
\newblock
\showISSN{0730-0301}
\urldef\tempurl%
\url{https://doi.org/10.1145/2010324.1964957}
\showDOI{\tempurl}


\bibitem[\protect\citeauthoryear{Mart\'{\i}nez, Dumas, Lefebvre, and
  Wei}{Mart\'{\i}nez et~al\mbox{.}}{2015}]%
        {Martinez:2015:SAO}
\bibfield{author}{\bibinfo{person}{Jon\`{a}s Mart\'{\i}nez},
  \bibinfo{person}{J{\'e}r{\'e}mie Dumas}, \bibinfo{person}{Sylvain Lefebvre},
  {and} \bibinfo{person}{Li-Yi Wei}.} \bibinfo{year}{2015}\natexlab{}.
\newblock \showarticletitle{Structure and Appearance Optimization for
  Controllable Shape Design}.
\newblock \bibinfo{journal}{\emph{ACM Trans. Graph.}} \bibinfo{volume}{34},
  \bibinfo{number}{6}, Article \bibinfo{articleno}{229} (\bibinfo{date}{Oct.}
  \bibinfo{year}{2015}), \bibinfo{numpages}{11}~pages.
\newblock
\showISSN{0730-0301}
\urldef\tempurl%
\url{https://doi.org/10.1145/2816795.2818101}
\showDOI{\tempurl}


\bibitem[\protect\citeauthoryear{Merrell and Manocha}{Merrell and
  Manocha}{2010}]%
        {Merrell:2010:ECS}
\bibfield{author}{\bibinfo{person}{Paul Merrell} {and} \bibinfo{person}{Dinesh
  Manocha}.} \bibinfo{year}{2010}\natexlab{}.
\newblock \showarticletitle{Example-based curve synthesis}.
\newblock \bibinfo{journal}{\emph{Computers \& Graphics}} \bibinfo{volume}{34},
  \bibinfo{number}{4} (\bibinfo{year}{2010}), \bibinfo{pages}{304--311}.
\newblock


\bibitem[\protect\citeauthoryear{Nancel and Cockburn}{Nancel and
  Cockburn}{2014}]%
        {Nancel:2014:CCM}
\bibfield{author}{\bibinfo{person}{Mathieu Nancel} {and} \bibinfo{person}{Andy
  Cockburn}.} \bibinfo{year}{2014}\natexlab{}.
\newblock \showarticletitle{Causality: A Conceptual Model of Interaction
  History}. In \bibinfo{booktitle}{\emph{CHI '14}}.
  \bibinfo{pages}{1777--1786}.
\newblock
\showISBNx{978-1-4503-2473-1}
\urldef\tempurl%
\url{https://doi.org/10.1145/2556288.2556990}
\showDOI{\tempurl}


\bibitem[\protect\citeauthoryear{Pedersen and Singh}{Pedersen and
  Singh}{2006}]%
        {Pedersen:2006:OLM}
\bibfield{author}{\bibinfo{person}{Hans Pedersen} {and} \bibinfo{person}{Karan
  Singh}.} \bibinfo{year}{2006}\natexlab{}.
\newblock \showarticletitle{Organic Labyrinths and Mazes}. In
  \bibinfo{booktitle}{\emph{NPAR '06}}. \bibinfo{pages}{79--86}.
\newblock
\showISBNx{1595933573}
\urldef\tempurl%
\url{https://doi.org/10.1145/1124728.1124742}
\showDOI{\tempurl}


\bibitem[\protect\citeauthoryear{Peng, Wei, Kazi, and Kim}{Peng
  et~al\mbox{.}}{2020}]%
        {Peng:2020:AAS}
\bibfield{author}{\bibinfo{person}{Mengqi Peng}, \bibinfo{person}{Li-Yi Wei},
  \bibinfo{person}{Rubaiat~Habib Kazi}, {and} \bibinfo{person}{Vladimir~G.
  Kim}.} \bibinfo{year}{2020}\natexlab{}.
\newblock \showarticletitle{Autocomplete Animated Sculpting}. In
  \bibinfo{booktitle}{\emph{UIST '20}}.
\newblock
\urldef\tempurl%
\url{https://doi.org/10.1145/3379337.3415884}
\showDOI{\tempurl}


\bibitem[\protect\citeauthoryear{Peng, Xing, and Wei}{Peng
  et~al\mbox{.}}{2018}]%
        {Peng:2018:A3S}
\bibfield{author}{\bibinfo{person}{Mengqi Peng}, \bibinfo{person}{Jun Xing},
  {and} \bibinfo{person}{Li-Yi Wei}.} \bibinfo{year}{2018}\natexlab{}.
\newblock \showarticletitle{Autocomplete {3D} Sculpting}.
\newblock \bibinfo{journal}{\emph{ACM Trans. Graph.}} \bibinfo{volume}{37},
  \bibinfo{number}{4}, Article \bibinfo{articleno}{132} (\bibinfo{date}{July}
  \bibinfo{year}{2018}), \bibinfo{numpages}{15}~pages.
\newblock
\showISSN{0730-0301}
\urldef\tempurl%
\url{https://doi.org/10.1145/3197517.3201297}
\showDOI{\tempurl}


\bibitem[\protect\citeauthoryear{Riesen and Bunke}{Riesen and Bunke}{2009}]%
        {Riesen:2009:AGE}
\bibfield{author}{\bibinfo{person}{Kaspar Riesen} {and} \bibinfo{person}{Horst
  Bunke}.} \bibinfo{year}{2009}\natexlab{}.
\newblock \showarticletitle{Approximate graph edit distance computation by
  means of bipartite graph matching}.
\newblock \bibinfo{journal}{\emph{Image and Vision computing}}
  \bibinfo{volume}{27}, \bibinfo{number}{7} (\bibinfo{year}{2009}),
  \bibinfo{pages}{950--959}.
\newblock


\bibitem[\protect\citeauthoryear{Roveri, {\"O}ztireli, Martin, Solenthaler, and
  Gross}{Roveri et~al\mbox{.}}{2015}]%
        {Roveri:2015:EBR}
\bibfield{author}{\bibinfo{person}{Riccardo Roveri}, \bibinfo{person}{A~Cengiz
  {\"O}ztireli}, \bibinfo{person}{Sebastian Martin}, \bibinfo{person}{Barbara
  Solenthaler}, {and} \bibinfo{person}{Markus Gross}.}
  \bibinfo{year}{2015}\natexlab{}.
\newblock \showarticletitle{Example based repetitive structure synthesis}.
\newblock \bibinfo{journal}{\emph{Computer Graphics Forum}}
  \bibinfo{volume}{34}, \bibinfo{number}{5} (\bibinfo{year}{2015}),
  \bibinfo{pages}{39--52}.
\newblock


\bibitem[\protect\citeauthoryear{Santoni and Pellacini}{Santoni and
  Pellacini}{2016}]%
        {Santoni:2016:GGP}
\bibfield{author}{\bibinfo{person}{Christian Santoni} {and}
  \bibinfo{person}{Fabio Pellacini}.} \bibinfo{year}{2016}\natexlab{}.
\newblock \showarticletitle{gTangle: A Grammar for the Procedural Generation of
  Tangle Patterns}.
\newblock \bibinfo{journal}{\emph{ACM Trans. Graph.}} \bibinfo{volume}{35},
  \bibinfo{number}{6}, Article \bibinfo{articleno}{182} (\bibinfo{date}{Nov.}
  \bibinfo{year}{2016}), \bibinfo{numpages}{11}~pages.
\newblock
\showISSN{0730-0301}
\urldef\tempurl%
\url{https://doi.org/10.1145/2980179.2982417}
\showDOI{\tempurl}


\bibitem[\protect\citeauthoryear{Schumacher, Thomaszewski, and
  Gross}{Schumacher et~al\mbox{.}}{2016}]%
        {Schumacher:2016:SDS}
\bibfield{author}{\bibinfo{person}{Christian Schumacher},
  \bibinfo{person}{Bernhard Thomaszewski}, {and} \bibinfo{person}{Markus
  Gross}.} \bibinfo{year}{2016}\natexlab{}.
\newblock \showarticletitle{Stenciling: Designing Structurally-Sound Surfaces
  with Decorative Patterns}.
\newblock \bibinfo{journal}{\emph{Computer Graphics Forum}}
  \bibinfo{volume}{35}, \bibinfo{number}{5} (\bibinfo{year}{2016}),
  \bibinfo{pages}{101--110}.
\newblock


\bibitem[\protect\citeauthoryear{Suzuki, Yeh, Yatani, and Gross}{Suzuki
  et~al\mbox{.}}{2017}]%
        {Suzuki:2017:ATP}
\bibfield{author}{\bibinfo{person}{Ryo Suzuki}, \bibinfo{person}{Tom Yeh},
  \bibinfo{person}{Koji Yatani}, {and} \bibinfo{person}{Mark~D Gross}.}
  \bibinfo{year}{2017}\natexlab{}.
\newblock \showarticletitle{Autocomplete Textures for 3D Printing}.
\newblock \bibinfo{journal}{\emph{arXiv preprint arXiv:1703.05700}}
  (\bibinfo{year}{2017}).
\newblock


\bibitem[\protect\citeauthoryear{Takayama, Sorkine, Nealen, and
  Igarashi}{Takayama et~al\mbox{.}}{2010}]%
        {Takayama:2010:VMD}
\bibfield{author}{\bibinfo{person}{Kenshi Takayama}, \bibinfo{person}{Olga
  Sorkine}, \bibinfo{person}{Andrew Nealen}, {and} \bibinfo{person}{Takeo
  Igarashi}.} \bibinfo{year}{2010}\natexlab{}.
\newblock \showarticletitle{Volumetric Modeling with Diffusion Surfaces}. In
  \bibinfo{booktitle}{\emph{SIGGRAPH ASIA '10}}. Article
  \bibinfo{articleno}{Article 180}, \bibinfo{numpages}{8}~pages.
\newblock
\urldef\tempurl%
\url{https://doi.org/10.1145/1866158.1866202}
\showDOI{\tempurl}


\bibitem[\protect\citeauthoryear{Tu}{Tu}{2020}]%
        {Tu:2020:CCTC}
\bibfield{author}{\bibinfo{person}{Peihan Tu}.}
  \bibinfo{year}{2020}\natexlab{}.
\newblock \bibinfo{title}{Continuous Curve Textures Source Code}.
\newblock
\newblock
\newblock
\shownote{\url{https://github.com/tph9608/continuous-curve-texture/}.}


\bibitem[\protect\citeauthoryear{Tu, Lischinski, and Huang}{Tu
  et~al\mbox{.}}{2019}]%
        {Tu:2019:PPS}
\bibfield{author}{\bibinfo{person}{Peihan Tu}, \bibinfo{person}{Dani
  Lischinski}, {and} \bibinfo{person}{Hui Huang}.}
  \bibinfo{year}{2019}\natexlab{}.
\newblock \showarticletitle{Point Pattern Synthesis via Irregular Convolution}.
\newblock \bibinfo{journal}{\emph{Computer Graphics Forum}}
  \bibinfo{volume}{38}, \bibinfo{number}{5} (\bibinfo{year}{2019}),
  \bibinfo{pages}{109--122}.
\newblock
\urldef\tempurl%
\url{https://doi.org/10.1111/cgf.13793}
\showDOI{\tempurl}


\bibitem[\protect\citeauthoryear{Wang, Yu, Zhou, and Guo}{Wang
  et~al\mbox{.}}{2011}]%
        {Wang:2011:MVV}
\bibfield{author}{\bibinfo{person}{Lvdi Wang}, \bibinfo{person}{Yizhou Yu},
  \bibinfo{person}{Kun Zhou}, {and} \bibinfo{person}{Baining Guo}.}
  \bibinfo{year}{2011}\natexlab{}.
\newblock \showarticletitle{Multiscale vector volumes}.
\newblock \bibinfo{journal}{\emph{ACM Transactions on Graphics (TOG)}}
  \bibinfo{volume}{30}, \bibinfo{number}{6} (\bibinfo{year}{2011}),
  \bibinfo{pages}{1--8}.
\newblock


\bibitem[\protect\citeauthoryear{Wang, Zhou, Yu, and Guo}{Wang
  et~al\mbox{.}}{2010}]%
        {Wang:2010:VST}
\bibfield{author}{\bibinfo{person}{Lvdi Wang}, \bibinfo{person}{Kun Zhou},
  \bibinfo{person}{Yizhou Yu}, {and} \bibinfo{person}{Baining Guo}.}
  \bibinfo{year}{2010}\natexlab{}.
\newblock \showarticletitle{Vector solid textures}.
\newblock \bibinfo{journal}{\emph{ACM Transactions on Graphics (TOG)}}
  \bibinfo{volume}{29}, \bibinfo{number}{4} (\bibinfo{year}{2010}),
  \bibinfo{pages}{1--8}.
\newblock


\bibitem[\protect\citeauthoryear{Wei}{Wei}{2016}]%
        {Wei:2016:TS}
\bibfield{author}{\bibinfo{person}{Li-Yi Wei}.}
  \bibinfo{year}{2016}\natexlab{}.
\newblock \bibinfo{title}{Texture Synthesis}.
\newblock
\newblock
\urldef\tempurl%
\url{https://github.com/1iyiwei/texture}
\showURL{%
\tempurl}


\bibitem[\protect\citeauthoryear{Wei, Lefebvre, Kwatra, and Turk}{Wei
  et~al\mbox{.}}{2009}]%
        {Wei:2009:SAE}
\bibfield{author}{\bibinfo{person}{Li-Yi Wei}, \bibinfo{person}{Sylvain
  Lefebvre}, \bibinfo{person}{Vivek Kwatra}, {and} \bibinfo{person}{Greg
  Turk}.} \bibinfo{year}{2009}\natexlab{}.
\newblock \showarticletitle{State of the Art in Example-based Texture
  Synthesis}. In \bibinfo{booktitle}{\emph{Eurographics 2009, State of the Art
  Report, EG-STAR}}. \bibinfo{publisher}{Eurographics Association}.
\newblock
\urldef\tempurl%
\url{http://www-sop.inria.fr/reves/Basilic/2009/WLKT09}
\showURL{%
\tempurl}


\bibitem[\protect\citeauthoryear{Wei and Levoy}{Wei and Levoy}{2000}]%
        {Wei:2000:FTS}
\bibfield{author}{\bibinfo{person}{Li-Yi Wei} {and} \bibinfo{person}{Marc
  Levoy}.} \bibinfo{year}{2000}\natexlab{}.
\newblock \showarticletitle{Fast Texture Synthesis Using Tree-structured Vector
  Quantization}. In \bibinfo{booktitle}{\emph{SIGGRAPH '00}}.
  \bibinfo{pages}{479--488}.
\newblock
\showISBNx{1-58113-208-5}
\urldef\tempurl%
\url{https://doi.org/10.1145/344779.345009}
\showDOI{\tempurl}


\bibitem[\protect\citeauthoryear{Wei and Levoy}{Wei and Levoy}{2001}]%
        {Wei:2001:TSO}
\bibfield{author}{\bibinfo{person}{Li-Yi Wei} {and} \bibinfo{person}{Marc
  Levoy}.} \bibinfo{year}{2001}\natexlab{}.
\newblock \showarticletitle{Texture Synthesis over Arbitrary Manifold
  Surfaces}. In \bibinfo{booktitle}{\emph{SIGGRAPH '01}}.
  \bibinfo{pages}{355--360}.
\newblock
\showISBNx{1-58113-374-X}
\urldef\tempurl%
\url{https://doi.org/10.1145/383259.383298}
\showDOI{\tempurl}


\bibitem[\protect\citeauthoryear{Xing, Chen, and Wei}{Xing
  et~al\mbox{.}}{2014}]%
        {Xing:2014:APR}
\bibfield{author}{\bibinfo{person}{Jun Xing}, \bibinfo{person}{Hsiang-Ting
  Chen}, {and} \bibinfo{person}{Li-Yi Wei}.} \bibinfo{year}{2014}\natexlab{}.
\newblock \showarticletitle{Autocomplete Painting Repetitions}.
\newblock \bibinfo{journal}{\emph{ACM Trans. Graph.}} \bibinfo{volume}{33},
  \bibinfo{number}{6}, Article \bibinfo{articleno}{172} (\bibinfo{date}{Nov.}
  \bibinfo{year}{2014}), \bibinfo{numpages}{11}~pages.
\newblock
\showISSN{0730-0301}
\urldef\tempurl%
\url{https://doi.org/10.1145/2661229.2661247}
\showDOI{\tempurl}


\bibitem[\protect\citeauthoryear{Xing, Wei, Shiratori, and Yatani}{Xing
  et~al\mbox{.}}{2015}]%
        {Xing:2015:AHA}
\bibfield{author}{\bibinfo{person}{Jun Xing}, \bibinfo{person}{Li-Yi Wei},
  \bibinfo{person}{Takaaki Shiratori}, {and} \bibinfo{person}{Koji Yatani}.}
  \bibinfo{year}{2015}\natexlab{}.
\newblock \showarticletitle{Autocomplete Hand-drawn Animations}.
\newblock \bibinfo{journal}{\emph{ACM Trans. Graph.}} \bibinfo{volume}{34},
  \bibinfo{number}{6}, Article \bibinfo{articleno}{169} (\bibinfo{date}{Oct.}
  \bibinfo{year}{2015}), \bibinfo{numpages}{11}~pages.
\newblock
\showISSN{0730-0301}
\urldef\tempurl%
\url{https://doi.org/10.1145/2816795.2818079}
\showDOI{\tempurl}


\bibitem[\protect\citeauthoryear{Zehnder, Coros, and Thomaszewski}{Zehnder
  et~al\mbox{.}}{2016}]%
        {Zehnder:2016:DSO}
\bibfield{author}{\bibinfo{person}{Jonas Zehnder}, \bibinfo{person}{Stelian
  Coros}, {and} \bibinfo{person}{Bernhard Thomaszewski}.}
  \bibinfo{year}{2016}\natexlab{}.
\newblock \showarticletitle{Designing Structurally-sound Ornamental Curve
  Networks}.
\newblock \bibinfo{journal}{\emph{ACM Trans. Graph.}} \bibinfo{volume}{35},
  \bibinfo{number}{4}, Article \bibinfo{articleno}{99} (\bibinfo{date}{July}
  \bibinfo{year}{2016}), \bibinfo{numpages}{10}~pages.
\newblock
\showISSN{0730-0301}
\urldef\tempurl%
\url{https://doi.org/10.1145/2897824.2925888}
\showDOI{\tempurl}


\bibitem[\protect\citeauthoryear{Zhou, Sun, Turk, and Rehg}{Zhou
  et~al\mbox{.}}{2007}]%
        {Zhou:2007:TSD}
\bibfield{author}{\bibinfo{person}{Howard Zhou}, \bibinfo{person}{Jie Sun},
  \bibinfo{person}{Greg Turk}, {and} \bibinfo{person}{James~M. Rehg}.}
  \bibinfo{year}{2007}\natexlab{}.
\newblock \showarticletitle{Terrain Synthesis from Digital Elevation Models}.
\newblock \bibinfo{journal}{\emph{IEEE Transactions on Visualization and
  Computer Graphics}} \bibinfo{volume}{13}, \bibinfo{number}{4}
  (\bibinfo{date}{July} \bibinfo{year}{2007}), \bibinfo{pages}{834–848}.
\newblock
\showISSN{1077-2626}
\urldef\tempurl%
\url{https://doi.org/10.1109/TVCG.2007.1027}
\showDOI{\tempurl}


\bibitem[\protect\citeauthoryear{Zhou, Huang, Wang, Tong, Desbrun, Guo, and
  Shum}{Zhou et~al\mbox{.}}{2006}]%
        {Zhou:2006:MQG}
\bibfield{author}{\bibinfo{person}{Kun Zhou}, \bibinfo{person}{Xin Huang},
  \bibinfo{person}{Xi Wang}, \bibinfo{person}{Yiying Tong},
  \bibinfo{person}{Mathieu Desbrun}, \bibinfo{person}{Baining Guo}, {and}
  \bibinfo{person}{Heung-Yeung Shum}.} \bibinfo{year}{2006}\natexlab{}.
\newblock \showarticletitle{Mesh Quilting for Geometric Texture Synthesis}.
\newblock \bibinfo{journal}{\emph{ACM Trans. Graph.}} \bibinfo{volume}{25},
  \bibinfo{number}{3} (\bibinfo{date}{July} \bibinfo{year}{2006}),
  \bibinfo{pages}{690--697}.
\newblock
\showISSN{0730-0301}
\urldef\tempurl%
\url{https://doi.org/10.1145/1141911.1141942}
\showDOI{\tempurl}


\bibitem[\protect\citeauthoryear{Zhou, Jiang, and Lefebvre}{Zhou
  et~al\mbox{.}}{2014}]%
        {Zhou:2014:TSV}
\bibfield{author}{\bibinfo{person}{Shizhe Zhou}, \bibinfo{person}{Changyun
  Jiang}, {and} \bibinfo{person}{Sylvain Lefebvre}.}
  \bibinfo{year}{2014}\natexlab{}.
\newblock \showarticletitle{Topology-constrained Synthesis of Vector Patterns}.
\newblock \bibinfo{journal}{\emph{ACM Trans. Graph.}} \bibinfo{volume}{33},
  \bibinfo{number}{6}, Article \bibinfo{articleno}{215} (\bibinfo{date}{Nov.}
  \bibinfo{year}{2014}), \bibinfo{numpages}{11}~pages.
\newblock
\showISSN{0730-0301}
\urldef\tempurl%
\url{https://doi.org/10.1145/2661229.2661238}
\showDOI{\tempurl}


\bibitem[\protect\citeauthoryear{Zitnick}{Zitnick}{2013}]%
        {Zitnick:2013:HBU}
\bibfield{author}{\bibinfo{person}{C.~Lawrence Zitnick}.}
  \bibinfo{year}{2013}\natexlab{}.
\newblock \showarticletitle{Handwriting Beautification Using Token Means}.
\newblock \bibinfo{journal}{\emph{ACM Trans. Graph.}} \bibinfo{volume}{32},
  \bibinfo{number}{4}, Article \bibinfo{articleno}{53} (\bibinfo{date}{July}
  \bibinfo{year}{2013}), \bibinfo{numpages}{8}~pages.
\newblock
\showISSN{0730-0301}
\urldef\tempurl%
\url{https://doi.org/10.1145/2461912.2461985}
\showDOI{\tempurl}


\end{thebibliography}
}

\clearpage
\appendix
\normalsize

\section{Existence Assignment}
\label{sec:appendix:existence_assignment}

The algorithm for generating additional samples is shown in \Cref{alg:generate_new_output_samples}.

\begin{algorithm}
	\begin{algorithmic}[1]
	\STATE $\mathbf{function}$
	 \funct{GenerateNewOutputSamples}($\{\neighinput, \neighoutput\}$)
	 \STATE $\{\samplecandidate\} \assign \emptyset$ \COMMENT{candidate sample set}
	 \STATE $\{\samplecluster\} \assign \emptyset$ 	 \COMMENT{$\samplecluster$ is a cluster that contains some candidate samples}
	 \STATE $\{\sampleoutput\} \assign \emptyset$ \COMMENT{new output sample set} 
    \FOR{$\neighinput, \neighoutput \in \{\neighinput, \neighoutput\}$}
	\FOR{$\sampleinputprime \in \neighinput$} 
	\IF{$\sampleinputprime$ is unmatched}
	\STATE Generate a candidate sample $\samplecandidate$ with $\samplespace(\samplecandidate)= \sampleposition(\sampleinputprime)-\sampleposition(\sampleinput)+\sampleposition(\sampleoutput)$ and other attributes are the same to $\sampleinputprime$
	\STATE $\{\samplecandidate\} \assign \{\samplecandidate\} \cup \samplecandidate $
	\ENDIF
	\ENDFOR
	\ENDFOR
	\FOR{$\samplecandidate \in \{\samplecandidate\}$} 	
	\STATE \COMMENT{Generate clusters from the candidate sample set by greedily looping over all candidates; more advanced clustering technique can be applied to replace this step}
	\STATE Find the $\samplecluster$ within $\{\samplecluster\}$ with nearest center to $\samplecandidate$
	\IF{\funct{Distance}($\samplecluster,\samplecandidate$) < $0.5 \samplingdistance$ }
	\STATE \COMMENT{\funct{Distance}($\samplecluster,\samplecandidate$) computes the spatial distance between the center of $\samplecluster$ and $\samplecandidate$}
	\STATE $\samplecluster \assign \samplecluster  \cup \samplecandidate$
	\ELSE
	\STATE $\samplecluster^{\prime} \assign \emptyset  $
	\STATE $\samplecluster^{\prime} \assign \samplecluster^{\prime}  \cup \samplecandidate $
	\STATE $\{\samplecluster\} \assign \{\samplecluster\}  \cup \samplecluster^{\prime} $
	\ENDIF
	\ENDFOR
	\FOR{$\samplecluster \in \{\samplecluster\} $}
	\STATE create a new $\sampleoutput$ with averaged sample positions and attributes by merging all $\{\samplecandidate\}$  within $\samplecluster$ 
	\STATE $\sampleexistence(\sampleoutput) \assign \frac{\#\samplecluster}{\#\text{overlapping}\ \sampleneigh\ \text{over}\ \sampleoutput} $
	\COMMENT{existence assignment}
	\IF{$\sampleexistence(\sampleoutput) > 0.5$}
	\STATE $\{\sampleoutput\} \assign \sampleoutput$
	\ENDIF
	\ENDFOR
	\RETURN $\{\sampleoutput\}$
	\end{algorithmic}
  \Caption{Generating new output samples in existence assignment.}
{%

}
\label{alg:generate_new_output_samples}
\end{algorithm}

The candidate samples are generated from pairs of $\neighinput$ and $\neighoutput$ (lines 5-12 in \Cref{alg:generate_new_output_samples}). 
In a pair of $\neighinput$ and $\neighoutput$, if there is an unmatched input sample $\sampleinputprime$ in $\neighinput$ (line 7), it will indicate the potential lack of an output sample, whose global position is $\samplespace(\samplecandidate)= \sampleposition(\sampleinputprime)-\sampleposition(\sampleinput)+\sampleposition(\sampleoutput)$  located within $\neighoutput$, and attributes are the same to  $\sampleinputprime$ (line 8).
The algorithm loops over all pairs of neighborhoods, each neighborhood pair may or may not produce new candidate samples. 
\nothing{
Note we only generate $\samplecandidate$ with $\samplespace(\samplecandidate)$ within the output patch.
}%
All these samples $\samplecandidate$ form a candidate sample set $\{\samplecandidate\}$.
We group $\{\samplecandidate\}$ into clusters $\{\samplecluster\}$ (line 13-24) by assigning a sample to its nearest cluster $\samplecluster$ with distance between $\samplecandidate$ and the center of $\samplecluster$ smaller than $0.5 \samplingdistance$ or otherwise create a new cluster $\samplecluster^{\prime}$ using the sample.
For each cluster $\samplecluster \in \{\samplecluster\}$  (line 25-31), we merge all its candidate samples $\{\samplecandidate\} \in \samplecluster$ as one output sample $\sampleoutput$ by averaging their position and attributes using the same way as in the assignment step (\Cref{subsubsec:assignment_step}).
The existence of $\sampleoutput$ is assigned as the ratio of the number of candidates within $\samplecluster$ over the number of overlapping $\sampleneigh$ over the position of $\sampleoutput$.
For the sake of explanation, assume $\sampleexistence(\sampleoutput) = 1$, it means all $\sampleneigh$ overlapping over $\sampleoutput$ produce one candidate sample on average, which suggests there could be missing samples, around $\sampleoutput$. 
Finally,  $\sampleoutput$ with $\sampleexistence > 0.5$ is added into the output sample distribution every iteration.

\section{Parameters}
\label{sec:appendix:parameters}
\nothing{
\Cref{fig:regular_brick_wall:automatic_synthesis} neighborhood sizes 50,40,30, sampling distance 20, 15, 10.
\Cref{fig:teaser:maze:output} neighborhood sizes 40,30,20, sampling distance 20, 15, 10.
\Cref{fig:distorted_blocks:automatic_synthesis} neigborhood sizes 50,40,30, sampling distance 30, 20, 10.
\Cref{fig:distorted_grid:automatic_synthesis}, neighborhood sizes 60, 50, sampling distance 40, 30.
\Cref{fig:prisma:auto} neighborhood sizes 60, 50, 40, sampling distance 40, 30, 30.
}%

\begin{table}[thb]
  \centering
  \Caption{Parameters.}
  {From left to right: neighborhood radii and sampling distances from lower to higher hierarchies. 
  The parameters in the bottom part of the table share default values.
  Input size is bounding box size of input exemplar.
  }
  \label{tab:parameters}
  \begin{tabular}{| l |c | c| c |} %
    \hline 
    & $\neighborhoodradius$ & $\samplingdistance$ & input size\\
    \hline
    \Cref{fig:regular_brick_wall:automatic_synthesis} & $\{50, 40, 30\}$ & $\{20, 15, 10\}$ & 350 $\times$ 200 \nothing{338 $\times$ 189} \\		
    \Cref{fig:distorted_grid:automatic_synthesis} & $\{60, 50\}$  & $\{40, 30\}$ & 300 $\times$ 300 \nothing{311 $\times$ 307} \\
    \Cref{fig:teaser:maze:output} &  $\{40, 30, 20\}$ & $\{20, 15, 10\}$ & 250 $\times$ 250 \nothing{248 $\times$ 236} \\
    \Cref{fig:prisma:auto} & $\{60, 50, 40\}$ & $\{40, 30, 30\}$ & 250 $\times$ 250 \nothing{255 $\times$ 273} \\
    \Cref{fig:distorted_blocks:automatic_synthesis} & $\{50, 40, 30\}$ & $\{30, 20, 10\}$ & 300 $\times$ 350 \nothing{303 $\times$ 330} \\
    \hline
    \Cref{fig:teaser:branch:output} & \multirow{4}{*}{$\vdots$} & \multirow{4}{*}{$\vdots$} & 400 $\times$ 400 \nothing{398 $\times$ 380}\\
    \Cref{fig:ablation:full:1} &  &  & 400 $\times$ 300 \nothing{411 $\times$ 298}\\
    \Cref{fig:circuit_board:automatic_synthesis} &  &  & 500 $\times$ 400 \nothing{504 $\times$ 375}\\
    \Cref{fig:curve_recon:ours:1} &  &  & 350 $\times$ 300 \nothing{358 $\times$ 282}\\
    \Cref{fig:diffusion:automatic_synthesis} & \multirow{4}{*}{$\vdots$}  & \multirow{4}{*}{$\vdots$} & 350 $\times$ 200 \nothing{366 $\times$ 224}\\
   \Cref{fig:fabric:automatic_synthesis} & &  & 300 $\times$ 300 \nothing{294 $\times$ 290}\\
    \Cref{fig:fence:automatic_synthesis} &  &  & 200 $\times$ 300 \nothing{189 $\times$ 282}\\
    \Cref{fig:curve_recon:ours:2} & \multirow{4}{*}{$\{60,50,40\}$} & \multirow{4}{*}{$\{40,30,25\}$}  & 300 $\times$ 300 \nothing{ 295 $\times$ 279 }\\
    \Cref{fig:heart:automatic_synthesis} &   &  & 200 $\times$ 300 \nothing{ 221 $\times$ 324 }\\
    \Cref{fig:topographic_map:automatic_synthesis} &  &  & 400 $\times$ 400
    \nothing{ 407 $\times$ 410 }\\         
    \Cref{fig:random_patch_copy_converged:tree} &  &  & 350 $\times$ 400 
    \nothing{ 362 $\times$ 398 }\\
    \Cref{fig:ablation:full:2} & \multirow{4}{*}{$\vdots$}  & \multirow{4}{*}{$\vdots$}  & 400 $\times$ 250
    \nothing{ 381 $\times$ 239 }\\
    \Cref{fig:teaser:string:output} &  &  & 300 $\times$ 250
    \nothing{ 293 $\times$ 240 }\\
    \Cref{fig:teaser:strip:output} &   &   & 350 $\times$ 300
    \nothing{ 361 $\times$ 312 }\\
    \Cref{fig:random_patch_copy_converged:1} &  &  & 350 $\times$ 250
    \nothing{ 370  $\times$ 245 }\\
    \Cref{fig:comparison_pattern_synthesis:wave:our} &  &  & 500 $\times$ 300
    \nothing{ 496  $\times$ 276 }\\
    \Cref{fig:waves:automatic_synthesis} & \multirow{4}{*}{$\vdots$} & \multirow{4}{*}{$\vdots$} & 400 $\times$ 200
    \nothing{ 375  $\times$ 196 }\\
    \Cref{fig:wood_ring:automatic_synthesis} &  &  & 250 $\times$  350\\
    \Cref{fig:chain:automatic_synthesis} &  &  & 250 $\times$  400
    \nothing{ 235 $\times$ 379 }\\
    \Cref{fig:zentangle:automatic_synthesis} &  &  & 300 $\times$ 300 \nothing{ 276 $\times$ 312 }\\
    \Cref{fig:zigzag:auto} &  &  & 300 $\times$  250
    \nothing{ 311  $\times$ 241 }\\
    \hline
  \end{tabular}
\end{table}

\Cref{tab:parameters} lists the parameters for the results shown in the paper.

\section{Performance}
\label{sec:performance}

Our current implementation in C++ is unoptimized. It takes about 160 seconds to synthesize a pattern with about 750, 1000, 1300 output samples and 30, 30, 20 samples on average within neighborhoods at each hierarchy, on a desktop with AMD Ryzen 9 3950 X 3.49 GHz 16-core processor and 32 GB RAM.
The major computational burden is on the neighborhood searching process (\Cref{subsubsec:search_step}). 
The computational complexity of neighborhood searching mainly depends on the number of samples and neighborhood radius. 
To compute the similarity and sample matching between a pair of neighborhoods, the Hungarian algorithm \cite{Kuhn:1955:HMA} has complexity $O(\numneighsample^3)$ where $\numneighsample$ is the number of samples within a neighborhoood (assume all input and output neighborhoods have same number of samples).
The patch match algorithm \cite{Barnes:2009:PRC} is composed of two alternating steps: propagation and random search. 
In an optimization step, there are the $a\numOutput \maxiter$ neighborhood matching computation where $a$ is a constant (in our implementation $a \approx 10$) which is related to the number of neighboring samples to a sample used in the propagation step and the range of random search, 
$\numOutput$ is the total number of output samples, and
$\maxiter$ is the maximum number of patch match iteration. 
The complexity of an optimization step  is thus $O(\numOutput \numneighsample^3)$.
Our algorithm has no more than 3 hierarchies and each hierarchy need about 7 steps. The lower hierarchy has less samples but a larger neighborhood radius (\Cref{sec:method:sample_synthesis:hierarchical_synthesis}).

\section{Texture Synthesis}
\label{sec:texture_synthesis}

\begin{figure}[tbh]
  \centering
  \captionsetup[subfigure]{labelformat=empty}
  \setlength{\tabcolsep}{0pt}
  \begin{tabular}{cccc}
    \centering
    \subfloat[]{
      \includegraphics[width=0.20\linewidth]{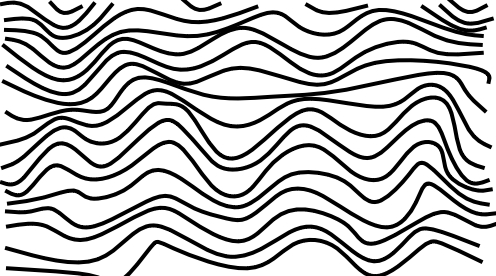}
    }%
    &\subfloat[]{
      \includegraphics[width=0.25\linewidth]{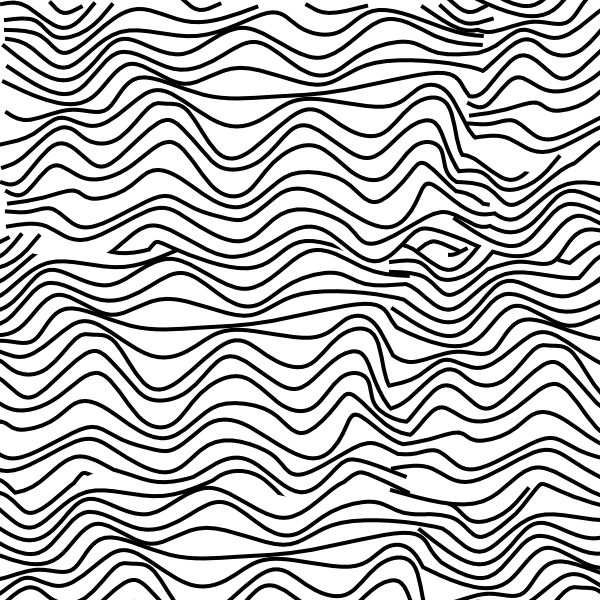}
    }%
    &\subfloat[]{
      \includegraphics[width=0.25\linewidth]{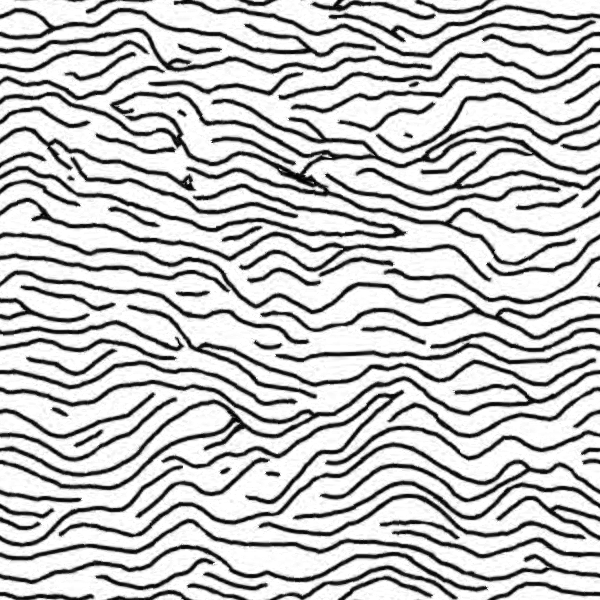}
    }%
    &\subfloat[]{
      \includegraphics[width=0.25\linewidth]{figs/results/vector_results/hier/g_curve_recon_global_waves_n0_60_n1_50_n2_40_level_2_iter_6.pdf}
    }%
    \vspace{-2em}
    \\
    \subfloat[]{
      \includegraphics[width=0.160\linewidth]{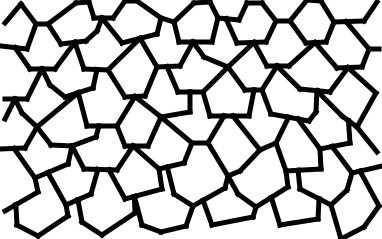}
    }%
    &\subfloat[]{
      \includegraphics[width=0.25\linewidth]{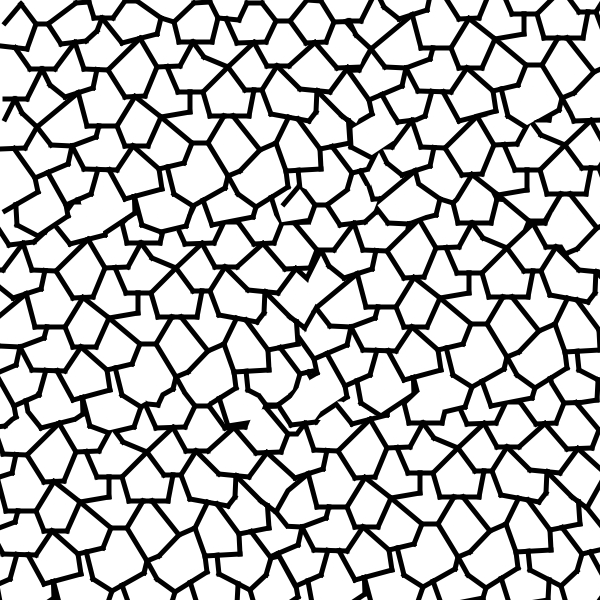}
    }%
    &\subfloat[]{
      \includegraphics[width=0.25\linewidth]{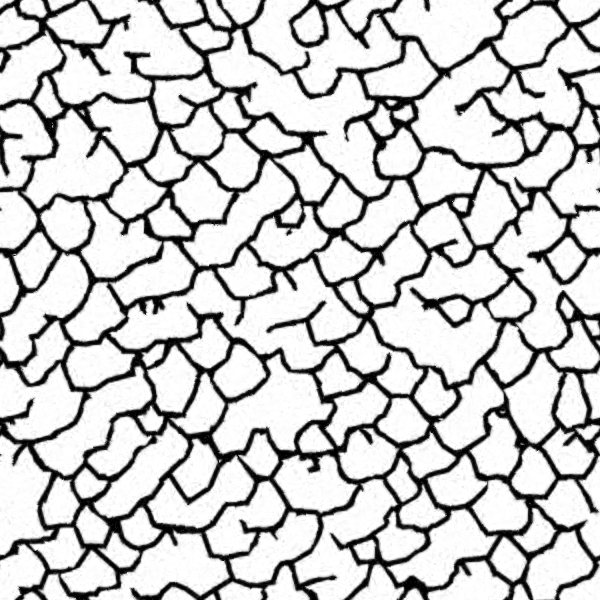}
    }%
    &\subfloat[]{
      \includegraphics[width=0.25\linewidth]{figs/results/vector_results/ours/g_curve_recon_roof_tiles_n0_60_n1_50_n2_40_level_2_iter_6.pdf}
    }%
    \vspace{-2em}
    \\
    \subfloat[\nothing{Exemplar}]{
      \includegraphics[width=0.17\linewidth]{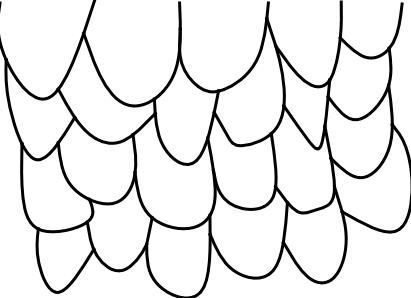}
    }%
    &\subfloat[\nothing{\cite{Cornet:2004:GTP}}]{
      \includegraphics[width=0.25\linewidth]{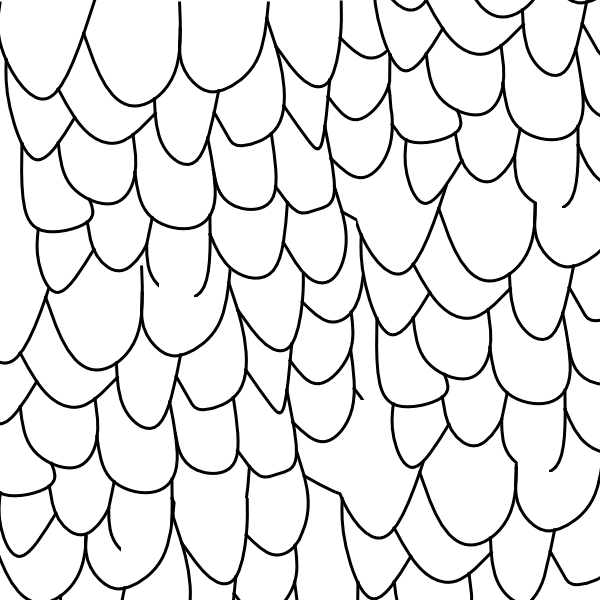}
    }%
    &\subfloat[\nothing{\cite{Wei:2016:TS}}]{
      \includegraphics[width=0.25\linewidth]{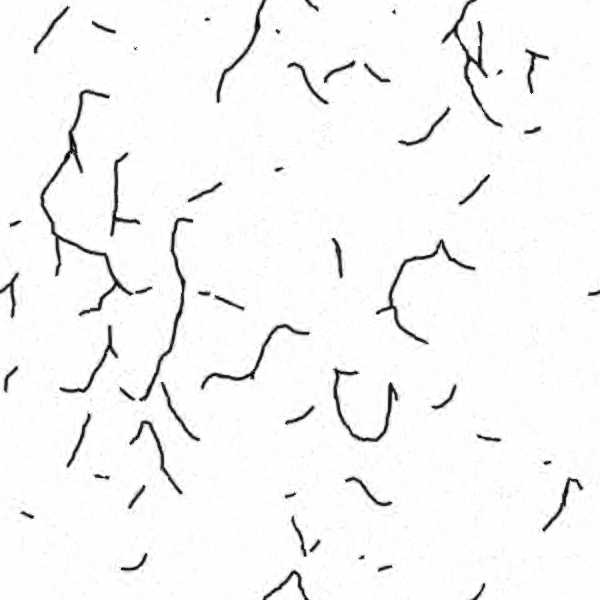}
    }%
    &\subfloat[\nothing{Ours}]{
      \includegraphics[width=0.25\linewidth]{figs/results/vector_results/ours/g_curve_recon_circles_pattern_corrected_n0_60_n1_45_n2_40_level_2_iter_10.pdf}
    }%
    \vspace{-2em}
    \\
    \subfloat[Exemplar]{
      \includegraphics[width=0.11\linewidth]{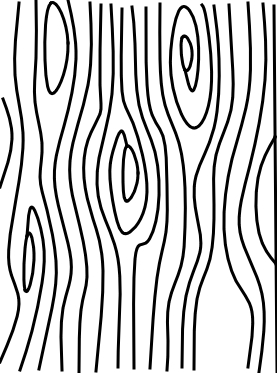}
    }%
    &\subfloat[\cite{Cornet:2004:GTP}]{
      \includegraphics[width=0.25\linewidth]{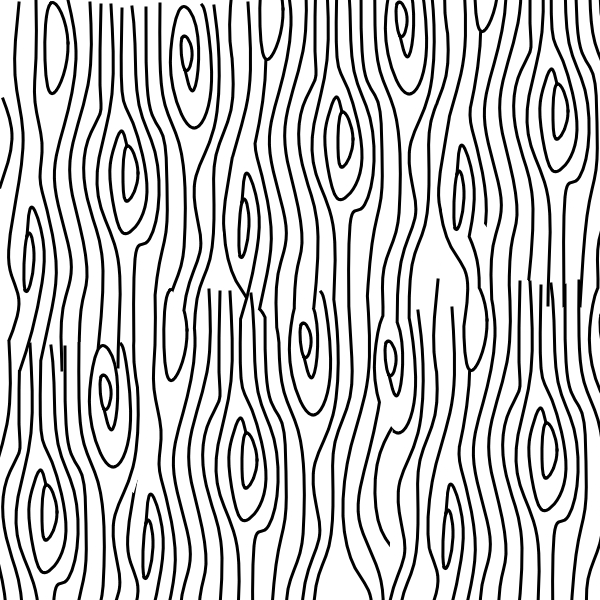}
    }%
    &\subfloat[\cite{Wei:2016:TS}]{
      \includegraphics[width=0.25\linewidth]{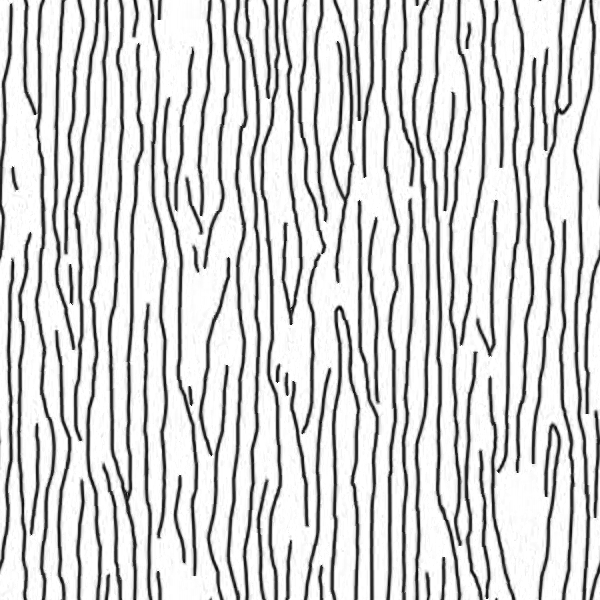}
    }%
    &\subfloat[Ours]{
      \includegraphics[width=0.25\linewidth]{figs/results/vector_results/ours/g_curve_recon_wood_ring_n0_60_n1_50_n2_40_level_2_iter_6.pdf}
    }%
  \end{tabular}
  \Caption{Comparison of our algorithm to texture synthesis.}
  {%
    Texture synthesis methods need additional vector-pixel conversions and need to process all pixels instead of just samples around patterns.
  }
  \label{fig:comparison_texture_synthesis}
\end{figure}

\nothing{
}%

\Cref{fig:comparison_texture_synthesis} shows the texture synthesis results by \nothing{the near state-of-the-art} graph cut \cite{Kwatra:2003:GTI} (using the implementation in the GIMP Texturize Plugin \cite{Cornet:2004:GTP}) and multi-resolution patch-match (using the implementation in \cite{Wei:2016:TS} with guidance channels \cite{Kaspar:2015:STT}).
As shown, pixel-based texture synthesis might not preserve continuous structures as well as our vector-based method.
These methods also need additional vector-pixel conversions and need to process all pixels instead of just samples around patterns.
\add{If standard texture synthesis methods are applied for vector patterns, the rasterization, synthesis, and vectorization process can introduce extra quality degradation and computation overhead, and thus might not be practical for interactive authoring as we present in the supplementary video.}%

\nothing{
}%

\ifthenelse{\equal{\final}{0}}
{
\clearpage
\pagenumbering{roman}
\section{Video}
\label{sec:video}

\nothing{
}%

\audio[background]
{
Repetitive patterns are fundamental for a variety of tasks in design and engineering.
Manually creating these patterns provides high degree of freedom but can also require significant technical and artistic expertise and manual labor.
}%
\video{
Show some examples of manual pattern authoring.
}%

\audio[prior_art]
{
Recent example-based methods can automate parts of the manual creation process, but are mainly tailored for discrete elements instead of more general continuous structures.
}%
\video{
Show discrete element and continuous patterns, selected from Pinterest.
}%

\paragraph{algorithm}

\audio[alg_intro]
{
We propose an example-based method to synthesize continuous curve patterns from exemplars.
}

\audio[alg_idea]
{

Our main idea is to extend prior example- and sample-based element synthesis to consider not only sample positions, or geometry, but also their connectivity, or topology.
}

\video
{
	Show (key differences between) previous and our pattern representation (\Cref{fig:pattern_representation}).
}

\audio[alg_details]
{
Our method takes an input exemplar, samples a graph representation, and synthesizes a larger output graph to reconstruct continuous curves.
}

\audio[alg_optim]{

Our algorithm can run at multiple resolutions.
Initially, we perform synthesis and reconstruction at a coarse level using sparse representation.
We continue this process with a denser and denser representation to obtain the final result.
}

\video {
	Show input exemplar, input graph, output graph, output reconstructed pattern
}

\paragraph{results}

\audio[results]
{
We show some automatic synthesis results.
}

\paragraph{UI}

\audio[ui_intro]
{
For further quality improvement and customization, we also present an autocomplete user interface which is built upon the core synthesis algorithm.

The interface allows users interactively create and edit patterns to maintain full control.
}%

\audio[ui_exemplar]
{
Users can first create a small exemplar using our system, or import vector patterns from external sources.
}

\audio[ui_autocomplete]
{
In the autocomplete mode, users can specify an output domain and our system generates predicted patterns that resembles what the users already draw.
}

\audio[ui_clone]
{
To give the users more control, in the clone mode, users can also specify a source region and clone it to a target region.
The synthesized patterns resemble the source patterns and seamlessly connect with what have already been drawn.
}

\audio[ui_manual]
{
Users can manually correct unsatisfactory parts within the predictions, or ask to resynthesize selected regions.
}

\video{
Show main features of our system, including autocomplete, workflow clone, and iterative editing.
}%

\video
{
Show results, including fully automatic synthesis and interactive authoring.
A key point here is to show that while automatic synthesis can produce reasonable results, it is not perfect, so enlisting user input can help, without burdening them with full manual authoring.
Show the number of autocomplete and manual strokes.
}

\audio[thank]
{
Thank you.
}%

\section{Fast-forward}
\label{sec:fast_forward}

\audio[background]
{
Repetitive curve patterns are ubiquitous, but can require high artistic expertise and manual labor to create.
}
\video{
Show some beautiful pattern images in our daily lives.
}%

\audio[prior]
{
These issues can be addressed by automatic synthesis, but prior methods mainly focus on discrete elements instead of continuous structures.
}%
\video{
Show discrete elements and our continuous structures.
}%

\audio[auto]
{
We propose an example-based method to synthesize continuous curve patterns from a variety of user exemplars. 
}%
\video{
Show more pattern synthesis input-output pairs.
}%

\audio[ui]
{
This method is integrated into our interactive authoring system, which has two major functions: autocomplete and clone, to generate predicted patterns.
For further quality improvement and customization, the users can  accept, partially accept or reject the predictions.
}
\video{
Show an interactive authoring video clip.
}%

\audio[idea]
{
Our main idea is to extend prior example- and sample-based element synthesis to consider not only sample positions, or geometry, but also their connectivity, or topology.
}%
\video{
Show key difference of representations between our and previous methods.
}%

\section{Submission Information}
\label{sec:sis}

\subsection{SIGGRAPH Asia 2020}

\paragraph{Authors}

Feel free to edit; the PhD adviser is usually listed as the last author.

\begin{enumerate}

\item
Peihan Tu

\item
Li-Yi Wei (on the committee)

\item
Koji Yatani

\item
Takeo Igarashi

\item
Matthias Zwicker
\end{enumerate}

\paragraph{Topic Areas}
\begin{description}
\item[Imaging/Video:] 1
\item[Interaction/VR:] 2
\end{description}

\paragraph{Keywords}
\begin{description}

\item[Imaging/Video:]
Texture Synthesis and Inpainting

\item[Interaction:]
Assistive Interfaces

\item[Modeling/Geometry:]
Procedural and Data-driven Modeling
\end{description}

\paragraph{Summary (maximum 30 words)}

We propose an example-based method to synthesize continuous curve patterns from user exemplars and an autocomplete user interface to facilitate interactive creation and iterative editing.

\section{Schedule}
\label{sec:todo}

\subsection{\href{https://sa2020.siggraph.org/submissions/technical-papers}{SIGGRAPH Asia 2020}}

\paragraph{Submission form deadline}

\todo{May 21, 2020 (22:00 GMT)}

\paragraph{Paper deadline}

\todo{May 22, 2020 (22:00 GMT)}

\section{Internship}
\label{sec:internship}

\paragraph{When}

May 26 to August 28 2020

\paragraph{Where}

Adobe San Jose

\paragraph{Who}

Peihan and Li-Yi, other collaborators include 
\begin{description}
\item[Rubaiat Habib]
UI expert and Li-Yi's long time collaborator with experience in pattern authoring, e.g. \cite{Kazi:2012:VIT}.

\item[Tarun Beri]
On the Illustrator product side who is a top expert on authoring vector graphics patterns).

\item[Paul Asente]
Who has done some related works in discrete elements \cite{AlMeraj:2013:TEE,AlMeraj:2013:PGT}.

\end{description}

\section{Discussion with Reviewers}

Dear reviewers:

Thank you for your comments.

We have modified our paper to address the two main required changes, as highlighted in color (orange for addition, blue for deletion): discussion and comparison to \cite{Roveri:2015:EBR}, and discussion about artifacts in \Cref{fig:comparison_sample_synthesis,subsec:examplepattern,sec:conclusion}.
Meanwhile, we are continuing to address other more detailed review suggestions.

Thank you,

The authors

\section{Blog}

\begin{figure}[htb]
  \centering
  \subfloat[raster]{
    \label{fig:example:raster}
    \includegraphics[width=0.4\linewidth]{161.jpg}
  }%
  \subfloat[vector]{
    \label{fig:example:svg}
    \includegraphics[width=0.58\linewidth]{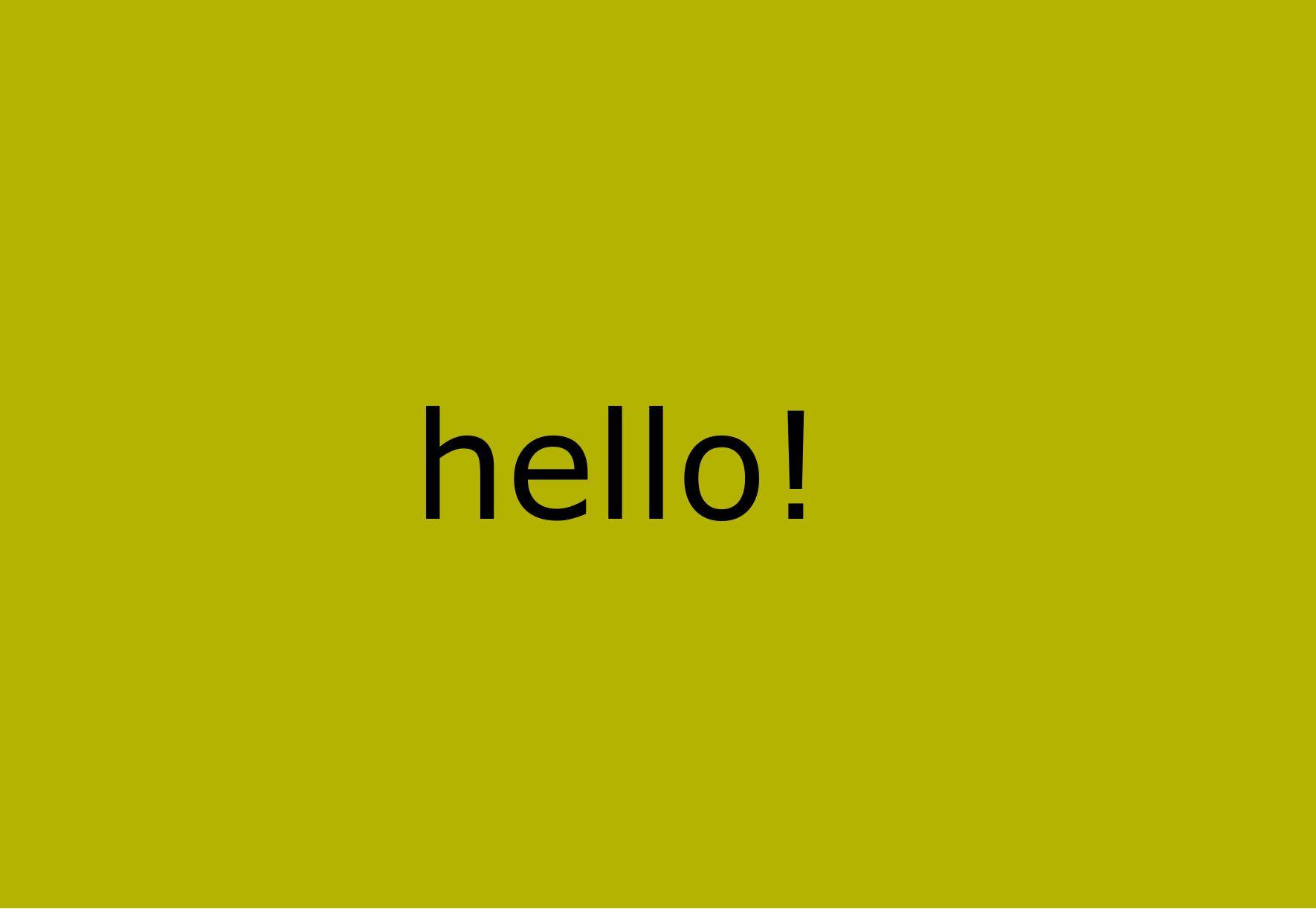}
  }%
  
  \subfloat[vector + raster]{
    \label{fig:example:svg_img}
    \includegraphics[width=0.9\linewidth]{svg_img_example.pdf}
  }%
 \Caption{Example figure.}
 {%
\subref{fig:example:raster} is a raster image and \subref{fig:example:svg} is a vector graphics.
Never, ever, rasterize vector graphics unless you want large size and low quality files.
We can also combine vector and raster graphics as in \subref{fig:example:svg_img}.
Use Inkscape to import  an image into an svg file, and draw over it using whatever stuff like texts or strokes.
Make sure the image is linked not embedded to avoid duplicating the image, and the link should be relative in the same path \cite{StackExchange:2011:HLI}.
 }
 \label{fig:example}
\end{figure}

\begin{description}
\item[December 12, 2020]
{

}
\item[October 17, 2020]
{
}
\item[October 10, 2020]
{

}

\item[October 8, 2020]
{

}

\item[October 3, 2020]
{

}

\item[September 23, 2020: log email]
{

}

\item[September 21, 2020]
{
}
\item[September 21, 2020]
{
}
\item[September 21, 2020]
{

}

\item[September 12, 2020]
{

}

\item[September 4, 2020]
{

}

\item[August 26, 2020]
{
}
\item[August 23, 2020]
{

}

\item[August 21, 2020]
{

}
\item[August 11, 2020]
{

}

\item[August 7, 2020]
{

}
\item[August 4, 2020]
{

}
\item[August 2, 2020]
{
}
\item[July 30, 2020]
{

}
\item[July 8, 2020]
{
}
\item[July 7, 2020]
{
}
\item[June 24,2020]
{

}

\item[June 22, 2020: animated vector textures]
{
[continuation from the previous post and today's meeting]

}

\item[June 20, 2020: element/pattern animation]
{
[continuation from the previous post]

}
\item[June 19, 2020]
{

}
\item[June 18, 2020]
{

}

\item[June 17, 2020]{

}

\item[June 16, 2020]
{

}

\item[June 15, 2020]
{

}

\item[June 12, 2020]
{
}
\item[June 11/12, 2020]{

}

\item[June 10, 2020]{

}

\item[June 9, 2020: patent]
{
}
\item[June 9, 2020]{

}

\item[May 30, 2020]{

}

\item[May 29, 2020]{

}

\item[May 28, 2020]{

}

\item[May 27, 2020]
{

}
\item[May 26, 2020]
{
}
\item[May 21, 2020]
{
}
\item[May 19, 2020]
{
}
\item[May 18, 2020]
{
}
\item[May 18, 2020]{

}

\item[May 17, 2020]
{
}
\item[May 11, 2020]
{
}
\item[May 10, 2020]
{

}
\item[April 29, 2020]{

}

\item[April 26, 2020]{

}

\item[April 24, 2020]{

}

\item[April 21, 2020]
{
[continuation from the previous post]

}
\item[April 19, 2020]{

}

\item[April 18, 2020]
{
}
\item[April 17, 2020]{

}

\item[April 16, 2020]
{
[continuation from the previous post]

}
\item[April 15, 2020]
{

}
\item[April 10, 2020]{

}

\item[April 9, 2020]{

}

\item[April  5, 2020]{

}

\item[April  3, 2020]{

}

\item[March 30, 2020]
{

}

\item[March 27, 2020]
{
}
\item[March 26, 2020]{

}

\item[March 24, 2020]{

}

\item[March 22, 2020]{
}

\item[March 18, 2020]
{

}

\item[March 17, 2020]
{
}

\item[March 6, 2020]{

}%

\item[March 3, 2020]
{
}

\item[March 2, 2020]
{
}
\item[March 2, 2020]{

}

\item[February 24, 2020]
{
[continuation from the previous post]

}
\item[February 23, 2020]{

}

\item[February 16, 2020]
{

}

\item[February 8, 2020]{

}

\item[January 20, 2020]
{
[continuation from the previous post]

}
\item[January 19, 2020]
{

}

\item[January 17, 2020]
{
}
\item[January 17, 2020]{

}

\item[January 16, 2020]
{
}
\item[January 14, 2020]
{
[continuation from the previous post]

}
\item[January 13, 2020]{
}

\item[January 7 ,2020]{
}
\item[January 2, 2020]
{
}

\item[January 2, 2020]
{

}
\item[January 2, 2020]
{

}

\item[January 1, 2020]{
}

\item[December 31, 2019]{

}

\item[December 28, 2019]
{
[continuation from the post before the previous one]

}
\item[December 27, 2019 - Pattern design: some results and some plans]
{
[log email to Koji, Takeo, and Matthias]

}
\item[December 25, 2019]
{
[continuation from the previous post]

}

\item[December 24, 2019]
{

}

\item[December 23, 2019]
{
[continuation from previous 3 posts]

}

\item[December 22, 2019]{
}

\item[December 21, 2019]{

}

\item[Decembr 20, 2019]{

}

\item[December 19, 2019]{
}

\item[December 11, 2019]
{
[continuation from the previous post]

}
\item[Decmber 9, 2019]{

}

\item[December 4, 2019]
{
}

\item[November 26, 2019]{

}

\item[November 25, 2019]
{
}
\item[November 24, 2019]{

}

\item[November 23, 2019]{

}

\item[November 20, 2019]
{

}

\item[November 11, 2019]
{
}
\item[November 10, 2019]
{
}
\item[November 7, 2019]
{

}

\item[October 30, 2019]
{	
}

\item[October 28, 2019]
{
[continuation from the previous post]

}
\item[October 26, 2019]
{

}
\item[October 25, 2019]
{
[continuation from the previous post]

}
\item[October 24, 2019]{

}

\item[October 23, 2019]
{
[continuation from the previous post]

}
\item[October 22, 2019: Tiling]{

}

\item[October 19, 2019: reconstruction]
{
[continuation from the previous post]

}
\item[October 18, 2019]{

}

\item[October 16, 2019]
{
[continuation from the previous post]

}
\item[October 15, 2019]{

}

\item[October 15, 2019]
{
[continuing discussions from the previous post]

}
\item[October 11, 2019]
{
[continuing discussions from the previous post]

}
\item[October 10, 2019]{

}

\item[September 30, 2019]
{
}
\item[September 26, 2019]{

}

\item[September 24, 2019]
{
[continuation from the previous post]

}
\item[September 23, 2019]{

}

\item[September 22, 2019]
{
}

\item[September 16, 2019]
{

}

\item[September 15, 2019]
{
}

\item[September 14, 2019]
{
}
\item[September 1, 2019]
{

\item[September 1, 2019]
{
}
}
\item[August 31, 2019]{

}

\item[August 31, 2019]
{ 
[continuation from prior posts from August 28, 2019]

}
\item[August 29, 2019]{

}

\item[August 28, 2019]{

}

\item[August 26, 2019]{

}

\item[August 25, 2019: Email: autocomplete pattern design]{

}

\item[August 22, 2019]
{

}

\item[August 20, 2019: Code: autocomplete pattern design]
{

}
\item[August 18, 2019]{

}

\item[August 16, 2019]{
}

\item[August 13, 2019]
{
}
\item[August 11, 2019]{

}

\item[July 23, 2019]
{
}

\item[July 21, 2019]
{
}
\item[July 19, 2019]{

}

\item[July 17, 2019]{

}

\item[July 2, 2019]{

}

\item[June 30, 2019]{

}

\item[June 22, 2019]
{
}

\item[June 19, 2019]{
}

\item[June 17, 2019]{

}

\item[June 2, 2019]{

}

\item[May 23, 2019]
{
}
\item[May 22, 2019]{
	
}

\item[May 20, 2019]{

}

\item[May 19, 2019]{

}

\item[May 16, 2019]{

}

\item[May 13, 2019]{
}

\item[May 9, 2019: hierarchical synthesis and interactive learning]{

}

\item[April 26, 2019: greedy algorithm for graph matching]{

}

\item[April 19, 2019: continous pattern synthesis with edge matching]{

}

\item[April  18, 2019: meeting note]{

}

\item[April 17, 2019]
{
}
\item[April 16, 2019]
{
}

\item[April 15, 2019: meeting note]{
}
\item[April 5, 2019]{

}

\item[April 1, 2019]
{
}
\item[March 22, 2019]{

}

\item[March 14, 2019]
{
}
\item[March 10, 2019]
{
}

\item[March 9 ,2019]
{
}

\item[March 8, 2019]
{

}

\item[March 4, 2019]{
}%
\item[February 28, 2019: meeting notes]
{
}
\item[February 13, 2019]
{
[continuation from the previous post]

}
\item[February 9, 2019]
{

}

\item[February 7, 2019]
{
[continuation from previous posts]

}
\item[February 5, 2019]
{

}

\item[February 1, 2019]
{

}

\item[January 31, 2019: hierarchical element synthesis]
{

}
\item[January 29, 2019]{
}%
\item[January 28, 2019]
{
}
\item[January 27, 2019]{
 
}

\item[January 18, 2019]
{
[continuation from the previous post]

}
\item[January 16, 2019]
{

}

\item[December 30, 2018]
{
[replying to the previous thread]

}
\item[December 27, 2018]
{
}
\item[July 7, 2018]
{

}

\item[July 6, 2018]
{

}
\item[July 2, 2018]
{
[continuation from the previous post]

}
\item[June 28, 2018]
{

}

\item[June 27, 2018: autocomplete element pattern design]
{
}
\item[June 26, 2018]
{
	
}

\item[June 22, 2018]
{

}
\item[June 17, 2018: example driven project development]
{
[continuation from prior posts]

}
\item[June 14, 2018]
{

}

\item[June 11, 2018]
{

}
\item[June 6, 2018]
{
		
}

\item[June 4, 2018]
{
	
}

\item[May 23, 2018]
{
[continuation from the previous post]

}
\item[May 21, 2018]{

}

\item[May 20, 2018]
{
}
\item[May 19, 2018]{

}

\item[May 13, 2018]{

}
\item[May 6, 2018]
{

}

\item[May 1, 2018]
{

}

\item[April 29, 2018]
{
}
\item[April 26, 2018]
{

}

\item[April 20, 2018]
{
}
\item[April 15, 2018]
{
	
}

\item[April 14, 2018]
{

}

\item[April 10, 2018]
{
}

\item[April 9, 2018]
{

}

\item[April 6, 2018]
{

}

\item[April 4, 2018]
{

}

\item[March 29, 2018]
{

}

\end{description}

\label{subsec:PET} %

\section{Texture Categories}
\label{sec:categories}

Element pattern is a large family. In principle, our system allows users to design any type of element textures. 
But a same category of element textures is always designed with similar authoring pipeline and supported by similar method internally.    

Before we discuss user interface and method, we first classify textures into two categories largely with several sub-categories. Because it is easy to design individual geometric element and the most important component is the element interactions or arrangements, the classification is mostly based on the arrangements.

\paragraph{Simple textures} 
There are three requirements for being a simple texture: 1) their elements can be \textit{compactly} enclosed by eclipse, 2) there are no precise element interactions such as contact and 3) the element arrangement is global without sub-arrangement. There are two types of simple textures with 1) no adjacency and 2) overlap between elements, as shown in \Cref{fig:simple_texture:no_adjacency,fig:simple_texture:overlap}.
		
\paragraph{Complex textures}
If the above three requirements are not completely satisfied, textures are complex textures. Complex textures not satisfying 1) and 3) consist of local simple textures. We give examples showing complex textures not satisfying one, two or three of the above three requirements. We term unsatisfied requirements 1) 2) and 3) as non-convex element (NCE), element contact (EC) and sub-arrangement (SA), respectively. \Cref{fig:texture_categories} shows examples with descriptions.
More examples are shown in the result section and supplement.

\begin{figure}[htb]
	\centering
	\subfloat[Simple texture: no adjacency]{
			\label{fig:simple_texture:no_adjacency}
			\includegraphics[width=0.33\linewidth]{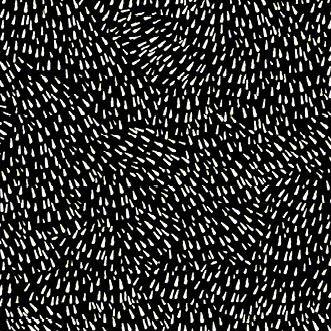}
	}%
	\subfloat[Simple texture: overlap]{
			\label{fig:simple_texture:overlap}
			\includegraphics[width=0.33\linewidth]{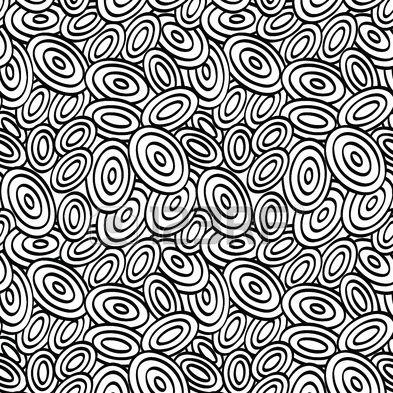}
	}%
	\subfloat[NCE]{
		\label{fig:complex_texture:NCE}
		\includegraphics[width=0.33\linewidth]{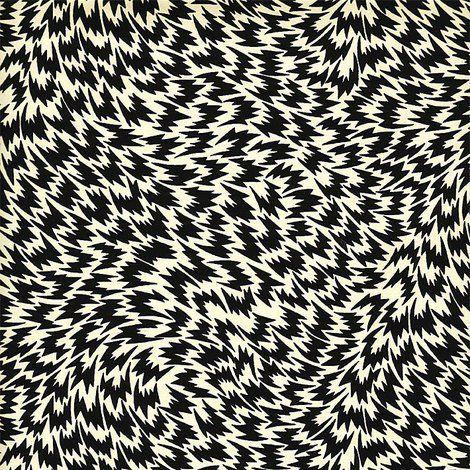}
	}%
	
	\subfloat[Curve-and-curve EC]{
			\label{fig:complex_texture:EC:curve_curve}
			\includegraphics[width=0.33\linewidth]{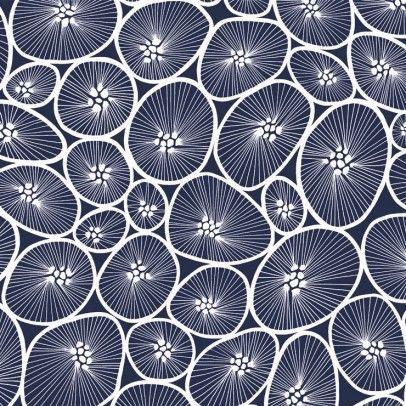}
	}%
	\nothing{
	\subfloat[Point-and-curve EC]{
		\label{fig:complex_texture:EC:point_curve}
		\includegraphics[width=0.33\linewidth]{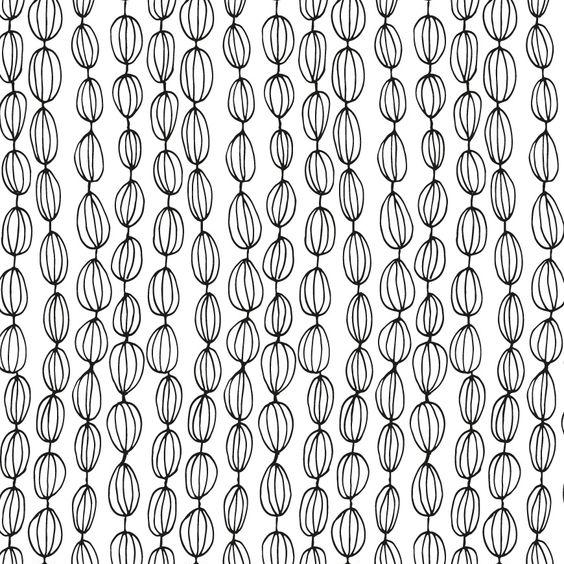}
	}%
}
	\subfloat[SA]{
			\label{fig:complex_texture:SA}
			\includegraphics[width=0.33\linewidth]{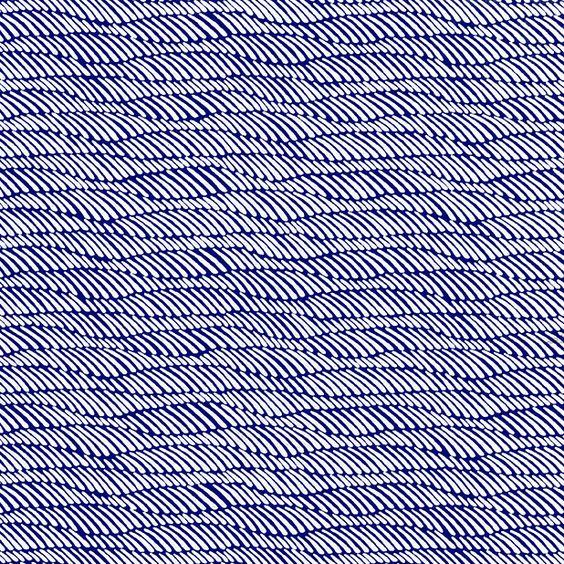}
	}%
	\subfloat[NCE and point-and-point EC]{
		\label{fig:complex_texture:NCE_EC}
		\includegraphics[width=0.33\linewidth]{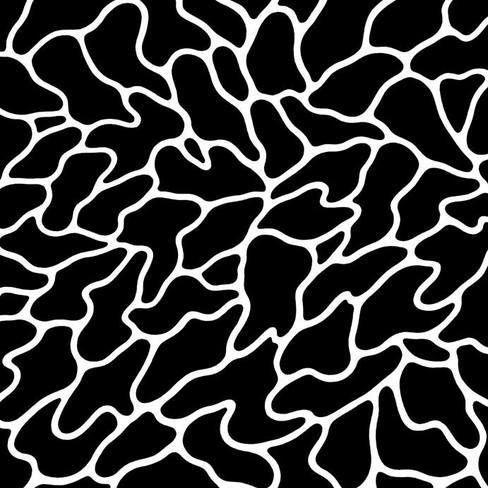}
	}%
	
	\subfloat[Point-and-curve EC and SA]{
			\label{fig:complex_texture:EC_SA}
			\includegraphics[width=0.33\linewidth]{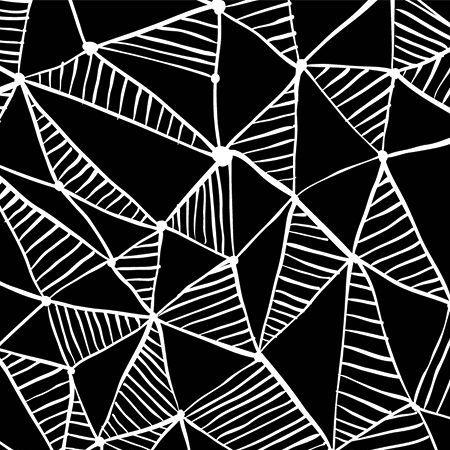}
	}%
	\subfloat[NCE and SA]{
			\label{fig:complex_texture:NCE_SA}
			\includegraphics[width=0.33\linewidth]{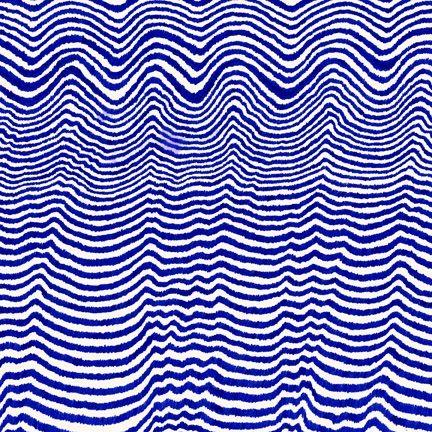}
	}%
	\subfloat[NCE, point-and-point EC and SA]{
			\label{fig:complex_texture:NCE_EC_SA}
			\includegraphics[width=0.33\linewidth]{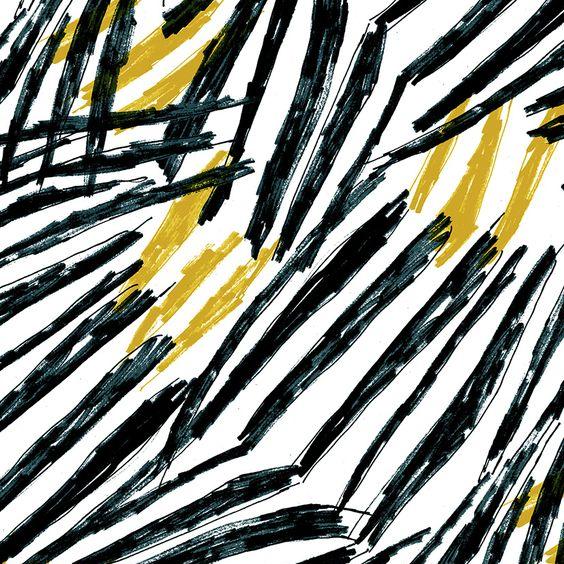}
	}%
	
	\Caption{Texture categories.}
	{%
		Simple textures with
		\subref{fig:simple_texture:no_adjacency} no adjacency  and  \subref{fig:simple_texture:overlap} overlap between elements.
		Complex textures with \subref{fig:complex_texture:NCE} non-convex elements,
		\subref{fig:complex_texture:EC:curve_curve} curve-and-curve element contacts where the irregular  circles are tangent with each other
		and \subref{fig:complex_texture:SA} sub-arrangement
		where elements are vertically and regularly distributed at large and compactly distributed along wave-line at small scale. 
		A complex texture with \subref{fig:complex_texture:NCE_EC} non-convex element and point-and-point element contacts where each element is an irregular short curve.
		Complex textures with
		\subref{fig:complex_texture:EC_SA} point-and-curve element contact and sub-arrangement, and 
		\subref{fig:complex_texture:NCE_SA} non-convex elements and sub-arrangements  where each element is a wave-line and the elements are non-uniformly and vertically distributed with compacter central section.
		A complex texture with
		\subref{fig:complex_texture:NCE_EC_SA} 
		non-convex element, point-and-point element contacts and sub-arrangements 
		where each element is a long and a little curvy brush stroke, the elements are regularly distributed locally and there are element contacts.		
	}
	\label{fig:texture_categories}
\end{figure}

\section{Method Old}

Unlike assigning sample locations and attributes which can be done individually for each sample, determining $\sampleedge$ during the assignment step is more constrained and thus more complicated.
This is because samples are isolated while $\sampleedge$ are directly connected with and thus constrained by nearby $\sampleedge$'s.

For each pair of output samples, we need to make a binary decision if $\sampleedgeoutput(\sampleoutput, \sampleoutputprime)$ is $1$ or $0$.
We formulate the edge assignment problem as labelling problem using Markov random field (MRF), where we want to find the optimal binary labellings of output edge set $\outputsampleedgeset$  so that the energy function $\edgeenergy$ defined using a MRF is minimized.
The energy function is defined as 
\begin{align}
\edgeenergy = \unaryenergy +  \binaryenergyweight\binaryenergy
\label{eq:edge_energy}
\end{align}
, where 
\begin{align}
\unaryenergy = \sum_{\sampleedgeoutput\in\outputsampleedgeset} \unarycost(\sampleedgeoutput) 
\end{align}
is the unary energy term. $\unarycost(\sampleedgeoutput)$ measures how appropriate a label (0 or 1) is for the edge $\sampleedgeoutput$,
and 
\begin{align}
\binaryenergy = \sum_{\{\sampleedgeoutput,\sampleedgeoutputprime\}\in \neighboringedgeset} \binarycost(\sampleedgeoutput,\sampleedgeoutputprime)
\label{eq:pairwise_energy}
\end{align}
is the pairwise energy term, and $\binarycost(\sampleedgeoutput,\sampleedgeoutputprime)$ evaluates the cost given by the neighboring edge pair $\sampleedgeoutput,\sampleedgeoutputprime$. $\binaryenergyweight$ is a weighting parameter.

We defined $\unarycost(\sampleedge)$ as 
\begin{equation}
\unarycost(\sampleedgeoutput)=\left\{
\begin{aligned}
-\log{\edgeconfidence}& , & \sampleedgeoutput=1, \\
-\log(1-\edgeconfidence) & , & \sampleedgeoutput=0.
\end{aligned}
\right.
\label{eq:unary_cost}
\end{equation}

$\binarycost(\sampleedgeoutput,\sampleedgeoutputprime)$ is defined as
\begin{align}
\binarycost(\sampleedgeoutput,\sampleedgeoutputprime)=
-\log{\jointedgeconfidence(\sampleedgeoutput,\sampleedgeoutputprime)}
\end{align}
where $\edgeconfidence$ and $\jointedgeconfidence$ are computed from a number of matched input neighborhoods over the output edge $\sampleedgeoutput$ or edge pair $\sampleoutput$ and $\sampleoutputprime$.
\begin{align}
\edgeconfidence=\frac{\sum{\sampleedgeinput}}{\nummatchedneighborhoods}
\end{align}
where $\sum{\sampleedgeinput}$ sums over all matched input edge of $\sampleedgeoutput$ (if there is no matched edge, $\sampleedgeinput=0$). 
$\nummatchedneighborhoods$ is the total number of matched input neighborhood overlapped over $\sampleedgeoutput$. 
If every input neighborhood indicates there is an input edge matched with  $\sampleedgeoutput$, $\edgeconfidence=1$ and thus $\unarycost(\sampleedgeoutput)=0$.
\begin{equation}
\jointedgeconfidence(\sampleedgeoutput,\sampleedgeoutputprime)=\frac{\sum{\mathbbm{1}(\sampleedgeoutput=\sampleedgeinput, \sampleedgeoutputprime=\sampleedgeinputprime)}}{\nummatchedneighborhoods}
\end{equation}
where $\mathbbm{1}(\cdot)$ is an indicator function that equals one if its condition $\cdot$ holds, and otherwise zero. $\sum$ sums over all matched input neighborhoods over the edge pair $\sampleedgeoutput$ and $\sampleedgeoutputprime$.

We optimize the \Cref{eq:edge_energy} using quadratic pseudo-boolean optimization (QPBO) methods \cite{Rother:2007:OBM}.

\subsubsection{Core Idea}
During user interaction, the system predict what the user might want to draw next. These predictions might or might not meet the user intention. 

User responses to the system predictions indicate whether the predictions are good or bad to the user. Our goal is to find a set of parameters that improves the quality of predictions. In other words, the future predictions should be similar to accepted predictions and existing user drawings but dissimilar to  bad predictions. 

\subsubsection{Optimization}
We calculate the set of parameters via minimizing the following optimization objective
\begin{align}
\max_{nsize,sampling} sim(exemplar, pred_{accept}) + dissim(examplar ,pred_{ignore})
\label{eqn:objective}
\end{align}
where $sim$ and $dissim$ indicates similarity and dissimilarity criteria. $pred$ indicates predictions. $nsize$ indicates neighborhood size.

\nothing{	
Our method includes two technical components.
\paragraph{Element extraction}
Users do not necessarily to specify each individual element (which is tedious).
Our system analyzes the user workflow and identify elements to enable subsequent synthesis.

\paragraph{Hierarchical Analysis and Synthesis}
In \cite{Ma:2011:DET}, each element has fixed sample representation during synthesis (e.g uses fixed number of samples with fixed locations relative to element object).
We extend \cite{Ma:2011:DET} to hierarchical synthesis for acceleration and improving the quality especially when dealing with structured pattern.
To this end, element is sampled hierarchically by analyzing its shape and interactive user responses.

\subsection{Hierarchical Synthesis}
\label{subsec:hierarchical_synthesis}
Here, we assume the hierarchical sample representations of element are already analyzed, which will be detailed in \Cref{subsec:hierarchical_analysis}.

\subsubsection{Neighborhood Metric}

We extend the patten optimization framework \cite{Ma:2011:DET} to hierarchical synthesis.
The pattern similarity is evaluated by summing up sample neighborhood similarities. Let $\sampleneigh(\samplesym)$ denote the spatial neighborhood of a sample $\samplesym$, consisting of samples within its spatial extent defined by a user specified size. The distance between two neighborhoods $\neighinput$ and $\neighoutput$ is defined as 
\begin{equation}
\label{eq:neighborhood_metric}
\euclidean{\neighinput-\neighoutput}
\end{equation}

\subsubsection{Pattern Metric}
Pattern metric is computed via comparing local neighborhoods defined in \Cref{eq:neighborhood_metric}.
\begin{equation}
\euclidean{\neighinput-\neighoutput}
\end{equation}

\subsubsection{Optimization}

\subsubsection{Graph similarity}

For continuous structures, both sample and edge similarity, namely graph similarity, should be taken into account. We define graph similarity using graph edit distance \cite{Sanfeliu:1983:DMA}, as follows
\begin{equation}
\differencesym{\editcost}(\graph,\graph\prime) = \min \sum_{\editop \in \editopset(\graph,\graph\prime)} \editcost(\editop)
\label{eq:graph_similarity}
\end{equation}
where $\editopset$ denotes the set of edit operations transforming $\graph$ to $\graph\prime$, which includes node/edge insertion, deletion and substitution. $\editcost(\editop)$ is the cost of the edit operation $\editop$.

Exact computation of \Cref{eq:graph_similarity} is intractable. We approximate \Cref{eq:graph_similarity} using the algorithm presented in \cite{Riesen:2009:AGE}. 
Riesen et al \shortcite{Riesen:2009:AGE} embeds edge information into samples, and reduce the quadratic assignment problem to linear one that can be readily solved via Hungarian algorithm \cite{Kuhn:1955:HMA}.
The sample similarity is appended as~\cite{Riesen:2009:AGE}
\nothing{
	\begin{equation}
	\differencesym{\samplevec}(\samplesym,\samplesymprime)=\big(\differencesym{\samplespace}(\samplesym,\samplesymprime),\weightappr\differencesym{\sampleappr}(\samplesym,\samplesymprime), \weighttemp\differencesym{\sampletemp}(\samplesym,\samplesymprime), \editcost(\sampleedge(\samplesym),\sampleedge(\samplesymprime))  \big)
	\end{equation}
}%
\begin{equation}
\differencesym{\samplevec}(\samplesym,\samplesymprime)=\big(\differencesym{\samplespace}(\samplesym,\samplesymprime),\weightappr\differencesym{\sampleid}(\samplesym,\samplesymprime), \editcost(\sampleedge(\samplesym),\sampleedge(\samplesymprime))  \big)
\label{eq:graph_sample_similarity}
\end{equation}

where $\editcost(\sampleedge(\samplesym),\sampleedge(\samplesymprime))$ is minimum edge edit cost implied by sample substitution between $\samplesym$ and $\samplesym\prime$ \cite{Riesen:2009:AGE}. Again, $\editcost(\sampleedge(\samplesym),\sampleedge(\samplesymprime))$ can be computed via Hungarian algorithm \cite{Kuhn:1955:HMA}.
Though such approximation gives only sub-optimal solutions,

\subsection{Hierarchical Analysis}
\label{subsec:hierarchical_analysis}
In some cases, it is sufficient to use a single sample (e.g. \Cref{fig:simple_texture:no_adjacency}) to encode the location and shape of an element object. 
In this case, it is acceptable to synthesize at a single resolution without hierarchy.
In other cases, it is required to apply hierarchical, multi-sample representation since the elements might have complex topology or interactions between each other (e.g. \Cref{fig:complex_texture:EC:curve_curve}).

To generate hierarchical sample representations, instead of defining it manually and statically via only analyzing patterns, our system also learns from interactive user responses.

During authoring, the generated predictions might or might not conform to user intentions. 
If the users tend to accept the predictions, it suggests that the synthesis quality is acceptable. Thus it's not necessary to use more samples. If the users tend to ignore the predictions, the pattern synthesizer needs to be improved either by changing its matching neighborhood size or using denser sample representation.
}%

\section{Method Overview}
\label{sec:overview}

\begin{figure*}[htb]
	\centering
	
	\includegraphics[width=0.9\linewidth]{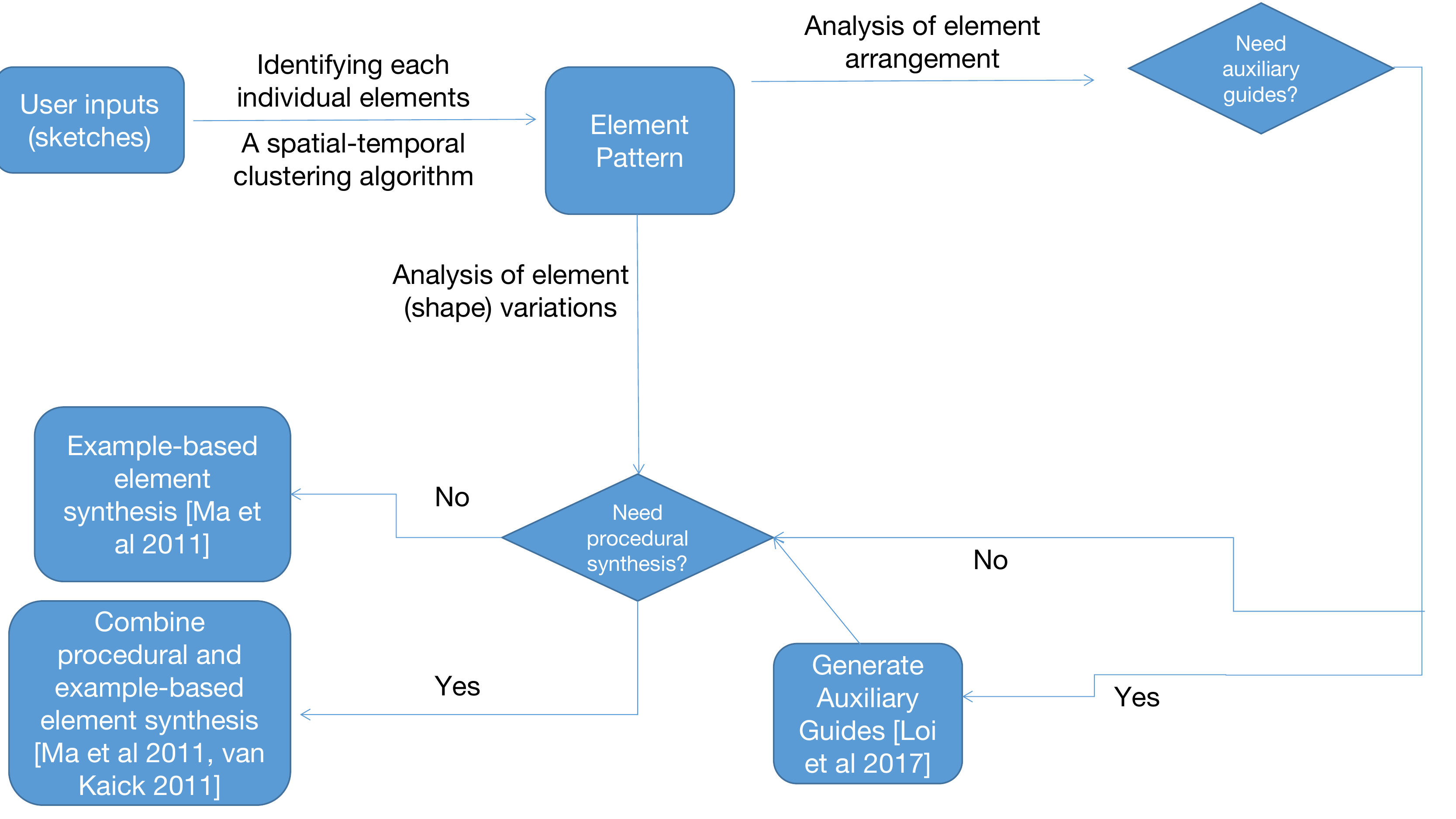}

	\Caption{Overview of our method.}
	{%
		Our method takes user input sketches. A spatial-temporal clustering algorithm is used to identify each individual elements.
		 Global arrangement and shape variations of the elements are analyzed. 
		 Specifically, if the elements are distributed regularly, our algorithm would generate auxiliary guides to help to preserve structures in the subsequent synthesis step.
		 If the element shape are dissimilar (have shape transformation among elements), our method combine example-based \cite{Ma:2011:DET} and procedural shape interpolation \cite{Van:2011:SSC} to generate predicted elements.
	
	}
	\label{fig:method_flowchart}
\end{figure*}

We represent each element and each auxiliary as a collection of samples that record global position and relative position within it --relevant information for example analysis and texture design. We also incorporate element orientation, size and other attributes deduced from inverse procedural modeling (section~\ref{sec:ElementDesign}) as additional information into each element explicitly. These information is intuitively well-understood by the users for better controls.
Here, we briefly overview the internal method of our system .
\paragraph{Element design} 
The design starts with generating geometric elements.

As the users draw elements, our system automatically create a procedural element, representing infinite number of fixed examples in parametric space, by inverse procedural modeling. The autocompleted elements by the system come from such procedural element space.

\nothing{
To make a visually pleasing texture, the distributed elements should be of slightly different appearance, e.g. undergoing shape transformations.}

\nothing{
Our system helps users to create \textit{procedural element}, representing infinite number of fixed examples in parametric space, by inverse procedural modeling. 
We interpolate multiple user-provided elements by finding shape correspondences \cite{Van:2011:SSC}.
}

\nothing{

}

\nothing{
\paragraph{Element interaction design} To describe interactions among elements and between elements and auxiliaries, the users can provide examples with distributions of elements and auxiliaries.
Our system also provides the users functions of designing \textit{procedural element interactions} by interpolating several examples \cite{Matusik:2005:TDU}.
}

\paragraph{Implicit auxiliaries generation}
When the users specify element arrangements via drawing or copy-and-paste, our system predicts the global arrangements. It is usually difficult to generate convincing global structure from local arrangements by example-based methods. 
Also, we find element arrangements can be commonly classified into regular and non-regular that can be expressed by concise grammar \cite{Loi:2017:PAE}. Therefore, our system predicts global auxiliary structures via predicting its grammars though which limit the search space, as in \cite{Nishida:2016:ISU}. These generated auxiliary guides help to determine the position of predicted element placement.

The auxiliaries prediction is made considering both user input and interactive user responses (e.g. accept or reject the system-generated predictions) via learning-based approach. The prediction is directly influenced and conditioned by user input, while interactive user responses reflect preferences \cite{Zsolnai:2018:GMS}. We train a neural network which receives user input and generate global auxiliary guides. A preference function is defined to re-score network outputs.

\nothing{
Generating auxiliary lines or points helps to determine and predict the position of element placement. 
We revise and extend partition operators in \cite{Loi:2017:PAE} to procedurally generate auxiliaries. Generated auxiliaries are marked with type attributes. 
}

\paragraph{Textures auto-completion} Textures auto-completion predicts what the textures might look like.
The predicts are synthesized by formulating a texture optimization problem as \cite{Ma:2011:DET} simultaneously considering previous workflows on element placement, generated auxiliary guides and explicit control maps. Our method provides good controls by manipulating weights among samples and energy terms. 

\paragraph{Merging} If necessary, some complex textures might be easier to design by merging two or multiple textures together \cite{Loi:2017:PAE}.

\section{Grammar}
\label{sec:grammar}

To simplify notations, we call both element $e$ and auxiliary line or point $al$ as object $obj$.
We represent each object with a collection of samples $s$ and various attributes $\Theta$.
Each sample $s$ records 1) its spatial position at $\mathbf{p(s)}$ 2) a sample-id $id(s)$ for the relative position within $e$ or $al$. Various attributes $\Theta$ include type-id1 $id_{t1}$ which indicates it belongs to element or  auxiliary line, type-id2 $id_{t2}$ for identifying which type of element or auxiliary line it is, group-id $id_g$ that records which group it affiliates with within a type of element or auxiliary line.
Additionally, for element object, the users might explicitly append orientation $\vec o$, size $k$ and extra attributes $\{\theta\}$ deduced from inverse procedural modeling to $obj$.

Formally, we identify an object as
\begin{equation*}
	obj=\langle \{\mathbf{u}(s)\},\Theta,\Phi \rangle
\end{equation*}
\begin{equation*}
	\mathbf{u}(s)=(\mathbf{p}(s),id_s(s))
\end{equation*}
\begin{equation*}
	\Theta=\langle id_{t1}, id_{t2}, id_g \rangle \qquad
	\Phi = \langle \vec o,k,\{\theta\} \rangle
\end{equation*}

$obj^0$ refers to element object, namely $id_{t1}=0$, and $obj^1$ refers to auxiliary. Specifically, auxiliary point only has one sample.

\section{Geometric Element Design by inverse procedural modeling}
\label{sec:ElementDesign}

\nothing{
\subsection{Procedural element from examples}
\label{subsec:ProcElem}
}

The users draw several element examples that are of extreme shape transformation (e.g. one is circle, the other is square), with user specified or system deduced sample points.
Our system can automatically deduce the shape transformation grammar and interpolate between example elements (inverse procedural modeling). This is done by finding shape correspondences and gradually warping from one example element to another \cite{Van:2011:SSC}. Formally, this process is expressed as

\begin{equation*}
	\begin{aligned} 
	proceduralElement():\Big\{\langle \{\mathbf{u}_{ij}(s)\}^{N_1}_{j=1},\Theta({id_{t1}=0}), \vec o, k \rangle\Big\}^n_{i=1} \rightarrow\\
	\{pEle\}=\Big\{\langle\{\mathbf{u}^{\prime}_{ij}(s)\}^{N_1}_{j=1},\Theta({id_{t1}=0}), \vec o, k, \{\theta_i\}  \rangle\Big\}^\infty_{i=1}
	\end{aligned} 
\end{equation*}
where $N_1$ is the number of samples that are used for shape matching within the element, and $n$ refers to the number of examples provided by the users.
To disentangle deduced $\{\theta\}$ from well-defined $\vec o$ and $k$, the example elements have equal $\vec o$ and $k$. The inverse procedural modeling gives the users a set with infinite number of elements controlled by parameter $\{\theta\}$.

\nothing{
If two elements are not sufficient for such interpolation, the users can provide additional example element. The element drawing process could be aided by ``Beautification" function provided by \cite{Xing:2015:AHA} (with some modifications).
}

\nothing{

}
\nothing{
\subsection{Procedural element interaction from examples}
\label{subsec:ProcElemDist}

The users can provide element interaction examples that are composed of several elements sampled from procedural element space as well as auxiliaries.  
Our system will automatically analyze the interactions and map them onto a space filled with procedurally generated auxiliaries (section~\ref{sec:AssistingOperator}). The users can design \textit{procedural element interaction} by interpolating several user provided examples. This process is formulated as 
\begin{equation}
\label{eq:ProceduralInteraction}
	\begin{split} 
	&proceduralInteraction(): \\
	&\Big\{\big\{\langle \{\mathbf{u}_{jm}(s)\}^{N_2}_{j=1},\Theta_m, \vec o_m, k_i,\{\theta_m\}\rangle\big\}^M_{m=1}\Big\}^n_{i=1} \rightarrow \\
	&\{pInct\} = \Big\{\big\{\langle \{\mathbf{u}^{\prime}_{jm}(s)\}^{N_2}_{j=1},\Theta_m, \vec o^\prime_m, k_i,\{\theta_m\}\rangle\big\}^M_{m=1},\{\eta_i\}\Big\}^\infty_{i=1}
	\end{split}	
\end{equation}
where $M$ is the total number of elements and auxiliaries within the provided example interaction. $N_2$ is the number of samples for encoding the interactions among elements and auxiliaries. 
Note that $N_2$ does not necessarily equal to $N_1$. 
$\{\eta\}$ is the deduced parameter controlling the interactions among elements and auxiliaries.
In the process of deducing procedural element interaction, the objects are rigid and only allowed to be translated or oriented (e.g. from $\vec o_i$ to $\vec o^\prime_i$ in Eq.~\eqref {eq:ProceduralInteraction} ).

}
\nothing{
Similar to \ref{subsec:ProcElem}, the users can also define example distributions themselves, and our system deduce its procedural parameters so that users can use a single or multiple parameters to control \textit{local} element distributions \cite{Emilien:2015:WIE}. The users may draw proxy \cite{Hurtut:2009:ASE} or sample points \cite{Ma:2011:DET} to represent interaction between elements. Also, the users might need to explicitly specify some special element distributions, such as contact and overlap. For instance, the users can assign particular mark to sample point indicating contact point.

}

\section{Texture Auto-completion}
\label{sec:autocompletion}
Our synthesis method is built upon the optimization framework \cite{Ma:2011:DET,Xing:2014:APR,Xing:2015:AHA}. However, different from \cite{Xing:2014:APR,Xing:2015:AHA}, 1) our method adopt batch synthesis instead of incremental synthesis for each element
2) although the design process is interactive and dynamic, we do not consider temporal differences in the similarity measure because the auxiliaries are labeled into different types by procedural generation, providing hard-core information on similarity
3) the similarity measure is extended for allowing element deformation
4) because of additional auxiliaries' labels, our method does not rely on search step for finding element placements with similar neighborhood

\nothing{
Given auxiliaries $\{\auxiliary\}$ (section~\ref{sec:AssistingOperator}), procedural element $\{pEle\}$ (section~\ref{sec:ElementDesign}), previous placement workflow, additional parameter maps $\{C\}$ controlling procedural objects and a interaction map $\text{H}$,   our system compute batch  by optimizing an energy function as follows.

\begin{equation}
\label{eq:energy}
E = E_{s} + E_{c} 
\end{equation}
where $E_s$ and $E_c$ represent similarity and constraint terms, respectively. 
}

\subsection{Similarity measure}

\subsubsection{Sample similarity}

\subsubsection{Element similarity}

\subsubsection{Neighborhood similarity}

\cite{Ma:2011:DET,Xing:2014:APR}

The neighborhood similarity is defined by summing up the distances of matched samples \cite{Ma:2011:DET,Xing:2014:APR,Peng:2018:A3S} as well as their connectivities within two neighborhoods $\neighoutput$, $\neighinput$:
\begin{equation}
\begin{aligned}
\distance{\neighoutput-\neighinput}=\sum_{
	\substack{\sampleoutputprime\in\neighoutput \\
		\sampleinputprime\in\neighinput \\
		\sampleinputprime=\match(\sampleoutputprime)}}
\distance{\differencesym{\samplevec}(\sampleoutput,\sampleoutputprime)-\differencesym{\samplevec}(\sampleinput,\sampleinputprime)}
+
\distance{\edgelength(\sampleoutput,\sampleoutputprime) - \edgelength(\sampleinput,\sampleinputprime)}
\end{aligned}
\label{eq:neighborhood_similarity}
\end{equation}
,
where $\match$ is the \nothing{one-to-one }sample matching function. In a neighborhood, there might be samples from both discrete elements and continuous structures. We only match samples with the same id, from the same type of patterns, and from the same type of elements within an element pattern. 
As in \cite{Ma:2011:DET,Ma:2013:DET}, we apply Hungarian algorithm \cite{Kuhn:1955:HMA} to build sample matches that minimizes \Cref{eq:neighborhood_similarity}.

\subsection{Constraint}
$E_s$ only measure the similarity between local element interactions and provided interaction examples. However, the users might also want to control the element shape, size when not violating the specified interactions. For example, in Fig.~\ref{fig:curve_curve_EC}, the users might want to deform the element while keeping it contact to the auxiliary lines. This is possible by finding a good trade off between $E_s$, and $E_c$ which is defined as the euclidean norm between specified  and actual parameters. 

\subsection{Optimization}
\subsubsection{Initialization}
We initialize the predicted element placement based on local similarity of the past placement by averaging.

\subsubsection{Assignment Step}

\subsubsection{Weighting}
If the users specify contact relation between objects, our system would assign higher weight to corresponding sample-pairs for keeping contact relation during synthesis.

\nothing{

}

\begin{figure}[htb]
	\centering
	\subfloat[before]{
	\label{subfig:workflow clone_before}
	\includegraphics[width=0.5\linewidth]{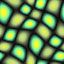}
	}
	\subfloat[after]{
	\label{subfig:workflow clone_after}
	\includegraphics[width=0.5\linewidth]{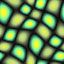}
	}

	\Caption{Workflow Clone}
	{
		With learning mode off, the user can clone the pattern to a larger region, our system can generates large amount of predictions.
	}
	\label{fig:ui_workflow_clone}
\end{figure}
Our system supports two types of functions  1)  workflow clone 2) interactive learning.

\subsection{Auto-complete}
\label{subsec:autocomplete}

	Our system allows the users to draw patterns as usual  \cite{Xing:2014:APR}. our system can \textit{autocomplete} what the users might want to draw next. 
	As will be discussed in \Cref{subsec:interactive_learning}, the system may generate multiple autocompleted predictions. For selected predictions, the users can accept, or reject the predictions.

	\nothing{
		Similar to \cite{Xing:2014:APR}, our system allow the users to draw patterns as usual.
		
		This process can also be applied to auxiliary structures which helps to determine elements arrangements. 
		We treat both elements and auxiliaries as basic component in the design process. The only difference is that elements are visible but auxiliary guides are not.
	}

	\subsection{Interactive Learning}
	
	\label{subsec:interactive_learning}
	
	The system can refine the autocompleted predictions by observing user responses. 
	In the interactive learning mode, the system generates two (more is also possible) predictions, the users can choose which one is better and perform further edits on it (\Cref{subsec:autocomplete}), and our system refine the synthesis parameters in the background.
	
	If the algorithm already produce sufficiently good predictions. The users can choose to disable the interactive learning mode. 
	If the users find the interactive learning fails to help to the system produce better and better results, the users can restart the learning process.

	\subsection{Workflow clone}
	
	Workflow clone tool allows the users to generate predictions over a specified large domain. This tool is useful especially when the interactive learning successfully helps the system to generate satisfied predictions.

\section{User Interface Old}
\label{sec:interfaceold}

\subsection{Scope}
\label{subsec:scopeold}
Our system provides great flexibility on designing \textit{2D} element textures that are composed of identifiable geometric elements. In principle, our system can design any element textures. 
The two key functions we provide are that 1) element placement auto-completion and 2) workflow clone -- both reduce the users effort on repetitive element placement. 
The more repetitions occur in expected textures, the more useful the system is. The worst case is that the users do not use any provided functions, and just place and tune element shape and locations by themselves.

The system currently only provides functions for designing \textit{black and white} element textures. But it can be easily extend to colorful textures within our framework.

Also, the design process only follows users' requirement on \textit{texture visual effect}. Our system does not consider any physical constraints \cite{Martinez:2015:SAO} required by fabrication. Such constraints should be explicitly provided by the users in the design process. We leave it as a future work.

\subsection{Basic mode}
\label{subsec:BasicMode}
We assume all inputs are in vector graphics format; if not, we vectorize raster inputs first.

Similar to \cite{Xing:2014:APR}, our system let the users to draw individual element and specify global and local arrangements via copy-and-paste as usual. 
This process can also be applied to auxiliary structures which helps to determine elements arrangements. 
We treat both elements and auxiliaries as basic component in the design process. The only difference is that elements are visible but auxiliary guides are not.

Our system can 1) \textit{autocomplete} what the users might want to do next, and the users can either accept or reject the predictions. 
We also let the users to clone previous workflows via \textit{workflow clone}. The users can also provide additional control map for manipulating the autocompletion process.

\subsection{Advanced mode}
\label{subsec:AdvancedMode}
As will be mentioned in \Cref{sec:method}, our system generates implicit procedural auxiliary guides for assisting basic mode (\Cref{subsec:BasicMode}). Because for most artists, they may not like to edit grammars as such workflow is very different from those in existing pattern design system. 

However, procedural grammar provides a great flexibility on editing. The users can optionally choose advanced mode and edit auxiliary guides explicitly. Advanced mode can let the users design more complex patterns.

\subsubsection{Advanced functions}
\label{subsubsec:AdvancedFunctions}
The advanced mode provides several additional functions for pattern design.

\nothing{
As briefly mentioned in \Cref{subsec:scope}, our system provides two key functions 1) global and local element arrangement generation and 2) element placement auto-completion.

For any type of textures, the design process is generally as follows

\paragraph{Specifying procedural elements}

Our system let the users create example elements from scratch, or they can import elements from external files. To improve element diversity, the users might draw several elements of the same type. These elements and their variants would form the final textures.
For both simple and complex textures, this step is the same.

Sometimes, element itself is a texture. Understandably, our system allows the users to design the element as a texture first. 

}%

\paragraph{Explicit auxiliaries generation}
The users select an enclosed region and generate auxiliary guides with grammars, instead of workflow clone and autocompletion during usual drawing process.

\nothing{
For simple textures, it is not necessary to recursively generate auxiliaries. A complex texture can consist of locally simple textures, which requires recursive auxiliaries generation.
Element can also be retagged as auxiliaries.

To generate auxiliaries (lines or points), \textit{Mask} defines enclosed regions where auxiliaries are generated. \textit{Auxiliaries Generator}, which includes many \textit{Assisting Operators} (please see section~\ref{sec:AssistingOperator}), is used to generate auxiliaries within a selected region. Auxiliaries are also represented by multiple samples with user specified spacing. 
 The generated auxiliaries by identical operator have identical attributes (e.g. type). The system lets the users to edit these attributes (section~\ref{sec:grammar}) conveniently via \textit{Intelligent Selector} and \textit{Intelligent Parameter Setter}.

\paragraph{Control map}
The auto-completion is controlled by the previous workflow as well as user specified control maps. These control maps might incorporate local orientation or size information.  They are generated by interpolating user's hints in 2D space.

\paragraph{Element placement and auto-completion}
Finally, the users can drag the designed elements onto the auxiliaries and tune the position. 
During placing, the system would predict final textures in real time. The users can either accept the predicts or reject them by continuing element placement.
For a particular element interaction -- contact, they should specify the contact relation between elements and auxiliaries.
}

\paragraph{Texture tuning}
If the users are not satisfied with what they get after texture creation, due to its procedure nature, it's easy for the users to modify the auxiliary structures and our system gives feedbacks about the modified textures to the users in real-time.

\subsubsection{Suggestions in advanced mode for different textures}
\label{subsubsec:SuggestedWorkflow}
We have classified different textures into categories in \Cref{sec:categories}. Our system is very flexible so that there are different procedures for designing an identical texture.
Here, we describe suggested workflow for the users to design different types of textures in \textit{advanced mode}.
\paragraph{Simple textures}
For designing simple textures, basic mode with explicit control maps is sufficient. Of course, the users can choose advanced mode if they like. If so, we suggest the users to generate auxiliary points.

\paragraph{Complex textures}
For complex textures, there are three points that are required to be considered.
\begin{description}
	\item[Non-convex element] We suggest the users to utilize auxiliary lines (straight or non-straight) instead of only points to assist its position characterization.
	\item[Element contact]  There are three types of contact: point and point, point and curve, curve and curve. 
	Understandably, only auxiliary points, both points and lines, and only lines are suggested to assist their interaction characterization, respectively.  
	\item[Sub-arrangments] Auxiliaries should be generated recursively. In each subregion, local textures might be simple and can be designed with workflow for simple textures. 
\end{description} 

\nothing{
\paragraph{Simple textures}
For simple textures, we suggest the users to generate auxiliary points regularly, or non-regularly with \cite{Li:2010:ABN,Wei:2010:MBN} and control maps for placing elements.

\paragraph{Complex textures}
Auxiliaries are used to assist describing element interactions. The suggested workflow for  complex textures are as follows.
\begin{description}
	\item[Non-convex element] We suggest the users to utilize auxiliary lines (straight or non-straight) instead of only points to assist its position characterization.
	\item[Element contact]  There are three types of contact: point and point, point and curve, curve and curve. 
	Understandably, only auxiliary points, both points and lines, and only lines are suggested to assist their interaction characterization, respectively.  
	\item[Sub-arrangments] Auxiliaries should be generated recursively. In each subregion, local textures might be simple and can be designed with workflow for simple textures. 
\end{description} 
}

\nothing{
\subsection{Examples}
\label{subsec:examples}
\Cref{fig:noadjacency:1,fig:noadjacency:2} show the texture grammar of simple textures. As described in \Cref{subsec:GeneralWorkflow,subsec:SuggestedWorkflow}, simple textures are generated with auxiliary points and control maps.

\Cref{fig:curve_curve_EC,fig:point_curve_EC_SA,fig:SA} show the design process of complex textures. 
In \Cref{fig:curve_curve_EC}, auxiliary points are used to determine the center of irregular circle and lines to assist characterizing the element contacts.
\Cref{fig:point_curve_EC_SA} shows a texture with point-and-curve EC and SA. As suggested in \Cref{subsec:SuggestedWorkflow}, both auxiliary lines and points are used to characterize such element contacts and the element arrangement is generated recursively.
In \Cref{fig:SA}, we retag the elements as auxiliary lines for specifying element arrangements.

More examples are included in the supplementary material.
}

\nothing{
\begin{figure}
	\centering
	\includegraphics[width=\linewidth]{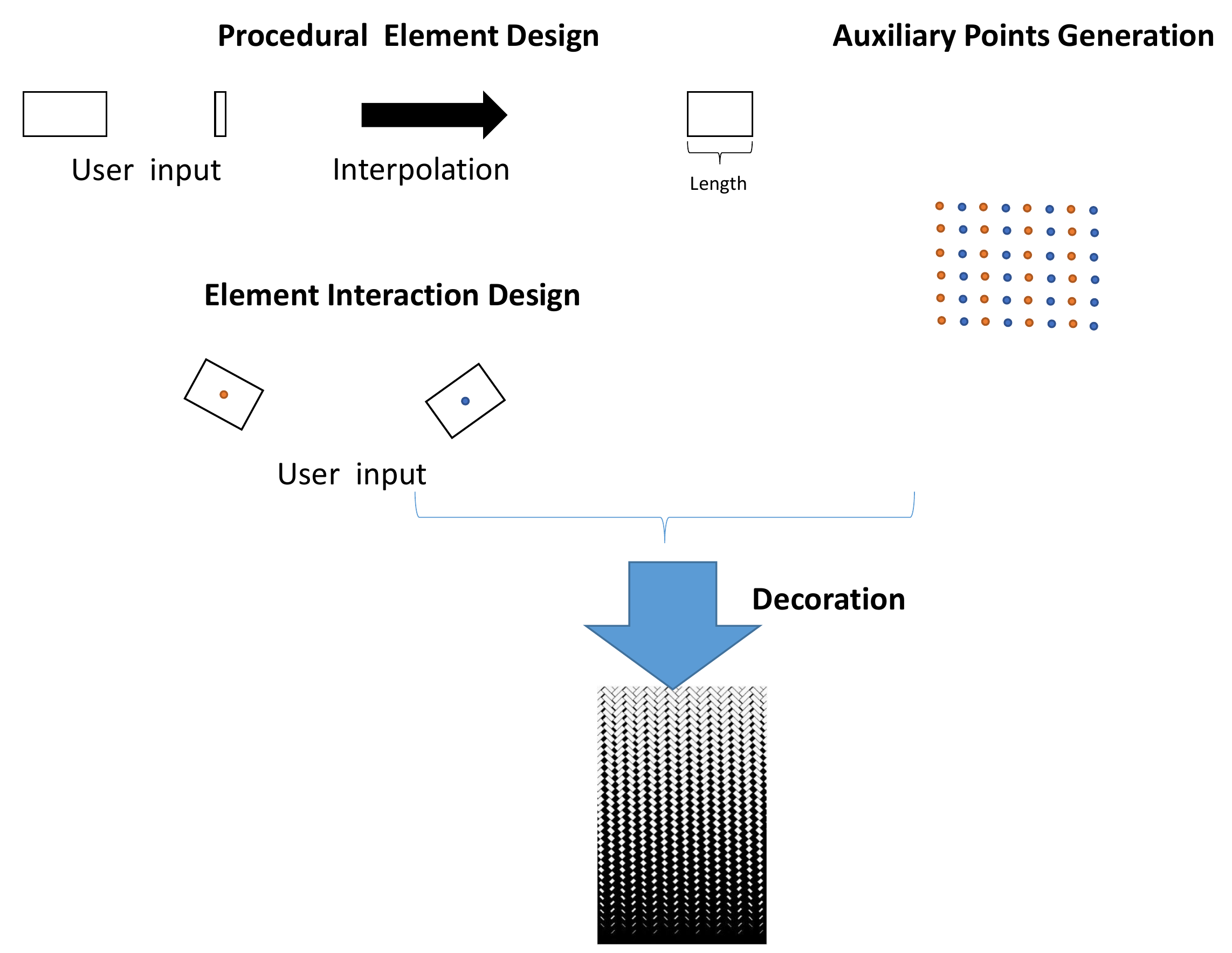}
	\Caption{No adjacency 1\cite{Tu:2018:NA1}}
	{
	}
	\label{fig:noadjacency:1}
\end{figure}
\begin{figure}
	\centering
	\includegraphics[width=\linewidth]{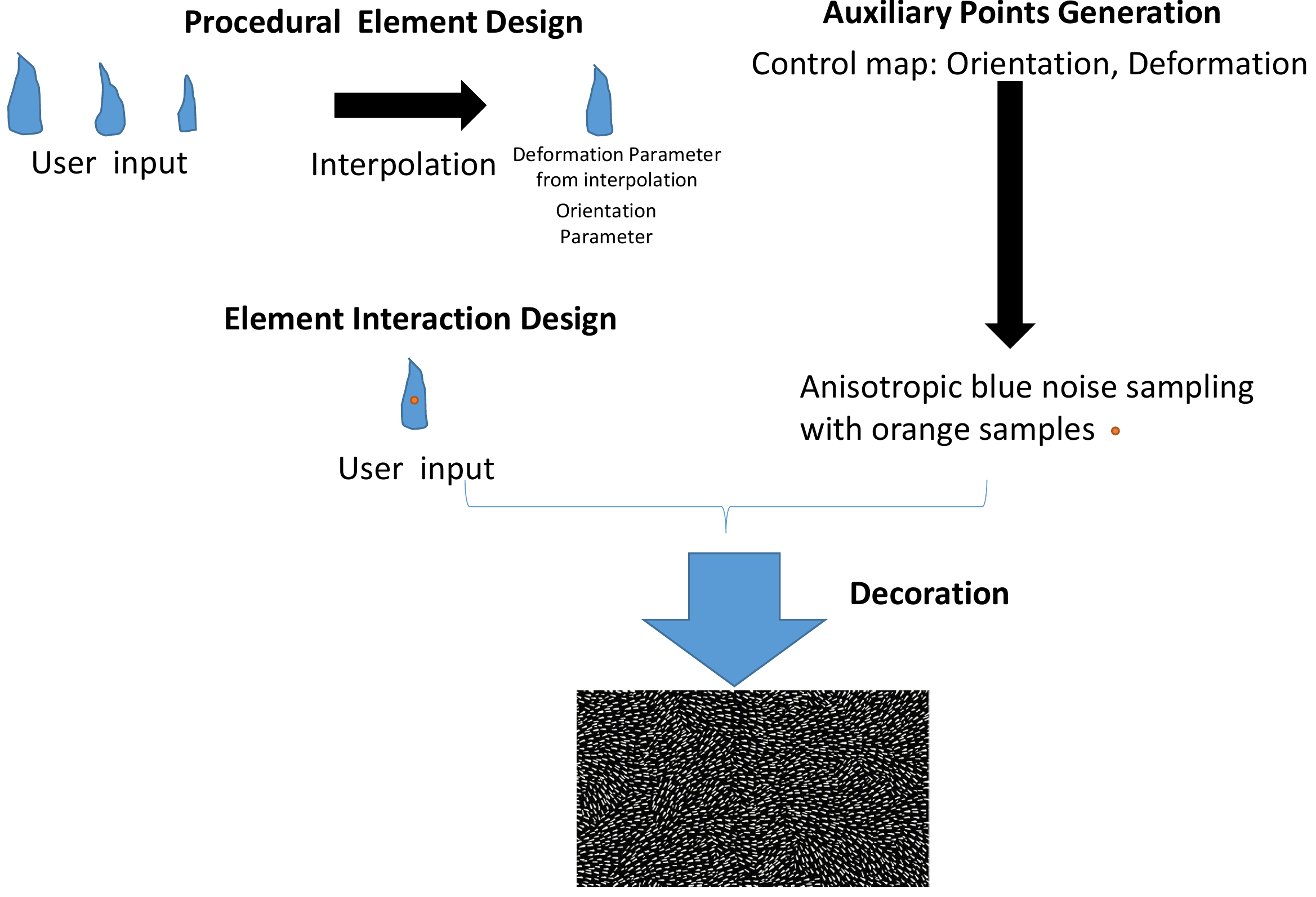}
	\Caption{No adjacency 2\cite{Tu:2018:NA2}}
	{
	}
	\label{fig:noadjacency:2}
\end{figure}
\begin{figure}
	\centering
	\includegraphics[width=\linewidth]{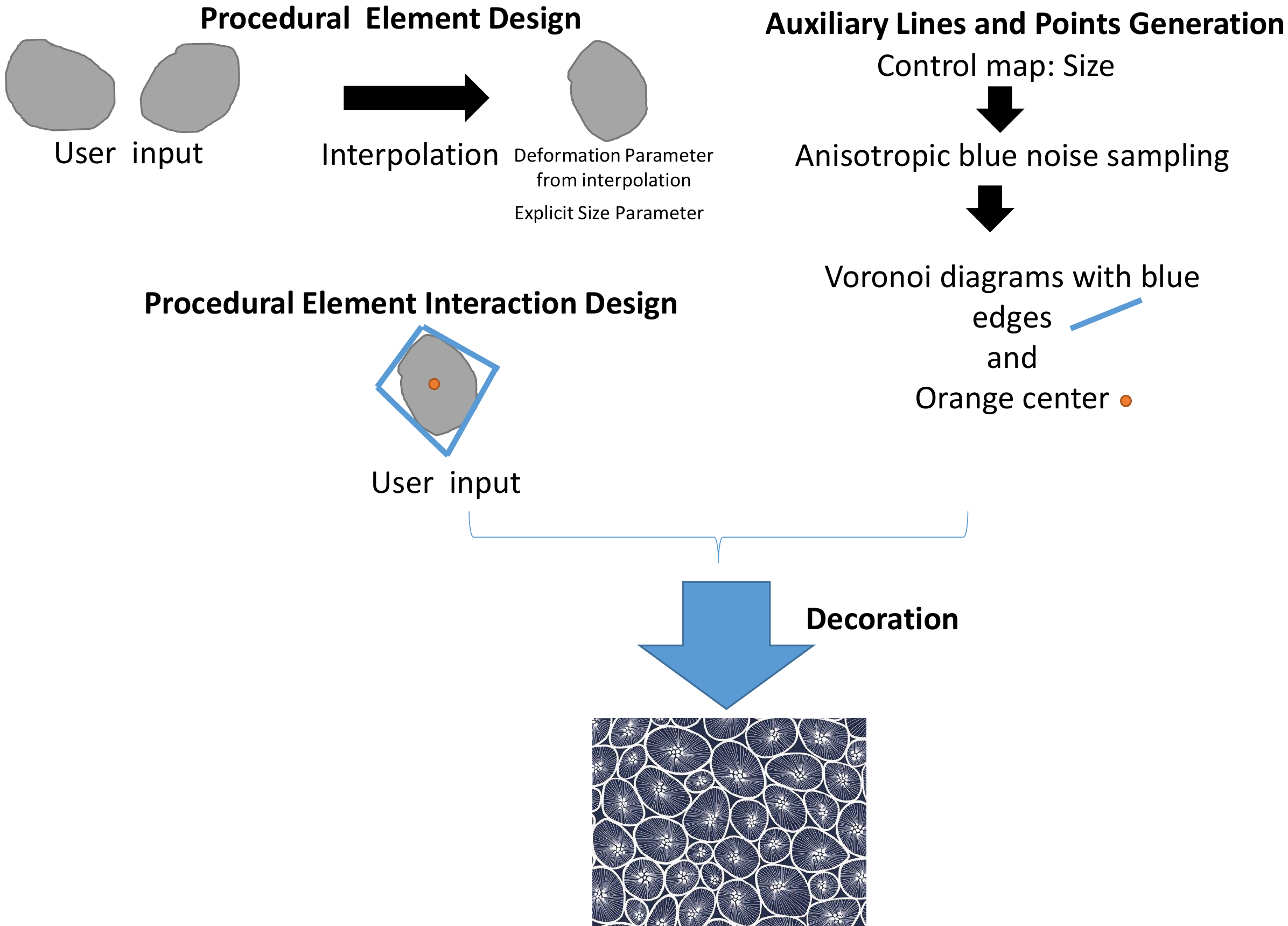}
	\Caption{Designing texture with curve-and-curve EC \cite{Tu:2018:WC1}}
	{Auxiliary points are used to determine the center of irregular circle and lines to assist characterizing the element contacts.
	}
	\label{fig:curve_curve_EC}
\end{figure}
\begin{figure}
	\centering
	\includegraphics[width=\linewidth]{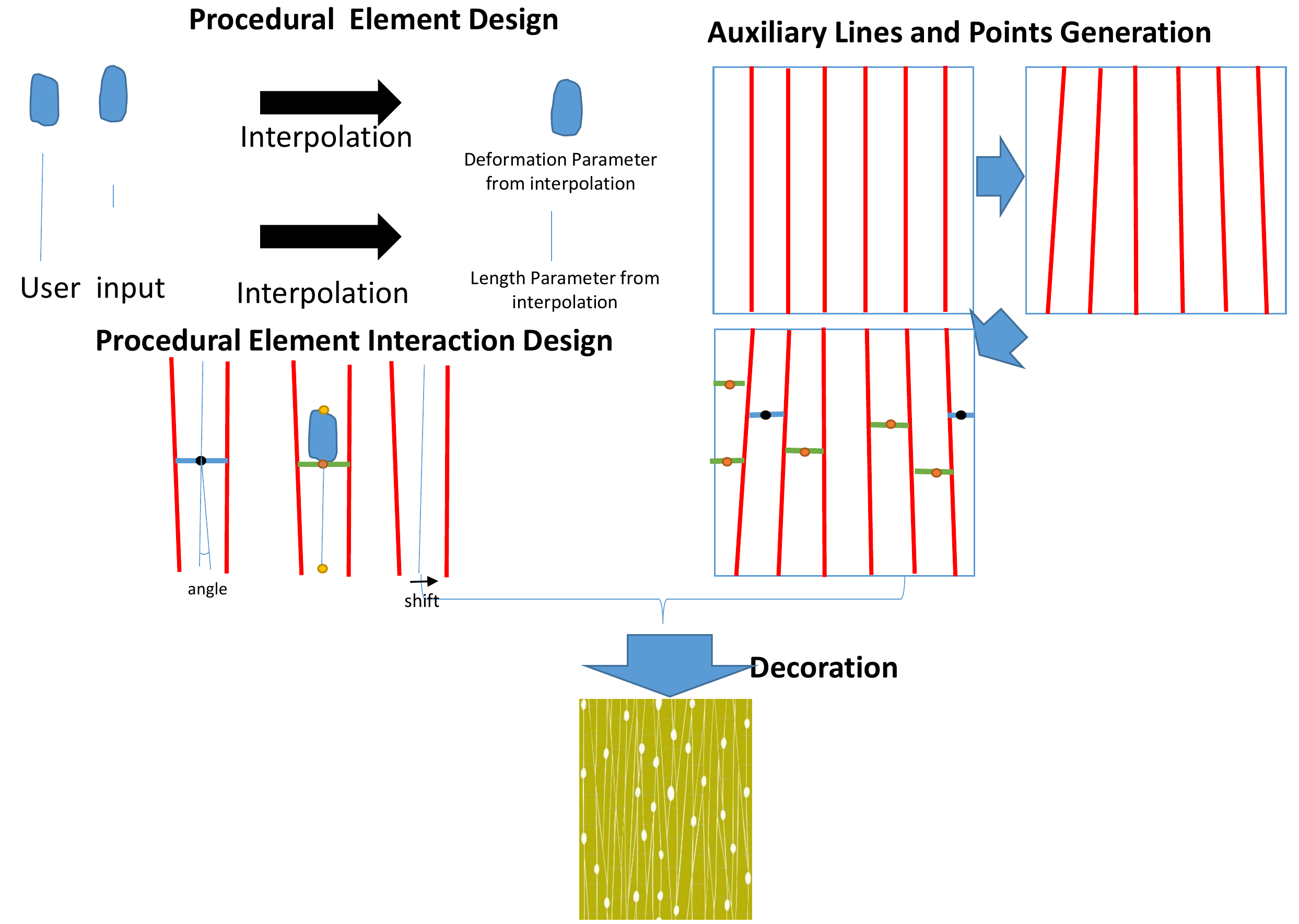}
	\Caption{Designing texture with point-and-curve EC and SA \cite{Tu:2018:WC2}}
	{Both auxiliary lines and points are used to characterize such element contacts and the element arrangement is generated recursively.
	}
	\label{fig:point_curve_EC_SA}
\end{figure}
\begin{figure}
	\centering
	\includegraphics[width=\linewidth]{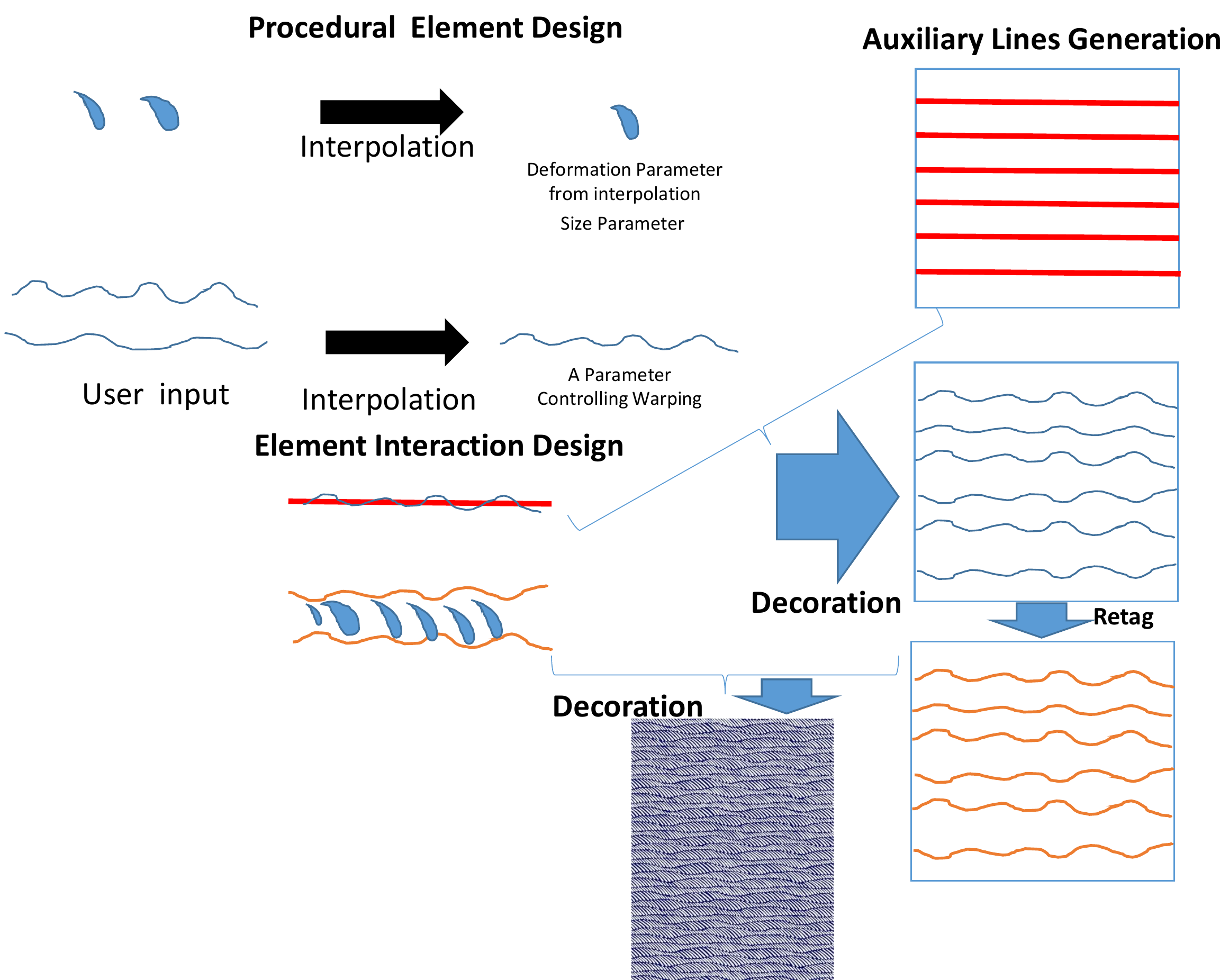}
	\Caption{Designing texture with SA \cite{Tu:2018:CT1}}
	{
		We retag the elements as auxiliary lines for specifying element arrangements.
	}
	\label{fig:SA}
\end{figure}
\begin{figure}[htb]
	\centering
	\subfloat[Procedural element]{
		\includegraphics[width=\linewidth]{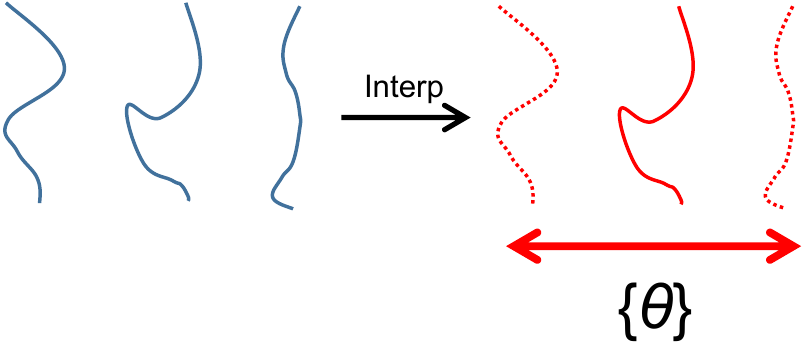}
	}%
	
	\subfloat[Auxiliaries generation]{
		\includegraphics[width=0.33\linewidth]{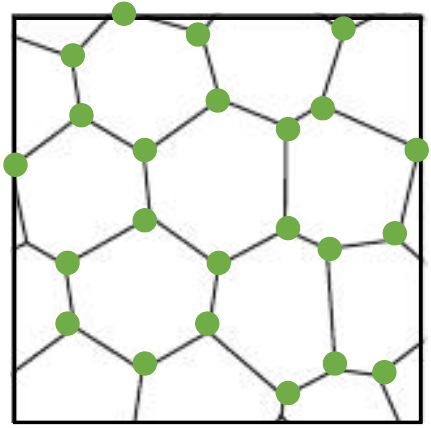}
	}%
	\subfloat[Element placement]{
		\includegraphics[width=0.33\linewidth]{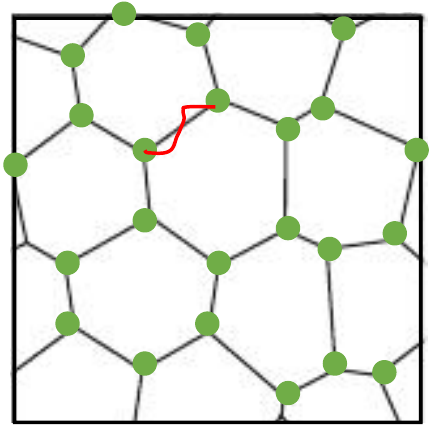}
	}%
	\subfloat[Created texture]{
			\includegraphics[width=0.33\linewidth]{figs/categories/crop_complex_texture_nce_ec.jpg}
	}%
	
	\Caption{Design textures with NCE and point-and-point EC.}
	{%
			The created texture has NCE and point-and-point EC.  As suggested in \Cref{subsubsec:SuggestedWorkflow}, we use auxiliary lines (black) to assist characterizing elements' position and points (green) to element contact. 
	}
	\label{fig:NCE_point_point_EC}
\end{figure}
}
\nothing{

}

\nothing{
}%

\subsection{Online Parameter Learning}

\subsubsection{Synthesis parameter sensitivity}

\nothing{
	\begin{figure*}[htb]
	\centering
		\subfloat[examplar]{
		\label{fig:param_sensitivity_exemplar1}
		\includegraphics[width=0.12\linewidth]{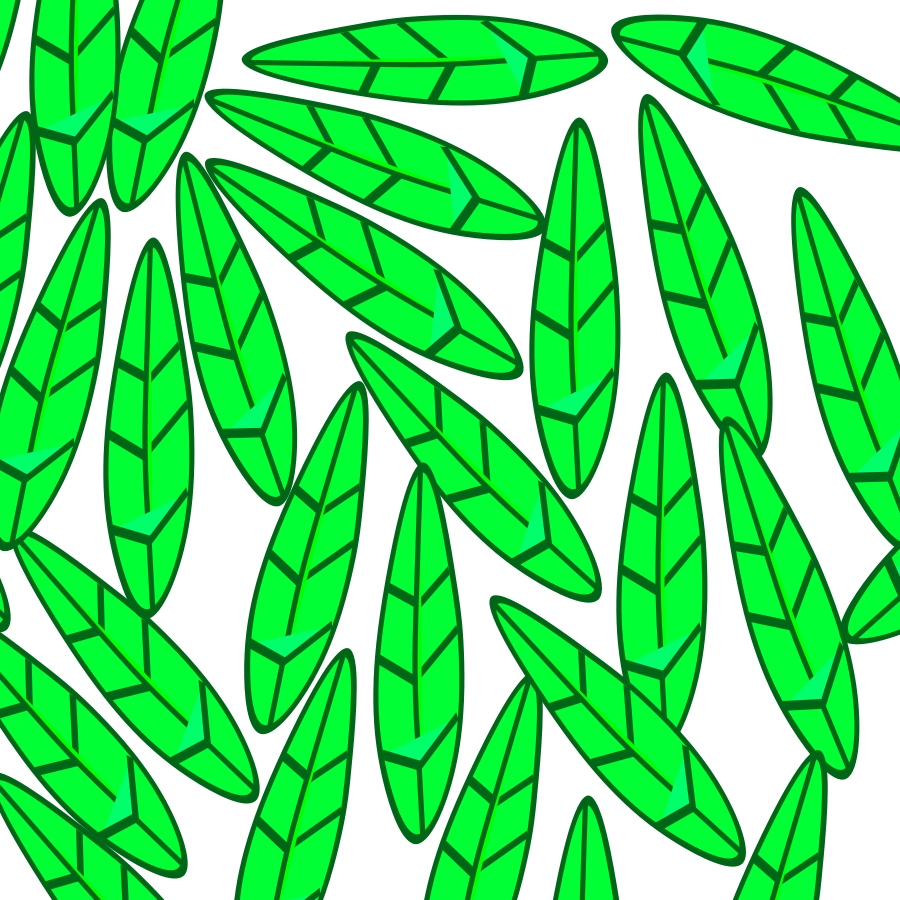}
	}%
		\subfloat[$\neighsize_1=50\ \neighsize_2=50\ \neighsize_3=50$]{
		\label{fig:param_sensitivity_1}
		\includegraphics[width=0.18\linewidth]{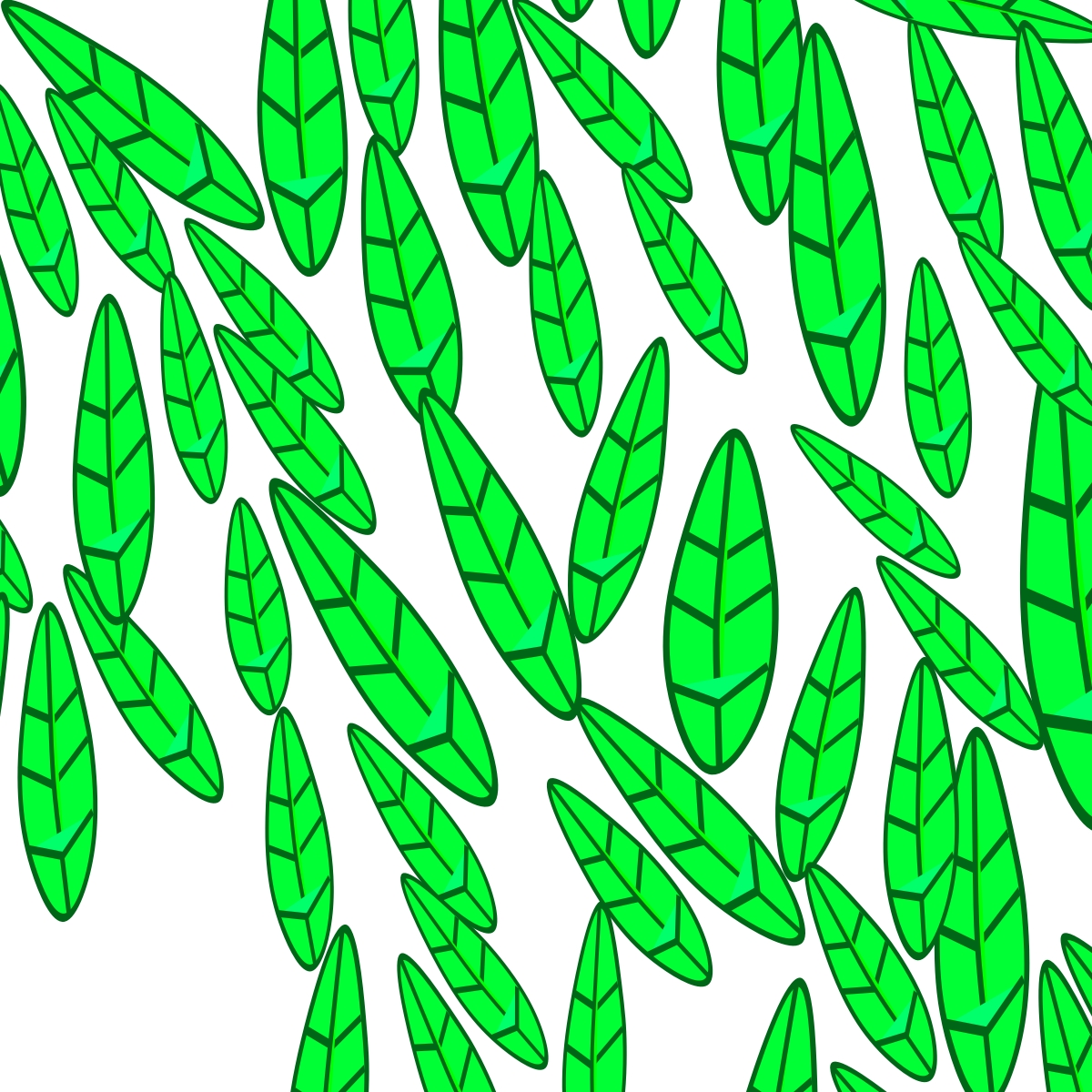}
	}%
	\subfloat[$\neighsize_1=150\ \neighsize_2=50\ \neighsize_3=50$]{
		\label{fig:param_sensitivity_2}
		\includegraphics[width=0.18\linewidth]{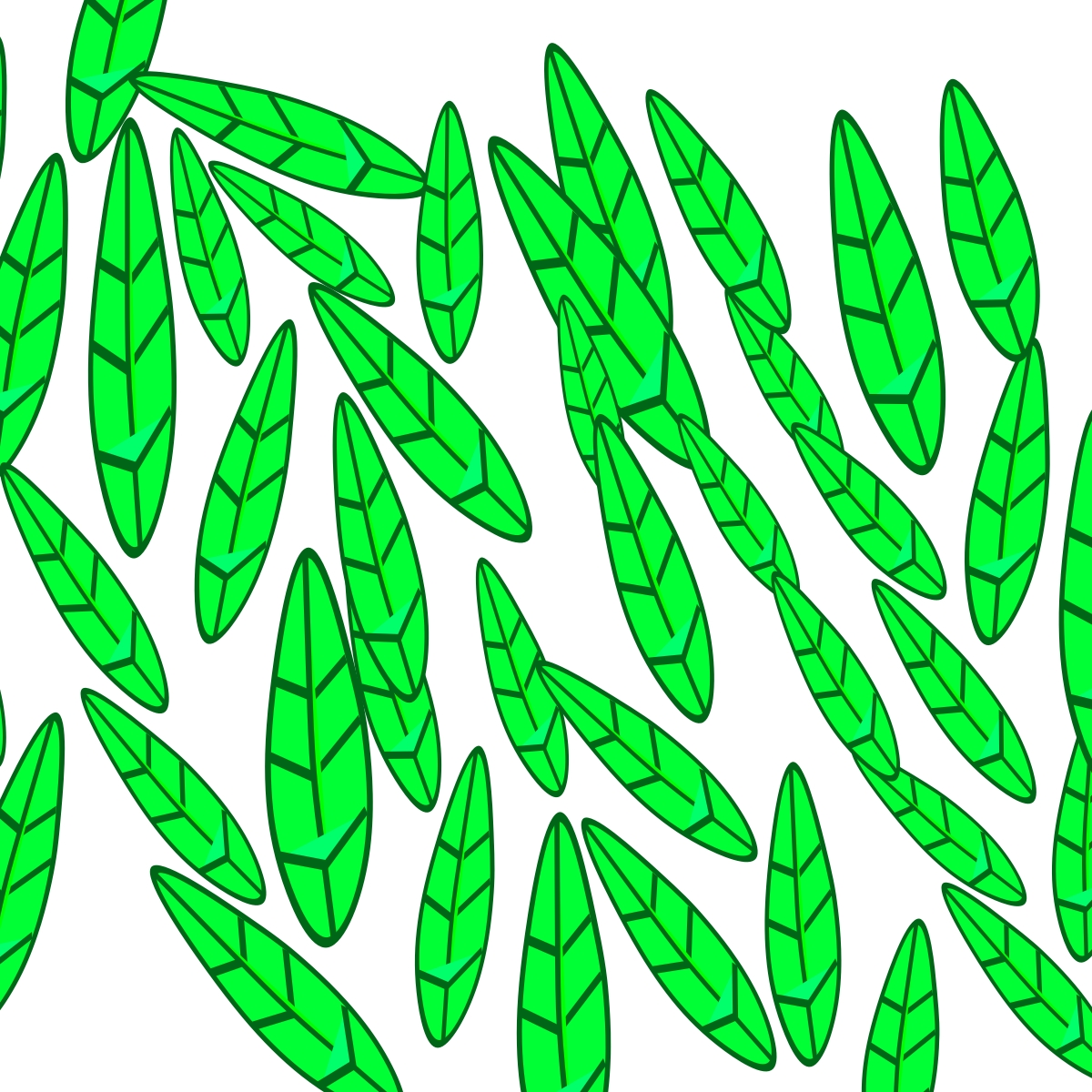}
	}%
	\subfloat[$\neighsize_1=50\ \neighsize_2=50\ \neighsize_3=25$]{
		\label{fig:param_sensitivity_3}
		\includegraphics[width=0.18\linewidth]{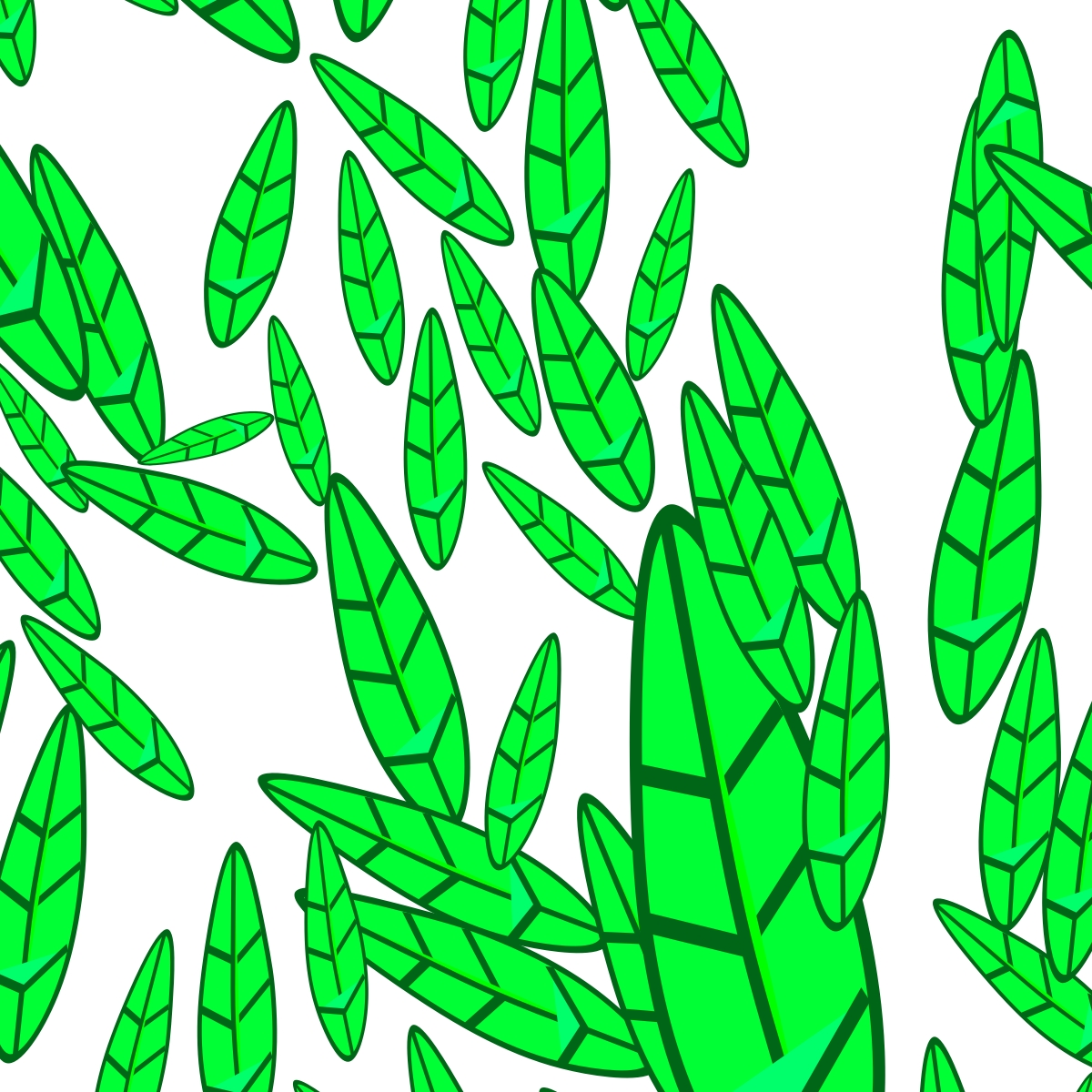}
	}%
	\subfloat[$\neighsize_1=50\ \neighsize_2=25\ \neighsize_3=25$]{
	\label{fig:param_sensitivity_4}
	\includegraphics[width=0.18\linewidth]{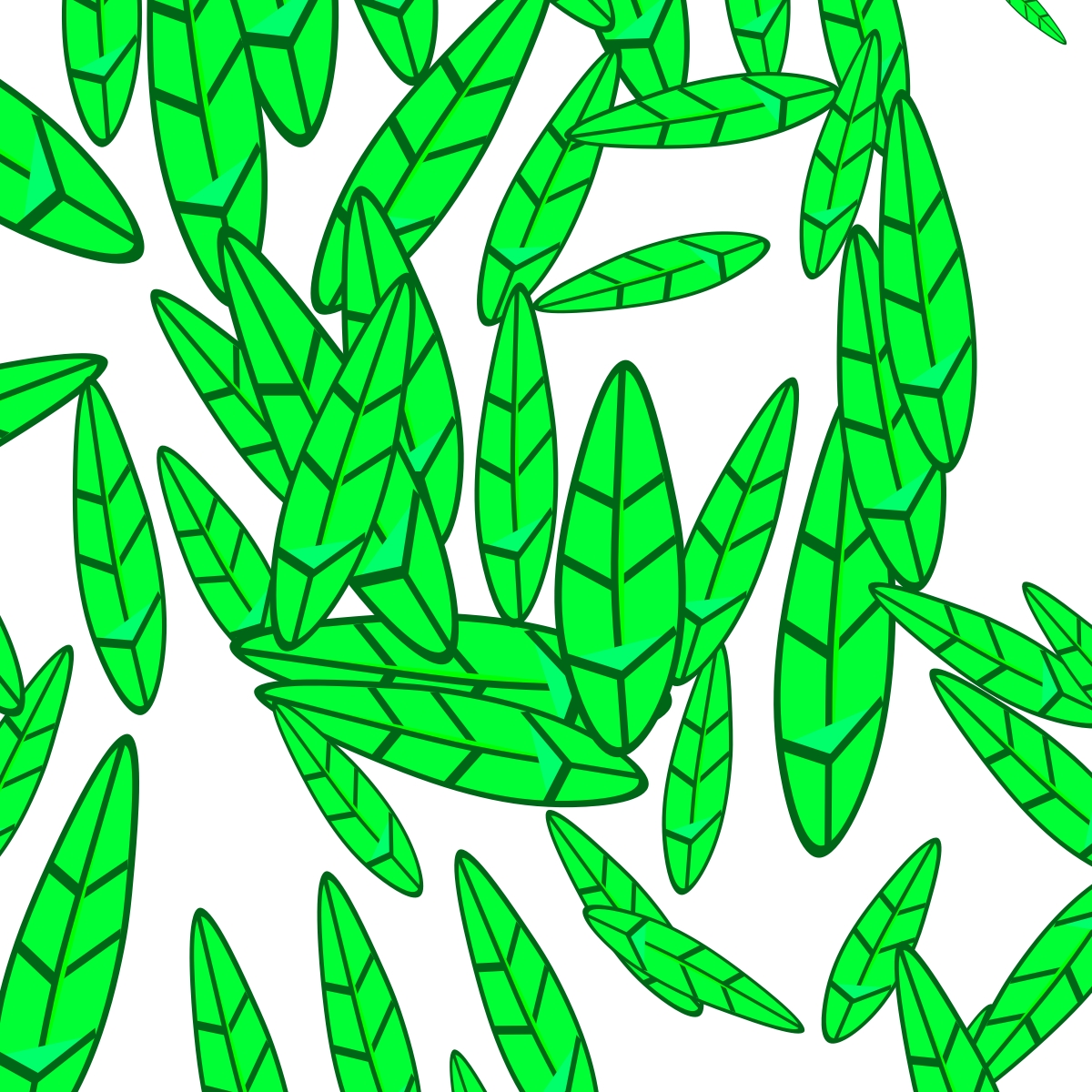}
}%

	\subfloat[examplar]{
	\label{fig:param_sensitivity_exemplar2}
	\includegraphics[width=0.12\linewidth]{figs/raster/161.jpg}
}%
\subfloat[$\neighsize=50$]{
	\includegraphics[width=0.18\linewidth]{figs/results/parameter_sensitivity/hier_synthesis_example_leaves2_n0_50_n1_50_n2_50_level_2.jpg}
}%
\subfloat[$\neighsize=75$]{
	\includegraphics[width=0.18\linewidth]{figs/results/parameter_sensitivity/hier_synthesis_example_leaves2_n0_150_n1_50_n2_50_level_2.jpg}
}%
\subfloat[$\neighsize=100$]{
	\includegraphics[width=0.18\linewidth]{figs/results/parameter_sensitivity/hier_synthesis_example_leaves2_n0_50_n1_50_n2_25_level_2.jpg}
}%
\subfloat[$\neighsize=125$]{
	\includegraphics[width=0.18\linewidth]{figs/results/parameter_sensitivity/hier_synthesis_example_leaves2_n0_50_n1_25_n2_25_level_2.jpg}
}%

	\Caption{Synthesis parameter sensitivity}
	{%
		The neighborhood sizes have substantial influence on synthesis quality.
	}
	
	\label{fig:param_sensitivity}
\end{figure*}

	The predictions are generated by example-based synthesis, which might be sensitive to algorithm parameters.
	Usually, these parameters are manually and heuristically selected and fixed during authoring. 
	For simple patterns that are targeted by \cite{Xing:2014:APR,Peng:2018:A3S}, such strategy might suffice in many cases.
	However, we target to synthesize complex, hierarchical patterns for which the synthesis quality is strongly influenced by parameters of synthesis algorithm \cite{Ma:2011:DET} (e.g. multi-level neighborhood sizes, pattern sampling).
	Too small neighborhood may not be able to capture large structures within patterns, while too large neighborhood may lose local ones due to averaging badly matched neighborhoods. Also, large neighborhood might produce distorted results and computationally expensive which is less favourable in interactive systems.
	
}%

\begin{figure}[htb]
	\centering
	\includegraphics[width=0.98\linewidth]{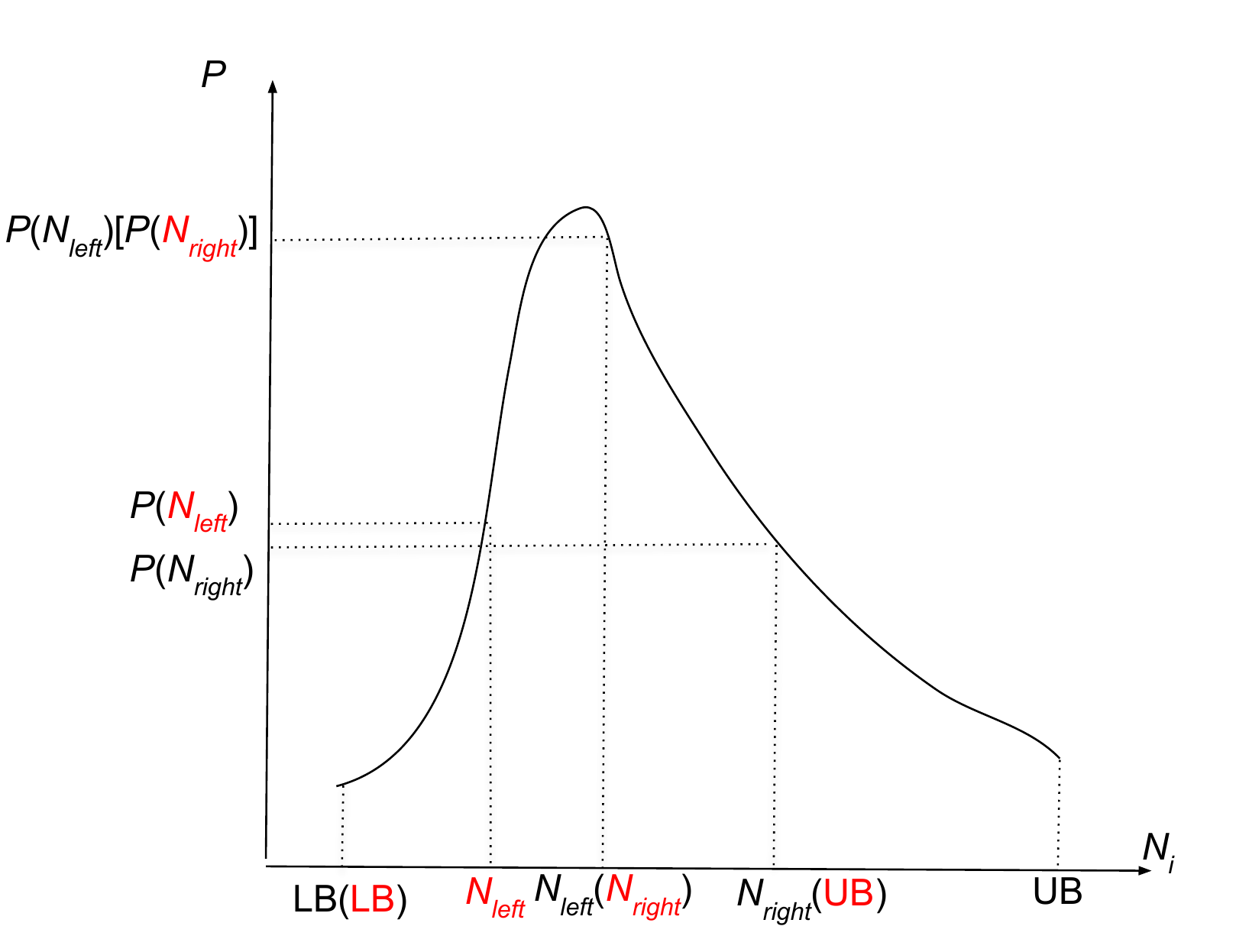}
	\Caption{Golden section search.} 
	{%
		We show the case where $\preferencefunction(\neighsize_{left}) > \preferencefunction(\neighsize_{right})$. In the next iteration, the golden section search narrows the search range by assigning the new bound  $\textcolor{red}{\boundconstraint} \assign [\lowerbound{\boundconstraint},\neighsize_{right}]$, left and right evaluation points $\textcolor{red}{\neighsize_{left}} \assign \lowerbound{\boundconstraint} + (1-\goldenratio)\cdot(\neighsize_{right} - \lowerbound{\boundconstraint})$, $\textcolor{red}{\neighsize_{right}} \assign \neighsize_{left}$
		(\textcolor{red}{red} indicates the next iteration)	
}
	\label{fig:GSS}
\end{figure}

\begin{algorithm}
  \begin{algorithmic}[1]
  	\REQUIRE noisy preference function $\preferencefunction$ without derivatives; maximum number of hierarchies $\maxhierlevel$; bound constraints $\boundconstraint_\hierlevel$ at level $\hierlevel$: $\boundconstraint_1$, $\boundconstraint_2$, $\boundconstraint_3$; termination threshold at level $\hierlevel$: $\goldensearchneightermin_1$, $\goldensearchneightermin_2$, $\goldensearchneightermin_3$
    \ENSURE $\neighvecsize=\{\neighsize_i\}^{\hierlevel}_{i=1}$
    \STATE  $\neighsize_1, \preferencescoremax  \assign \funct{GoldenSectionSearch}(\preferencefunction,\boundconstraint_1,\goldensearchneightermin_1)$, $\neighsize_2 \assign 0$,  $\neighsize_3 \assign 0$
    \IF{$\preferencescoremax<\preferencescorethreshold$}
   	\STATE $\neighsize_2, \preferencescoremax \assign \funct{GoldenSectionSearch}(\preferencefunction, \boundconstraint_2,\goldensearchneightermin_2)$
   	\IF {$\preferencescoremax<\preferencescorethreshold$}
  	\STATE $\neighsize_3, \preferencescoremax \assign \funct{GoldenSectionSearch}(\preferencefunction,\boundconstraint_3,\goldensearchneightermin_3)$
  	\ENDIF
    \ENDIF
    \RETURN{} $\neighvecsize=\{\neighsize_1,\neighsize_2,\neighsize_3\}$
  \end{algorithmic}
  \Caption{Online parameter learning.}
  {%
  	Our method decomposes a MD optimization problem into several 1D problems. Each 1D problem is solved by golden section search (\Cref{alg:goldensection}) via boolean-valued function comparison. When necessary, the function can be evaluated by prediction acceptance rate.
  }
  \label{alg:learning}
\end{algorithm}

\begin{algorithm}
	\begin{algorithmic}[1]
	\STATE $\mathbf{function}$ \funct{GoldenSectionSearch}($\preferencefunction$,$\boundconstraint$,$\goldensearchneightermin$)
	\STATE $\neighsize_{left} \assign \lowerbound{\boundconstraint} + (1-\goldenratio)\cdot(\upperbound{\boundconstraint} - \lowerbound{\boundconstraint})$ \COMMENT {$\upperbound{\cdot}$,$\lowerbound{\cdot}$ indicates lower and upper bound of $\boundconstraint$, $\goldenratio$ is the inverse of golden ratio }
	\STATE $\neighsize_{right} \assign \lowerbound{\boundconstraint} + \goldenratio\cdot(\upperbound{\boundconstraint} - \lowerbound{\boundconstraint})$ 
	\IF{$\abs{\neighsize_{left}-\neighsize_{right}}<\goldensearchneightermin$}
	\STATE $\neighsizeoptim \assign (\neighsize_{right}+\neighsize_{left})/2$
	\RETURN $\neighsizeoptim$, $\preferencefunction({\neighsizeoptim})$  
	\ENDIF
	 \STATE \COMMENT{It is not necessary to evaluate $\preferencefunction$}
	\IF{$\preferencefunction(\neighsize_{left}) \approx \preferencefunction(\neighsize_{right})$} 
	\STATE  $\boundconstraint \assign [\neighsize_{left},\neighsize_{right}]$
	\RETURN \funct{GoldenSectionSearch}($\preferencefunction$,$\boundconstraint$,$\goldensearchneightermin$) 
	\ENDIF 
	\IF{$\preferencefunction(\neighsize_{left})<\preferencefunction(\neighsize_{right})$ }
	\STATE  $\boundconstraint \assign [\neighsize_{left},\upperbound{\boundconstraint}]$
	\RETURN \funct{GoldenSectionSearch}($\preferencefunction$,$\boundconstraint$,$\goldensearchneightermin$) 
	\ENDIF
	\IF{$\preferencefunction(\neighsize_{left})>\preferencefunction(\neighsize_{right})$}
	\STATE  $\boundconstraint \assign [\lowerbound{\boundconstraint},\neighsize_{right}]$
	\RETURN \funct{GoldenSectionSearch}($\preferencefunction$,$\boundconstraint$,$\goldensearchneightermin$) 
	\ENDIF 

	\end{algorithmic}
  \Caption{Golden section search.}
{%
	In lines 9, 13, 17, it's not necessary to evaluate $\preferencefunction$. In line 6, $\preferencefunction$ can be evaluated by acceptance rate.
}
\label{alg:goldensection}
\end{algorithm}

\subsubsection{Problem formulation}
Our goal is to maximize an underlying preference function $\preferencefunction(\neighvecsize)$ with respect to the neighborhood size  at each level $\neighvecsize=\{\neighsize_i\}^{\hierlevel}_{i=1}$ . We find usually two or three hierarchies can suffice in most cases. 
\begin{equation}
\argmax_{\neighvecsize}\preferencefunction(\neighvecsize)
\label{eq:optim_obj}
\end{equation}
The key difficulty is that we don't know the explicit form of the function, and it can vary depending on the individual patterns and users.
Although the function may be evaluated by analyzing interactive user responses, due to variations of user inputs, the analyzed function values might be noisy prevent us to calculate numerical derivatives (e.g. by using finite difference approximations).

There are two observations that help us to design the optimization algorithm
\begin{itemize}
	\item Based on prior practices about parameter tuning in texture synthesis \cite{Wei:2009:SAE}, we know that, with the increase of neighborhood size, the synthesis quality first increases and then decreases. 
	\item When we synthesize the pattern hierarchically, the synthesis in the next level is always initialized with the result from the previous level. A better initialization will lead to better synthesized results.
\end{itemize}
Based on these two observations, 1) we can assume the preference function is unimodal with respect to the neighborhood size; 2) the multi-dimensional (MD) optimization problem can be decomposed into several one-dimensional (1D) problem in a greedy fashion; our algorithm first optimize the neighborhood size at the first level and then proceed to higher levels.

\nothing{

	In general, $\preferencefunction(\neighvecsize)$ is expensive to evaluate, noisy and its derivative information is unavailable. To optimize such functions, we apply a derivate-free optimization algorithm \cite{Conn:2009:IDF,Rios:2013:DFO}, , which targets to optimize convex but noisy functions without derivatives, as will be detailed in \Cref{subsubsec:optimization}.
	
	Next, we discuss how to evaluate the preference function by analyzing user responses.

}

\subsubsection{Learning by optimization}
\label{subsubsec:optimization}

Here, we introduce our optimization algorithm to \Cref{eq:optim_obj} for finding the optimum synthesis parameter settings. Our algorithm automatically determines the number of hierarchies as well as neighborhood size at each level.  Our algorithm is summarized in \Cref{alg:learning}.

The MD problem is decomposed into several 1D problems. We first optimize $\neighsize_1$ while assuming $\neighsize_2=0$ and $\neighsize_3=0$. 
The 1D function is unimodal, thus we apply golden section search (GSS) \cite{Kiefer:1953:SMS} for optimizing it. 
GSS can be performed by solely using boolean-valued function comparisons. It's not necessary to evaluate the function values $\preferencefunction$. The comparison is made when the users  choose which prediction they prefer as the system generates two set of predictions. 

After the optimal $\neighsize_1$ is found, the system uses it to synthesize predictions. If the quality of the predictions is sufficiently good, it means one level suffices to generate desired patterns. The quality of predictions or preference $\preferencefunction$ is calculated as the {\em prediction acceptance rate}.
Otherwise, we proceed to optimize $\neighsize_3$ while keeping $\neighsize_2=0$  and the hierarchical synthesis has two levels. If two levels are not sufficient, the algorithm further augment the number of hierarchies. 

In our algorithm, we set $\boundconstraint_1= [0, 0.25 \canvassize ]$, $\boundconstraint_2=[0, 0.125 \canvassize]$, $\boundconstraint_3=[0, 0.125  \canvassize]$, $\goldensearchneightermin_1 = 0.025  \canvassize$, $\goldensearchneightermin_2 = 0.0125,  \canvassize$, $\goldensearchneightermin_3 = 0.0125  \canvassize$, where $\canvassize$ is the canvas size.

\nothing{
	\subsubsection{Evaluation of preference function}
	\label{subsubsec:eval_preference_function}
	
	Assume once the system generates an element set $\predelementset=\{\predelement^i\}^{\predelementsetsize}_{i=1}$  as \textit{p}redictions. Users can accept, modify or reject the generated predictions. Through these operations, the resulting set by \textit{u}sers is $\userelementset=\{\userelement^i\}^{\userelementsetsize}_{i=1}$.
	$\userelementset$ excludes rejected elements. $\predelementsetsize$ and $\userelementsetsize$ are the number of elements within $\predelementset$ and $\userelementset$, and $\predelementsetsize \geq \userelementsetsize$ . We can quantitatively evaluate the user preference $\preferencescore$ to $\predelementset$ by comparing it to $\userelementset$.
	
	The preference $\preferencescore$ is evaluated by computing the similarity between 
	$\userelementset$ and $\predelementset$, which is defined as
	\begin{equation}
	\preferencescore=\frac{1}{\predelementsetsize}\sum_{\substack{\userelement \in \userelementset \\ \predelement \in \predelementset }}^{} \shapesimilarity(\predelement,\userelement)
	\end{equation}

	where $\shapesimilarity(\predelement,\userelement)$ returns the similarity between $\predelement$ and $\userelement$. $\predelement$ and $\userelement$ are elements within $\predelementset$
	and $\userelementset$. The users interactively generate elements within $\userelementset$ one by one, during which we match the element pairs $\userelement$ and $\predelement$ one-by-one by finding the element within $\predelementset$ that minimizes $\shapesimilarity$.
	
	The similarity $\shapesimilarity$ is defined as
	\begin{equation}
	\shapesimilarity(\predelement,\userelement)=\frac{1}{\matchedsetsize} \sum_{\substack{ \usersample \in \userelement\\  \predsample \in \predelement } }^{}\exp(-\frac{\norm{\predsample-\usersample} }{\sigma^2} )
	\end{equation}
	where $\usersample$ and $\predsample$ are samples that belong to $\userelement$ and $\predelement$, respectively. $\sigma$ is a hyperparameter. $\usersample$ and $\predsample$ are matched via a point set registration algorithm \cite{Myronenko:2010:PSR}.  $\matchedsetsize$
	The above formula equals one if an element is accepted in which case the $\userelement$ and $\predelement$ entirely overlap.
	
}

\nothing{
	\subsubsection{Fundamentals of Bayesian Optimization}
	Bayesian optimization is a technique of finding global optimum for black-box functions that does not require derivatives \cite{Brochu:2010:TBO,Shahriari:2015:THL}. 
	
	Bayesian optimization is designed to minimize the number of samples needed in the optimization. This is especially useful for functions that are expensive to evaluate (e.g. user preference function). 
}%

\begin{figure}[htb]
	\centering

	\subfloat[output graph]{
	\label{fig:output_graph}
	\includegraphics[width=0.48\linewidth]{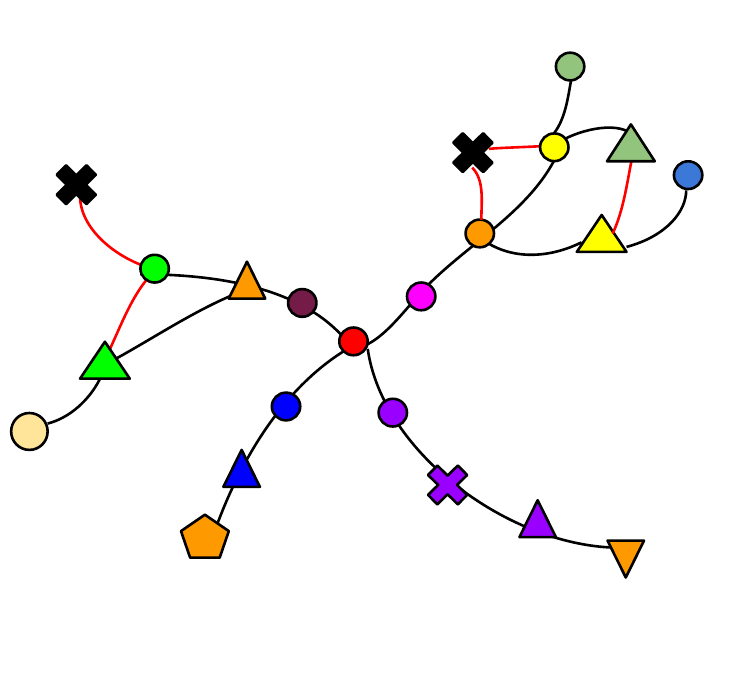}
}%
\subfloat[input graph]{
	\label{fig:input_graph}
	\includegraphics[width=0.48\linewidth]{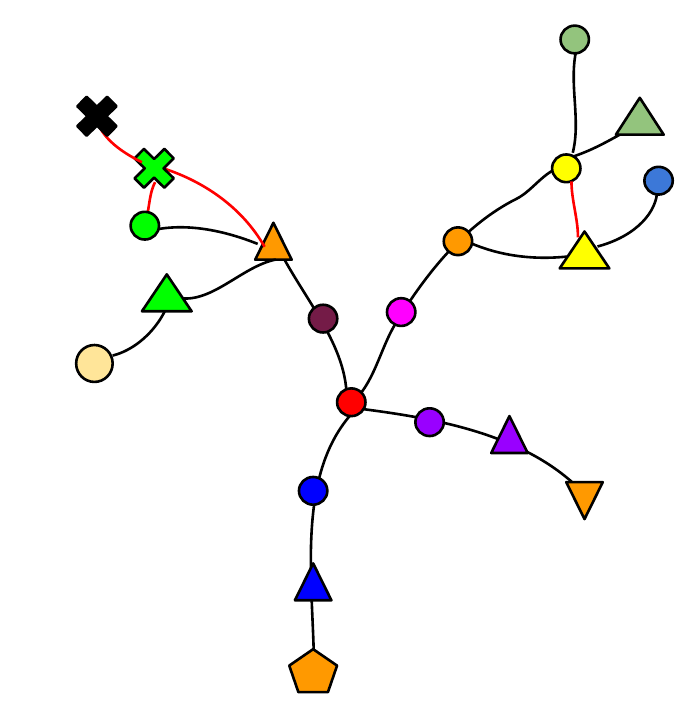}
}%

	\Caption{Graph matching.}
	{%
		First, we match two central  (\textcolor{red}{red}) nodes from the input and output neighborhoods.
		Excluding path nodes, we search for neighboring, unmatched  (\textcolor{orange}{orange}) nodes of most recently matched pair of (\textcolor{red}{red}) nodes.
		These nodes from the input and output are matched via Hungarian algorithm \cite{Kuhn:1955:HMA}.
		Again, for each most recently matched pair of  (\textcolor{orange}{orange}) nodes, we find its neighbors (\textcolor{green}{green}, \textcolor{yellow}{yellow}) and matched via Hungarian algorithm. This process is repeated until the input or output run out of \textit{joint} and \textit{end} nodes. Finally, \textit{path} nodes (e.g. \textcolor{blue}{blue} and \textcolor{pink}{pink}) are matched via Hungarian algorithm. In this illustration, according to the above descriptions, only nodes with the same color are matched via Hungarian algorithm. Shape indicates matched pair. Cross shape indicates there is no matched node. Our algorithm does not enforce edge constraints among neighbors of a matched node  (e.g. \textcolor{green}{green} and \textcolor{yellow}{yellow} ) 
		 
	}
	\label{fig:graph_matching_algo}
\end{figure}

\label{fig:curve_curve_EC} %

\begin{figure}[htb]
  \centering
  	\captionsetup[subfigure]{labelformat=empty}
  	
  	\label{fig:example:edge_assignment:exemplar:1}
  	\frame{\includegraphics[width=0.18\linewidth]{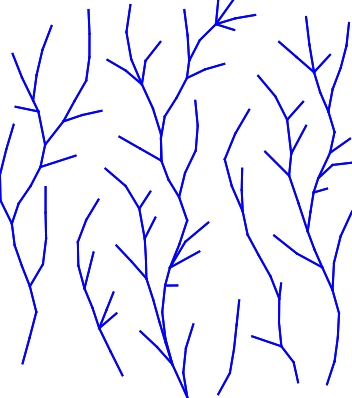}}
	\label{fig:example:edge_assignment:inital:1}
	\frame{\includegraphics[width=0.38\linewidth]{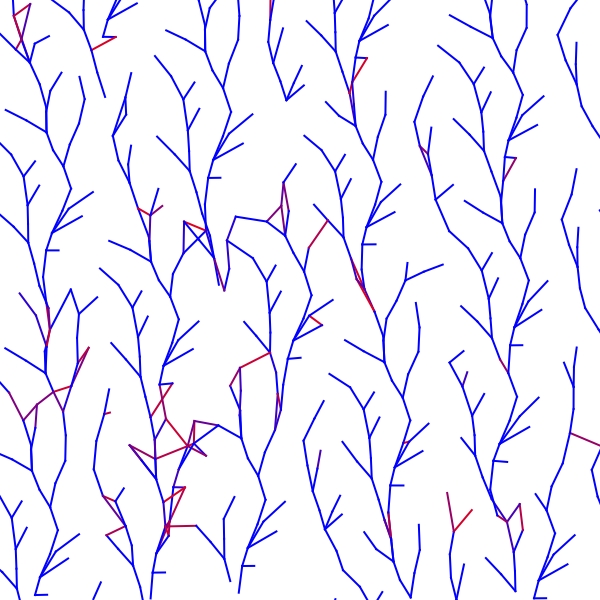}}
	\label{fig:example:edge_assignment:clean:1}
	\frame{\includegraphics[width=0.38\linewidth]{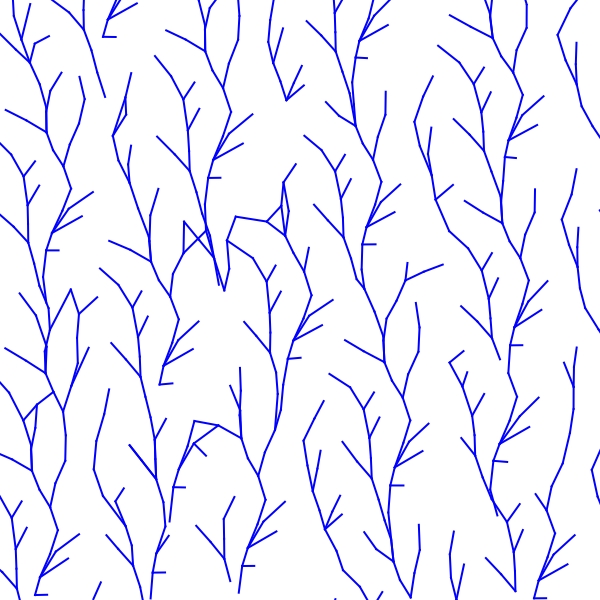}}

   \subfloat[exemplar]{
  	\label{fig:example:edge_assignment:exemplar:2}
	\frame{\includegraphics[width=0.18\linewidth]{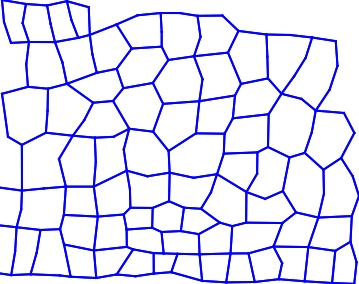}}
}%
  \subfloat[edge confidence]{
	\label{fig:example:edge_assignment:inital:2}
	\frame{\includegraphics[width=0.38\linewidth]{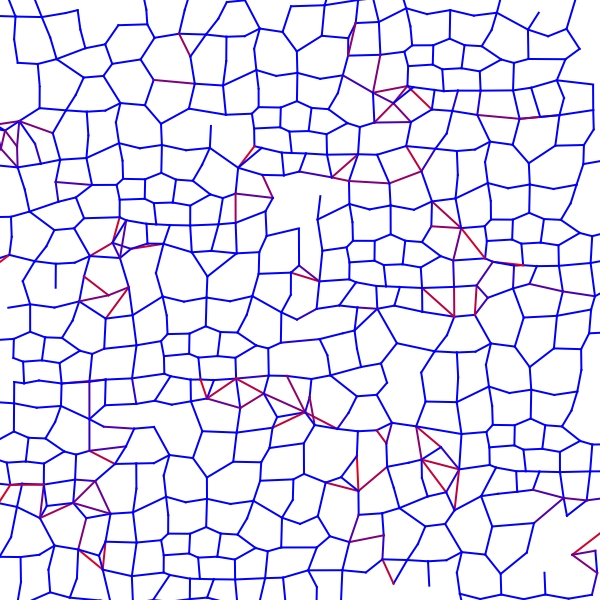}}
}%
\subfloat[binarized]{
	\label{fig:example:edge_assignment:clean:2}
	\frame{\includegraphics[width=0.38\linewidth]{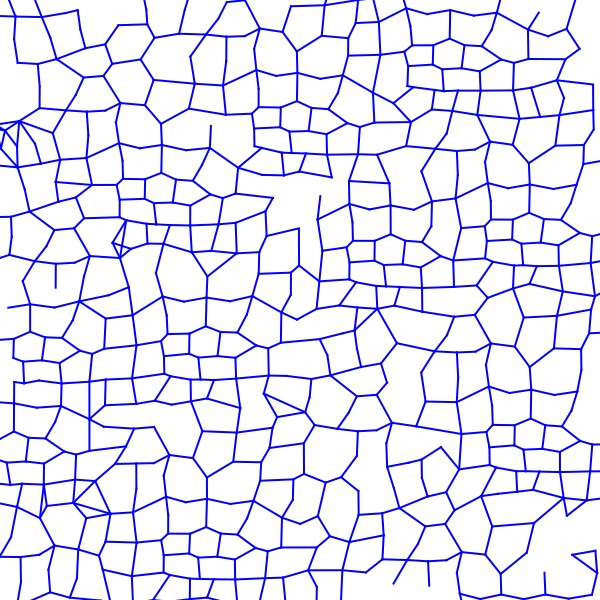}}
}%
  
 \Caption{Edge assignment visualization.}
 {%
 		The value of edge $\sampleedge(\sampleoutput,\sampleoutputprime)$ computed from \Cref{eqn:assignment:edge}  is visualized.
 		Input graph exemplars are shown on the left. 
 		In the middle, red indicates the edge is less likely to exist; blue indicates the edge is more likely to exist. 
 		On the right, the edge values are (binarized) rounded to  $0/1$. 
 		In other words, the edges with values smaller than 0.5 are removed and those large than 0.5 are kept.
 		For better visualization, we only show edges but nodes. 
 		Note the algorithm is not perfect. The simple strategy occasionally removes good and keep bad edges.
 }
 \label{fig:explain_edge_assignment}
\end{figure}

\nothing{

}

\subsubsection{Probabilistic neighborhood matching}
\label{sec:neighborhood_matching:probabilistic}

In \cite{Ma:2011:DET}, each output sample is exactly matched with another one sample in the input, which is problematic, since some input samples should not be matched to any output samples and some output samples are missing. 
But it allows to easily define sample similarity for various sample attributes, as shown in \Cref{subsubsec:sample_similarity}.
We term it {\em hard neighborhood matching} (\Cref{fig:hard_neighborhood_matching}).

In \cite{Roveri:2015:EBR}, the samples are matched via matching their transformation transformed by density kernels. {\em There is no explicit matching between samples}. 
We call it {\em soft neighborhood matching} (\Cref{fig:soft_neighborhood_matching}) (actually the samples are not matched).
This similarity criterion is not easy to integrate with various sample attributes. 
The attributes must be continuous and can be summed together (kernels perform local sum of attributes), which is also ambiguous.

We propose a robust neigborhood similarity that explicitly considers outliers and missing points to address these issues (\Cref{fig:semi_soft_neighborhood_matching}).
We consider cost of an output not matching with any input samples. 

We define the neighborhood $\sampleneigh(\samplesym)$ around $\samplesym$ as a set of samples within its spatial neighborhood.

An output sample $\sampleoutputprime$ within $\neighoutput$  is probabilistically matched with samples within $\neighinput$. The probability of matching $\sampleoutputprime$ to $\sampleinputprime$  is computed as 
\begin{equation}
\probability(\sampleoutputprime,\sampleinputprime)=\frac{1}{\normalizationfactor}
\exp\left(
-\frac{\norm{ \differencesym{\samplevec}(\sampleoutput,\sampleoutputprime)-\differencesym{\samplevec}(\sampleinput,\sampleinputprime)}}{2\softmatchingsigma^2}
\right)
\label{eq:matching_probability}
\end{equation}
where $\normalizationfactor=\sum_{\samplesym \in \{\neighinput, \sampleinput\}}^{}\exp\left(-\frac{\norm{ \differencesym{\samplevec}(\sampleoutput,\sampleoutputprime)-\differencesym{\samplevec}(\sampleinput,\samplesym)}}{2\softmatchingsigma^2}\right)$ is the normalization factor to make sure the sum of probability for matching $\sampleoutputprime$ to every $\sampleinputprime$ and center $\sampleoutput$ equals 1. 
$\softmatchingsigma$ is a parameter controlling the Gaussian function, which can be readily adapted by pattern sampling rate.
In a neighborhood, there might be samples from both discrete elements and continuous structures. We only match samples with the same id, from the same type of patterns, and from the same type of elements within an element pattern. 
For central samples, we also use \Cref{eq:matching_probability} to compute the probability, where $\differencesym{\samplevec}(\sampleoutput,\sampleoutput)=0$, $\differencesym{\samplevec}(\sampleinput,\sampleinput)=0$.

\subsubsection{Bidirectional neighborhood similarity}
\label{subsubsec:neighborhood_similarity}

We can exchange the output and input in \Cref{eq:matching_probability} to compute the probability for matching $\sampleinputprime$ to $\sampleoutputprime$.
The bidirectional neighborhood similarity is defined as
\begin{align}
&\distance{\neighoutput-\neighinput}= \\
&\sum_{
	\substack{\sampleoutputprime\in\neighoutput}}\sum_{\sampleinputprime\in\neighinput}
\probabilitybi(\sampleoutputprime,\sampleinputprime)\left(\distance{\differencesym{\samplevec}(\sampleoutput,\sampleoutputprime)-\differencesym{\samplevec}(\sampleinput,\sampleinputprime)}
+ \distance{\edgelength(\sampleoutput,\sampleoutputprime) - \edgelength(\sampleinput,\sampleinputprime)}\right)
\label{eq:bidirectional_neighborhood_similarity}
\end{align}
,
where 
\begin{align}
\probabilitybi(\sampleoutputprime,\sampleinputprime)=\frac{\probability(\sampleoutputprime,\sampleinputprime)+\probability(\sampleinputprime,\sampleoutputprime)}{2}
\label{eq:probabilitybi}
\end{align}
is a symmetric version of $\probability$ with $\probabilitybi(\sampleoutputprime,\sampleinputprime)=\probabilitybi(\sampleinputprime,\sampleoutputprime)$.

\nothing{
	Matching samples from continuous structures are constrained by graph edges which is a classic, NP-hard (for global optimal solution) graph matching problem \cite{West:1996:IGT}. We propose a greedy algorithm for deciding the matchings. 
	For purposes of algorithm clarification, graph nodes with one, two and more than two neighbors are called \textit{end node}, \textit{path node}, \textit{joint node}, as illustrated in \Cref{fig:graph_representation}.
	The basic idea is as follows, as illustrated in 1) We match each pair of nodes in a greedy fashion, the next one or several matchings are constrained by existing matchings and graph edge.  2)  We first match two central nodes $\sampleinput$ and $\sampleoutput$, and then proceed to match joint and end nodes in an order of traversing the graph (which excludes path nodes) by breadth-first search, and finally match path nodes.
	For matching neighbors of a pair of matched node, we apply Hungarian algorithm \cite{Kuhn:1955:HMA}. This process is repeated until there is no joint or end node that can be matched in $\neighoutput$. 3) Finally, we only match path nodes from the same graph edge (a graph excluding path nodes).
}

\nothing{
}%

\begin{figure*}[htb]
	\centering

\captionsetup[subfigure]{labelformat=empty}

	\subfloat[exemplar]{
	\frame{\includegraphics[width=0.13\linewidth]{figs/results/vector_results/exemplars/g_exemplar_example_leaves.pdf}}
}
\subfloat[non-hierarchical (small)]{
	\frame{\includegraphics[width=0.20\linewidth]{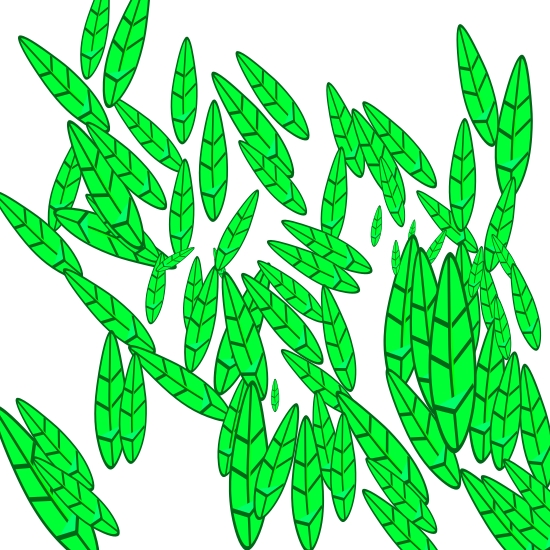}}
}
\subfloat[non-hierarchical (median)]{
	\frame{\includegraphics[width=0.20\linewidth]{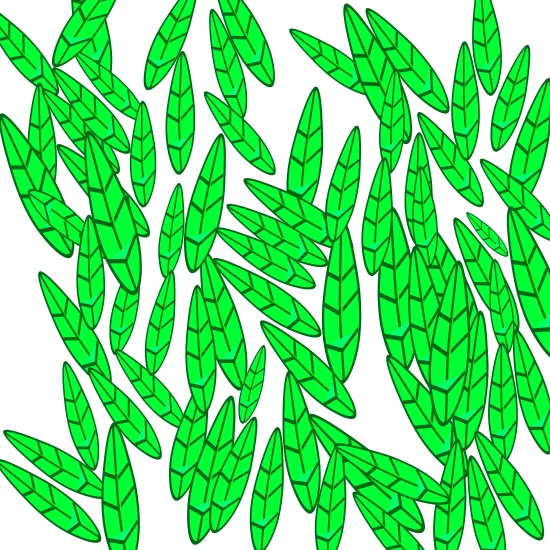}}
}
\subfloat[non-hierarchical (large)]{
	\frame{\includegraphics[width=0.20\linewidth]{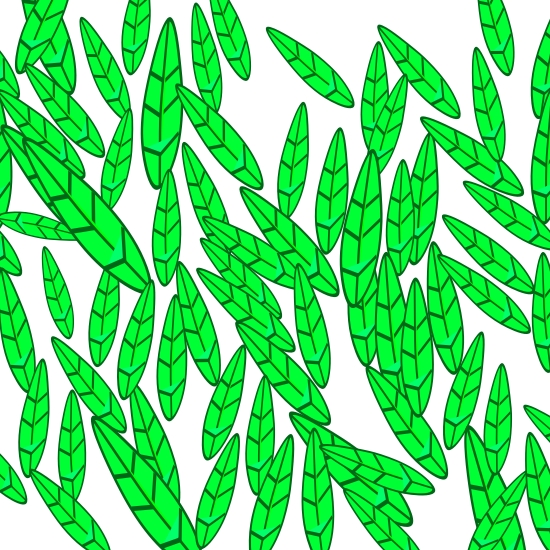}}
}
\subfloat[Hierarchical]{
	\frame{\includegraphics[width=0.20\linewidth]{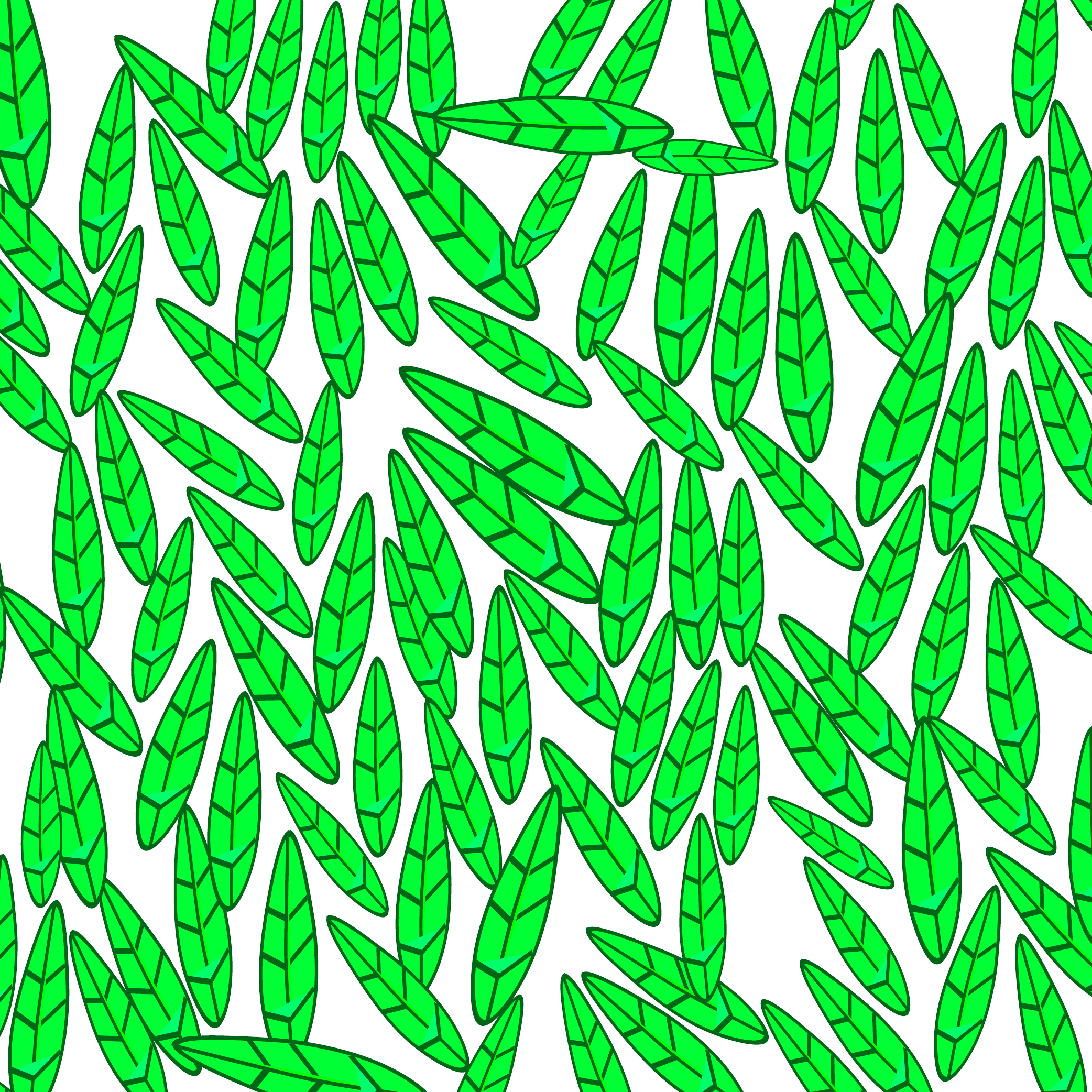}}
}

	\Caption{Hierarchical versus non-hierarchical synthesis.}
	{%
		The first column shows the exemplar; the second and third columns show non-hierarchical synthesis results with small and large neighbourhoods, respectively. 
		A small neighborhood has limited information, using which the synthesis cannot reproduce larger structures (There are unwanted overlap between elements).
		A large neighborhood may be not well matched. The least squares solvers will average unmatched neighborhoods and result in similar-size elements and kind of uniform distribution.
		A median neighborhoood blends two types of artifacts.
		The rightmost column shows the hierarchical synthesis generates higher-quality results. 
	}
	\label{fig:compare_synthesis}
\end{figure*}

\begin{figure*}[tb!p]
	\centering

	\Caption{Continuation from \protect\Cref{fig:full_auto_outputs}.}
	{%
		\subref{fig:distorted_grid:automatic_synthesis} is generated with only first two hierarchies.
		 is of size about $250 \times 250$.
		 \subref{fig:distorted_blocks:automatic_synthesis} are generated with three hierarchies with neighborhood sizes 40, 30, 20 and sample distances 30, 20, 10.
	}
	\label{fig:full_auto_outputs_extra}
\end{figure*}

\begin{figure}[htb]
	\centering
	\subfloat[non-hierarchical representation]{
		\label{fig:non_hier_repre}
		\includegraphics[width=0.45\linewidth]{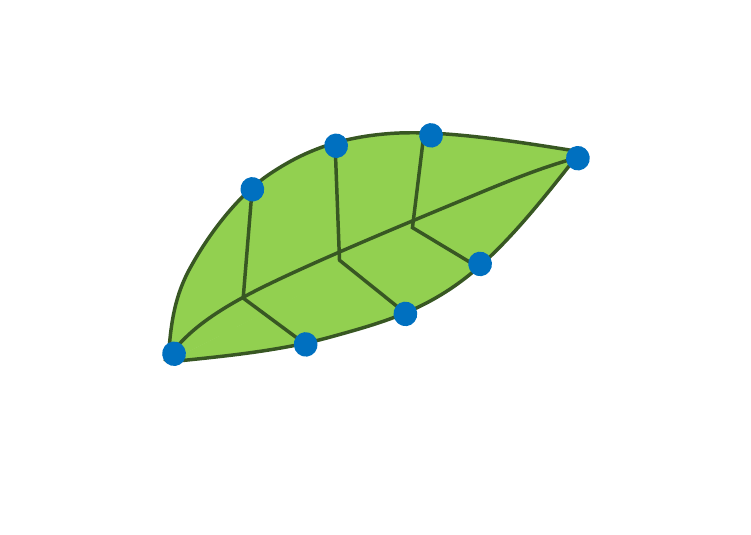}
	}%
	\subfloat[hierarchical representation]{
		\label{fig:hier_repre}
		\includegraphics[width=0.45\linewidth]{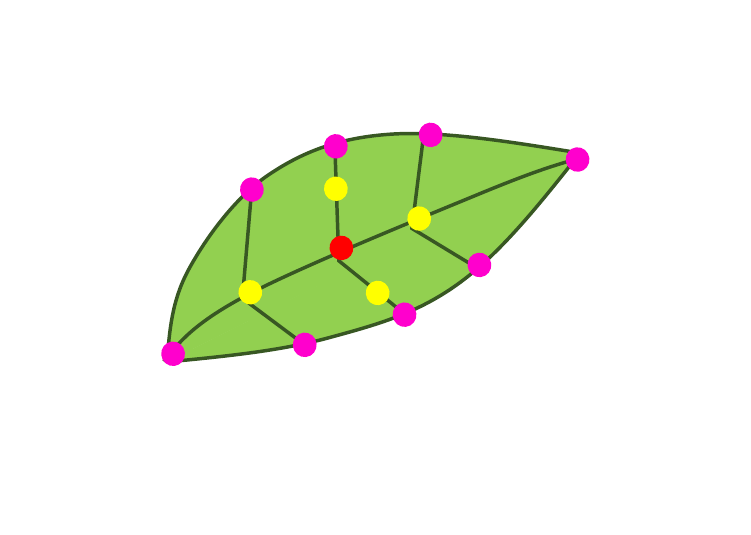}
	}%

	\Caption{Representations.}
	{%
		The non-hierarchical representation  \subref{fig:non_hier_repre} is less expressive than our hierarchical representation \subref{fig:hier_repre} that can generate more complex patterns. In \subref{fig:hier_repre}, the synthesis is employed from \textcolor{red}{red} $\rightarrow$ \textcolor[rgb]{0.78,0.78,0}{yellow}$\rightarrow$  \textcolor{magenta}{magenta} samples.

	}
	\label{fig:hier_representation}
\end{figure}
\label{fig:NCE_point_point_EC} 
\label{fig:noadjacency:1}
\label{fig:noadjacency:2}
\label{fig:point_curve_EC_SA}
\label{fig:SA}
\label{fig:branches:exemplar}
\label{subsec:proceduralpattern}
\label{subsec:analysis}
\label{eq:graph_similarity}
\label{fig:semi_soft_neighborhood_matching}
\label{sub@fig:semi_soft_neighborhood_matching}
\label{subsubsec:adaptive_sampling}
\label{subsubsec:recon_continuous_structures:init}
\label{subsec:preferencelearning}
\label{fig:hier_synthesis_2}
\label{fig:cells:automatic_synthesis}
\label{fig:sampling_rate}

\begin{figure}[htb]
	\centering
	\subfloat[before]{
		\label{fig:continuous_pattern_recon_before}
		\includegraphics[width=0.48\linewidth]{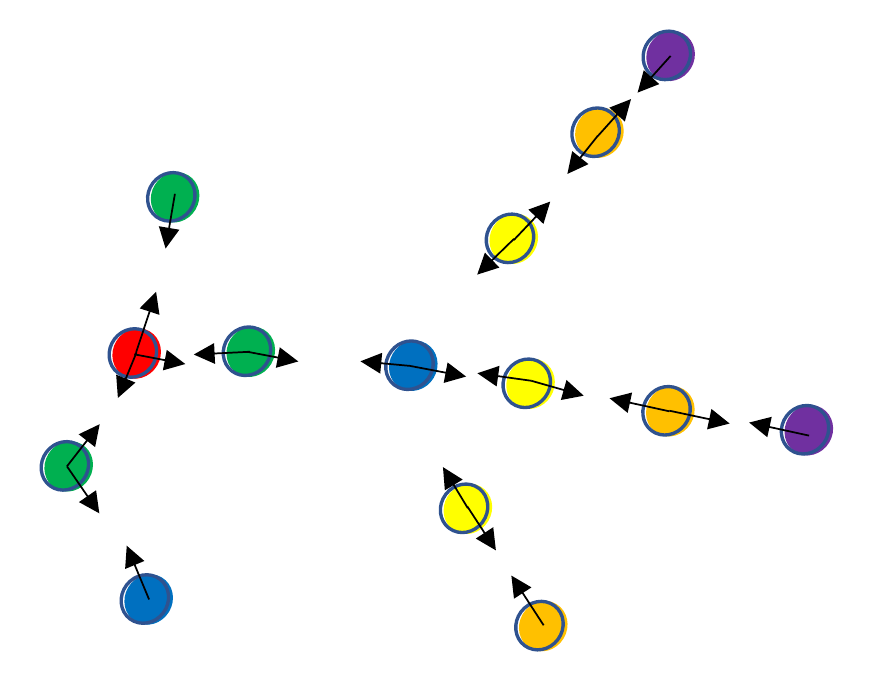}
	}%
	\subfloat[after]{
		\label{fig:continuous_pattern_recon_after}
		\includegraphics[width=0.48\linewidth]{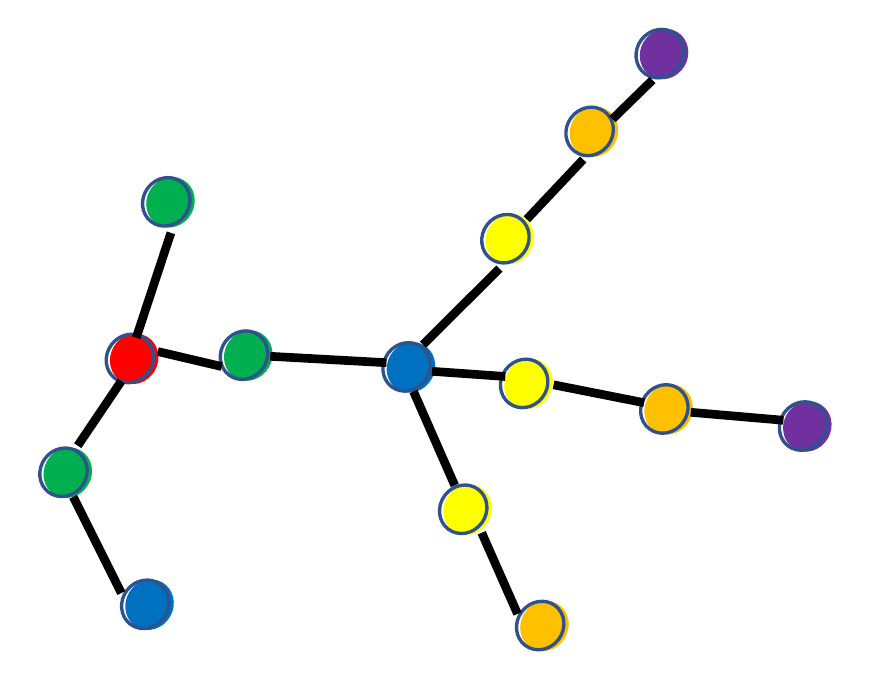}
	}%
	\Caption{Continuous pattern reconstruction.}
	{%
		\subref{fig:continuous_pattern_recon_before} shows the sample distribution we have. 
		Each sample has an orientation vector that indicates the number of edges and their orientations. 
		Each sample also records the expected geodesic distance to other samples on the reconstructed graph. In \subref{fig:continuous_pattern_recon_before}, we visualize the estimated distance between the red to other samples. Green, blue, yellow, orange and purple indicates the geodesic distance 1, 2, 3, 4, 5.
		Our reconstruction algorithm generates \subref{fig:continuous_pattern_recon_after} with edges connecting those samples. 
	}
	\label{fig:continuous_pattern_recon}
\end{figure}
\begin{figure*}[htb]
	\centering
		\subfloat[examplar]{
		\label{fig:continuous_exemplar}
		\includegraphics[width=0.2\linewidth]{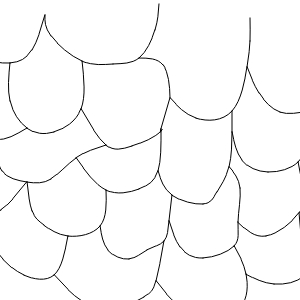}
	}%
	\subfloat[example graph]{
	\label{fig:example_graph_init}
	\includegraphics[width=0.2\linewidth]{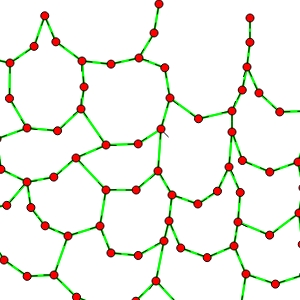}
}%
	\subfloat[initialization]{
	\label{fig:initialization}
	\includegraphics[width=0.2\linewidth]{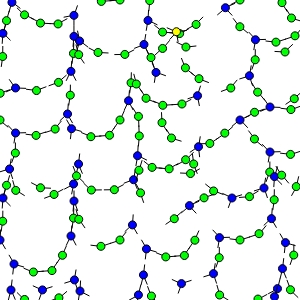}
}%
	\subfloat[sample optimization]{
		\label{fig:sample_optimization}
		\includegraphics[width=0.2\linewidth]{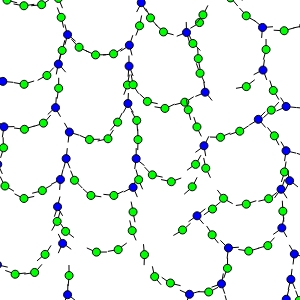}
	}%

	\subfloat[initial graph reconstructionl]{
	\label{fig:initial_graph_reconstruction}
	\includegraphics[width=0.2\linewidth]{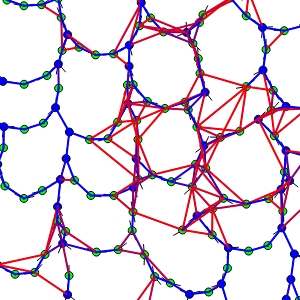}
}%
	\subfloat[Denoising]{
	\label{fig:denoising}
	\includegraphics[width=0.2\linewidth]{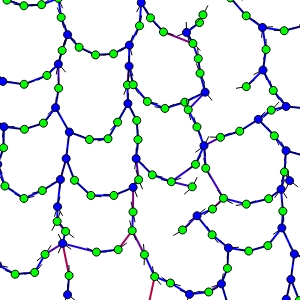}
}
	\subfloat[Curve reconstruction from graph]{
	\label{fig:curve_reconstruction}
	\includegraphics[width=0.22\linewidth]{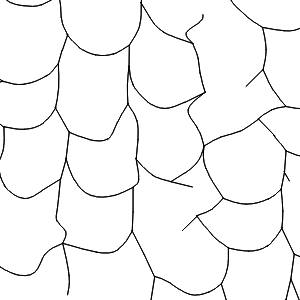}
}%

	\Caption{Continuous pattern synthesis.}
	{%
			Our algorithm proceed from sample synthesis to curve reconstruction. For the input exemplar \subref{fig:continuous_exemplar}, we use a graph to represent it \subref{fig:example_graph_init}. 
			During synthesis, instead of synthesizing graph directly, we synthesize samples (graph node) where each sample record attributes including number of connections and edge orientations. 
			The synthesized sample distribution \subref{fig:sample_optimization}  is used to construct an initial output graph \subref{fig:initial_graph_reconstruction}. In \subref{fig:example_graph_init} and \subref{fig:sample_optimization},
			the color of each sample indicates the number of edges on it (green = 2, blue =3 yellow = 4). 
			The initial graph is noisy with bad connections. We denoise the graph with MST \subref{fig:denoising}. 
			Finally, the pattern curves are reconstructed \subref{fig:curve_reconstruction}  from the graph by interpolating graph paths that are free of branches.
			
	}
	\label{fig:continuous_pattern_synthesis}
\end{figure*}
\begin{algorithm}
  \begin{algorithmic}[1]
  	\REQUIRE Output samples $\{\sampleoutput^k\}_{k=1}^{\numoutputsamples}$, input samples $\{\sampleinput^k\}_{k=1}^{\numinputsamples}$, edge connections between $\{\sampleinput\}$ recorded in adjacency matrix $\inputadjacencymatrix$ matched neigborhood pairs $\{\neighoutput,\neighinput,\match\}$
  	\ENSURE Graph edges connecting  $\{\sampleoutput\}$, connections recorded in adjacency matrix $\outputadjacencymatrix$
  	\STATE $\forall \outputadjacencymatrix^{ij}\assign 0$ 
  	\STATE $\forall \weightedadjacencymatrix^{ij}\assign 0$
  	\COMMENT{all entries are initialized with 0, $\weightedadjacencymatrix$ is a weighted adjacency matrix, where a large weight indicates the edge is more likely to exist.}
  	\FOR{$\neighoutput, \neighinput, \match \in \{\neighoutput,\neighinput,\match\}$ }
    \FOR{$\sampleoutput^1 \in \neighoutput$ }
    \FOR{$\sampleoutput^2 \in \neighoutput$ }
 	\IF{$\sampleoutput^1 \neq \sampleoutput^2$}
   	\STATE $\sampleinput^1 \assign \match(\sampleoutput^1)$ \ $\sampleinput^2 \assign \match(\sampleoutput^2)$
 	\STATE $\inputpath \assign$ \funct{FindShortestPath}($\sampleinput^1,\sampleinput^2,\inputadjacencymatrix$) 
    \COMMENT{Find the shortest path between $\sampleinput^1,\sampleinput^2$ on input graph using breadth-first search}
    \FOR{$\sampleinput \in \inputpath$}
    \STATE $\sampleoutput \assign \match(\sampleinput)$
    \IF{$\sampleoutput \neq NULL$ }
     \STATE $\outputpath$ {\em append} $\sampleoutput$
    \ELSE
    \STATE  $\outputpath \assign \emptyset$
    \STATE \textbf{break}
    \ENDIF
    \ENDFOR
    \FOR{edges $ \sampleedge \in \outputpath$}
    \STATE $\indexstart \assign \funct{GetStartIndex}(\sampleedge) $
    \STATE $\indexend = \funct{GetEndIndex}(\sampleedge) $
   \STATE $\weightedadjacencymatrix[\indexstart][\indexend]\pluseq1$
    \STATE $\weightedadjacencymatrix[\indexend][\indexstart]\pluseq1$
    \COMMENT{$\weightedadjacencymatrix$ is symmetric}
    \ENDFOR
	\ENDIF
    \ENDFOR
    \ENDFOR
     \ENDFOR
     \FOR{$\indexstart = i\ to\ \numoutputsamples$}
     \STATE  $t  \assign$ $\sampleconnections(\sampleoutput^{\indexstart})$ \STATE $\{\indexend^i\}_{i=1}^{t} \assign$ \funct{FindtLargestEntryIndexes}($\weightedadjacencymatrix[\indexstart],t$)
     \FOR{$\indexend \in\{\indexend^i\}_{i=1}^{t}$}
   	\STATE $\outputadjacencymatrix[\indexstart][\indexend]=1$\ $\outputadjacencymatrix[\indexend][\indexstart]=1$
   	    \COMMENT{$\outputadjacencymatrix$ is symmetric}
   	\ENDFOR
     \ENDFOR     \COMMENT{Edge filtering}
    \RETURN{} $\outputadjacencymatrix$
     
  \end{algorithmic}
  \Caption{Continuous pattern reconstruction.}
  {%

  }
  \label{alg:continuous_recon}
\end{algorithm}

\section{User Study}

\subsection{Participants and setup}

Our recruit 12 participants. 

participants include  professional artist and  novices. 

\subsection{Procedure}

We ask the users to evaluate the quality of synthesized predictions by looking at how much portion of elements you can accept.

\subsubsection{Warm-up session}
The goal is to help the participants to familiarize the system.  

\subsubsection{Target session}
We ask our participants to draw target patterns with or without assisted functions. There are three conditions 1) with parameter learning, 2) without parameter learning 3) without autocomplete. Tasks are assigned in counter-balance order across the conditions.
 
The sketching time is measured. In particular, the acceptance rate of predictions is also measured with respect to time when the interactive learning function is enabled. 

We also compare two interface designs for integrating interactive learning module into the system.

\subsubsection{Open session}
The goal is to observe participants behavior and identify potential issues of the system. 
The only requirement is to create repetitive patterns.

\subsubsection{Interview}
We collect feedbacks from the participants, including interface design, overall satisfaction.

\subsubsection{Performance}

\subsubsection{Subjective evaluation}

\section{Meta}
\label{sec:meta}

\paragraph{Google site}
\url{https://sites.google.com/site/peihanturesearch/}

}
{}

\end{document}